\documentclass[a4paper,12pt]{article}
\pdfoutput=1
\usepackage{epsfig,amsmath,amsfonts,amsthm}
\usepackage[figuresright]{rotating}
\usepackage[left=28mm,right=20mm,top=30mm,bottom=20mm]{geometry}
\usepackage{subfig}
\usepackage{graphicx}
\usepackage{rotating}
\usepackage{booktabs}
\usepackage{moreverb}

\DeclareGraphicsExtensions{.pdf,.png,.jpg}
\usepackage{natbib}
\numberwithin{equation}{section}

\def\cov{\hbox{Cov}}

\def\var{\hbox{Var}}

\def\cov{\hbox{Cov}}
\def\Bin{\hbox{Bin}}

\def\var{\hbox{Var}}

\def\logit{\hbox{logit}}

\usepackage[nokeyprefix]{refstyle}
\usepackage{varioref}
\usepackage{xr}
\usepackage{hyperref}

\begin{document}
\title{
Exploring  Consequences of Simulation Design for  Apparent Performance of Statistical Methods.\\
1: Results from simulations with constant sample sizes
}
\author{Elena Kulinskaya, David C. Hoaglin, and Ilyas Bakbergenuly}
\date{\today}

\maketitle
\abstract{
Contemporary statistical publications rely on simulation to evaluate performance of new methods and compare them with established methods. In the context of meta-analysis of log-odds-ratios, we investigate how the ways in which simulations are implemented affect such conclusions. Choices of distributions for sample sizes and/or control probabilities considerably affect conclusions about statistical methods.  Here we report on the results for constant sample sizes. Our two subsequent publications will cover normally and uniformly distributed sample sizes.}

\section{Introduction}
Many methodological publications in applied statistics develop a new method, illustrate it in examples, and evaluate its performance by simulation. Our interest lies in methods for meta-analysis (MA). For meta-analysis of odds ratios, we demonstrate how researchers' choices of simulation design can affect conclusions on the comparative merits of various methods.

The meta-analysis of odds ratios from $K$ studies involves $2 K$ binomial variables, $X_{ij}\sim \Bin(n_{ij}, p_{ij})$ for $i = 1,\ldots, K$  and $j = C$ or $T$ (for the Control or Treatment arm).
The conventional random-effects model assumes that $\logit(p_{ij}) = \alpha_i + \theta_{i} z_{ij}$ for $\theta_{i} \sim N(\theta, \tau^2)$ and an indicator $z_{ij}$ taking values $0$ (for Control) and 1 (for Treatment). In this notation, $\alpha_i = \logit(p_{iC})$ and $\alpha_i + \theta_{i} = \logit(p_{iT})$.

A design specifies a systematic collection of situations involving the number of studies, $K$; the sample sizes, $n_{ij}$; the control-arm probabilities, $p_{iC}$, or, equivalently, their logits, $\alpha_i$; the overall log-odds-ratio, $\theta$; and the between-study variance, $\tau^2$. For each situation the simulation uses $M$ replications, where $M$ is typically large, say 10,000.

For simplicity, we consider equal arm-level sample sizes, $n_{iC} = n_{iT} = n_i $. Studies vary in how they specify the $n_i$. Choices include setting $n_1 = \cdots = n_K$ in all $M$ replications (our choice here), using a constant $n_i$, and using some distribution (typically normal or uniform) to generate a new set of $n_i$ in each replication. We consider these two latter choices in companion reports.

Similarly, the $p_{iC}$ or their logits $\alpha_i$ can be constant or generated from some distribution. Again, normal and uniform distributions are the typical choices.

\section{Generation of log-odds-ratios and control-arm probabilities} \label{sec:GenLORandpC}

Consider $K$ studies that used a particular individual-level binary outcome.
Each study reports $X_{iT}$ and $X_{iC}$, the numbers of events in the $n_{iT}$ subjects in the Treatment arm and the $n_{iC}$ subjects in the Control arm, for $i = 1, \ldots ,K$. It is customary to treat $X_{iT}$ and $X_{iC}$ as independent binomial variables:
\begin{equation} \label{eq:binomialXs}
X_{iT}\sim {\Bin}(n_{iT},p_{iT})\qquad \text{and}\qquad X_{iC}\sim {\Bin}(n_{iC},p_{iC}).
\end{equation}
The log-odds-ratio for Study $i$ is
\begin{equation}\label{eq:psi}
\theta_{i}=\log_{e}\left(\frac{p_{iT}(1-p_{iC})}{p_{iC}(1-p_{iT})}\right)\qquad\text{estimated by} \qquad
\hat\theta_{i}=\log_{e}\left(\frac{\hat p_{iT}(1-\hat p_{iC})}{\hat p_{iC}(1-\hat p_{iT})}\right).
\end{equation}
The (conditional, given $p_{ij}$ and $n_{ij}$) variance of $\hat{\theta}_i$, derived by the delta method, is
\begin{equation} \label{eq:sigma}
v_{i}^2 = {\var}(\hat{\theta}_{i}) = \frac{1} {n_{iT} {p}_{iT} (1 - {p}_{iT})} + \frac{1}{n_{iC} {p}_{iC} (1 - {p}_{iC})},
\end{equation}
estimated by substituting $\hat{p}_{ij}$ for $p_{ij}$.  (We take the $\hat{p}_{ij}$ as given by the particular method.)

Under the binomial-normal random-effects model (REM),  the true study-level effects, $\theta_{i}$, follow a normal distribution:
\begin{equation} \label{standardREM}
 \theta_{i} \sim N(\theta, \tau^2).
\end{equation}
The resulting logistic mixed-effects model belongs to the class of generalized linear mixed models (GLMMs) (\cite{turner2000multilevel}, \cite{stijnen2010random}). \cite{kuss2015statistical}, \cite{jackson2018comparison}, and \cite{bakbergenuly2018GLMM} review these GLMM methods.

In practice $p_{iC}$ and $p_{iT}$ vary among studies in a variety of ways, not necessarily described by any particular distribution.  
 Simulations can treat the $p_{iC}$ as constant (e.g., at a sequence of values) or sample them from a distribution, either directly (usually from a uniform distribution) or indirectly, by generating $\logit(p_{iC})$ (usually from a Gaussian distribution).

\subsection{Models with fixed and random intercepts} \label{sec:interceptmodels}
We consider two fixed-intercept random-effects models (FIM1 and FIM2, Section~\ref{sec:FIM}) and two random-intercept random-effects models (RIM1 and RIM2, Section~\ref{sec:RIM}) as in \cite{bakbergenuly2018GLMM}. These models are  equivalent to Models 2 and 4 (for FIM)  and Models 3 and 5 (for RIM), respectively, of \cite{jackson2018comparison}.
Briefly, the FIMs include fixed control-arm effects (log-odds of the control-arm probabilities), and the RIMs replace these fixed effects with random effects.

Under the fixed-effect (common-effect) model, $\tau^2 = 0$ and  $\theta_i \equiv \theta$. Still, the control-arm effects can be either fixed or random, resulting in two fixed-effect models: the fixed-intercept fixed-effect model FIM1F, and the random-intercept fixed-effect model RIM1F. Random-intercept fixed-effect models were considered by \cite{kuss2015statistical} and \cite{PiagetRossel2019}. However, GLMMs with random $\theta_i$ are traditional in meta-analysis.

\subsubsection{Fixed-intercept models (FIM1 and FIM2)} \label{sec:FIM}
The fixed-intercept models fit fixed effects for the studies' control arms and account for heterogeneity in odds ratios among studies. Given the binomial distributions in the two arms (Equation~(\ref{eq:binomialXs})),
the model is ($i = 1, \ldots ,K$)
\begin{equation} \label{FIM}
\begin{array}{ll}
\log \left(\frac{p_{iT}}{1 - p_{iT}}\right) &= \alpha_{i} + \theta + (1-c)b_{i} \\
\log \left(\frac{p_{iC}}{1 - p_{iC}}\right) &= \alpha_{i} - cb_{i},
\end{array}
\end{equation}
where the $\alpha_{i}$ are the fixed control-arm effects (usually regarded as nuisance parameters), $\theta$ is the overall log-odds-ratio, and the $b_{i} \sim N(0,\tau^2)$ are random effects. Under FIM1, $c = 0$, resulting in higher variance in the treatment group. Under FIM2, $c = 1/2$,  splitting the random effect $b_i$ equally  between the two equations and yielding equal variance in the two arms.
The fixed study-specific intercepts $\alpha_i$ have to be estimated, along with $\theta$ and $\tau^2$.
In a logistic mixed-effects regression, these $K+2$ parameters are estimated iteratively, using marginal quasi-likelihood, penalized quasi-likelihood, or a first- or second-order-expansion approximation. \cite{jackson2018comparison} demonstrate that inference using FIM2 is preferable, even though they generate data from FIM1.
When $\tau^2 \equiv 0$, these two models become a fixed-intercept fixed-effect model, FIM1F.

\subsubsection{Random-intercept models (RIM1 and RIM2)} \label{sec:RIM}
As $K$ becomes large, it may be inconvenient, even problematic, to have a separate $\alpha_i$ for each study.  One can replace those fixed effects with random effects $\alpha + u_{i}$, centered at $\alpha$:
\begin{equation}\label{RIM}
\begin{array}{ll}
\log \left(\frac{p_{iT}}{1 - p_{iT}}\right) &= \alpha + u_{i} + \theta + (1-c)b_{i} \\
\log \left(\frac{p_{iC}}{1 - p_{iC}}\right) &= \alpha + u_{i} - cb_{i}.
\end{array}
\end{equation}
As  before, $\theta$ is the overall log-odds-ratio, and $b_{i} \sim N(0,\tau^2)$. RIM1 and RIM2 correspond to $c = 0$ and $1/2$, respectively. Now the $u_{i} \sim N(0,\sigma^2)$, and $u_i$ and $b_i$ can be correlated: $\cov(u_{i}, b_{i}) = \rho \sigma \tau$.
Typically,  $\rho$ is taken as zero in simulation and in estimation. Again, RIM2 is preferable to RIM1 for inference.

When $\tau^2\equiv 0$, these two models become a random-intercept fixed-effect model, denoted by RIM1F.

The vast majority of simulation studies use FIM1 or RIM1 for data generation, both for standard two-stage methods of MA and when studying performance of GLMMs, even when they use FIM2 or RIM2 for inference.

\subsection{Non-Gaussian random-intercept models} \label{sec:NGRIM}

Other distributions besides the Gaussian yield a mixture of control-arm probabilities.

We are not aware of any simulation studies that intentionally used a beta distribution for control-arm probabilities. However, the Beta(1,1) distribution is the same as $U(0,\;1)$, and a popular choice is a uniform distribution on an interval, $(p_l,\;p_u) \subset [0,1]$. \cite{viechtbauer2007confidence}, \cite{sidik2007}, and  \cite{Nagashima} (set iii)
generated the $p_{iC}$ from $U(0.05,\;0.65)$ in combination with the Gaussian REM. Similarly, \cite{jackson2018comparison} (setting 13) generated the $p_{iC}$ from $U(0.1,\;0.3)$. All these studies add a uniform distribution of control-arm probabilities to the FIM1 setting, to arrive, unintentionally,  at a random-intercept model that we denote by URIM1. This model retains the normal distribution of the $\theta_i$.

\section{Design of simulations} \label{sec:DesignSims}
Our simulations keep the arm-level sample sizes equal in the $K$ (= 5, 10, 30) studies. The control-arm probability $p_{iC} =  0.1, \;0.4$. For the log-odds-ratios $\theta_i$, we use Equation~(\ref{standardREM}) with $\theta$ = 0, 0.5, 1, 1.5, and 2 and  $\tau^2=0 (0.1)1$. We vary two components of the data-generating mechanism: the model (at five levels: FIM1, FIM2, RIM1, RIM2, and URIM1) and the arm-level sample sizes, $n$, centered at 40, 100, 250, and 1000 (constant, normally distributed, or uniformly distributed). Here we provide simulation results only for  constant sample sizes. 
We also vary the variance $\sigma^2=0.1,\;0.4$ for RIM.

We keep the control-arm probabilities $p_{iC}$ and the log-odds-ratios $\theta_i$ independent (i.e., $\rho = 0$ in the RIMs).

For control-arm probabilities, when using a normal (on the logit scale) or a uniform distribution, we can have approximately the same variance on the probability scale by taking $\Delta_p = \sqrt{12 [p_C^0 (1 - p_C^0)]^2 \sigma^2}$ in comparator simulations (the expression for $\Delta_p$ is an approximation based on the delta method).

For each generated dataset, we use two-stage methods of MA for log-odds-ratio. We also use the GLMM methods based on FIM2 and RIM2 as implemented in \texttt{metafor}  \cite{bakbergenuly2018GLMM, viechtbauer2015package}.

For each combination of the parameters and a data-generating mechanism, we generated data for 1000 simulated meta-analyses.

Table~\ref{tab:components} shows the components of the simulations. We included the DerSimonian-Laird (DL), restricted maximum-likelihood (REML), Mandel-Paule (MP), and Kulinskaya-Dollinger (KD) estimators { of $\tau^2$  with corresponding inverse-variance-weighted estimators of $\theta$ and confidence intervals with critical values from the normal distribution}. \cite{Bakbergenuly2020} studied those inverse-variance-weighted estimators in detail. We also included the SSW point estimator of $\theta$, whose weights depend only on the studies' arm-level sample sizes, and a corresponding confidence interval, which uses $\hat{\theta}_{SSW}$ as the midpoint, $\hat{\tau}^{2}_{KD}$ in the estimate of its variance, and critical values from the $t$ distribution on $K - 1$ degrees of freedom.  Among the estimators, FIM2 and RIM2 denote the estimators in the corresponding GLMMs.

\begin{table}[ht]
	\caption{ \label{tab:components} Components of the simulations for log-odds-ratio}
	\begin{tabular}
		{|l|l|}
		\hline
		Parameter & Values \\
                	\hline
		$K$              & 5, 10, 30 \\
		$n$              & 40, 100, 250, 1000  \\
		$\theta$       & 0, 0.5, 1, 1.5, 2 \\
		$\tau^{2}$    & 0(0.1)1 \\
		$p_C$         & 0.1, 0.4 \\
		$\sigma^2$  & 0.1, 0.4 \\
		\hline
		Data-generation mechanisms & \\
		\hline
		FIM1   & Section~\ref{sec:FIM} \\
		FIM2   & Section~\ref{sec:FIM} \\
		RIM1   & Section~\ref{sec:RIM} \\
		RIM2   & Section~\ref{sec:RIM} \\
		URIM1 & Section~\ref{sec:NGRIM}, $p_{iC} \sim U(p_{C} - \sigma\sqrt{3}p_{C}(1 - p_{C}), p_{C} + \sigma\sqrt{3}p_{C}(1 - p_{C}) )$ \\
		\hline
		Estimation targets & Estimators \\
		\hline
		bias in estimating $\tau^{2}$ & DL, REML, MP, KD, FIM2. RIM2 \\
		bias in estimating $\theta$    & DL, REML, MP, KD, FIM2, RIM2, SSW \\
		coverage of $\theta$             & DL, REML, MP, KD, FIM2, RIM2, \\
		                                              & SSW (with $\hat{\tau}^{2}_{KD}$ and $t_{K - 1}$ critical values) \\
		\hline
	\end{tabular}
\end{table}

\section{Summary of the results}

Our simulations explored two main components of design: the data-generation mechanism and the distribution of study-level sample sizes. The second of these had essentially no impact on bias of estimators of $\tau^2$, bias of estimators of $\theta$, or coverage of confidence intervals for $\theta$.

The five data-generation mechanisms often produced different results for at least one of those measures of performance. In the most frequent pattern FIM2 and RIM2 yield similar results, and FIM1, RIM1, and URIM1 also yield results that are similar but different from those of FIM2 and RIM2. In some situations URIM1 stands apart (e.g., for the bias of $\hat{\tau}^2_{MP}$ and the bias of $\hat{\theta}_{SSW}$), and so does FIM1 (for the bias of $\hat{\tau}^2_{RIM2}$ and the bias of $\hat{\theta}_{RIM2}$). For $K = 30$ some figures show a particularly unusual pattern, in which the traces for the five data-generation mechanisms are mostly separate.

In summary, except for the coverage of the SSW confidence interval and, in most situations, the bias of $\hat{\theta}_{SSW}$, the choice of data-generation mechanism affects the results. These differences can complicate the process of integrating results from separate simulation studies.

\bibliographystyle{plainnat}
\bibliography{simul_20Feb20}

\clearpage
\renewcommand{\thesection}{A1.\arabic{section}}
\setcounter{section}{0}

\section*{A1: Plots of bias of  estimators of $\tau^2$ for log-odds-ratio}
Each figure corresponds to a value of $\theta$ (= 0, 0.5, 1, 1.5, 2), a value of $p_C$ (= .1, .4), and a value of $\sigma^2$ (= 0.1, 0.4). \\
Each panel of a figure corresponds to a value of $n$ (= 40, 100, 250, 1000) and a value of $K$ (= 5, 10, 30) and has $\tau^2$ = 0.0(0.1)1.0 on the horizontal axis. \\
The data-generation mechanisms are
\begin{itemize}
	\item FIM1 - Fixed-intercept model with $c = 0$
	\item FIM2 - Fixed-intercept model with $c = 1/2$
	\item RIM1 - Random-intercept model with $c = 0$
	\item RIM2 - Random-intercept model with $c = 1/2$
	\item URIM1 - Random-intercept model with $c = 0$ and $p_{iC}$ uniformly distributed on [$p_{C} - \sigma\sqrt{3}p_{C}(1 - p_{C})$, $p_{C} + \sigma\sqrt{3}p_{C}(1-p_{C})$]
\end{itemize}
The point estimators of $\tau^2$ are
\begin{itemize}
\item $\hat{\tau}^2_{DL}$ - DerSimonian-Laird
\item $\hat{\tau}^2_{REML}$ - Restricted maximum-likelihood
\item $\hat{\tau}^2_{MP}$ - Mandel-Paule
\item $\hat{\tau}^2_{KD}$ - Kulinskaya-Dollinger
\item $\hat{\tau}^2_{FIM2}$ - Estimator of $\tau^2$ in the FIM2 GLMM
\item $\hat{\tau}^2_{RIM2}$ - Estimator of $\tau^2$ in the RIM2 GLMM
\end{itemize}

\clearpage
\subsection*{A1.1 Bias of $\hat{\tau}_{DL}^2$}
\renewcommand{\thefigure}{A1.1.\arabic{figure}}
\setcounter{figure}{0}

\begin{figure}[t]
	\centering
	\includegraphics[scale=0.33]{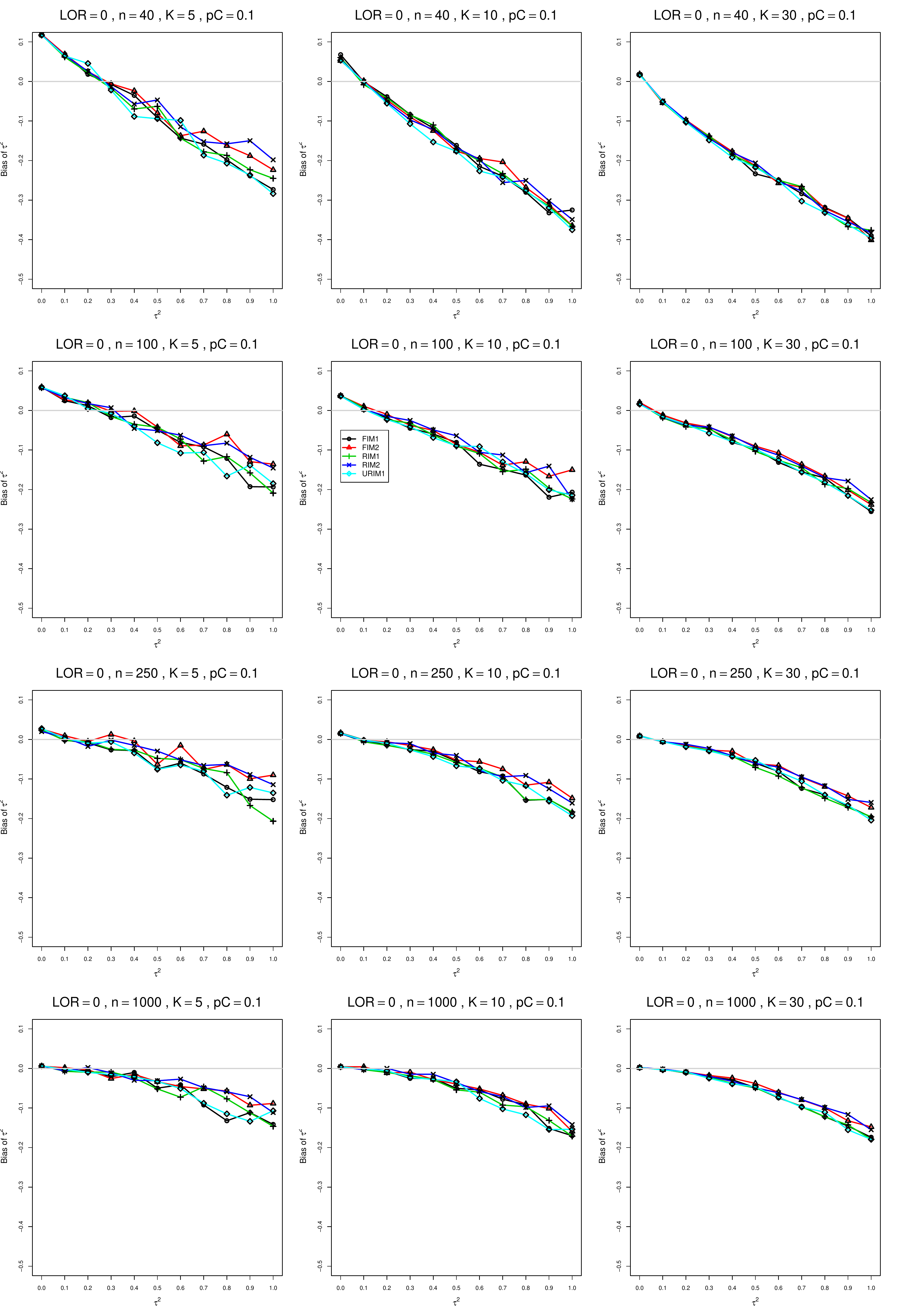}
	\caption{Bias of between-studies variance $\hat{\tau}_{DL}^2$ for $\theta=0$, $p_{C}=0.1$, $\sigma^2=0.1$, constant sample sizes $n=40,\;100,\;250,\;1000$.
The data-generation mechanisms are FIM1 ($\circ$), FIM2 ($\triangle$), RIM1 (+), RIM2 ($\times$), and URIM1 ($\diamond$).	
		\label{PlotBiasTau2mu0andpC01LOR_DLsigma01}}
\end{figure}
\begin{figure}[t]
	\centering
	\includegraphics[scale=0.33]{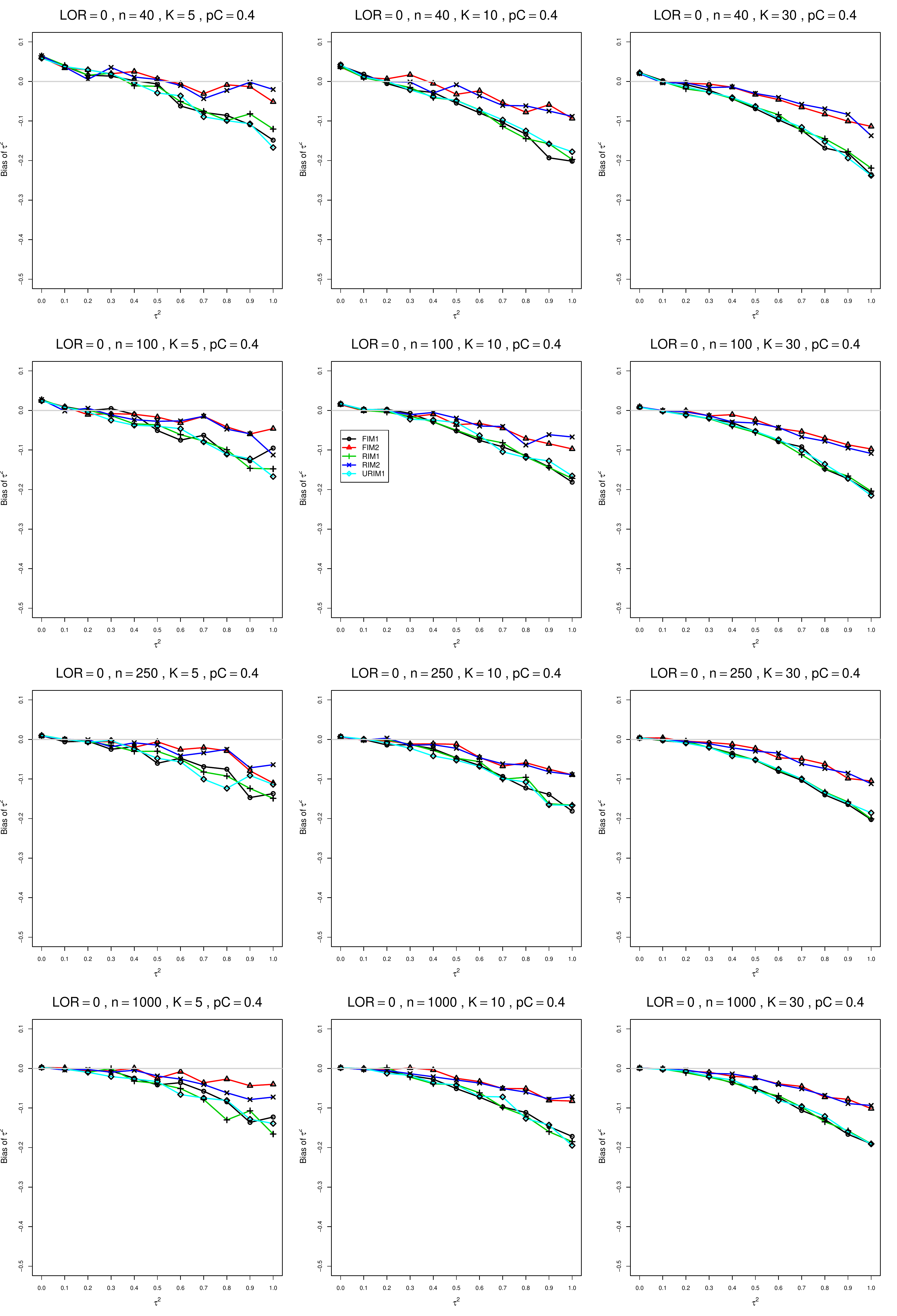}
	\caption{Bias of  between-studies variance $\hat{\tau}_{DL}^2$ for $\theta=0$, $p_{C}=0.4$, $\sigma^2=0.1$, constant sample sizes $n=40,\;100,\;250,\;1000$.
The data-generation mechanisms are FIM1 ($\circ$), FIM2 ($\triangle$), RIM1 (+), RIM2 ($\times$), and URIM1 ($\diamond$).
		\label{PlotBiasTau2mu0andpC04LOR_DLsigma01}}
\end{figure}
\begin{figure}[t]
	\centering
	\includegraphics[scale=0.33]{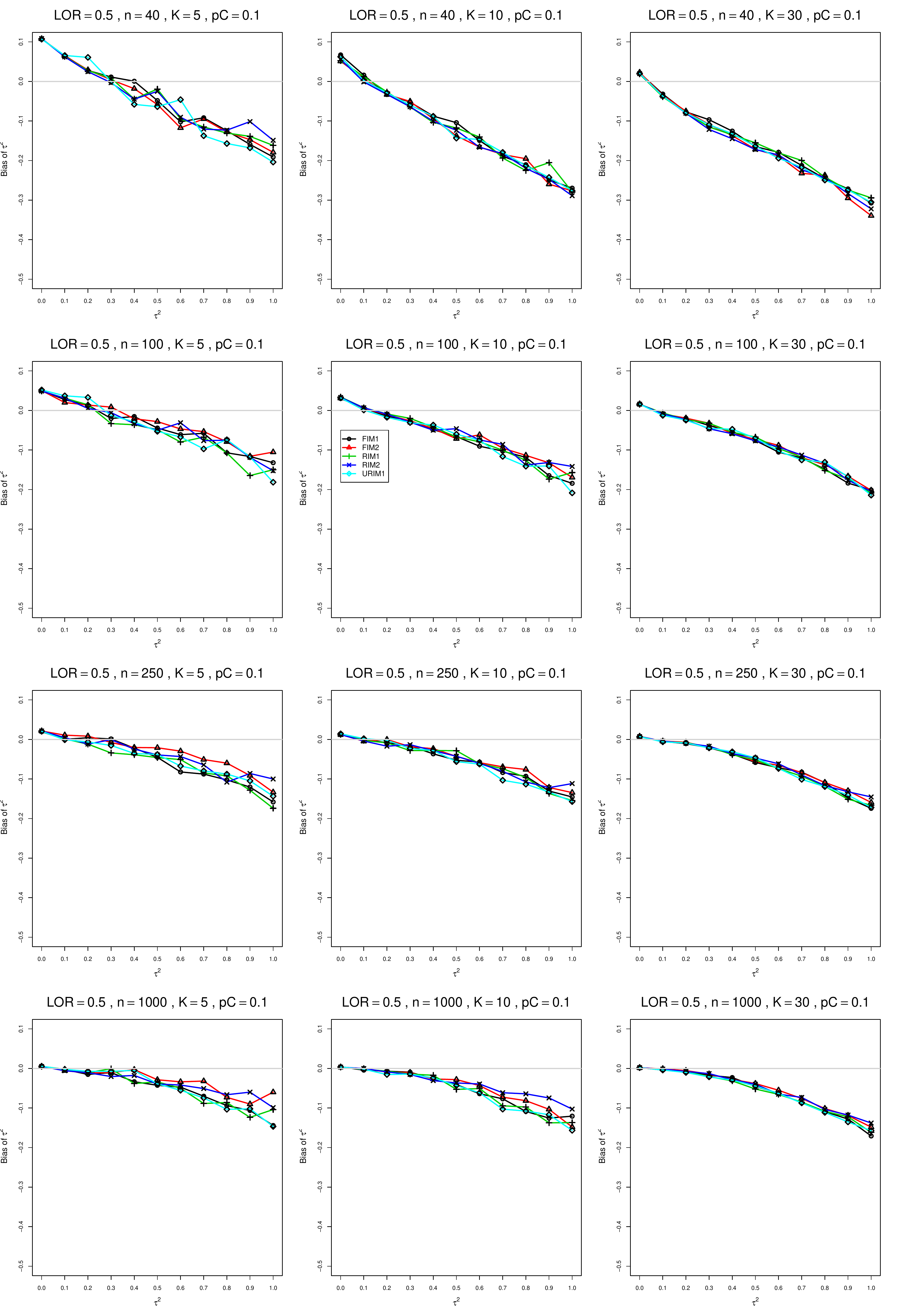}
	\caption{Bias of  between-studies variance $\hat{\tau}_{DL}^2$ for $\theta=0.5$, $p_{C}=0.1$, $\sigma^2=0.1$, constant sample sizes $n=40,\;100,\;250,\;1000$.
The data-generation mechanisms are FIM1 ($\circ$), FIM2 ($\triangle$), RIM1 (+), RIM2 ($\times$), and URIM1 ($\diamond$).
		\label{PlotBiasTau2mu05andpC01LOR_DLsigma01}}
\end{figure}
\begin{figure}[t]
	\centering
	\includegraphics[scale=0.33]{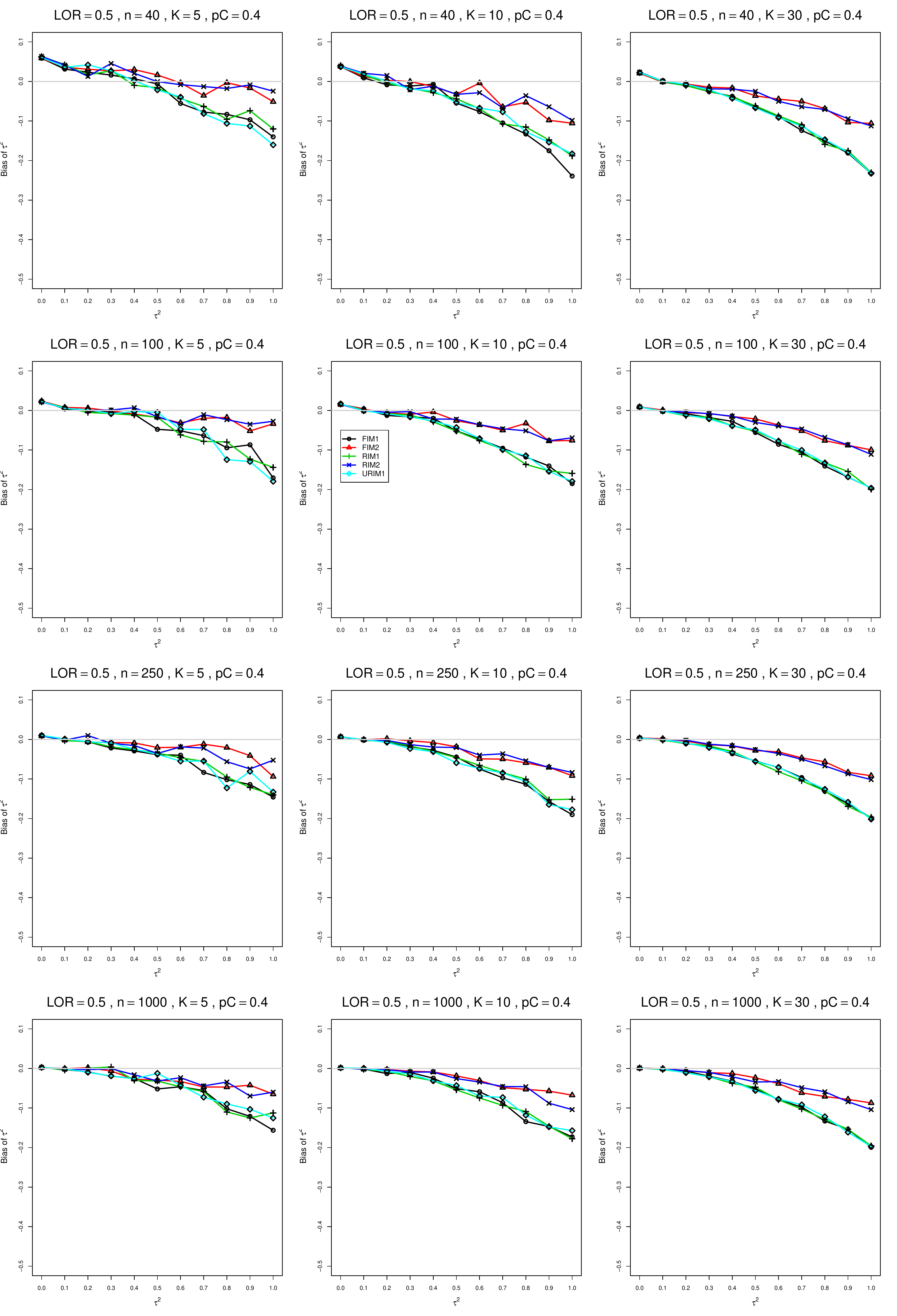}
	\caption{Bias of  between-studies variance $\hat{\tau}_{DL}^2$ for $\theta=0.5$, $p_{C}=0.4$, $\sigma^2=0.1$, constant sample sizes $n=40,\;100,\;250,\;1000$.
The data-generation mechanisms are FIM1 ($\circ$), FIM2 ($\triangle$), RIM1 (+), RIM2 ($\times$), and URIM1 ($\diamond$).
		\label{PlotBiasTau2mu05andpC04LOR_DLsigma01}}
\end{figure}
\begin{figure}[t]
	\centering
	\includegraphics[scale=0.33]{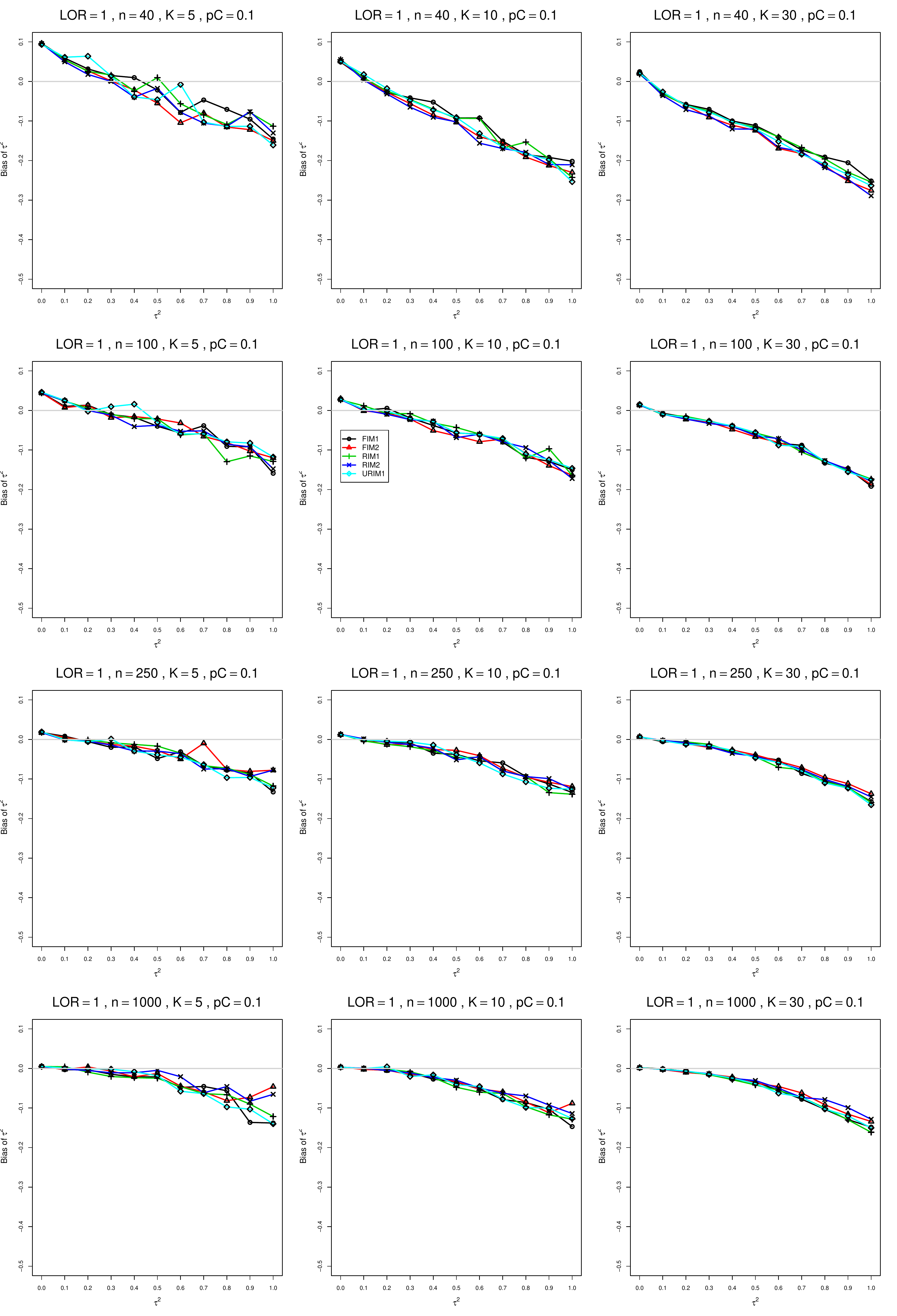}
	\caption{Bias of  between-studies variance $\hat{\tau}_{DL}^2$ for $\theta=1$, $p_{C}=0.1$, $\sigma^2=0.1$, constant sample sizes $n=40,\;100,\;250,\;1000$.
The data-generation mechanisms are FIM1 ($\circ$), FIM2 ($\triangle$), RIM1 (+), RIM2 ($\times$), and URIM1 ($\diamond$).
		\label{PlotBiasTau2mu1andpC01LOR_DLsigma01}}
\end{figure}
\begin{figure}[t]
	\centering
	\includegraphics[scale=0.33]{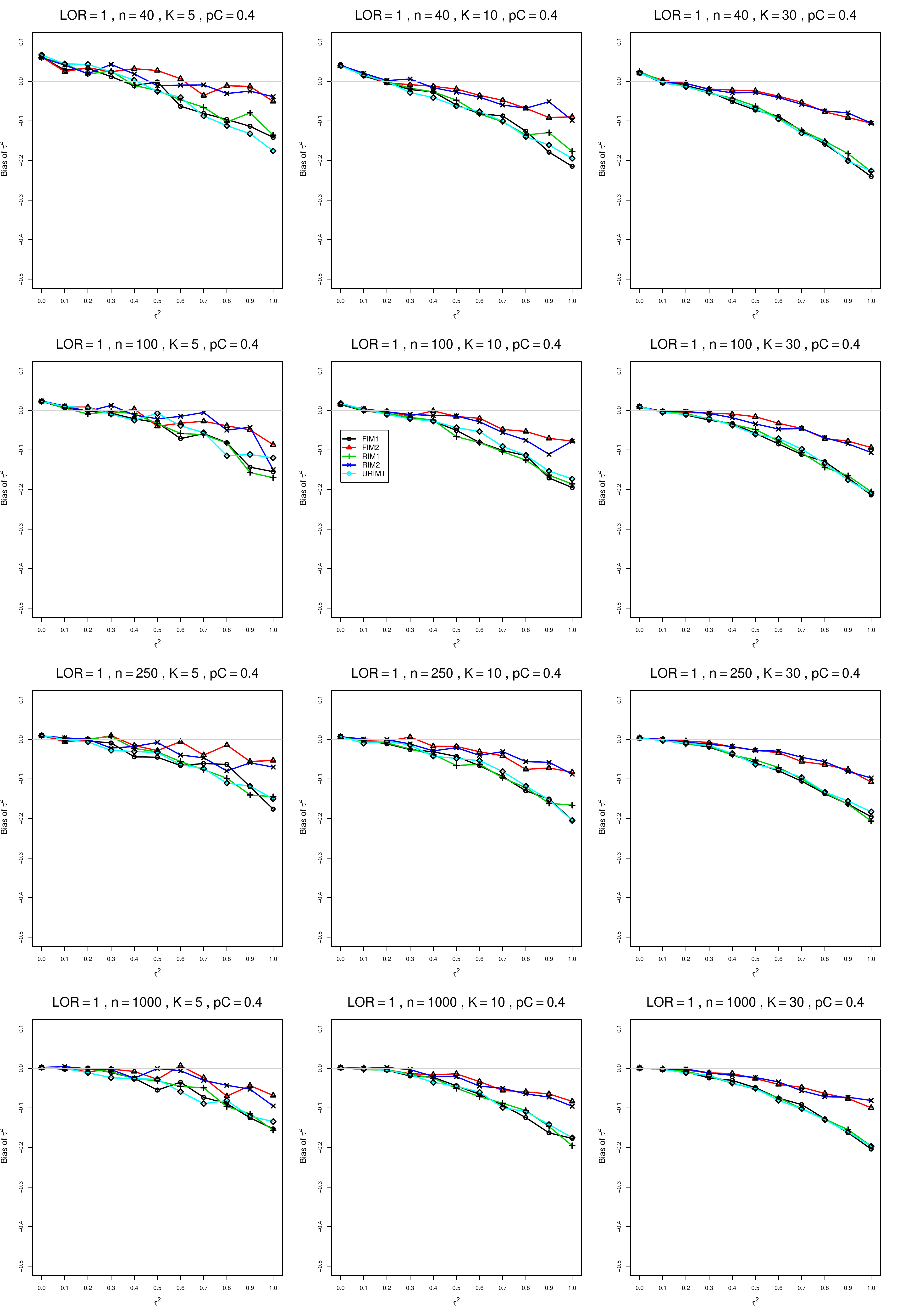}
	\caption{Bias of  between-studies variance $\hat{\tau}_{DL}^2$ for $\theta=1$, $p_{C}=0.4$, $\sigma^2=0.1$, constant sample sizes $n=40,\;100,\;250,\;1000$.
The data-generation mechanisms are FIM1 ($\circ$), FIM2 ($\triangle$), RIM1 (+), RIM2 ($\times$), and URIM1 ($\diamond$).
		\label{PlotBiasTau2mu1andpC04LOR_DLsigma01}}
\end{figure}
\begin{figure}[t]
	\centering
	\includegraphics[scale=0.33]{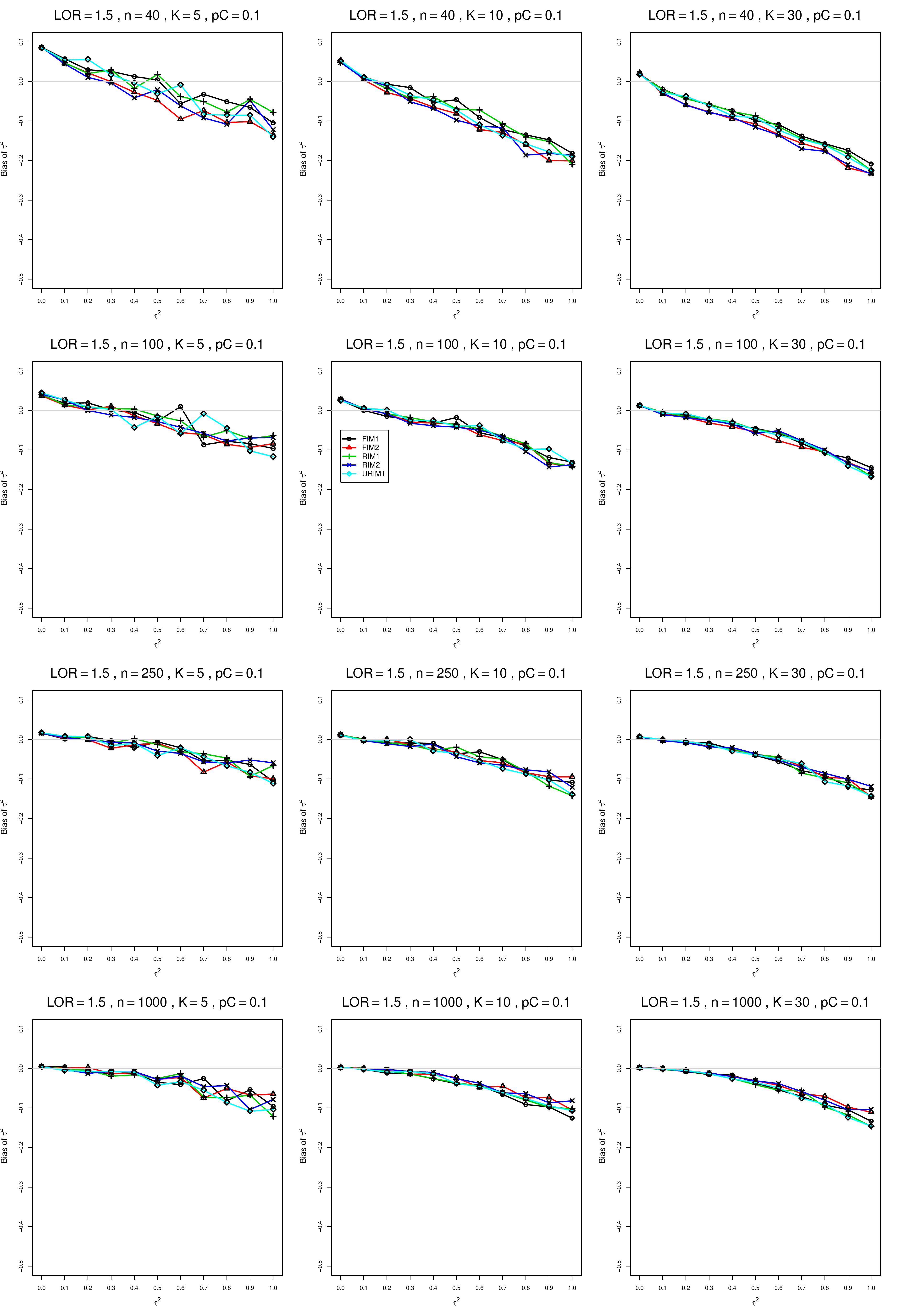}
	\caption{Bias of  between-studies variance $\hat{\tau}_{DL}^2$ for $\theta=1.5$, $p_{C}=0.1$, $\sigma^2=0.1$, constant sample sizes $n=40,\;100,\;250,\;1000$.
The data-generation mechanisms are FIM1 ($\circ$), FIM2 ($\triangle$), RIM1 (+), RIM2 ($\times$), and URIM1 ($\diamond$).
		\label{PlotBiasTau2mu15andpC01LOR_DLsigma01}}
\end{figure}
\begin{figure}[t]
	\centering
	\includegraphics[scale=0.33]{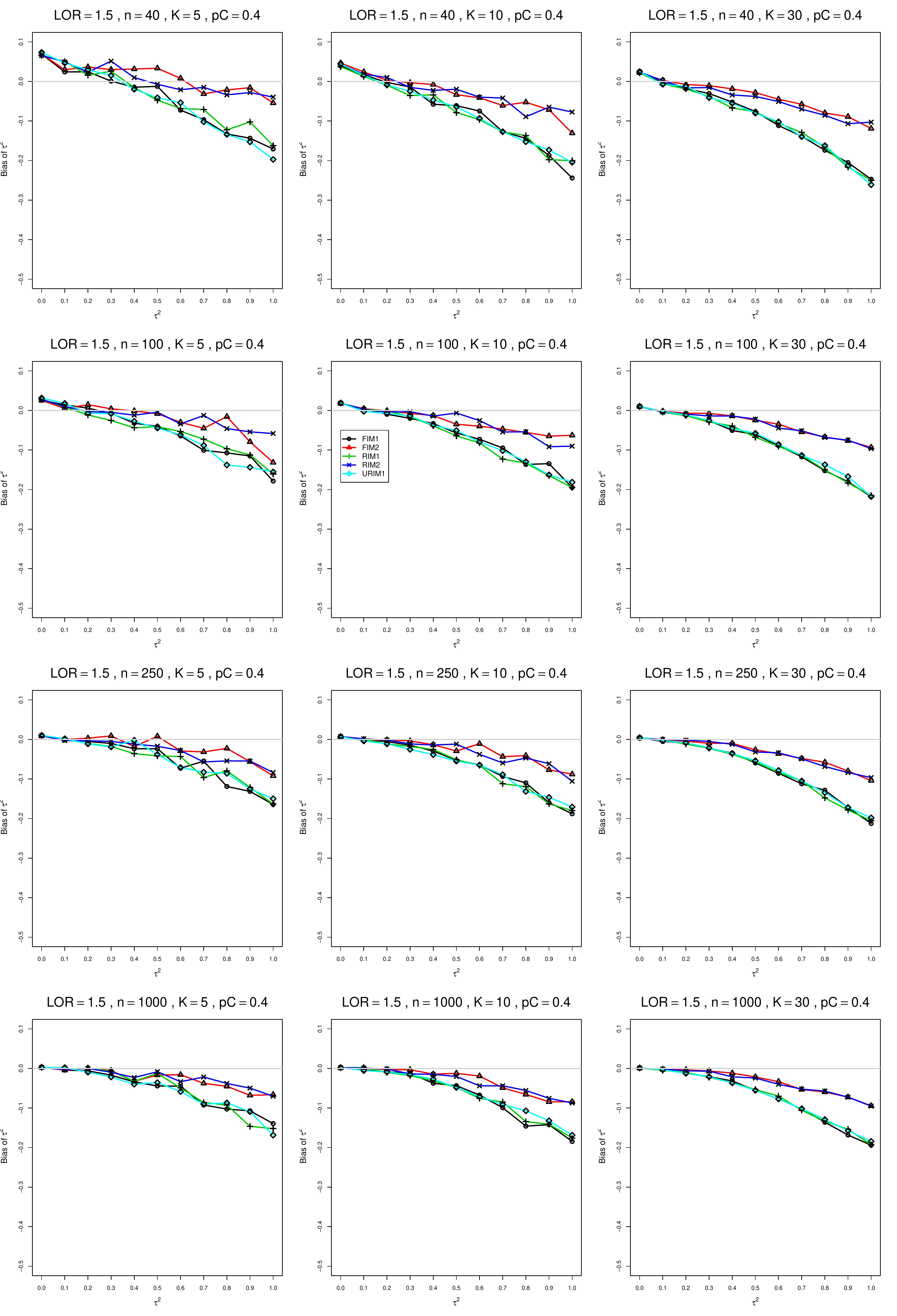}
	\caption{Bias of  between-studies variance $\hat{\tau}_{DL}^2$ for $\theta=1.5$, $p_{C}=0.4$, $\sigma^2=0.1$, constant sample sizes $n=40,\;100,\;250,\;1000$.
The data-generation mechanisms are FIM1 ($\circ$), FIM2 ($\triangle$), RIM1 (+), RIM2 ($\times$), and URIM1 ($\diamond$).
		\label{PlotBiasTau2mu15andpC04LOR_DLsigma01}}
\end{figure}
\begin{figure}[t]
	\centering
	\includegraphics[scale=0.33]{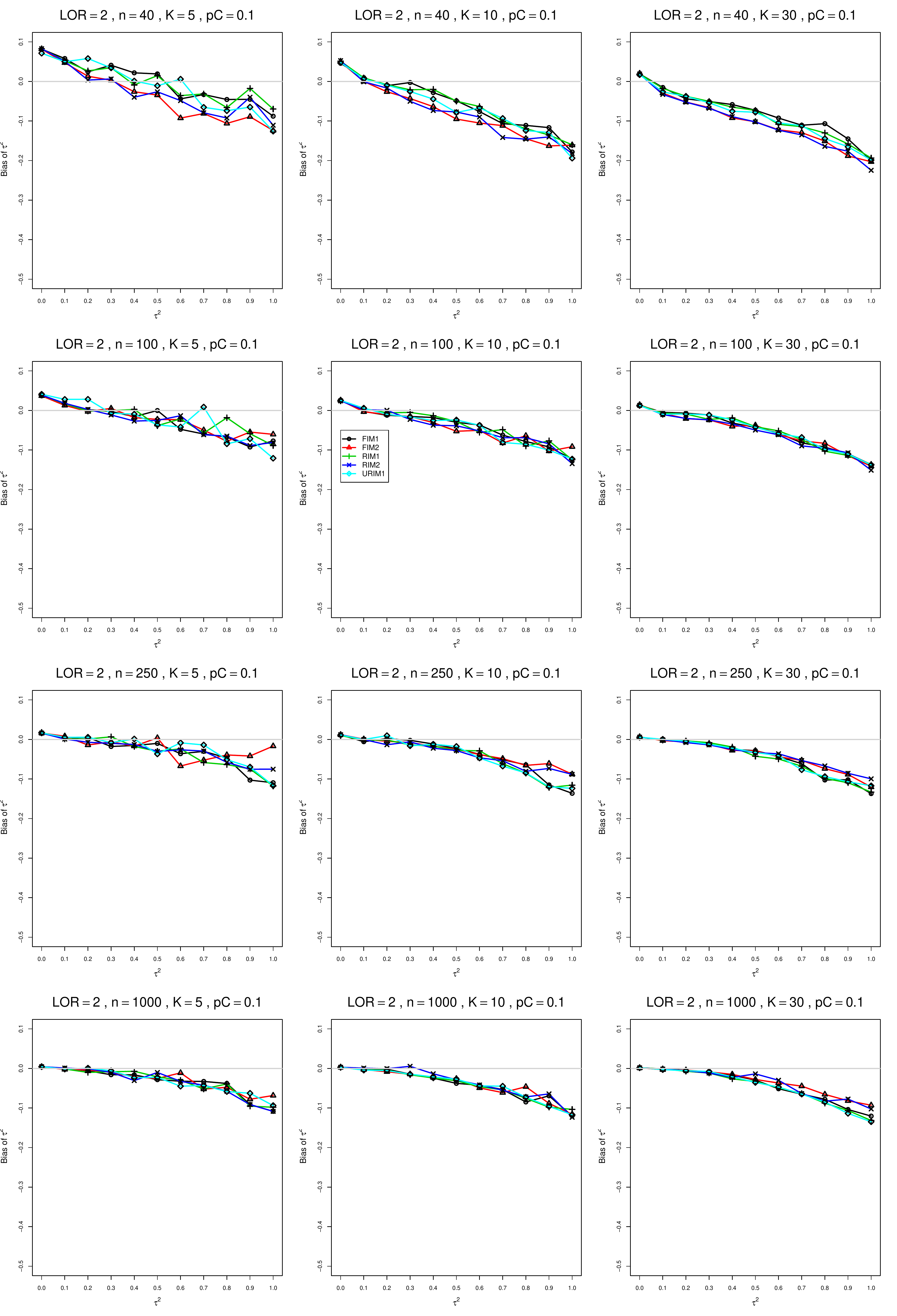}
	\caption{Bias of  between-studies variance $\hat{\tau}_{DL}^2$ for $\theta=2$, $p_{C}=0.1$, $\sigma^2=0.1$, constant sample sizes $n=40,\;100,\;250,\;1000$.
The data-generation mechanisms are FIM1 ($\circ$), FIM2 ($\triangle$), RIM1 (+), RIM2 ($\times$), and URIM1 ($\diamond$).
		\label{PlotBiasTau2mu2andpC01LOR_DLsigma01}}
\end{figure}
\begin{figure}[t]
	\centering
	\includegraphics[scale=0.33]{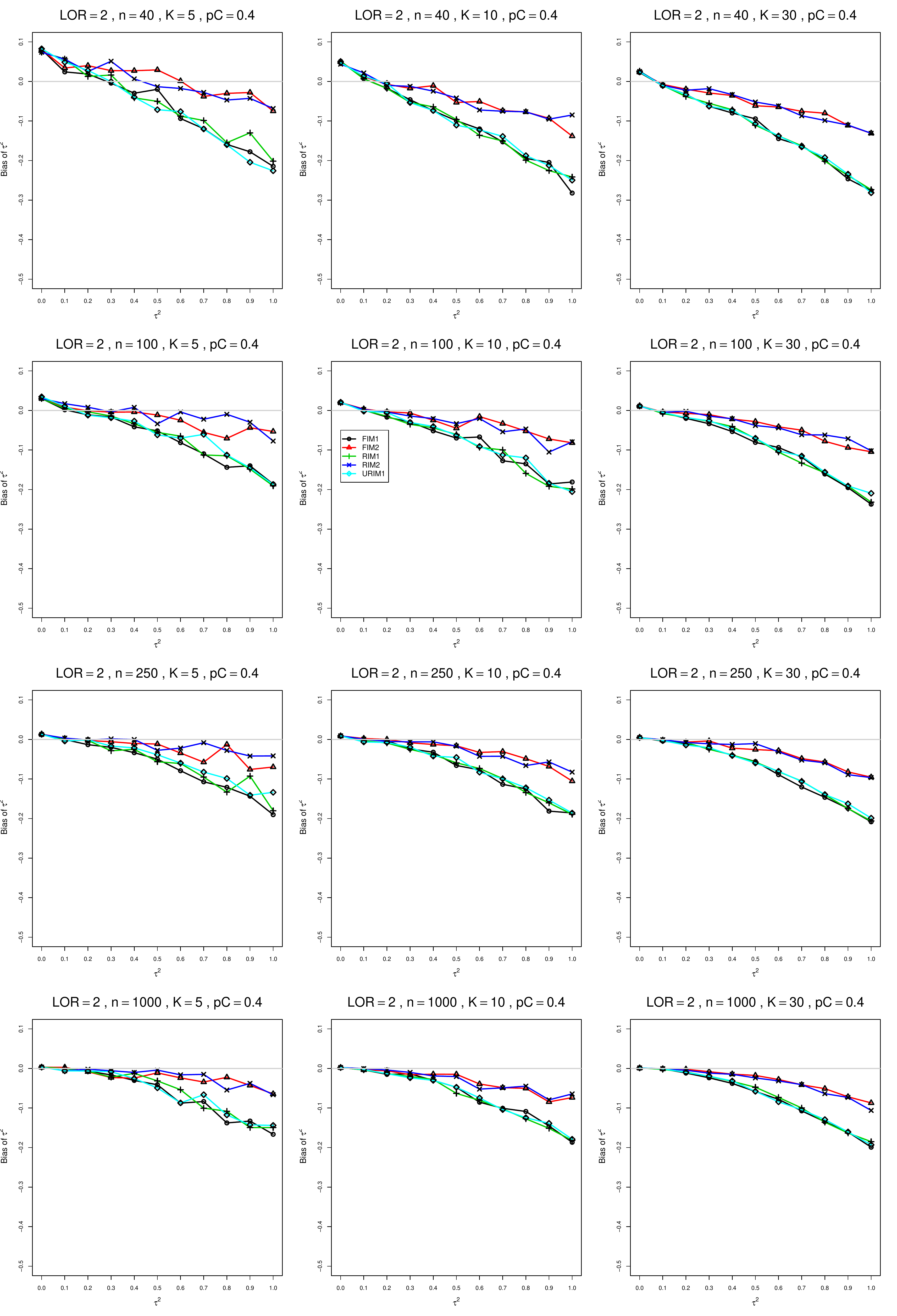}
	\caption{Bias of  between-studies variance $\hat{\tau}_{DL}^2$ for $\theta=2$, $p_{C}=0.4$, $\sigma^2=0.1$, constant sample sizes $n=40,\;100,\;250,\;1000$.
The data-generation mechanisms are FIM1 ($\circ$), FIM2 ($\triangle$), RIM1 (+), RIM2 ($\times$), and URIM1 ($\diamond$).
		\label{PlotBiasTau2mu2andpC04LOR_DLsigma01}}
\end{figure}
\begin{figure}[t]
	\centering
	\includegraphics[scale=0.33]{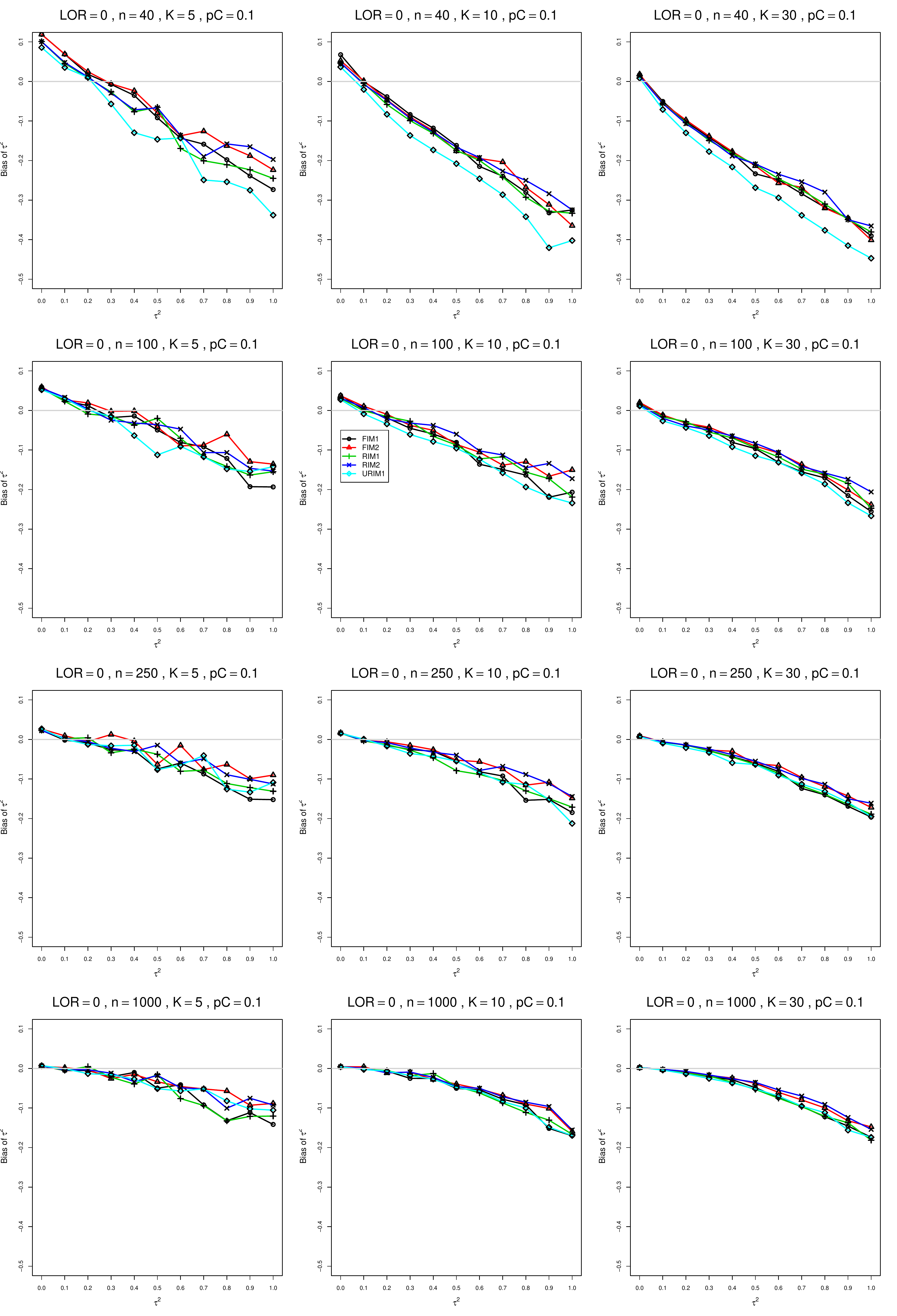}
	\caption{Bias of  between-studies variance $\hat{\tau}_{DL}^2$ for $\theta=0$, $p_{C}=0.1$, $\sigma^2=0.4$, constant sample sizes $n=40,\;100,\;250,\;1000$.
The data-generation mechanisms are FIM1 ($\circ$), FIM2 ($\triangle$), RIM1 (+), RIM2 ($\times$), and URIM1 ($\diamond$).
		\label{PlotBiasTau2mu0andpC01LOR_DLsigma04}}
\end{figure}
\begin{figure}[t]
	\centering
	\includegraphics[scale=0.33]{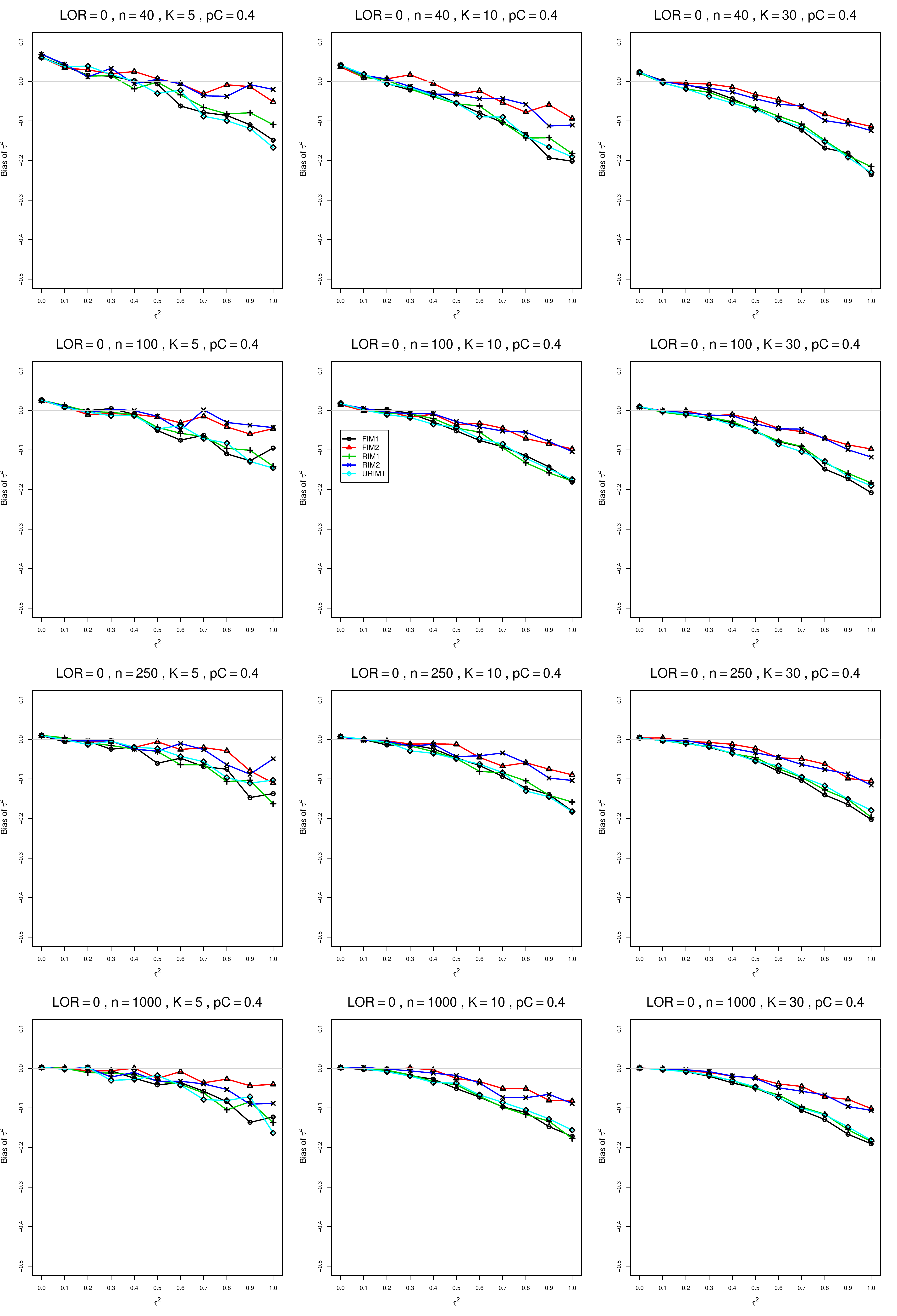}
	\caption{Bias of  between-studies variance $\hat{\tau}_{DL}^2$ for $\theta=0$, $p_{C}=0.4$, $\sigma^2=0.4$, constant sample sizes $n=40,\;100,\;250,\;1000$.
The data-generation mechanisms are FIM1 ($\circ$), FIM2 ($\triangle$), RIM1 (+), RIM2 ($\times$), and URIM1 ($\diamond$).
		\label{PlotBiasTau2mu0andpC04LOR_DLsigma04}}
\end{figure}
\begin{figure}[t]
	\centering
	\includegraphics[scale=0.33]{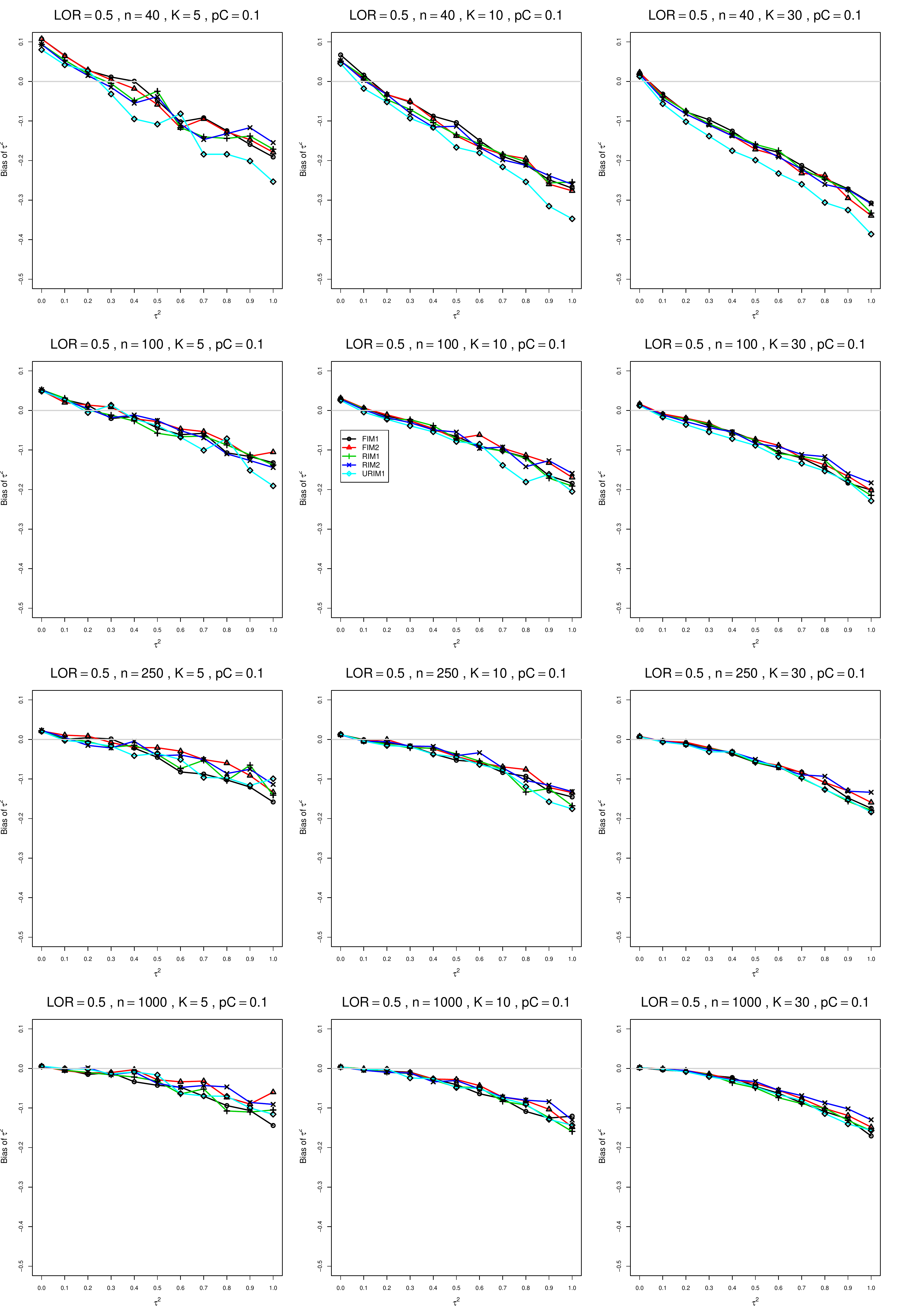}
	\caption{Bias of  between-studies variance $\hat{\tau}_{DL}^2$ for $\theta=0.5$, $p_{C}=0.1$, $\sigma^2=0.4$, constant sample sizes $n=40,\;100,\;250,\;1000$.
The data-generation mechanisms are FIM1 ($\circ$), FIM2 ($\triangle$), RIM1 (+), RIM2 ($\times$), and URIM1 ($\diamond$).
		\label{PlotBiasTau2mu05andpC01LOR_DLsigma04}}
\end{figure}
\begin{figure}[t]
	\centering
	\includegraphics[scale=0.33]{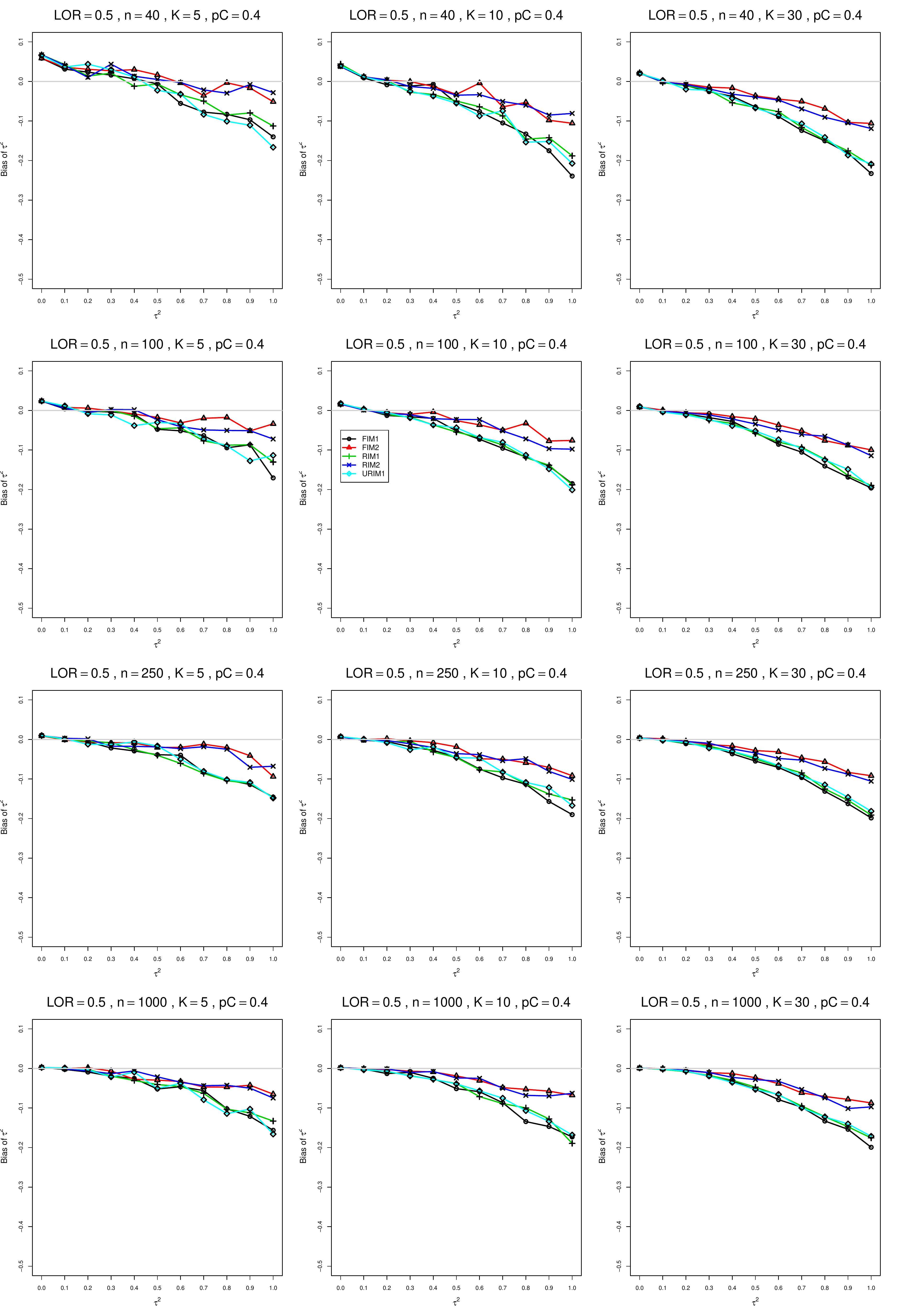}
	\caption{Bias of  between-studies variance $\hat{\tau}_{DL}^2$ for $\theta=0.5$, $p_{C}=0.4$, $\sigma^2=0.4$, constant sample sizes $n=40,\;100,\;250,\;1000$.
The data-generation mechanisms are FIM1 ($\circ$), FIM2 ($\triangle$), RIM1 (+), RIM2 ($\times$), and URIM1 ($\diamond$).
		\label{PlotBiasTau2mu05andpC04LOR_DLsigma04}}
\end{figure}
\begin{figure}[t]
	\centering
	\includegraphics[scale=0.33]{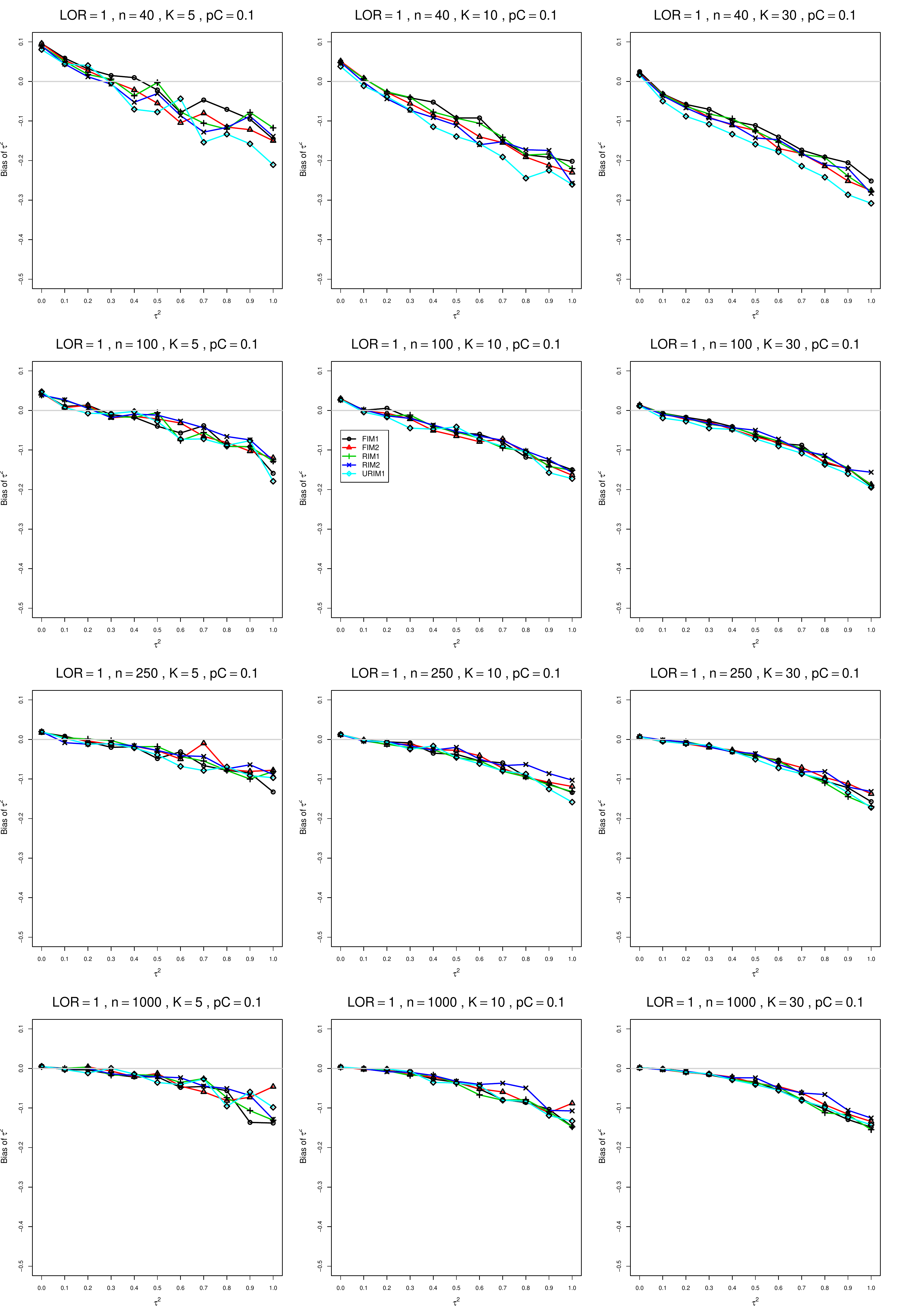}
	\caption{Bias of  between-studies variance $\hat{\tau}_{DL}^2$ for $\theta=1$, $p_{C}=0.1$, $\sigma^2=0.4$, constant sample sizes $n=40,\;100,\;250,\;1000$.
The data-generation mechanisms are FIM1 ($\circ$), FIM2 ($\triangle$), RIM1 (+), RIM2 ($\times$), and URIM1 ($\diamond$).
		\label{PlotBiasTau2mu1andpC01LOR_DLsigma04}}
\end{figure}
\begin{figure}[t]
	\centering
	\includegraphics[scale=0.33]{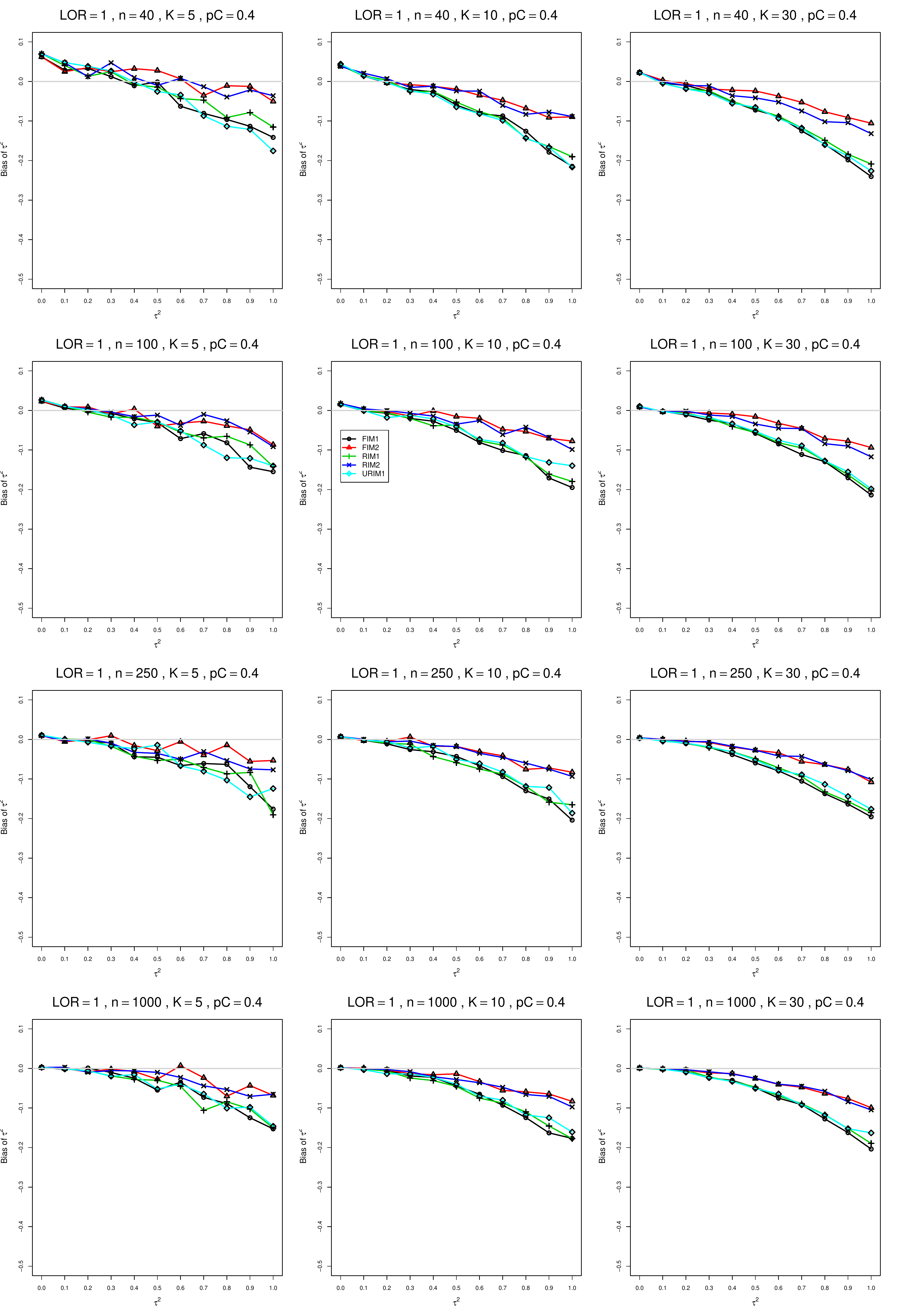}
	\caption{Bias of  between-studies variance $\hat{\tau}_{DL}^2$ for $\theta=1$, $p_{C}=0.4$, $\sigma^2=0.4$, constant sample sizes $n=40,\;100,\;250,\;1000$.
The data-generation mechanisms are FIM1 ($\circ$), FIM2 ($\triangle$), RIM1 (+), RIM2 ($\times$), and URIM1 ($\diamond$).
		\label{PlotBiasTau2mu1andpC04LOR_DLsigma04}}
\end{figure}
\begin{figure}[t]
	\centering
	\includegraphics[scale=0.33]{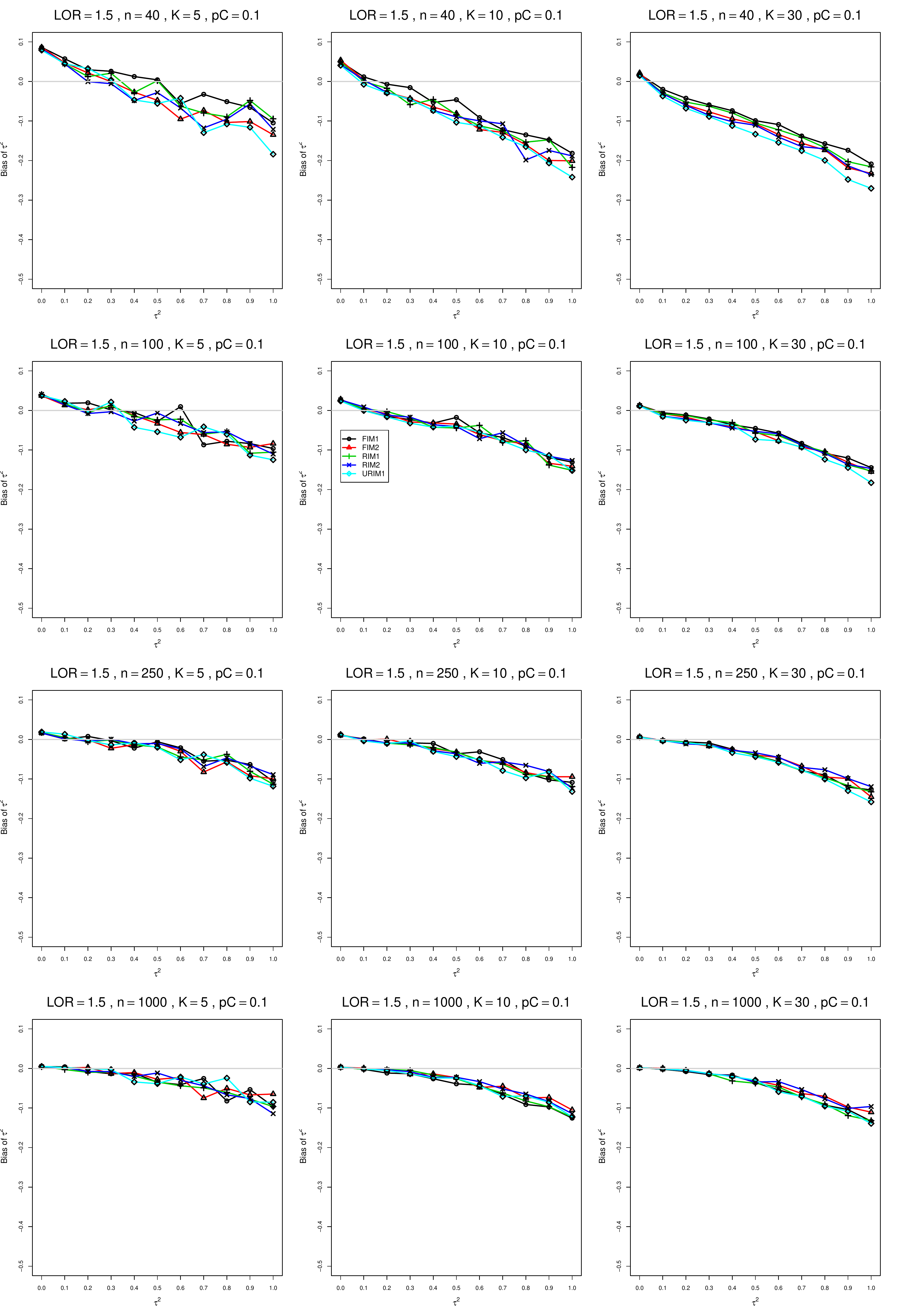}
	\caption{Bias of  between-studies variance $\hat{\tau}_{DL}^2$ for $\theta=1.5$, $p_{C}=0.1$, $\sigma^2=0.4$, constant sample sizes $n=40,\;100,\;250,\;1000$.
The data-generation mechanisms are FIM1 ($\circ$), FIM2 ($\triangle$), RIM1 (+), RIM2 ($\times$), and URIM1 ($\diamond$).
		\label{PlotBiasTau2mu15andpC01LOR_DLsigma04}}
\end{figure}
\begin{figure}[t]
	\centering
	\includegraphics[scale=0.33]{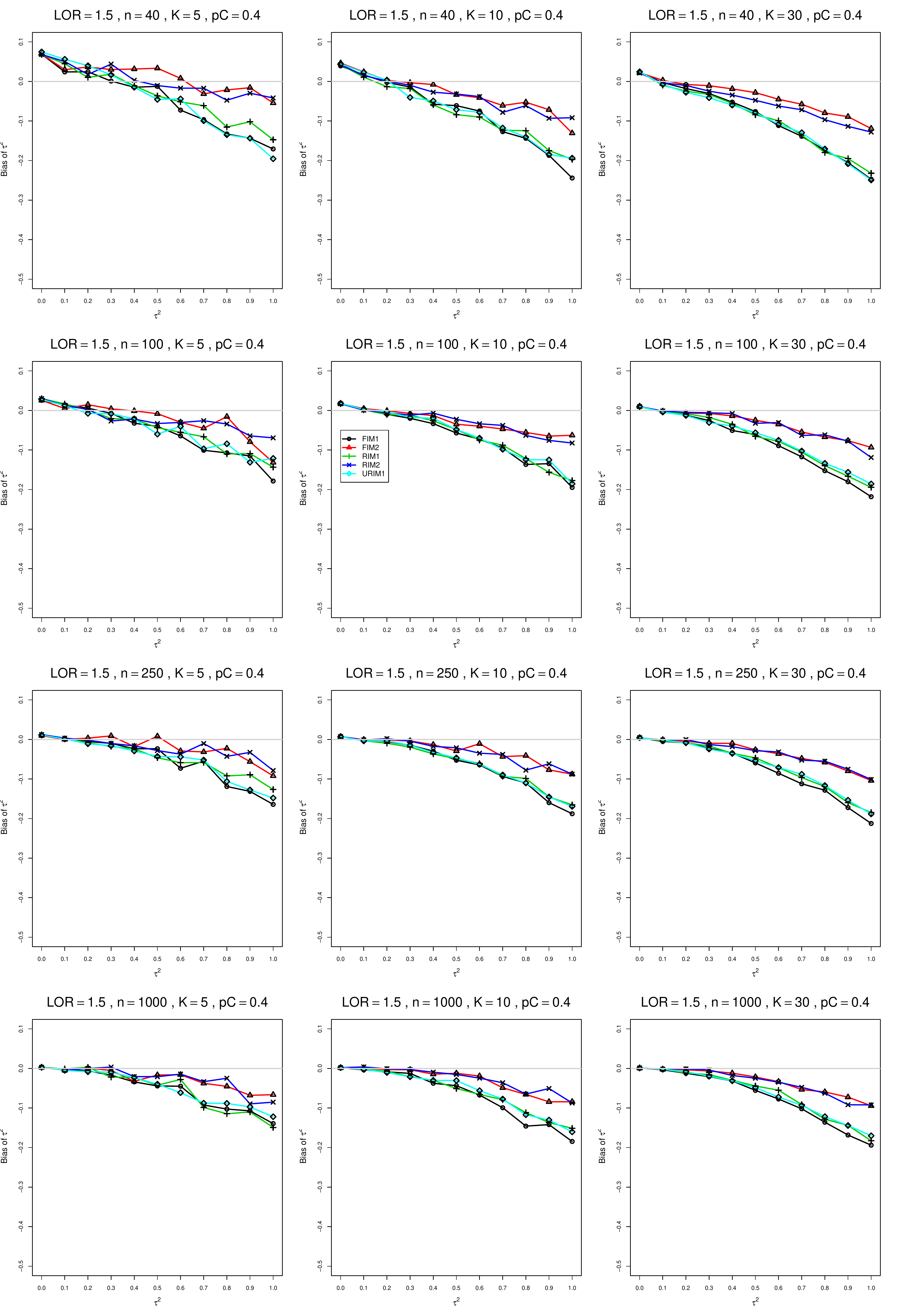}
	\caption{Bias of  between-studies variance $\hat{\tau}_{DL}^2$ for $\theta=1.5$, $p_{C}=0.4$, $\sigma^2=0.4$, constant sample sizes $n=40,\;100,\;250,\;1000$.
The data-generation mechanisms are FIM1 ($\circ$), FIM2 ($\triangle$), RIM1 (+), RIM2 ($\times$), and URIM1 ($\diamond$).
		\label{PlotBiasTau2mu15andpC04LOR_DLsigma04}}
\end{figure}
\begin{figure}[t]
	\centering
	\includegraphics[scale=0.33]{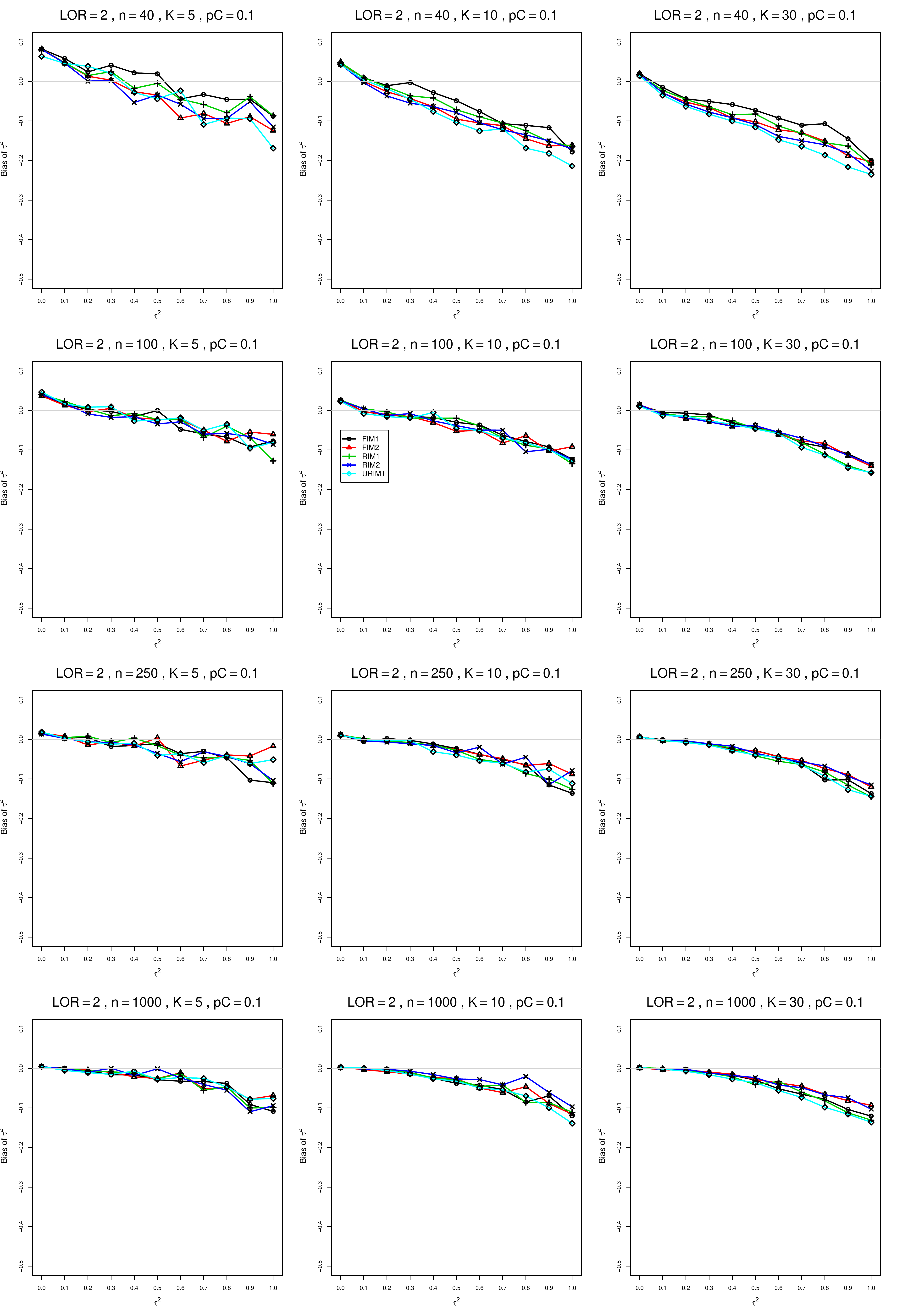}
	\caption{Bias of  between-studies variance $\hat{\tau}_{DL}^2$ for $\theta=2$, $p_{C}=0.1$, $\sigma^2=0.4$, constant sample sizes $n=40,\;100,\;250,\;1000$.
The data-generation mechanisms are FIM1 ($\circ$), FIM2 ($\triangle$), RIM1 (+), RIM2 ($\times$), and URIM1 ($\diamond$).
		\label{PlotBiasTau2mu2andpC01LOR_DLsigma04}}
\end{figure}
\begin{figure}[t]
	\centering
	\includegraphics[scale=0.33]{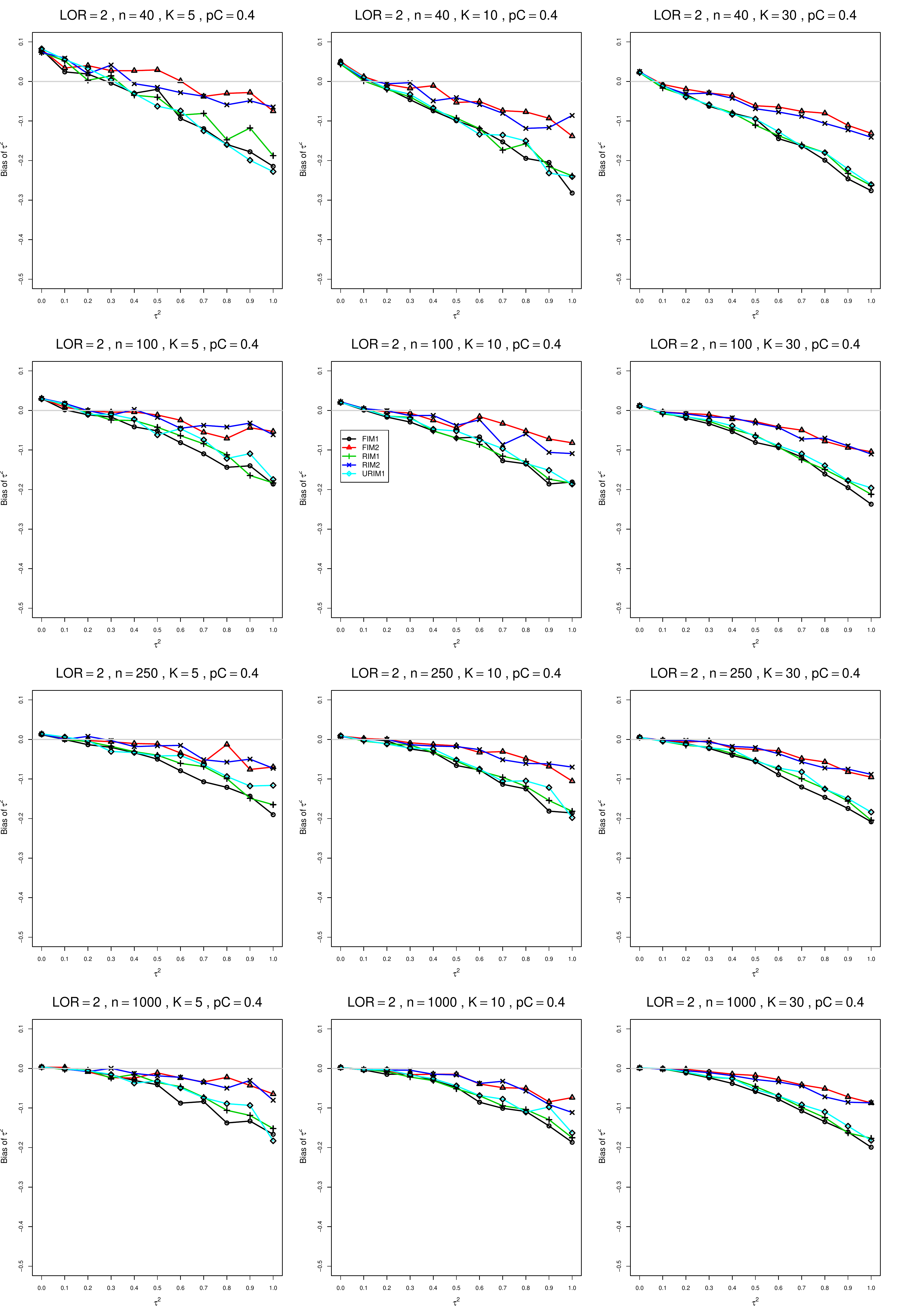}
	\caption{Bias of  between-studies variance $\hat{\tau}_{DL}^2$ for $\theta=2$, $p_{C}=0.4$, $\sigma^2=0.4$, constant sample sizes $n=40,\;100,\;250,\;1000$.
The data-generation mechanisms are FIM1 ($\circ$), FIM2 ($\triangle$), RIM1 (+), RIM2 ($\times$), and URIM1 ($\diamond$).
		\label{PlotBiasTau2mu2andpC04LOR_DLsigma04}}
\end{figure}

\clearpage
\subsection*{A1.2 Bias of $\hat{\tau}_{REML}^2$}
\renewcommand{\thefigure}{A1.2.\arabic{figure}}
\setcounter{figure}{0}

\begin{figure}[t]
	\centering
	\includegraphics[scale=0.33]{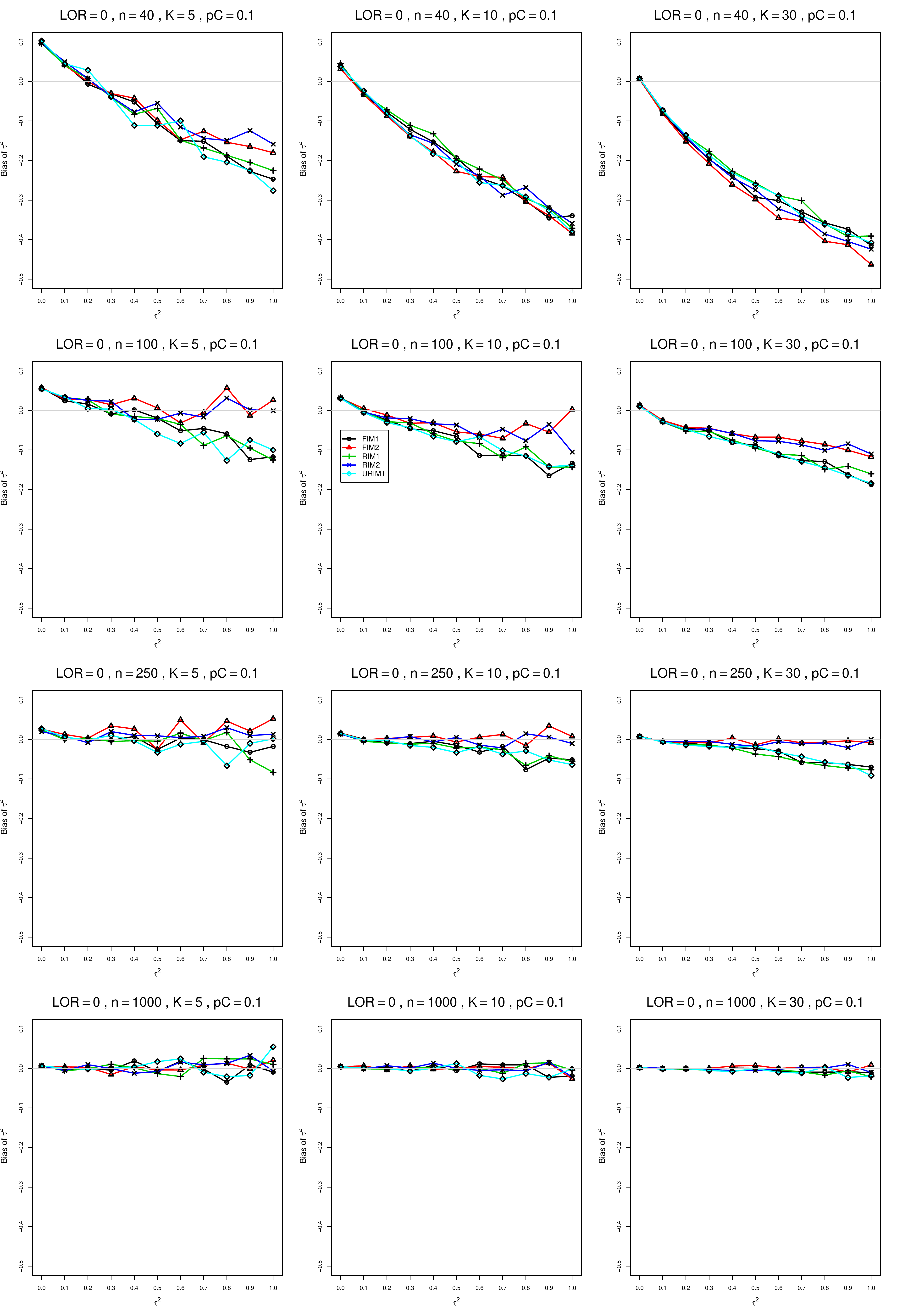}
	\caption{Bias of  between-studies variance $\hat{\tau}_{REML}^2$ for $\theta=0$, $p_{C}=0.1$, $\sigma^2=0.1$, constant sample sizes $n=40,\;100,\;250,\;1000$.
The data-generation mechanisms are FIM1 ($\circ$), FIM2 ($\triangle$), RIM1 (+), RIM2 ($\times$), and URIM1 ($\diamond$).
		\label{PlotBiasTau2mu0andpC01LOR_REMLsigma01}}
\end{figure}
\begin{figure}[t]
	\centering
	\includegraphics[scale=0.33]{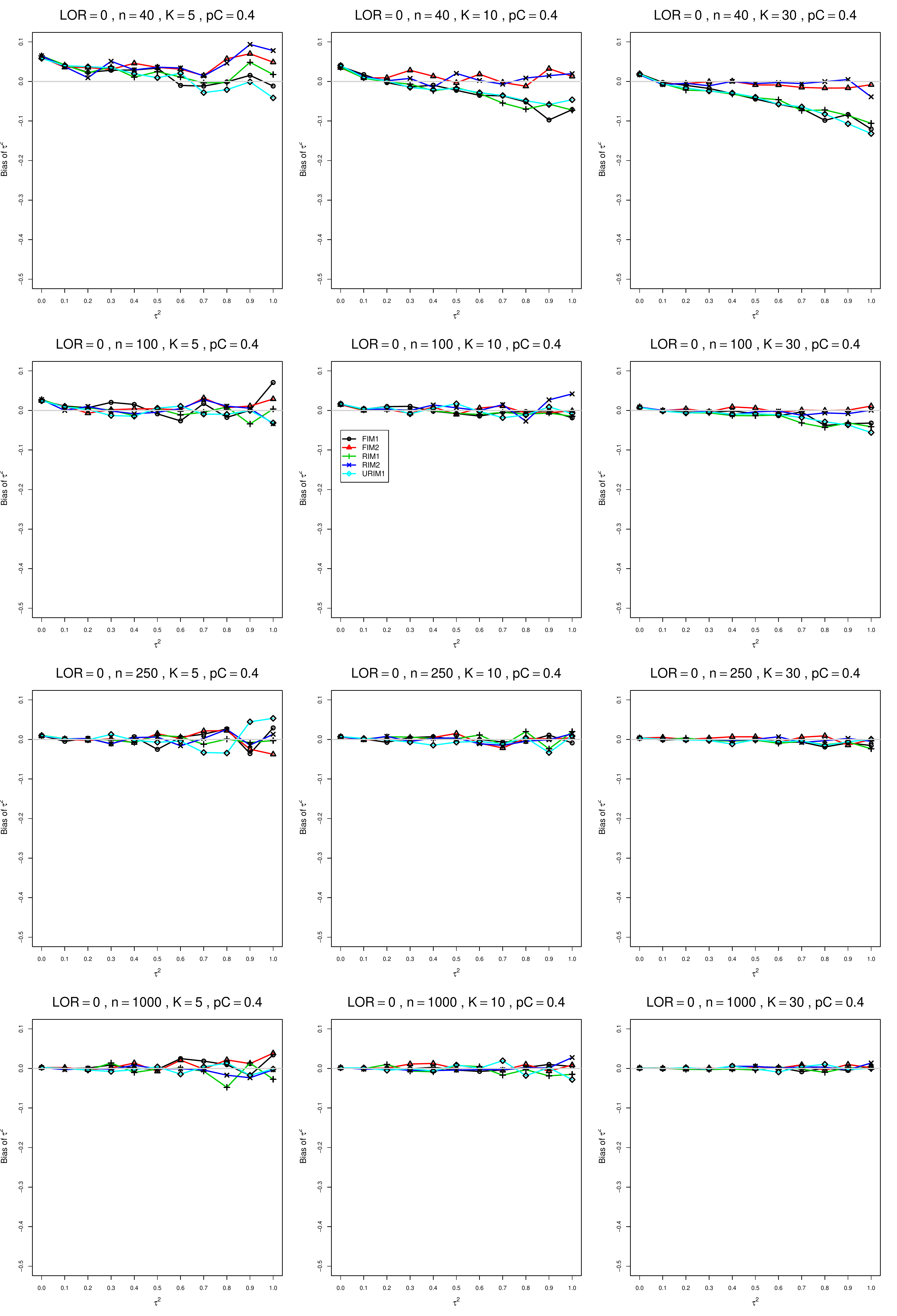}
	\caption{Bias of  between-studies variance $\hat{\tau}_{REML}^2$ for $\theta=0$, $p_{C}=0.4$, $\sigma^2=0.1$, constant sample sizes $n=40,\;100,\;250,\;1000$.
The data-generation mechanisms are FIM1 ($\circ$), FIM2 ($\triangle$), RIM1 (+), RIM2 ($\times$), and URIM1 ($\diamond$).
		\label{PlotBiasTau2mu0andpC04LOR_REMLsigma01}}
\end{figure}
\begin{figure}[t]
	\centering
	\includegraphics[scale=0.33]{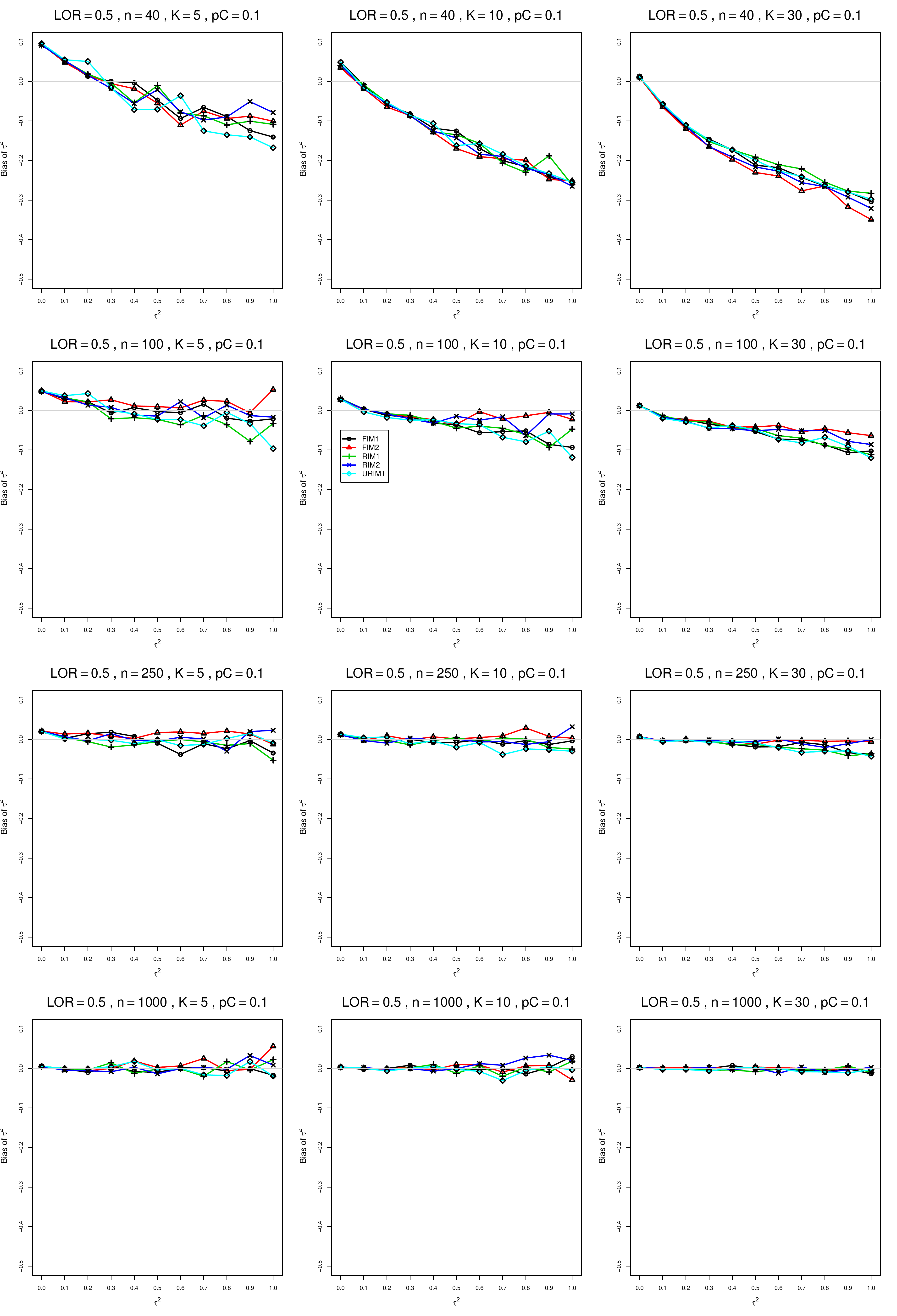}
	\caption{Bias of  between-studies variance $\hat{\tau}_{REML}^2$ for $\theta=0.5$, $p_{C}=0.1$, $\sigma^2=0.1$, constant sample sizes $n=40,\;100,\;250,\;1000$.
The data-generation mechanisms are FIM1 ($\circ$), FIM2 ($\triangle$), RIM1 (+), RIM2 ($\times$), and URIM1 ($\diamond$).
		\label{PlotBiasTau2mu05andpC01LOR_REMLsigma01}}
\end{figure}
\begin{figure}[t]
	\centering
	\includegraphics[scale=0.33]{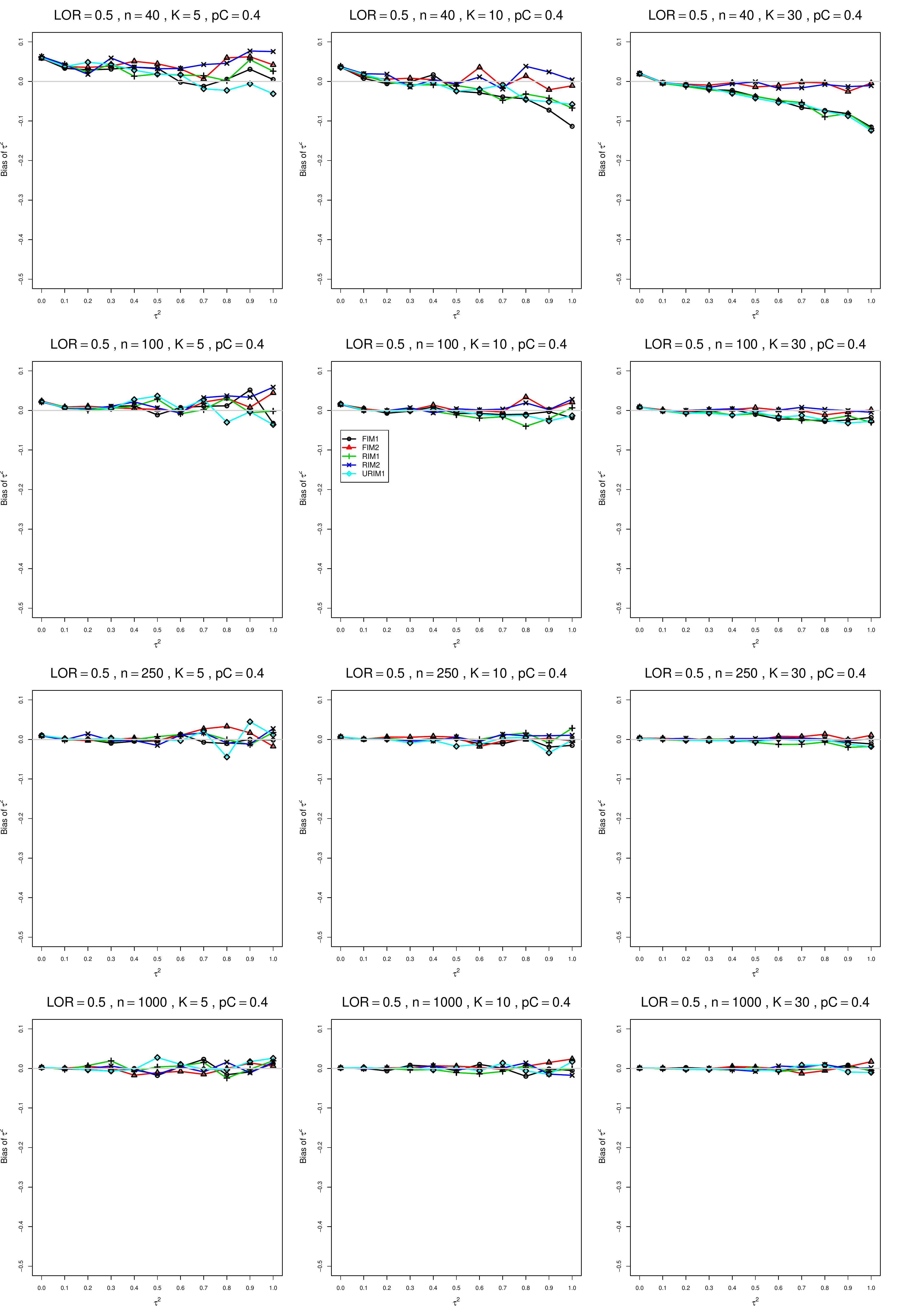}
	\caption{Bias of  between-studies variance $\hat{\tau}_{REML}^2$ for $\theta=0.5$, $p_{C}=0.4$, $\sigma^2=0.1$, constant sample sizes $n=40,\;100,\;250,\;1000$.
The data-generation mechanisms are FIM1 ($\circ$), FIM2 ($\triangle$), RIM1 (+), RIM2 ($\times$), and URIM1 ($\diamond$).
		\label{PlotBiasTau2mu05andpC04LOR_REMLsigma01}}
\end{figure}
\begin{figure}[t]
	\centering
	\includegraphics[scale=0.33]{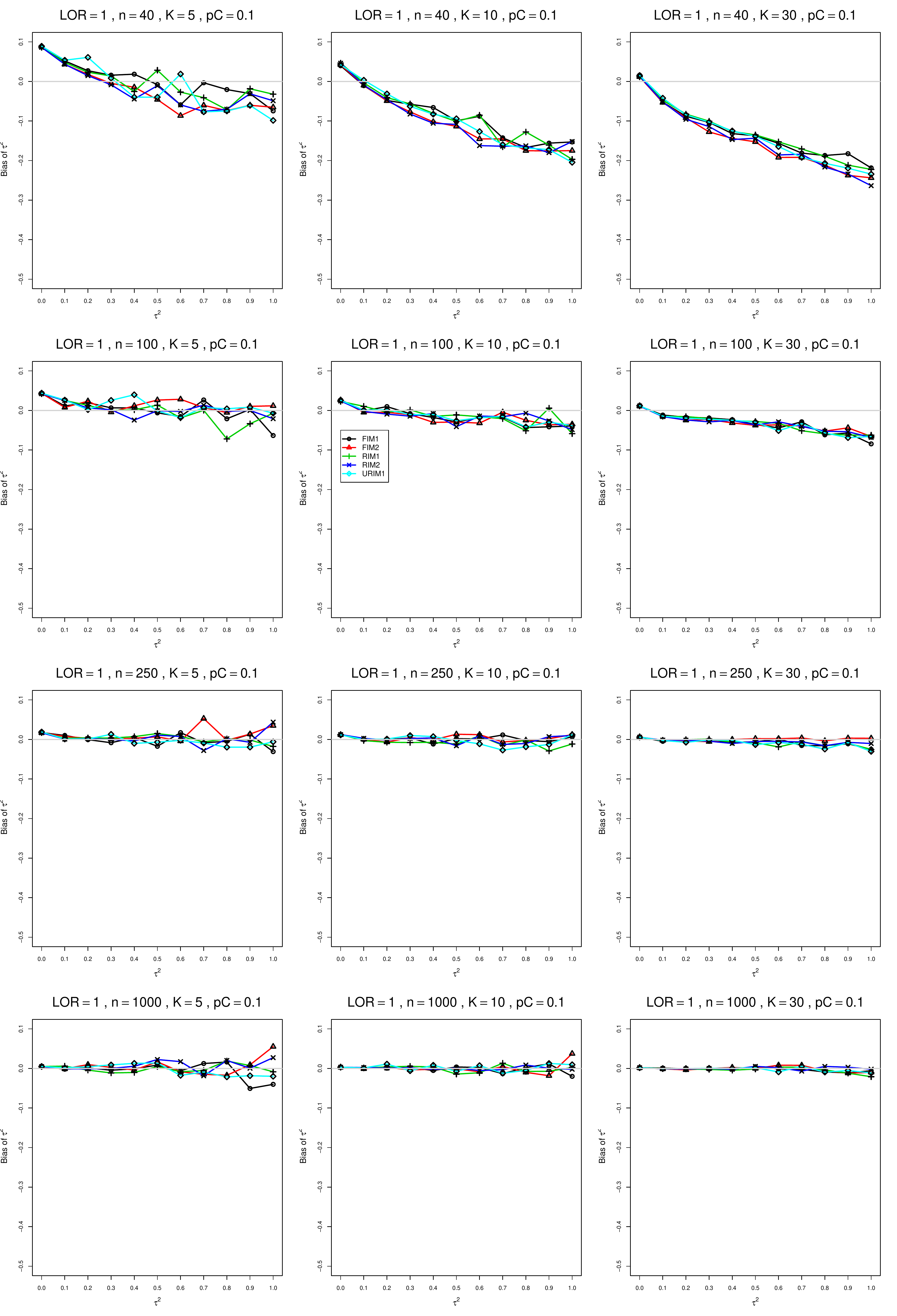}
	\caption{Bias of  between-studies variance $\hat{\tau}_{REML}^2$ for $\theta=1$, $p_{C}=0.1$, $\sigma^2=0.1$, constant sample sizes $n=40,\;100,\;250,\;1000$.
The data-generation mechanisms are FIM1 ($\circ$), FIM2 ($\triangle$), RIM1 (+), RIM2 ($\times$), and URIM1 ($\diamond$).
		\label{PlotBiasTau2mu1andpC01LOR_REMLsigma01}}
\end{figure}
\begin{figure}[t]
	\centering
	\includegraphics[scale=0.33]{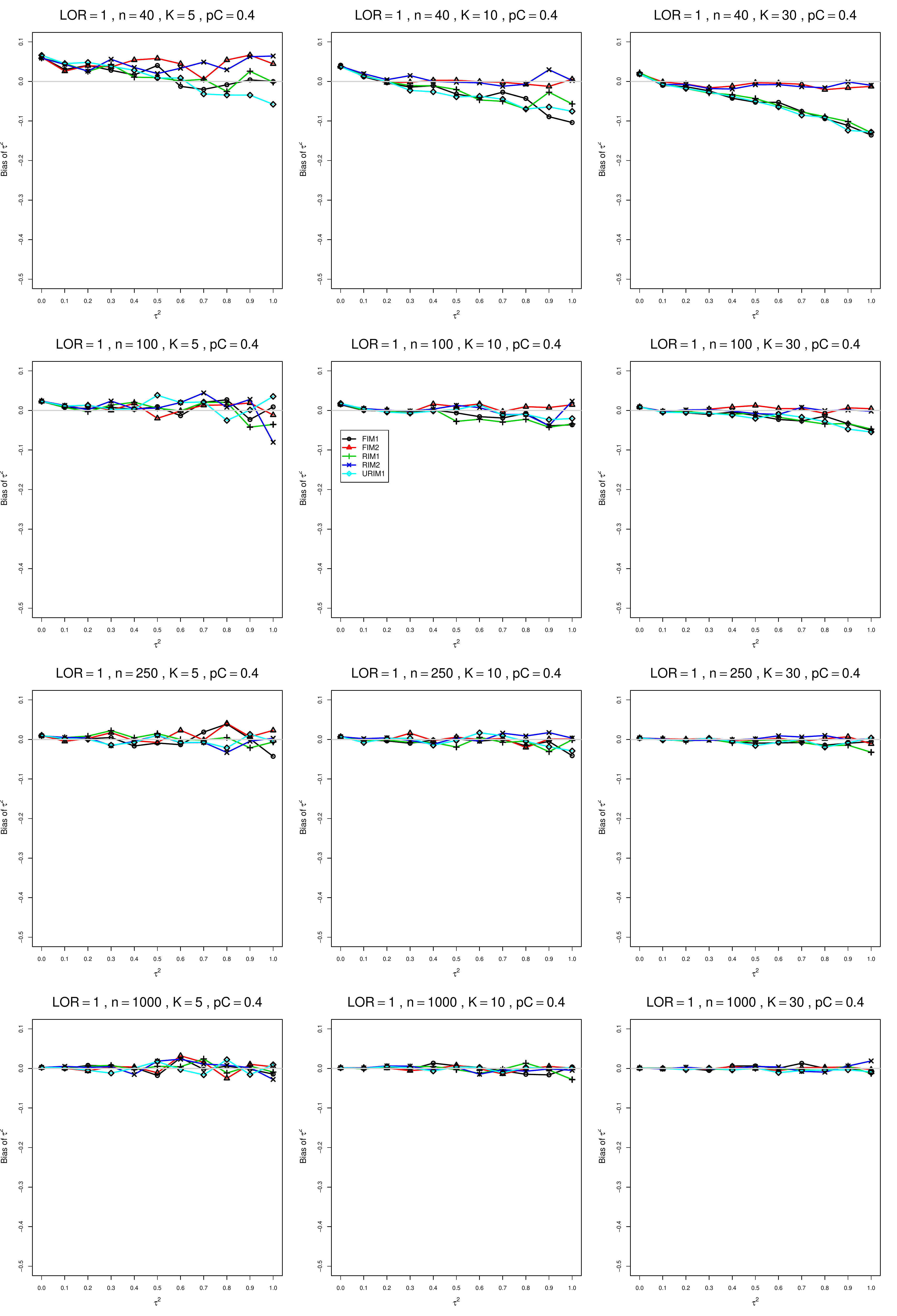}
	\caption{Bias of  between-studies variance $\hat{\tau}_{REML}^2$ for $\theta=1$, $p_{C}=0.4$, $\sigma^2=0.1$, constant sample sizes $n=40,\;100,\;250,\;1000$.
The data-generation mechanisms are FIM1 ($\circ$), FIM2 ($\triangle$), RIM1 (+), RIM2 ($\times$), and URIM1 ($\diamond$).
		\label{PlotBiasTau2mu1andpC04LOR_REMLsigma01}}
\end{figure}
\begin{figure}[t]
	\centering
	\includegraphics[scale=0.33]{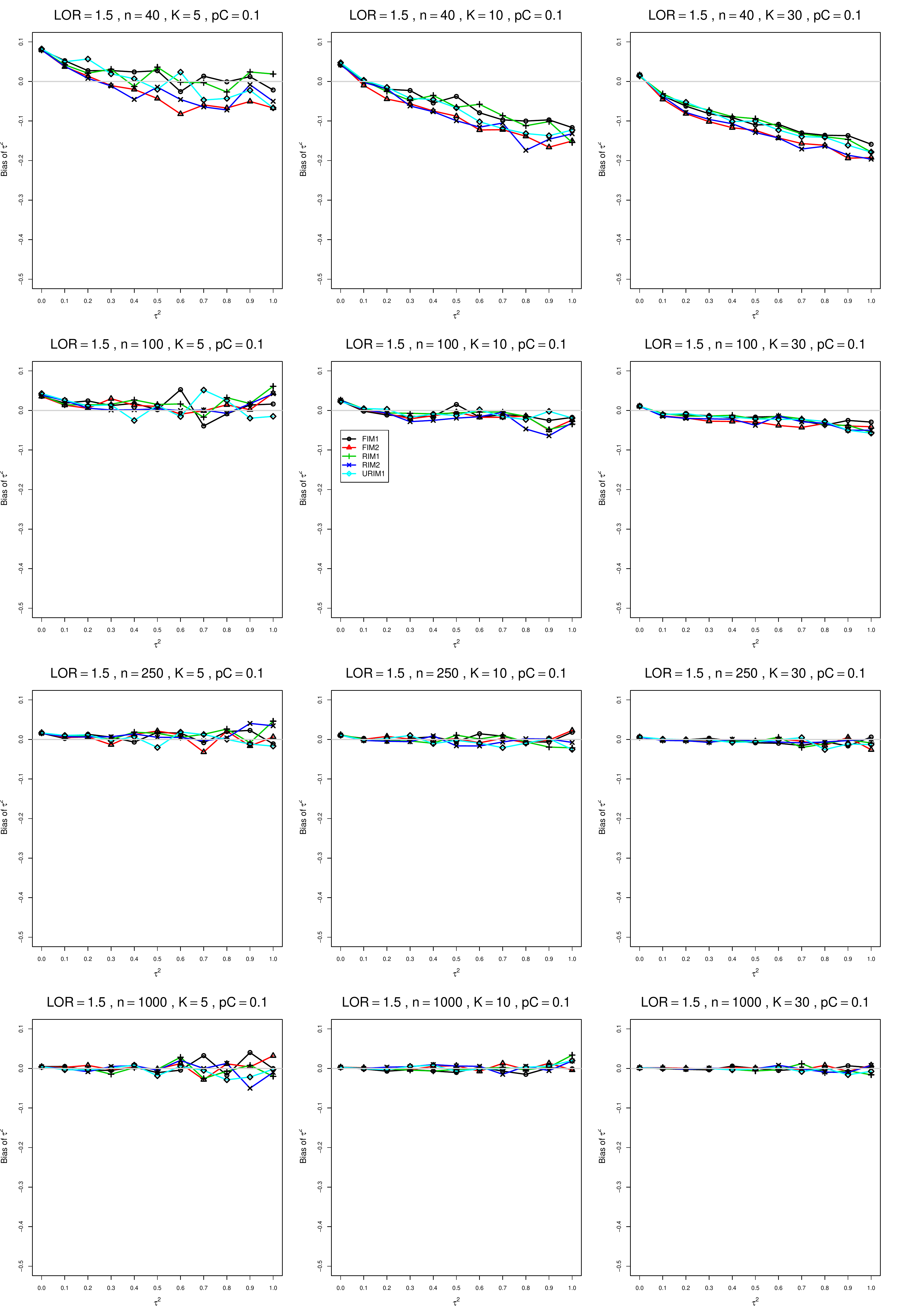}
	\caption{Bias of  between-studies variance $\hat{\tau}_{REML}^2$ for $\theta=1.5$, $p_{C}=0.1$, $\sigma^2=0.1$, constant sample sizes $n=40,\;100,\;250,\;1000$.
The data-generation mechanisms are FIM1 ($\circ$), FIM2 ($\triangle$), RIM1 (+), RIM2 ($\times$), and URIM1 ($\diamond$).
		\label{PlotBiasTau2mu15andpC01LOR_REMLsigma01}}
\end{figure}
\begin{figure}[t]
	\centering
	\includegraphics[scale=0.33]{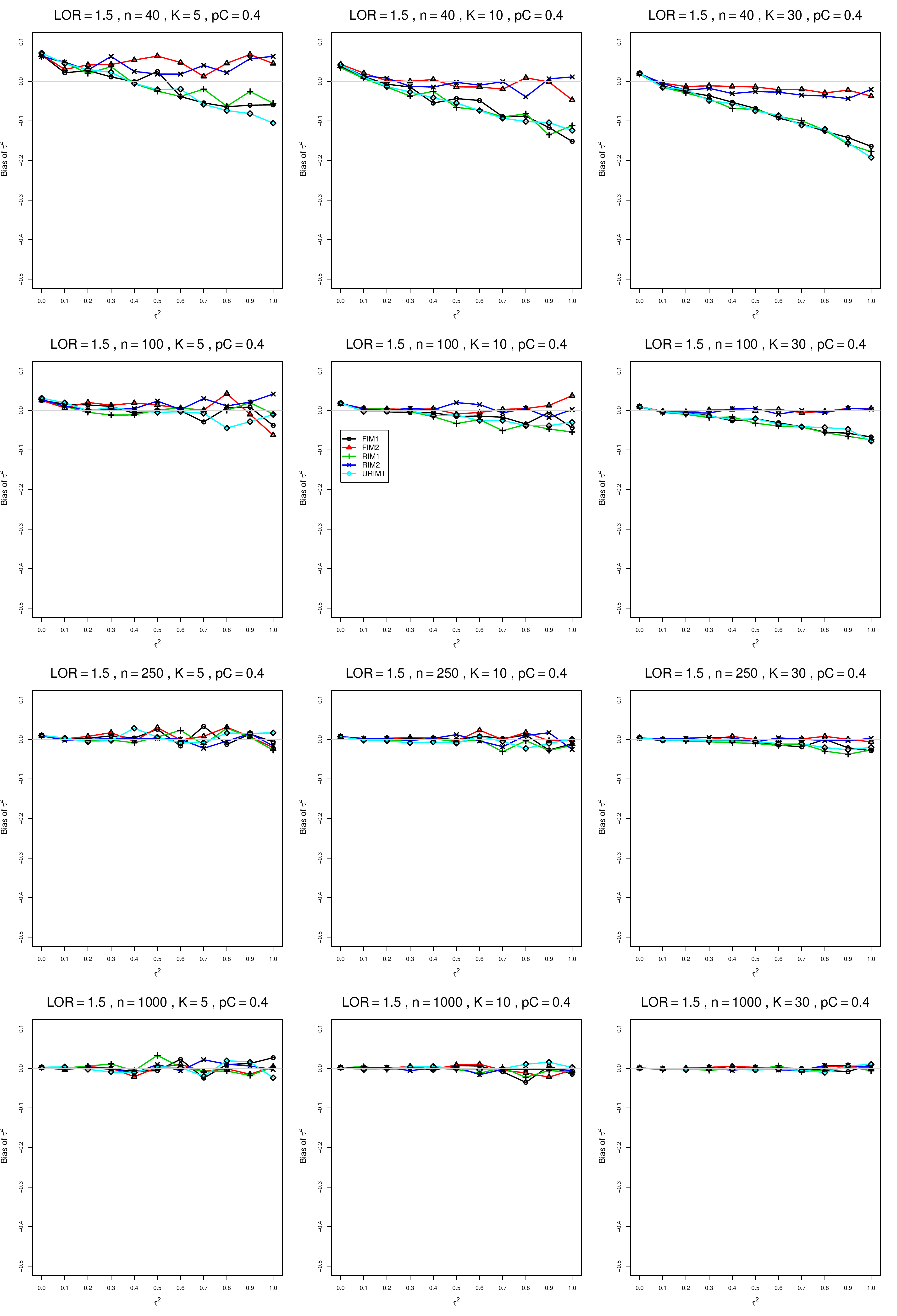}
	\caption{Bias of  between-studies variance $\hat{\tau}_{REML}^2$ for $\theta=1.5$, $p_{C}=0.4$, $\sigma^2=0.1$, constant sample sizes $n=40,\;100,\;250,\;1000$.
The data-generation mechanisms are FIM1 ($\circ$), FIM2 ($\triangle$), RIM1 (+), RIM2 ($\times$), and URIM1 ($\diamond$).
		\label{PlotBiasTau2mu15andpC04LOR_REMLsigma01}}
\end{figure}
\begin{figure}[t]
	\centering
	\includegraphics[scale=0.33]{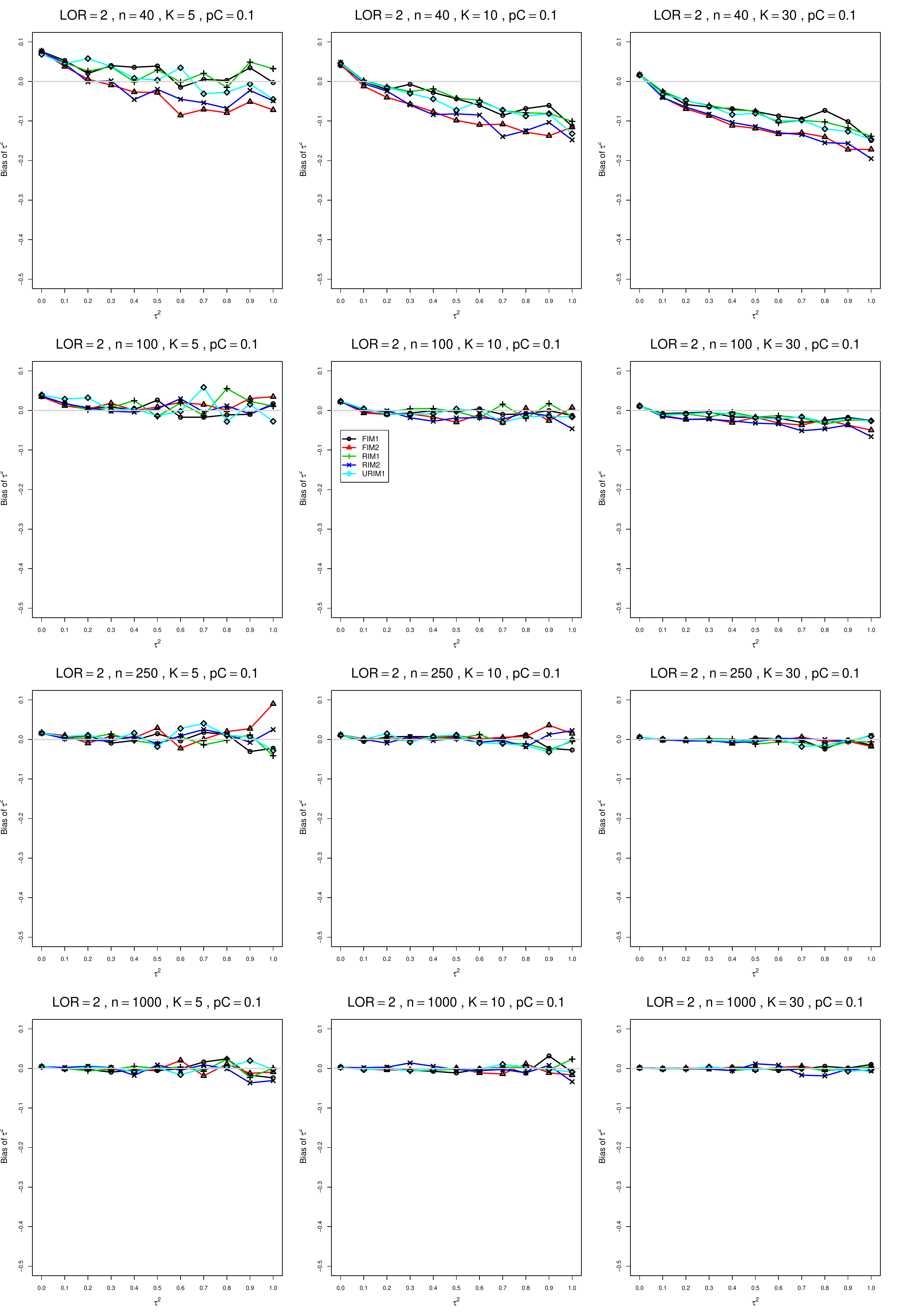}
	\caption{Bias of  between-studies variance $\hat{\tau}_{REML}^2$ for $\theta=2$, $p_{C}=0.1$, $\sigma^2=0.1$, constant sample sizes $n=40,\;100,\;250,\;1000$.
The data-generation mechanisms are FIM1 ($\circ$), FIM2 ($\triangle$), RIM1 (+), RIM2 ($\times$), and URIM1 ($\diamond$).
		\label{PlotBiasTau2mu2andpC01LOR_REMLsigma01}}
\end{figure}
\begin{figure}[t]
	\centering
	\includegraphics[scale=0.33]{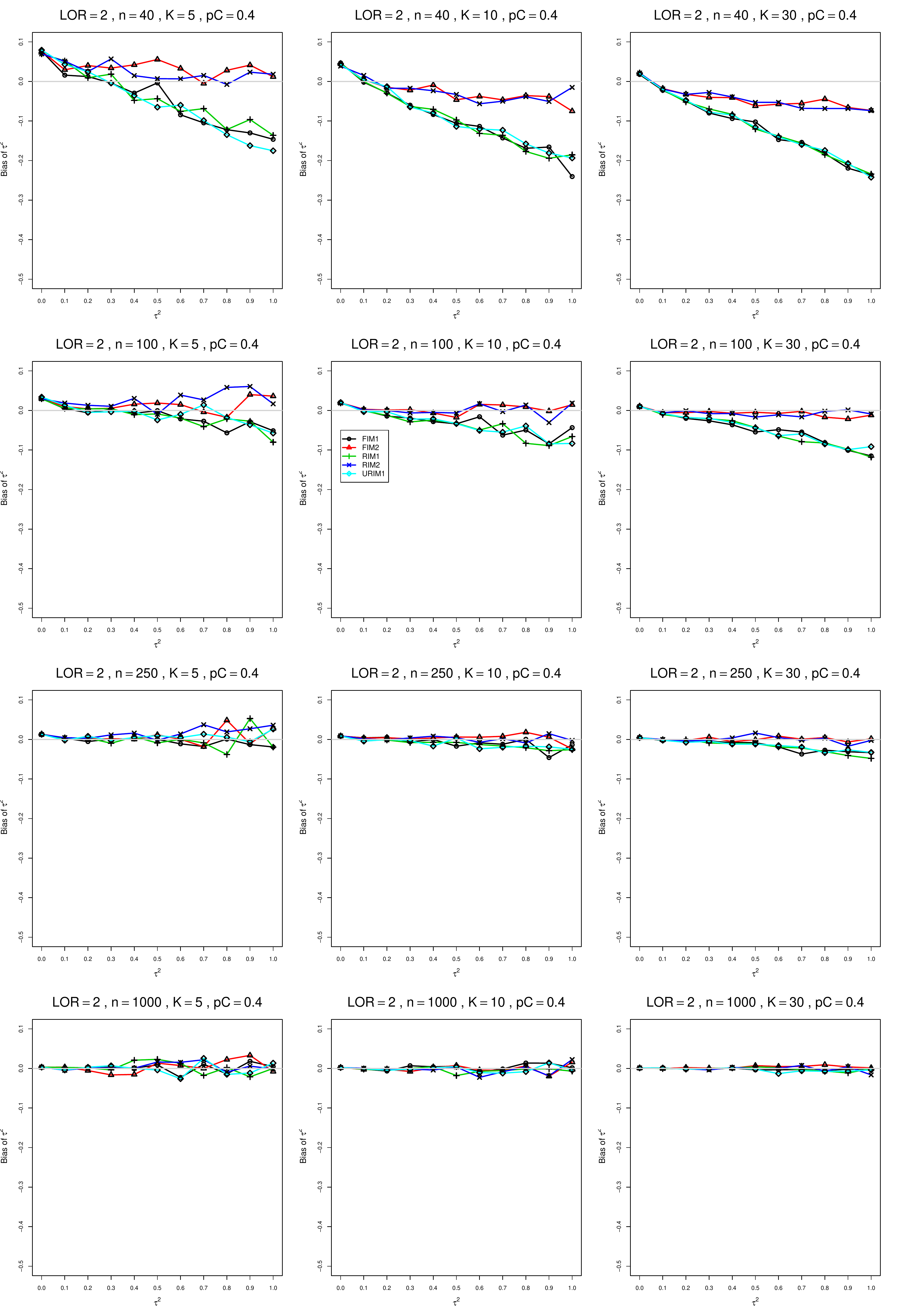}
	\caption{Bias of  between-studies variance $\hat{\tau}_{REML}^2$ for $\theta=2$, $p_{C}=0.4$, $\sigma^2=0.1$, constant sample sizes $n=40,\;100,\;250,\;1000$.
The data-generation mechanisms are FIM1 ($\circ$), FIM2 ($\triangle$), RIM1 (+), RIM2 ($\times$), and URIM1 ($\diamond$).
		\label{PlotBiasTau2mu2andpC04LOR_REMLsigma01}}
\end{figure}
\begin{figure}[t]
	\centering
	\includegraphics[scale=0.33]{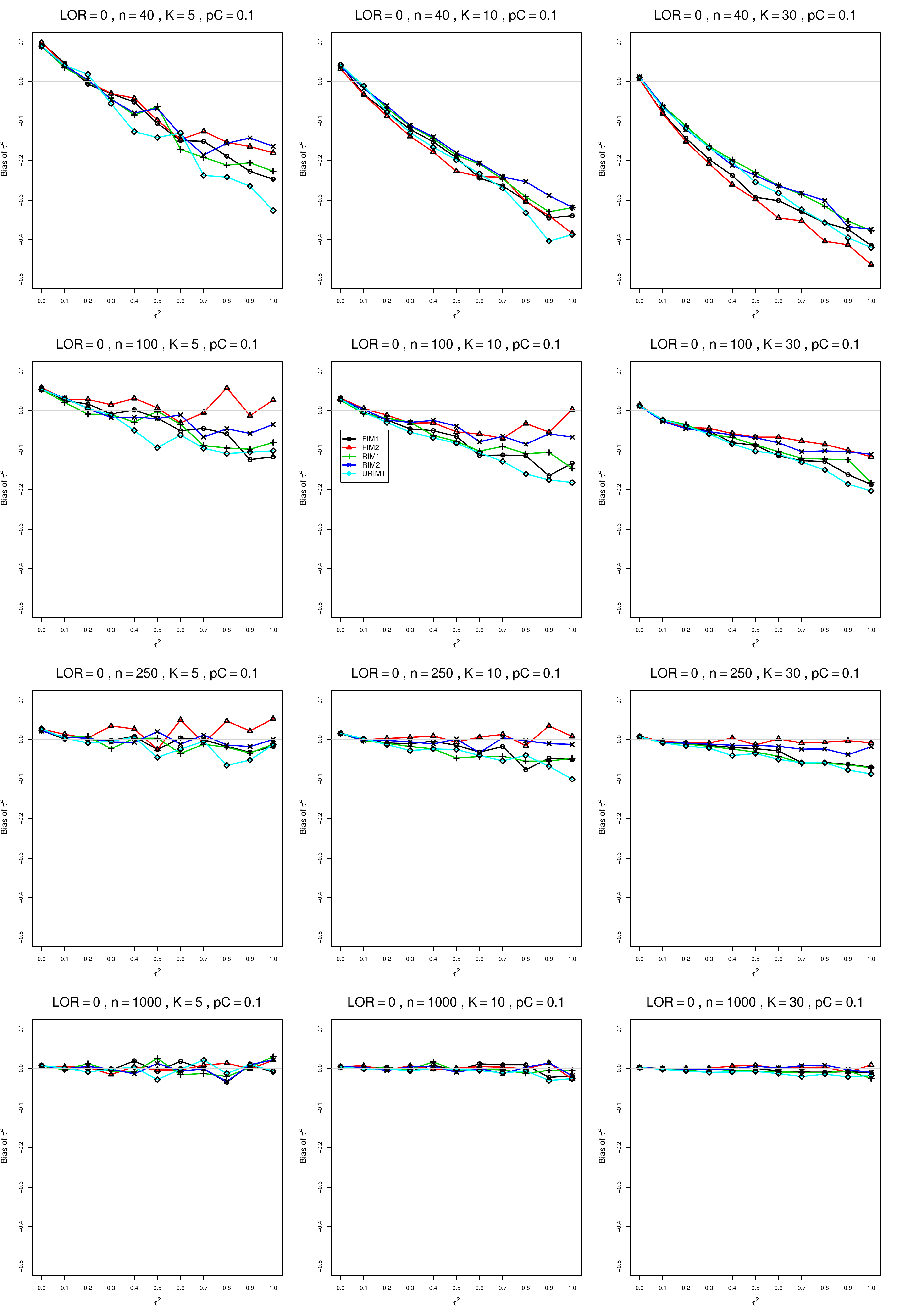}
	\caption{Bias of  between-studies variance $\hat{\tau}_{REML}^2$ for $\theta=0$, $p_{C}=0.1$, $\sigma^2=0.4$, constant sample sizes $n=40,\;100,\;250,\;1000$.
The data-generation mechanisms are FIM1 ($\circ$), FIM2 ($\triangle$), RIM1 (+), RIM2 ($\times$), and URIM1 ($\diamond$).
		\label{PlotBiasTau2mu0andpC01LOR_REMLsigma04}}
\end{figure}
\begin{figure}[t]
	\centering
	\includegraphics[scale=0.33]{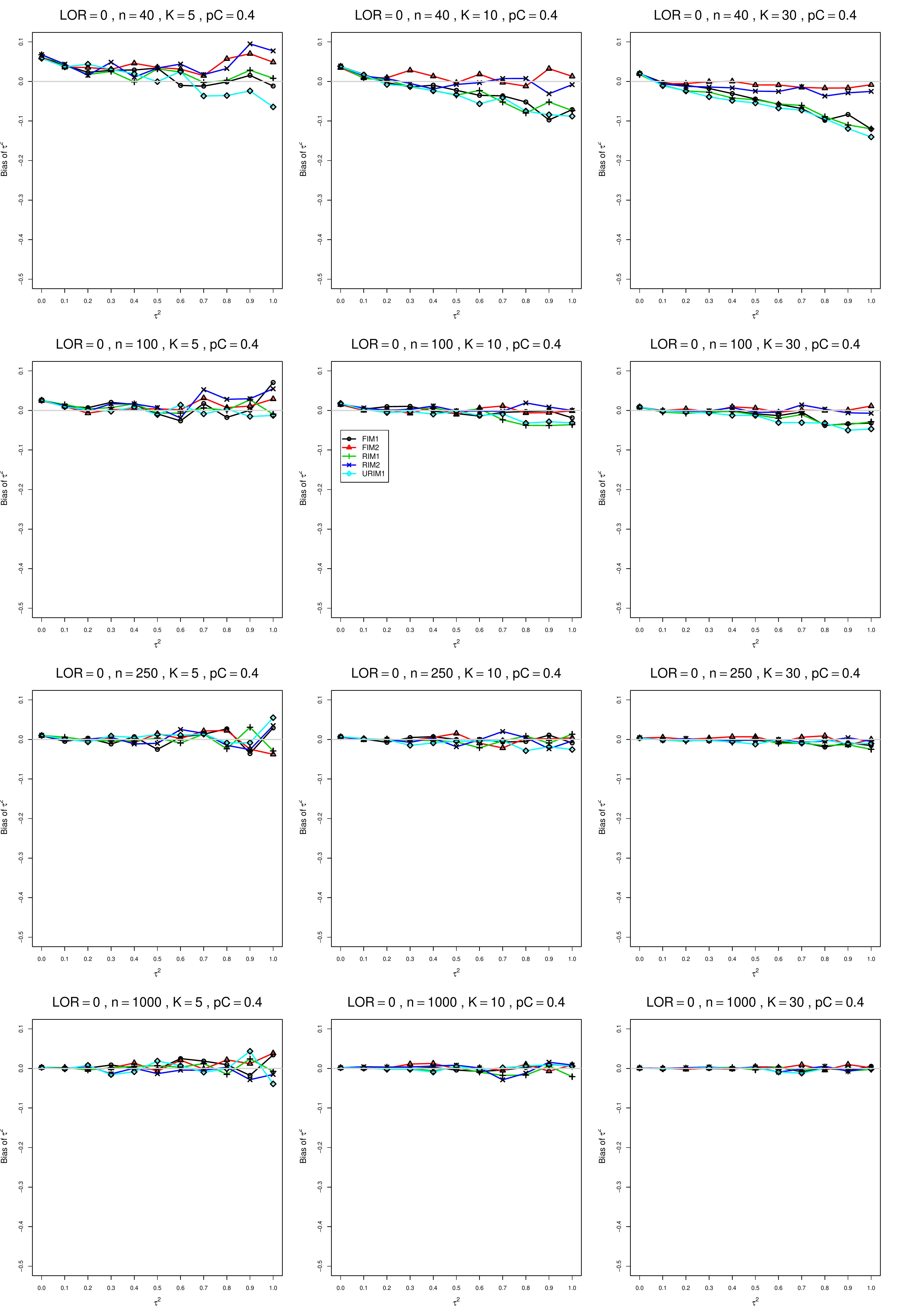}
	\caption{Bias of  between-studies variance $\hat{\tau}_{REML}^2$ for $\theta=0$, $p_{C}=0.4$, $\sigma^2=0.4$, constant sample sizes $n=40,\;100,\;250,\;1000$.
The data-generation mechanisms are FIM1 ($\circ$), FIM2 ($\triangle$), RIM1 (+), RIM2 ($\times$), and URIM1 ($\diamond$).
		\label{PlotBiasTau2mu0andpC04LOR_REMLsigma04}}
\end{figure}
\begin{figure}[t]
	\centering
	\includegraphics[scale=0.33]{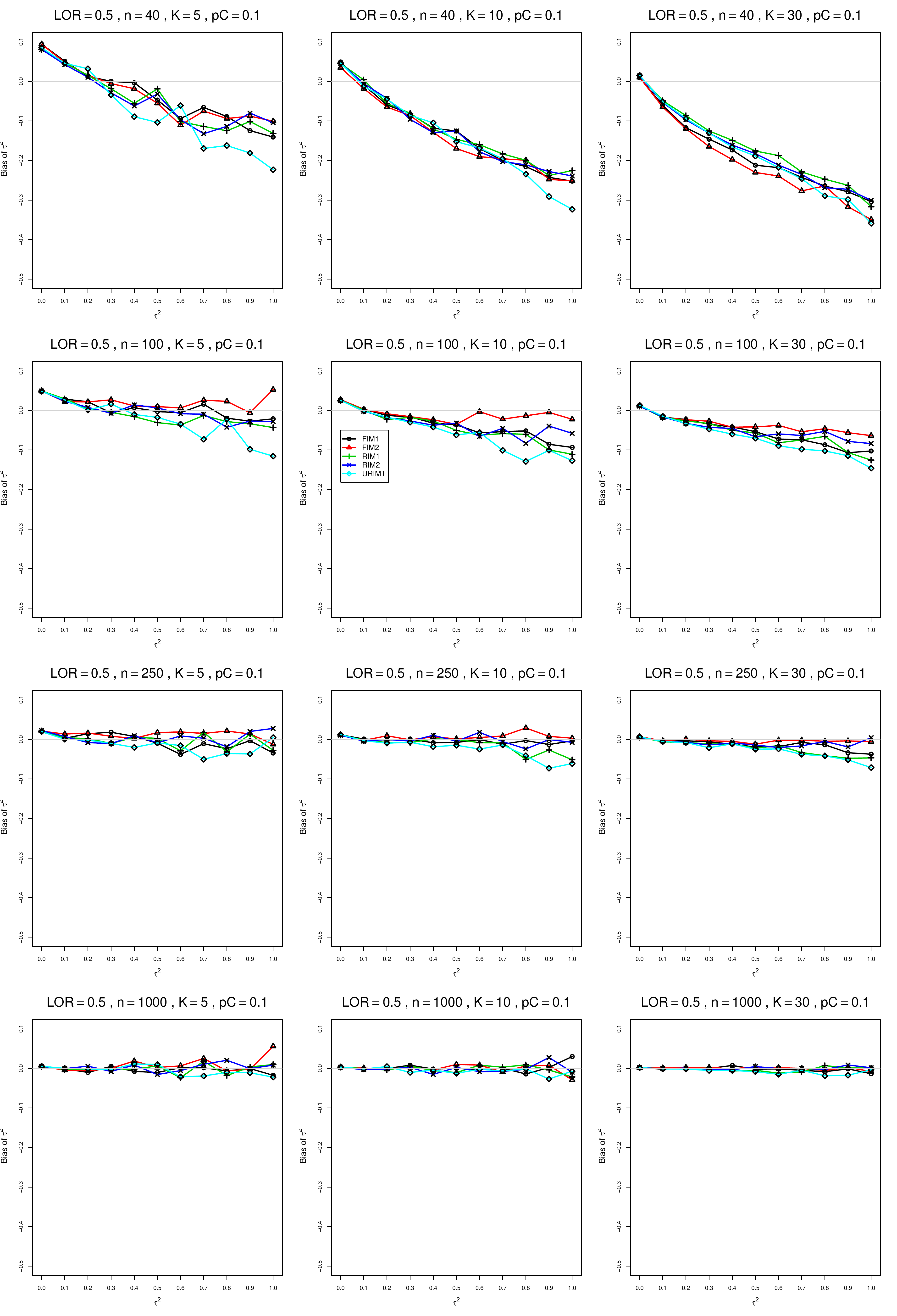}
	\caption{Bias of  between-studies variance $\hat{\tau}_{REML}^2$ for $\theta=0.5$, $p_{C}=0.1$, $\sigma^2=0.4$, constant sample sizes $n=40,\;100,\;250,\;1000$.
The data-generation mechanisms are FIM1 ($\circ$), FIM2 ($\triangle$), RIM1 (+), RIM2 ($\times$), and URIM1 ($\diamond$).
		\label{PlotBiasTau2mu05andpC01LOR_REMLsigma04}}
\end{figure}
\begin{figure}[t]
	\centering
	\includegraphics[scale=0.33]{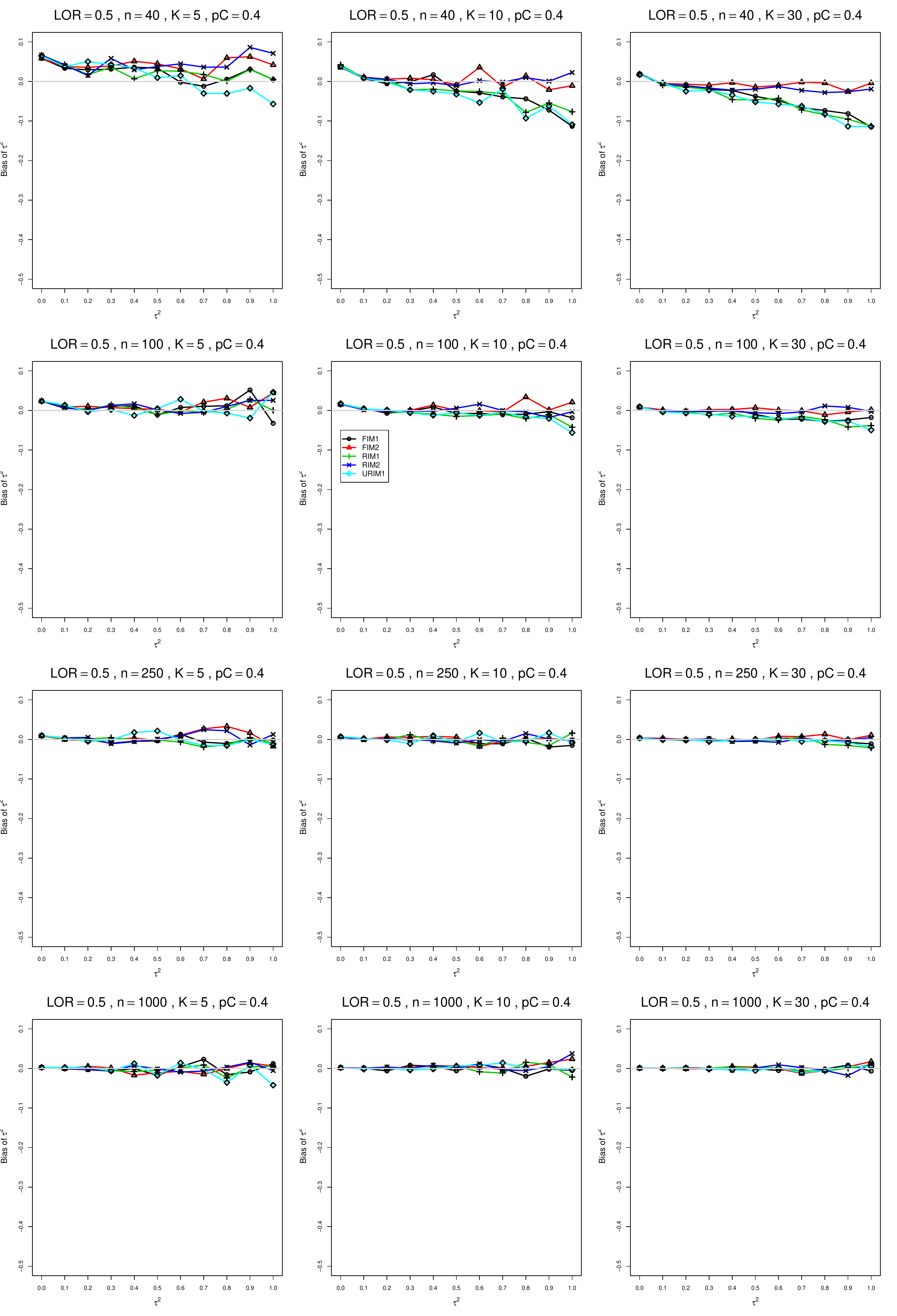}
	\caption{Bias of  between-studies variance $\hat{\tau}_{REML}^2$ for $\theta=0.5$, $p_{C}=0.4$, $\sigma^2=0.4$, constant sample sizes $n=40,\;100,\;250,\;1000$.
The data-generation mechanisms are FIM1 ($\circ$), FIM2 ($\triangle$), RIM1 (+), RIM2 ($\times$), and URIM1 ($\diamond$).
		\label{PlotBiasTau2mu05andpC04LOR_REMLsigma04}}
\end{figure}
\begin{figure}[t]
	\centering
	\includegraphics[scale=0.33]{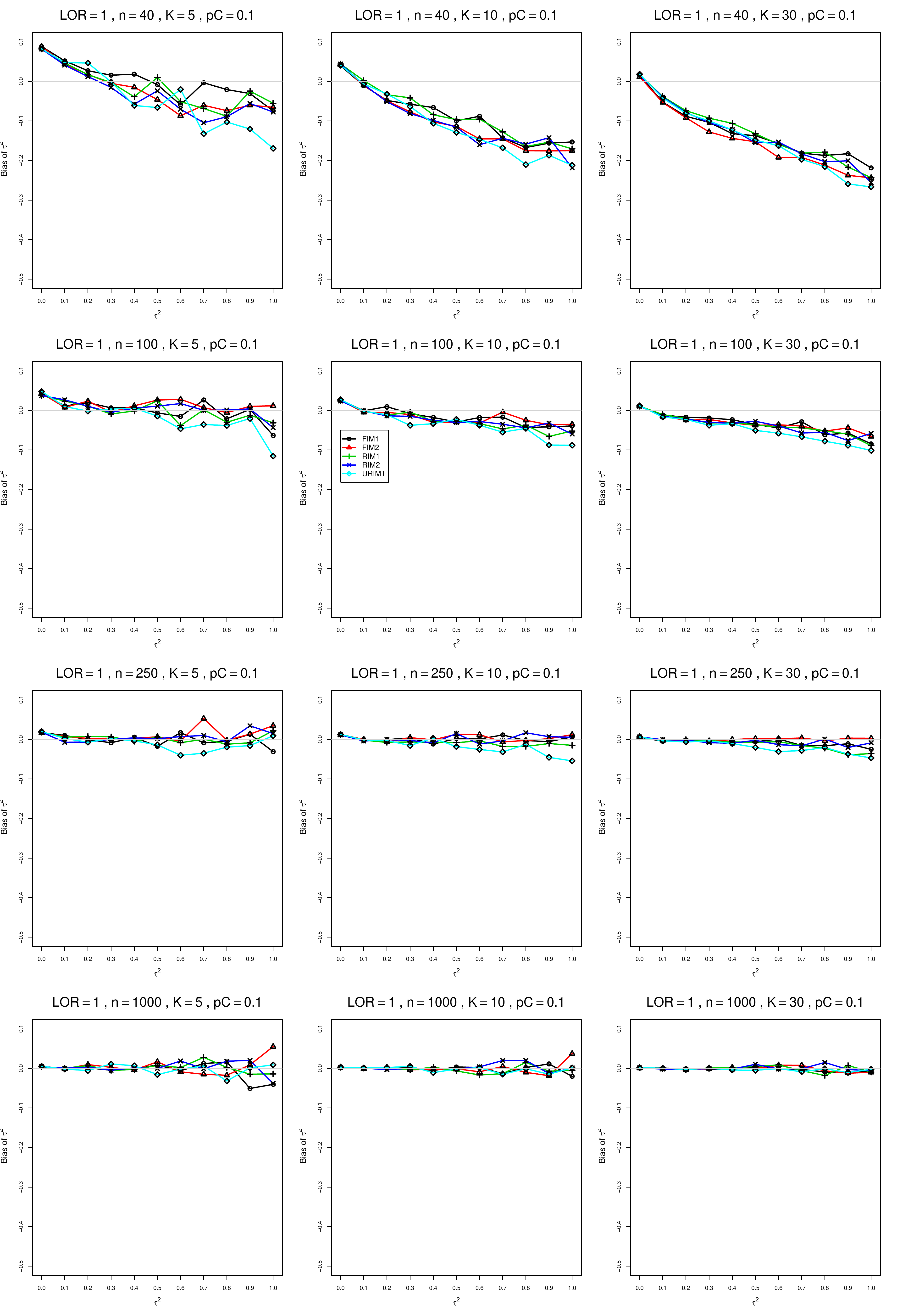}
	\caption{Bias of  between-studies variance $\hat{\tau}_{REML}^2$ for $\theta=1$, $p_{C}=0.1$, $\sigma^2=0.4$, constant sample sizes $n=40,\;100,\;250,\;1000$.
The data-generation mechanisms are FIM1 ($\circ$), FIM2 ($\triangle$), RIM1 (+), RIM2 ($\times$), and URIM1 ($\diamond$).
		\label{PlotBiasTau2mu1andpC01LOR_REMLsigma04}}
\end{figure}
\begin{figure}[t]
	\centering
	\includegraphics[scale=0.33]{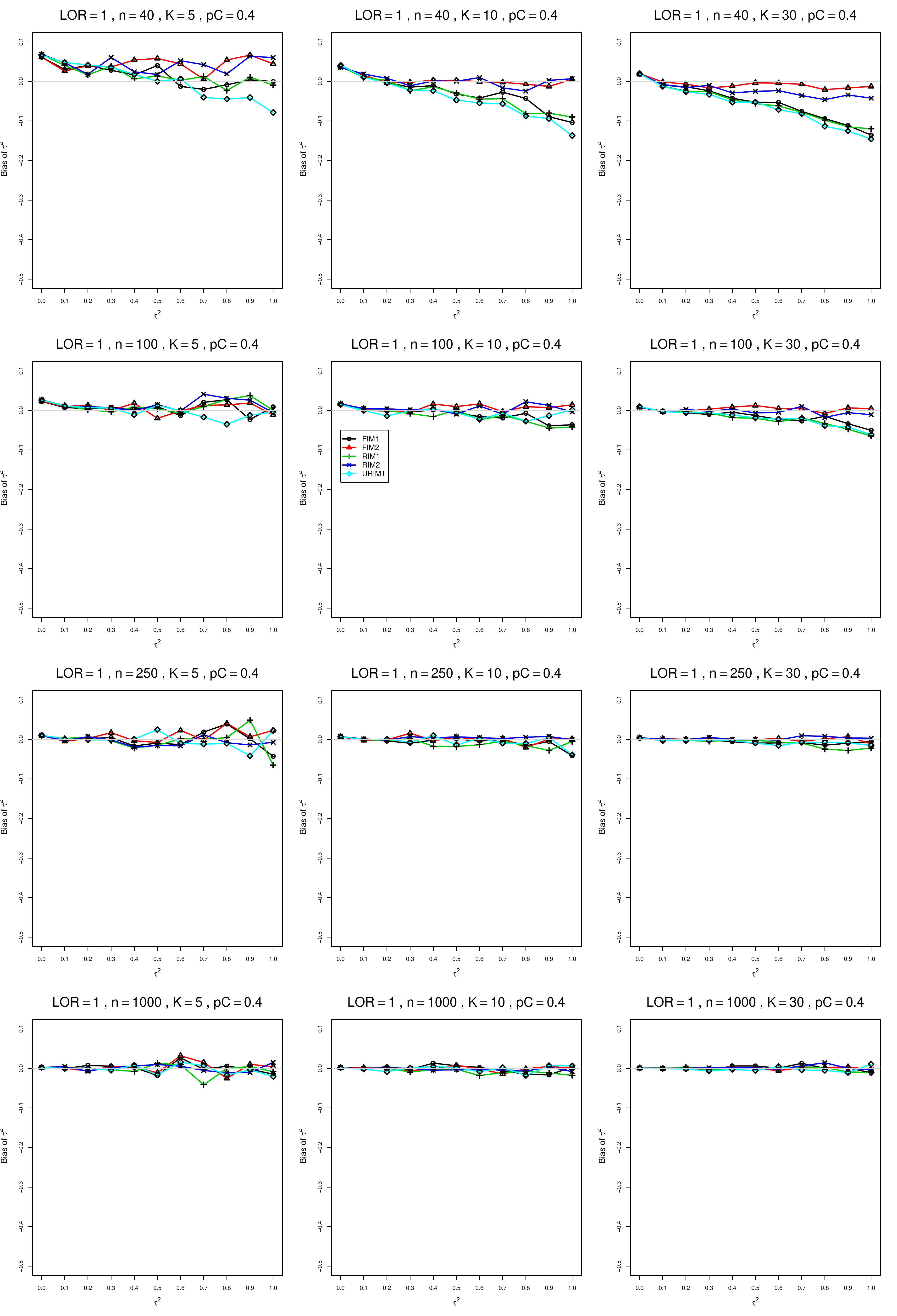}
	\caption{Bias of  between-studies variance $\hat{\tau}_{REML}^2$ for $\theta=1$, $p_{C}=0.4$, $\sigma^2=0.4$, constant sample sizes $n=40,\;100,\;250,\;1000$.
The data-generation mechanisms are FIM1 ($\circ$), FIM2 ($\triangle$), RIM1 (+), RIM2 ($\times$), and URIM1 ($\diamond$).
		\label{PlotBiasTau2mu1andpC04LOR_REMLsigma04}}
\end{figure}
\begin{figure}[t]
	\centering
	\includegraphics[scale=0.33]{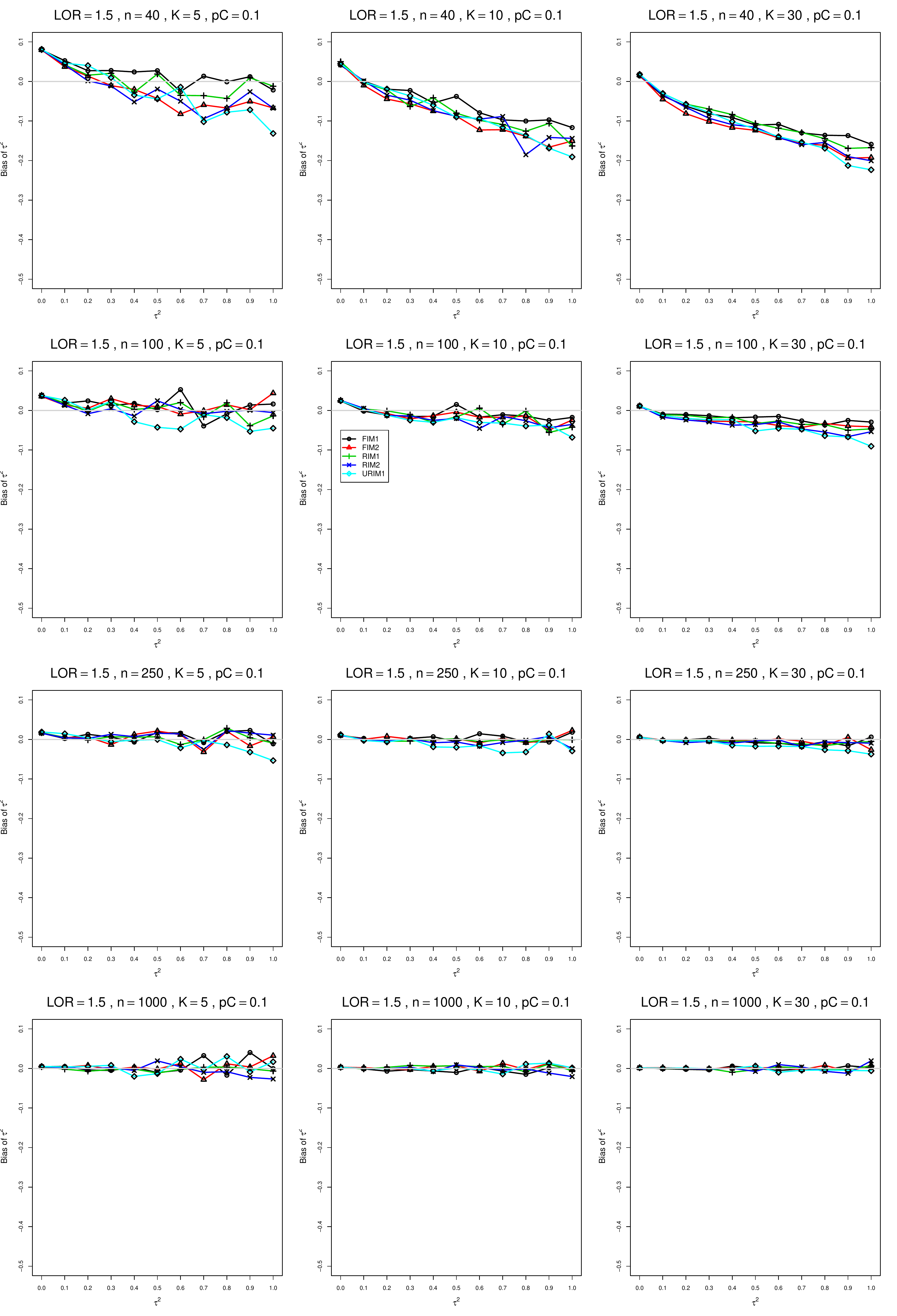}
	\caption{Bias of  between-studies variance $\hat{\tau}_{REML}^2$ for $\theta=1.5$, $p_{C}=0.1$, $\sigma^2=0.4$, constant sample sizes $n=40,\;100,\;250,\;1000$.
The data-generation mechanisms are FIM1 ($\circ$), FIM2 ($\triangle$), RIM1 (+), RIM2 ($\times$), and URIM1 ($\diamond$).
		\label{PlotBiasTau2mu15andpC01LOR_REMLsigma04}}
\end{figure}
\begin{figure}[t]
	\centering
	\includegraphics[scale=0.33]{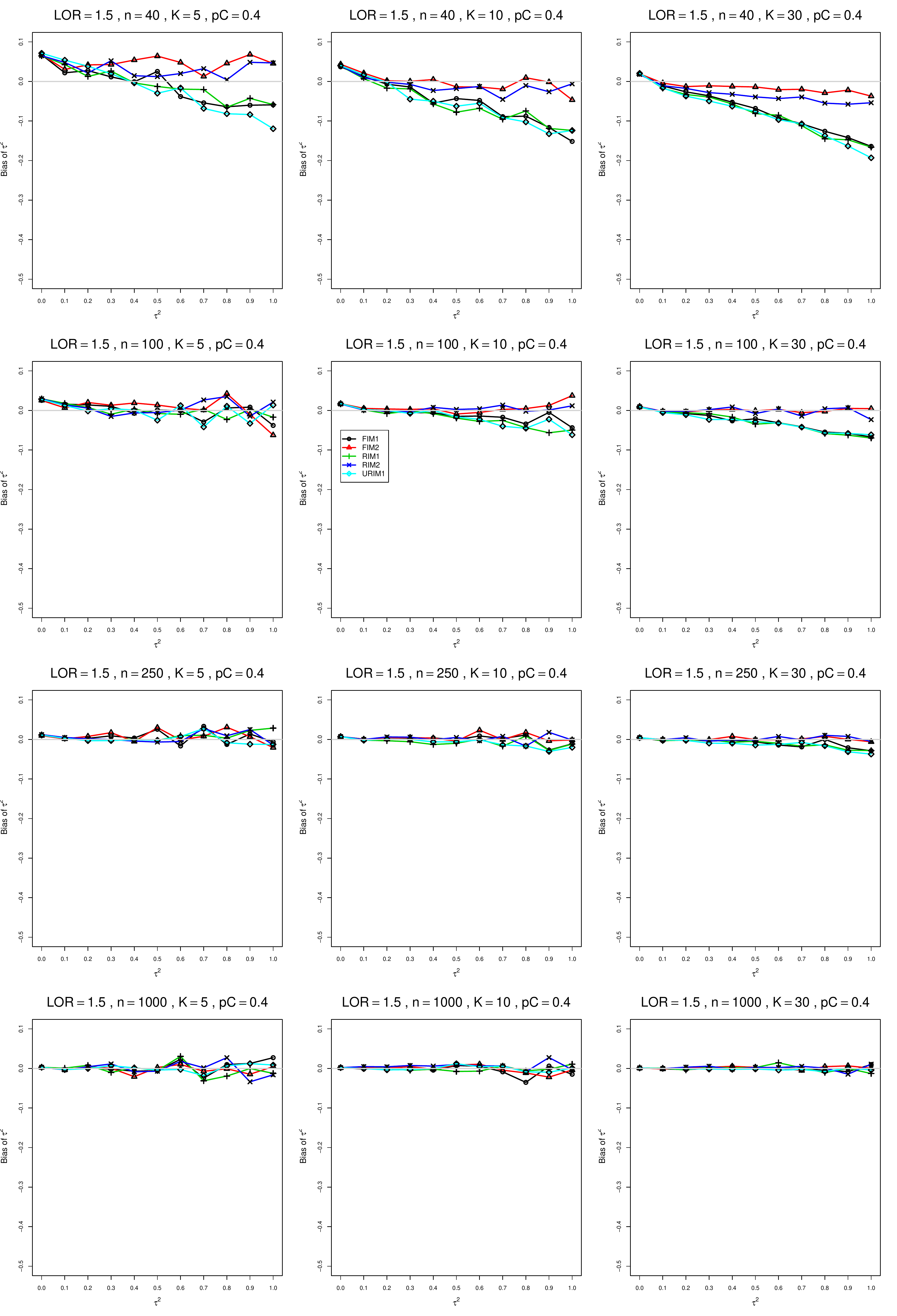}
	\caption{Bias of  between-studies variance $\hat{\tau}_{REML}^2$ for $\theta=1.5$, $p_{C}=0.4$, $\sigma^2=0.4$, constant sample sizes $n=40,\;100,\;250,\;1000$.
The data-generation mechanisms are FIM1 ($\circ$), FIM2 ($\triangle$), RIM1 (+), RIM2 ($\times$), and URIM1 ($\diamond$).
		\label{PlotBiasTau2mu15andpC04LOR_REMLsigma04}}
\end{figure}
\begin{figure}[t]
	\centering
	\includegraphics[scale=0.33]{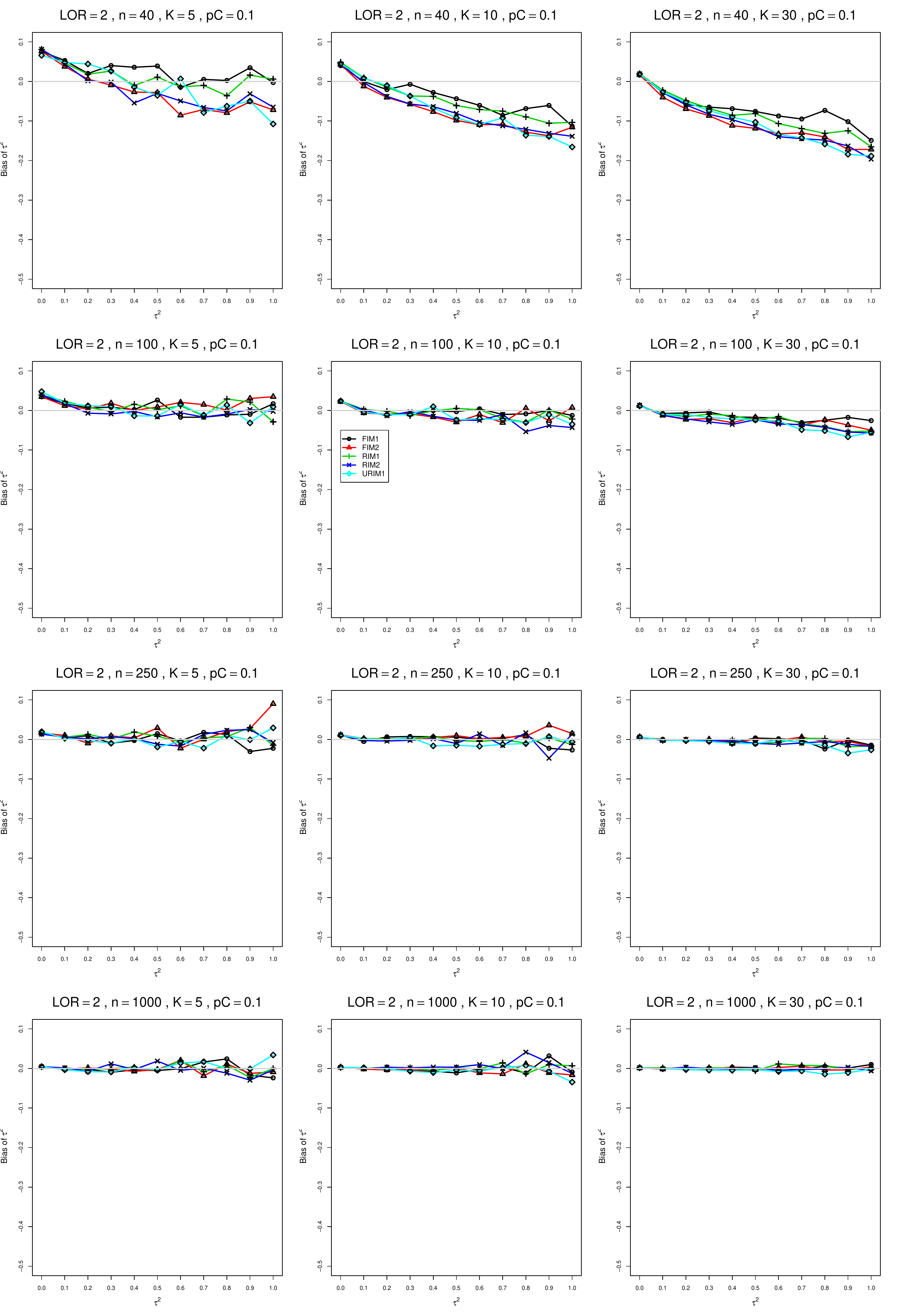}
	\caption{Bias of  between-studies variance $\hat{\tau}_{REML}^2$ for $\theta=2$, $p_{C}=0.1$, $\sigma^2=0.4$, constant sample sizes $n=40,\;100,\;250,\;1000$.
The data-generation mechanisms are FIM1 ($\circ$), FIM2 ($\triangle$), RIM1 (+), RIM2 ($\times$), and URIM1 ($\diamond$).
		\label{PlotBiasTau2mu2andpC01LOR_REMLsigma04}}
\end{figure}
\begin{figure}[t]
	\centering
	\includegraphics[scale=0.33]{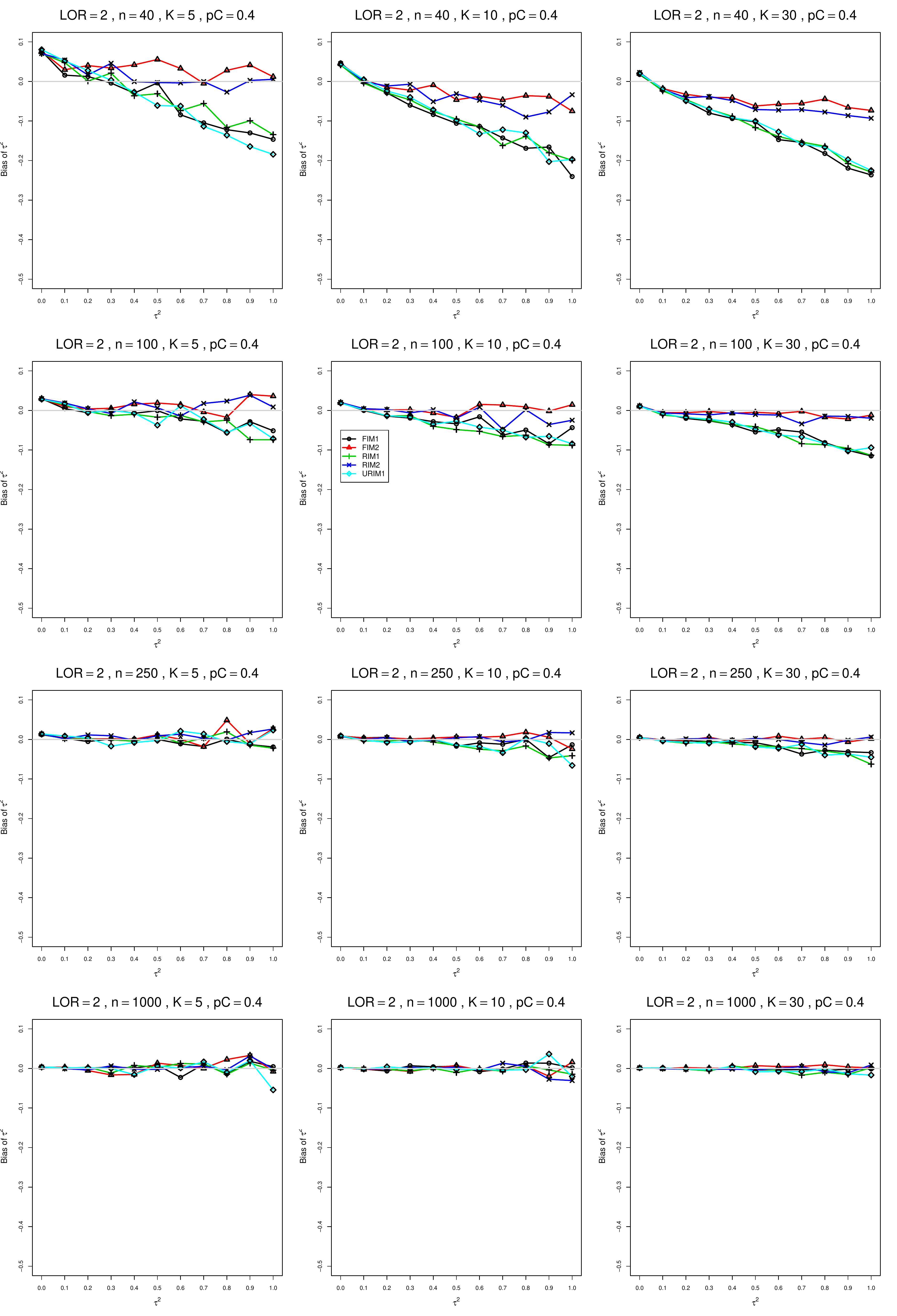}
	\caption{Bias of  between-studies variance $\hat{\tau}_{REML}^2$ for $\theta=2$, $p_{C}=0.4$, $\sigma^2=0.4$, constant sample sizes $n=40,\;100,\;250,\;1000$.
The data-generation mechanisms are FIM1 ($\circ$), FIM2 ($\triangle$), RIM1 (+), RIM2 ($\times$), and URIM1 ($\diamond$).
		\label{PlotBiasTau2mu2andpC04LOR_REMLsigma04}}
\end{figure}

\clearpage
\subsection*{A1.3 Bias of $\hat{\tau}_{MP}^2$}
\renewcommand{\thefigure}{A1.3.\arabic{figure}}
\setcounter{figure}{0}

\begin{figure}[t]
	\centering
	\includegraphics[scale=0.33]{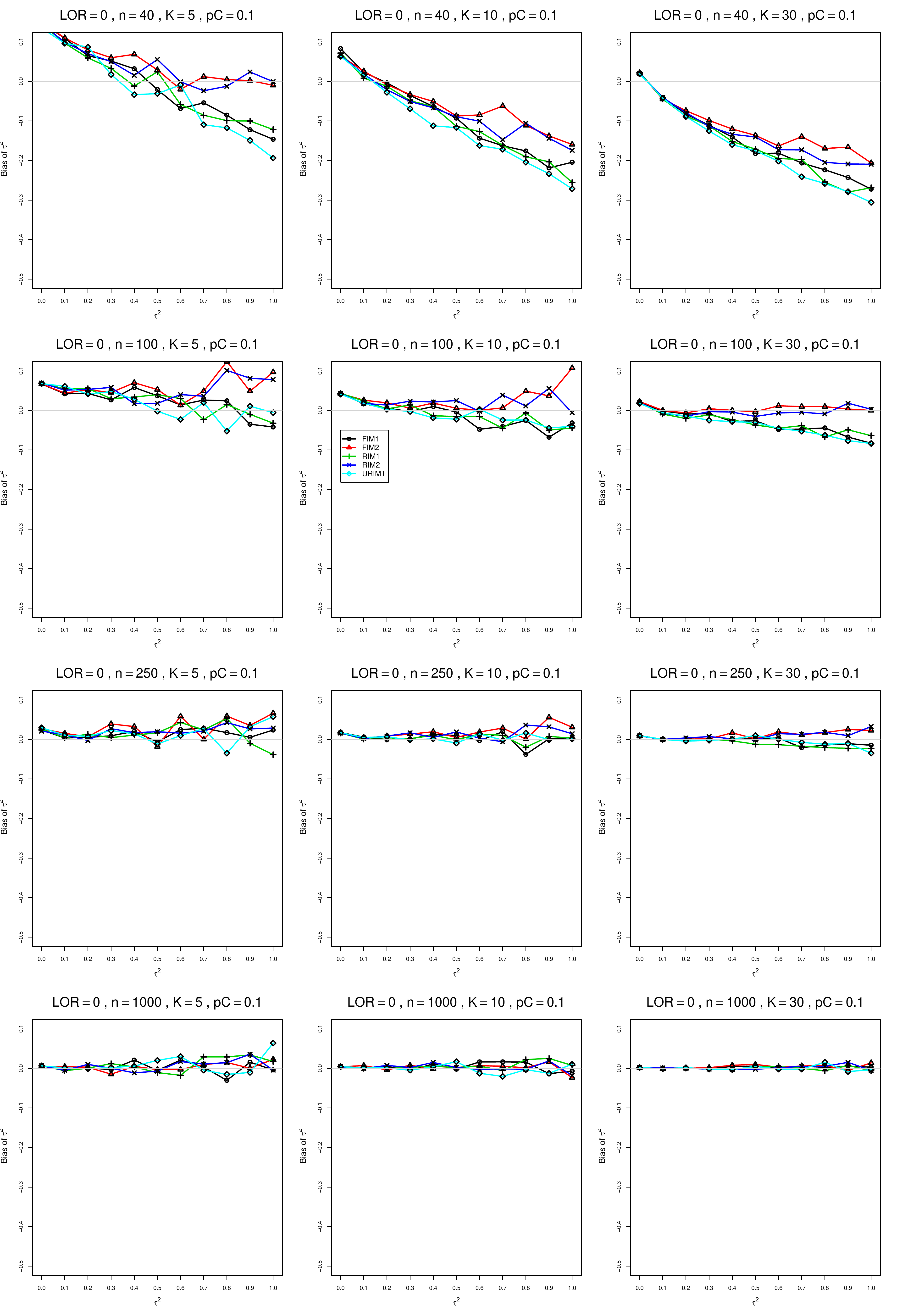}
	\caption{Bias of  between-studies variance $\hat{\tau}_{MP}^2$ for $\theta=0$, $p_{C}=0.1$, $\sigma^2=0.1$, constant sample sizes $n=40,\;100,\;250,\;1000$.
The data-generation mechanisms are FIM1 ($\circ$), FIM2 ($\triangle$), RIM1 (+), RIM2 ($\times$), and URIM1 ($\diamond$).
		\label{PlotBiasTau2mu0andpC01LOR_MPsigma01}}
\end{figure}
\begin{figure}[t]
	\centering
	\includegraphics[scale=0.33]{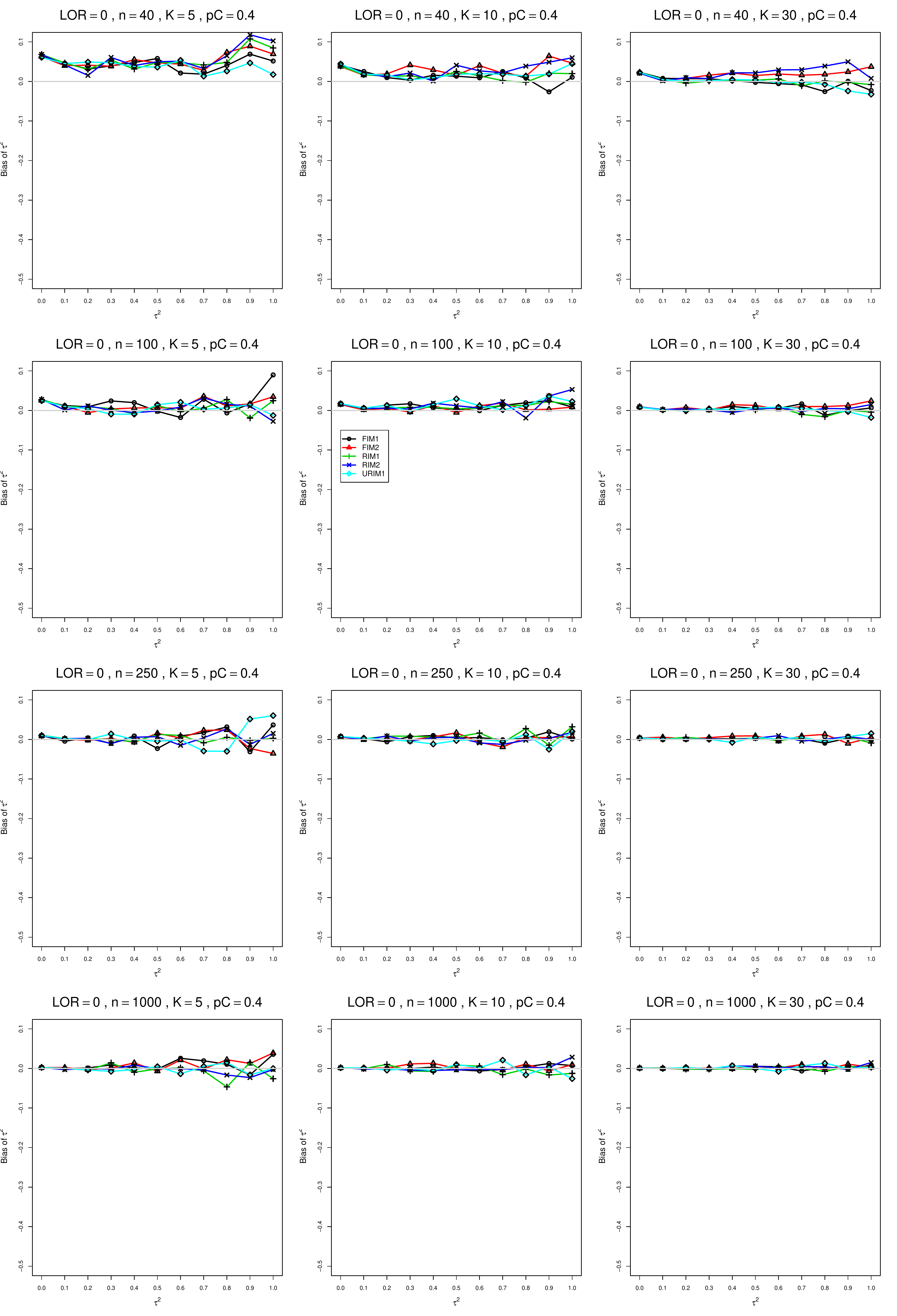}
	\caption{Bias of  between-studies variance $\hat{\tau}_{MP}^2$ for $\theta=0$, $p_{C}=0.4$, $\sigma^2=0.1$, constant sample sizes $n=40,\;100,\;250,\;1000$.
The data-generation mechanisms are FIM1 ($\circ$), FIM2 ($\triangle$), RIM1 (+), RIM2 ($\times$), and URIM1 ($\diamond$).
		\label{PlotBiasTau2mu0andpC04LOR_MPsigma01}}
\end{figure}
\begin{figure}[t]
	\centering
	\includegraphics[scale=0.33]{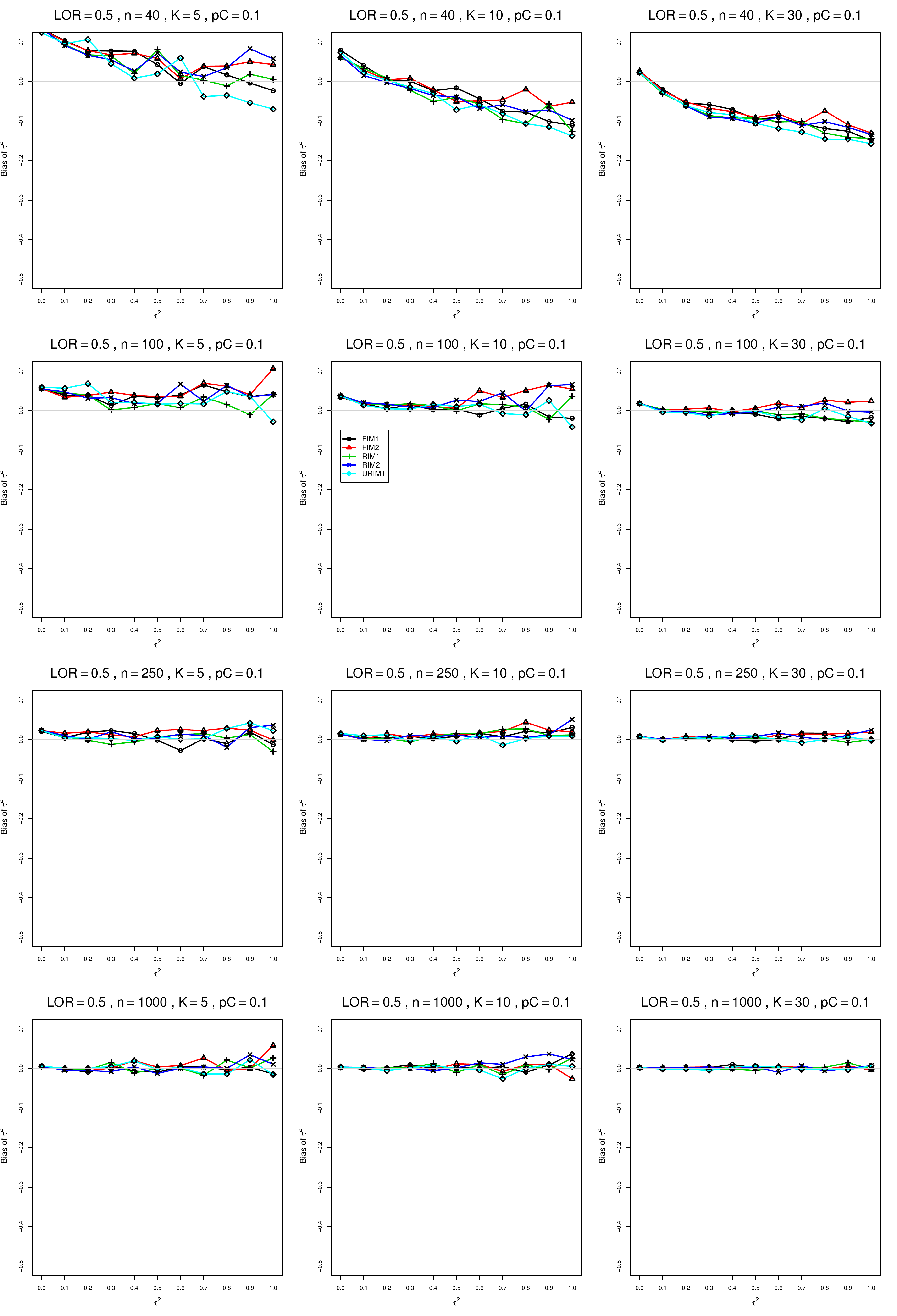}
	\caption{Bias of  between-studies variance $\hat{\tau}_{MP}^2$ for $\theta=0.5$, $p_{C}=0.1$, $\sigma^2=0.1$, constant sample sizes $n=40,\;100,\;250,\;1000$.
The data-generation mechanisms are FIM1 ($\circ$), FIM2 ($\triangle$), RIM1 (+), RIM2 ($\times$), and URIM1 ($\diamond$).
		\label{PlotBiasTau2mu05andpC01LOR_MPsigma01}}
\end{figure}
\begin{figure}[t]
	\centering
	\includegraphics[scale=0.33]{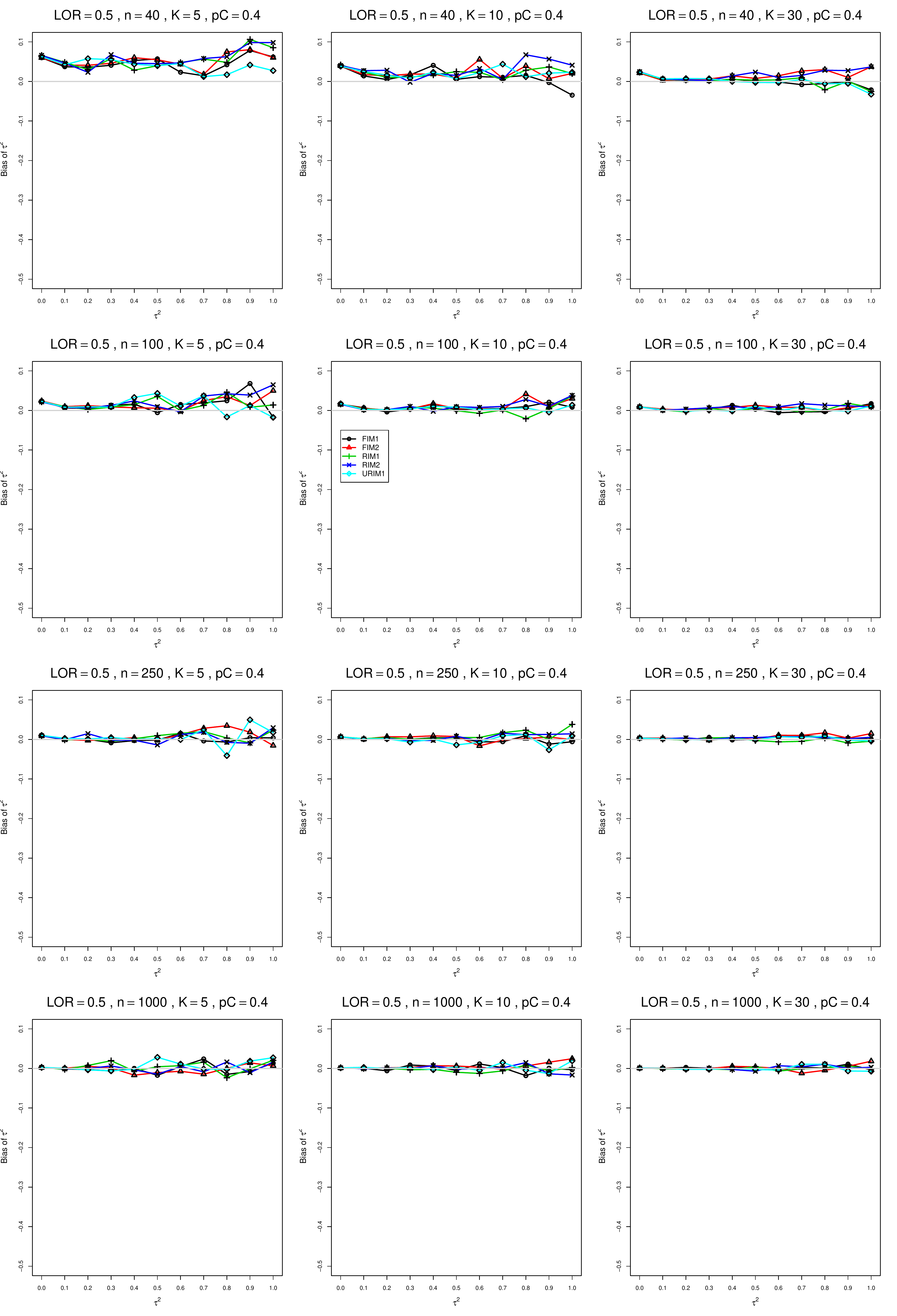}
	\caption{Bias of  between-studies variance $\hat{\tau}_{MP}^2$ for $\theta=0.5$, $p_{C}=0.4$, $\sigma^2=0.1$, constant sample sizes $n=40,\;100,\;250,\;1000$.
The data-generation mechanisms are FIM1 ($\circ$), FIM2 ($\triangle$), RIM1 (+), RIM2 ($\times$), and URIM1 ($\diamond$).
		\label{PlotBiasTau2mu05andpC04LOR_MPsigma01}}
\end{figure}
\begin{figure}[t]
	\centering
	\includegraphics[scale=0.33]{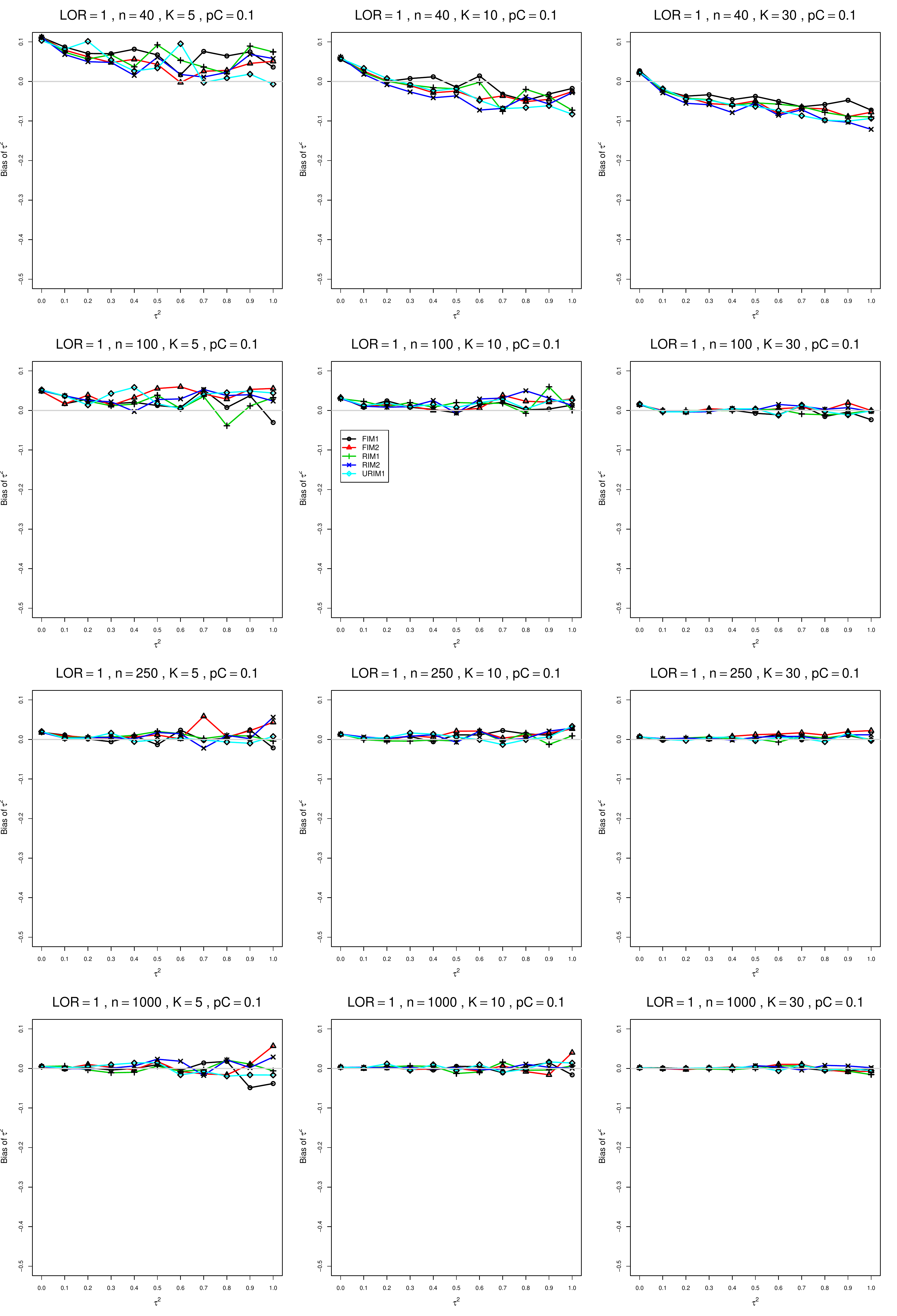}
	\caption{Bias of  between-studies variance $\hat{\tau}_{MP}^2$ for $\theta=1$, $p_{C}=0.1$, $\sigma^2=0.1$, constant sample sizes $n=40,\;100,\;250,\;1000$.
The data-generation mechanisms are FIM1 ($\circ$), FIM2 ($\triangle$), RIM1 (+), RIM2 ($\times$), and URIM1 ($\diamond$).
		\label{PlotBiasTau2mu1andpC01LOR_MPsigma01}}
\end{figure}
\begin{figure}[t]
	\centering
	\includegraphics[scale=0.33]{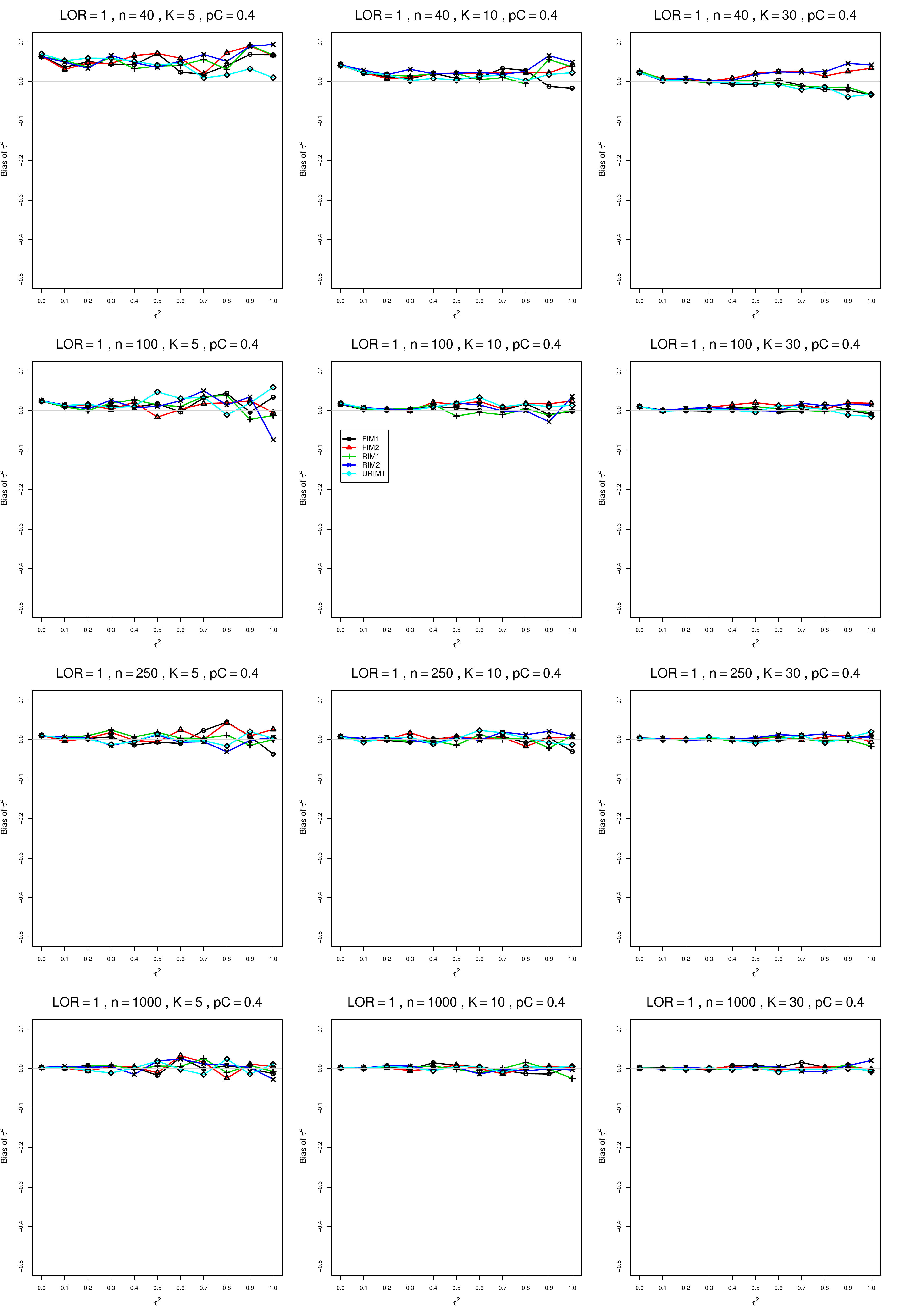}
	\caption{Bias of  between-studies variance $\hat{\tau}_{MP}^2$ for $\theta=1$, $p_{C}=0.4$, $\sigma^2=0.1$, constant sample sizes $n=40,\;100,\;250,\;1000$.
The data-generation mechanisms are FIM1 ($\circ$), FIM2 ($\triangle$), RIM1 (+), RIM2 ($\times$), and URIM1 ($\diamond$).
		\label{PlotBiasTau2mu1andpC04LOR_MPsigma01}}
\end{figure}
\begin{figure}[t]
	\centering
	\includegraphics[scale=0.33]{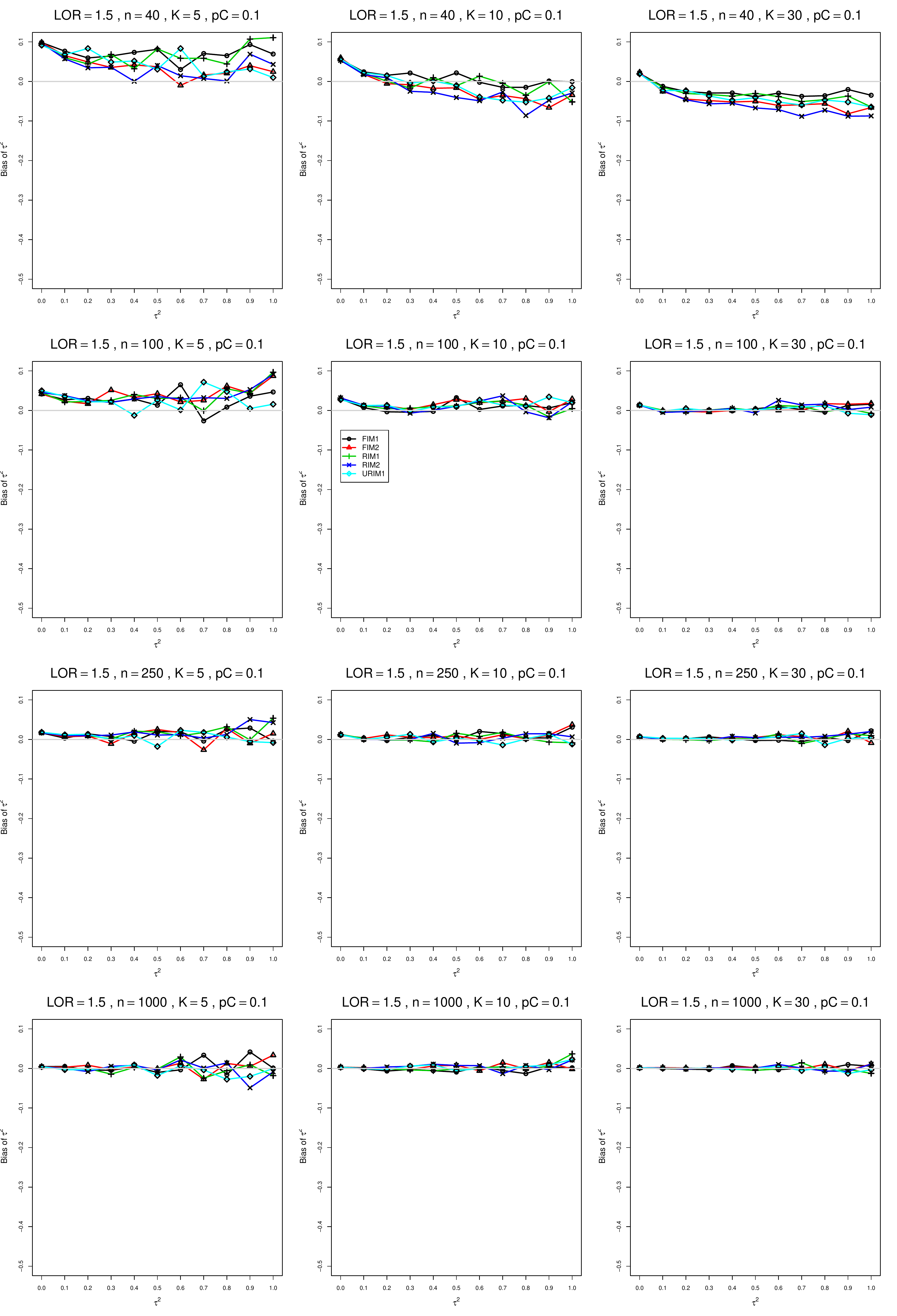}
	\caption{Bias of  between-studies variance $\hat{\tau}_{MP}^2$ for $\theta=1.5$, $p_{C}=0.1$, $\sigma^2=0.1$, constant sample sizes $n=40,\;100,\;250,\;1000$.
The data-generation mechanisms are FIM1 ($\circ$), FIM2 ($\triangle$), RIM1 (+), RIM2 ($\times$), and URIM1 ($\diamond$).
		\label{PlotBiasTau2mu15andpC01LOR_MPsigma01}}
\end{figure}
\begin{figure}[t]
	\centering
	\includegraphics[scale=0.33]{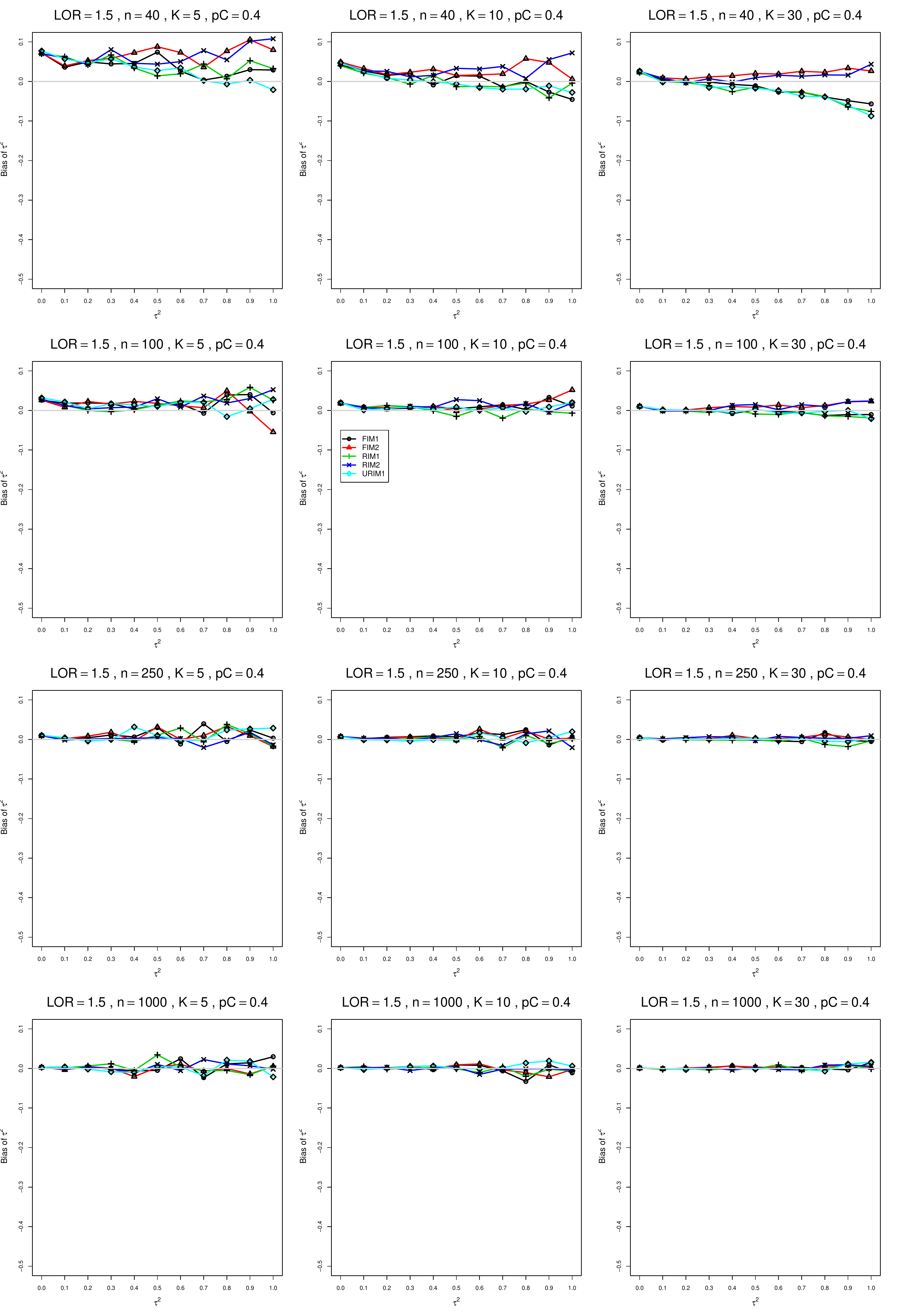}
	\caption{Bias of  between-studies variance $\hat{\tau}_{MP}^2$ for $\theta=1.5$, $p_{C}=0.4$, $\sigma^2=0.1$, constant sample sizes $n=40,\;100,\;250,\;1000$.
The data-generation mechanisms are FIM1 ($\circ$), FIM2 ($\triangle$), RIM1 (+), RIM2 ($\times$), and URIM1 ($\diamond$).
		\label{PlotBiasTau2mu15andpC04LOR_MPsigma01}}
\end{figure}
\begin{figure}[t]
	\centering
	\includegraphics[scale=0.33]{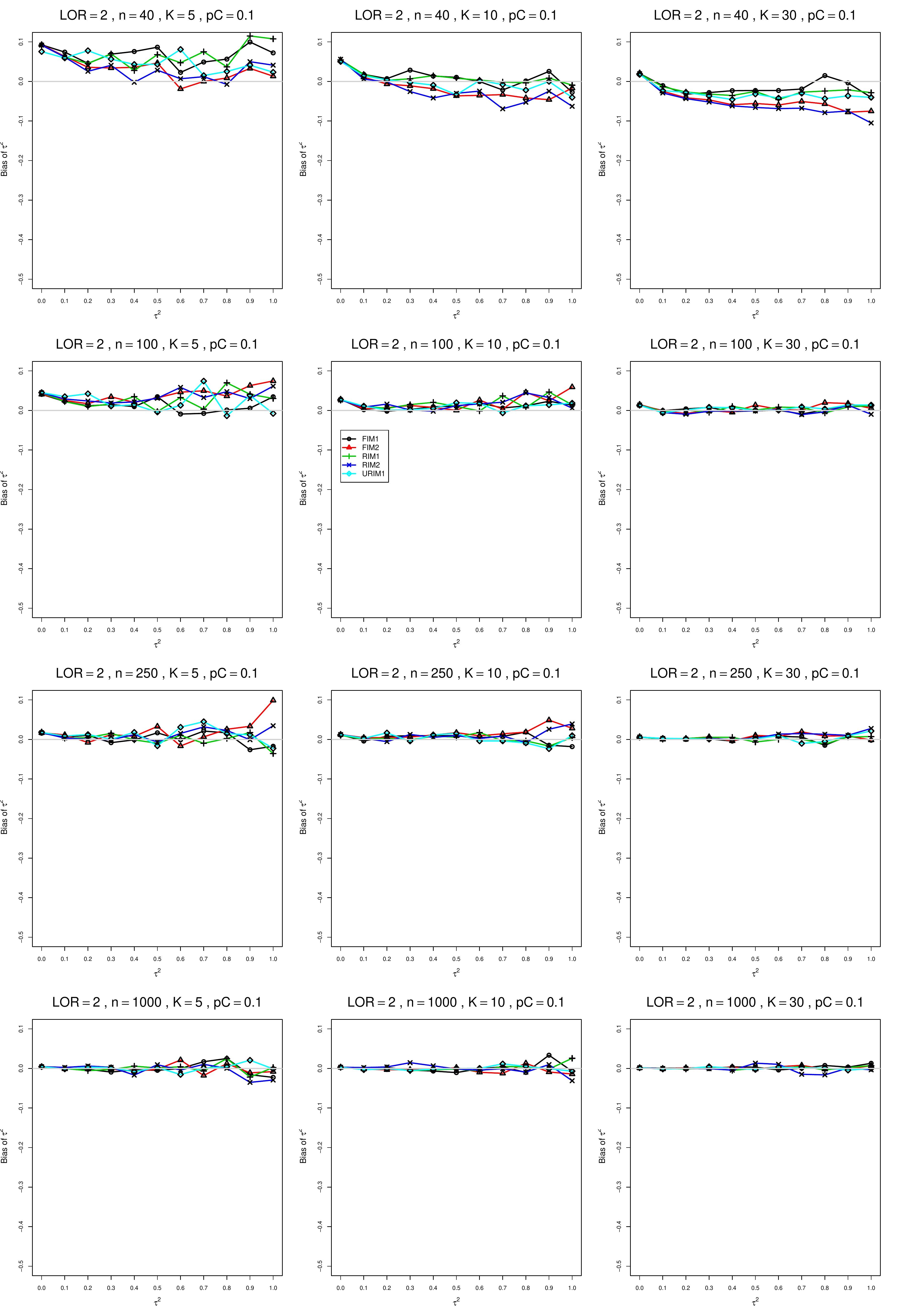}
	\caption{Bias of  between-studies variance $\hat{\tau}_{MP}^2$ for $\theta=2$, $p_{C}=0.1$, $\sigma^2=0.1$, constant sample sizes $n=40,\;100,\;250,\;1000$.
The data-generation mechanisms are FIM1 ($\circ$), FIM2 ($\triangle$), RIM1 (+), RIM2 ($\times$), and URIM1 ($\diamond$).
		\label{PlotBiasTau2mu2andpC01LOR_MPsigma01}}
\end{figure}
\begin{figure}[t]
	\centering
	\includegraphics[scale=0.33]{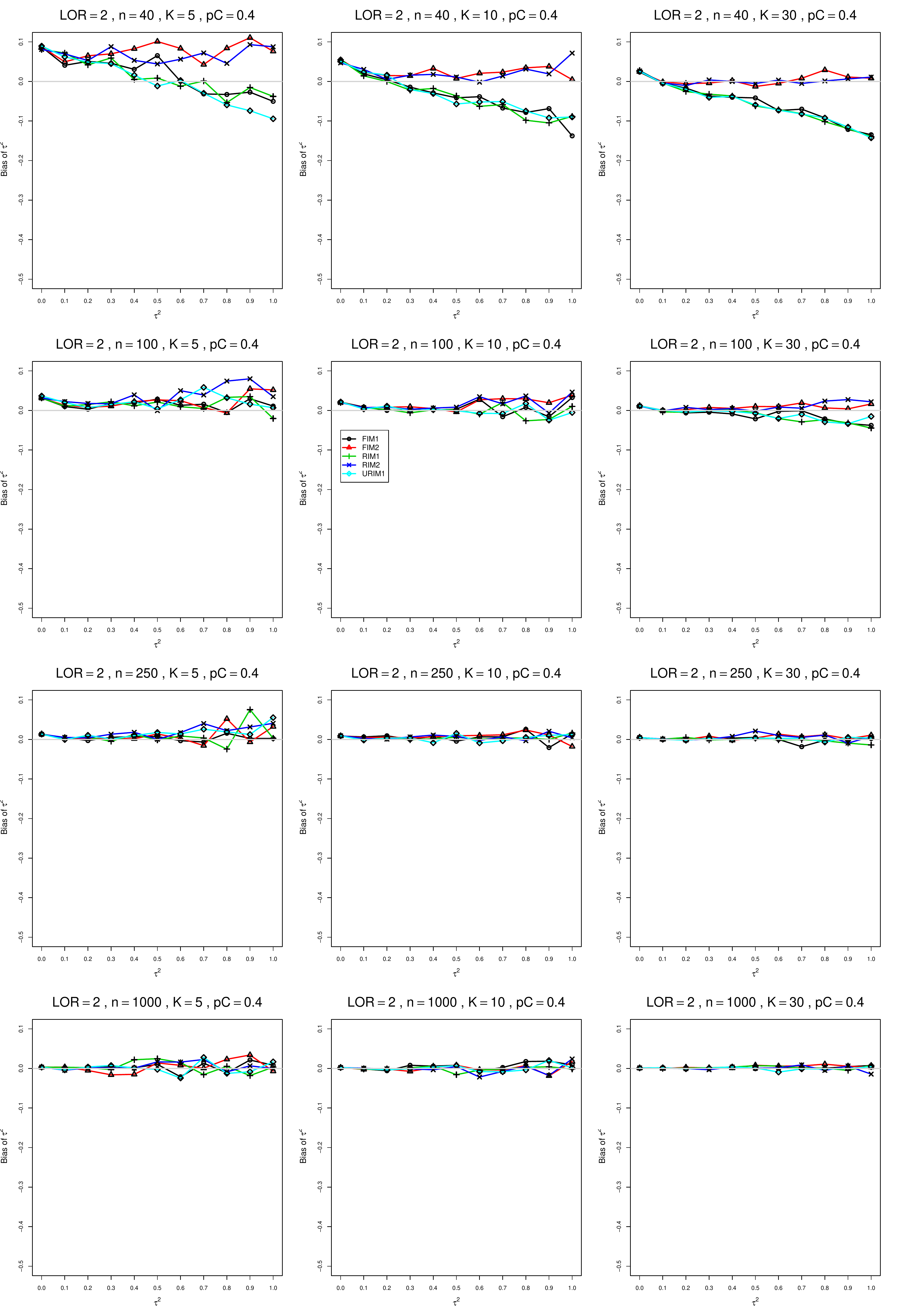}
	\caption{Bias of  between-studies variance $\hat{\tau}_{MP}^2$ for $\theta=2$, $p_{C}=0.4$, $\sigma^2=0.1$, constant sample sizes $n=40,\;100,\;250,\;1000$.
The data-generation mechanisms are FIM1 ($\circ$), FIM2 ($\triangle$), RIM1 (+), RIM2 ($\times$), and URIM1 ($\diamond$).
		\label{PlotBiasTau2mu2andpC04LOR_MPsigma01}}
\end{figure}
\begin{figure}[t]
	\centering
	\includegraphics[scale=0.33]{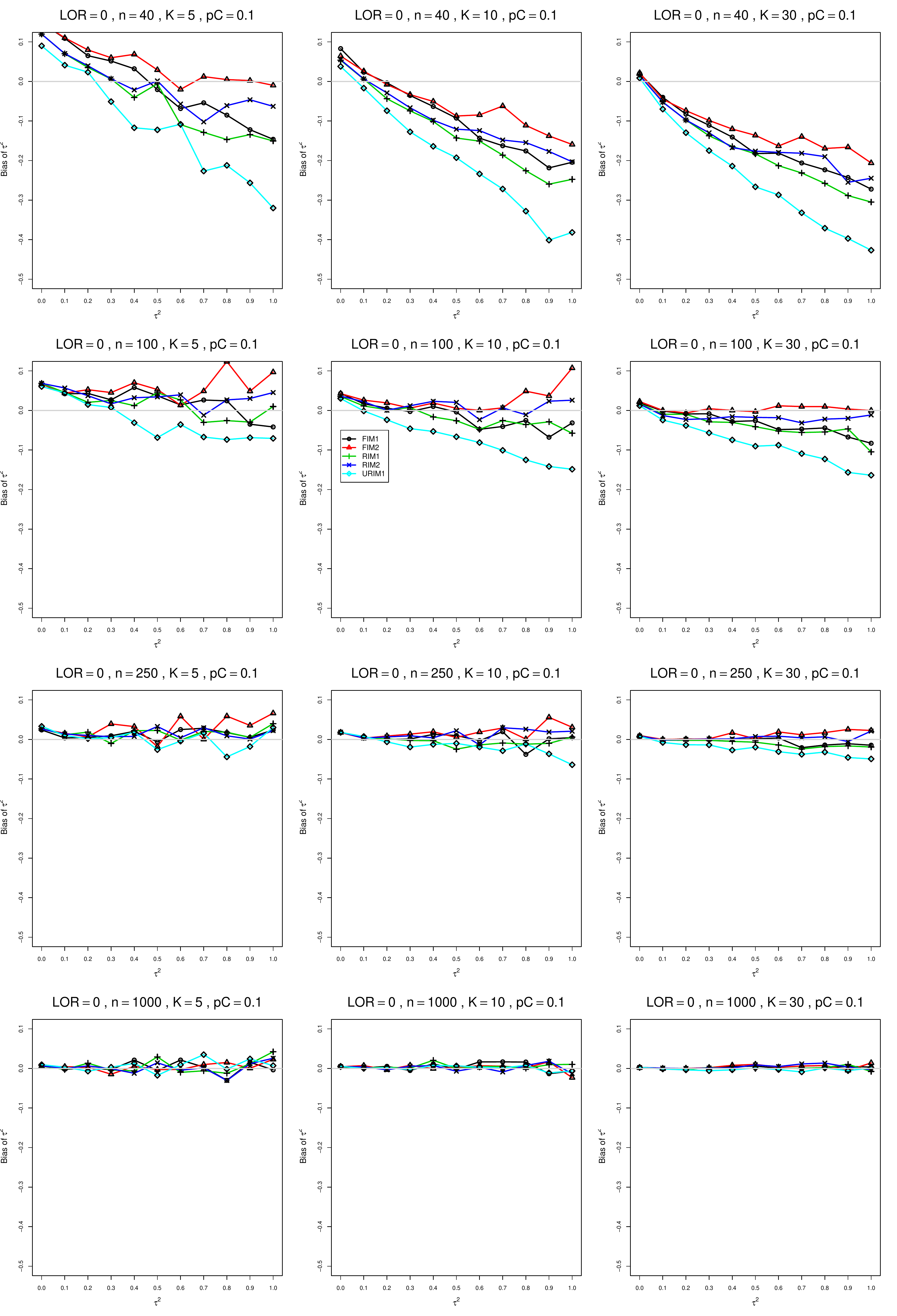}
	\caption{Bias of  between-studies variance $\hat{\tau}_{MP}^2$ for $\theta=0$, $p_{C}=0.1$, $\sigma^2=0.4$, constant sample sizes $n=40,\;100,\;250,\;1000$.
The data-generation mechanisms are FIM1 ($\circ$), FIM2 ($\triangle$), RIM1 (+), RIM2 ($\times$), and URIM1 ($\diamond$).
		\label{PlotBiasTau2mu0andpC01LOR_MPsigma04}}
\end{figure}
\begin{figure}[t]
	\centering
	\includegraphics[scale=0.33]{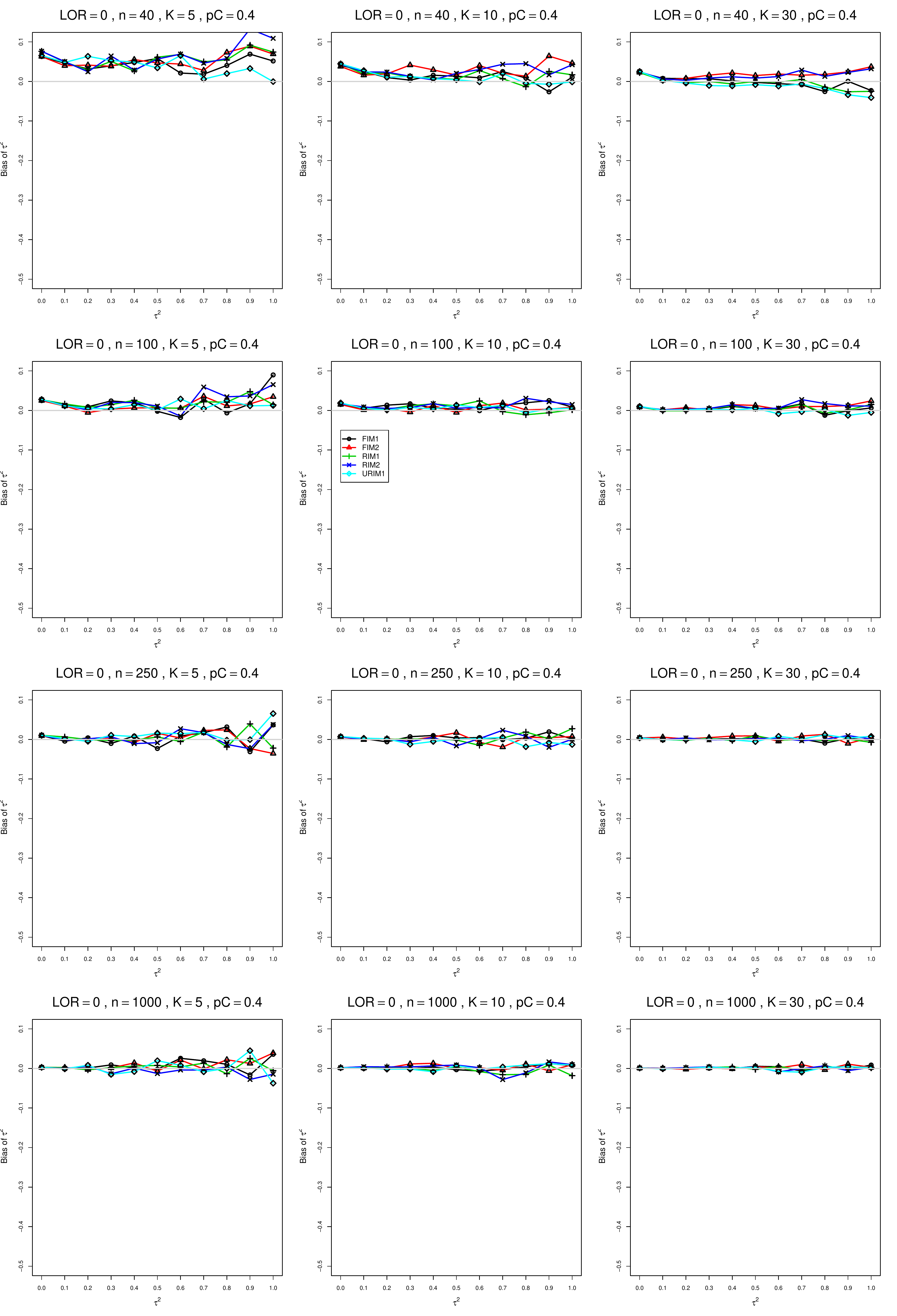}
	\caption{Bias of  between-studies variance $\hat{\tau}_{MP}^2$ for $\theta=0$, $p_{C}=0.4$, $\sigma^2=0.4$, constant sample sizes $n=40,\;100,\;250,\;1000$.
The data-generation mechanisms are FIM1 ($\circ$), FIM2 ($\triangle$), RIM1 (+), RIM2 ($\times$), and URIM1 ($\diamond$).
		\label{PlotBiasTau2mu0andpC04LOR_MPsigma04}}
\end{figure}
\begin{figure}[t]
	\centering
	\includegraphics[scale=0.33]{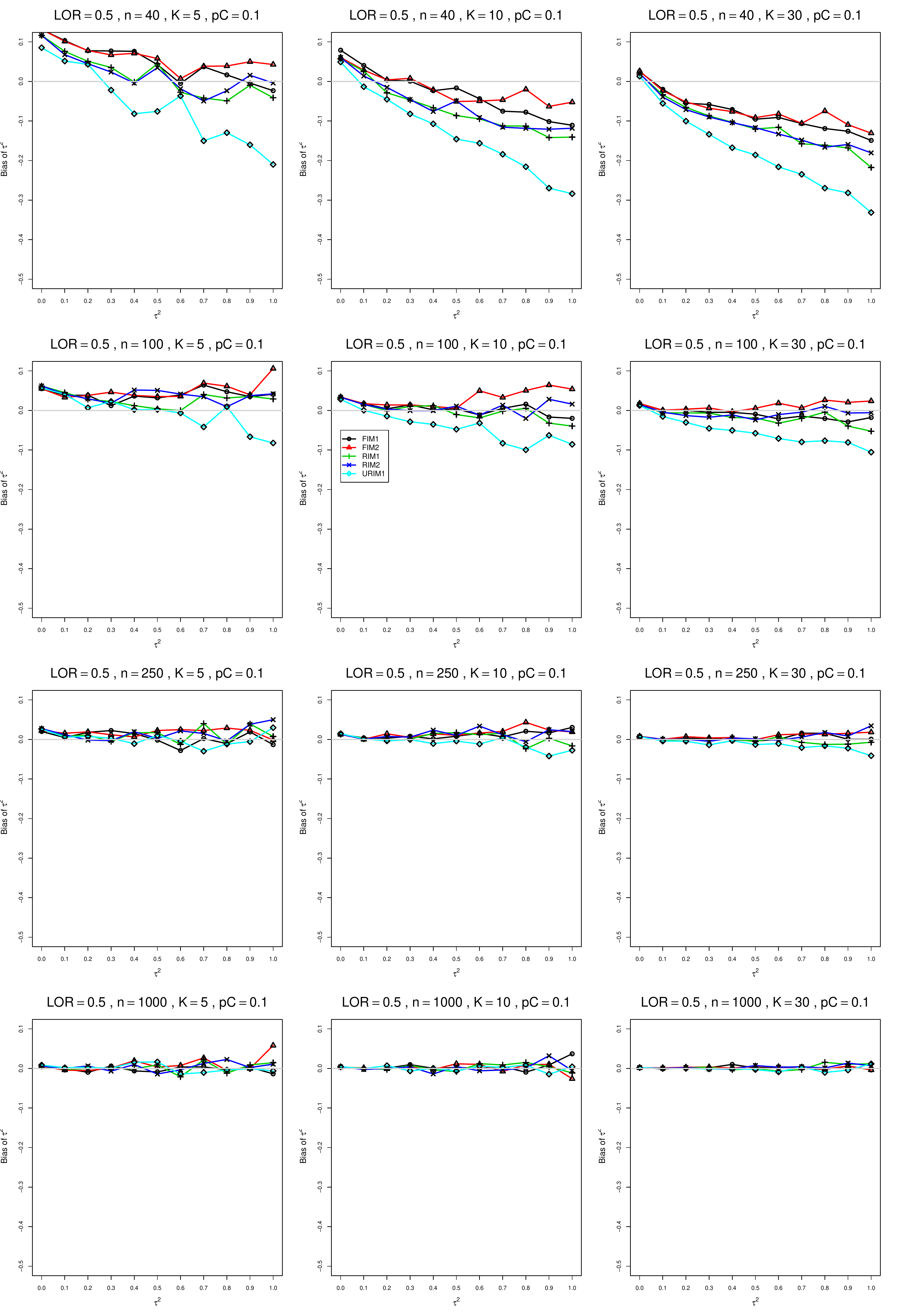}
	\caption{Bias of  between-studies variance $\hat{\tau}_{MP}^2$ for $\theta=0.5$, $p_{C}=0.1$, $\sigma^2=0.4$, constant sample sizes $n=40,\;100,\;250,\;1000$.
The data-generation mechanisms are FIM1 ($\circ$), FIM2 ($\triangle$), RIM1 (+), RIM2 ($\times$), and URIM1 ($\diamond$).
		\label{PlotBiasTau2mu05andpC01LOR_MPsigma04}}
\end{figure}
\begin{figure}[t]
	\centering
	\includegraphics[scale=0.33]{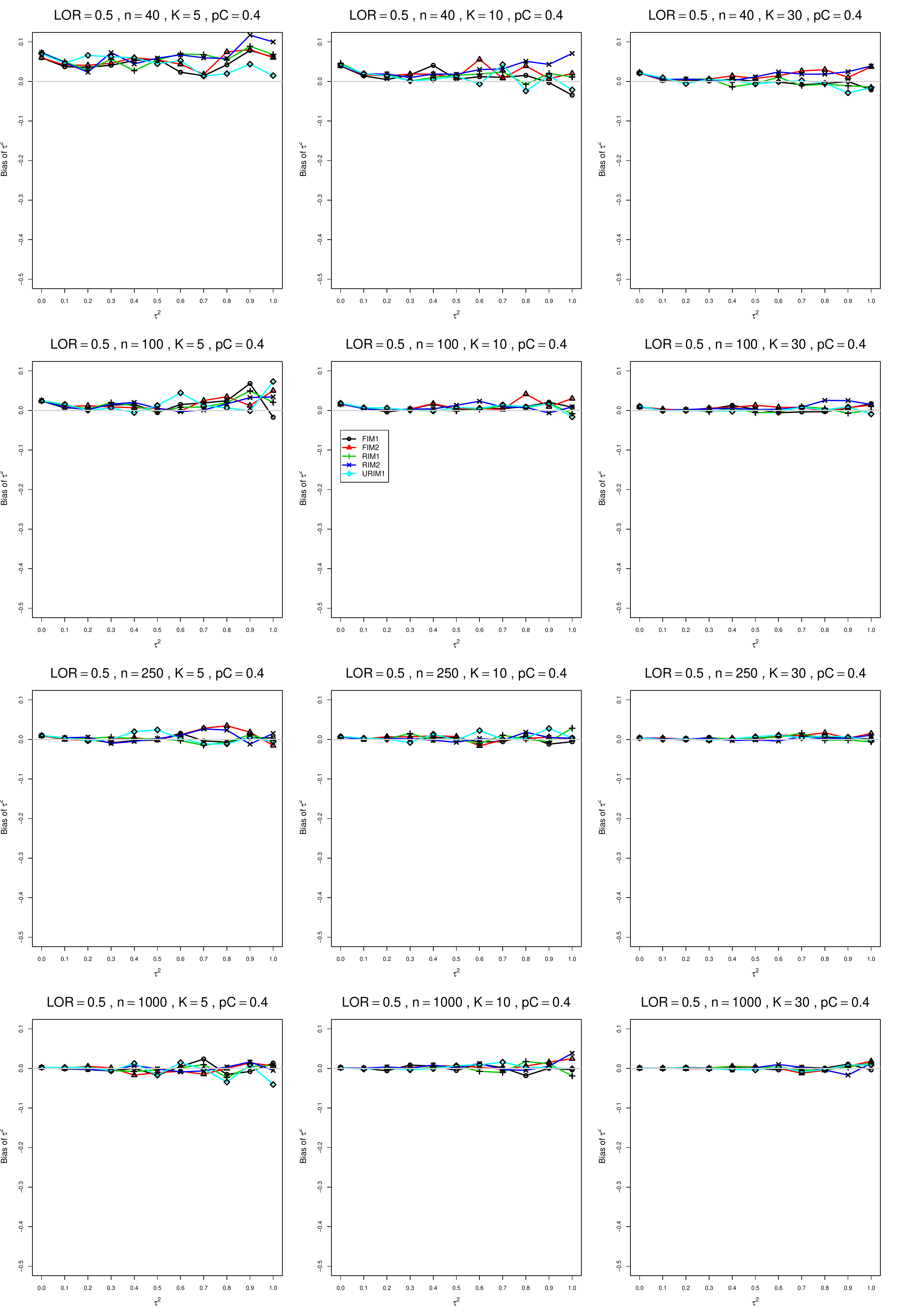}
	\caption{Bias of  between-studies variance $\hat{\tau}_{MP}^2$ for $\theta=0.5$, $p_{C}=0.4$, $\sigma^2=0.4$, constant sample sizes $n=40,\;100,\;250,\;1000$.
The data-generation mechanisms are FIM1 ($\circ$), FIM2 ($\triangle$), RIM1 (+), RIM2 ($\times$), and URIM1 ($\diamond$).
		\label{PlotBiasTau2mu05andpC04LOR_MPsigma04}}
\end{figure}
\begin{figure}[t]
	\centering
	\includegraphics[scale=0.33]{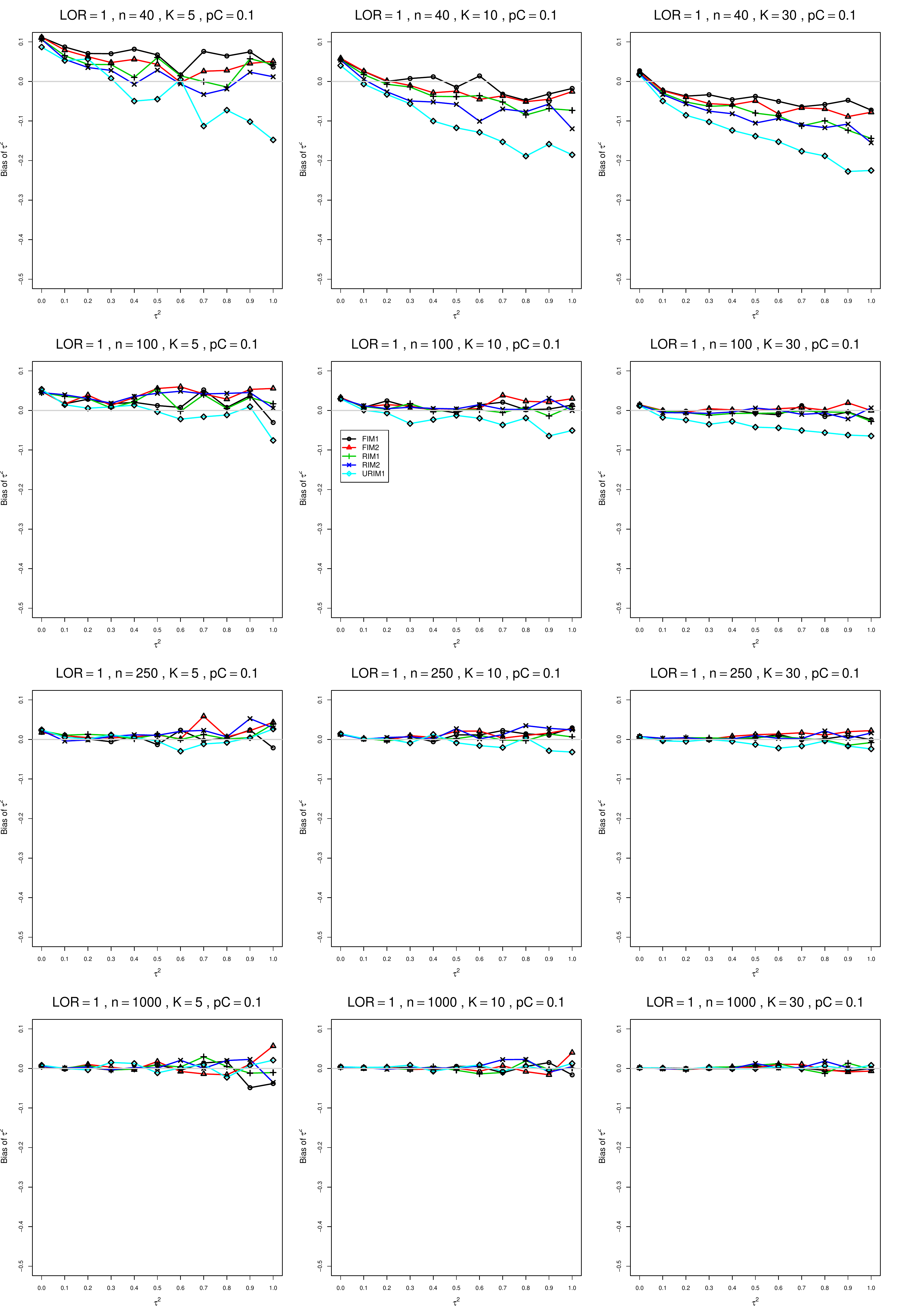}
	\caption{Bias of  between-studies variance $\hat{\tau}_{MP}^2$ for $\theta=1$, $p_{C}=0.1$, $\sigma^2=0.4$, constant sample sizes $n=40,\;100,\;250,\;1000$.
The data-generation mechanisms are FIM1 ($\circ$), FIM2 ($\triangle$), RIM1 (+), RIM2 ($\times$), and URIM1 ($\diamond$).
		\label{PlotBiasTau2mu1andpC01LOR_MPsigma04}}
\end{figure}
\begin{figure}[t]
	\centering
	\includegraphics[scale=0.33]{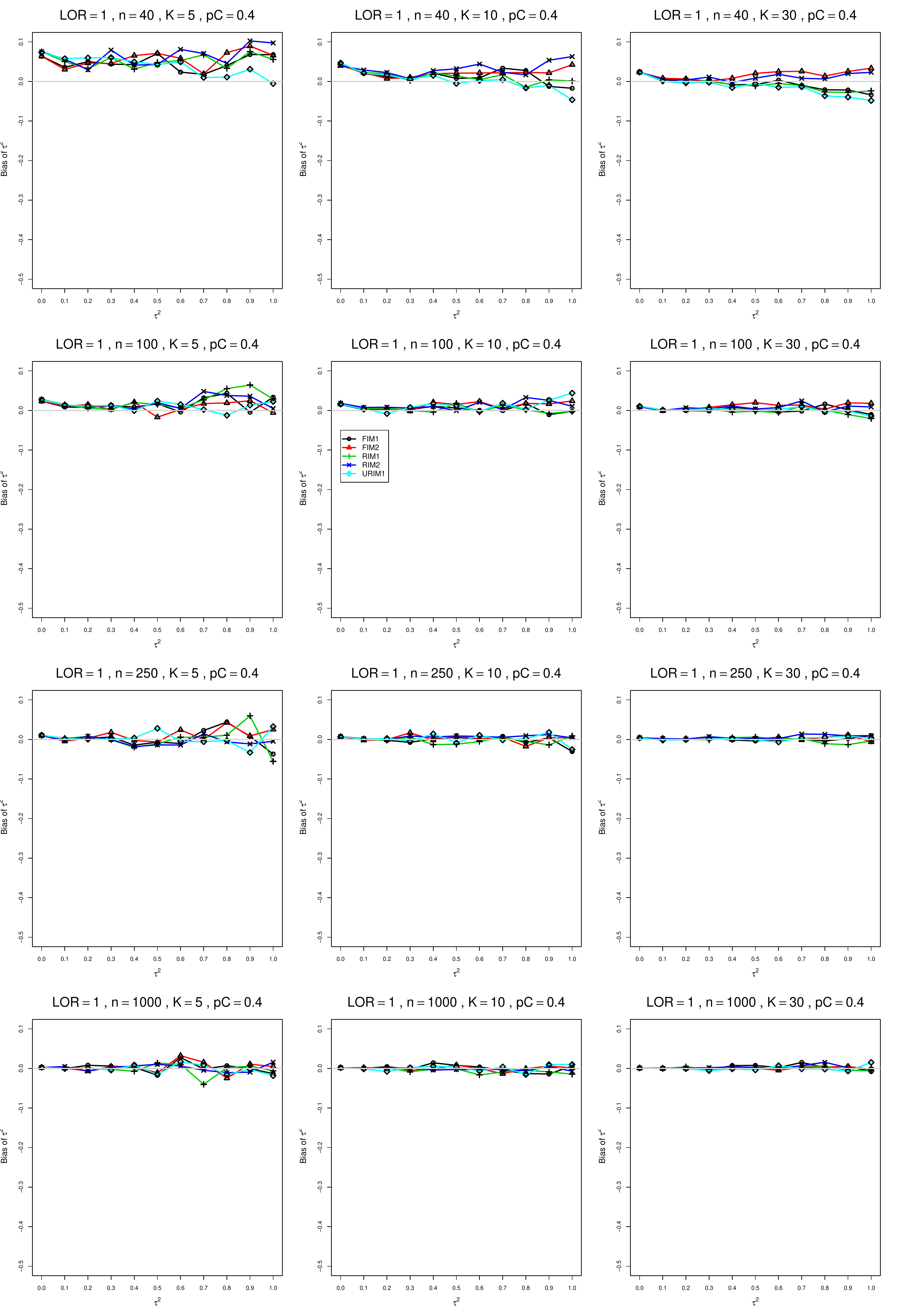}
	\caption{Bias of  between-studies variance $\hat{\tau}_{MP}^2$ for $\theta=1$, $p_{C}=0.4$, $\sigma^2=0.4$, constant sample sizes $n=40,\;100,\;250,\;1000$.
The data-generation mechanisms are FIM1 ($\circ$), FIM2 ($\triangle$), RIM1 (+), RIM2 ($\times$), and URIM1 ($\diamond$).
		\label{PlotBiasTau2mu1andpC04LOR_MPsigma04}}
\end{figure}
\begin{figure}[t]
	\centering
	\includegraphics[scale=0.33]{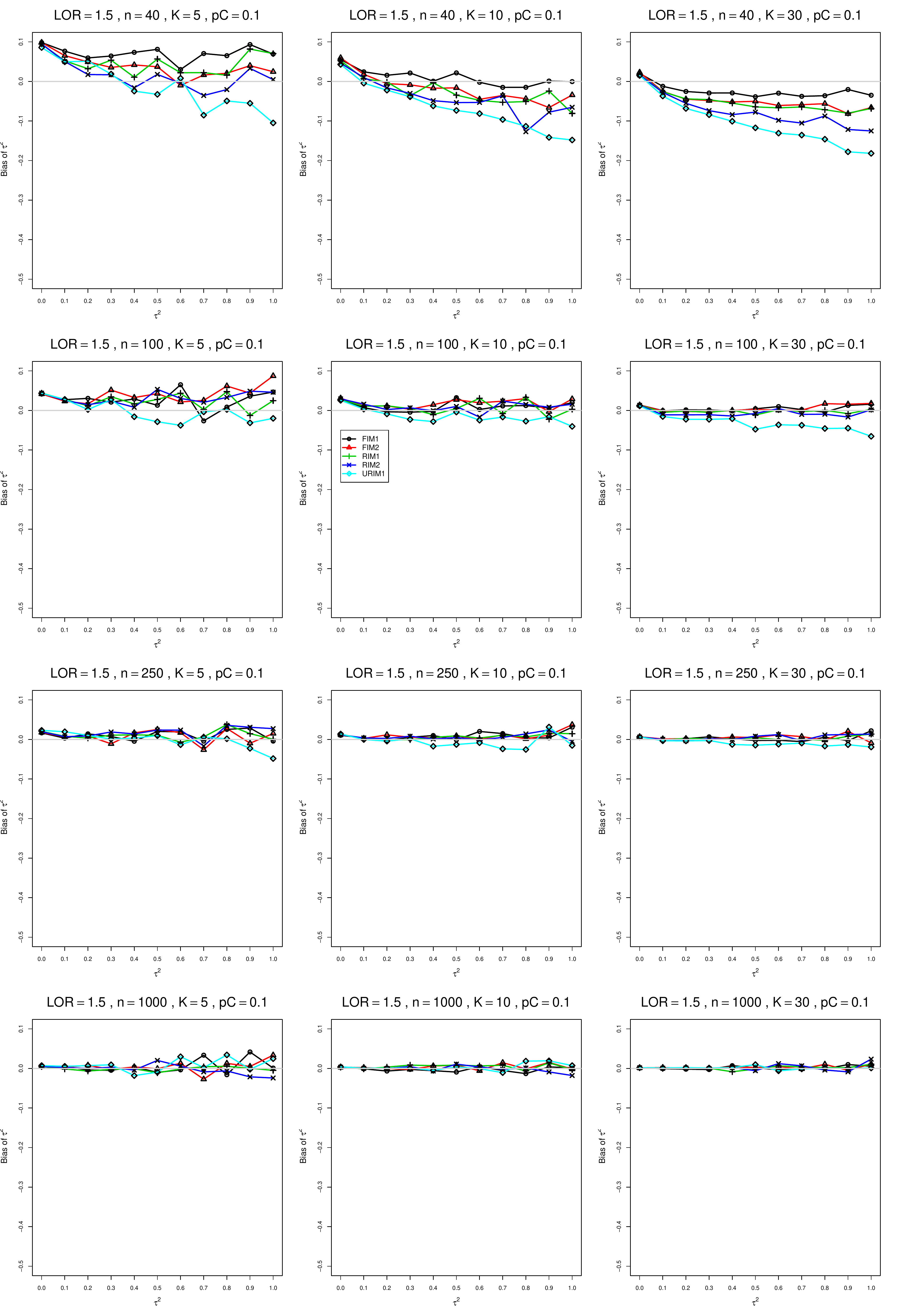}
	\caption{Bias of  between-studies variance $\hat{\tau}_{MP}^2$ for $\theta=1.5$, $p_{C}=0.1$, $\sigma^2=0.4$, constant sample sizes $n=40,\;100,\;250,\;1000$.
The data-generation mechanisms are FIM1 ($\circ$), FIM2 ($\triangle$), RIM1 (+), RIM2 ($\times$), and URIM1 ($\diamond$).
		\label{PlotBiasTau2mu15andpC01LOR_MPsigma04}}
\end{figure}
\begin{figure}[t]
	\centering
	\includegraphics[scale=0.33]{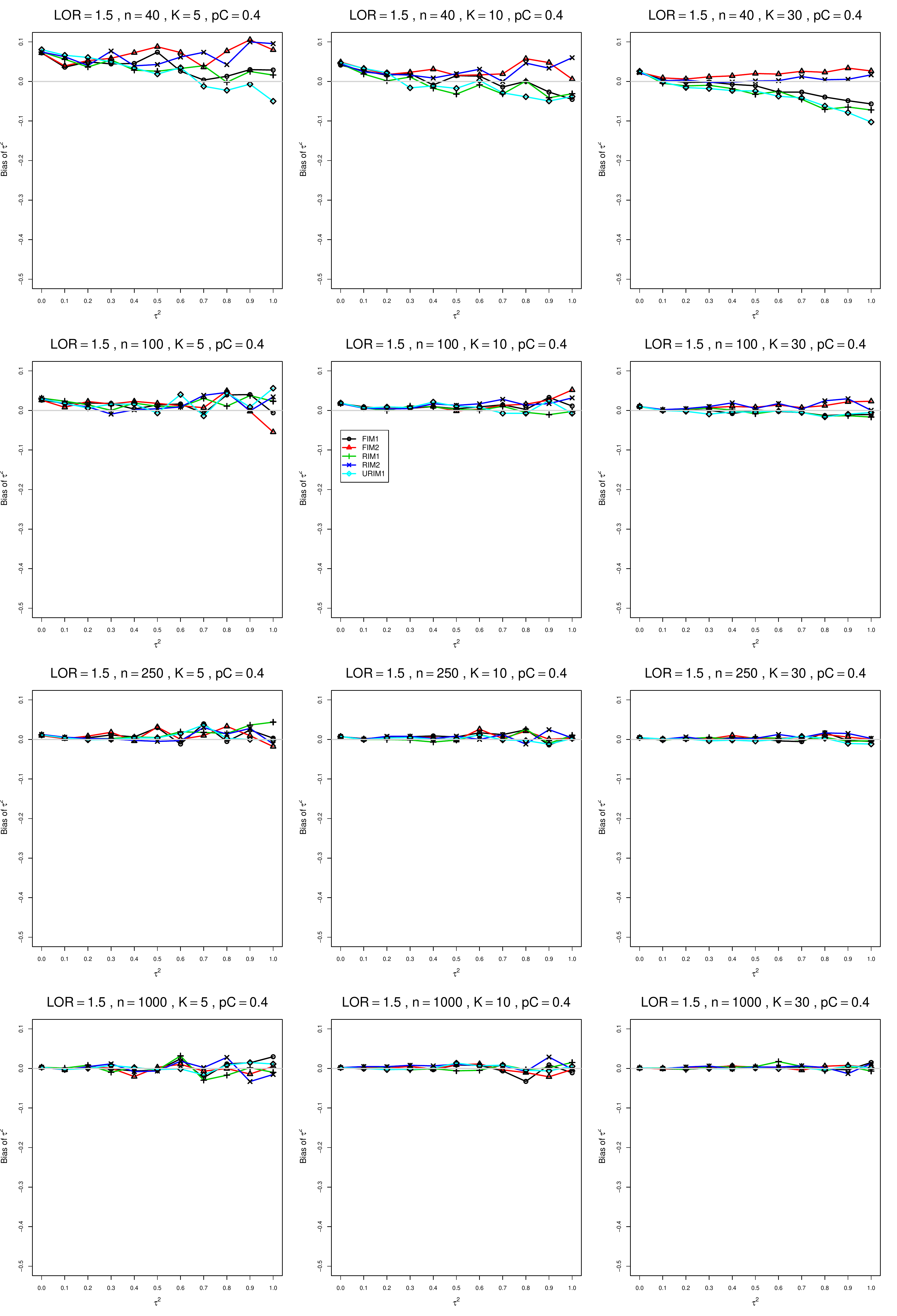}
	\caption{Bias of  between-studies variance $\hat{\tau}_{MP}^2$ for $\theta=1.5$, $p_{C}=0.4$, $\sigma^2=0.4$, constant sample sizes $n=40,\;100,\;250,\;1000$.
The data-generation mechanisms are FIM1 ($\circ$), FIM2 ($\triangle$), RIM1 (+), RIM2 ($\times$), and URIM1 ($\diamond$).
		\label{PlotBiasTau2mu15andpC04LOR_MPsigma04}}
\end{figure}
\begin{figure}[t]
	\centering
	\includegraphics[scale=0.33]{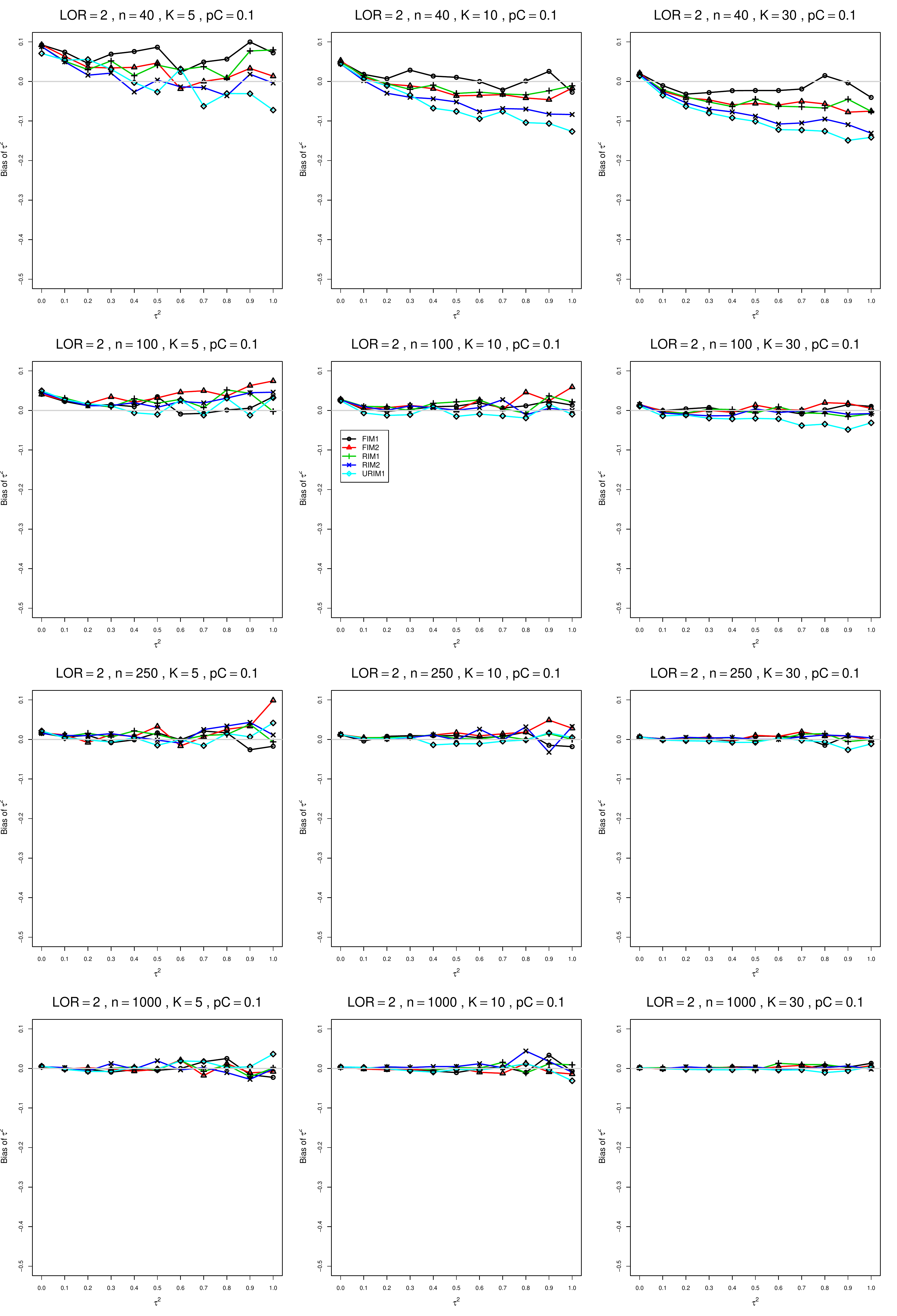}
	\caption{Bias of  between-studies variance $\hat{\tau}_{MP}^2$ for $\theta=2$, $p_{C}=0.1$, $\sigma^2=0.4$, constant sample sizes $n=40,\;100,\;250,\;1000$.
The data-generation mechanisms are FIM1 ($\circ$), FIM2 ($\triangle$), RIM1 (+), RIM2 ($\times$), and URIM1 ($\diamond$).
		\label{PlotBiasTau2mu2andpC01LOR_MPsigma04}}
\end{figure}
\begin{figure}[t]
	\centering
	\includegraphics[scale=0.33]{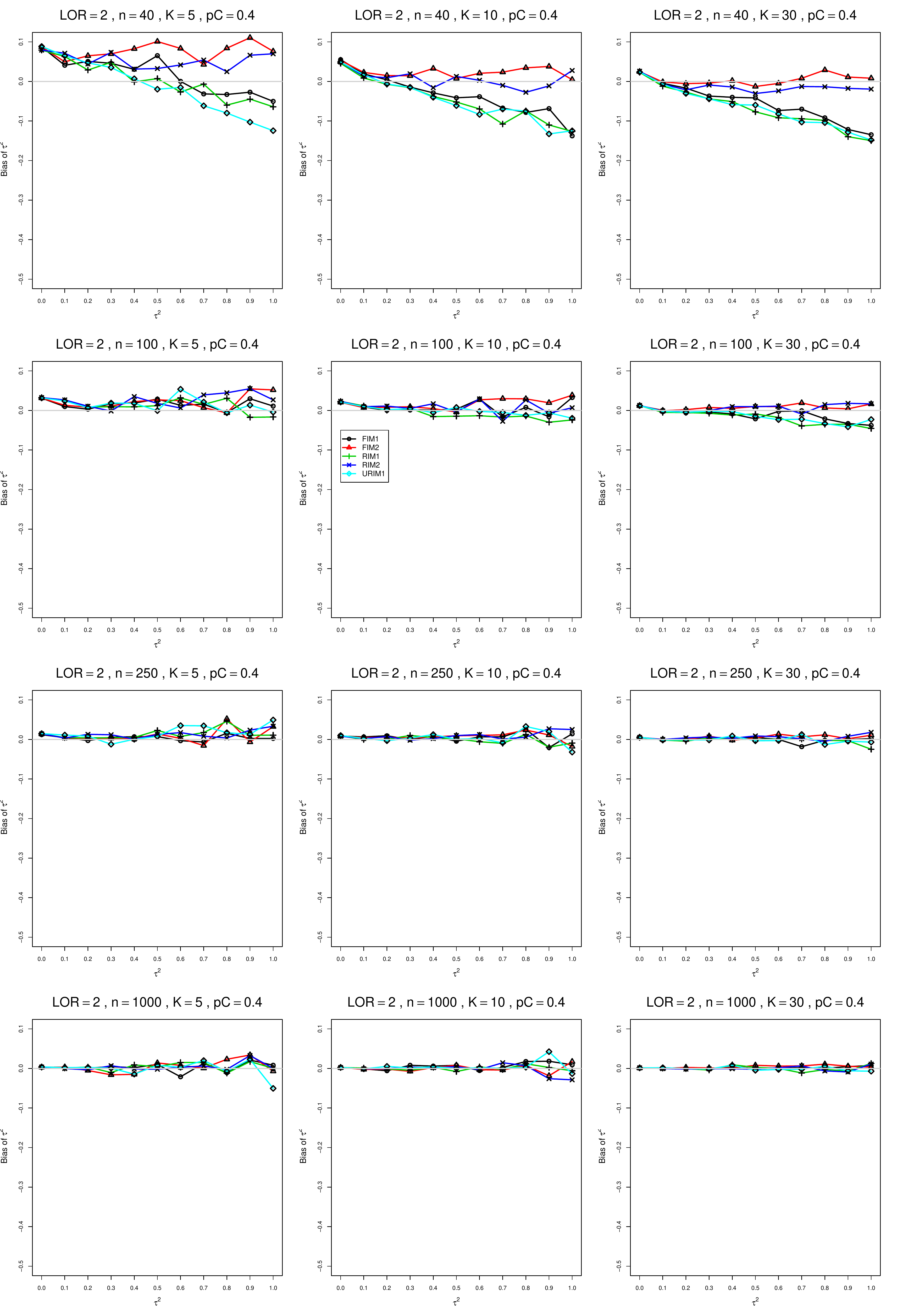}
	\caption{Bias of  between-studies variance $\hat{\tau}_{MP}^2$ for $\theta=2$, $p_{C}=0.4$, $\sigma^2=0.4$, constant sample sizes $n=40,\;100,\;250,\;1000$.
The data-generation mechanisms are FIM1 ($\circ$), FIM2 ($\triangle$), RIM1 (+), RIM2 ($\times$), and URIM1 ($\diamond$).
		\label{PlotBiasTau2mu2andpC04LOR_MPsigma04}}
\end{figure}

\clearpage
\subsection*{A1.4 Bias of $\hat{\tau}_{KD}^2$}
\renewcommand{\thefigure}{A1.4.\arabic{figure}}
\setcounter{figure}{0}

\begin{figure}[t]
	\centering
	\includegraphics[scale=0.33]{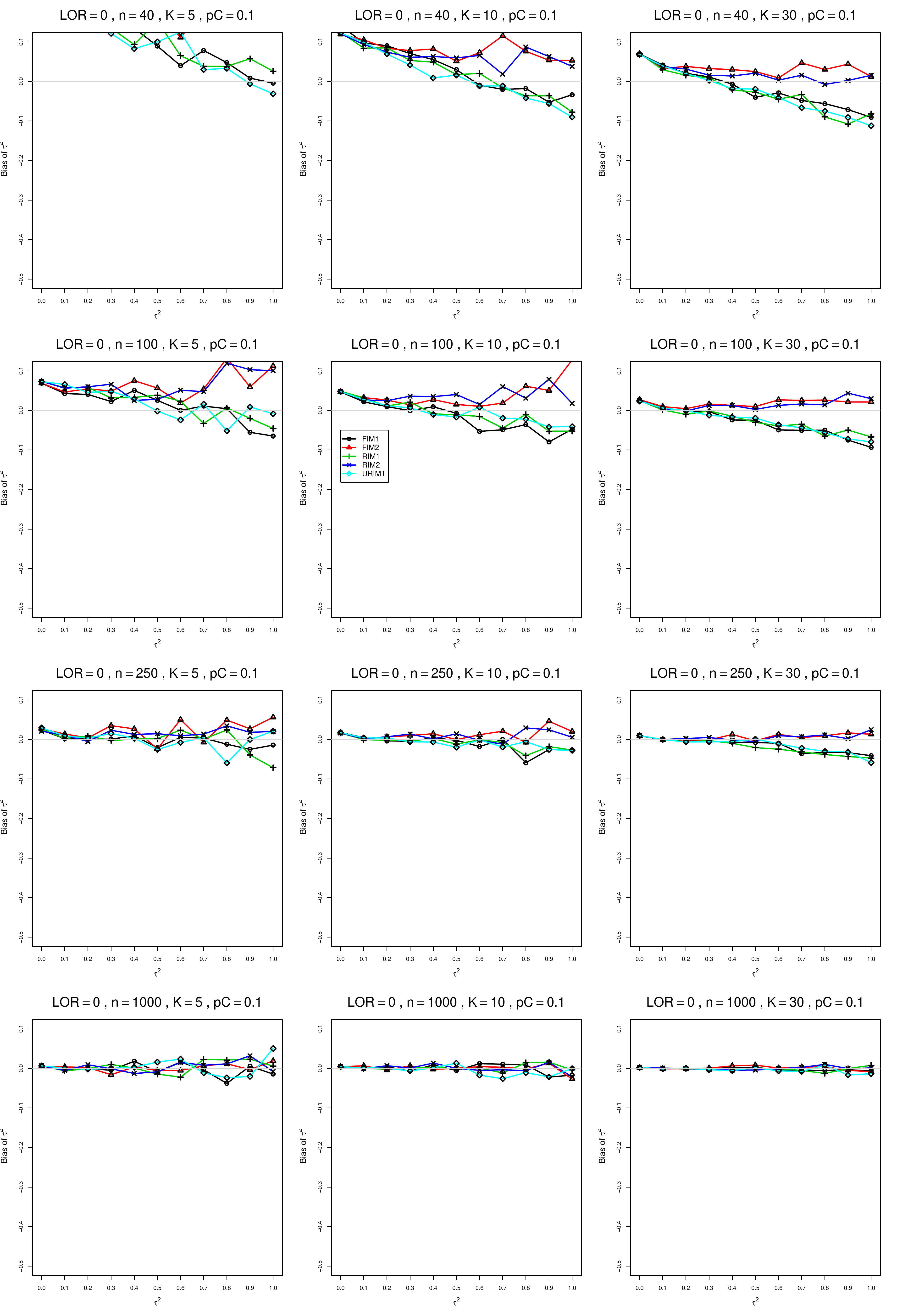}
	\caption{Bias of  between-studies variance $\hat{\tau}_{KD}^2$ for $\theta=0$, $p_{C}=0.1$, $\sigma^2=0.1$, constant sample sizes $n=40,\;100,\;250,\;1000$.
The data-generation mechanisms are FIM1 ($\circ$), FIM2 ($\triangle$), RIM1 (+), RIM2 ($\times$), and URIM1 ($\diamond$).
		\label{PlotBiasTau2mu0andpC01LOR_KDsigma01}}
\end{figure}
\begin{figure}[t]
	\centering
	\includegraphics[scale=0.33]{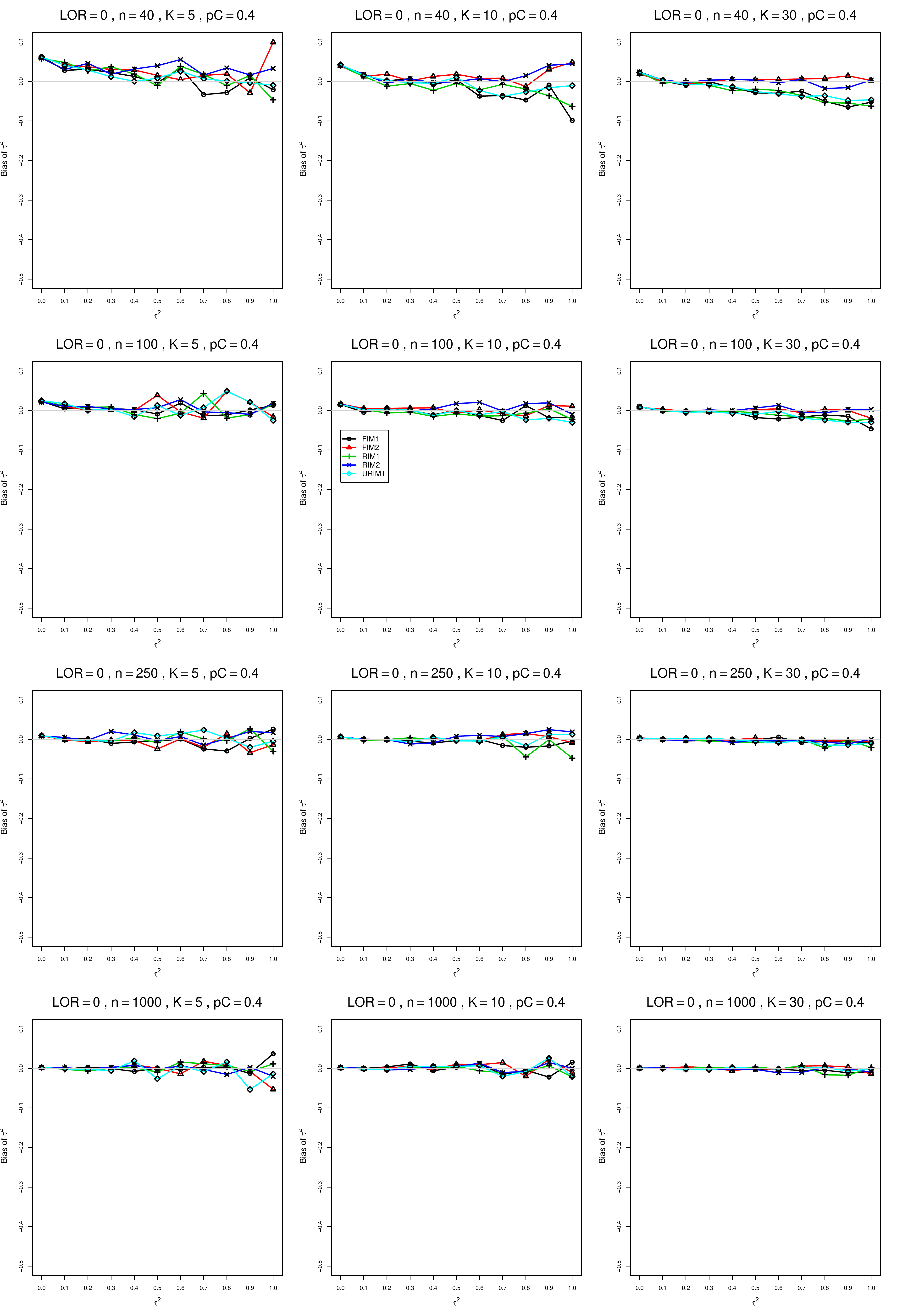}
	\caption{Bias of  between-studies variance $\hat{\tau}_{KD}^2$ for $\theta=0$, $p_{C}=0.4$, $\sigma^2=0.1$, constant sample sizes $n=40,\;100,\;250,\;1000$.
The data-generation mechanisms are FIM1 ($\circ$), FIM2 ($\triangle$), RIM1 (+), RIM2 ($\times$), and URIM1 ($\diamond$).
		\label{PlotBiasTau2mu0andpC04LOR_KDsigma01}}
\end{figure}
\begin{figure}[t]
	\centering
	\includegraphics[scale=0.33]{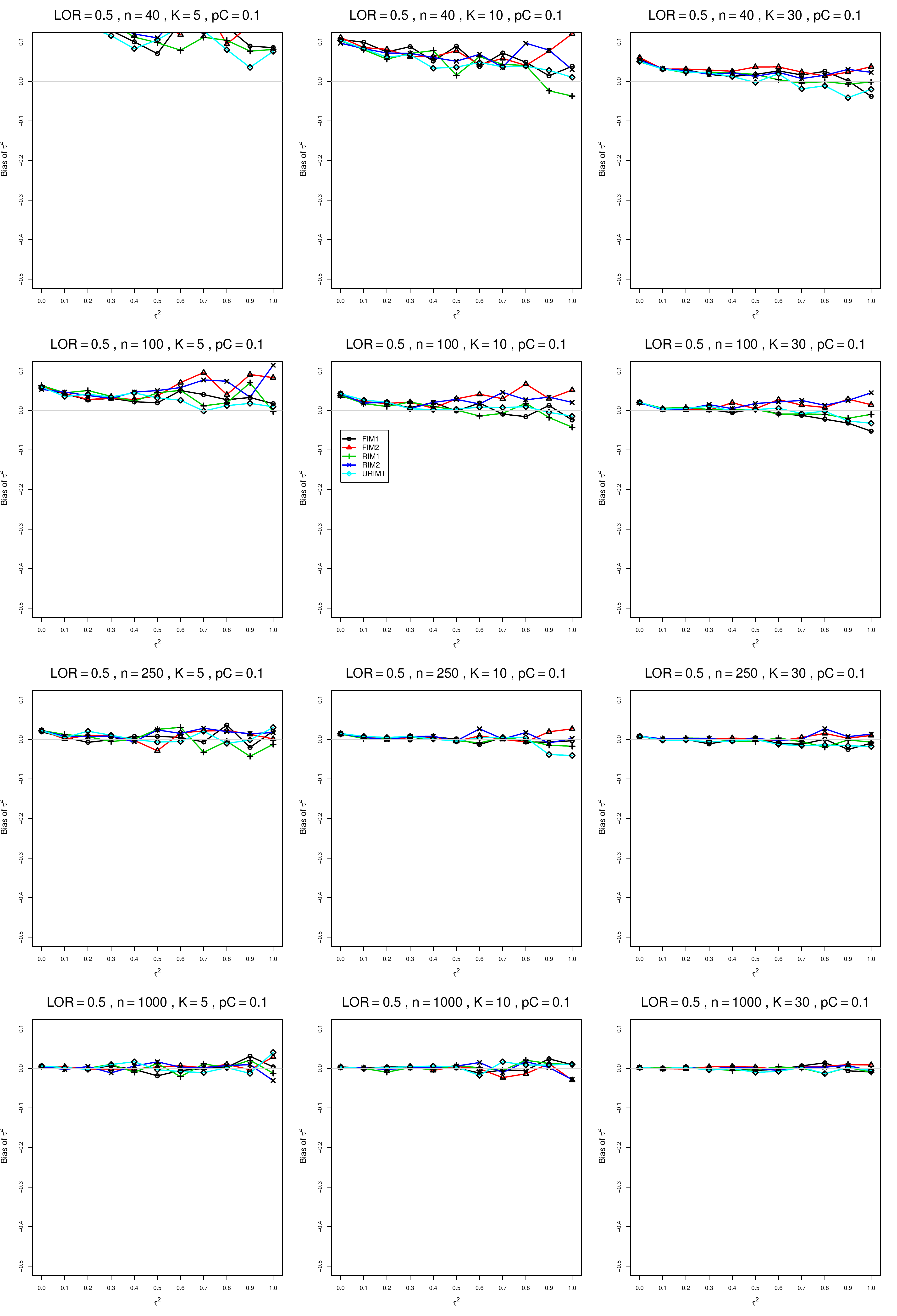}
	\caption{Bias of  between-studies variance $\hat{\tau}_{KD}^2$ for $\theta=0.5$, $p_{C}=0.1$, $\sigma^2=0.1$, constant sample sizes $n=40,\;100,\;250,\;1000$.
The data-generation mechanisms are FIM1 ($\circ$), FIM2 ($\triangle$), RIM1 (+), RIM2 ($\times$), and URIM1 ($\diamond$).
		\label{PlotBiasTau2mu05andpC01LOR_KDsigma01}}
\end{figure}
\begin{figure}[t]
	\centering
	\includegraphics[scale=0.33]{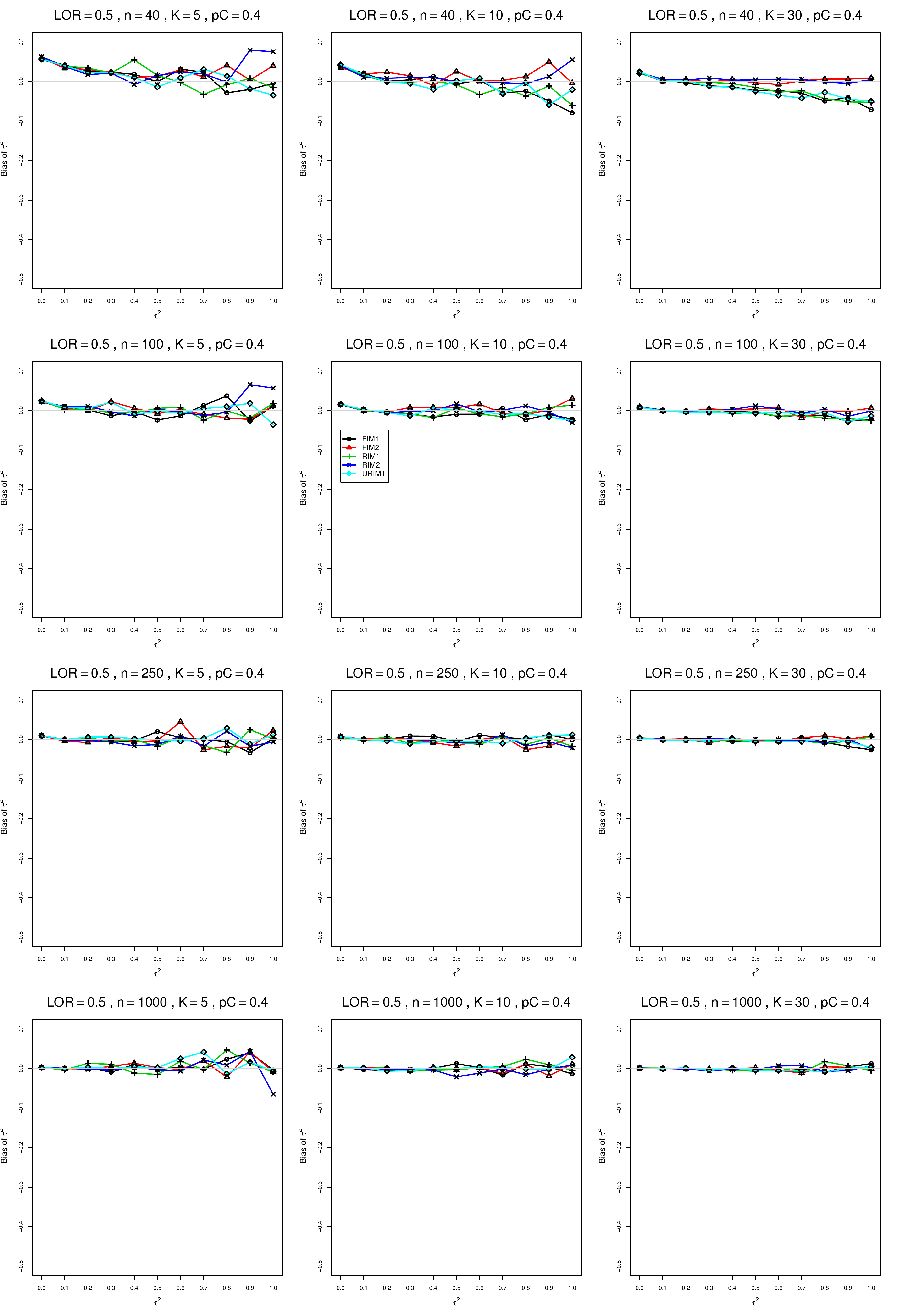}
	\caption{Bias of  between-studies variance $\hat{\tau}_{KD}^2$ for $\theta=0.5$, $p_{C}=0.4$, $\sigma^2=0.1$, constant sample sizes $n=40,\;100,\;250,\;1000$.
The data-generation mechanisms are FIM1 ($\circ$), FIM2 ($\triangle$), RIM1 (+), RIM2 ($\times$), and URIM1 ($\diamond$).
		\label{PlotBiasTau2mu05andpC04LOR_KDsigma01}}
\end{figure}
\begin{figure}[t]
	\centering
	\includegraphics[scale=0.33]{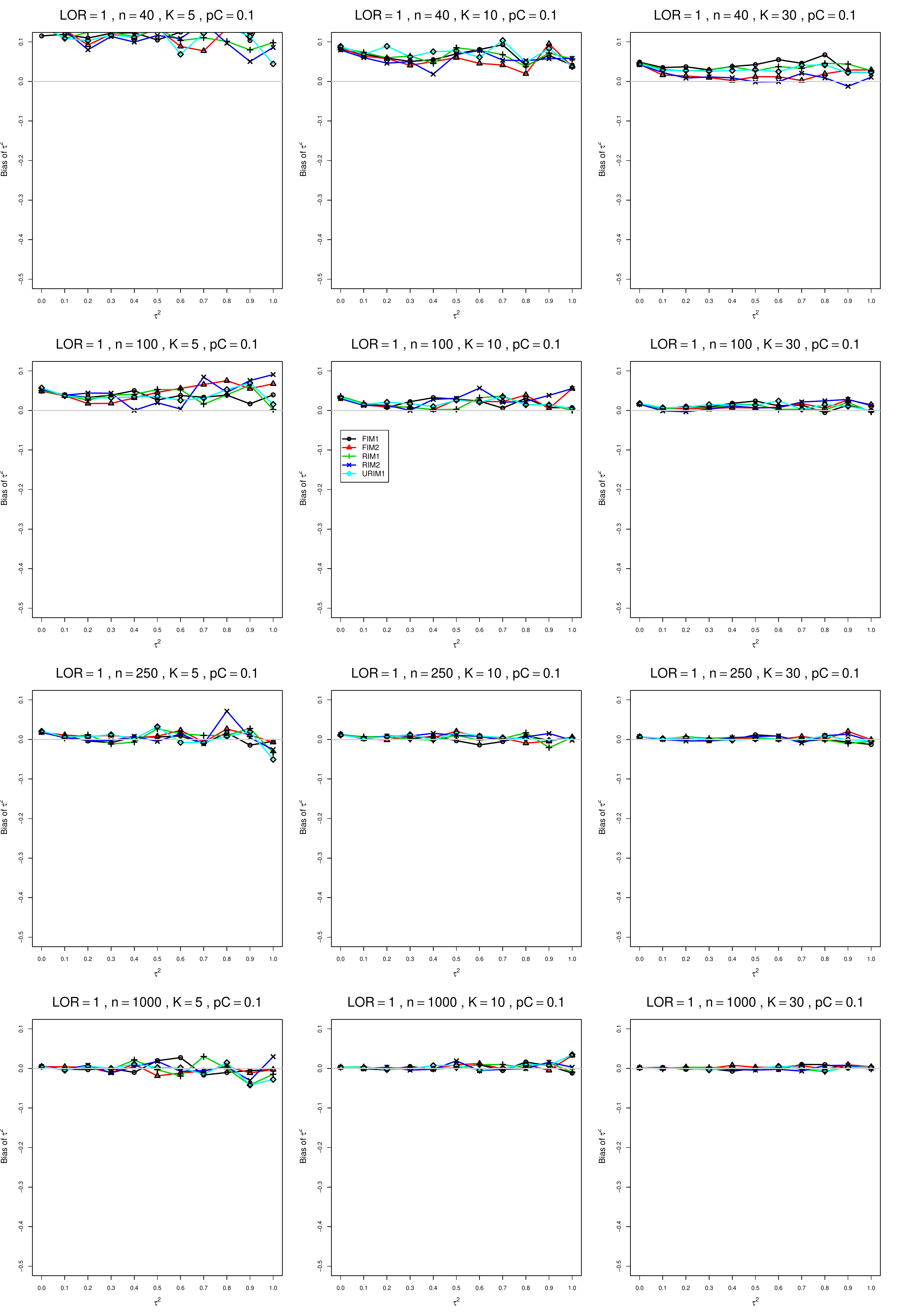}
	\caption{Bias of  between-studies variance $\hat{\tau}_{KD}^2$ for $\theta=1$, $p_{C}=0.1$, $\sigma^2=0.1$, constant sample sizes $n=40,\;100,\;250,\;1000$.
The data-generation mechanisms are FIM1 ($\circ$), FIM2 ($\triangle$), RIM1 (+), RIM2 ($\times$), and URIM1 ($\diamond$).
		\label{PlotBiasTau2mu1andpC01LOR_KDsigma01}}
\end{figure}
\begin{figure}[t]
	\centering
	\includegraphics[scale=0.33]{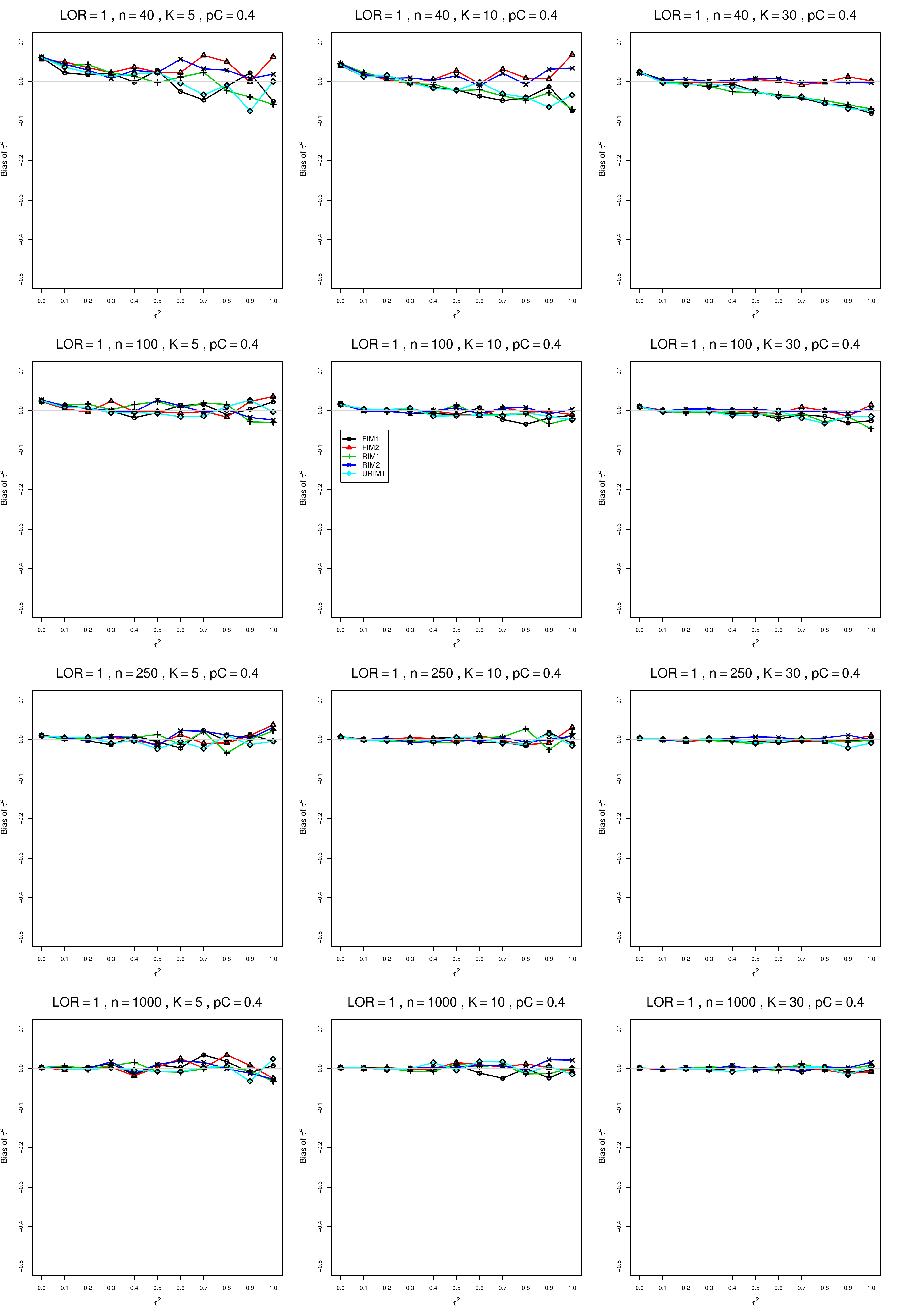}
	\caption{Bias of  between-studies variance $\hat{\tau}_{KD}^2$ for $\theta=1$, $p_{C}=0.4$, $\sigma^2=0.1$, constant sample sizes $n=40,\;100,\;250,\;1000$.
The data-generation mechanisms are FIM1 ($\circ$), FIM2 ($\triangle$), RIM1 (+), RIM2 ($\times$), and URIM1 ($\diamond$).
		\label{PlotBiasTau2mu1andpC04LOR_KDsigma01}}
\end{figure}
\begin{figure}[t]
	\centering
	\includegraphics[scale=0.33]{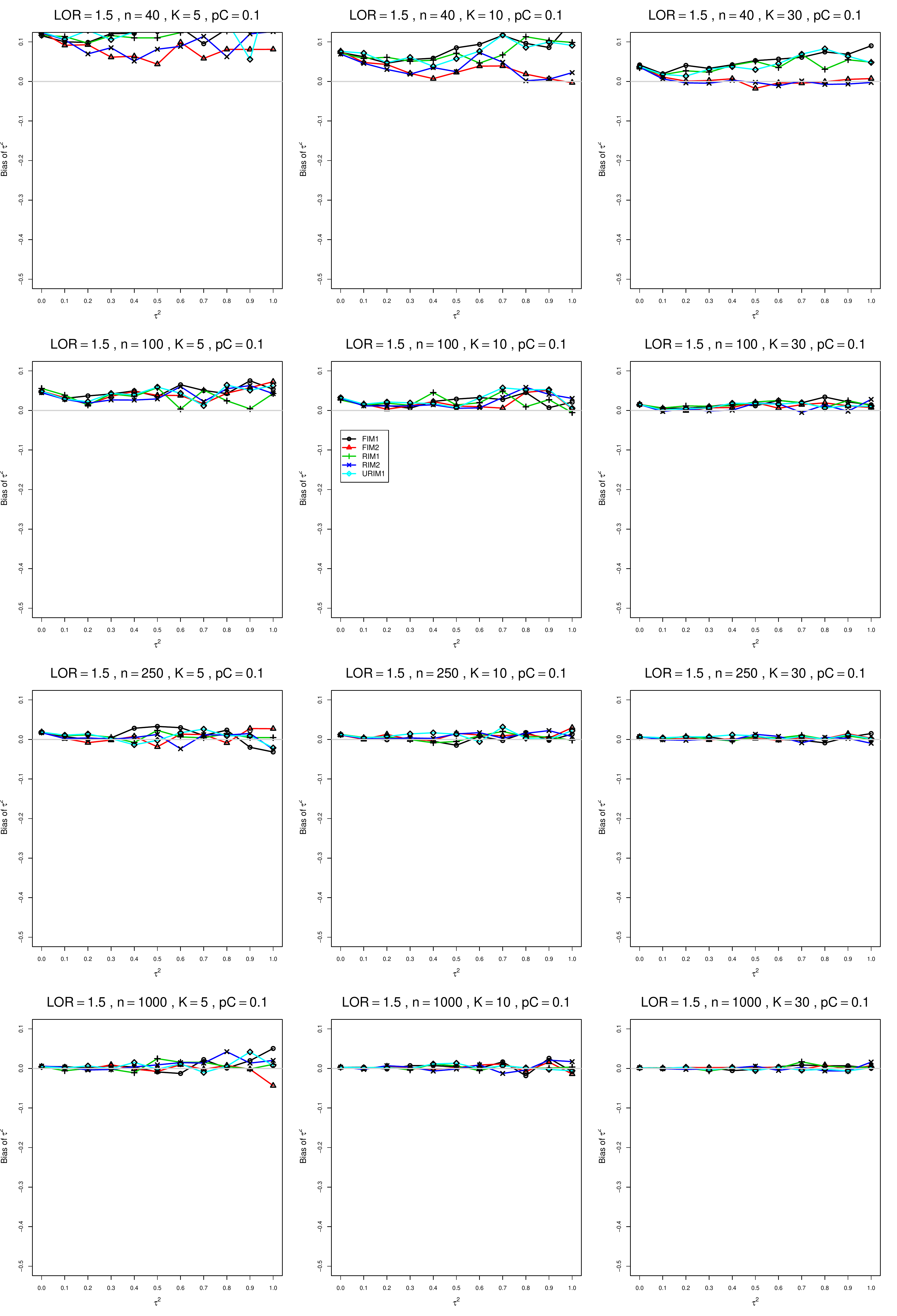}
	\caption{Bias of  between-studies variance $\hat{\tau}_{KD}^2$ for $\theta=1.5$, $p_{C}=0.1$, $\sigma^2=0.1$, constant sample sizes $n=40,\;100,\;250,\;1000$.
The data-generation mechanisms are FIM1 ($\circ$), FIM2 ($\triangle$), RIM1 (+), RIM2 ($\times$), and URIM1 ($\diamond$).
		\label{PlotBiasTau2mu15andpC01LOR_KDsigma01}}
\end{figure}
\begin{figure}[t]
	\centering
	\includegraphics[scale=0.33]{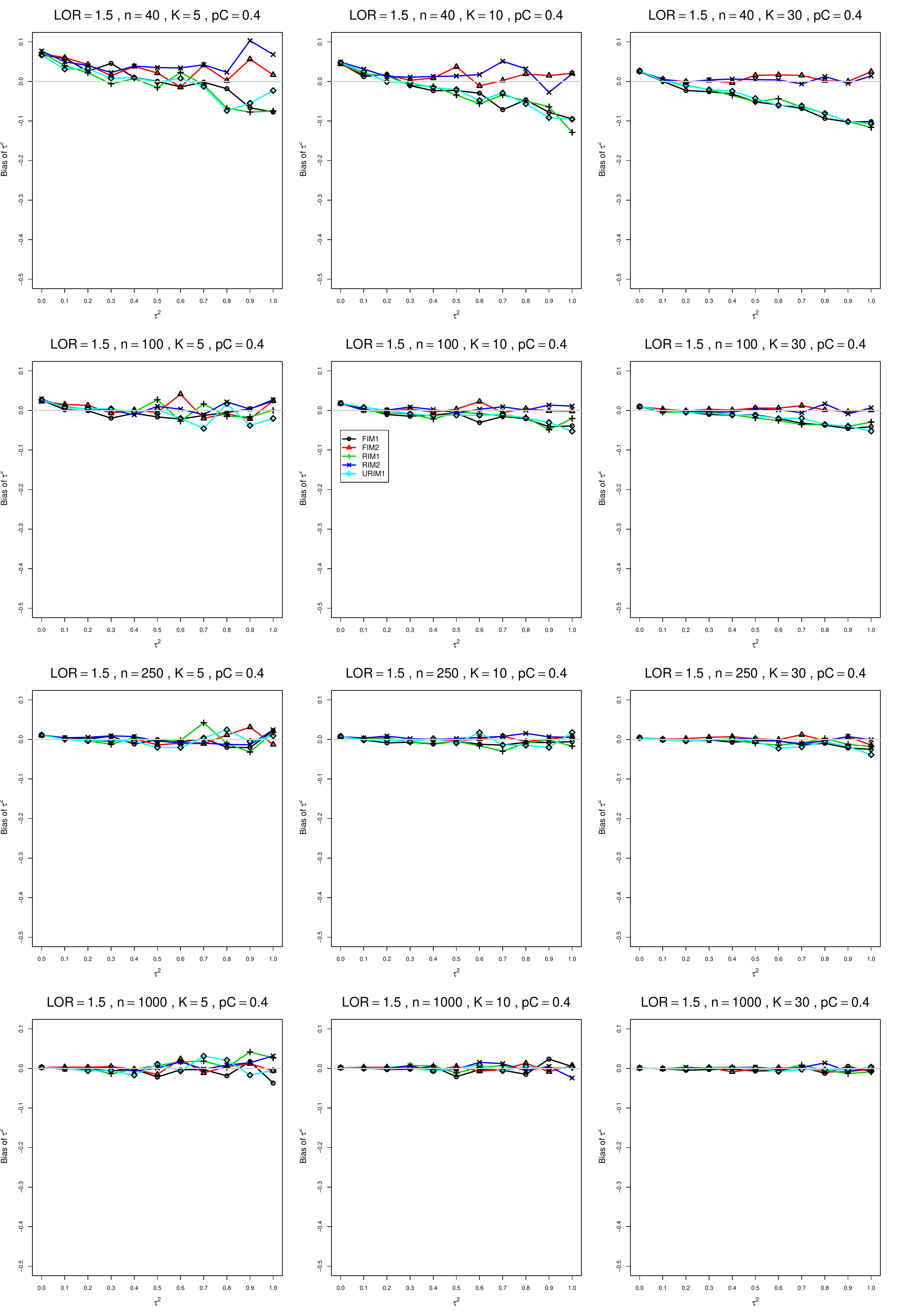}
	\caption{Bias of  between-studies variance $\hat{\tau}_{KD}^2$ for $\theta=1.5$, $p_{C}=0.4$, $\sigma^2=0.1$, constant sample sizes $n=40,\;100,\;250,\;1000$.
The data-generation mechanisms are FIM1 ($\circ$), FIM2 ($\triangle$), RIM1 (+), RIM2 ($\times$), and URIM1 ($\diamond$).
		\label{PlotBiasTau2mu15andpC04LOR_KDsigma01}}
\end{figure}
\begin{figure}[t]
	\centering
	\includegraphics[scale=0.33]{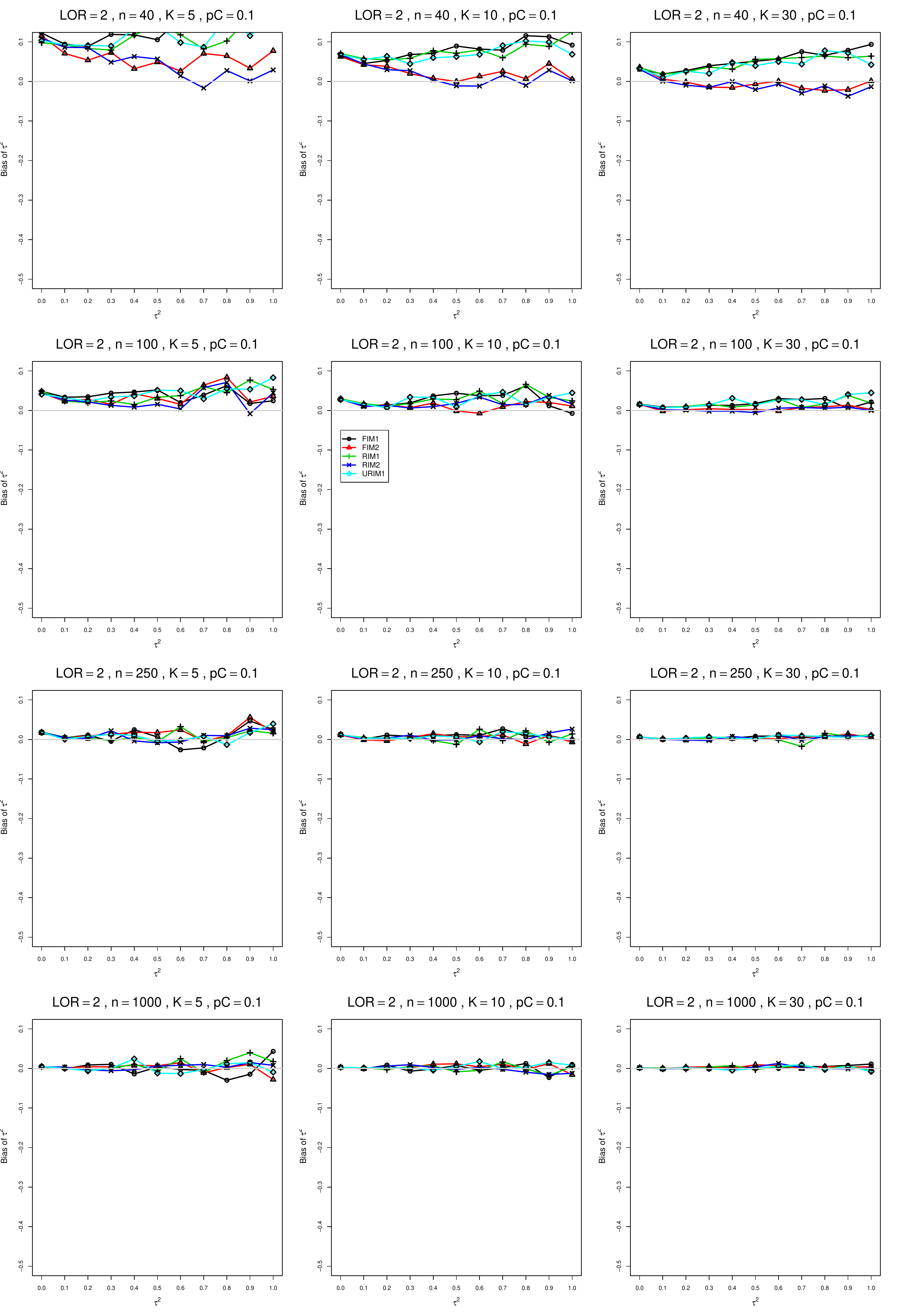}
	\caption{Bias of  between-studies variance $\hat{\tau}_{KD}^2$ for $\theta=2$, $p_{C}=0.1$, $\sigma^2=0.1$, constant sample sizes $n=40,\;100,\;250,\;1000$.
The data-generation mechanisms are FIM1 ($\circ$), FIM2 ($\triangle$), RIM1 (+), RIM2 ($\times$), and URIM1 ($\diamond$).
		\label{PlotBiasTau2mu2andpC01LOR_KDsigma01}}
\end{figure}
\begin{figure}[t]
	\centering
	\includegraphics[scale=0.33]{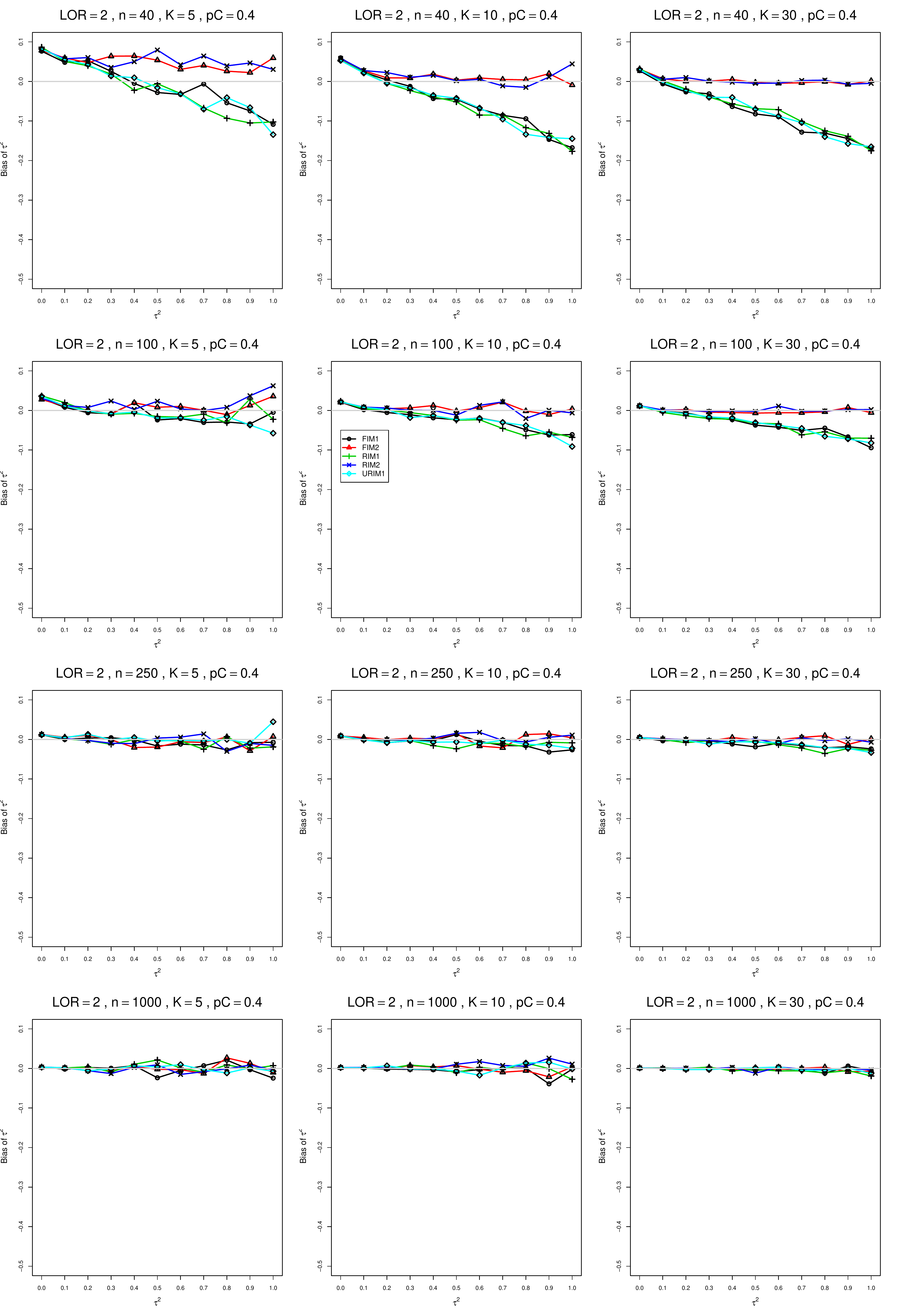}
	\caption{Bias of  between-studies variance $\hat{\tau}_{KD}^2$ for $\theta=2$, $p_{C}=0.4$, $\sigma^2=0.1$, constant sample sizes $n=40,\;100,\;250,\;1000$.
The data-generation mechanisms are FIM1 ($\circ$), FIM2 ($\triangle$), RIM1 (+), RIM2 ($\times$), and URIM1 ($\diamond$).
		\label{PlotBiasTau2mu2andpC04LOR_KDsigma01}}
\end{figure}
\begin{figure}[t]
	\centering
	\includegraphics[scale=0.33]{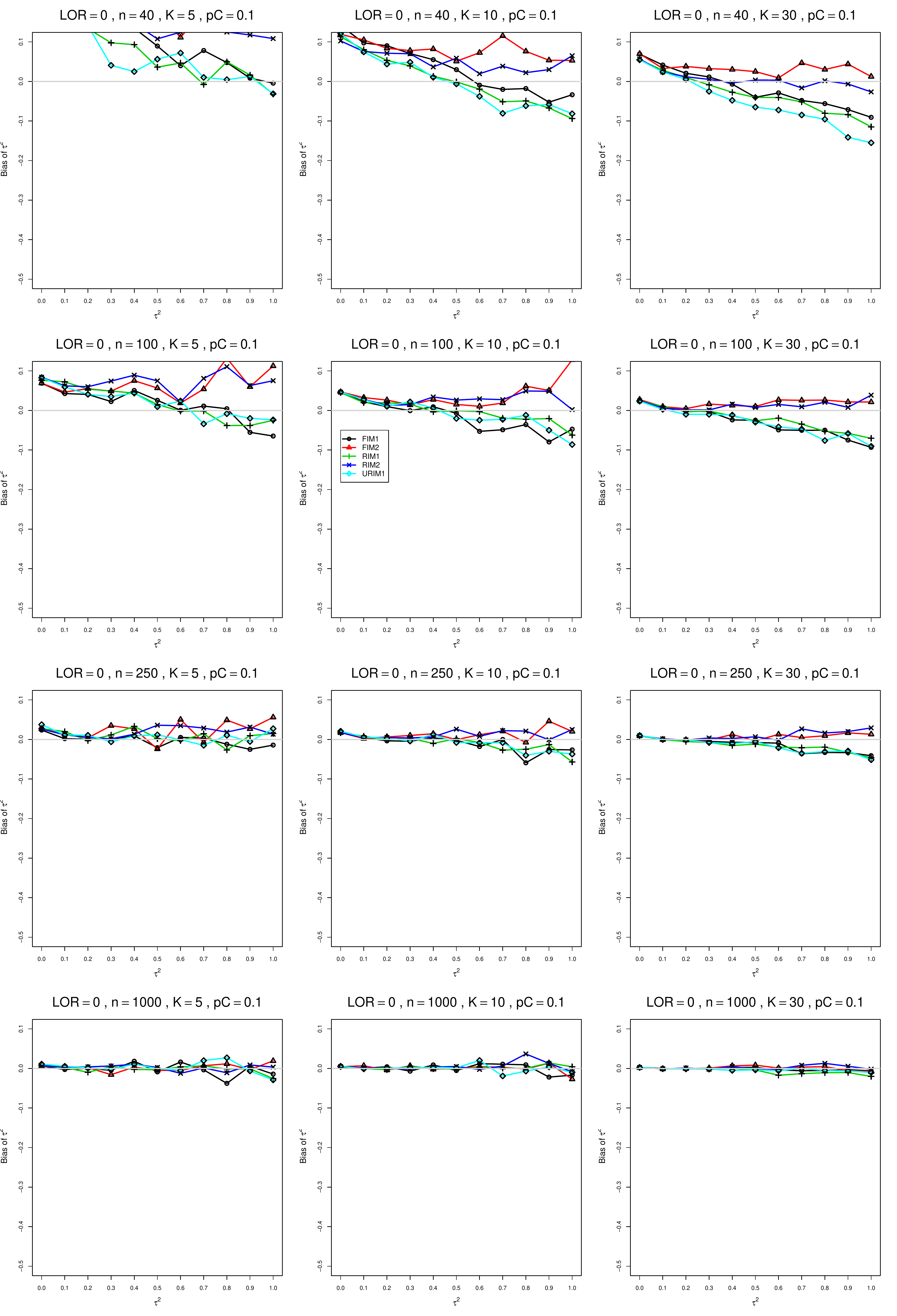}
	\caption{Bias of  between-studies variance $\hat{\tau}_{KD}^2$ for $\theta=0$, $p_{C}=0.1$, $\sigma^2=0.4$, constant sample sizes $n=40,\;100,\;250,\;1000$.
The data-generation mechanisms are FIM1 ($\circ$), FIM2 ($\triangle$), RIM1 (+), RIM2 ($\times$), and URIM1 ($\diamond$).
		\label{PlotBiasTau2mu0andpC01LOR_KDsigma04}}
\end{figure}
\begin{figure}[t]
	\centering
	\includegraphics[scale=0.33]{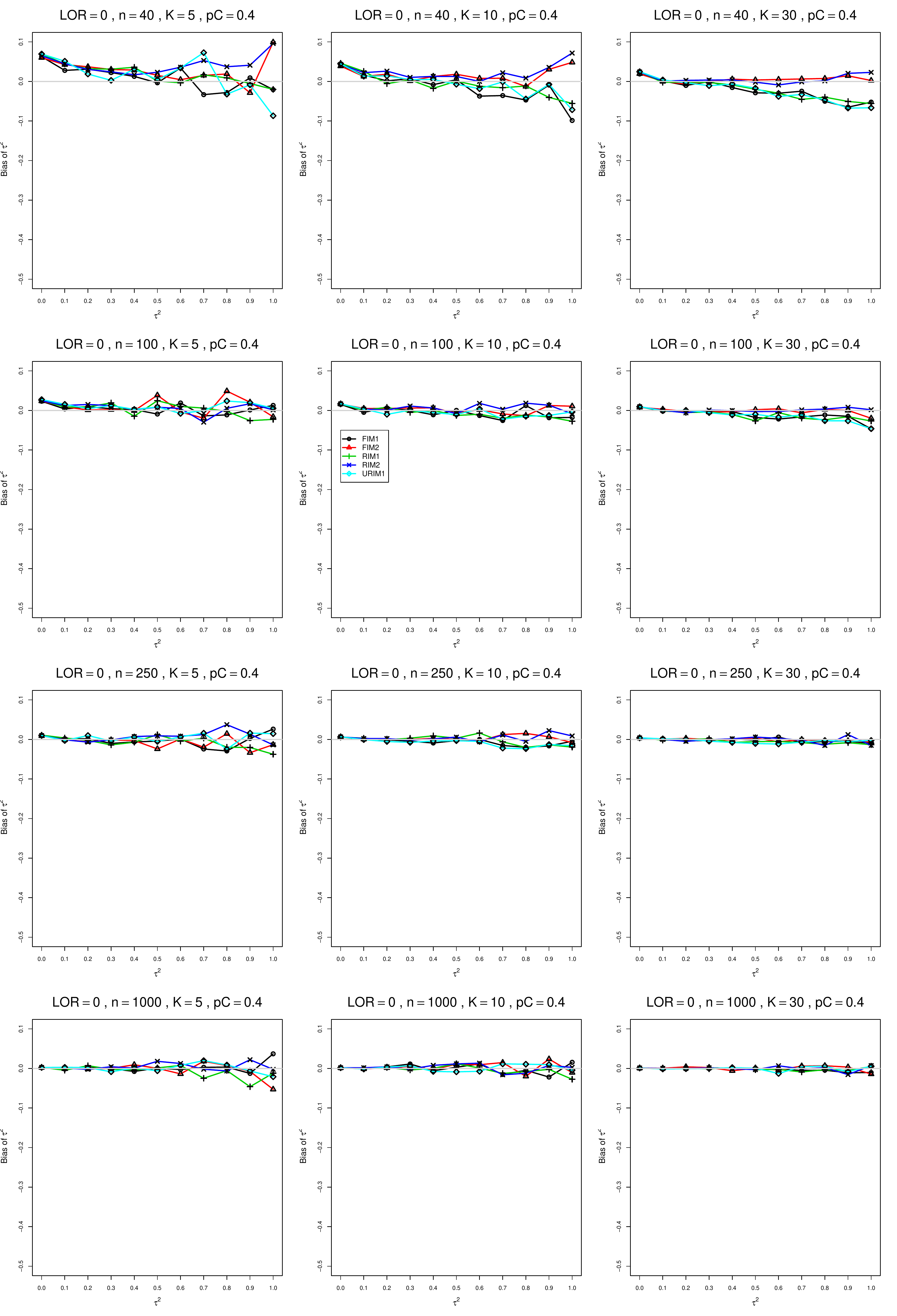}
	\caption{Bias of  between-studies variance $\hat{\tau}_{KD}^2$ for $\theta=0$, $p_{C}=0.4$, $\sigma^2=0.4$, constant sample sizes $n=40,\;100,\;250,\;1000$.
The data-generation mechanisms are FIM1 ($\circ$), FIM2 ($\triangle$), RIM1 (+), RIM2 ($\times$), and URIM1 ($\diamond$).
		\label{PlotBiasTau2mu0andpC04LOR_KDsigma04}}
\end{figure}
\begin{figure}[t]
	\centering
	\includegraphics[scale=0.33]{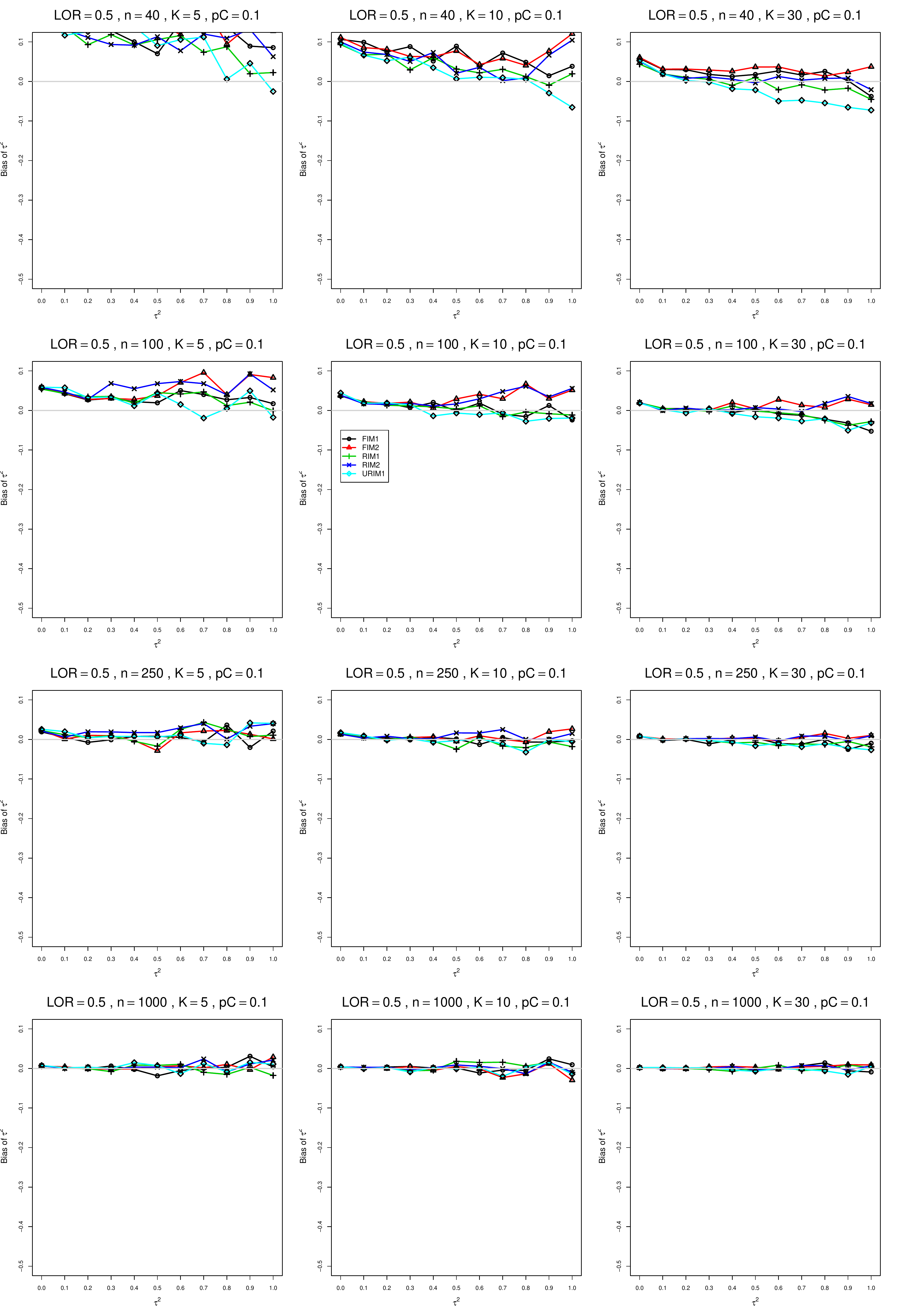}
	\caption{Bias of  between-studies variance $\hat{\tau}_{KD}^2$ for $\theta=0.5$, $p_{C}=0.1$, $\sigma^2=0.4$, constant sample sizes $n=40,\;100,\;250,\;1000$.
The data-generation mechanisms are FIM1 ($\circ$), FIM2 ($\triangle$), RIM1 (+), RIM2 ($\times$), and URIM1 ($\diamond$).
		\label{PlotBiasTau2mu05andpC01LOR_KDsigma04}}
\end{figure}
\begin{figure}[t]
	\centering
	\includegraphics[scale=0.33]{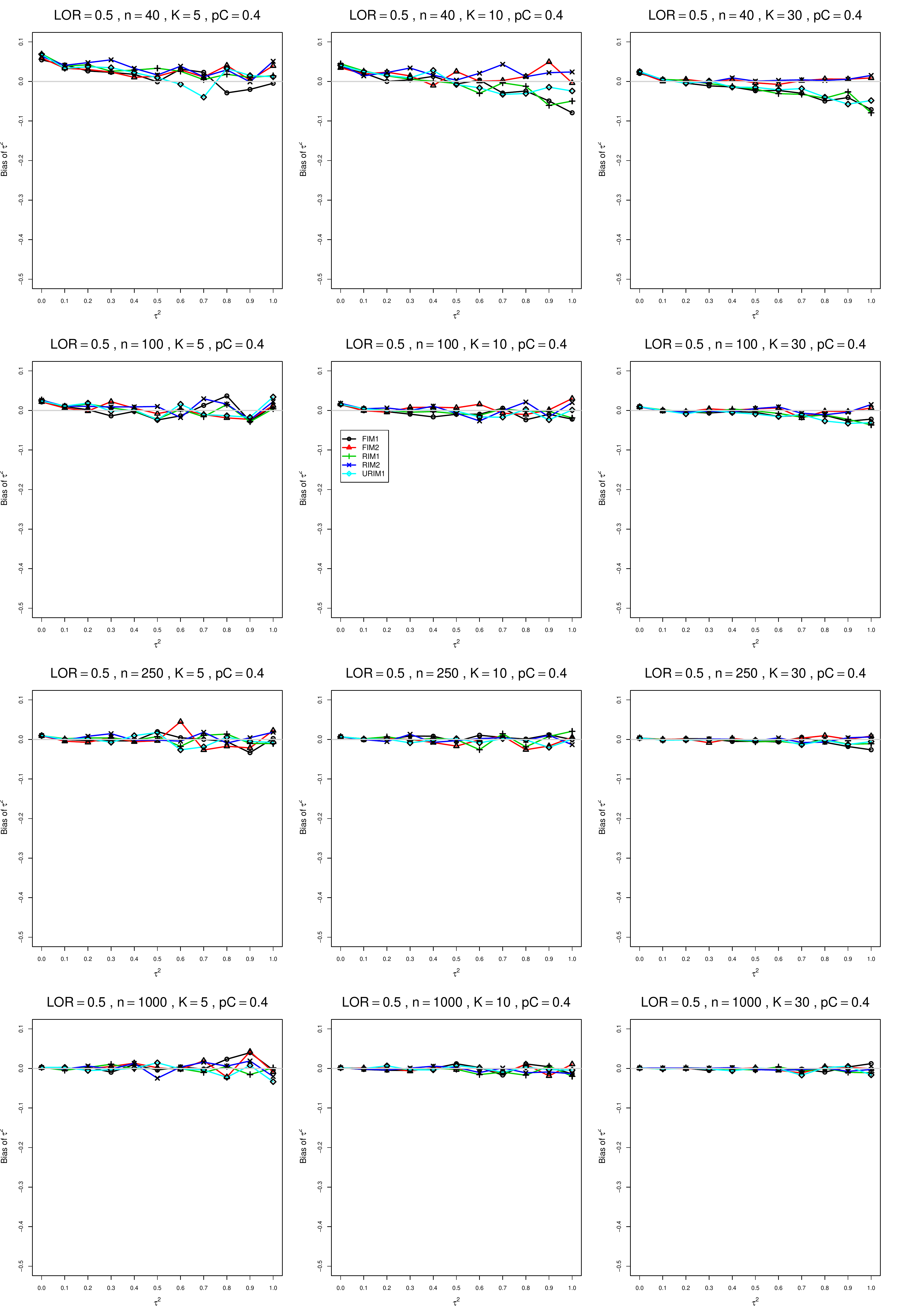}
	\caption{Bias of  between-studies variance $\hat{\tau}_{KD}^2$ for $\theta=0.5$, $p_{C}=0.4$, $\sigma^2=0.4$, constant sample sizes $n=40,\;100,\;250,\;1000$.
The data-generation mechanisms are FIM1 ($\circ$), FIM2 ($\triangle$), RIM1 (+), RIM2 ($\times$), and URIM1 ($\diamond$).
		\label{PlotBiasTau2mu05andpC04LOR_KDsigma04}}
\end{figure}
\begin{figure}[t]
	\centering
	\includegraphics[scale=0.33]{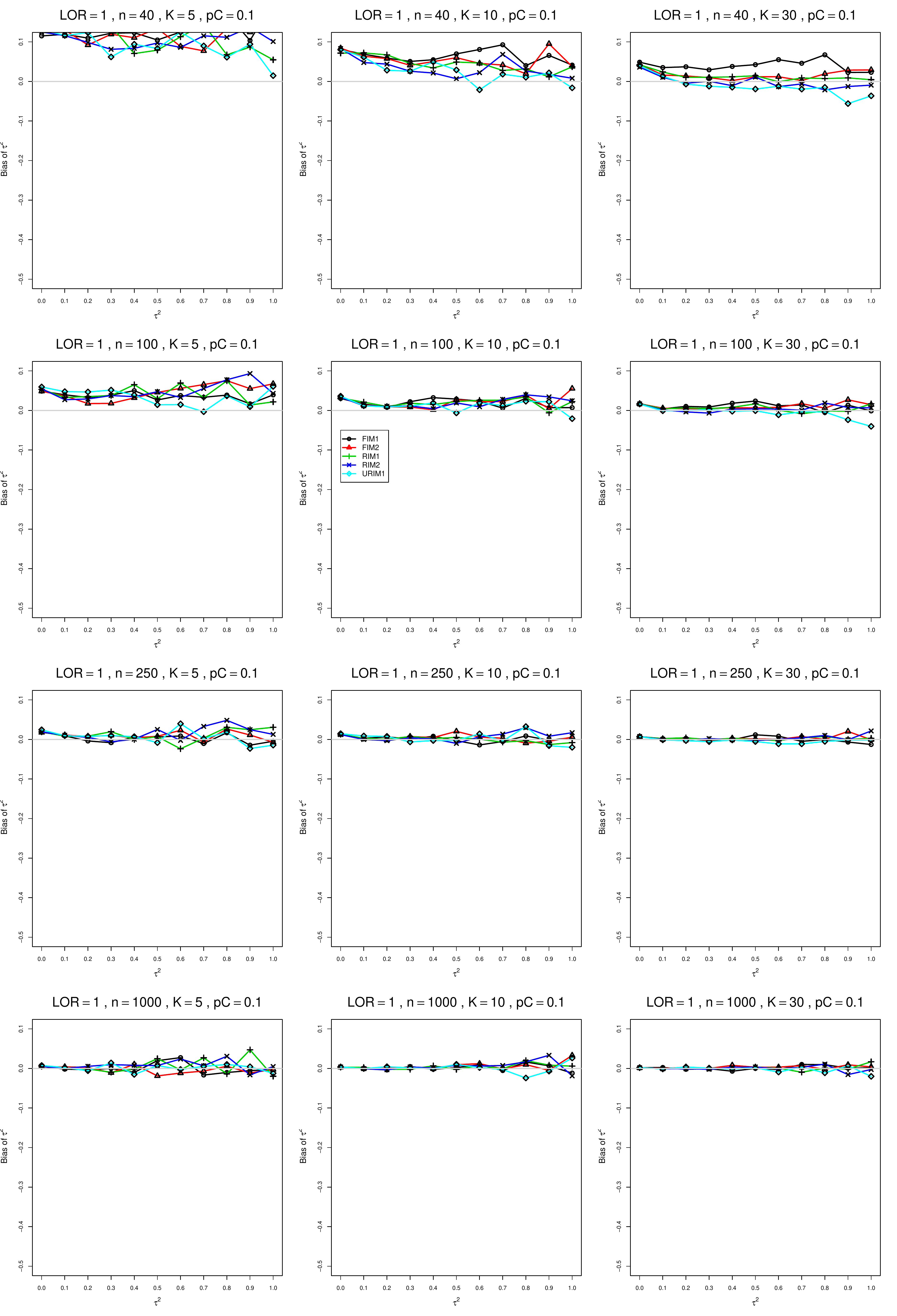}
	\caption{Bias of  between-studies variance $\hat{\tau}_{KD}^2$ for $\theta=1$, $p_{C}=0.1$, $\sigma^2=0.4$, constant sample sizes $n=40,\;100,\;250,\;1000$.
The data-generation mechanisms are FIM1 ($\circ$), FIM2 ($\triangle$), RIM1 (+), RIM2 ($\times$), and URIM1 ($\diamond$).
		\label{PlotBiasTau2mu1andpC01LOR_KDsigma04}}
\end{figure}
\begin{figure}[t]
	\centering
	\includegraphics[scale=0.33]{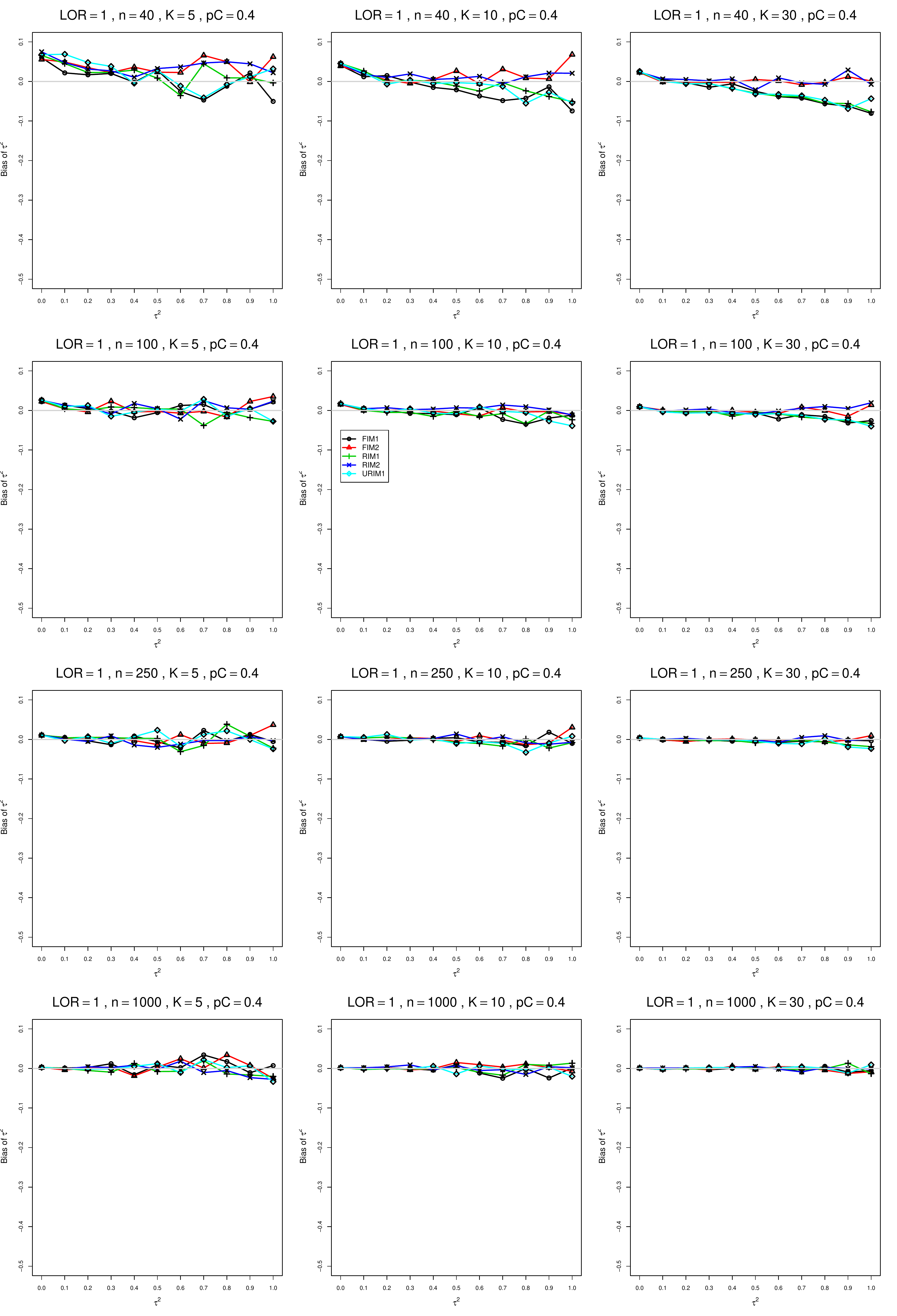}
	\caption{Bias of  between-studies variance $\hat{\tau}_{KD}^2$ for $\theta=1$, $p_{C}=0.4$, $\sigma^2=0.4$, constant sample sizes $n=40,\;100,\;250,\;1000$.
The data-generation mechanisms are FIM1 ($\circ$), FIM2 ($\triangle$), RIM1 (+), RIM2 ($\times$), and URIM1 ($\diamond$).
		\label{PlotBiasTau2mu1andpC04LOR_KDsigma04}}
\end{figure}
\begin{figure}[t]
	\centering
	\includegraphics[scale=0.33]{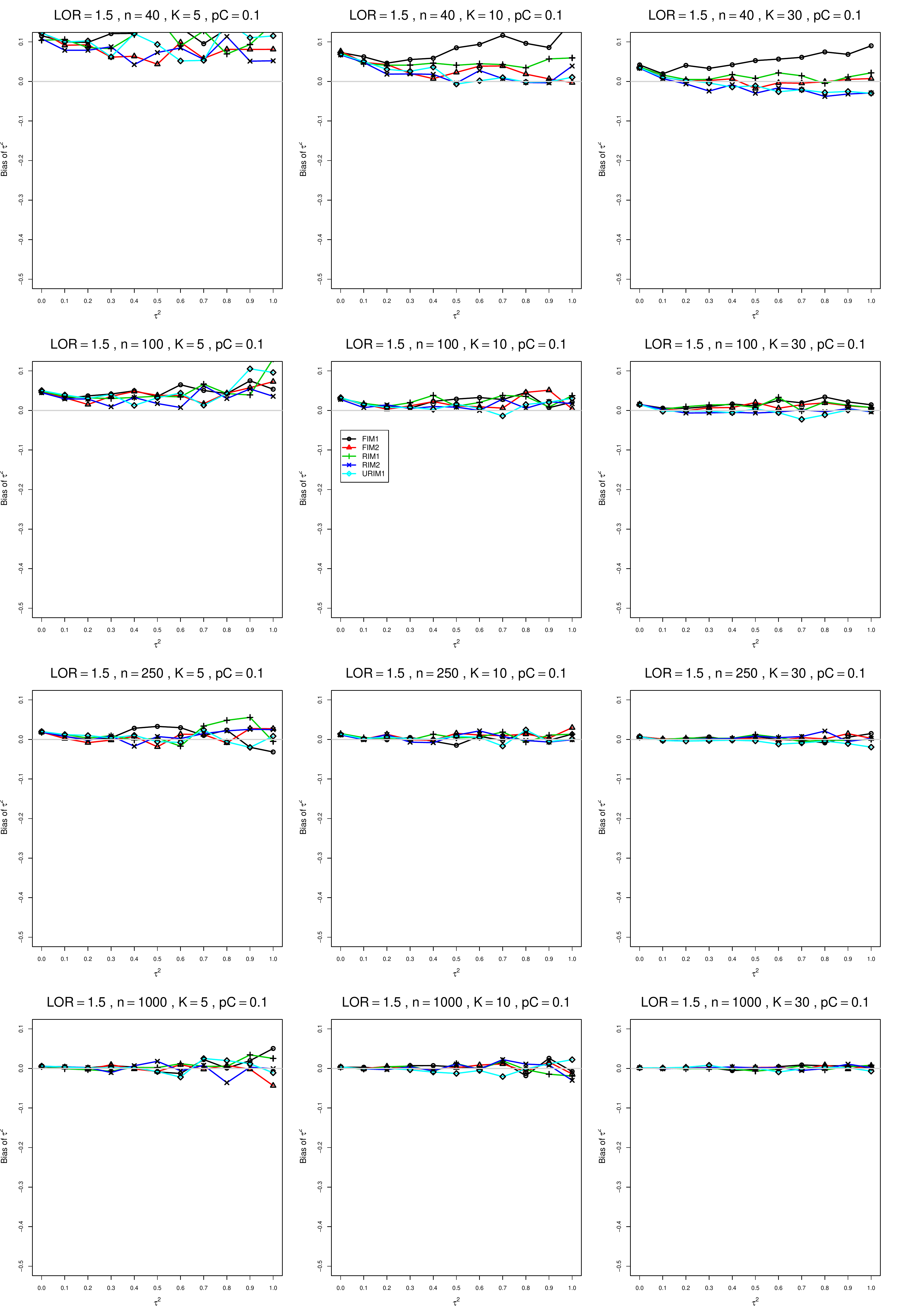}
	\caption{Bias of  between-studies variance $\hat{\tau}_{KD}^2$ for $\theta=1.5$, $p_{C}=0.1$, $\sigma^2=0.4$, constant sample sizes $n=40,\;100,\;250,\;1000$.
The data-generation mechanisms are FIM1 ($\circ$), FIM2 ($\triangle$), RIM1 (+), RIM2 ($\times$), and URIM1 ($\diamond$).
		\label{PlotBiasTau2mu15andpC01LOR_KDsigma04}}
\end{figure}
\begin{figure}[t]
	\centering
	\includegraphics[scale=0.33]{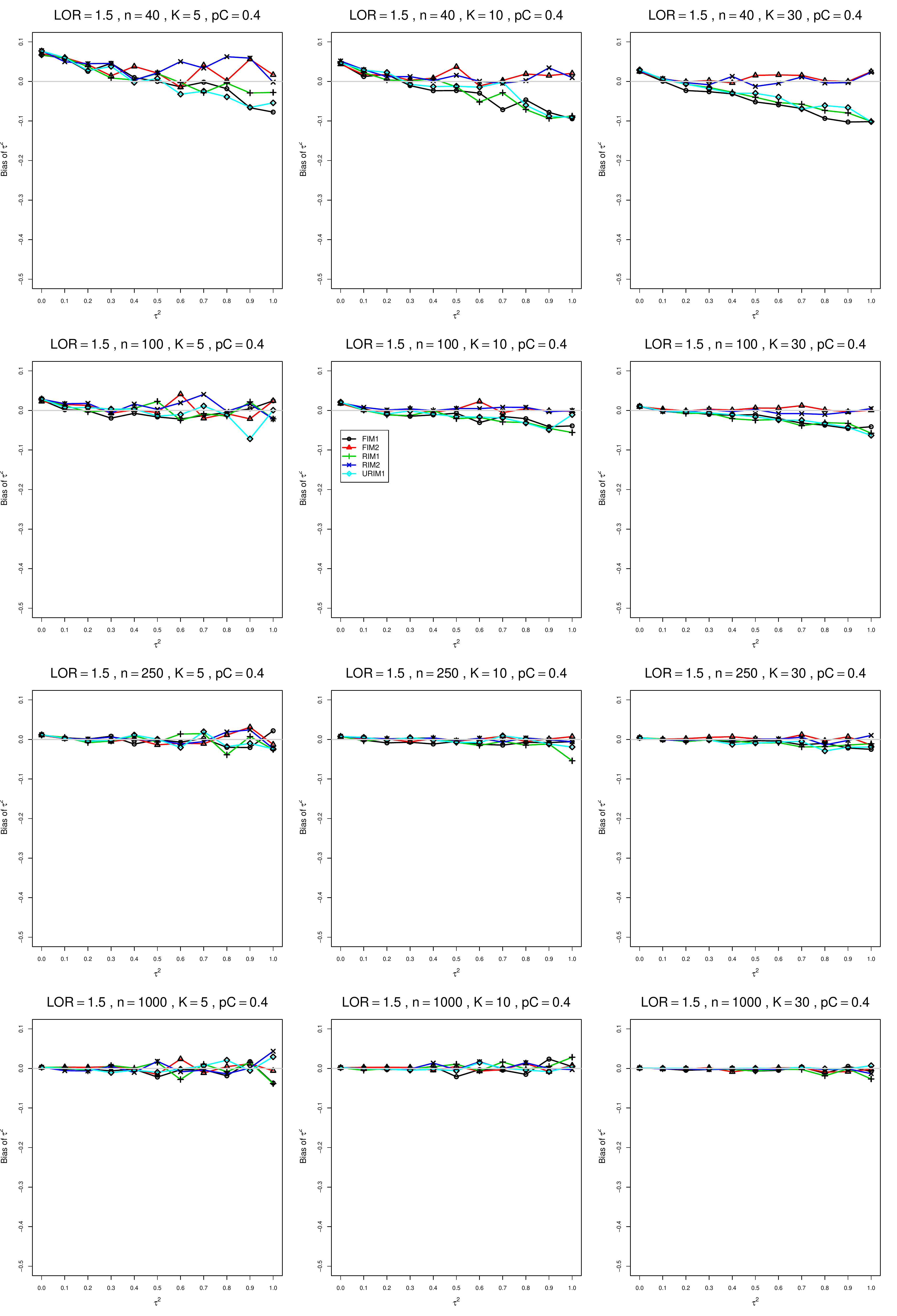}
	\caption{Bias of  between-studies variance $\hat{\tau}_{KD}^2$ for $\theta=1.5$, $p_{C}=0.4$, $\sigma^2=0.4$, constant sample sizes $n=40,\;100,\;250,\;1000$.
The data-generation mechanisms are FIM1 ($\circ$), FIM2 ($\triangle$), RIM1 (+), RIM2 ($\times$), and URIM1 ($\diamond$).
		\label{PlotBiasTau2mu15andpC04LOR_KDsigma04}}
\end{figure}
\begin{figure}[t]
	\centering
	\includegraphics[scale=0.33]{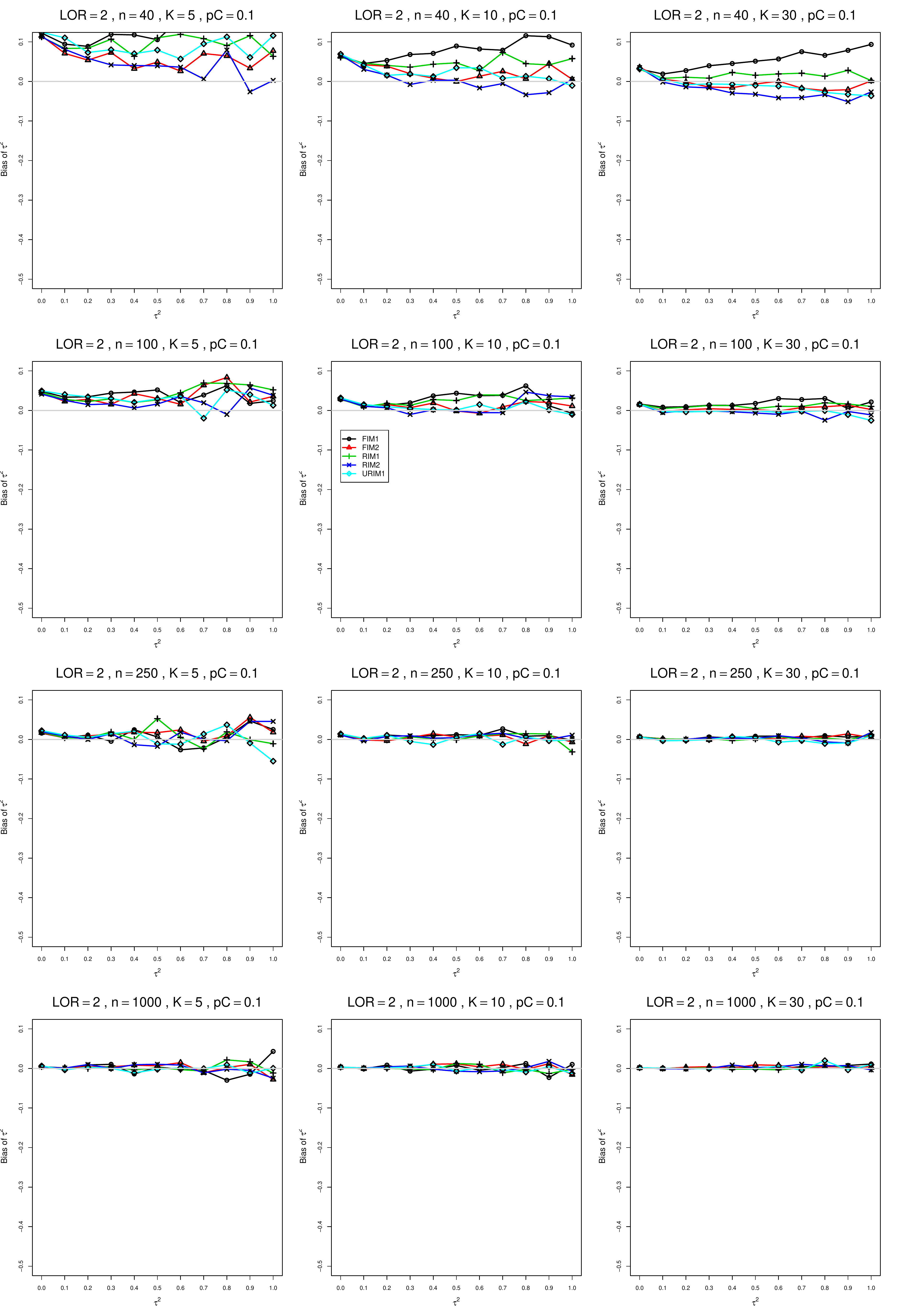}
	\caption{Bias of  between-studies variance $\hat{\tau}_{KD}^2$ for $\theta=2$, $p_{C}=0.1$, $\sigma^2=0.4$, constant sample sizes $n=40,\;100,\;250,\;1000$.
The data-generation mechanisms are FIM1 ($\circ$), FIM2 ($\triangle$), RIM1 (+), RIM2 ($\times$), and URIM1 ($\diamond$).
		\label{PlotBiasTau2mu2andpC01LOR_KDsigma04}}
\end{figure}
\begin{figure}[t]
	\centering
	\includegraphics[scale=0.33]{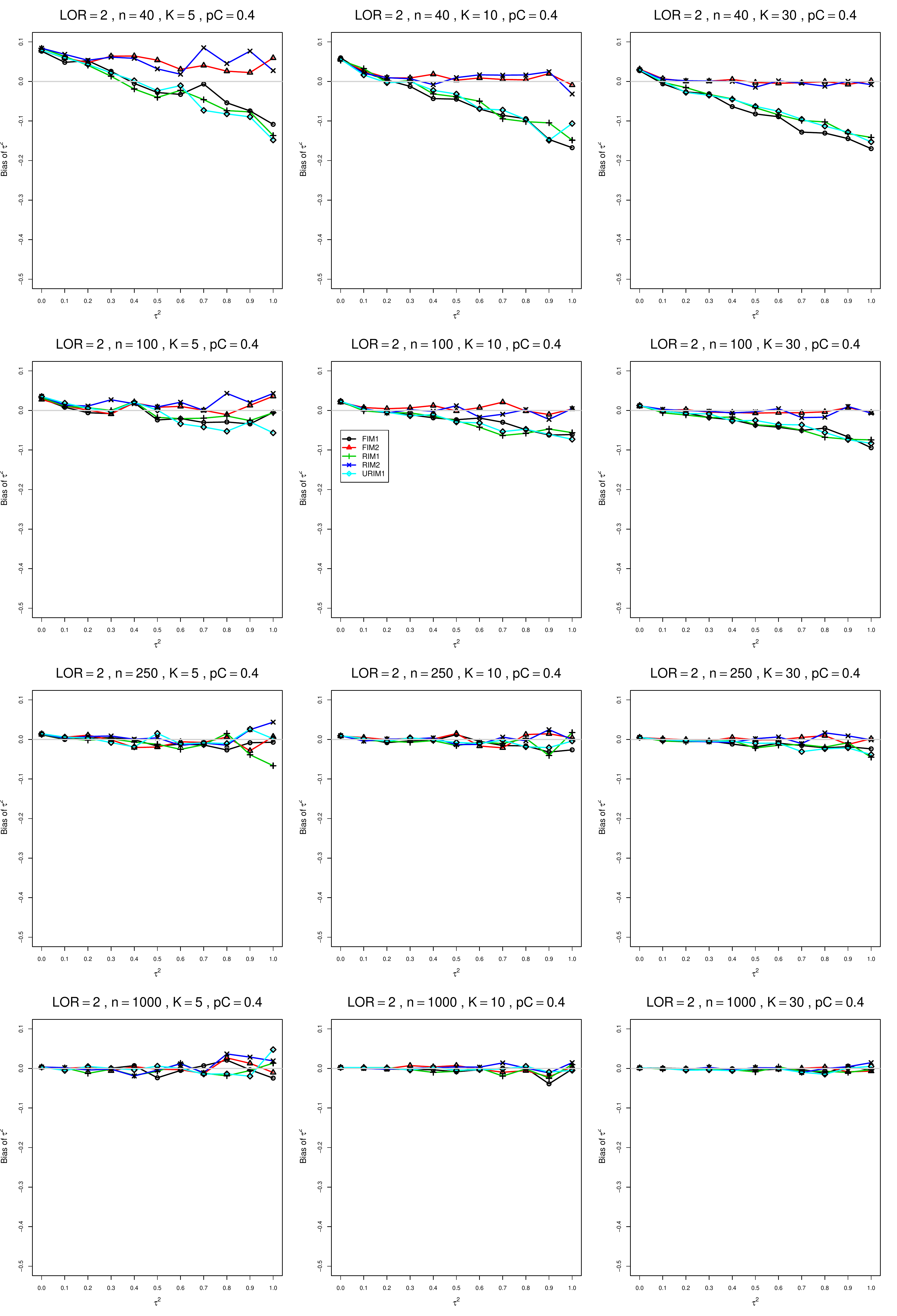}
	\caption{Bias of  between-studies variance $\hat{\tau}_{KD}^2$ for $\theta=2$, $p_{C}=0.4$, $\sigma^2=0.4$, constant sample sizes $n=40,\;100,\;250,\;1000$.
The data-generation mechanisms are FIM1 ($\circ$), FIM2 ($\triangle$), RIM1 (+), RIM2 ($\times$), and URIM1 ($\diamond$).
		\label{PlotBiasTau2mu2andpC04LOR_KDsigma04}}
\end{figure}

\clearpage
\subsection*{A1.5 Bias of $\hat{\tau}_{FIM2}^2$}
\renewcommand{\thefigure}{A1.5.\arabic{figure}}
\setcounter{figure}{0}

\begin{figure}[t]
	\centering
	\includegraphics[scale=0.33]{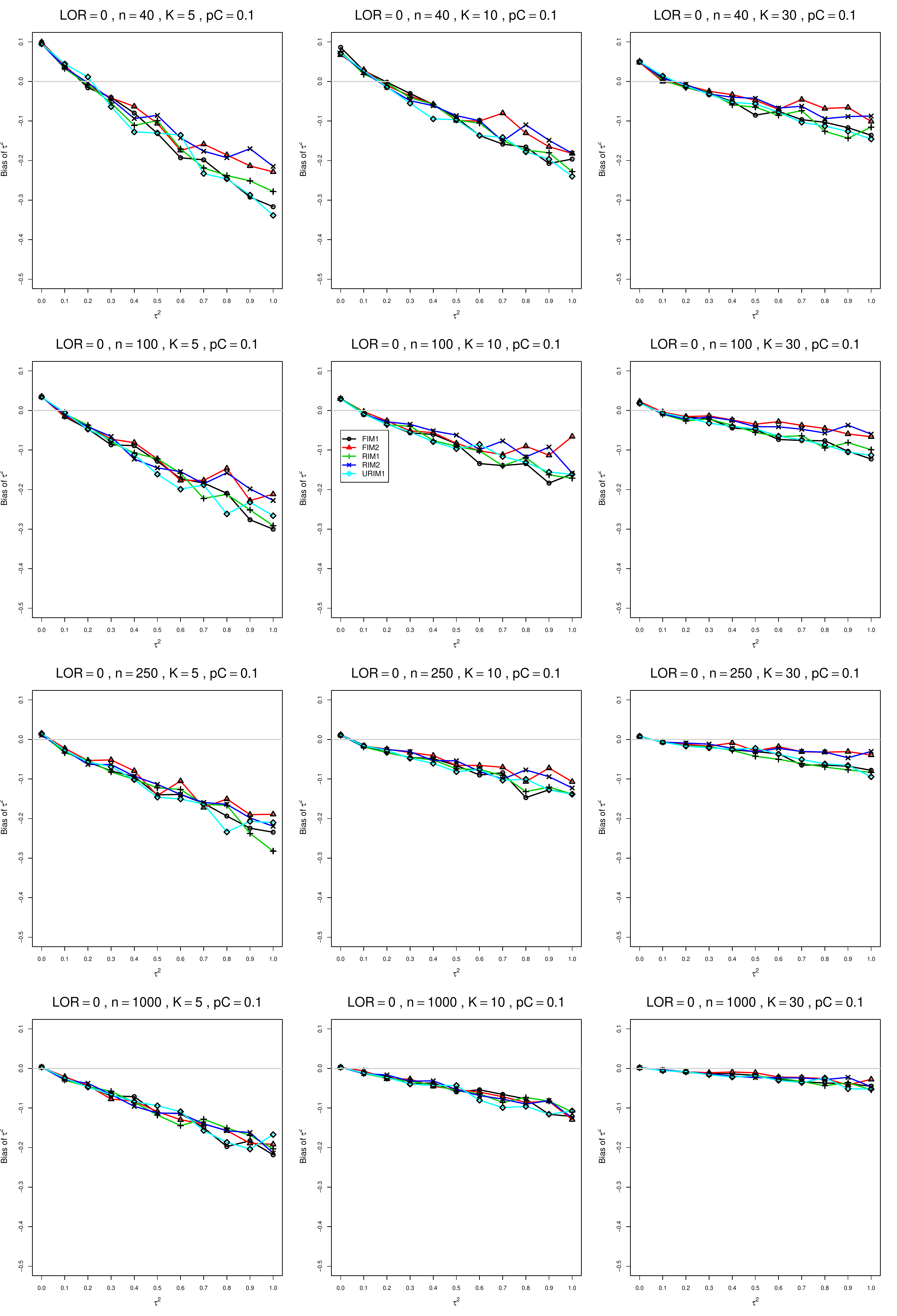}
	\caption{Bias of  between-studies variance $\hat{\tau}_{FIM2}^2$ for $\theta=0$, $p_{C}=0.1$, $\sigma^2=0.1$, constant sample sizes $n=40,\;100,\;250,\;1000$.
The data-generation mechanisms are FIM1 ($\circ$), FIM2 ($\triangle$), RIM1 (+), RIM2 ($\times$), and URIM1 ($\diamond$).
		\label{PlotBiasTau2mu0andpC01LOR_UMFSsigma01}}
\end{figure}
\begin{figure}[t]
	\centering
	\includegraphics[scale=0.33]{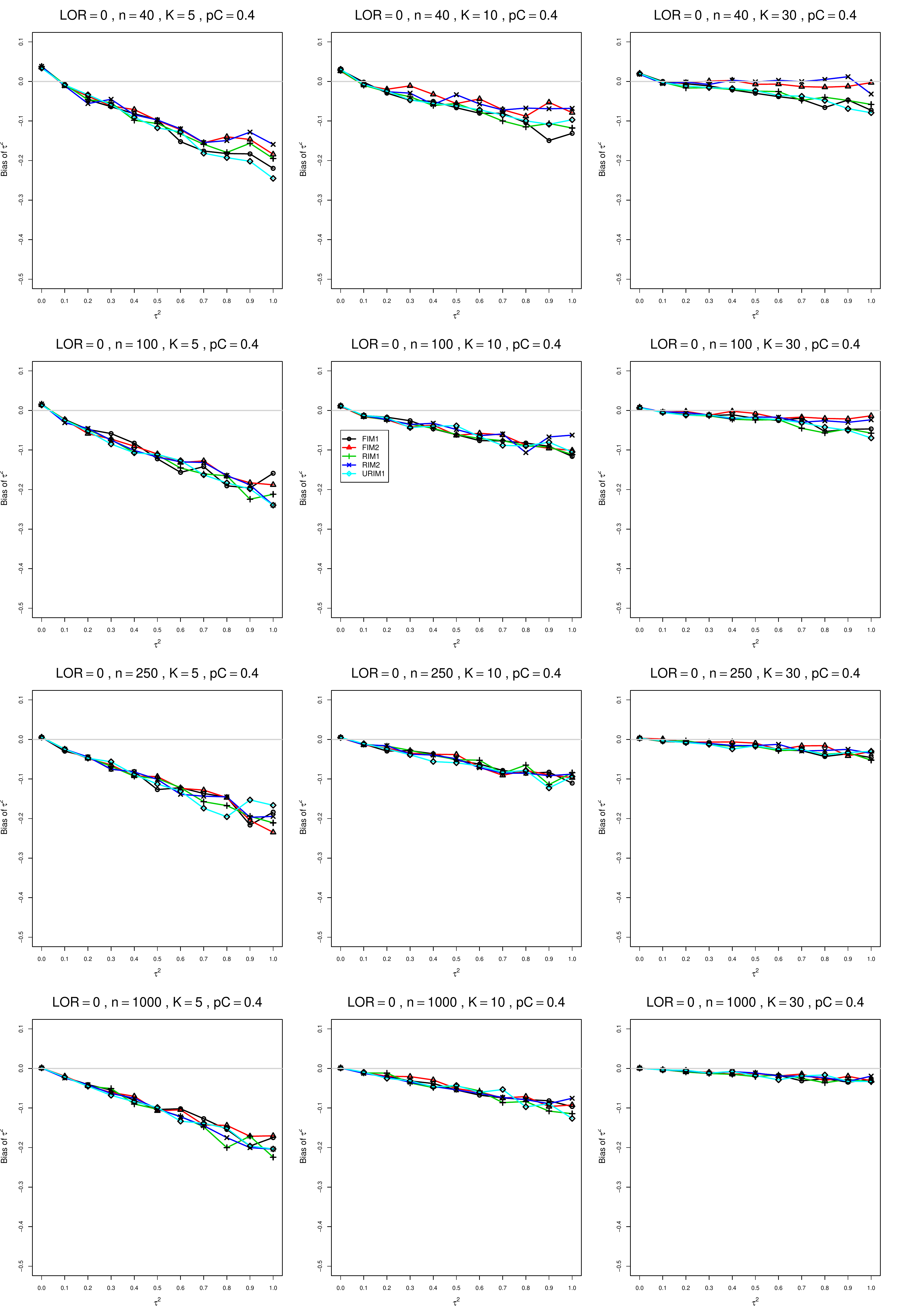}
	\caption{Bias of  between-studies variance $\hat{\tau}_{FIM2}^2$ for $\theta=0$, $p_{C}=0.4$, $\sigma^2=0.1$, constant sample sizes $n=40,\;100,\;250,\;1000$.
The data-generation mechanisms are FIM1 ($\circ$), FIM2 ($\triangle$), RIM1 (+), RIM2 ($\times$), and URIM1 ($\diamond$).
		\label{PlotBiasTau2mu0andpC04LOR_UMFSsigma01}}
\end{figure}
\begin{figure}[t]
	\centering
	\includegraphics[scale=0.33]{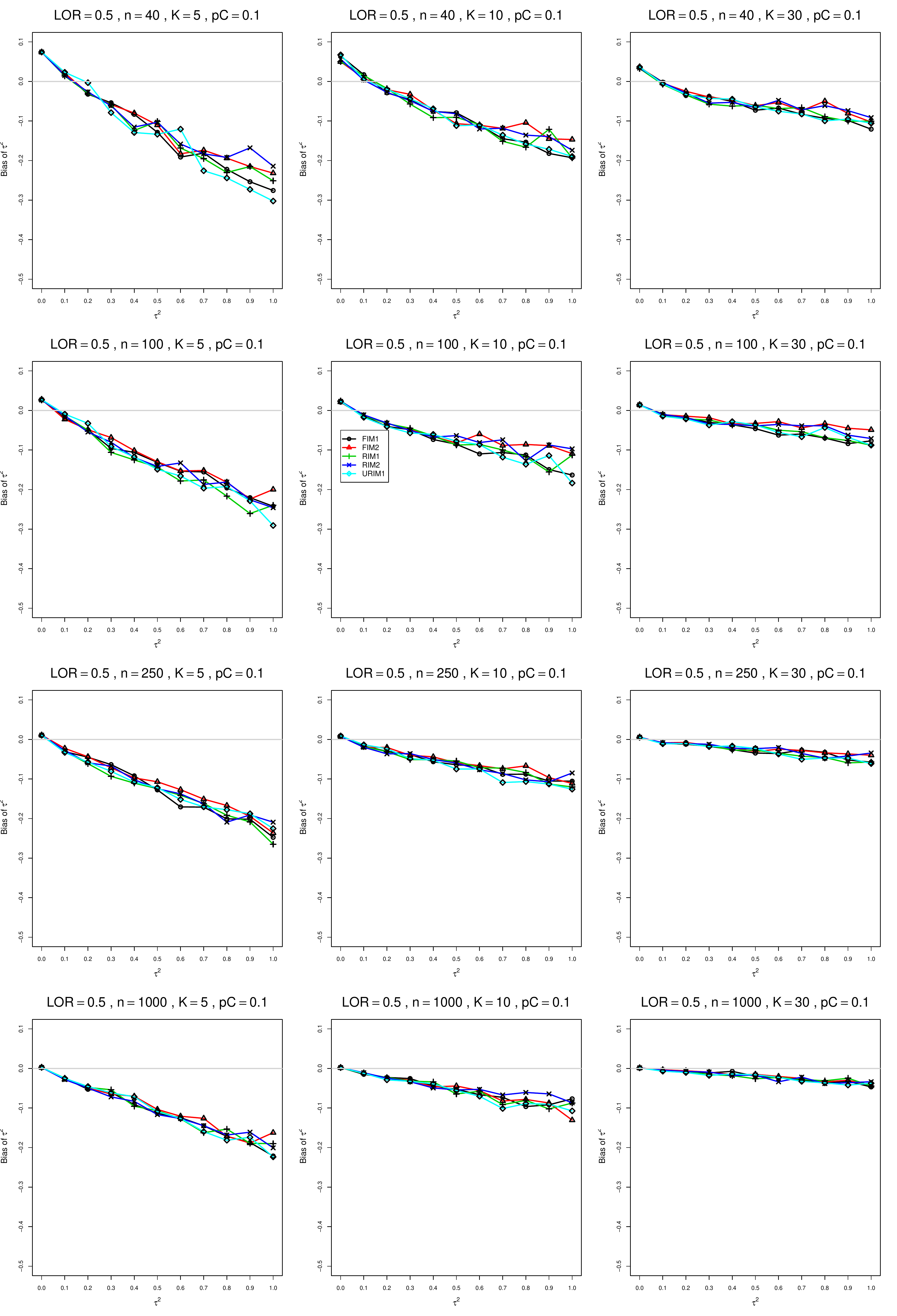}
	\caption{Bias of  between-studies variance $\hat{\tau}_{FIM2}^2$ for $\theta=0.5$, $p_{C}=0.1$, $\sigma^2=0.1$, constant sample sizes $n=40,\;100,\;250,\;1000$.
The data-generation mechanisms are FIM1 ($\circ$), FIM2 ($\triangle$), RIM1 (+), RIM2 ($\times$), and URIM1 ($\diamond$).
		\label{PlotBiasTau2mu05andpC01LOR_UMFSsigma01}}
\end{figure}
\begin{figure}[t]
	\centering
	\includegraphics[scale=0.33]{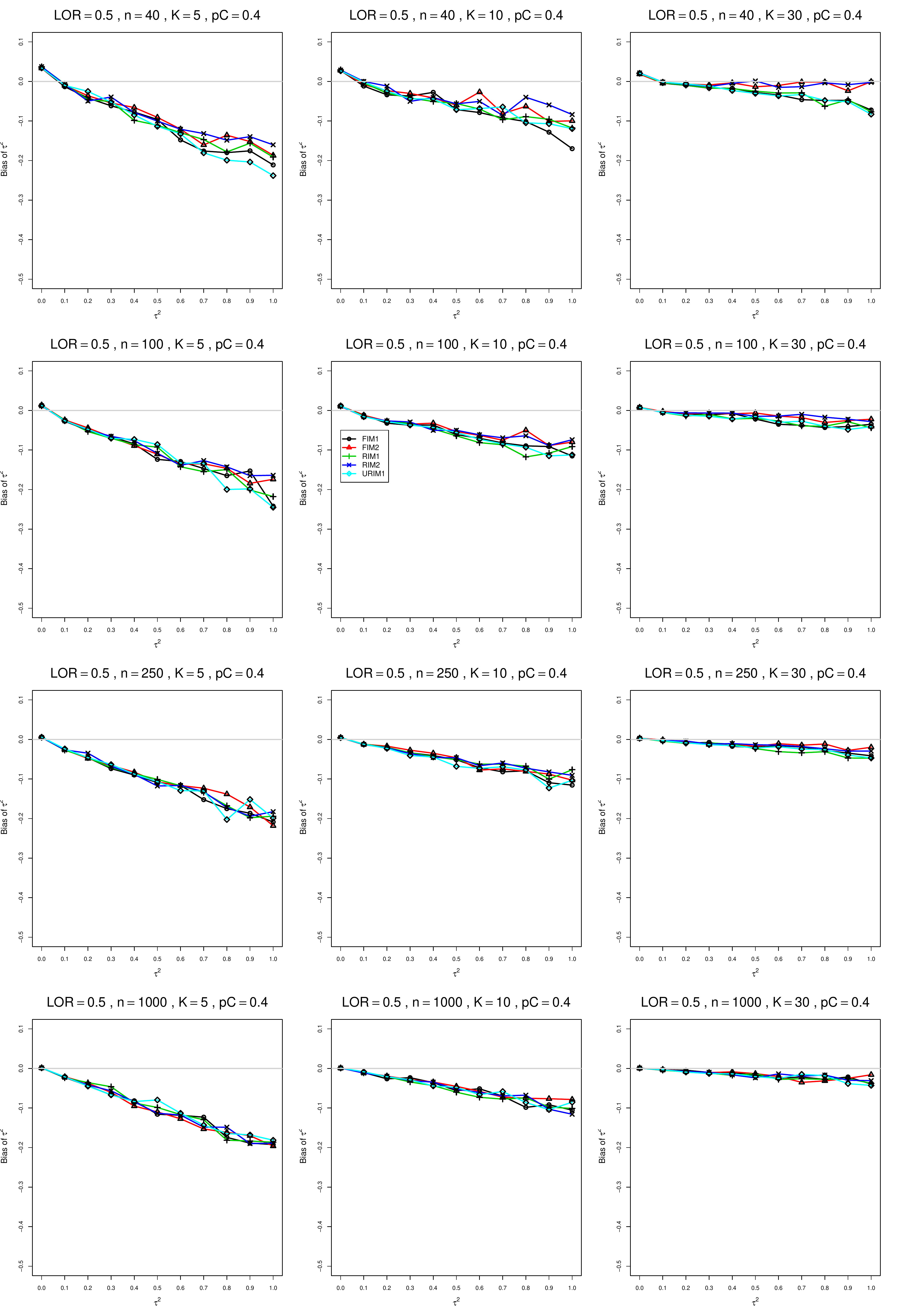}
	\caption{Bias of  between-studies variance $\hat{\tau}_{FIM2}^2$ for $\theta=0.5$, $p_{C}=0.4$, $\sigma^2=0.1$, constant sample sizes $n=40,\;100,\;250,\;1000$.
The data-generation mechanisms are FIM1 ($\circ$), FIM2 ($\triangle$), RIM1 (+), RIM2 ($\times$), and URIM1 ($\diamond$).
		\label{PlotBiasTau2mu05andpC04LOR_UMFSsigma01}}
\end{figure}
\begin{figure}[t]
	\centering
	\includegraphics[scale=0.33]{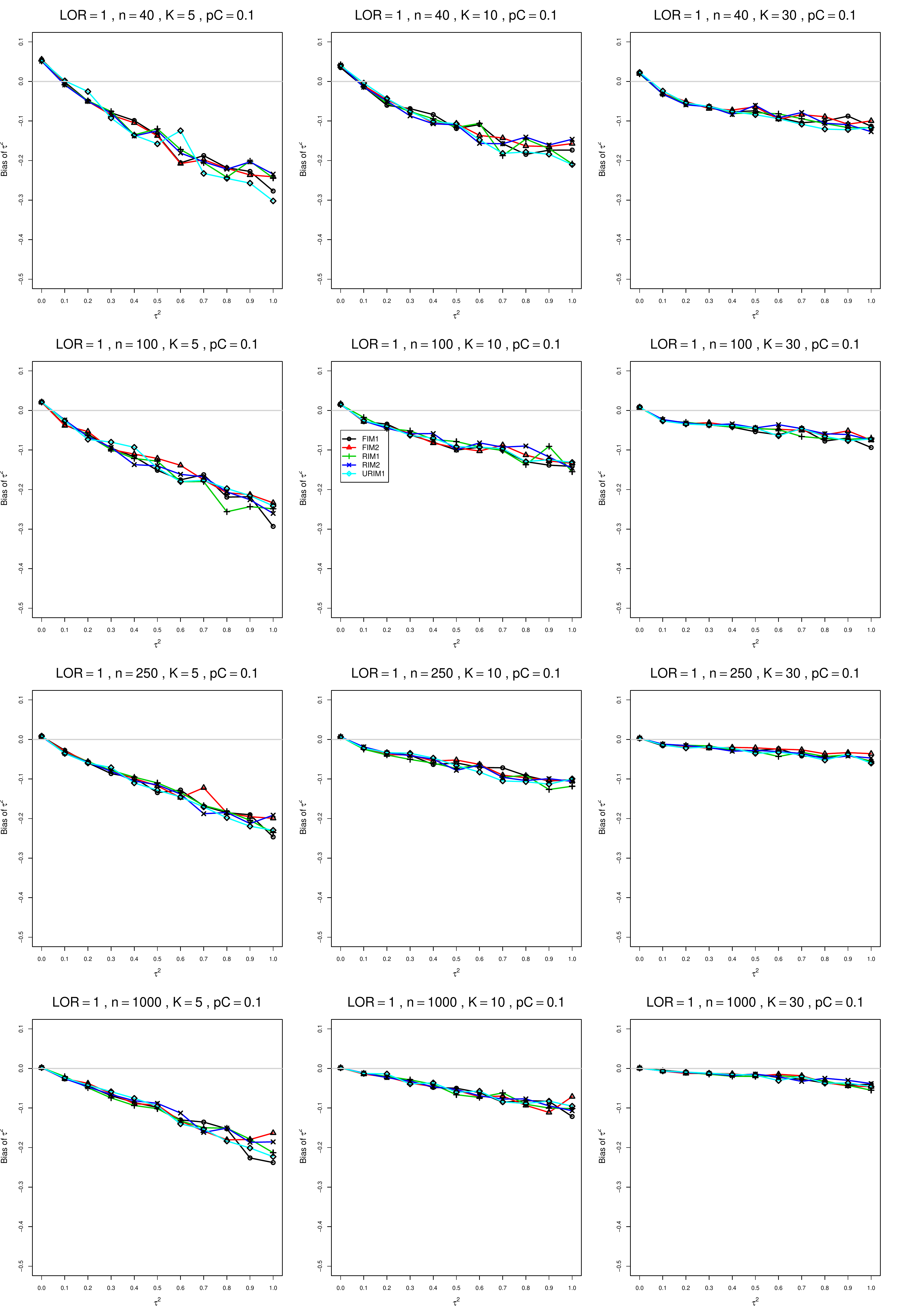}
	\caption{Bias of  between-studies variance $\hat{\tau}_{FIM2}^2$ for $\theta=1$, $p_{C}=0.1$, $\sigma^2=0.1$, constant sample sizes $n=40,\;100,\;250,\;1000$.
The data-generation mechanisms are FIM1 ($\circ$), FIM2 ($\triangle$), RIM1 (+), RIM2 ($\times$), and URIM1 ($\diamond$).
		\label{PlotBiasTau2mu1andpC01LOR_UMFSsigma01}}
\end{figure}
\begin{figure}[t]
	\centering
	\includegraphics[scale=0.33]{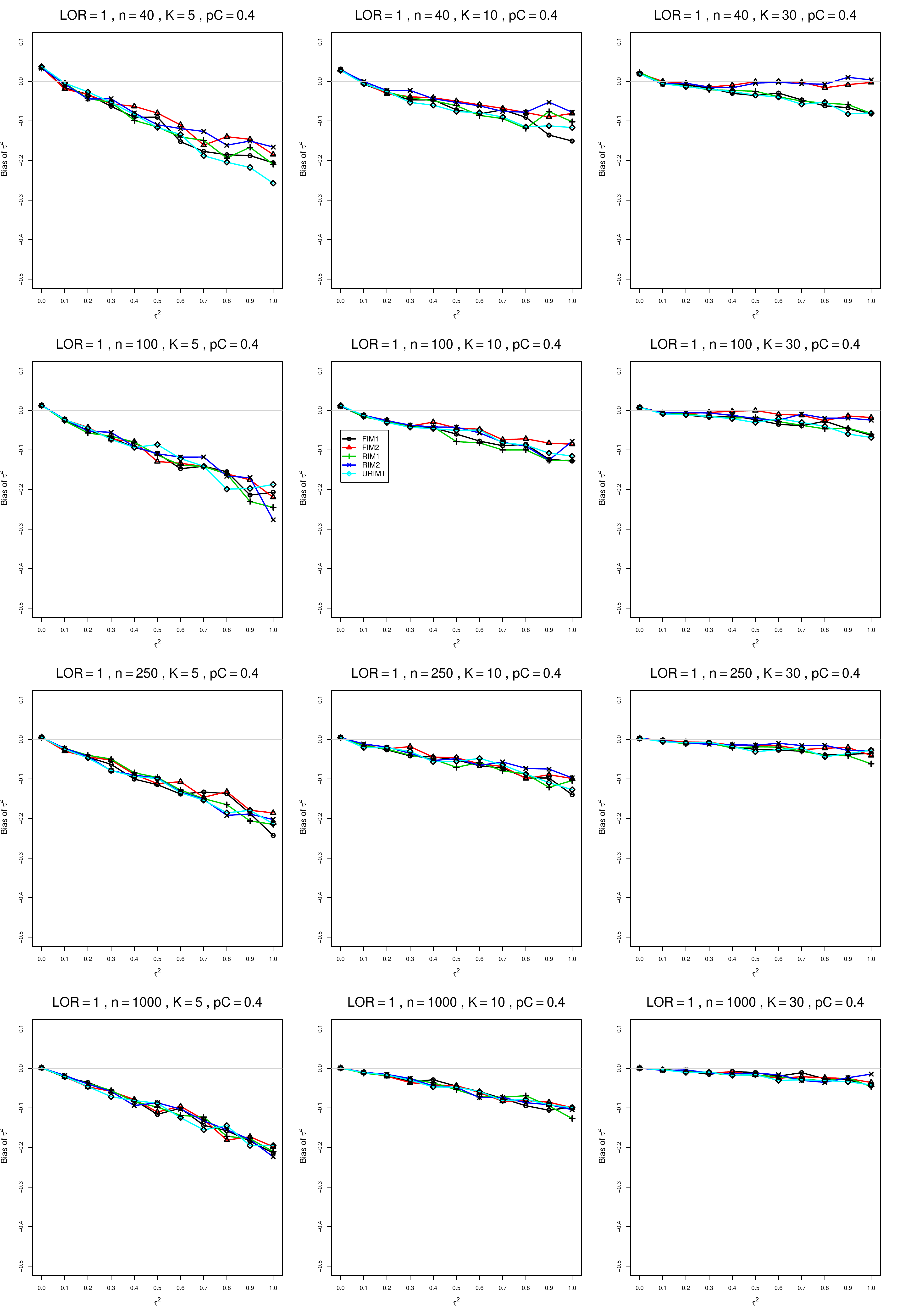}
	\caption{Bias of  between-studies variance $\hat{\tau}_{FIM2}^2$ for $\theta=1$, $p_{C}=0.4$, $\sigma^2=0.1$, constant sample sizes $n=40,\;100,\;250,\;1000$.
The data-generation mechanisms are FIM1 ($\circ$), FIM2 ($\triangle$), RIM1 (+), RIM2 ($\times$), and URIM1 ($\diamond$).
		\label{PlotBiasTau2mu1andpC04LOR_UMFSsigma01}}
\end{figure}
\begin{figure}[t]
	\centering
	\includegraphics[scale=0.33]{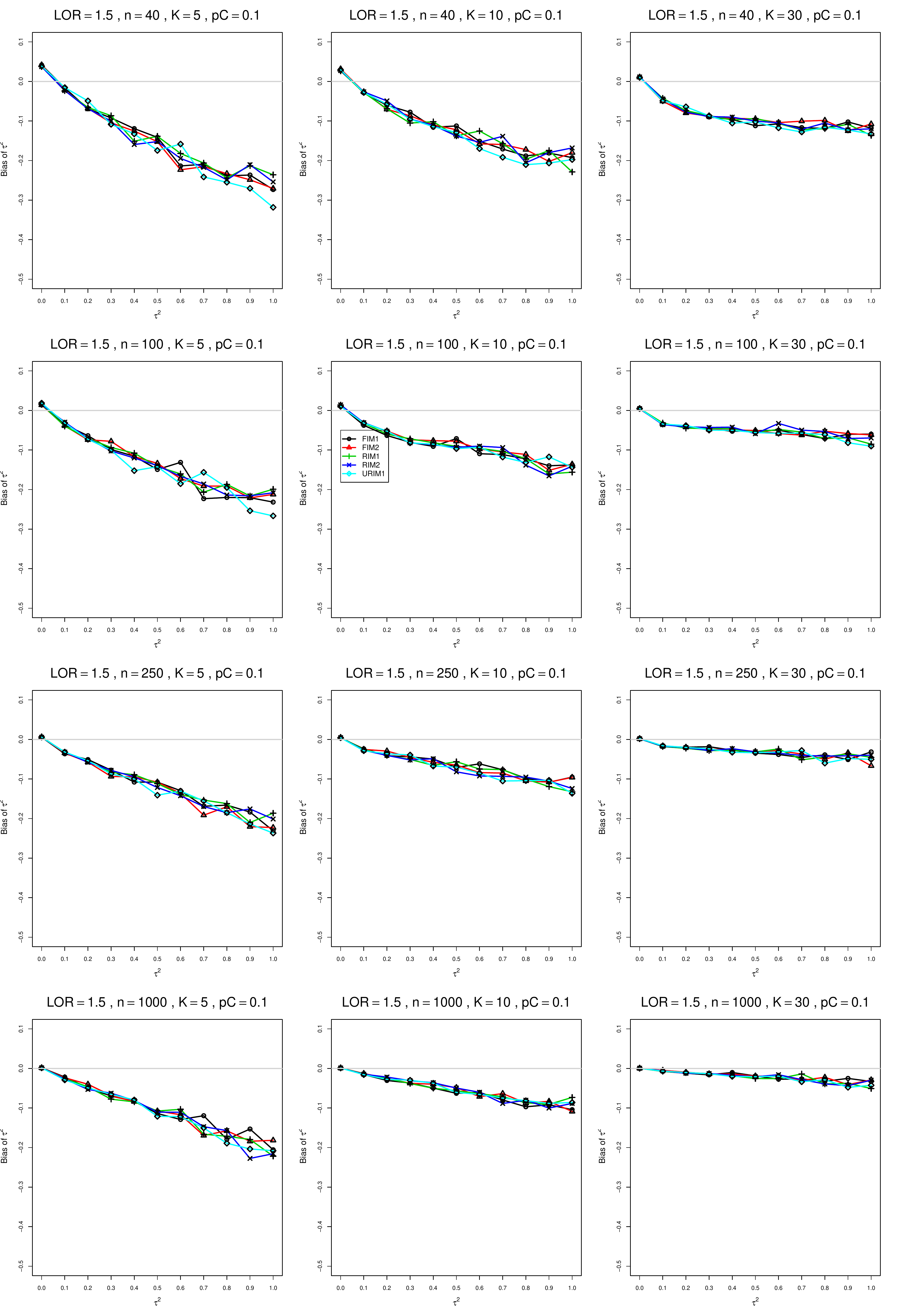}
	\caption{Bias of  between-studies variance $\hat{\tau}_{FIM2}^2$ for $\theta=1.5$, $p_{C}=0.1$, $\sigma^2=0.1$, constant sample sizes $n=40,\;100,\;250,\;1000$.
The data-generation mechanisms are FIM1 ($\circ$), FIM2 ($\triangle$), RIM1 (+), RIM2 ($\times$), and URIM1 ($\diamond$).
		\label{PlotBiasTau2mu15andpC01LOR_UMFSsigma01}}
\end{figure}
\begin{figure}[t]
	\centering
	\includegraphics[scale=0.33]{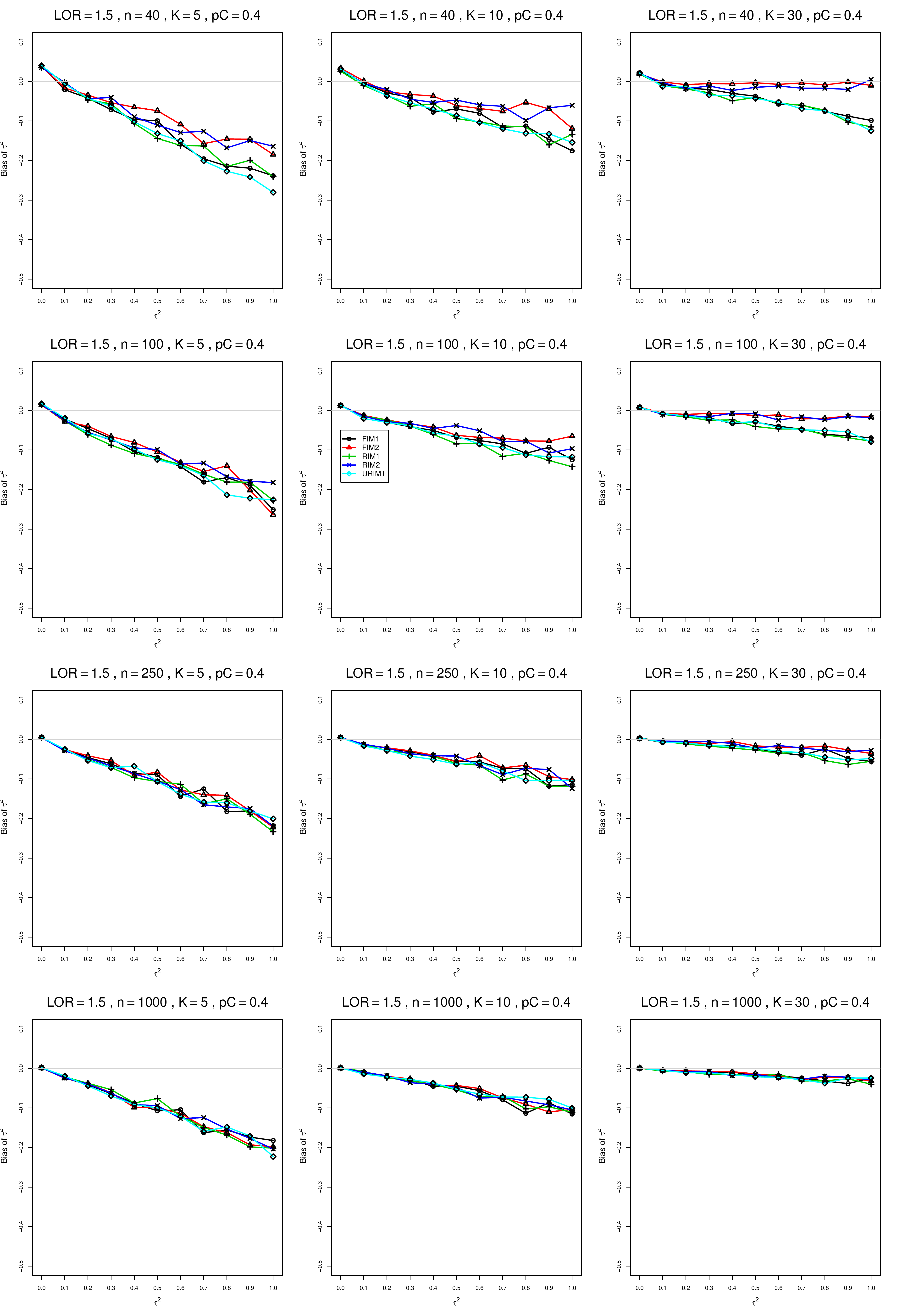}
	\caption{Bias of  between-studies variance $\hat{\tau}_{FIM2}^2$ for $\theta=1.5$, $p_{C}=0.4$, $\sigma^2=0.1$, constant sample sizes $n=40,\;100,\;250,\;1000$.
The data-generation mechanisms are FIM1 ($\circ$), FIM2 ($\triangle$), RIM1 (+), RIM2 ($\times$), and URIM1 ($\diamond$).
		\label{PlotBiasTau2mu15andpC04LOR_UMFSsigma01}}
\end{figure}
\begin{figure}[t]
	\centering
	\includegraphics[scale=0.33]{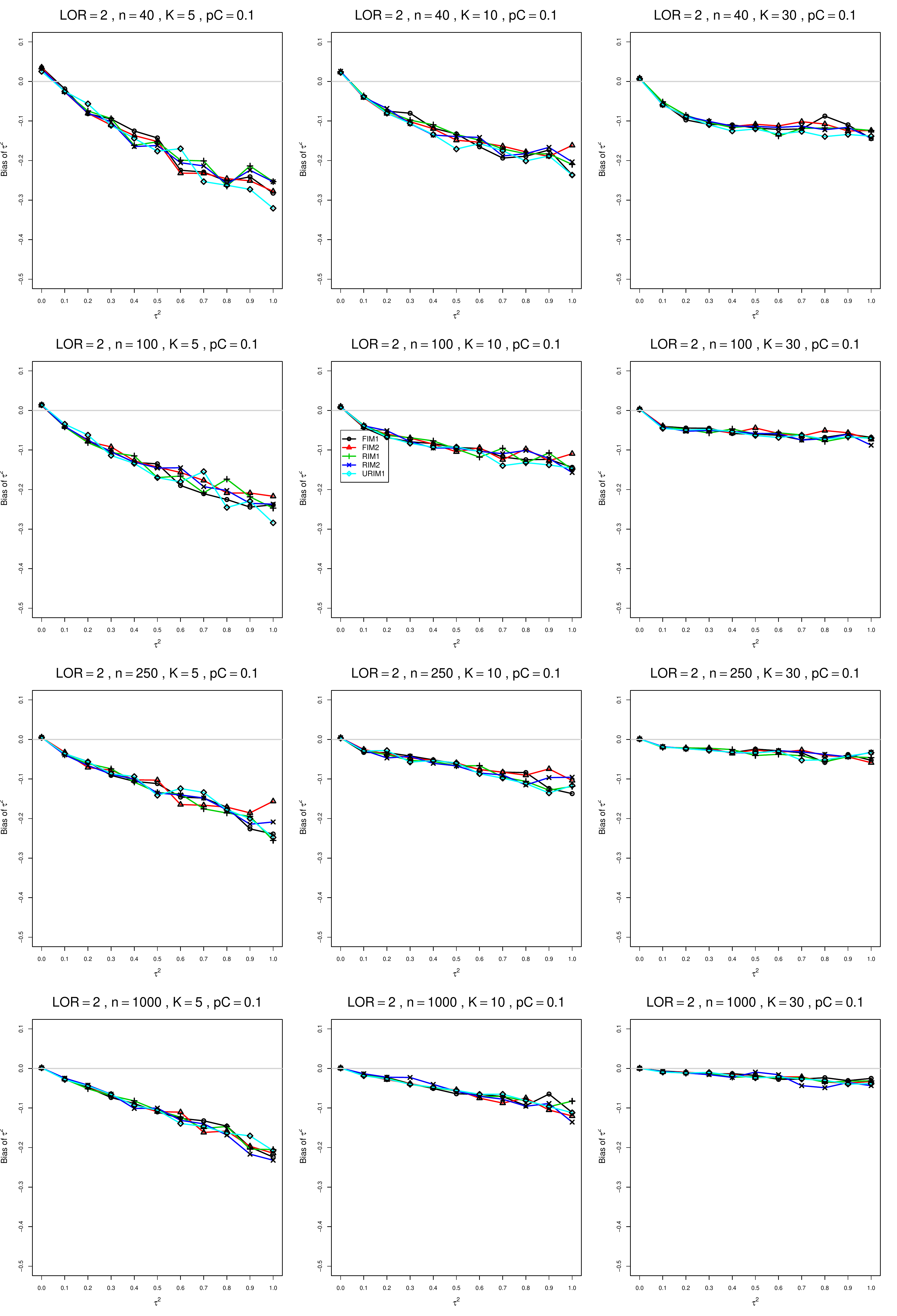}
	\caption{Bias of  between-studies variance $\hat{\tau}_{FIM2}^2$ for $\theta=2$, $p_{C}=0.1$, $\sigma^2=0.1$, constant sample sizes $n=40,\;100,\;250,\;1000$.
The data-generation mechanisms are FIM1 ($\circ$), FIM2 ($\triangle$), RIM1 (+), RIM2 ($\times$), and URIM1 ($\diamond$).
		\label{PlotBiasTau2mu2andpC01LOR_UMFSsigma01}}
\end{figure}
\begin{figure}[t]
	\centering
	\includegraphics[scale=0.33]{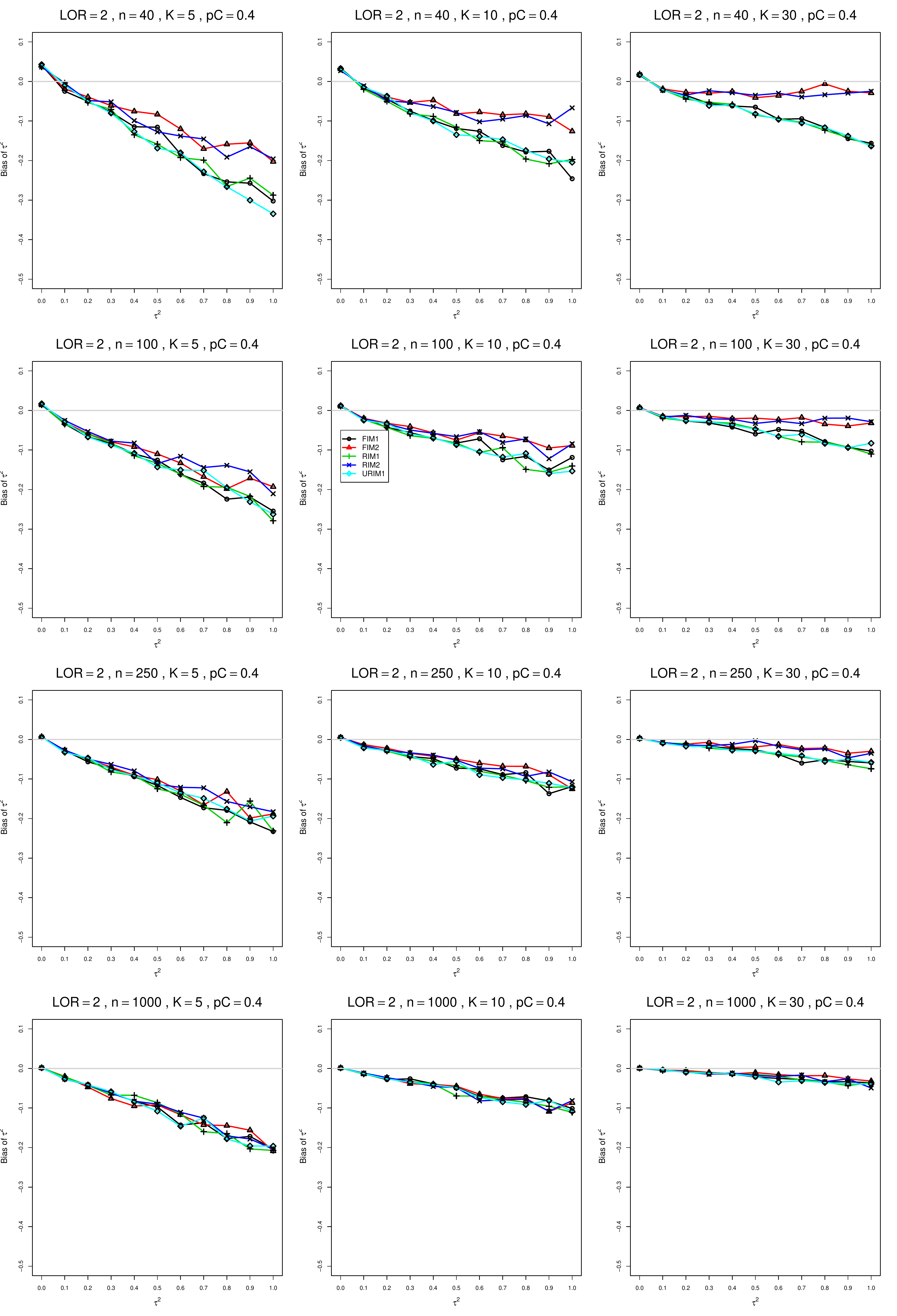}
	\caption{Bias of  between-studies variance $\hat{\tau}_{FIM2}^2$ for $\theta=2$, $p_{C}=0.4$, $\sigma^2=0.1$, constant sample sizes $n=40,\;100,\;250,\;1000$.
The data-generation mechanisms are FIM1 ($\circ$), FIM2 ($\triangle$), RIM1 (+), RIM2 ($\times$), and URIM1 ($\diamond$).
		\label{PlotBiasTau2mu2andpC04LOR_UMFSsigma01}}
\end{figure}
\begin{figure}[t]
	\centering
	\includegraphics[scale=0.33]{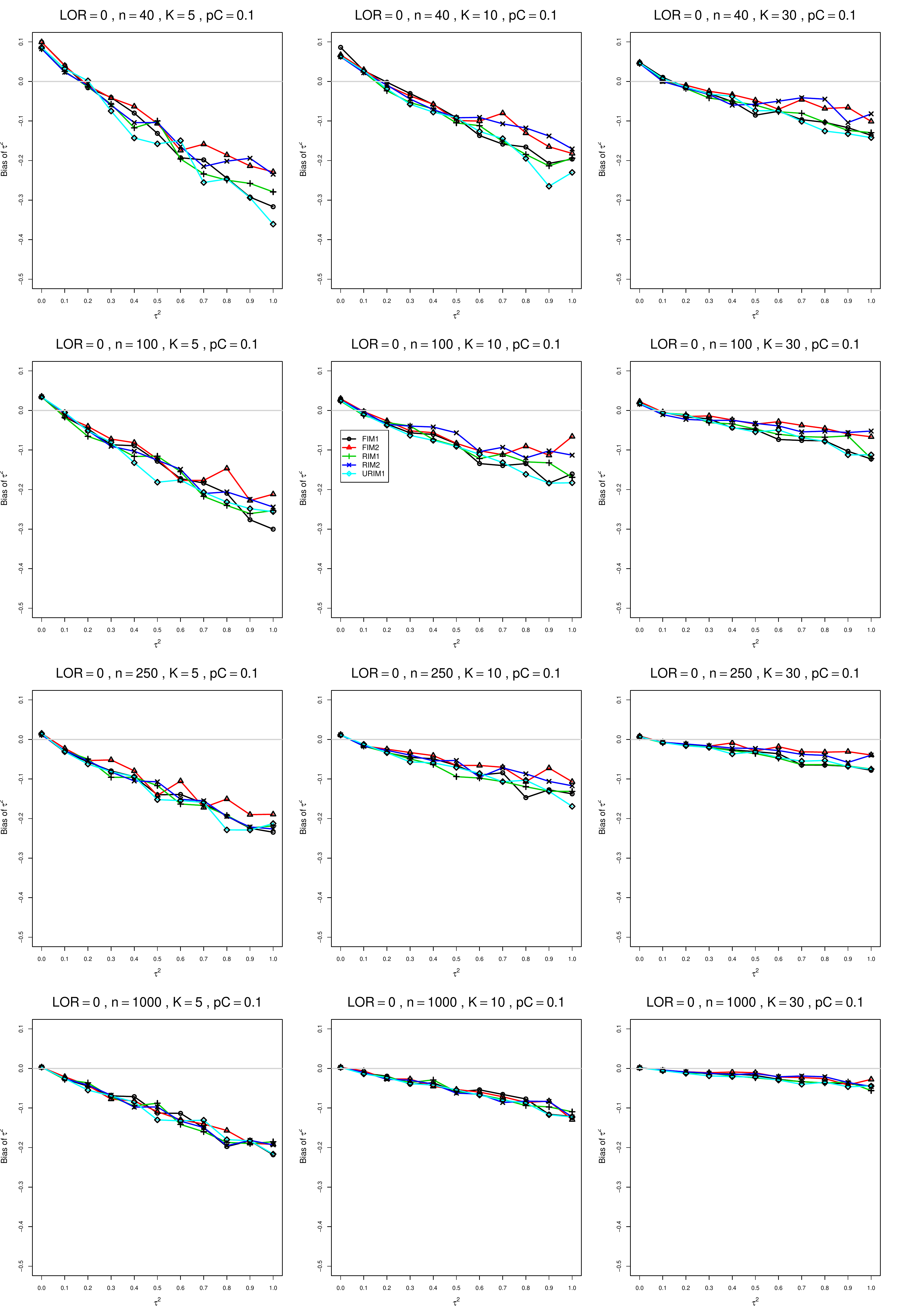}
	\caption{Bias of  between-studies variance $\hat{\tau}_{FIM2}^2$ for $\theta=0$, $p_{C}=0.1$, $\sigma^2=0.4$, constant sample sizes $n=40,\;100,\;250,\;1000$.
The data-generation mechanisms are FIM1 ($\circ$), FIM2 ($\triangle$), RIM1 (+), RIM2 ($\times$), and URIM1 ($\diamond$).
		\label{PlotBiasTau2mu0andpC01LOR_UMFSsigma04}}
\end{figure}
\begin{figure}[t]
	\centering
	\includegraphics[scale=0.33]{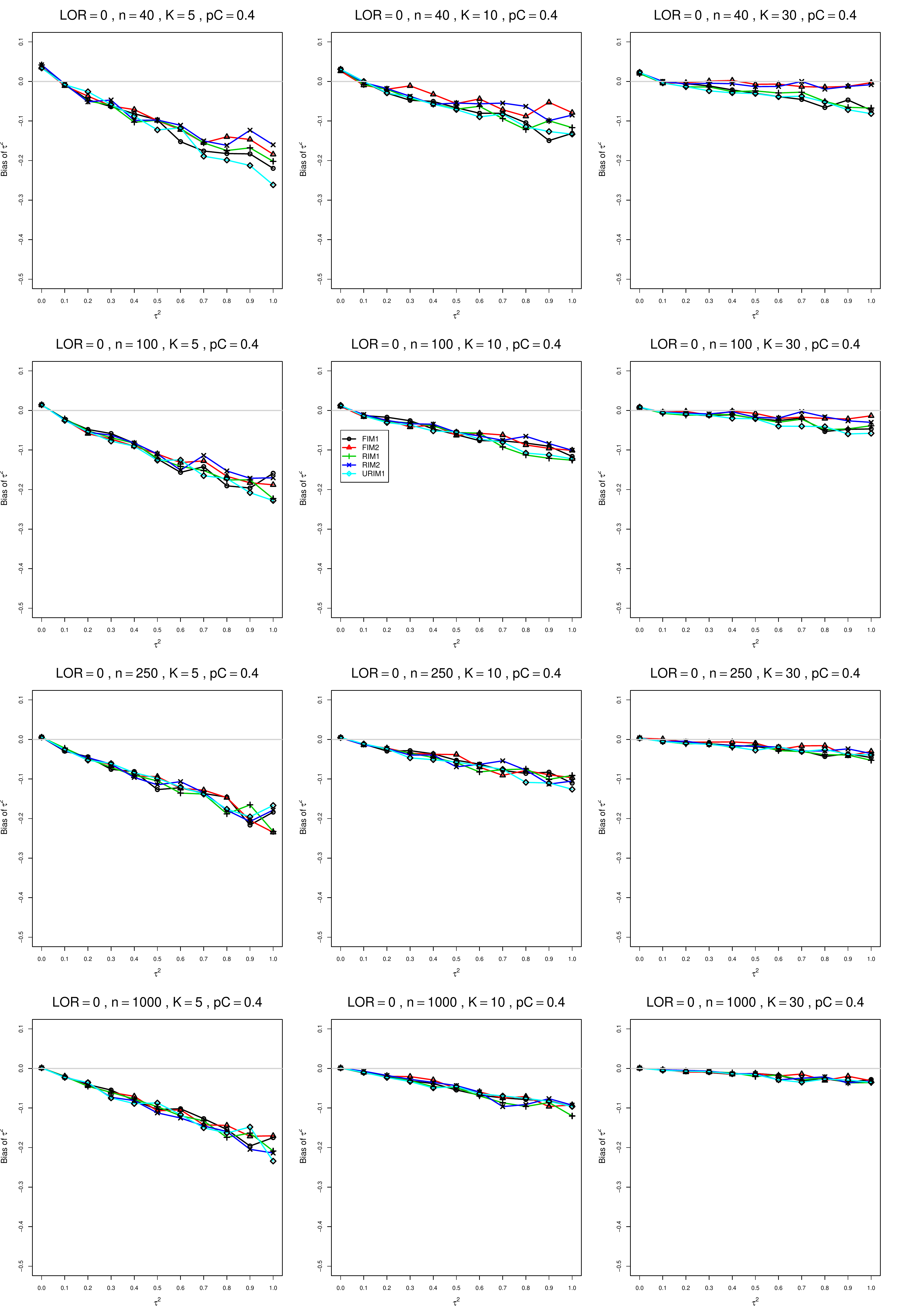}
	\caption{Bias of  between-studies variance $\hat{\tau}_{FIM2}^2$ for $\theta=0$, $p_{C}=0.4$, $\sigma^2=0.4$, constant sample sizes $n=40,\;100,\;250,\;1000$.
The data-generation mechanisms are FIM1 ($\circ$), FIM2 ($\triangle$), RIM1 (+), RIM2 ($\times$), and URIM1 ($\diamond$).
		\label{PlotBiasTau2mu0andpC04LOR_UMFSsigma04}}
\end{figure}
\begin{figure}[t]
	\centering
	\includegraphics[scale=0.33]{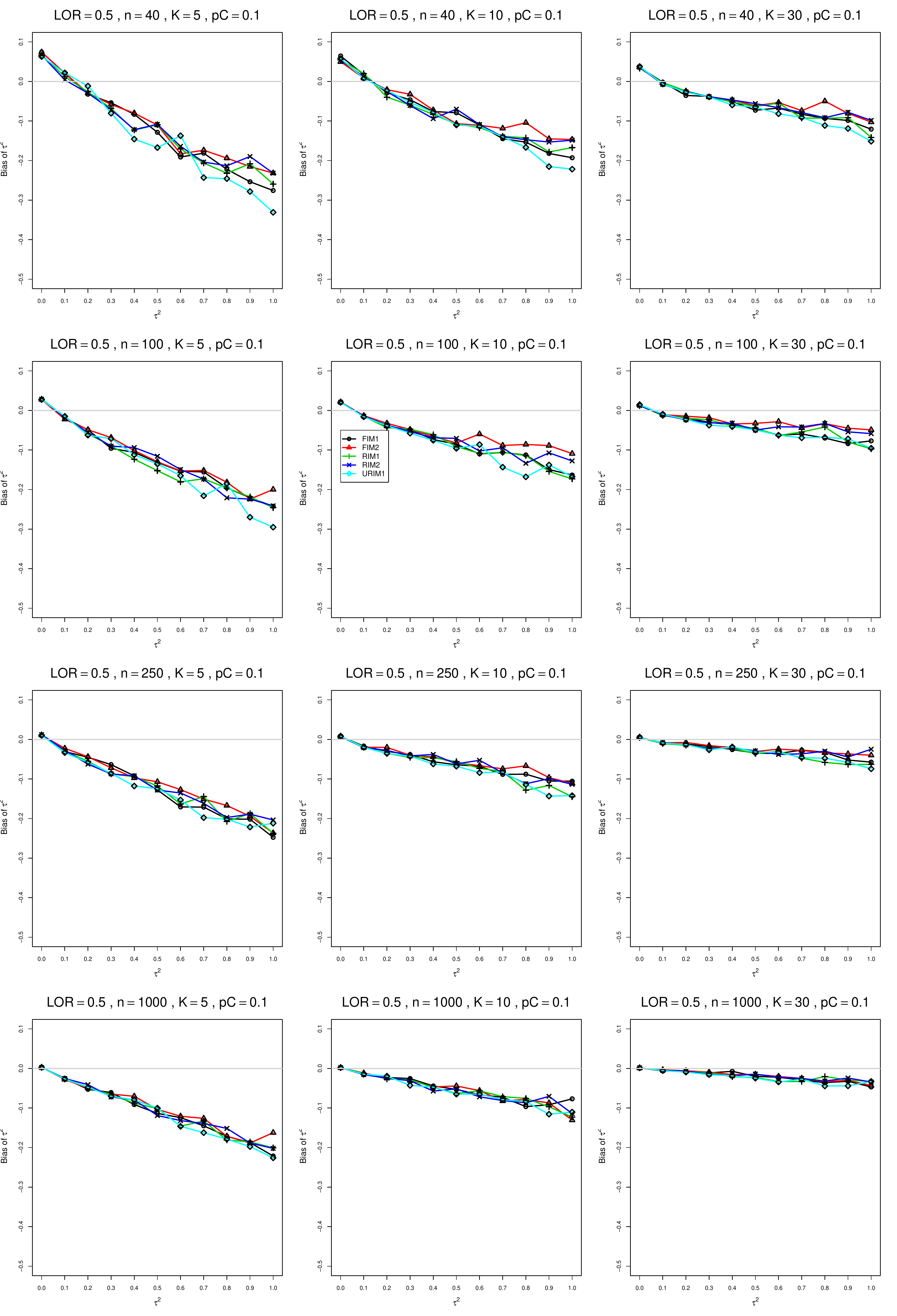}
	\caption{Bias of  between-studies variance $\hat{\tau}_{FIM2}^2$ for $\theta=0.5$, $p_{C}=0.1$, $\sigma^2=0.4$, constant sample sizes $n=40,\;100,\;250,\;1000$.
The data-generation mechanisms are FIM1 ($\circ$), FIM2 ($\triangle$), RIM1 (+), RIM2 ($\times$), and URIM1 ($\diamond$).
		\label{PlotBiasTau2mu05andpC01LOR_UMFSsigma04}}
\end{figure}
\begin{figure}[t]
	\centering
	\includegraphics[scale=0.33]{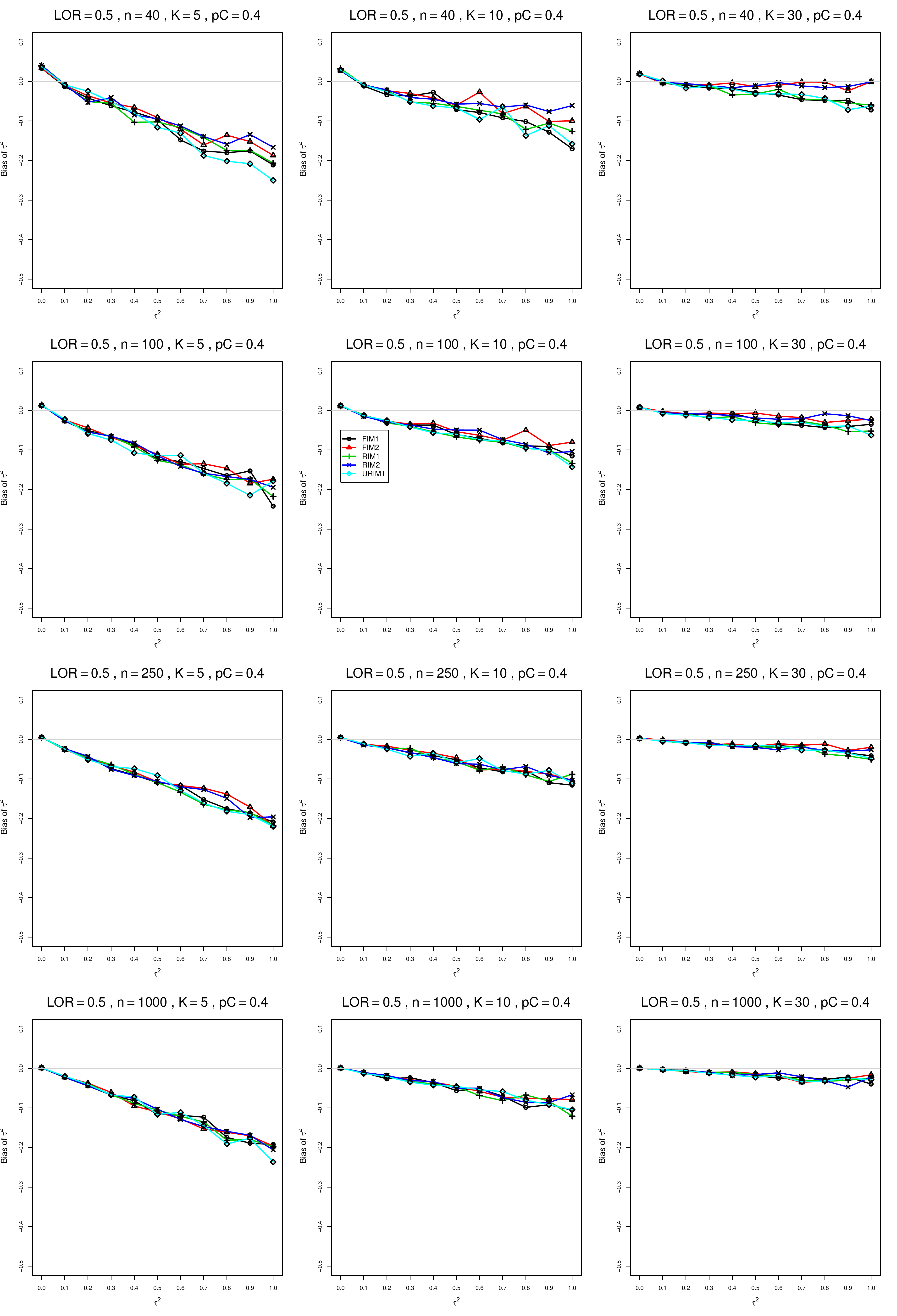}
	\caption{Bias of  between-studies variance $\hat{\tau}_{FIM2}^2$ for $\theta=0.5$, $p_{C}=0.4$, $\sigma^2=0.4$, constant sample sizes $n=40,\;100,\;250,\;1000$.
The data-generation mechanisms are FIM1 ($\circ$), FIM2 ($\triangle$), RIM1 (+), RIM2 ($\times$), and URIM1 ($\diamond$).
		\label{PlotBiasTau2mu05andpC04LOR_UMFSsigma04}}
\end{figure}
\begin{figure}[t]
	\centering
	\includegraphics[scale=0.33]{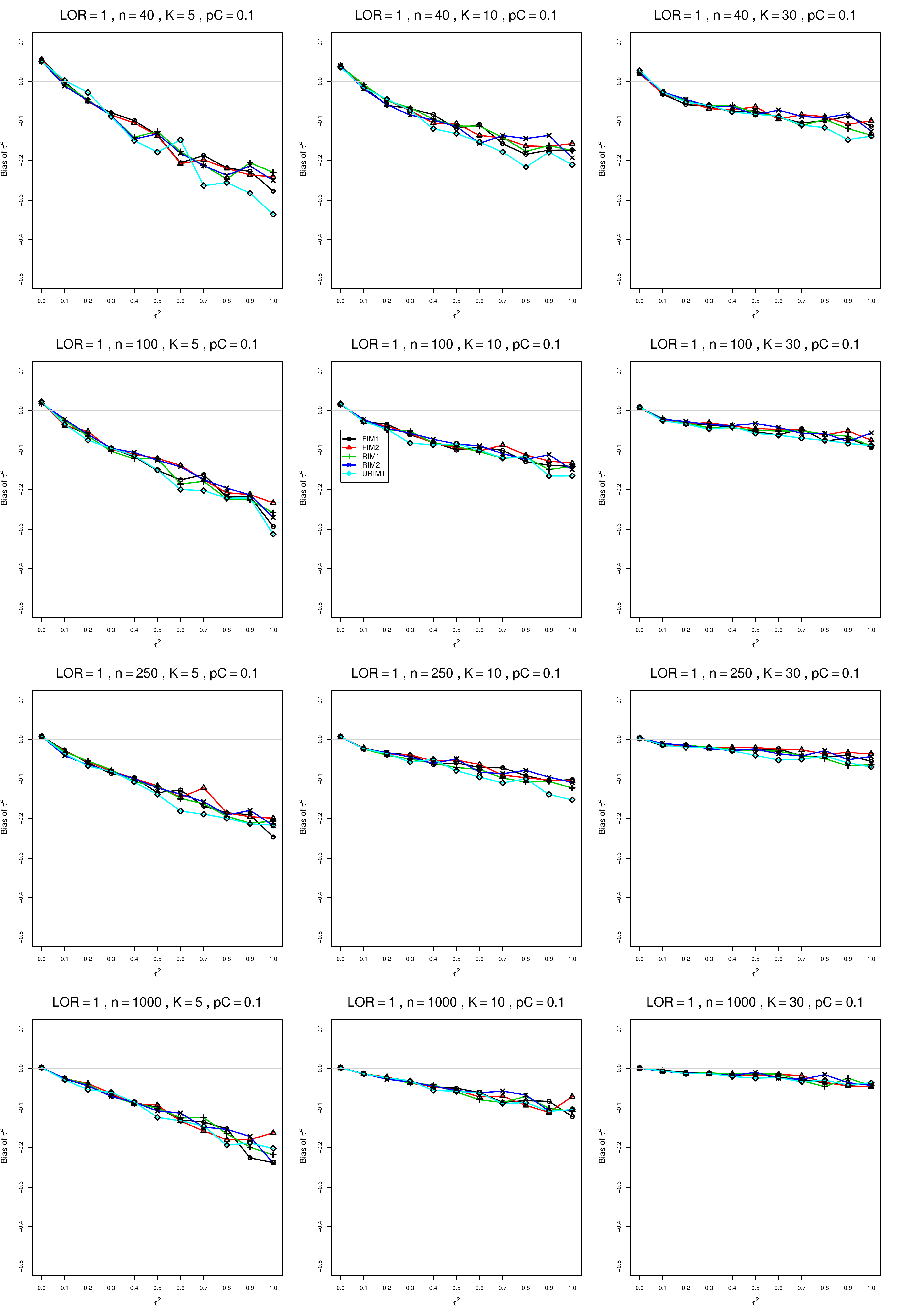}
	\caption{Bias of  between-studies variance $\hat{\tau}_{FIM2}^2$ for $\theta=1$, $p_{C}=0.1$, $\sigma^2=0.4$, constant sample sizes $n=40,\;100,\;250,\;1000$.
The data-generation mechanisms are FIM1 ($\circ$), FIM2 ($\triangle$), RIM1 (+), RIM2 ($\times$), and URIM1 ($\diamond$).
		\label{PlotBiasTau2mu1andpC01LOR_UMFSsigma04}}
\end{figure}
\begin{figure}[t]
	\centering
	\includegraphics[scale=0.33]{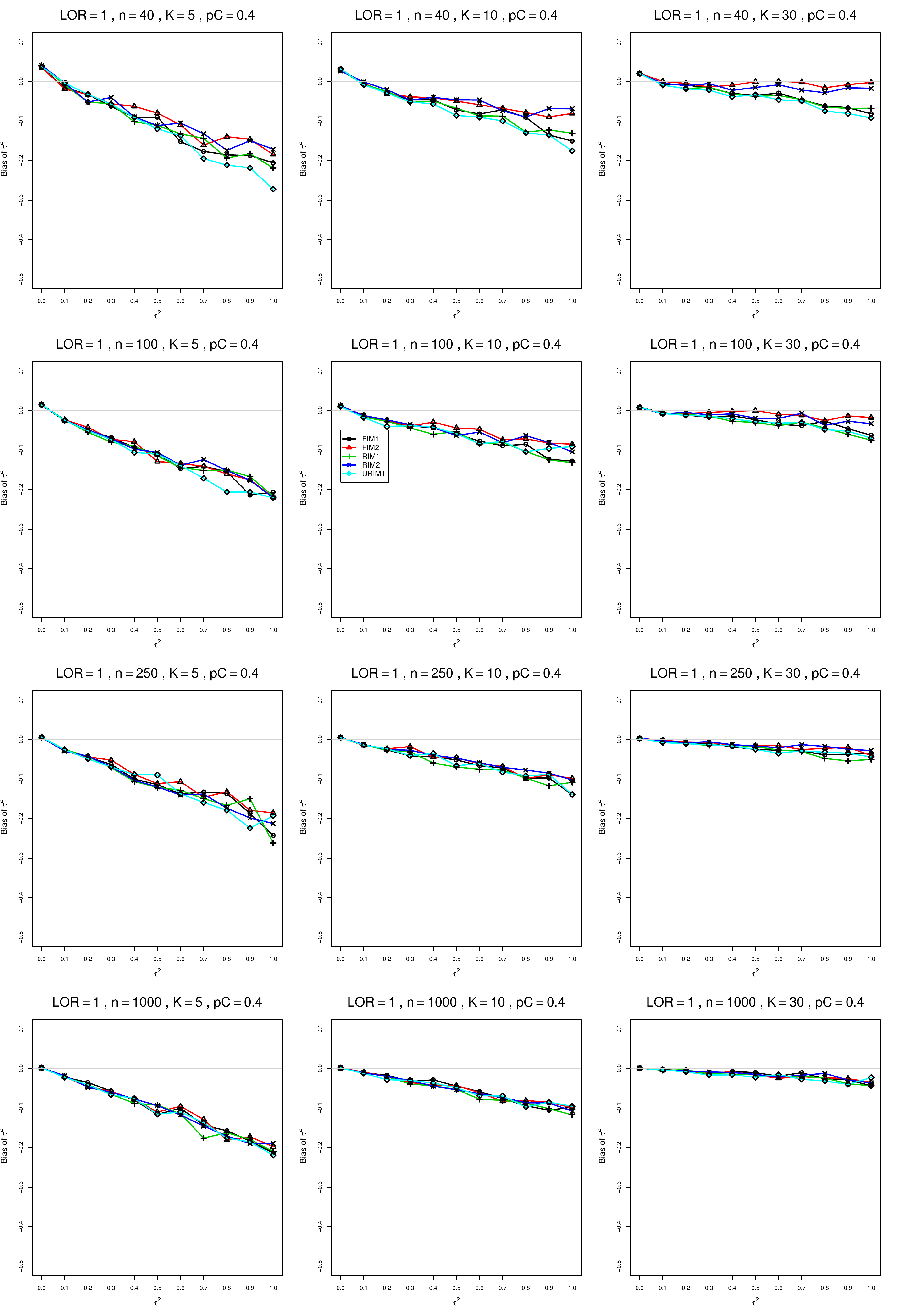}
	\caption{Bias of  between-studies variance $\hat{\tau}_{FIM2}^2$ for $\theta=1$, $p_{C}=0.4$, $\sigma^2=0.4$, constant sample sizes $n=40,\;100,\;250,\;1000$.
The data-generation mechanisms are FIM1 ($\circ$), FIM2 ($\triangle$), RIM1 (+), RIM2 ($\times$), and URIM1 ($\diamond$).
		\label{PlotBiasTau2mu1andpC04LOR_UMFSsigma04}}
\end{figure}
\begin{figure}[t]
	\centering
	\includegraphics[scale=0.33]{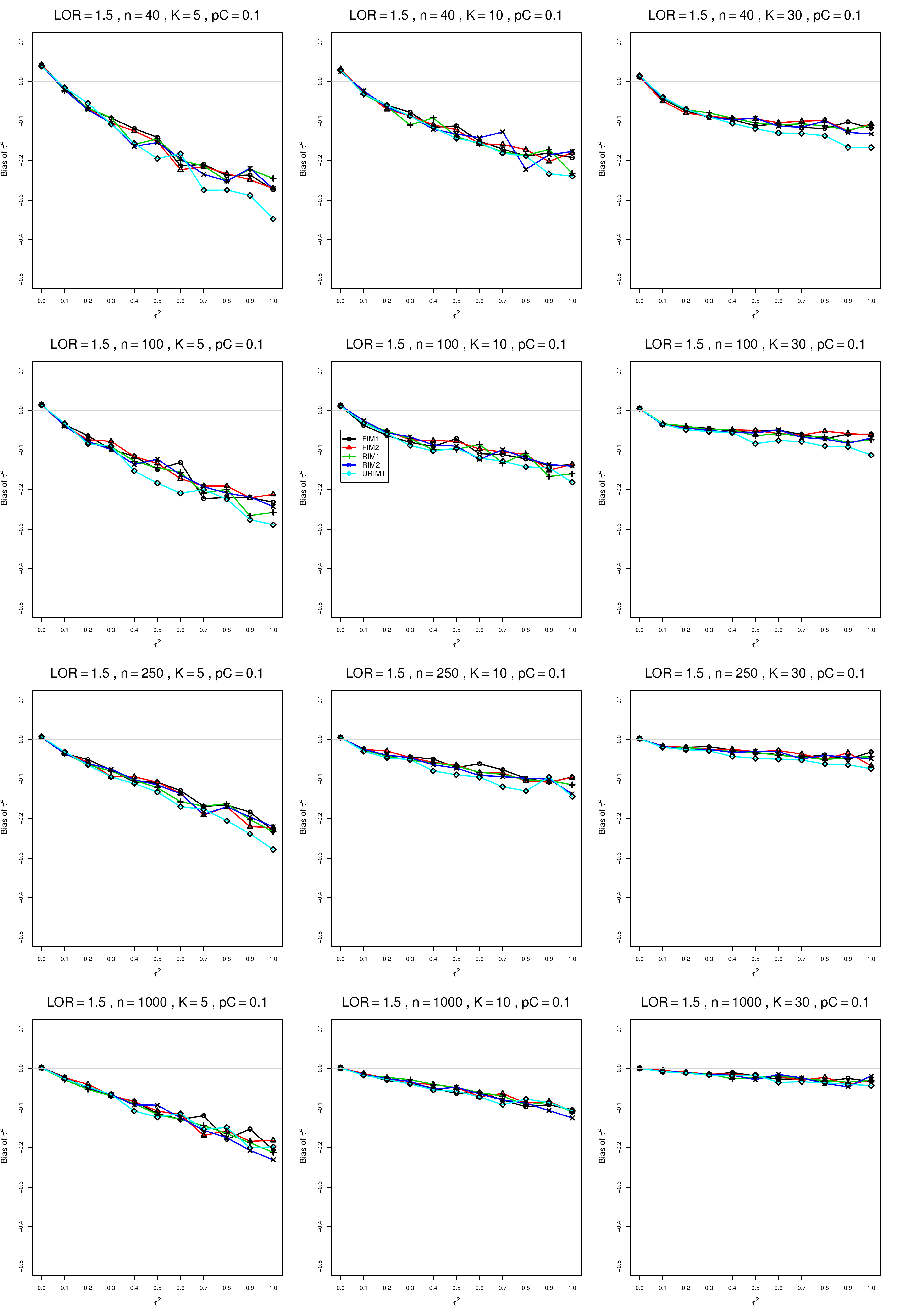}
	\caption{Bias of  between-studies variance $\hat{\tau}_{FIM2}^2$ for $\theta=1.5$, $p_{C}=0.1$, $\sigma^2=0.4$, constant sample sizes $n=40,\;100,\;250,\;1000$.
The data-generation mechanisms are FIM1 ($\circ$), FIM2 ($\triangle$), RIM1 (+), RIM2 ($\times$), and URIM1 ($\diamond$).
		\label{PlotBiasTau2mu15andpC01LOR_UMFSsigma04}}
\end{figure}
\begin{figure}[t]
	\centering
	\includegraphics[scale=0.33]{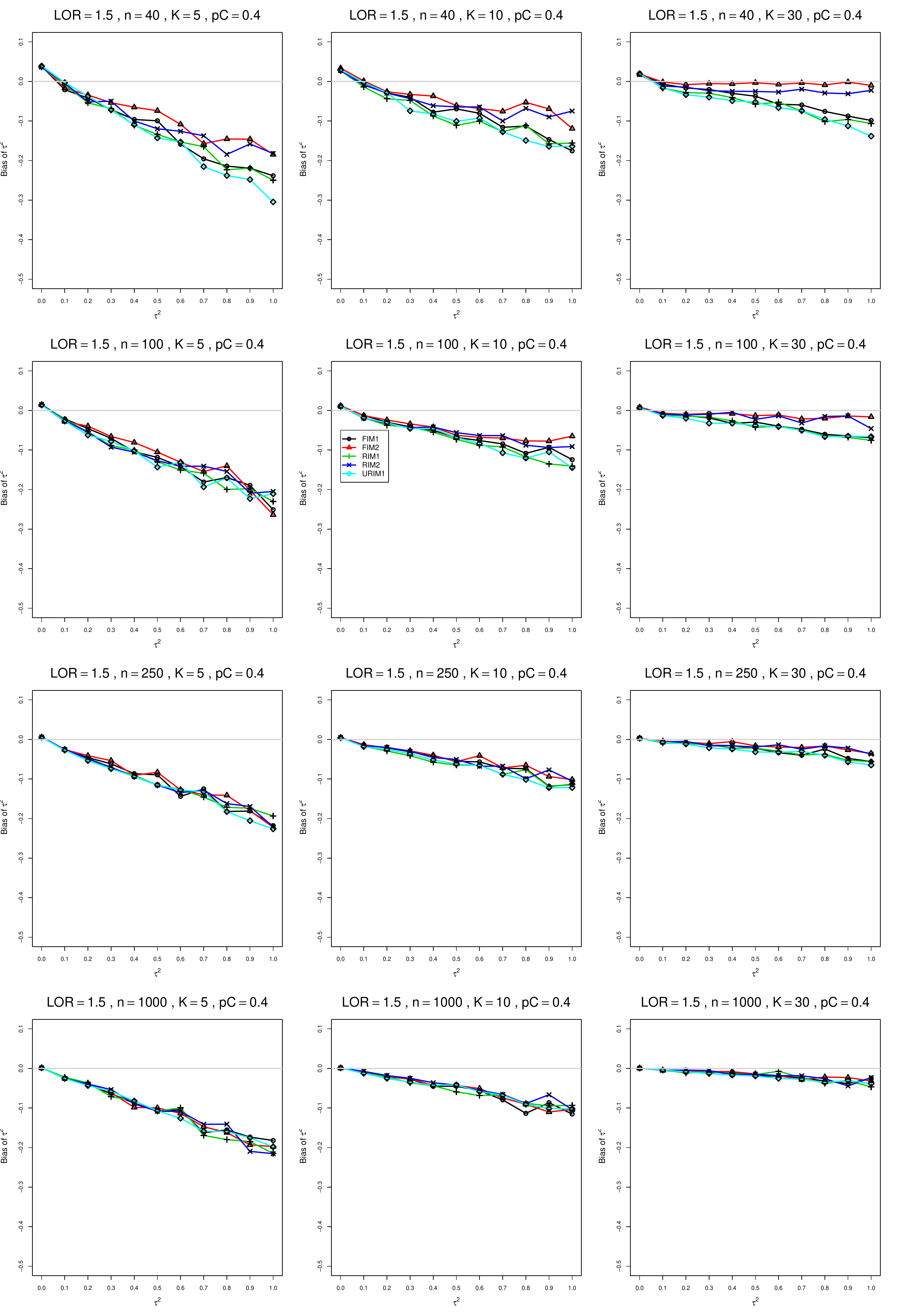}
	\caption{Bias of  between-studies variance $\hat{\tau}_{FIM2}^2$ for $\theta=1.5$, $p_{C}=0.4$, $\sigma^2=0.4$, constant sample sizes $n=40,\;100,\;250,\;1000$.
The data-generation mechanisms are FIM1 ($\circ$), FIM2 ($\triangle$), RIM1 (+), RIM2 ($\times$), and URIM1 ($\diamond$).
		\label{PlotBiasTau2mu15andpC04LOR_UMFSsigma04}}
\end{figure}
\begin{figure}[t]
	\centering
	\includegraphics[scale=0.33]{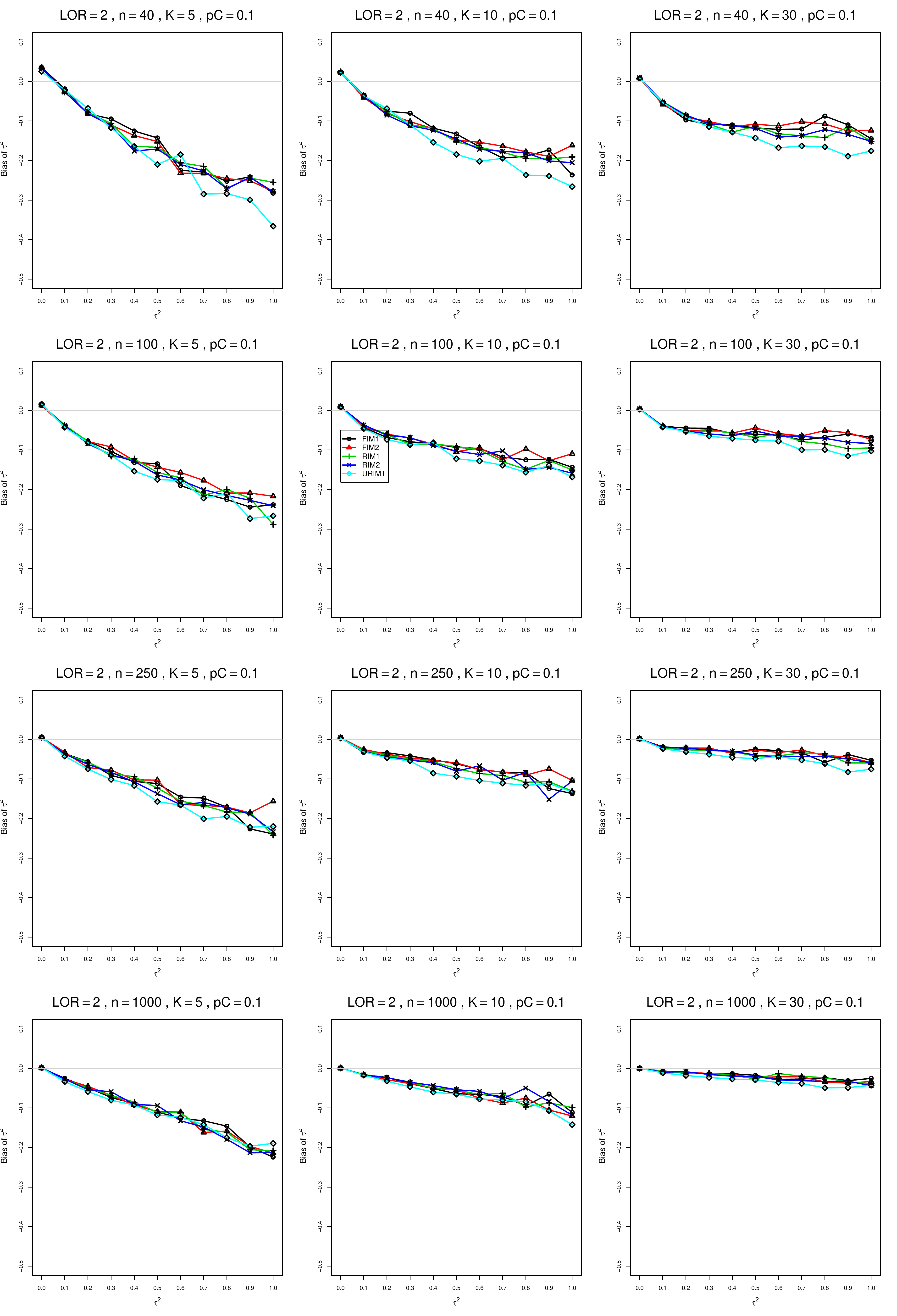}
	\caption{Bias of  between-studies variance $\hat{\tau}_{FIM2}^2$ for $\theta=2$, $p_{C}=0.1$, $\sigma^2=0.4$, constant sample sizes $n=40,\;100,\;250,\;1000$.
The data-generation mechanisms are FIM1 ($\circ$), FIM2 ($\triangle$), RIM1 (+), RIM2 ($\times$), and URIM1 ($\diamond$).
		\label{PlotBiasTau2mu2andpC01LOR_UMFSsigma04}}
\end{figure}
\begin{figure}[t]
	\centering
	\includegraphics[scale=0.33]{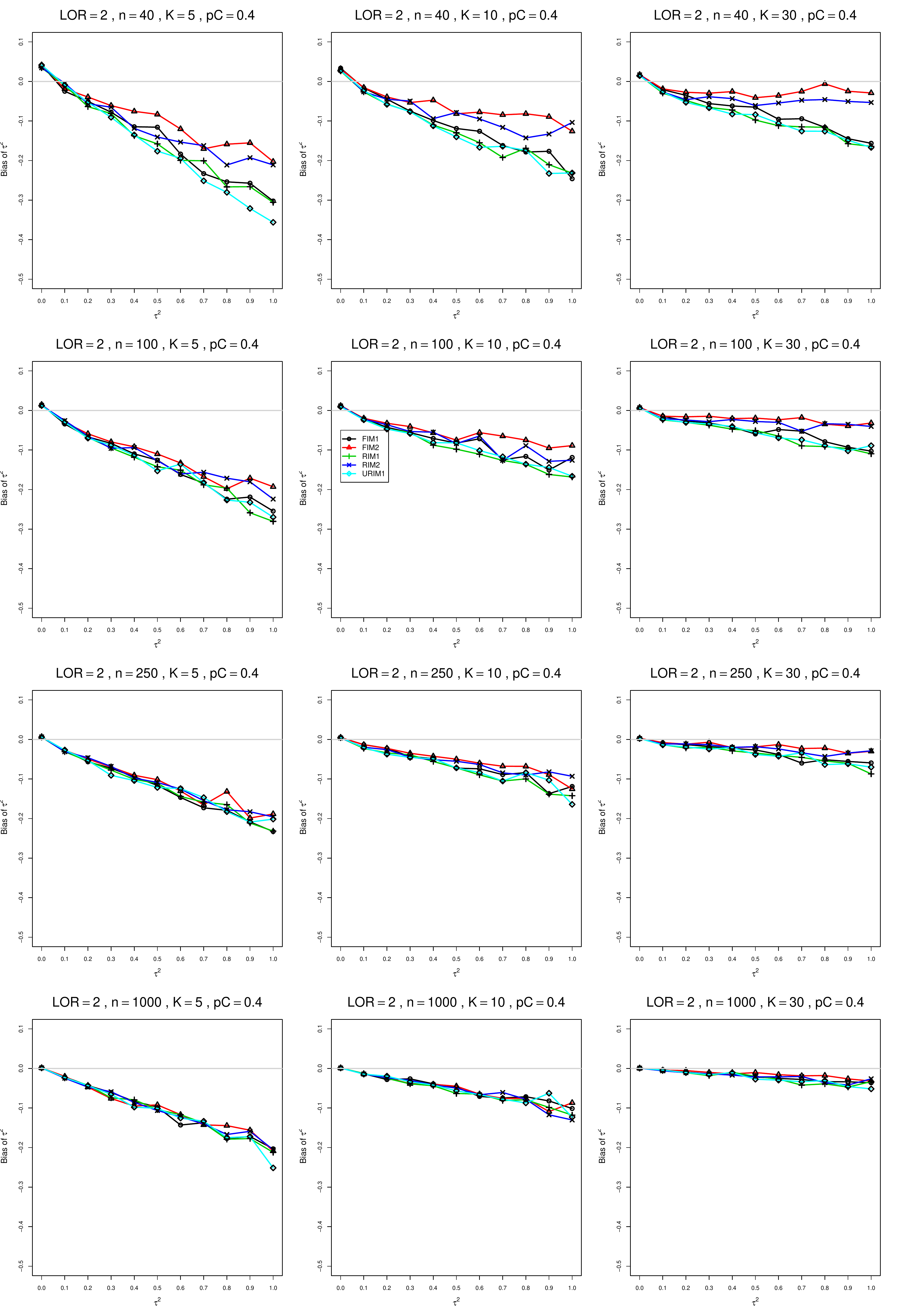}
	\caption{Bias of  between-studies variance $\hat{\tau}_{FIM2}^2$ for $\theta=2$, $p_{C}=0.4$, $\sigma^2=0.4$, constant sample sizes $n=40,\;100,\;250,\;1000$.
The data-generation mechanisms are FIM1 ($\circ$), FIM2 ($\triangle$), RIM1 (+), RIM2 ($\times$), and URIM1 ($\diamond$).
		\label{PlotBiasTau2mu2andpC04LOR_UMFSsigma04}}
\end{figure}

\clearpage
\subsection*{A1.6 Bias of $\hat{\tau}_{RIM2}^2$}
\renewcommand{\thefigure}{A1.6.\arabic{figure}}
\setcounter{figure}{0}

\begin{figure}[t]
	\centering
	\includegraphics[scale=0.33]{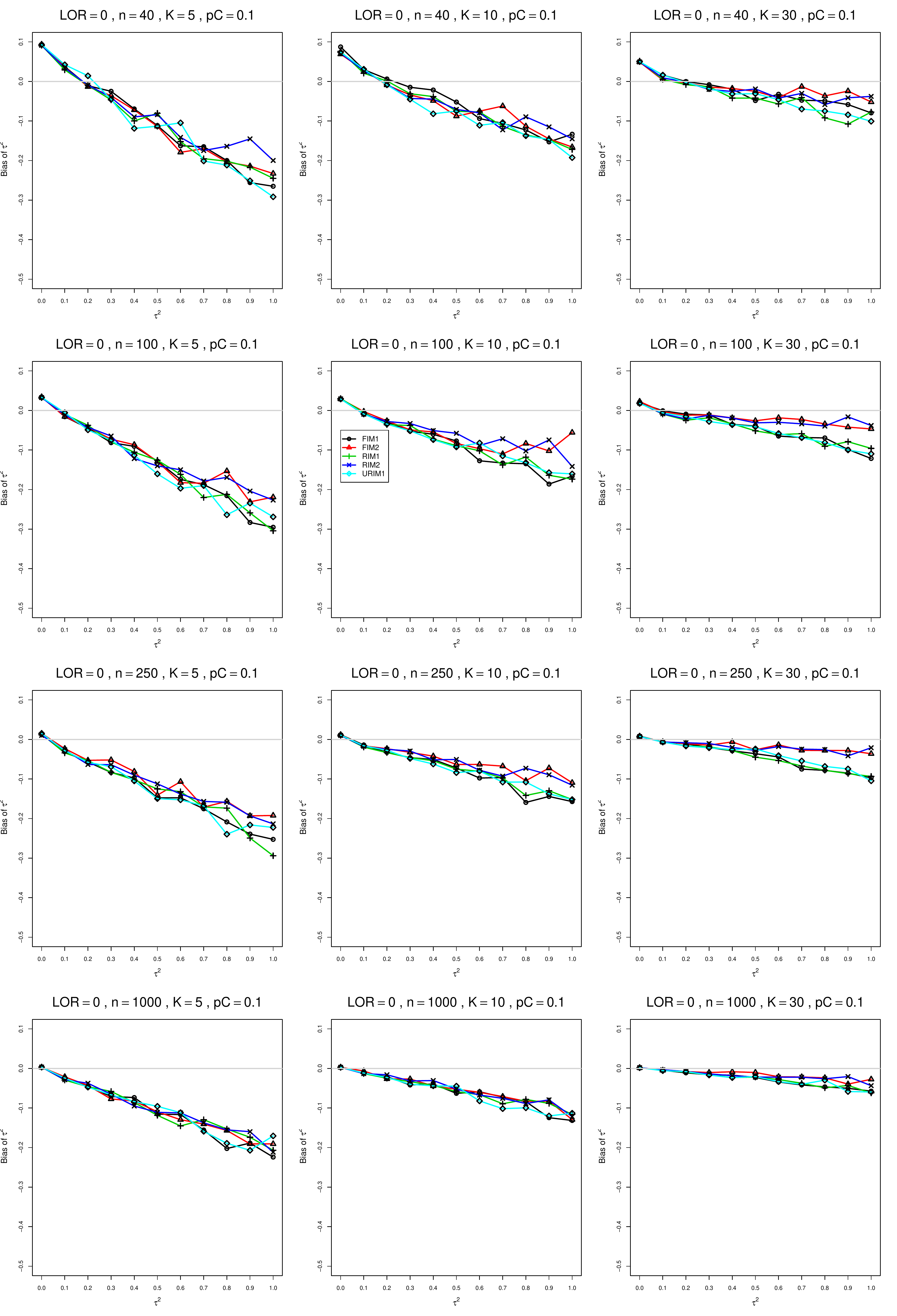}
	\caption{Bias of  between-studies variance $\hat{\tau}_{RIM2}^2$ for $\theta=0$, $p_{C}=0.1$, $\sigma^2=0.1$, constant sample sizes $n=40,\;100,\;250,\;1000$.
The data-generation mechanisms are FIM1 ($\circ$), FIM2 ($\triangle$), RIM1 (+), RIM2 ($\times$), and URIM1 ($\diamond$).
		\label{PlotBiasTau2mu0andpC01LOR_UMRSsigma01}}
\end{figure}
\begin{figure}[t]
	\centering
	\includegraphics[scale=0.33]{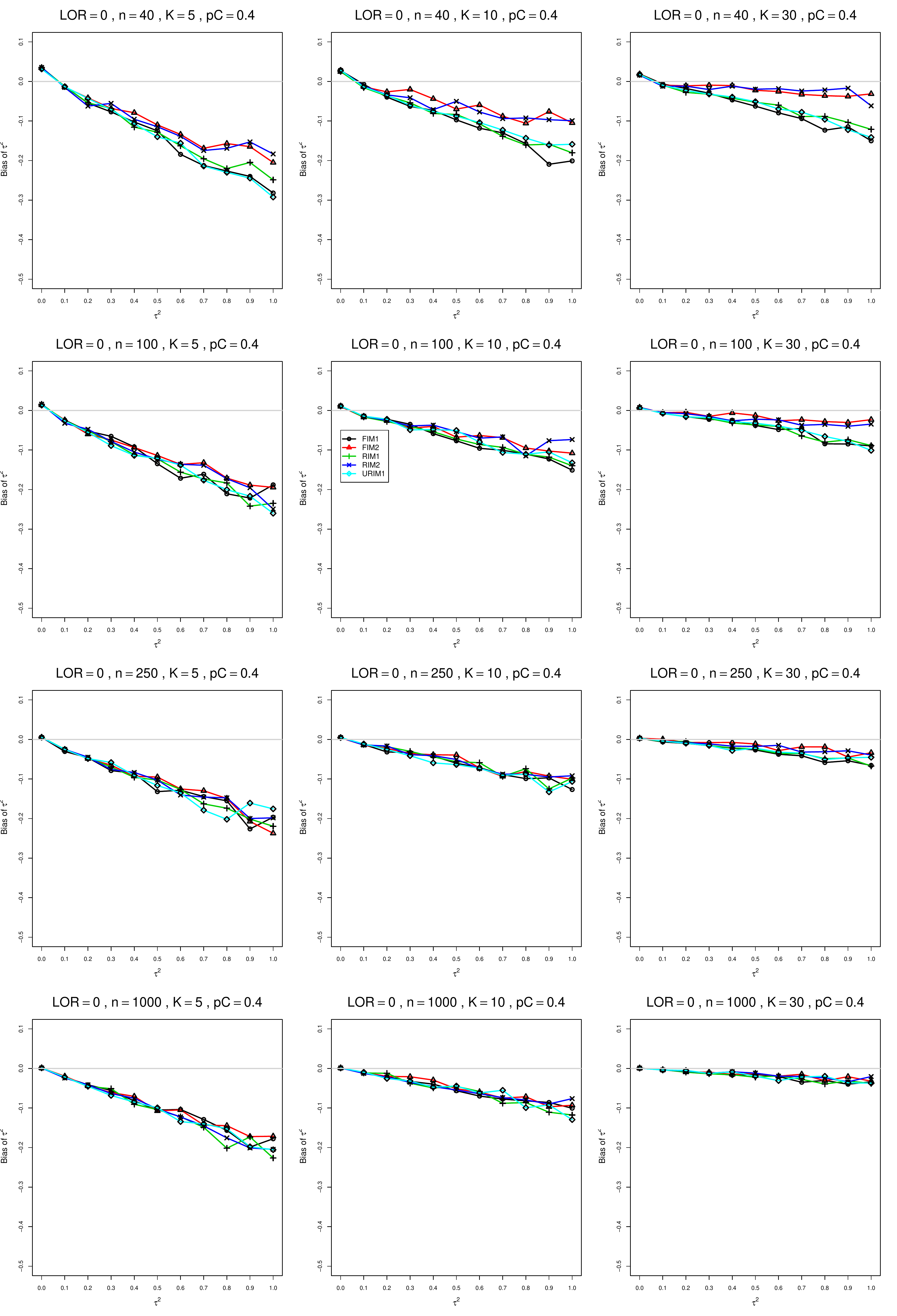}
	\caption{Bias of  between-studies variance $\hat{\tau}_{RIM2}^2$ for $\theta=0$, $p_{C}=0.4$, $\sigma^2=0.1$, constant sample sizes $n=40,\;100,\;250,\;1000$.
The data-generation mechanisms are FIM1 ($\circ$), FIM2 ($\triangle$), RIM1 (+), RIM2 ($\times$), and URIM1 ($\diamond$).
		\label{PlotBiasTau2mu0andpC04LOR_UMRSsigma01}}
\end{figure}
\begin{figure}[t]
	\centering
	\includegraphics[scale=0.33]{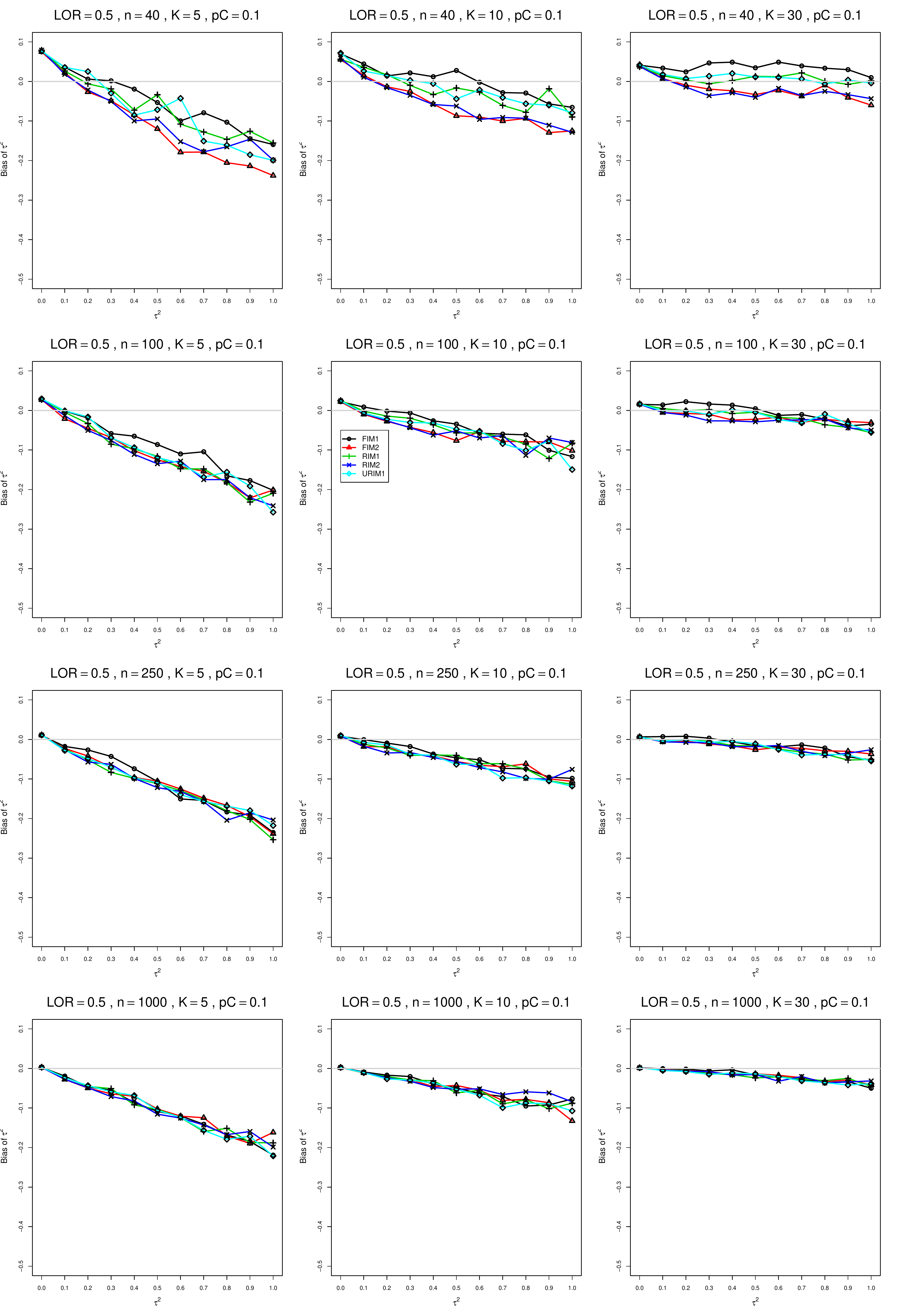}
	\caption{Bias of  between-studies variance $\hat{\tau}_{RIM2}^2$ for $\theta=0.5$, $p_{C}=0.1$, $\sigma^2=0.1$, constant sample sizes $n=40,\;100,\;250,\;1000$.
The data-generation mechanisms are FIM1 ($\circ$), FIM2 ($\triangle$), RIM1 (+), RIM2 ($\times$), and URIM1 ($\diamond$).
		\label{PlotBiasTau2mu05andpC01LOR_UMRSsigma01}}
\end{figure}
\begin{figure}[t]
	\centering
	\includegraphics[scale=0.33]{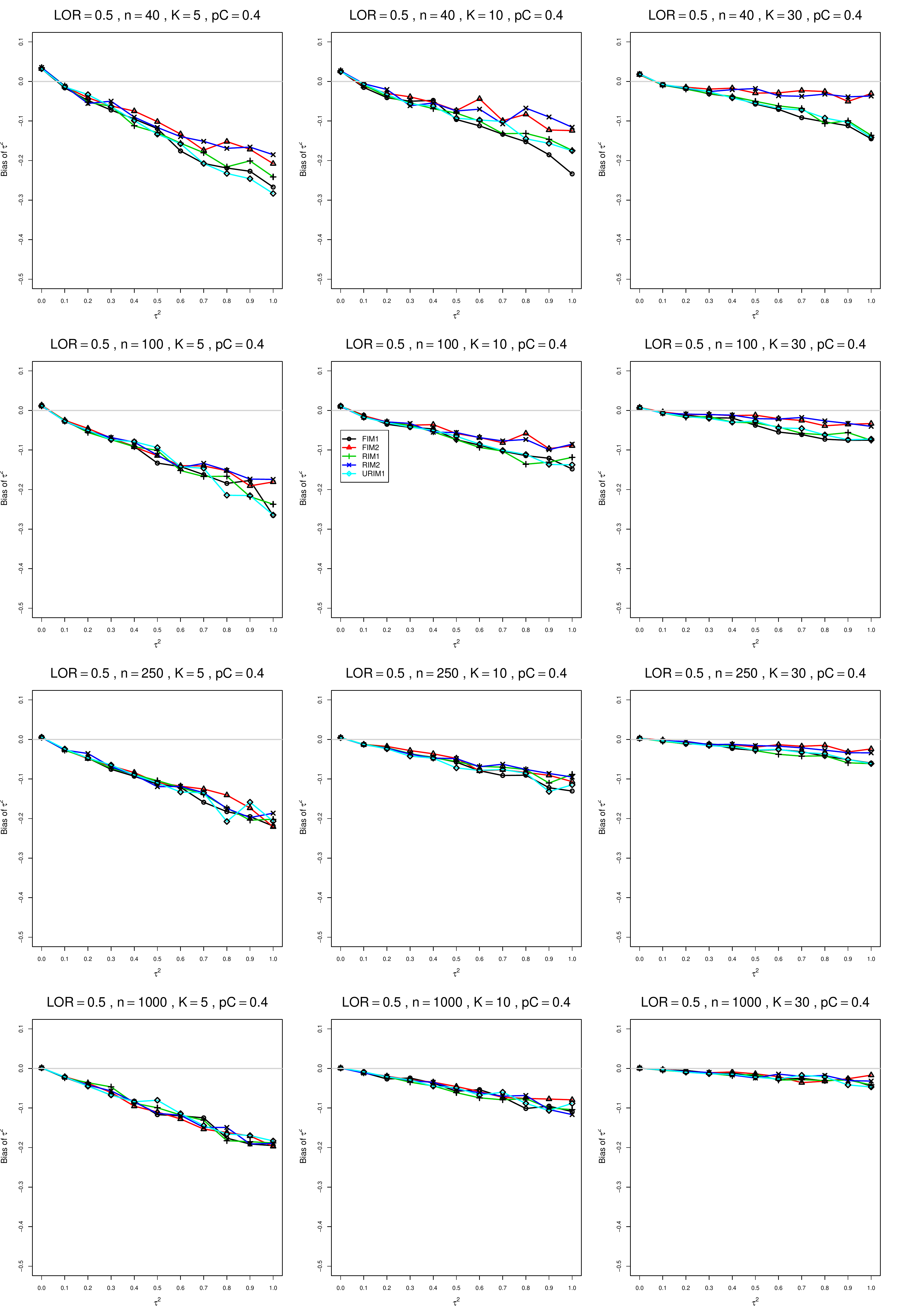}
	\caption{Bias of  between-studies variance $\hat{\tau}_{RIM2}^2$ for $\theta=0.5$, $p_{C}=0.4$, $\sigma^2=0.1$, constant sample sizes $n=40,\;100,\;250,\;1000$.
The data-generation mechanisms are FIM1 ($\circ$), FIM2 ($\triangle$), RIM1 (+), RIM2 ($\times$), and URIM1 ($\diamond$).
		\label{PlotBiasTau2mu05andpC04LOR_UMRSsigma01}}
\end{figure}
\begin{figure}[t]
	\centering
	\includegraphics[scale=0.33]{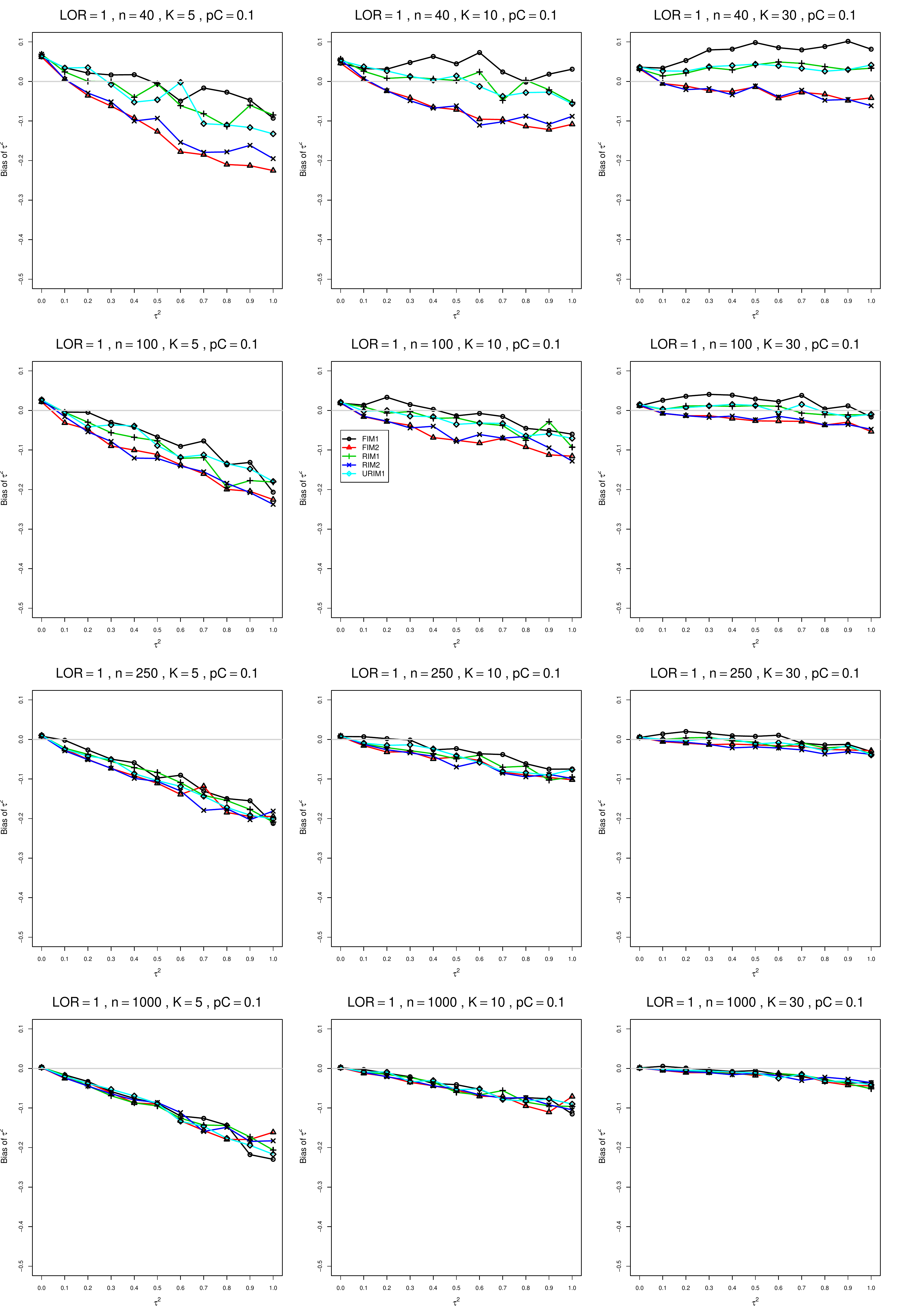}
	\caption{Bias of  between-studies variance $\hat{\tau}_{RIM2}^2$ for $\theta=1$, $p_{C}=0.1$, $\sigma^2=0.1$, constant sample sizes $n=40,\;100,\;250,\;1000$.
The data-generation mechanisms are FIM1 ($\circ$), FIM2 ($\triangle$), RIM1 (+), RIM2 ($\times$), and URIM1 ($\diamond$).
		\label{PlotBiasTau2mu1andpC01LOR_UMRSsigma01}}
\end{figure}
\begin{figure}[t]
	\centering
	\includegraphics[scale=0.33]{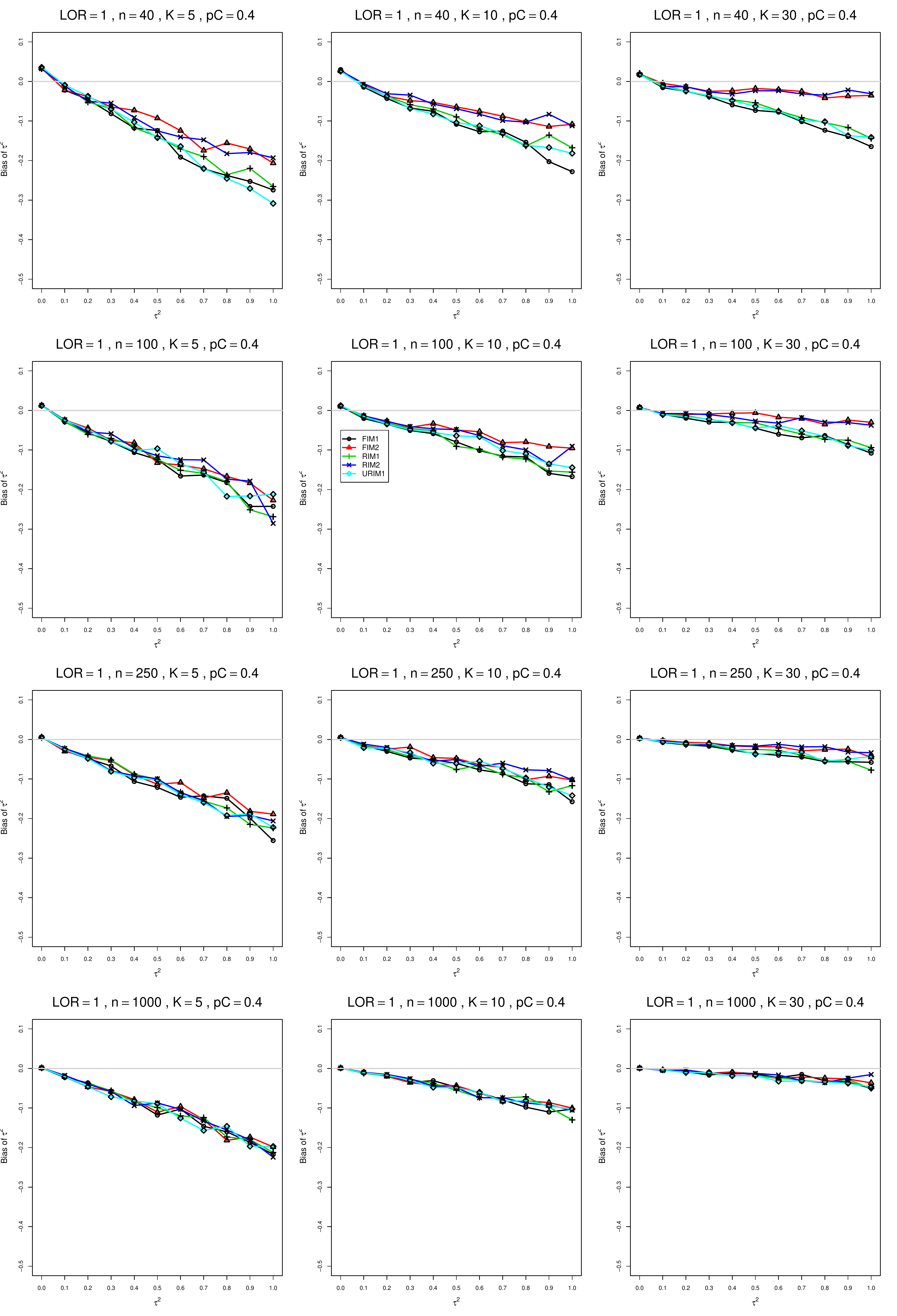}
	\caption{Bias of  between-studies variance $\hat{\tau}_{RIM2}^2$ for $\theta=1$, $p_{C}=0.4$, $\sigma^2=0.1$, constant sample sizes $n=40,\;100,\;250,\;1000$.
The data-generation mechanisms are FIM1 ($\circ$), FIM2 ($\triangle$), RIM1 (+), RIM2 ($\times$), and URIM1 ($\diamond$).
		\label{PlotBiasTau2mu1andpC04LOR_UMRSsigma01}}
\end{figure}
\begin{figure}[t]
	\centering
	\includegraphics[scale=0.33]{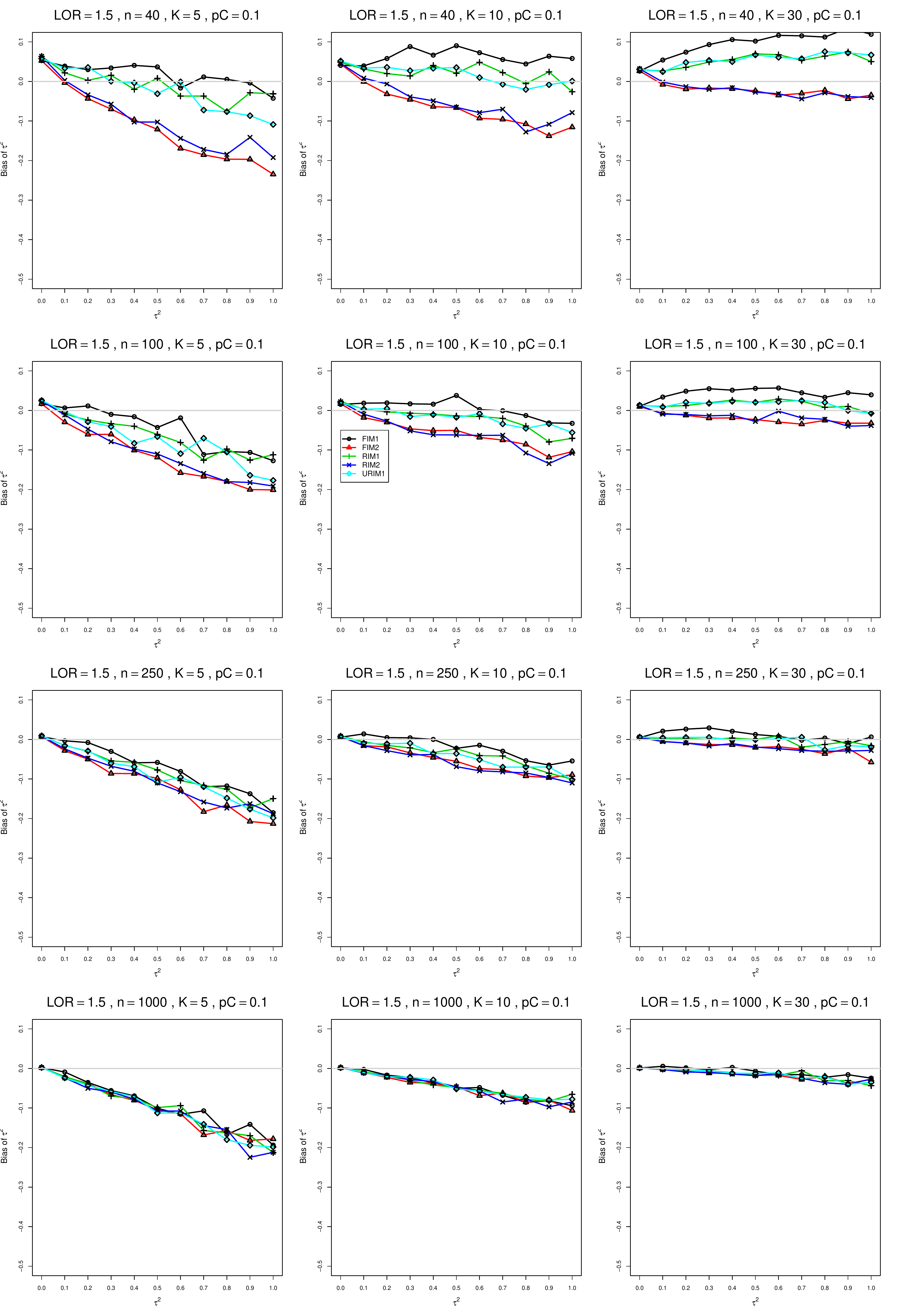}
	\caption{Bias of  between-studies variance $\hat{\tau}_{RIM2}^2$ for $\theta=1.5$, $p_{C}=0.1$, $\sigma^2=0.1$, constant sample sizes $n=40,\;100,\;250,\;1000$.
The data-generation mechanisms are FIM1 ($\circ$), FIM2 ($\triangle$), RIM1 (+), RIM2 ($\times$), and URIM1 ($\diamond$).
		\label{PlotBiasTau2mu15andpC01LOR_UMRSsigma01}}
\end{figure}
\begin{figure}[t]
	\centering
	\includegraphics[scale=0.33]{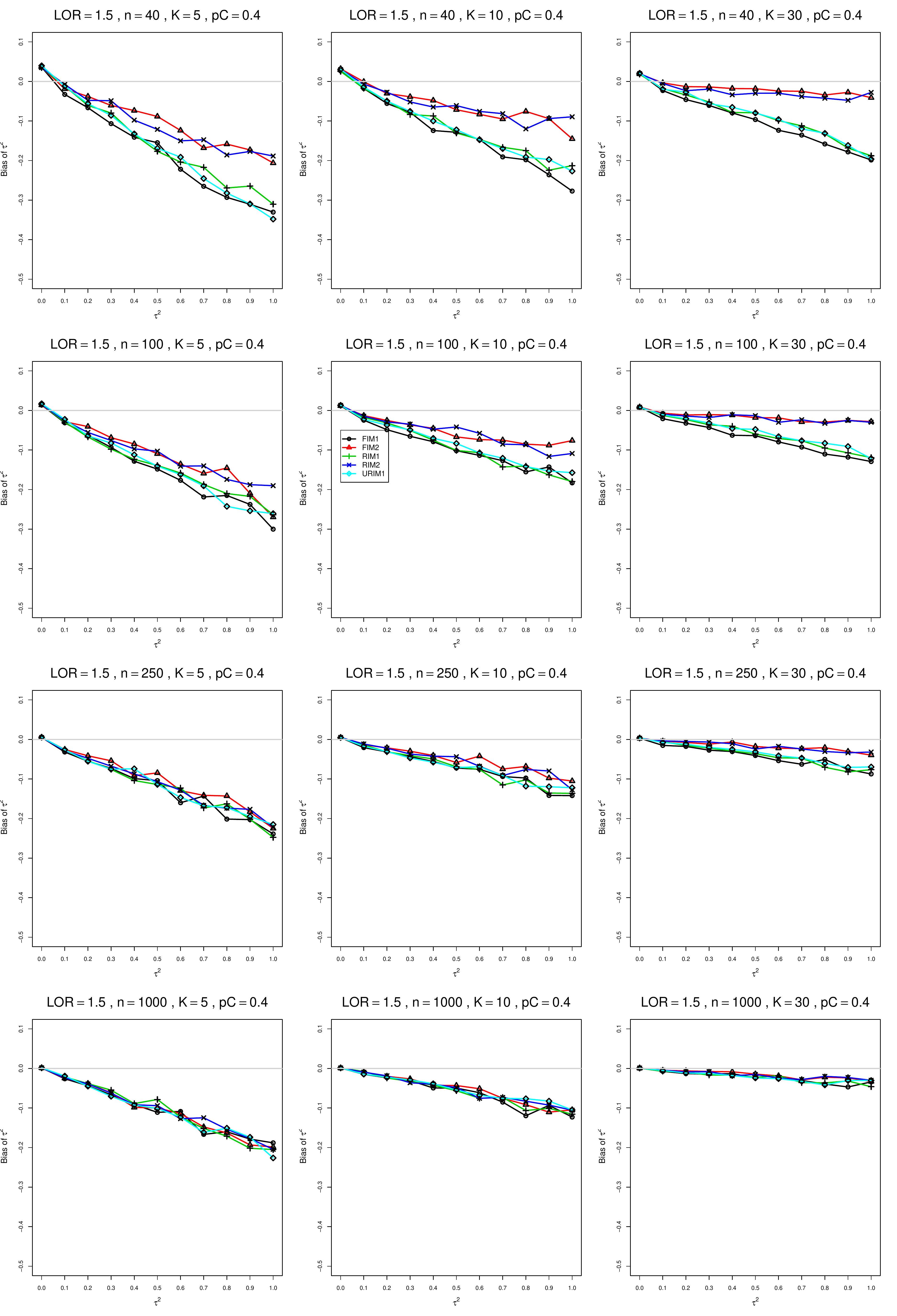}
	\caption{Bias of  between-studies variance $\hat{\tau}_{RIM2}^2$ for $\theta=1.5$, $p_{C}=0.4$, $\sigma^2=0.1$, constant sample sizes $n=40,\;100,\;250,\;1000$.
The data-generation mechanisms are FIM1 ($\circ$), FIM2 ($\triangle$), RIM1 (+), RIM2 ($\times$), and URIM1 ($\diamond$).
		\label{PlotBiasTau2mu15andpC04LOR_UMRSsigma01}}
\end{figure}
\begin{figure}[t]
	\centering
	\includegraphics[scale=0.33]{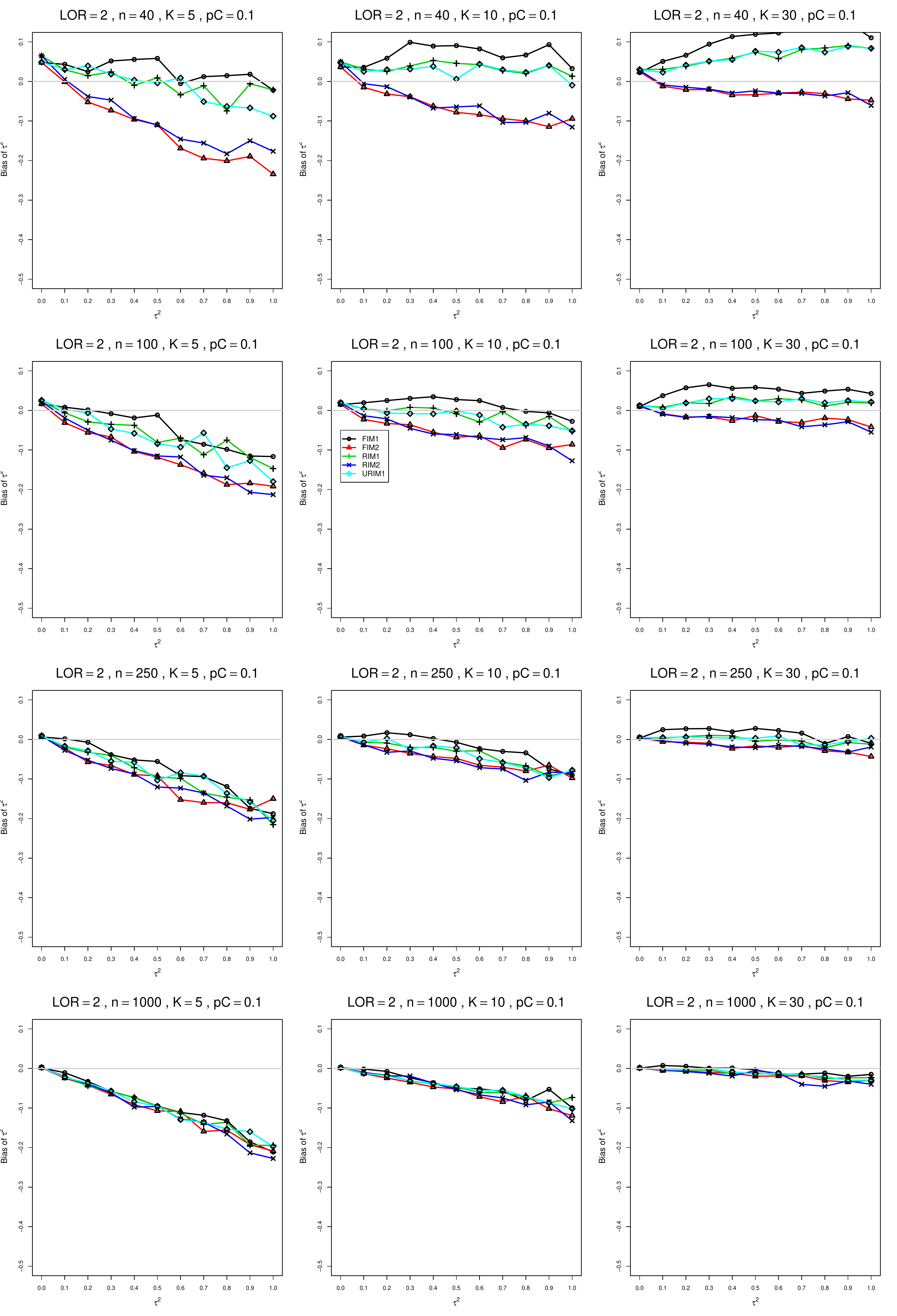}
	\caption{Bias of  between-studies variance $\hat{\tau}_{RIM2}^2$ for $\theta=2$, $p_{C}=0.1$, $\sigma^2=0.1$, constant sample sizes $n=40,\;100,\;250,\;1000$.
The data-generation mechanisms are FIM1 ($\circ$), FIM2 ($\triangle$), RIM1 (+), RIM2 ($\times$), and URIM1 ($\diamond$).
		\label{PlotBiasTau2mu2andpC01LOR_UMRSsigma01}}
\end{figure}
\begin{figure}[t]
	\centering
	\includegraphics[scale=0.33]{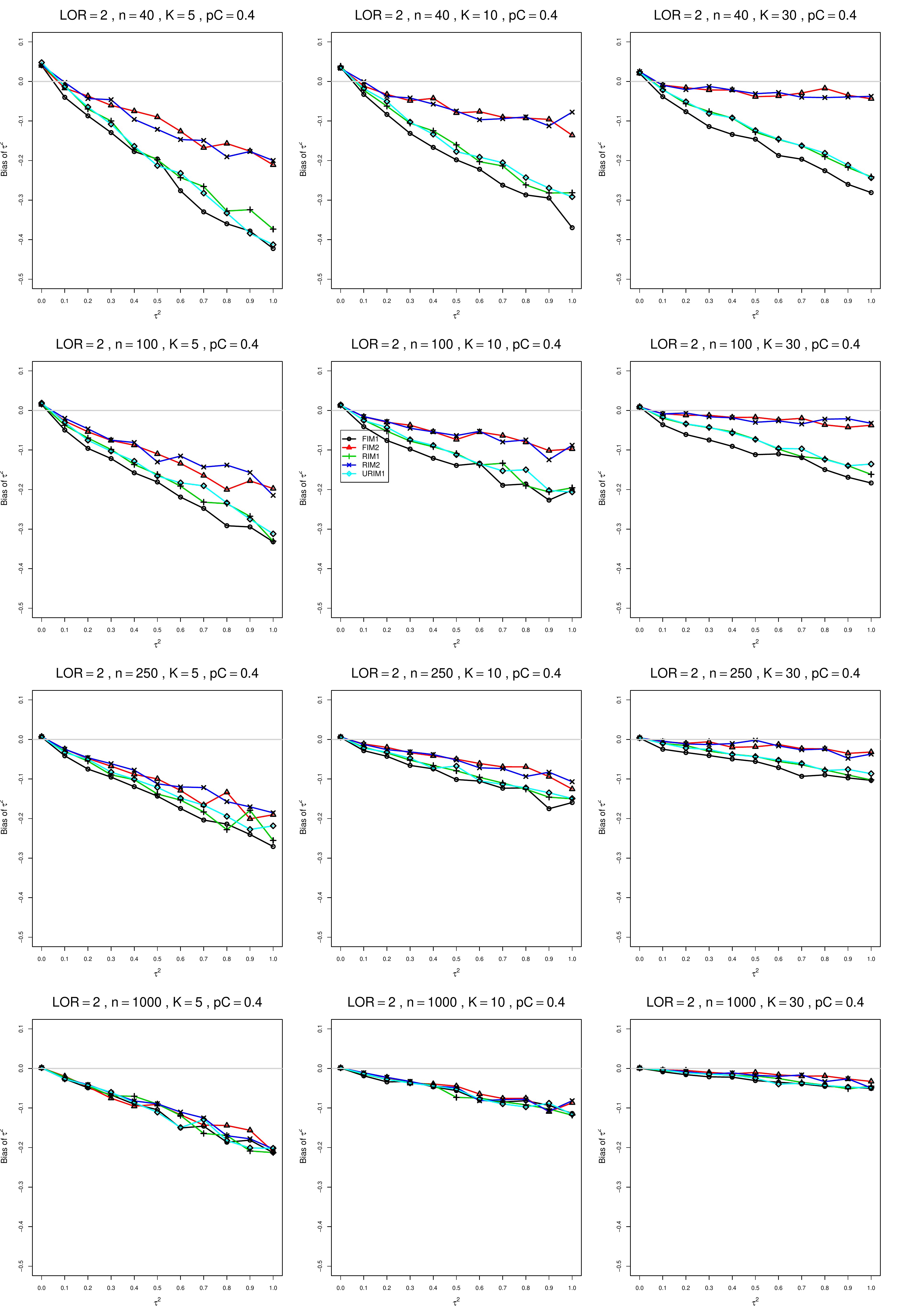}
	\caption{Bias of  between-studies variance $\hat{\tau}_{RIM2}^2$ for $\theta=2$, $p_{C}=0.4$, $\sigma^2=0.1$, constant sample sizes $n=40,\;100,\;250,\;1000$.
The data-generation mechanisms are FIM1 ($\circ$), FIM2 ($\triangle$), RIM1 (+), RIM2 ($\times$), and URIM1 ($\diamond$).
		\label{PlotBiasTau2mu2andpC04LOR_UMRSsigma01}}
\end{figure}
\begin{figure}[t]
	\centering
	\includegraphics[scale=0.33]{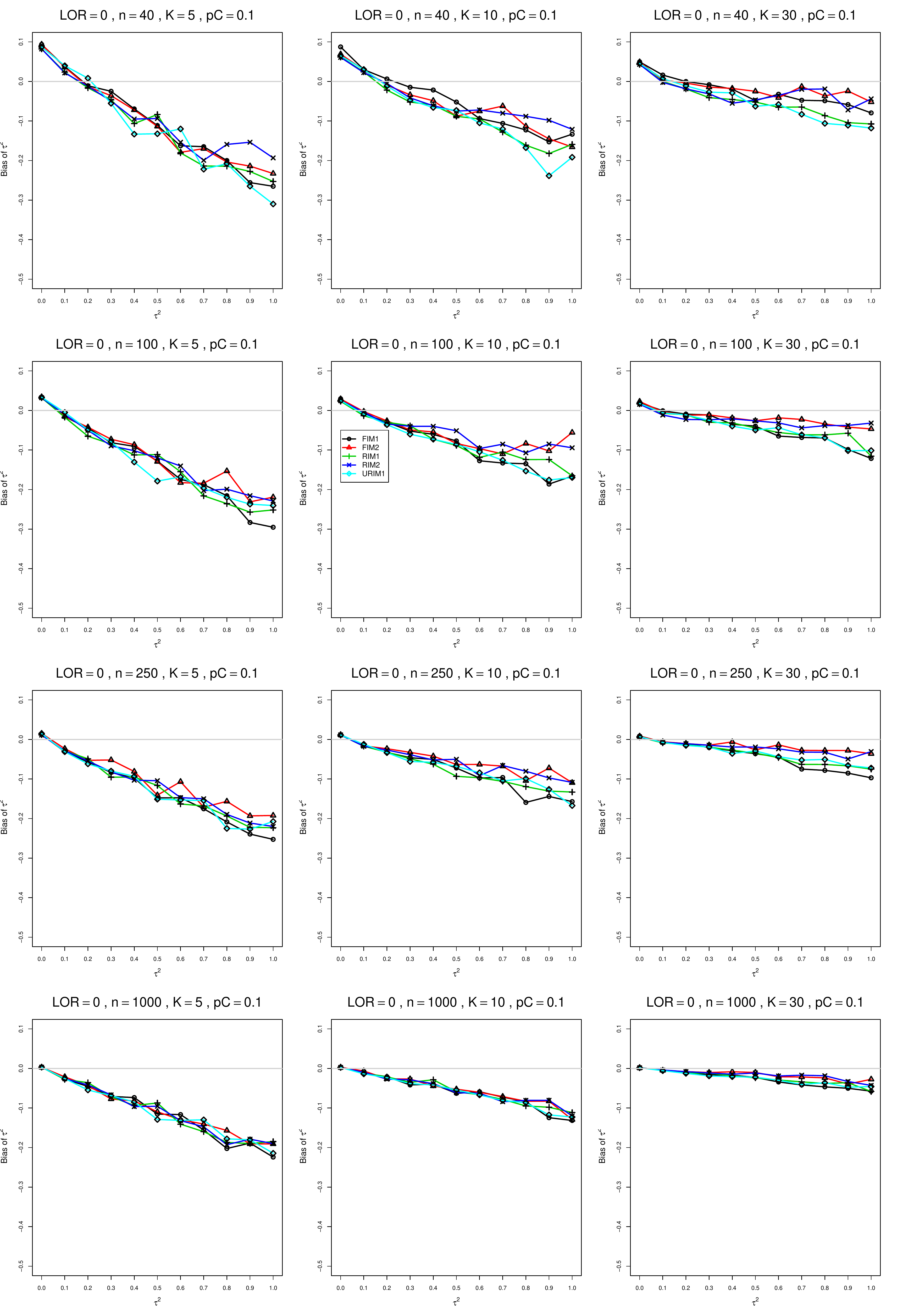}
	\caption{Bias of  between-studies variance $\hat{\tau}_{RIM2}^2$ for $\theta=0$, $p_{C}=0.1$, $\sigma^2=0.4$, constant sample sizes $n=40,\;100,\;250,\;1000$.
The data-generation mechanisms are FIM1 ($\circ$), FIM2 ($\triangle$), RIM1 (+), RIM2 ($\times$), and URIM1 ($\diamond$).
		\label{PlotBiasTau2mu0andpC01LOR_UMRSsigma04}}
\end{figure}
\begin{figure}[t]
	\centering
	\includegraphics[scale=0.33]{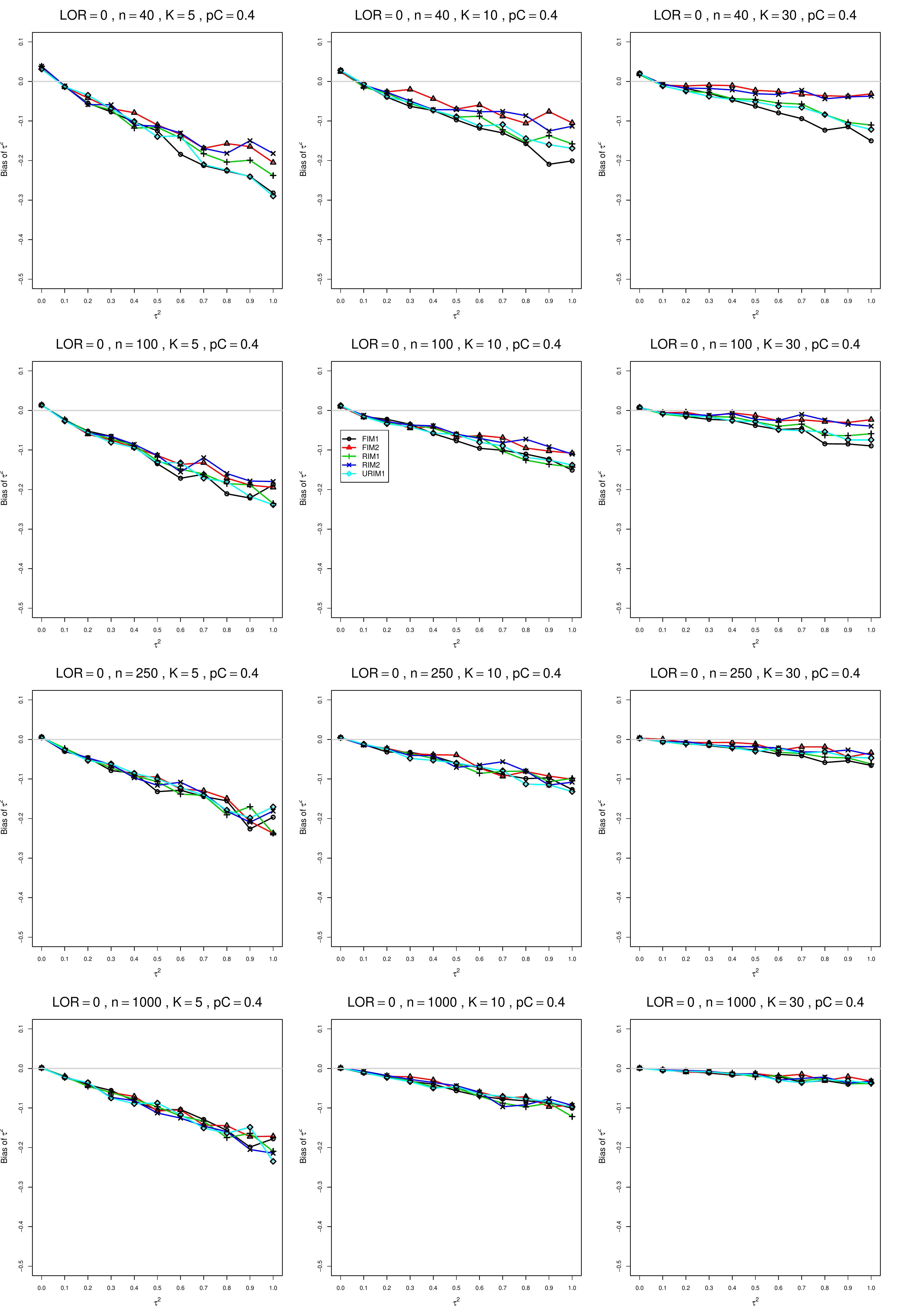}
	\caption{Bias of  between-studies variance $\hat{\tau}_{RIM2}^2$ for $\theta=0$, $p_{C}=0.4$, $\sigma^2=0.4$, constant sample sizes $n=40,\;100,\;250,\;1000$.
The data-generation mechanisms are FIM1 ($\circ$), FIM2 ($\triangle$), RIM1 (+), RIM2 ($\times$), and URIM1 ($\diamond$).
		\label{PlotBiasTau2mu0andpC04LOR_UMRSsigma04}}
\end{figure}
\begin{figure}[t]
	\centering
	\includegraphics[scale=0.33]{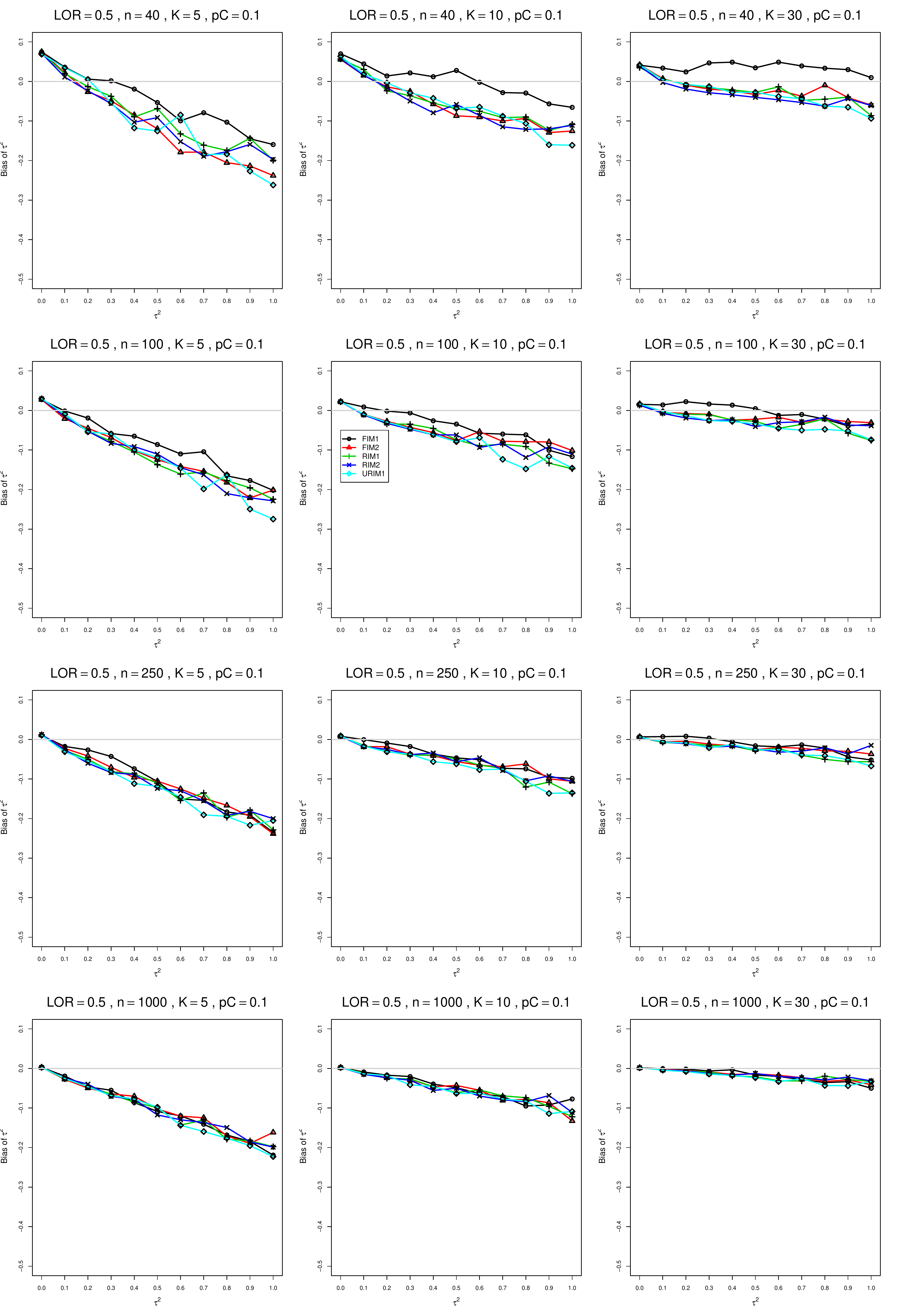}
	\caption{Bias of  between-studies variance $\hat{\tau}_{RIM2}^2$ for $\theta=0.5$, $p_{C}=0.1$, $\sigma^2=0.4$, constant sample sizes $n=40,\;100,\;250,\;1000$.
The data-generation mechanisms are FIM1 ($\circ$), FIM2 ($\triangle$), RIM1 (+), RIM2 ($\times$), and URIM1 ($\diamond$).
		\label{PlotBiasTau2mu05andpC01LOR_UMRSsigma04}}
\end{figure}
\begin{figure}[t]
	\centering
	\includegraphics[scale=0.33]{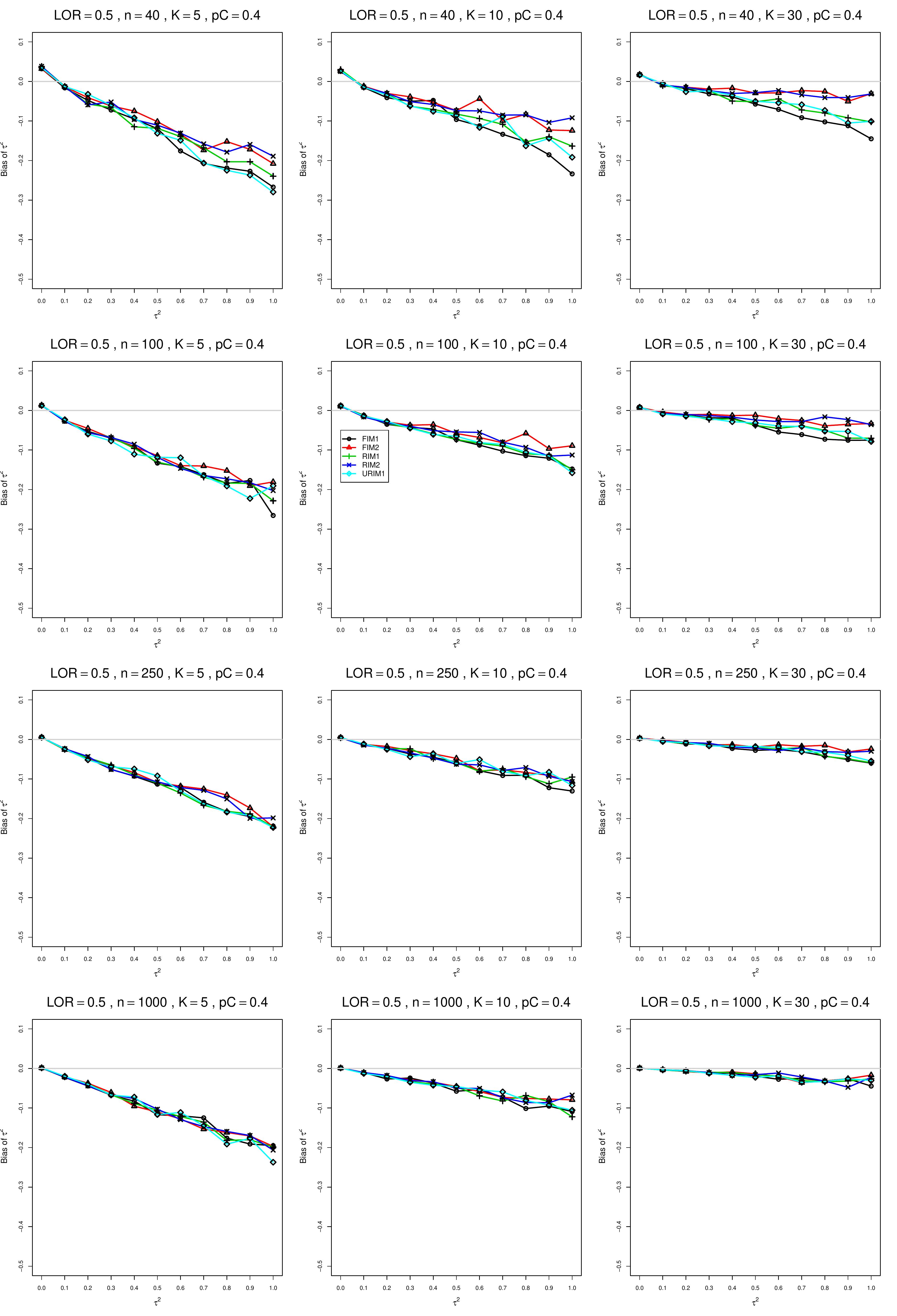}
	\caption{Bias of  between-studies variance $\hat{\tau}_{RIM2}^2$ for $\theta=0.5$, $p_{C}=0.4$, $\sigma^2=0.4$, constant sample sizes $n=40,\;100,\;250,\;1000$.
The data-generation mechanisms are FIM1 ($\circ$), FIM2 ($\triangle$), RIM1 (+), RIM2 ($\times$), and URIM1 ($\diamond$).
		\label{PlotBiasTau2mu05andpC04LOR_UMRSsigma04}}
\end{figure}
\begin{figure}[t]
	\centering
	\includegraphics[scale=0.33]{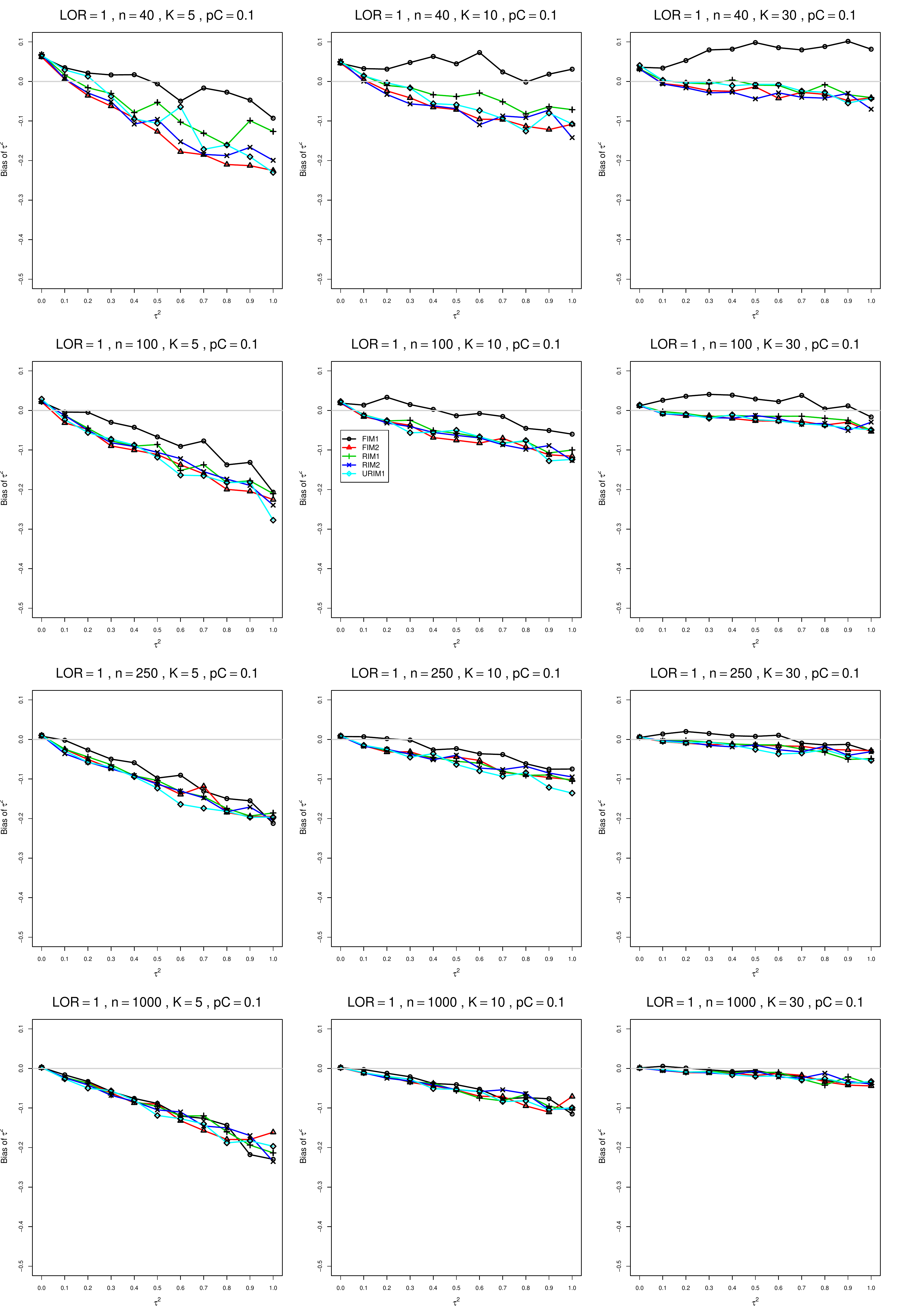}
	\caption{Bias of  between-studies variance $\hat{\tau}_{RIM2}^2$ for $\theta=1$, $p_{C}=0.1$, $\sigma^2=0.4$, constant sample sizes $n=40,\;100,\;250,\;1000$.
The data-generation mechanisms are FIM1 ($\circ$), FIM2 ($\triangle$), RIM1 (+), RIM2 ($\times$), and URIM1 ($\diamond$).
		\label{PlotBiasTau2mu1andpC01LOR_UMRSsigma04}}
\end{figure}
\begin{figure}[t]
	\centering
	\includegraphics[scale=0.33]{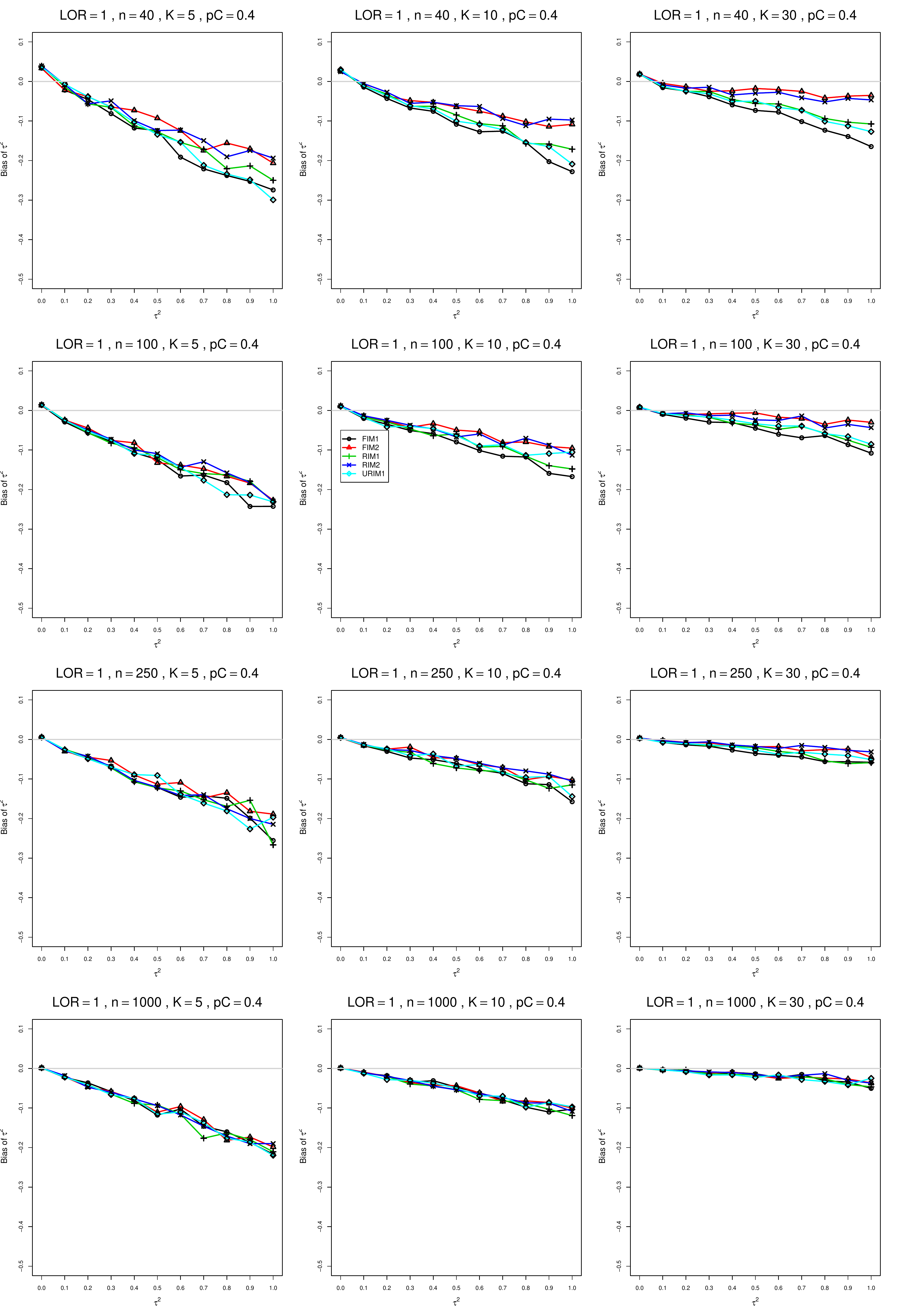}
	\caption{Bias of  between-studies variance $\hat{\tau}_{RIM2}^2$ for $\theta=1$, $p_{C}=0.4$, $\sigma^2=0.4$, constant sample sizes $n=40,\;100,\;250,\;1000$.
The data-generation mechanisms are FIM1 ($\circ$), FIM2 ($\triangle$), RIM1 (+), RIM2 ($\times$), and URIM1 ($\diamond$).
		\label{PlotBiasTau2mu1andpC04LOR_UMRSsigma04}}
\end{figure}
\begin{figure}[t]
	\centering
	\includegraphics[scale=0.33]{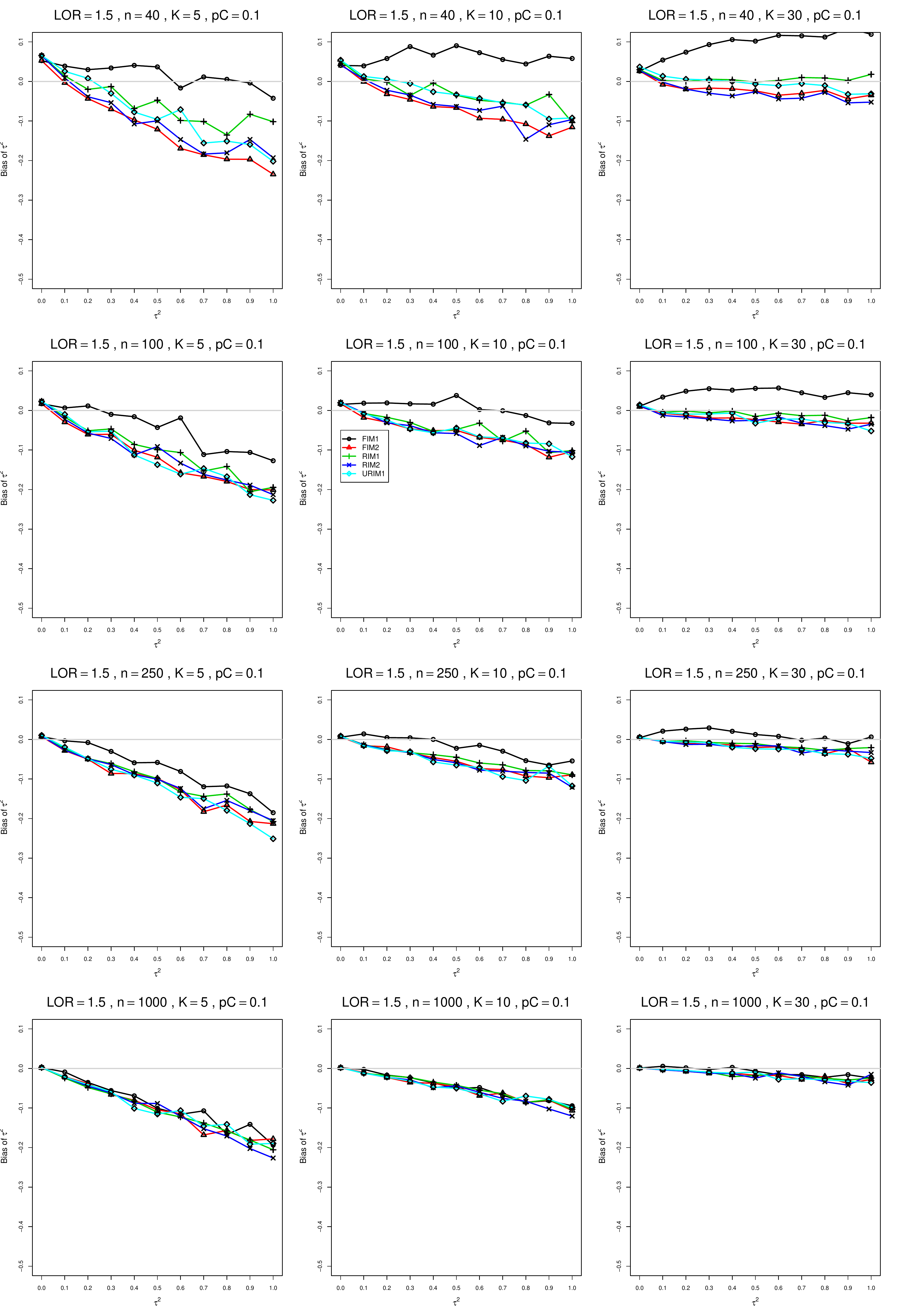}
	\caption{Bias of  between-studies variance $\hat{\tau}_{RIM2}^2$ for $\theta=1.5$, $p_{C}=0.1$, $\sigma^2=0.4$, constant sample sizes $n=40,\;100,\;250,\;1000$.
The data-generation mechanisms are FIM1 ($\circ$), FIM2 ($\triangle$), RIM1 (+), RIM2 ($\times$), and URIM1 ($\diamond$).
		\label{PlotBiasTau2mu15andpC01LOR_UMRSsigma04}}
\end{figure}
\begin{figure}[t]
	\centering
	\includegraphics[scale=0.33]{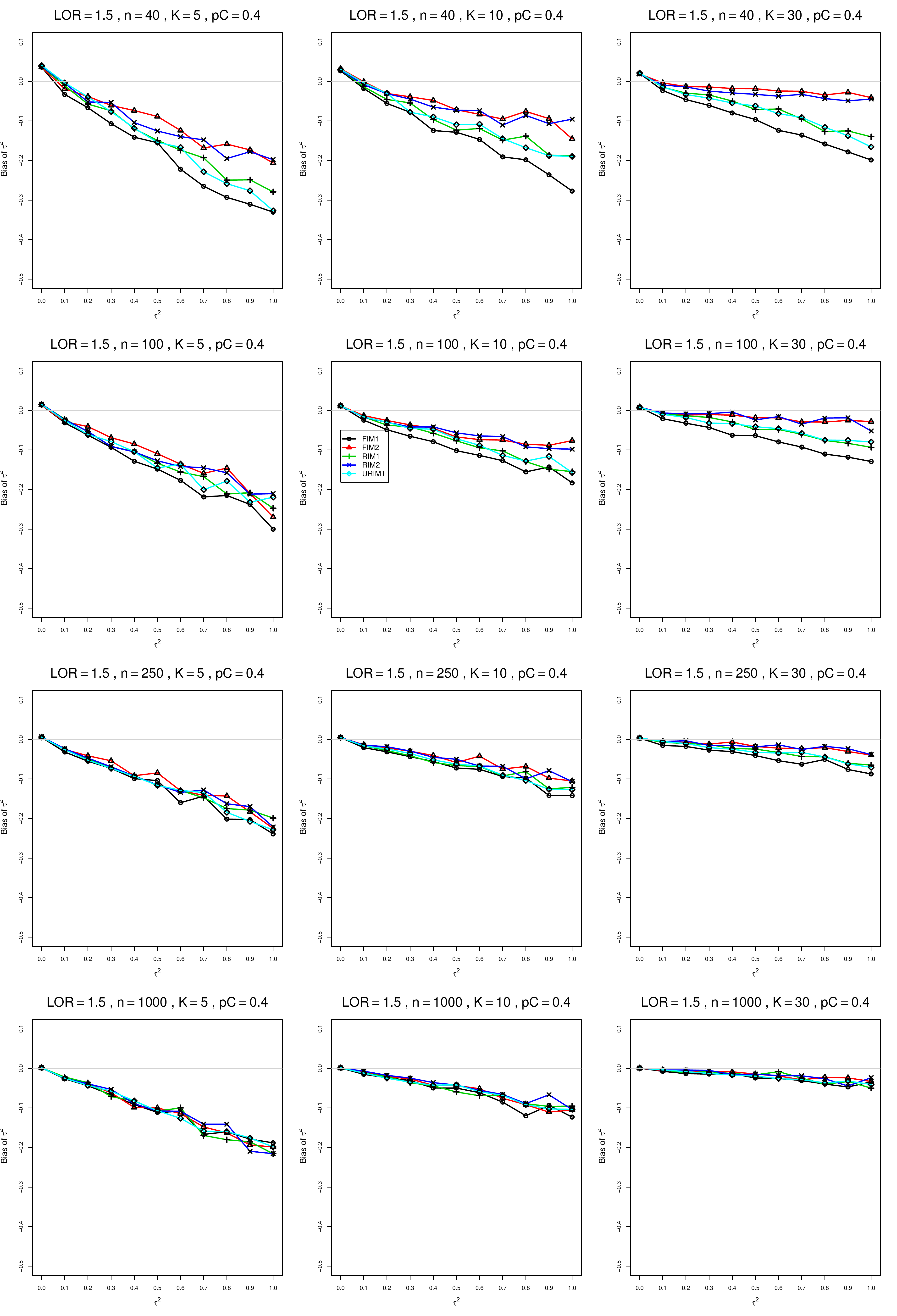}
	\caption{Bias of  between-studies variance $\hat{\tau}_{RIM2}^2$ for $\theta=1.5$, $p_{C}=0.4$, $\sigma^2=0.4$, constant sample sizes $n=40,\;100,\;250,\;1000$.
The data-generation mechanisms are FIM1 ($\circ$), FIM2 ($\triangle$), RIM1 (+), RIM2 ($\times$), and URIM1 ($\diamond$).
		\label{PlotBiasTau2mu15andpC04LOR_UMRSsigma04}}
\end{figure}
\begin{figure}[t]
	\centering
	\includegraphics[scale=0.33]{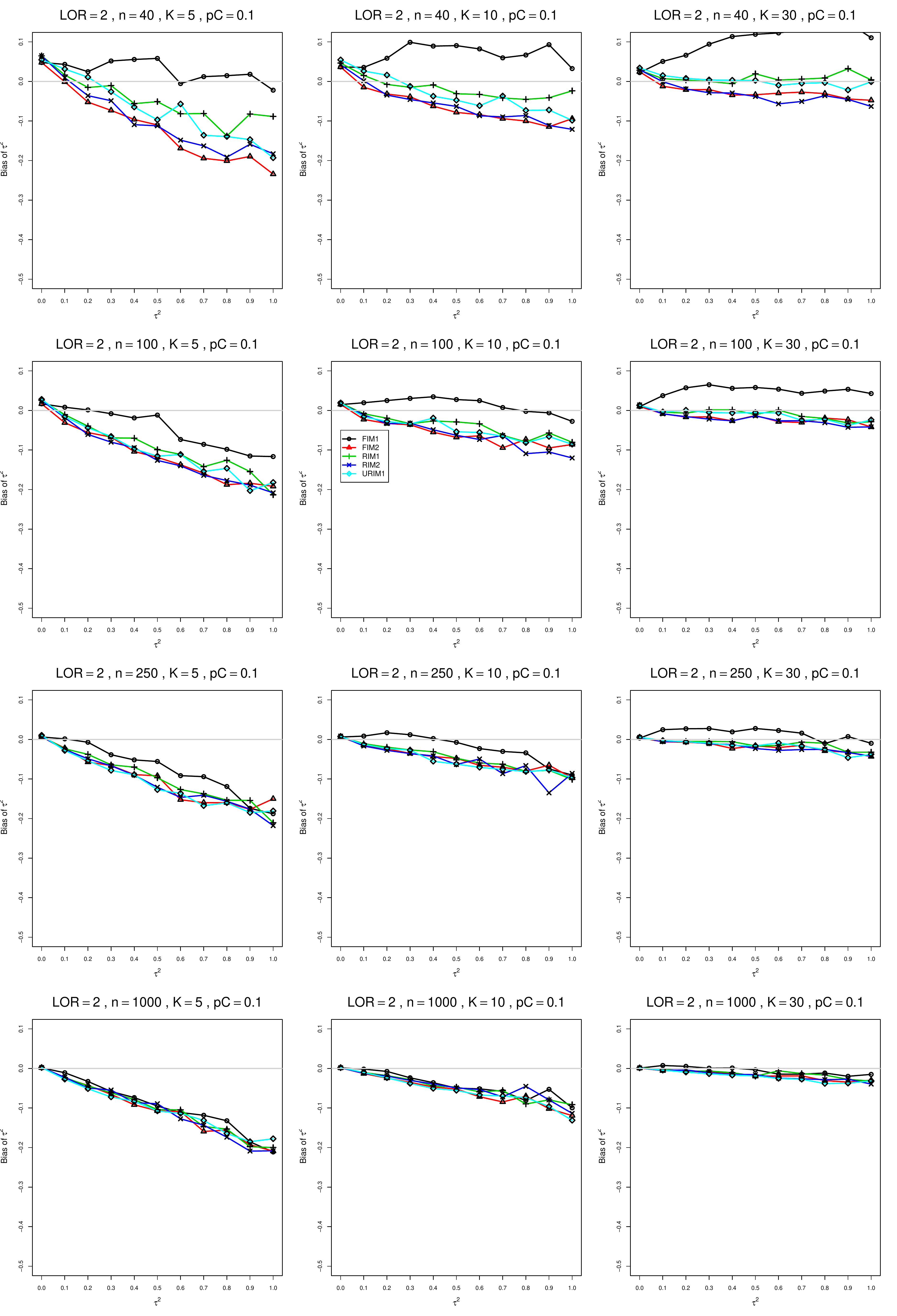}
	\caption{Bias of  between-studies variance $\hat{\tau}_{RIM2}^2$ for $\theta=2$, $p_{C}=0.1$, $\sigma^2=0.4$, constant sample sizes $n=40,\;100,\;250,\;1000$.
The data-generation mechanisms are FIM1 ($\circ$), FIM2 ($\triangle$), RIM1 (+), RIM2 ($\times$), and URIM1 ($\diamond$).
		\label{PlotBiasTau2mu2andpC01LOR_UMRSsigma04}}
\end{figure}
\begin{figure}[t]
	\centering
	\includegraphics[scale=0.33]{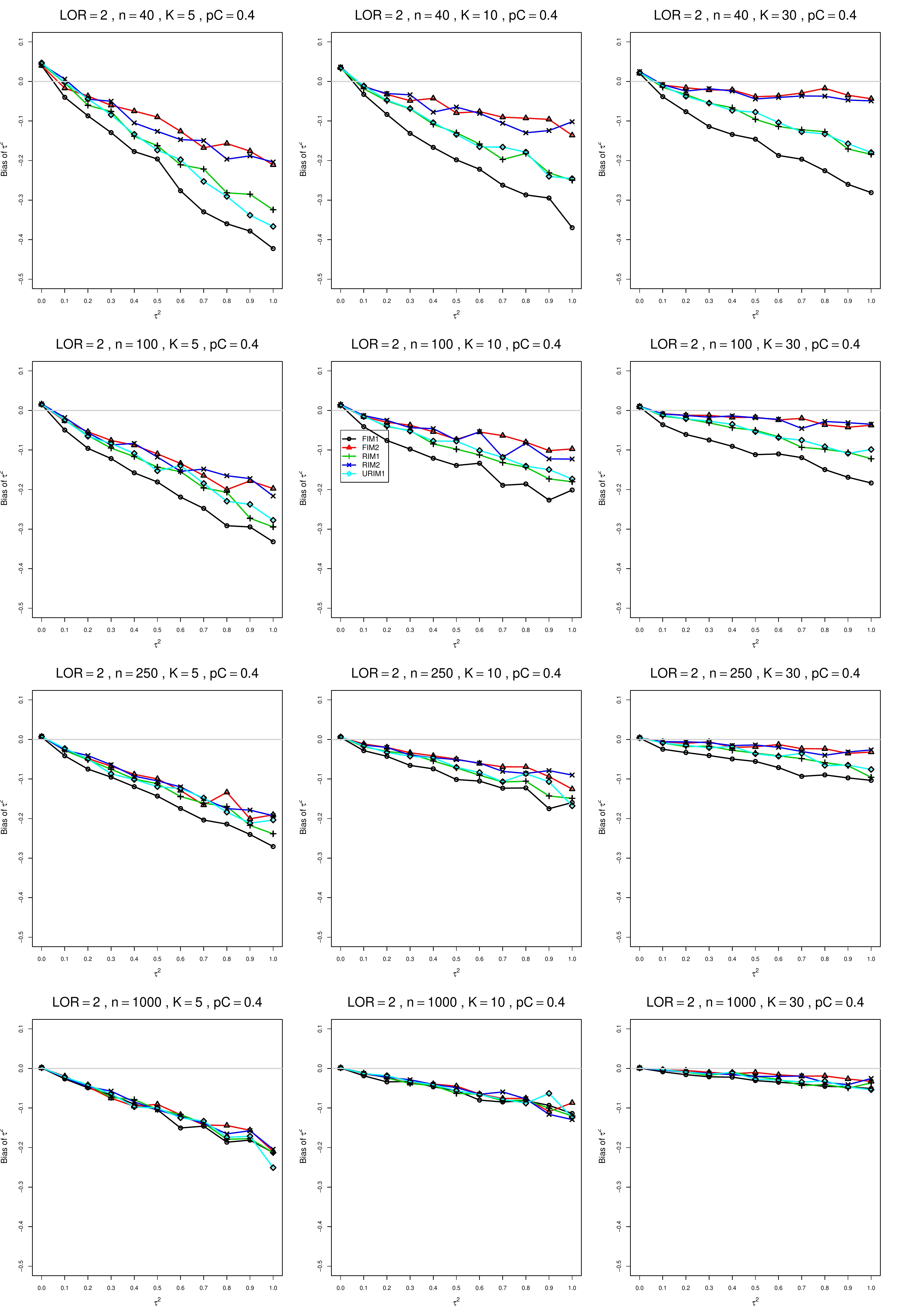}
	\caption{Bias of  between-studies variance $\hat{\tau}_{RIM2}^2$ for $\theta=2$, $p_{C}=0.4$, $\sigma^2=0.4$, constant sample sizes $n=40,\;100,\;250,\;1000$.
The data-generation mechanisms are FIM1 ($\circ$), FIM2 ($\triangle$), RIM1 (+), RIM2 ($\times$), and URIM1 ($\diamond$).
		\label{PlotBiasTau2mu2andpC04LOR_UMRSsigma04}}
\end{figure}

\clearpage
\section*{A2: Plots of bias of estimators of $\theta$ for log-odds-ratio}
Each panel of a figure corresponds to a value of n (= 40, 100, 250, 1000) and a value of K (= 5, 10, 30) and has $\tau^2$ = 0.0(0.1)1.0 on the horizontal axis. \\
The data-generation mechanisms are
\begin{itemize}
	\item FIM1 - Fixed-intercept model with $c = 0$
	\item FIM2 - Fixed-intercept model with $c = 1/2$
	\item RIM1 - Random-intercept model with $c = 0$
	\item RIM2 - Random-intercept model with $c = 1/2$
	\item URIM1 - Random-intercept model with $c = 0$ and $p_{iC}$ is uniformly distributed on [$p_{iC}-\sigma\sqrt{3}p_{iC}(1-p_{iC})$, $p_{iC}+\sigma\sqrt{3}p_{iC}(1-p_{iC})$]
\end{itemize}
The point estimators of $\theta$ are
\begin{itemize}
	\item $\hat{\theta}_{DL}$ - DerSimonian-Laird
	\item $\hat{\theta}_{REML}$ - Restricted maximum-likelihood
	\item $\hat{\theta}_{MP}$ - Mandel-Paule
	\item $\hat{\theta}_{KD}$ - Kulinskaya-Dollinger
	\item $\hat{\theta}_{FIM2}$ - Estimator of $\theta$ in the FIM2 GLMM
	\item $\hat{\theta}_{RIM2}$ - Estimator of $\theta$ in the RIM2 GLMM
    \item $\hat{\theta}_{SSW}$ -  sample-size-weighted estimator of $\theta$
\end{itemize}

\clearpage
\subsection*{A2.1 Bias of $\hat{\theta}_{DL}$}
\renewcommand{\thefigure}{A2.1.\arabic{figure}}
\setcounter{figure}{0}

\begin{figure}[t]
	\centering
	\includegraphics[scale=0.33]{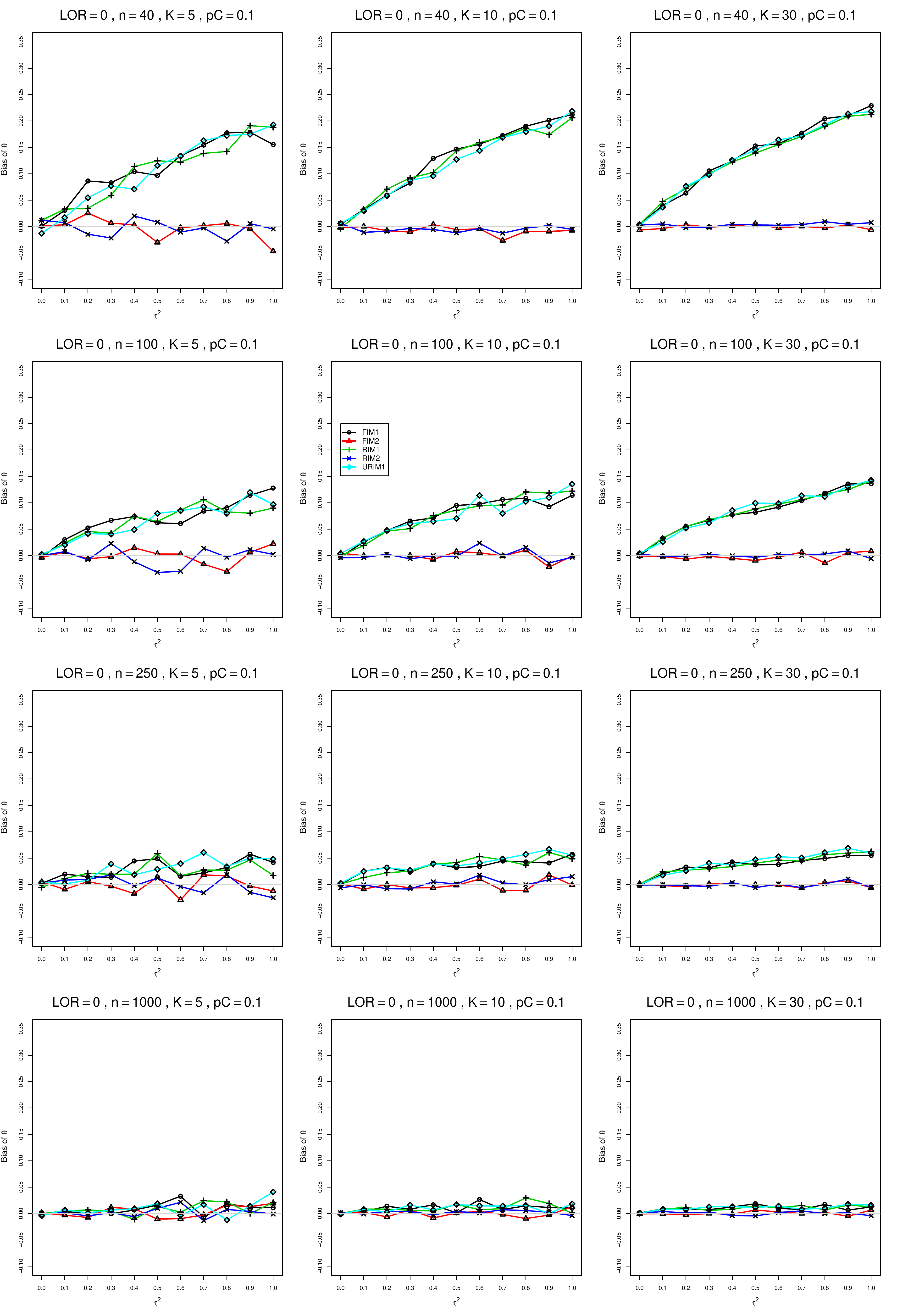}
	\caption{Bias of  overall log-odds ratio $\hat{\theta}_{DL}$ for $\theta=0$, $p_{C}=0.1$, $\sigma^2=0.1$, constant sample sizes $n=40,\;100,\;250,\;1000$.
The data-generation mechanisms are FIM1 ($\circ$), FIM2 ($\triangle$), RIM1 (+), RIM2 ($\times$), and URIM1 ($\diamond$).
		\label{PlotBiasThetamu0andpC01LOR_DLsigma01}}
\end{figure}
\begin{figure}[t]
	\centering
	\includegraphics[scale=0.33]{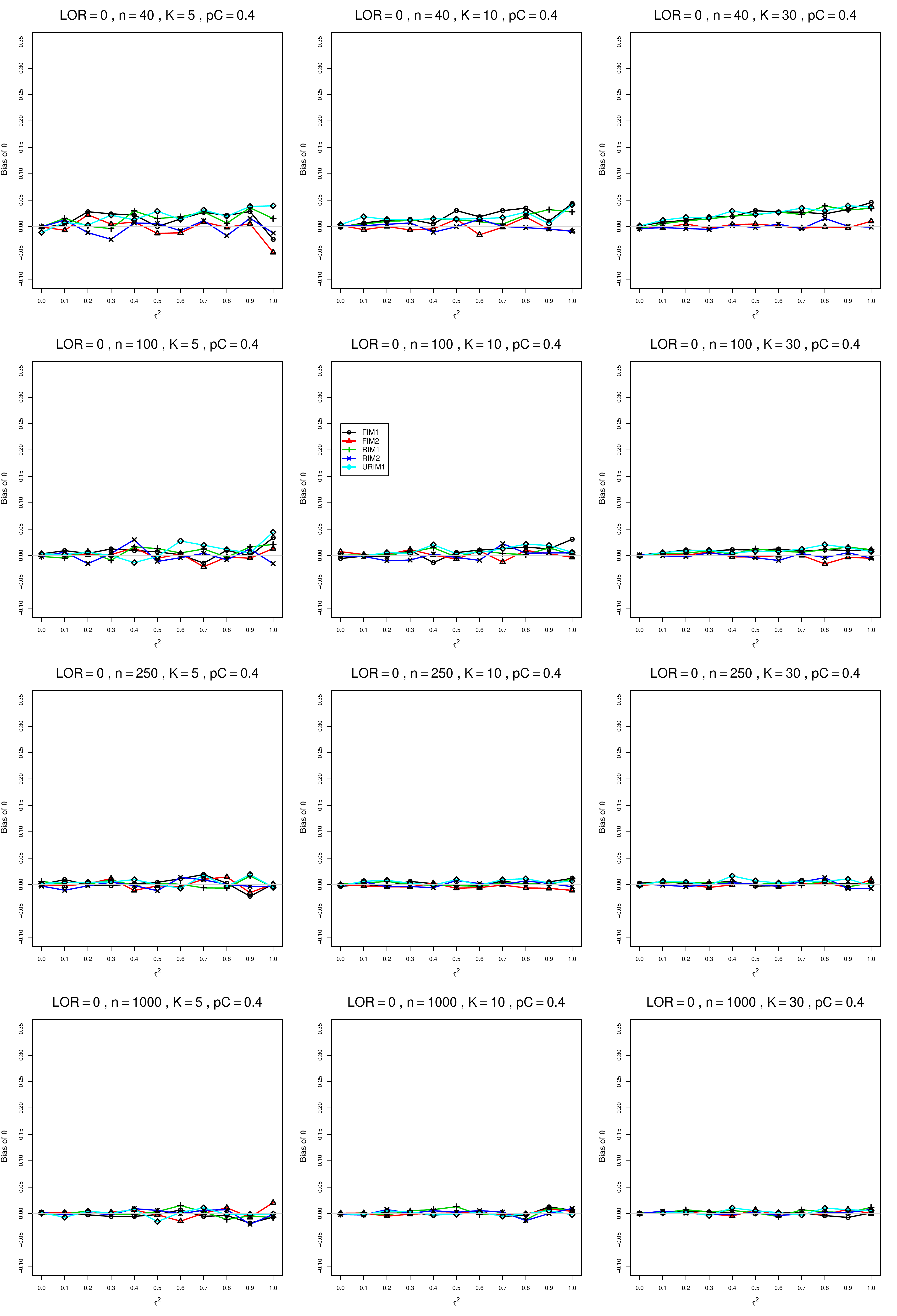}
	\caption{Bias of  overall log-odds ratio $\hat{\theta}_{DL}$ for $\theta=0$, $p_{C}=0.4$, $\sigma^2=0.1$, constant sample sizes $n=40,\;100,\;250,\;1000$.
The data-generation mechanisms are FIM1 ($\circ$), FIM2 ($\triangle$), RIM1 (+), RIM2 ($\times$), and URIM1 ($\diamond$).
		\label{PlotBiasThetamu0andpC04LOR_DLsigma01}}
\end{figure}
\begin{figure}[t]
	\centering
	\includegraphics[scale=0.33]{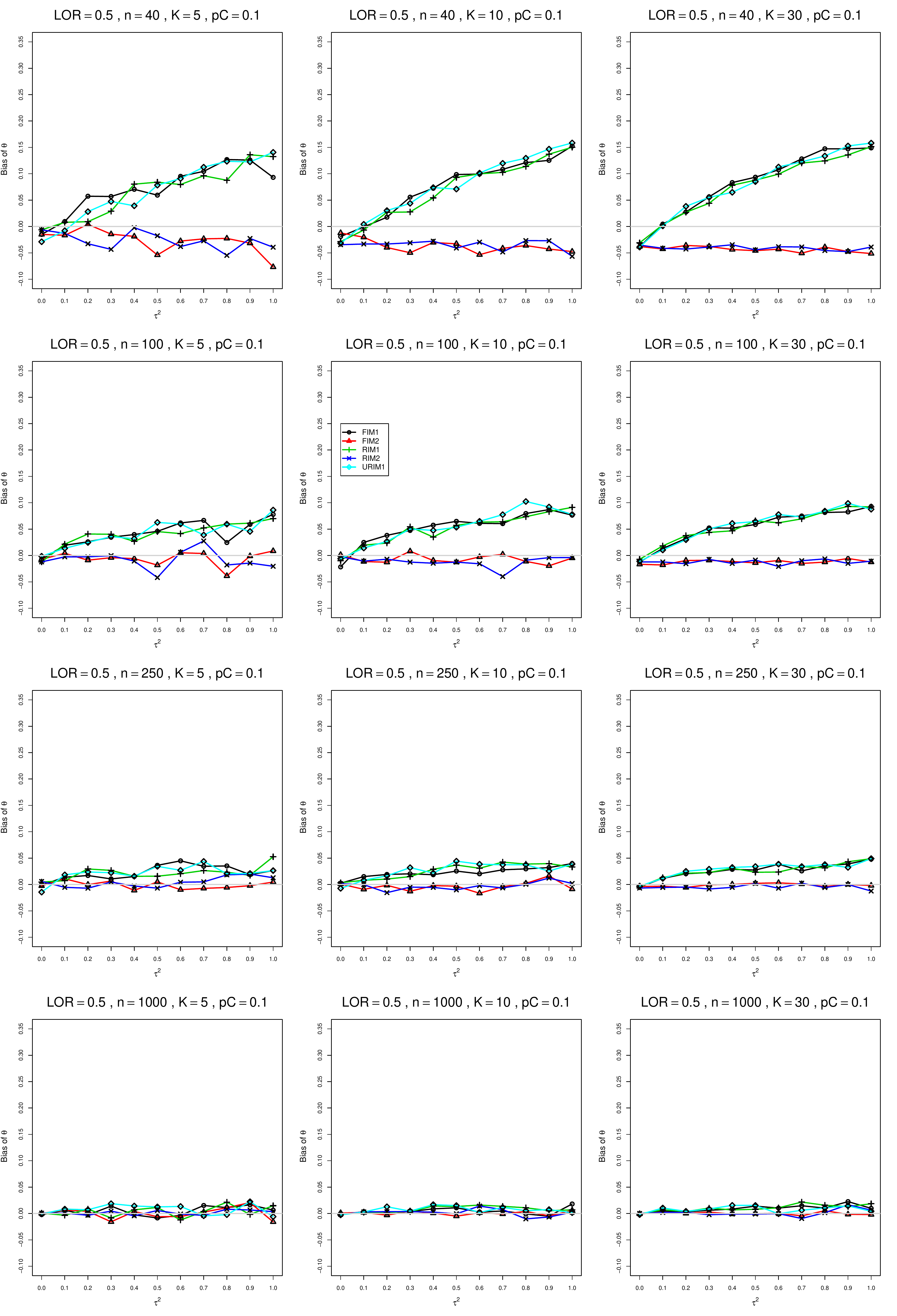}
	\caption{Bias of  overall log-odds ratio $\hat{\theta}_{DL}$ for $\theta=0.5$, $p_{C}=0.1$, $\sigma^2=0.1$, constant sample sizes $n=40,\;100,\;250,\;1000$.
The data-generation mechanisms are FIM1 ($\circ$), FIM2 ($\triangle$), RIM1 (+), RIM2 ($\times$), and URIM1 ($\diamond$).
		\label{PlotBiasThetamu05andpC01LOR_DLsigma01}}
\end{figure}
\begin{figure}[t]
	\centering
	\includegraphics[scale=0.33]{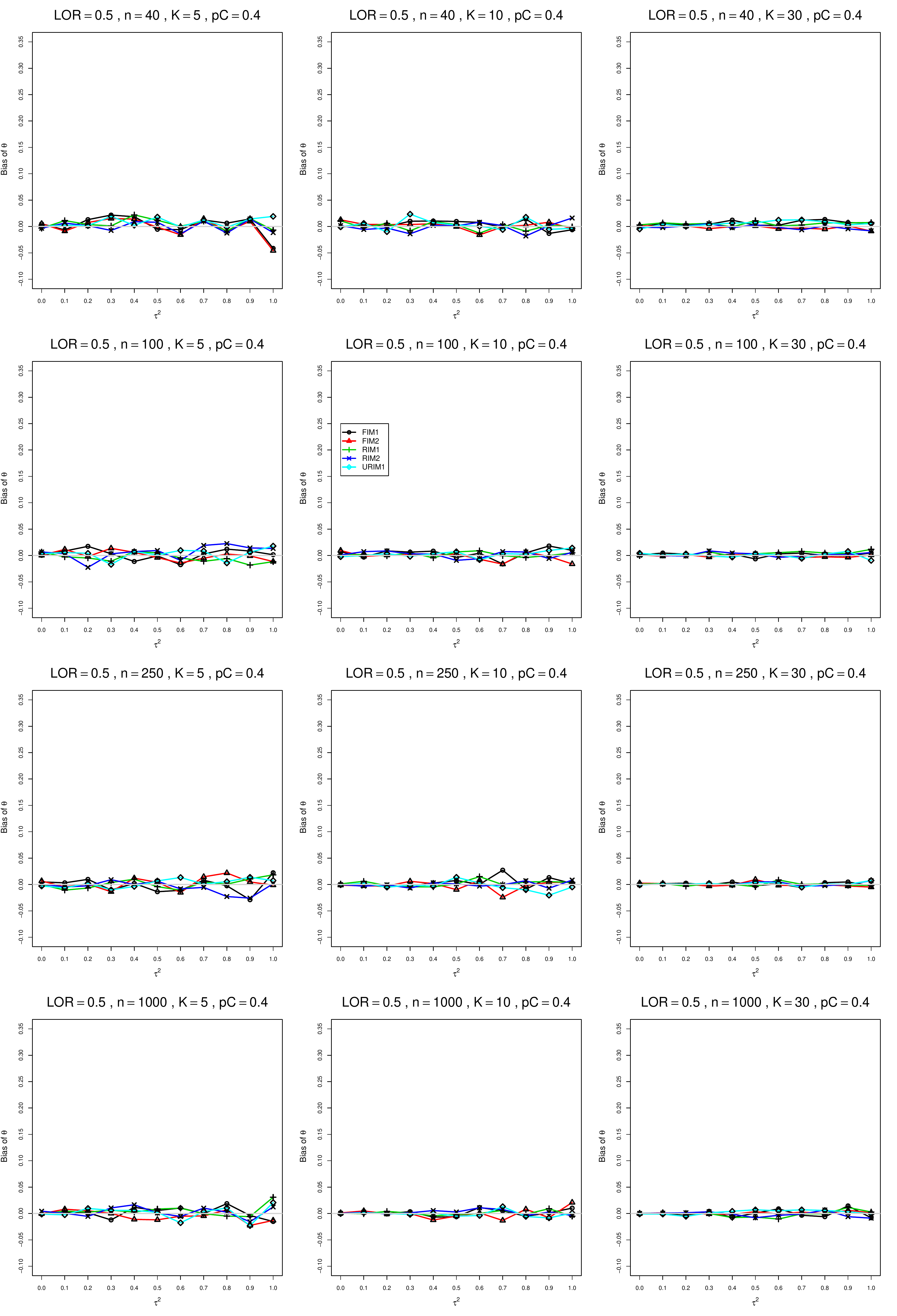}
	\caption{Bias of  overall log-odds ratio $\hat{\theta}_{DL}$ for $\theta=0.5$, $p_{C}=0.4$, $\sigma^2=0.1$, constant sample sizes $n=40,\;100,\;250,\;1000$.
The data-generation mechanisms are FIM1 ($\circ$), FIM2 ($\triangle$), RIM1 (+), RIM2 ($\times$), and URIM1 ($\diamond$).
		\label{PlotBiasThetamu05andpC04LOR_DLsigma01}}
\end{figure}
\begin{figure}[t]
	\centering
	\includegraphics[scale=0.33]{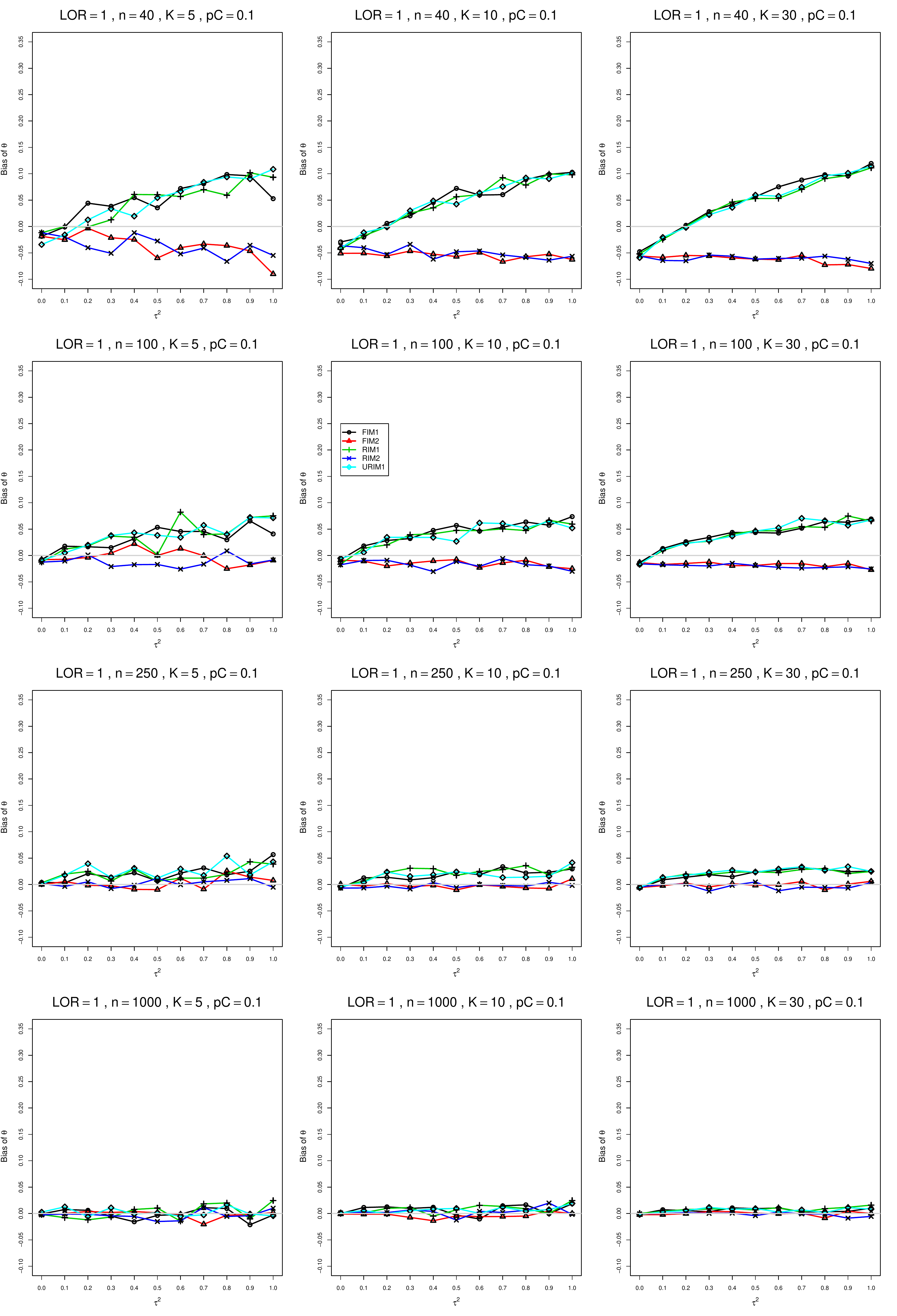}
	\caption{Bias of  overall log-odds ratio $\hat{\theta}_{DL}$ for $\theta=1$, $p_{C}=0.1$, $\sigma^2=0.1$, constant sample sizes $n=40,\;100,\;250,\;1000$.
The data-generation mechanisms are FIM1 ($\circ$), FIM2 ($\triangle$), RIM1 (+), RIM2 ($\times$), and URIM1 ($\diamond$).
		\label{PlotBiasThetamu1andpC01LOR_DLsigma01}}
\end{figure}
\begin{figure}[t]
	\centering
	\includegraphics[scale=0.33]{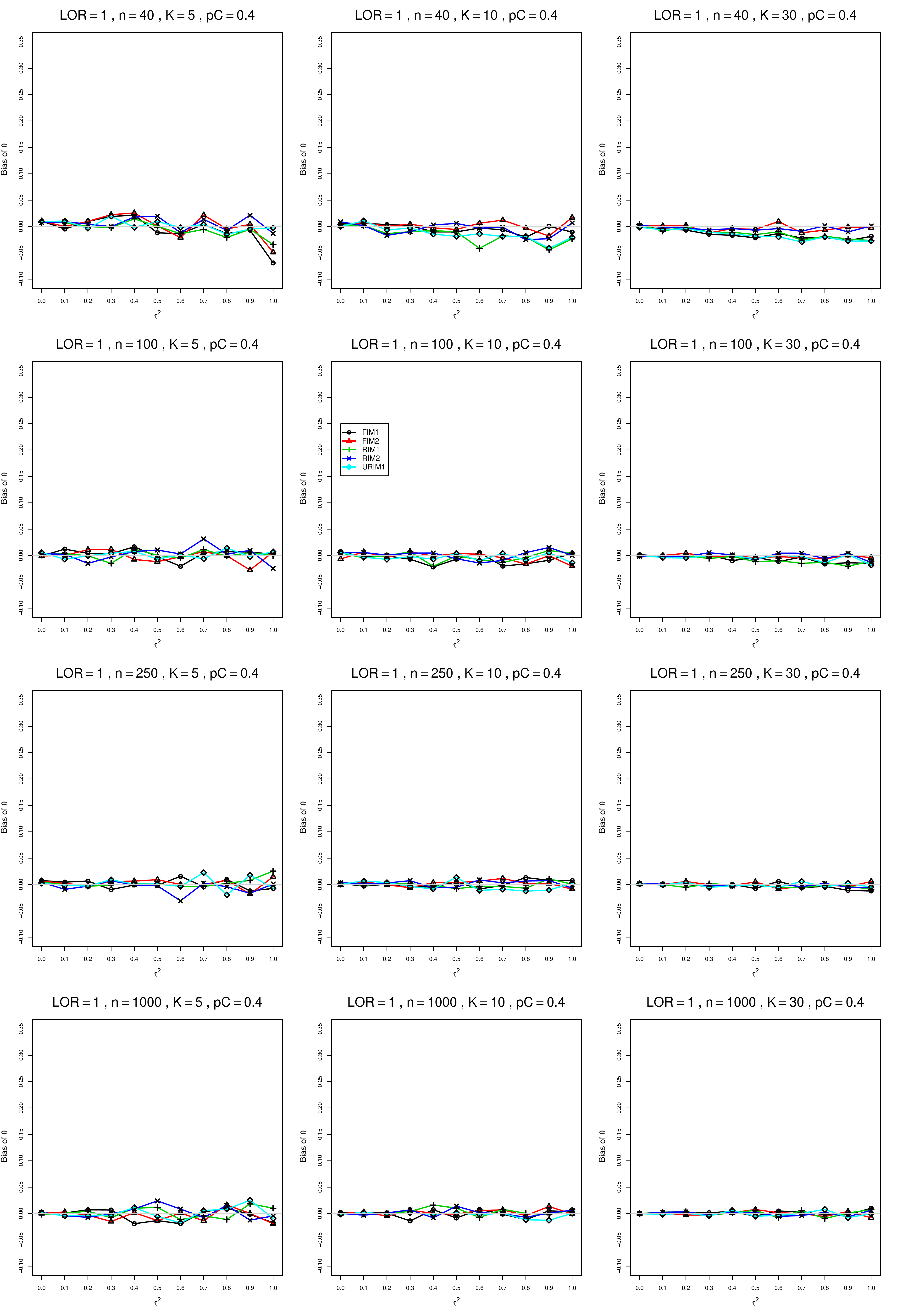}
	\caption{Bias of  overall log-odds ratio $\hat{\theta}_{DL}$ for $\theta=1$, $p_{C}=0.4$, $\sigma^2=0.1$, constant sample sizes $n=40,\;100,\;250,\;1000$.
The data-generation mechanisms are FIM1 ($\circ$), FIM2 ($\triangle$), RIM1 (+), RIM2 ($\times$), and URIM1 ($\diamond$).
		\label{PlotBiasThetamu1andpC04LOR_DLsigma01}}
\end{figure}
\begin{figure}[t]
	\centering
	\includegraphics[scale=0.33]{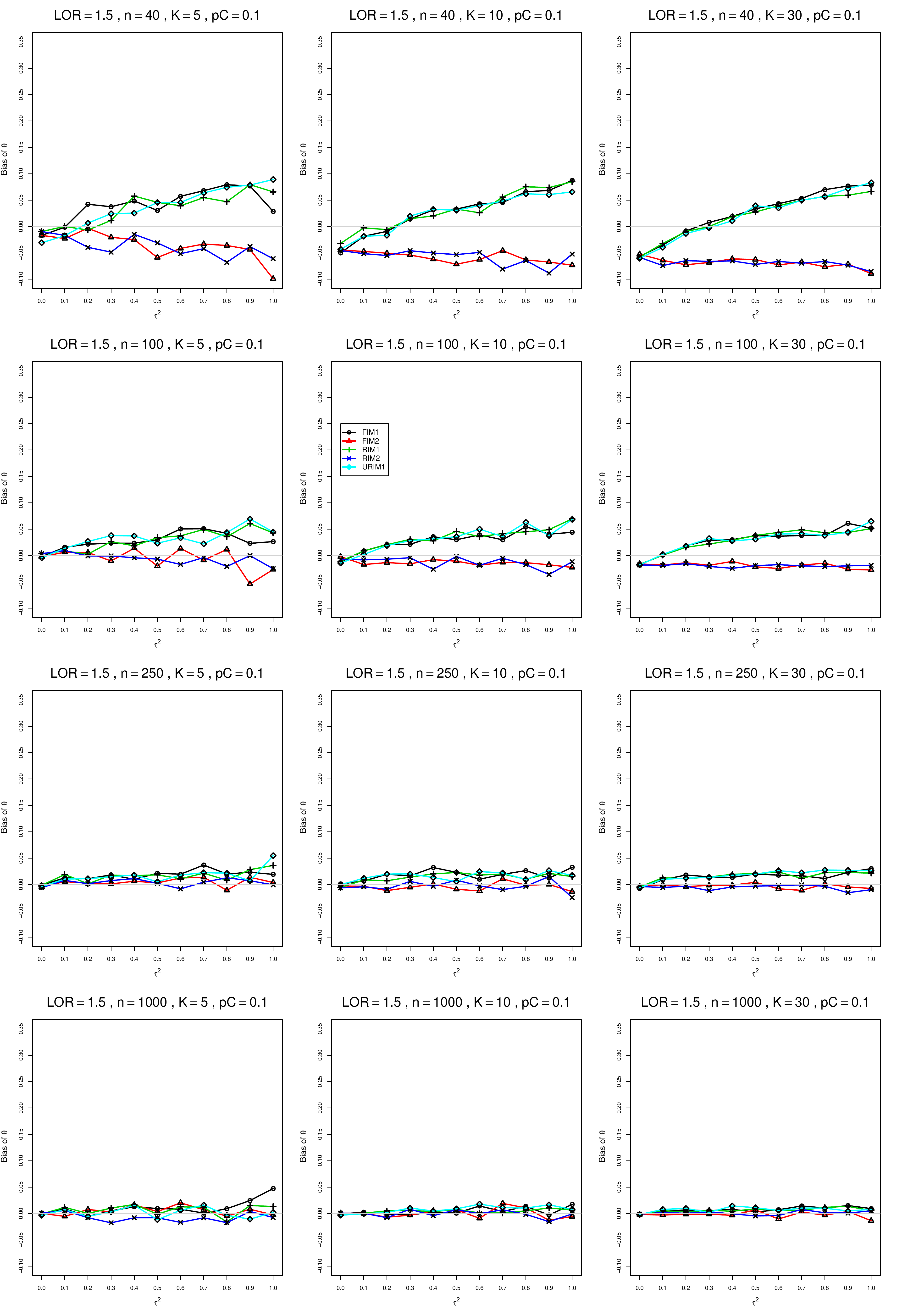}
	\caption{Bias of  overall log-odds ratio $\hat{\theta}_{DL}$ for $\theta=1.5$, $p_{C}=0.1$, $\sigma^2=0.1$, constant sample sizes $n=40,\;100,\;250,\;1000$.
The data-generation mechanisms are FIM1 ($\circ$), FIM2 ($\triangle$), RIM1 (+), RIM2 ($\times$), and URIM1 ($\diamond$).
		\label{PlotBiasThetamu15andpC01LOR_DLsigma01}}
\end{figure}
\begin{figure}[t]
	\centering
	\includegraphics[scale=0.33]{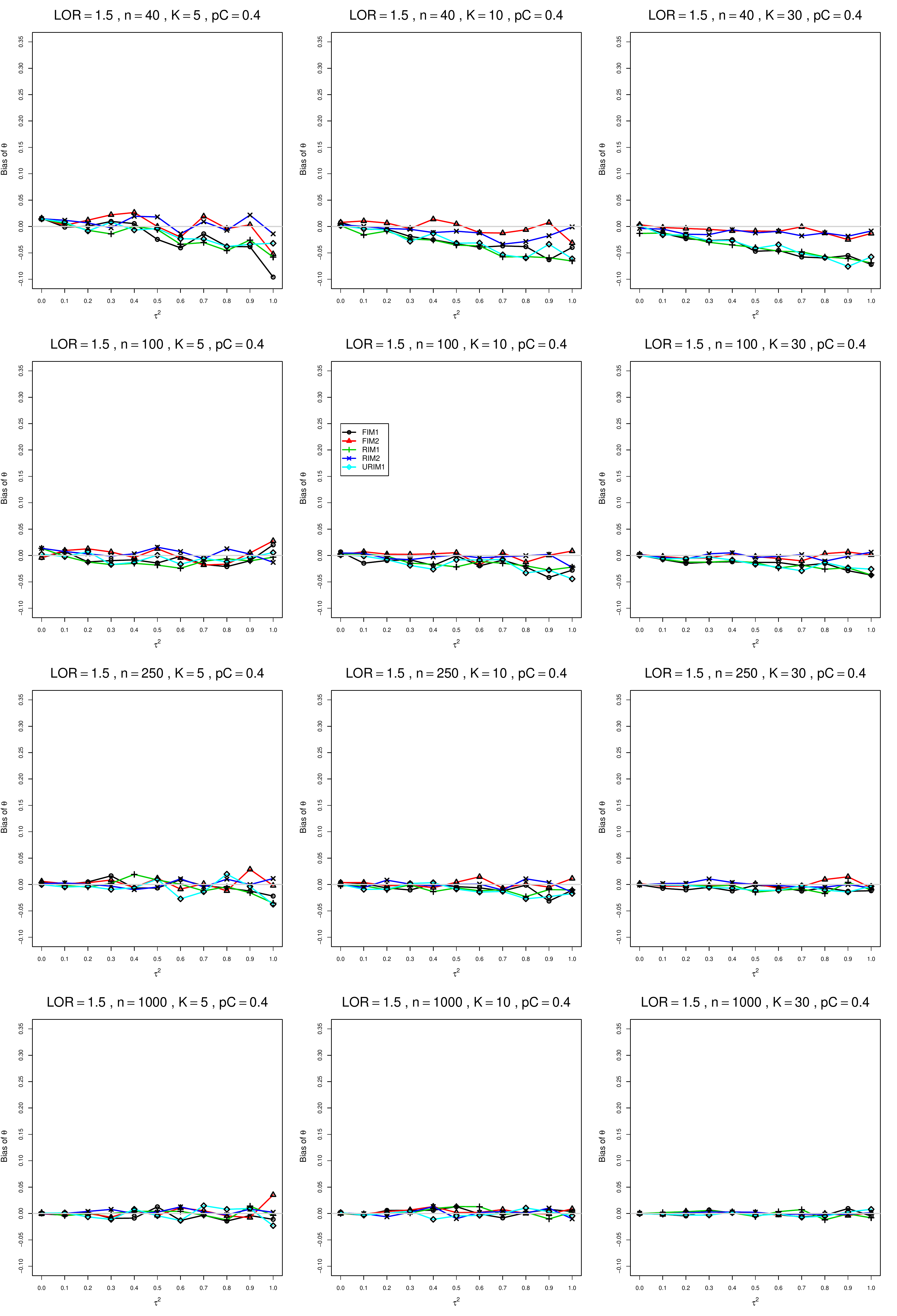}
	\caption{Bias of  overall log-odds ratio $\hat{\theta}_{DL}$ for $\theta=1.5$, $p_{C}=0.4$, $\sigma^2=0.1$, constant sample sizes $n=40,\;100,\;250,\;1000$.
The data-generation mechanisms are FIM1 ($\circ$), FIM2 ($\triangle$), RIM1 (+), RIM2 ($\times$), and URIM1 ($\diamond$).
		\label{PlotBiasThetamu15andpC04LOR_DLsigma01}}
\end{figure}
\begin{figure}[t]
	\centering
	\includegraphics[scale=0.33]{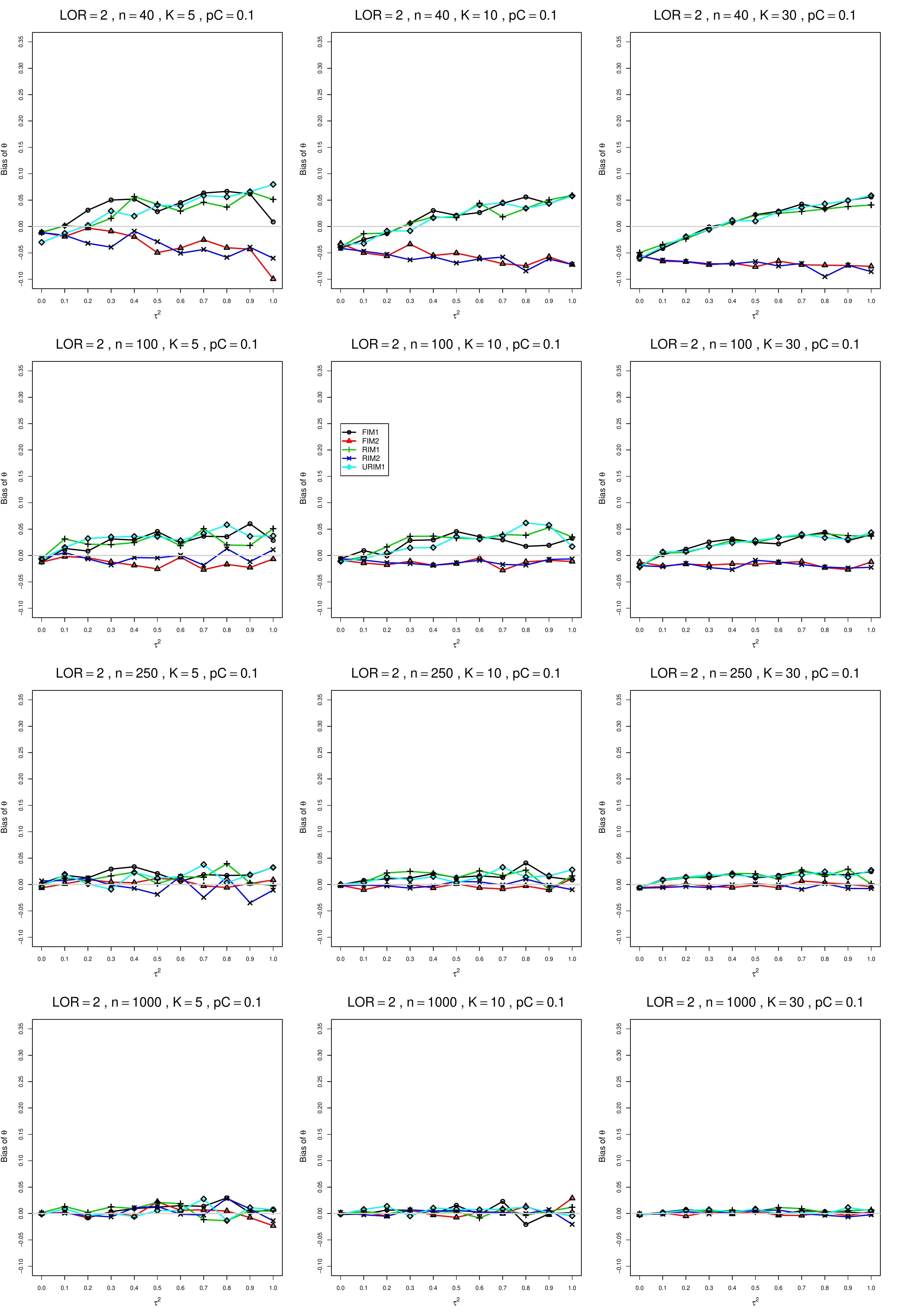}
	\caption{Bias of  overall log-odds ratio $\hat{\theta}_{DL}$ for $\theta=2$, $p_{C}=0.1$, $\sigma^2=0.1$, constant sample sizes $n=40,\;100,\;250,\;1000$.
The data-generation mechanisms are FIM1 ($\circ$), FIM2 ($\triangle$), RIM1 (+), RIM2 ($\times$), and URIM1 ($\diamond$).
		\label{PlotBiasThetamu2andpC01LOR_DLsigma01}}
\end{figure}
\begin{figure}[t]
	\centering
	\includegraphics[scale=0.33]{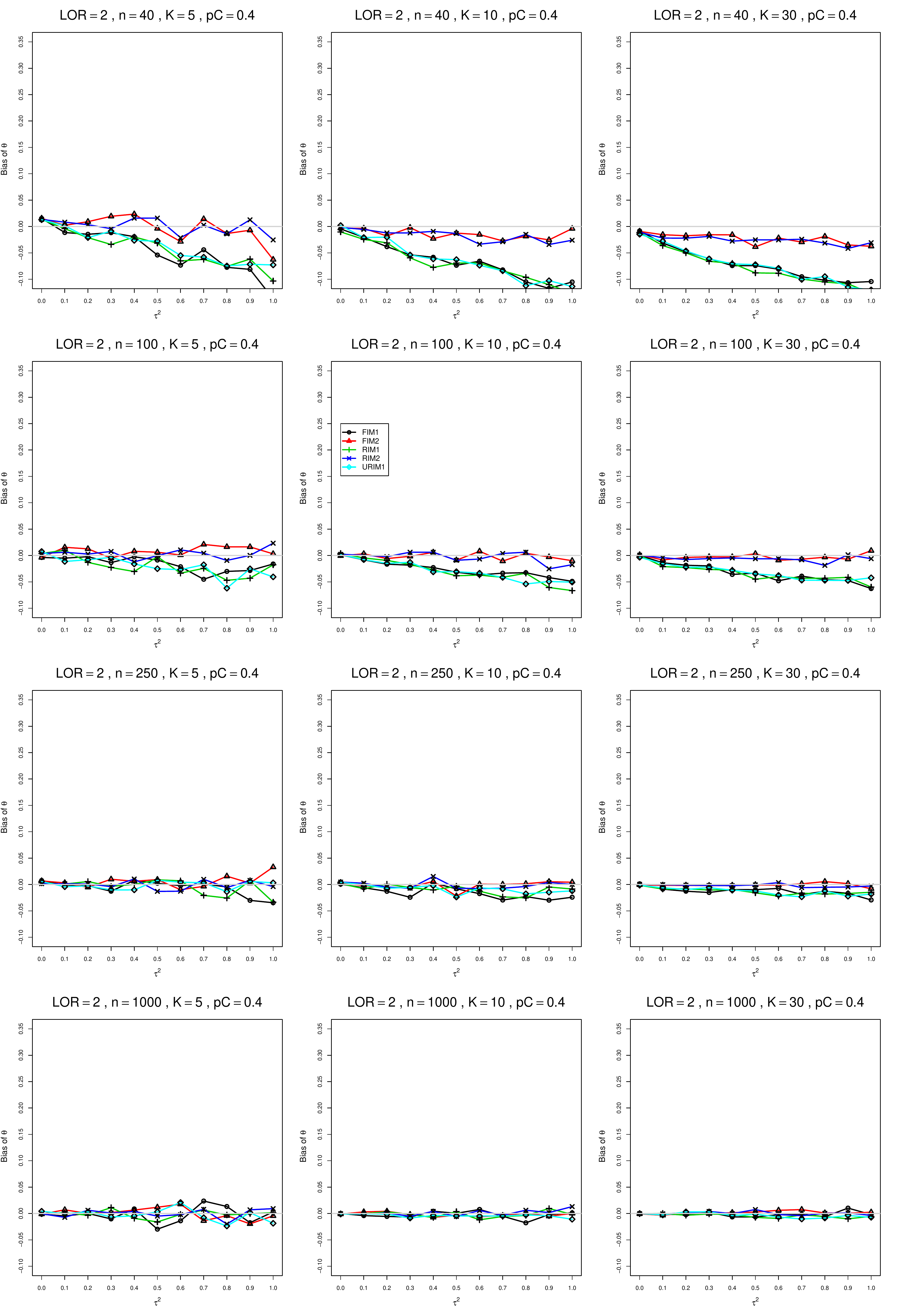}
	\caption{Bias of  overall log-odds ratio $\hat{\theta}_{DL}$ for $\theta=2$, $p_{C}=0.4$, $\sigma^2=0.1$, constant sample sizes $n=40,\;100,\;250,\;1000$.
The data-generation mechanisms are FIM1 ($\circ$), FIM2 ($\triangle$), RIM1 (+), RIM2 ($\times$), and URIM1 ($\diamond$).
		\label{PlotBiasThetamu2andpC04LOR_DLsigma01}}
\end{figure}
\begin{figure}[t]
	\centering
	\includegraphics[scale=0.33]{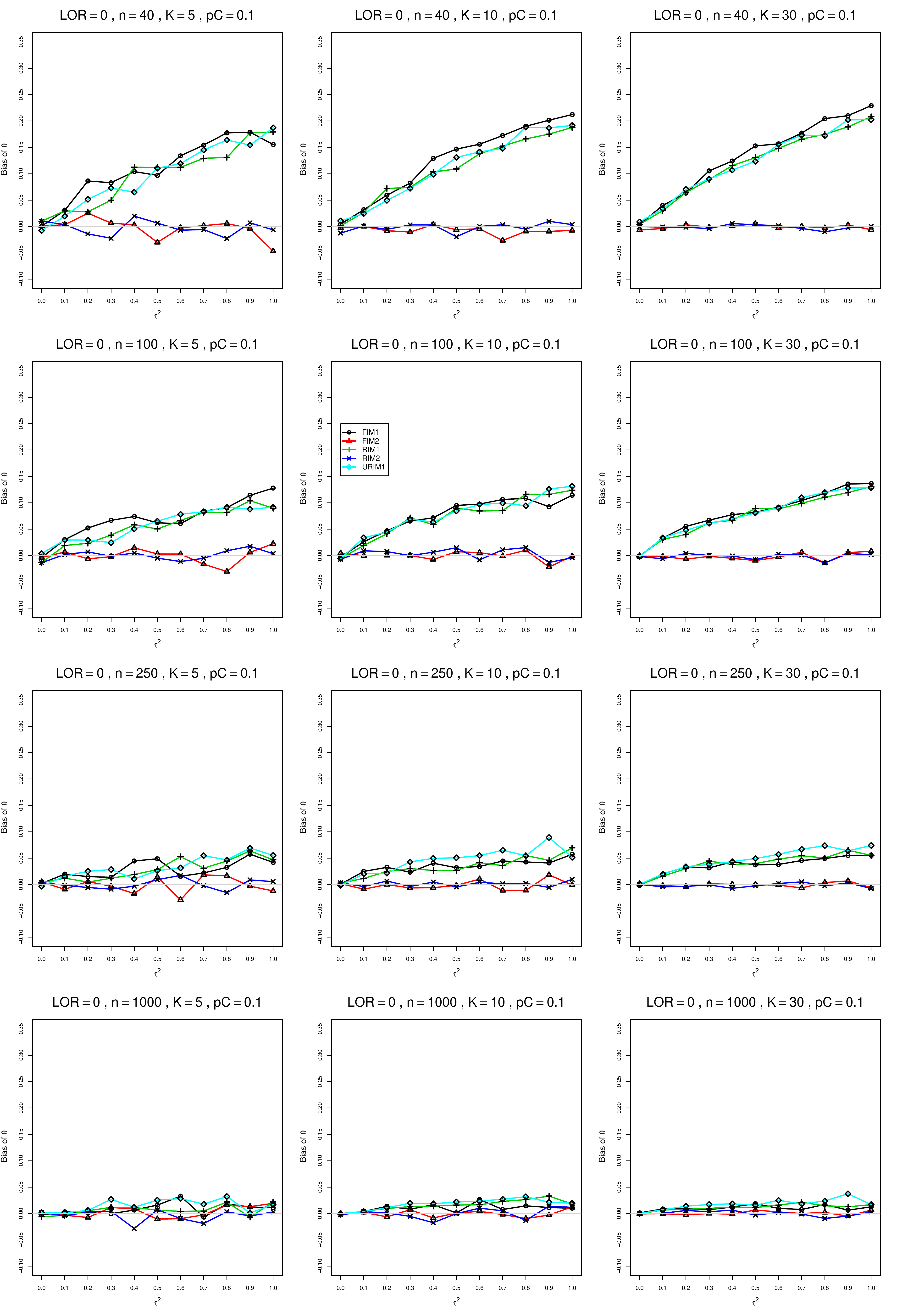}
	\caption{Bias of  overall log-odds ratio $\hat{\theta}_{DL}$ for $\theta=0$, $p_{C}=0.1$, $\sigma^2=0.4$, constant sample sizes $n=40,\;100,\;250,\;1000$.
The data-generation mechanisms are FIM1 ($\circ$), FIM2 ($\triangle$), RIM1 (+), RIM2 ($\times$), and URIM1 ($\diamond$).
		\label{PlotBiasThetamu0andpC01LOR_DLsigma04}}
\end{figure}
\begin{figure}[t]
	\centering
	\includegraphics[scale=0.33]{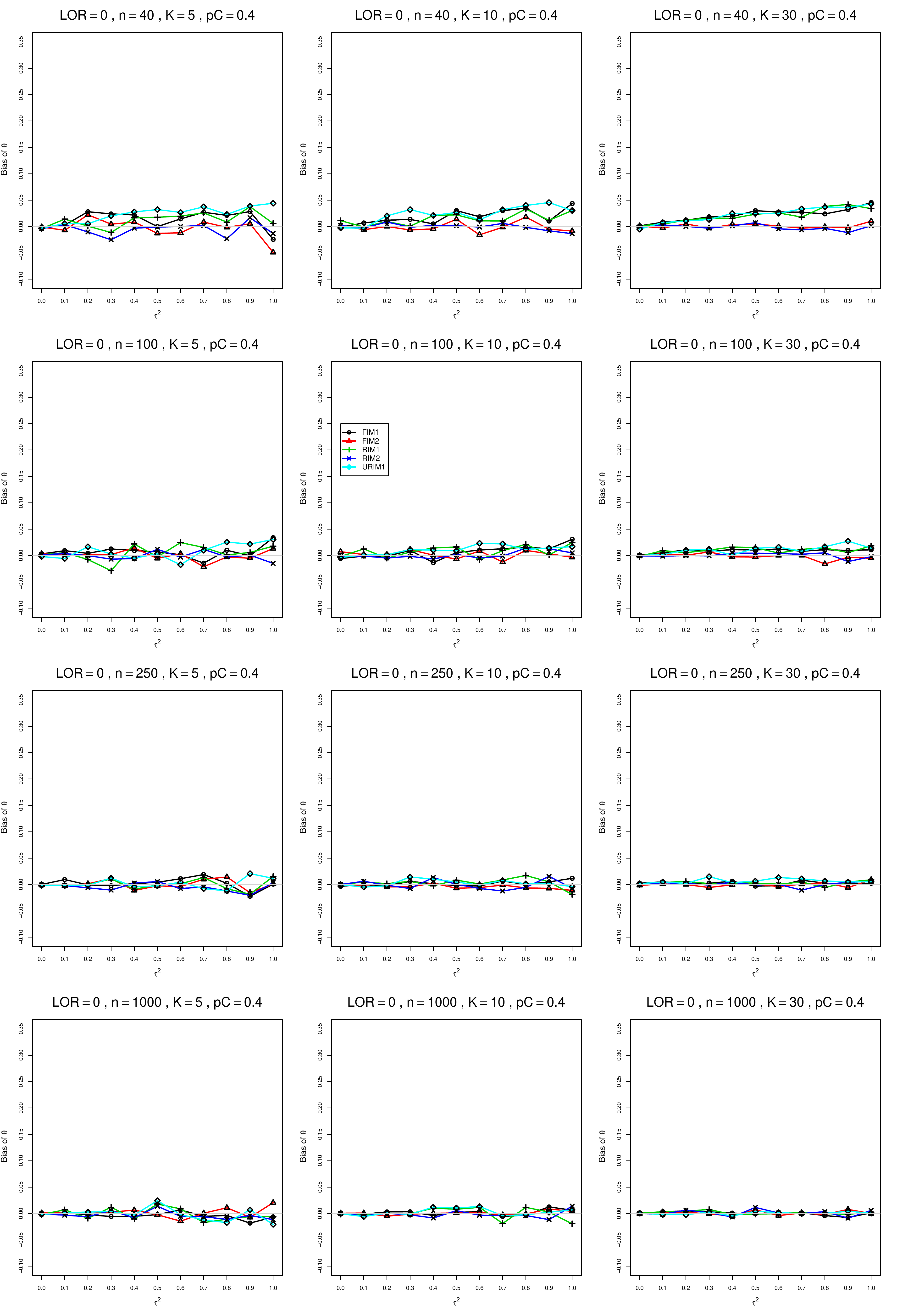}
	\caption{Bias of  overall log-odds ratio $\hat{\theta}_{DL}$ for $\theta=0$, $p_{C}=0.4$, $\sigma^2=0.4$, constant sample sizes $n=40,\;100,\;250,\;1000$.
The data-generation mechanisms are FIM1 ($\circ$), FIM2 ($\triangle$), RIM1 (+), RIM2 ($\times$), and URIM1 ($\diamond$).
		\label{PlotBiasThetamu0andpC04LOR_DLsigma04}}
\end{figure}
\begin{figure}[t]
	\centering
	\includegraphics[scale=0.33]{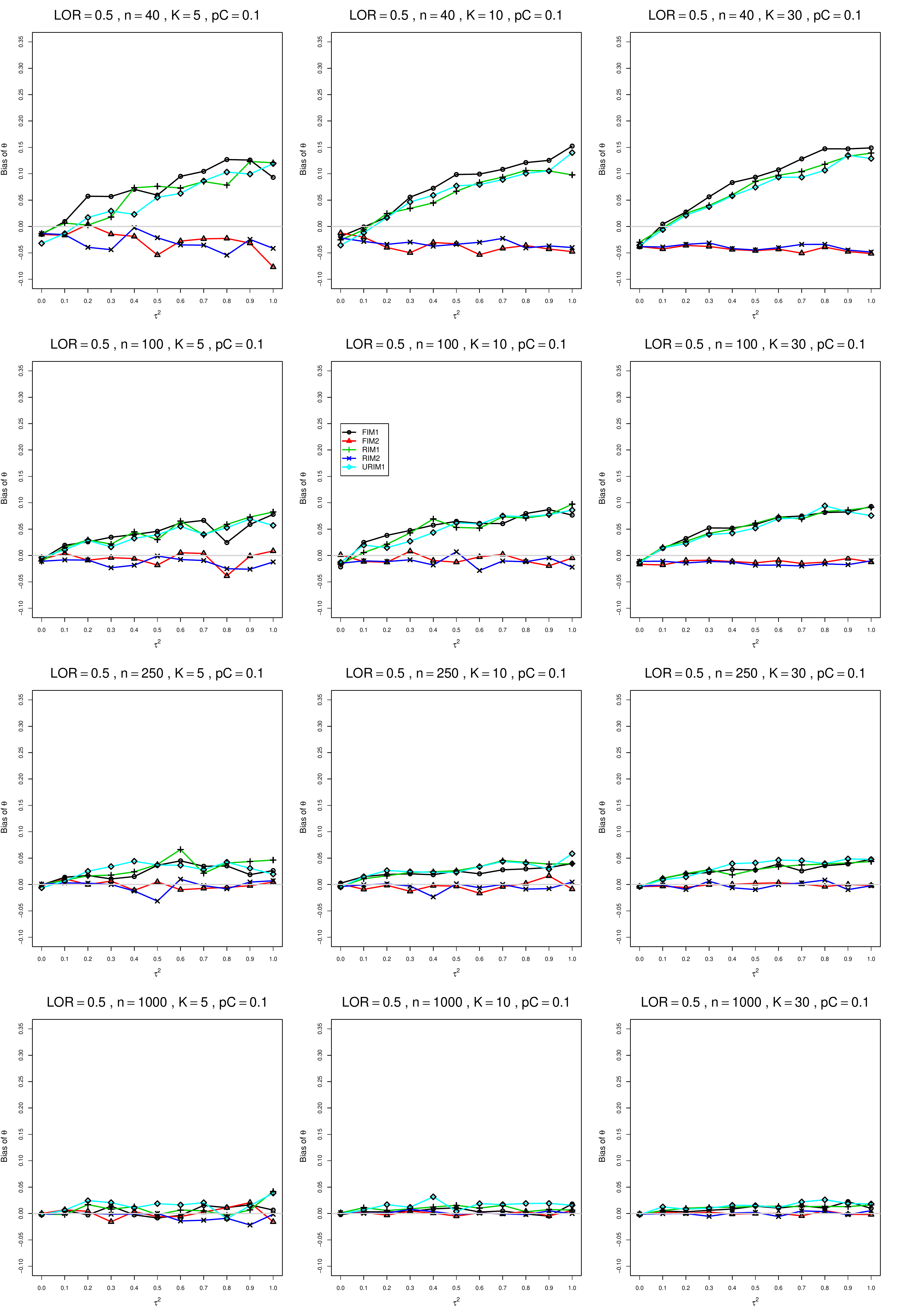}
	\caption{Bias of  overall log-odds ratio $\hat{\theta}_{DL}$ for $\theta=0.5$, $p_{C}=0.1$, $\sigma^2=0.4$, constant sample sizes $n=40,\;100,\;250,\;1000$.
The data-generation mechanisms are FIM1 ($\circ$), FIM2 ($\triangle$), RIM1 (+), RIM2 ($\times$), and URIM1 ($\diamond$).
		\label{PlotBiasThetamu05andpC01LOR_DLsigma04}}
\end{figure}
\begin{figure}[t]
	\centering
	\includegraphics[scale=0.33]{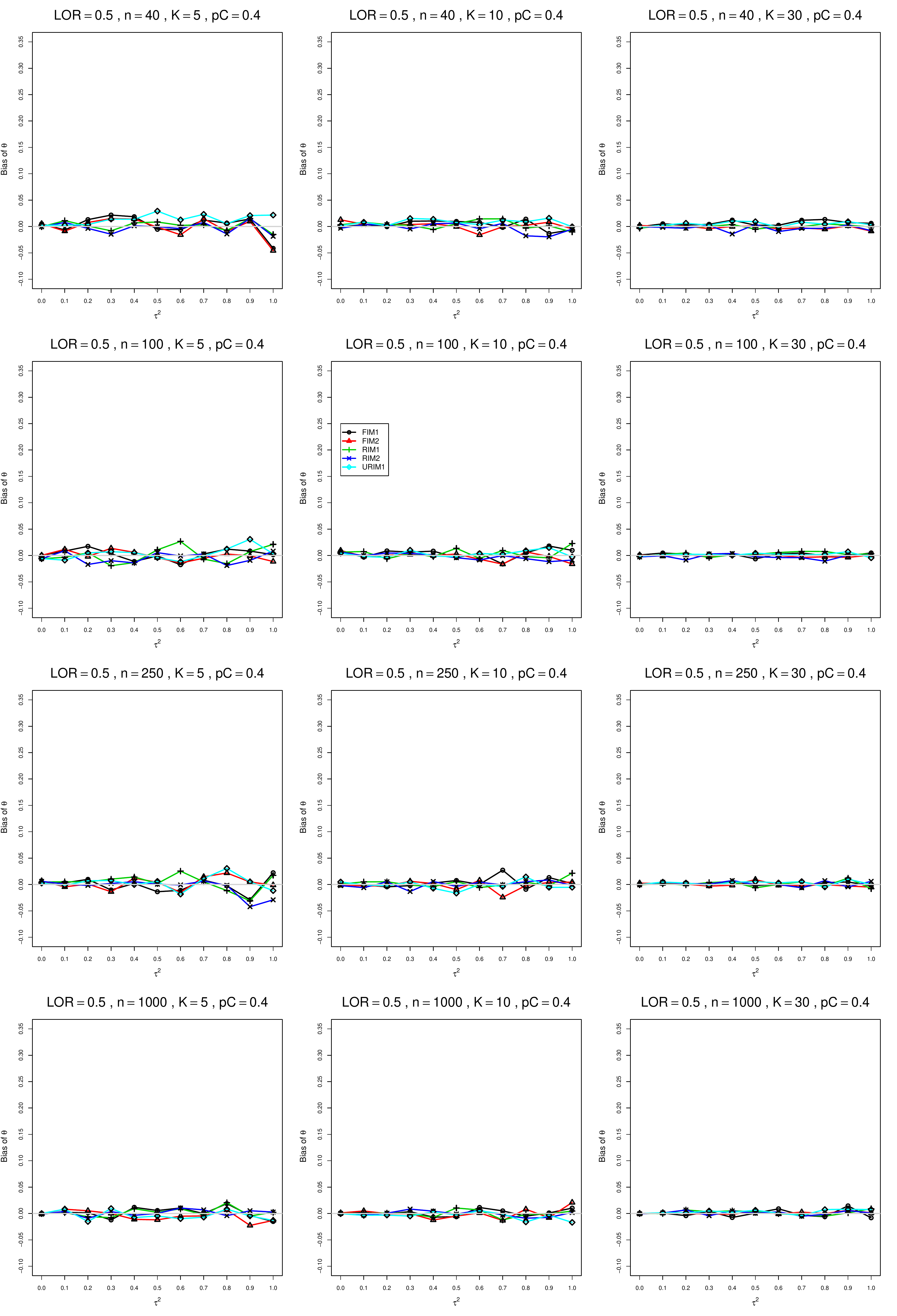}
	\caption{Bias of  overall log-odds ratio $\hat{\theta}_{DL}$ for $\theta=0.5$, $p_{C}=0.4$, $\sigma^2=0.4$, constant sample sizes $n=40,\;100,\;250,\;1000$.
The data-generation mechanisms are FIM1 ($\circ$), FIM2 ($\triangle$), RIM1 (+), RIM2 ($\times$), and URIM1 ($\diamond$).
		\label{PlotBiasThetamu05andpC04LOR_DLsigma04}}
\end{figure}
\begin{figure}[t]
	\centering
	\includegraphics[scale=0.33]{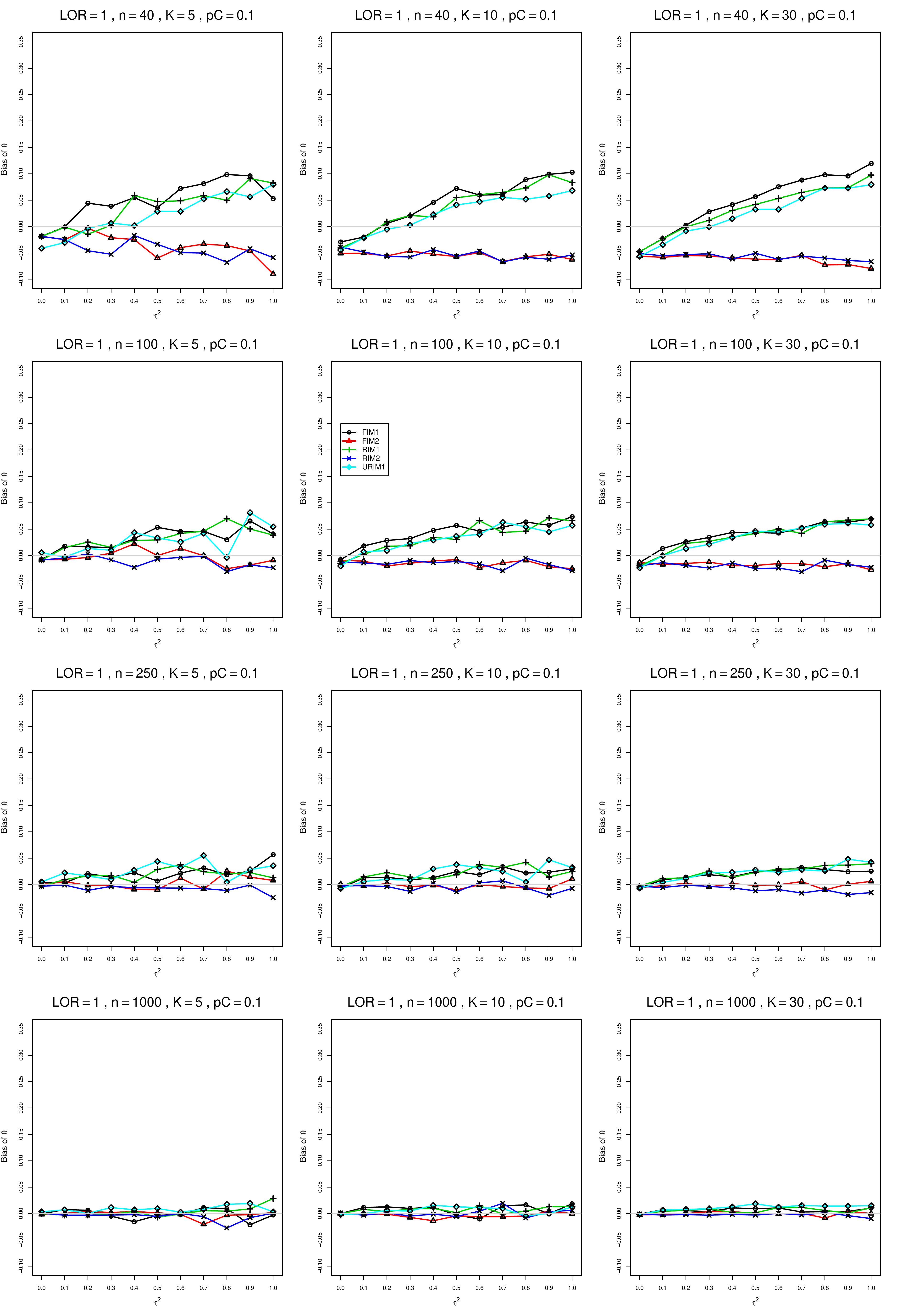}
	\caption{Bias of  overall log-odds ratio $\hat{\theta}_{DL}$ for $\theta=1$, $p_{C}=0.1$, $\sigma^2=0.4$, constant sample sizes $n=40,\;100,\;250,\;1000$.
The data-generation mechanisms are FIM1 ($\circ$), FIM2 ($\triangle$), RIM1 (+), RIM2 ($\times$), and URIM1 ($\diamond$).
		\label{PlotBiasThetamu1andpC01LOR_DLsigma04}}
\end{figure}
\begin{figure}[t]
	\centering
	\includegraphics[scale=0.33]{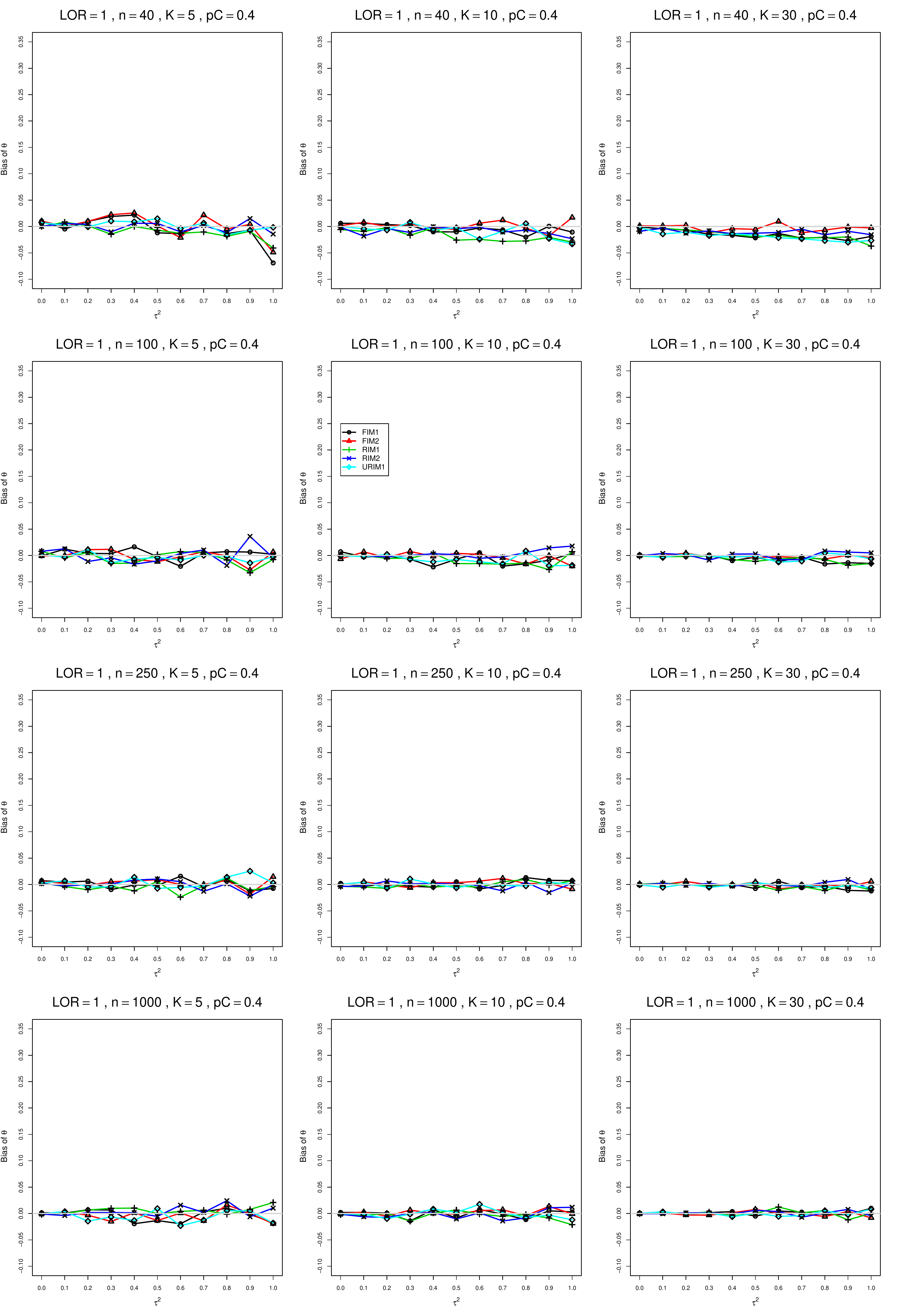}
	\caption{Bias of  overall log-odds ratio $\hat{\theta}_{DL}$ for $\theta=1$, $p_{C}=0.4$, $\sigma^2=0.4$, constant sample sizes $n=40,\;100,\;250,\;1000$.
The data-generation mechanisms are FIM1 ($\circ$), FIM2 ($\triangle$), RIM1 (+), RIM2 ($\times$), and URIM1 ($\diamond$).
		\label{PlotBiasThetamu1andpC04LOR_DLsigma04}}
\end{figure}
\begin{figure}[t]
	\centering
	\includegraphics[scale=0.33]{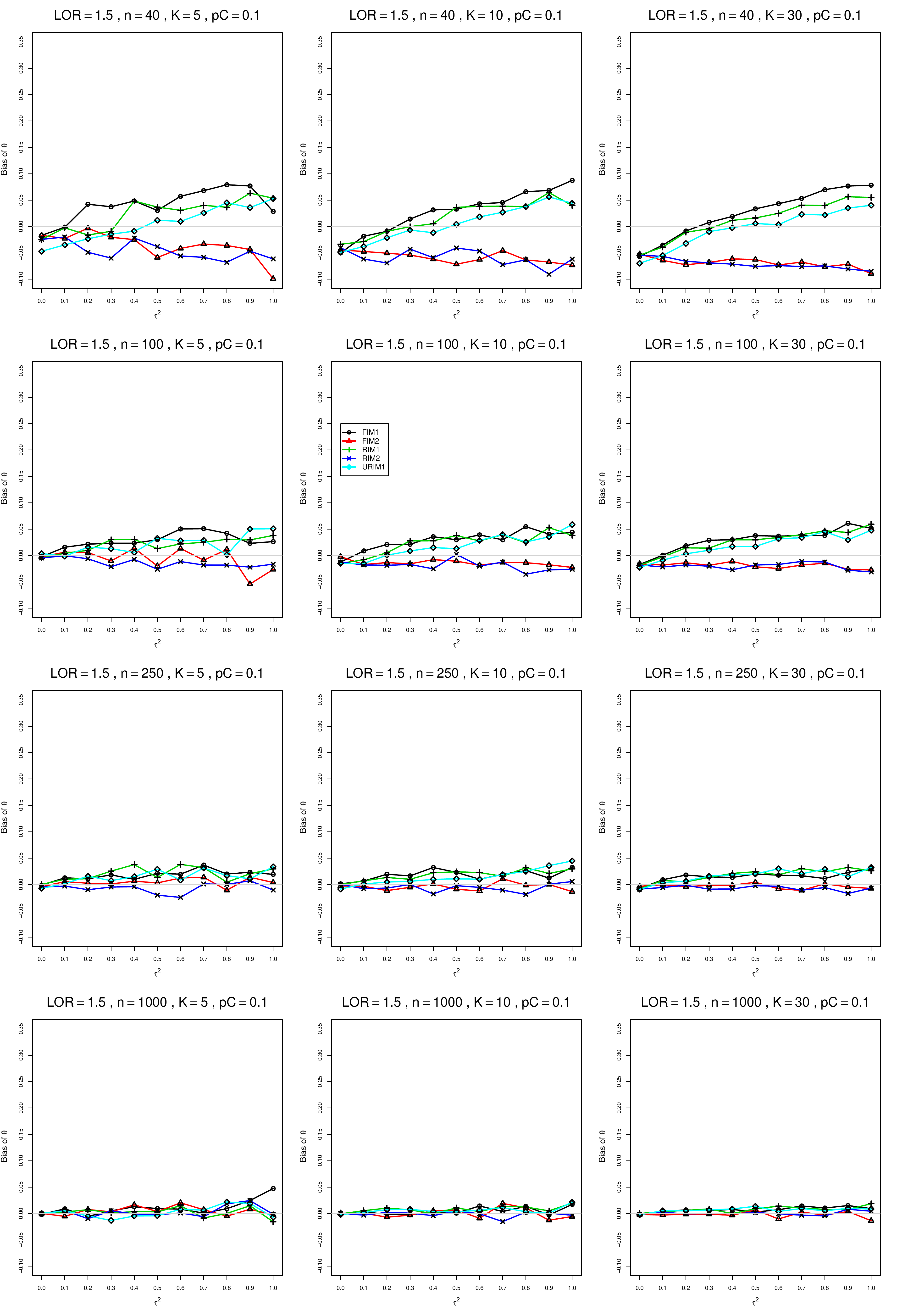}
	\caption{Bias of  overall log-odds ratio $\hat{\theta}_{DL}$ for $\theta=1.5$, $p_{C}=0.1$, $\sigma^2=0.4$, constant sample sizes $n=40,\;100,\;250,\;1000$.
The data-generation mechanisms are FIM1 ($\circ$), FIM2 ($\triangle$), RIM1 (+), RIM2 ($\times$), and URIM1 ($\diamond$).
		\label{PlotBiasThetamu15andpC01LOR_DLsigma04}}
\end{figure}
\begin{figure}[t]
	\centering
	\includegraphics[scale=0.33]{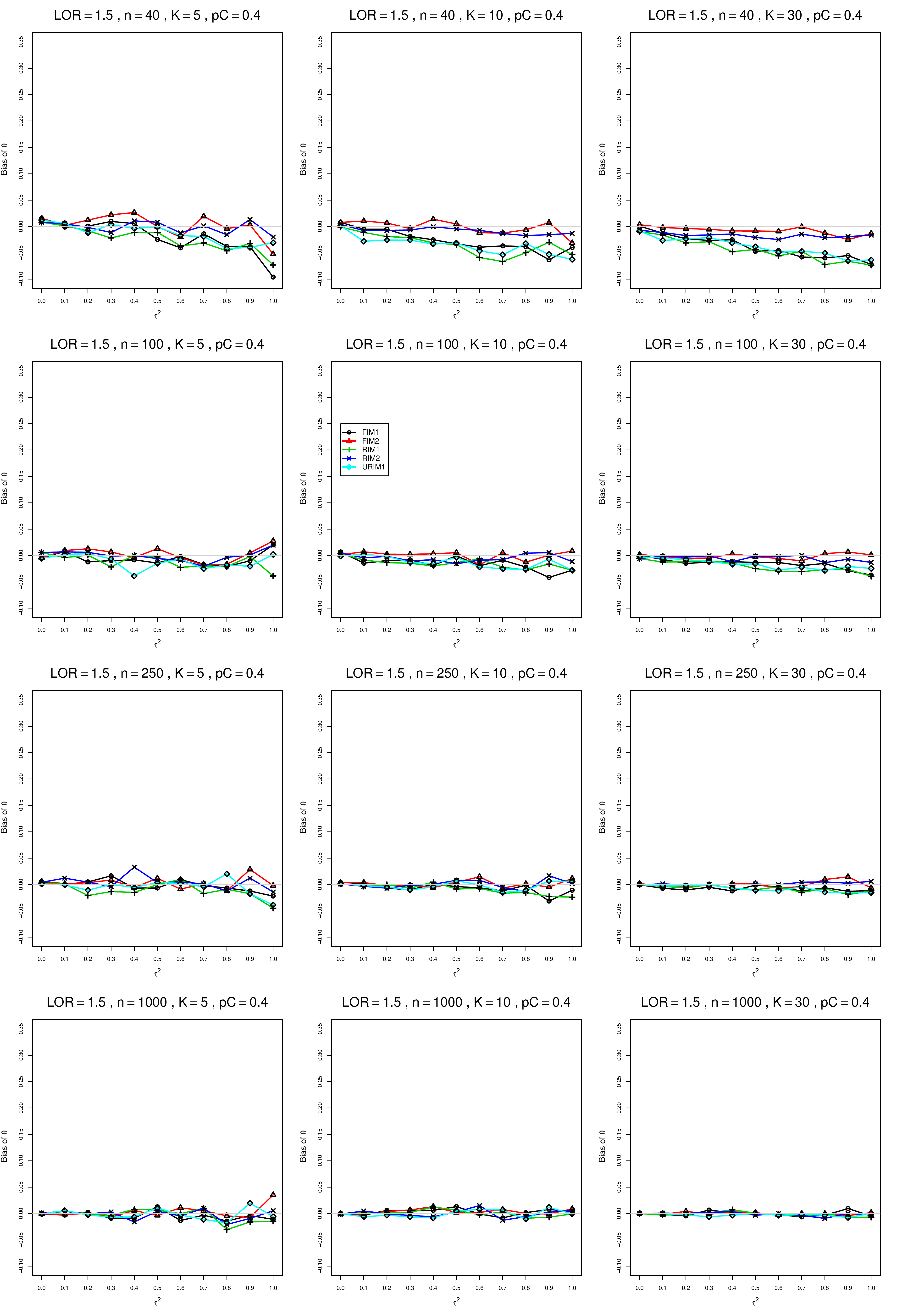}
	\caption{Bias of  overall log-odds ratio $\hat{\theta}_{DL}$ for $\theta=1.5$, $p_{C}=0.4$, $\sigma^2=0.4$, constant sample sizes $n=40,\;100,\;250,\;1000$.
The data-generation mechanisms are FIM1 ($\circ$), FIM2 ($\triangle$), RIM1 (+), RIM2 ($\times$), and URIM1 ($\diamond$).
		\label{PlotBiasThetamu15andpC04LOR_DLsigma04}}
\end{figure}
\begin{figure}[t]
	\centering
	\includegraphics[scale=0.33]{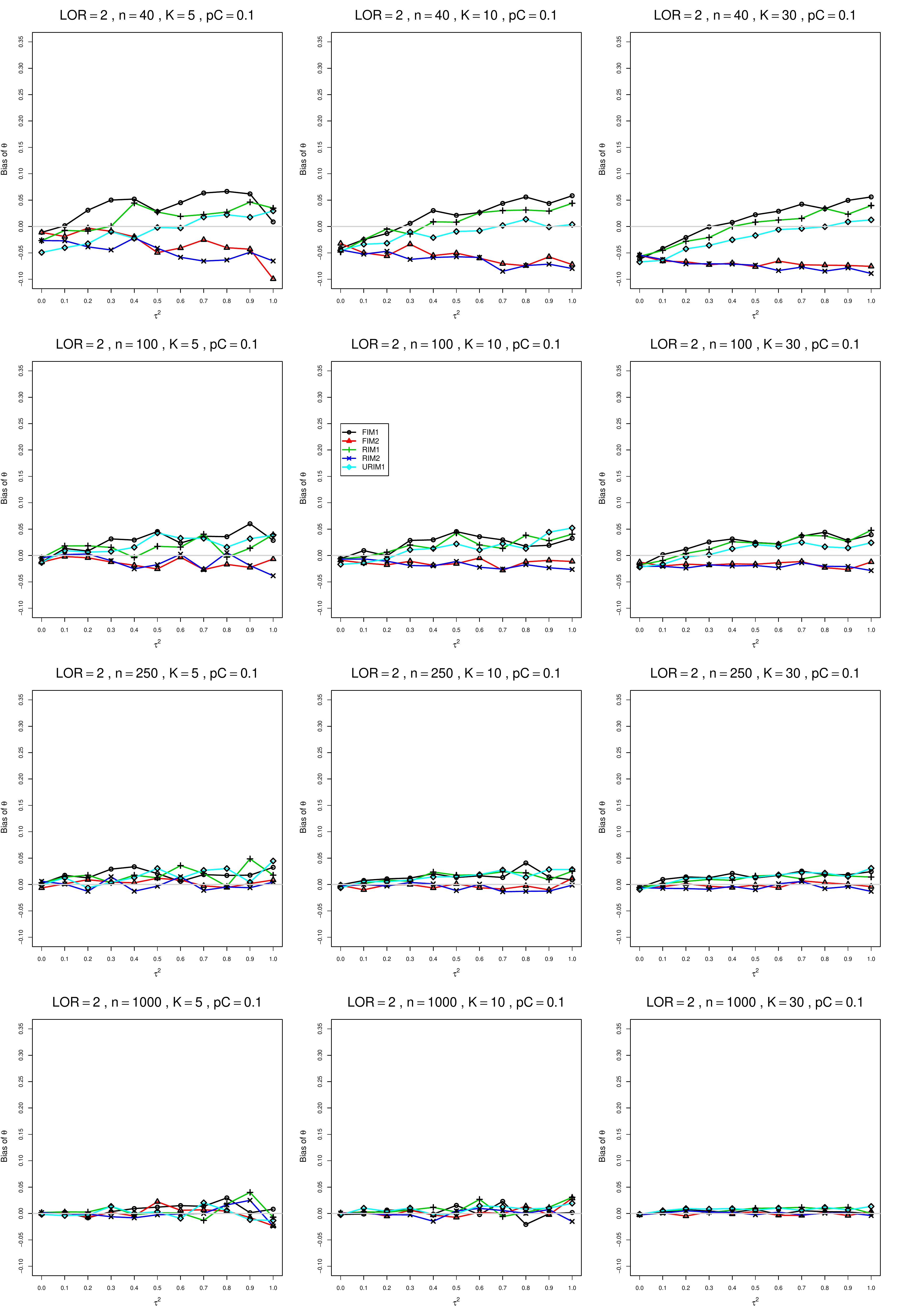}
	\caption{Bias of  overall log-odds ratio $\hat{\theta}_{DL}$ for $\theta=2$, $p_{C}=0.1$, $\sigma^2=0.4$, constant sample sizes $n=40,\;100,\;250,\;1000$.
The data-generation mechanisms are FIM1 ($\circ$), FIM2 ($\triangle$), RIM1 (+), RIM2 ($\times$), and URIM1 ($\diamond$).
		\label{PlotBiasThetamu2andpC01LOR_DLsigma04}}
\end{figure}
\begin{figure}[t]
	\centering
	\includegraphics[scale=0.33]{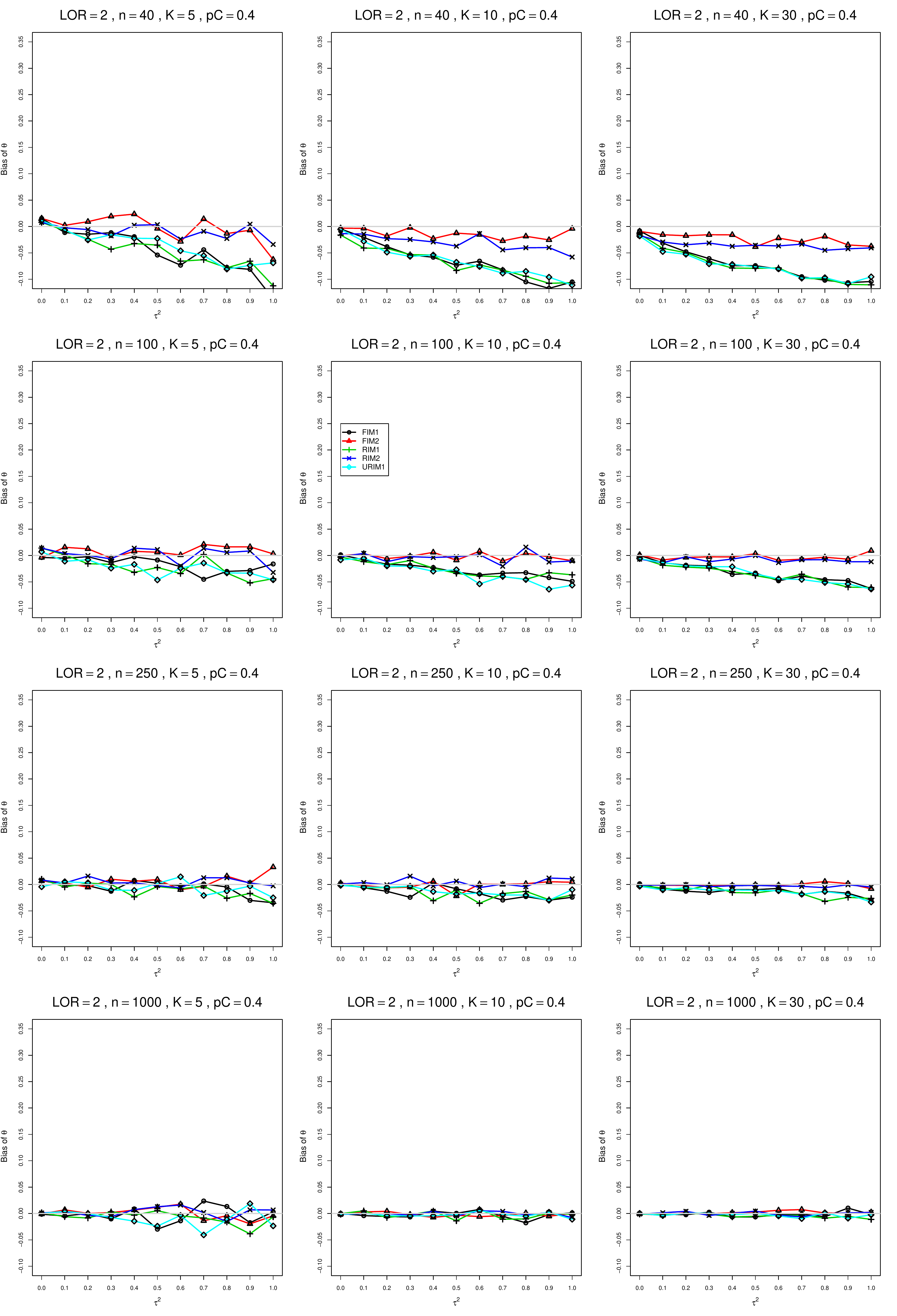}
	\caption{Bias of  overall log-odds ratio $\hat{\theta}_{DL}$ for $\theta=2$, $p_{C}=0.4$, $\sigma^2=0.4$, constant sample sizes $n=40,\;100,\;250,\;1000$.
The data-generation mechanisms are FIM1 ($\circ$), FIM2 ($\triangle$), RIM1 (+), RIM2 ($\times$), and URIM1 ($\diamond$).
		\label{PlotBiasThetamu2andpC04LOR_DLsigma04}}
\end{figure}

\clearpage
\subsection*{A2.2 Bias of $\hat{\theta}_{REML}$}
\renewcommand{\thefigure}{A2.2.\arabic{figure}}
\setcounter{figure}{0}

\begin{figure}[t]
	\centering
	\includegraphics[scale=0.33]{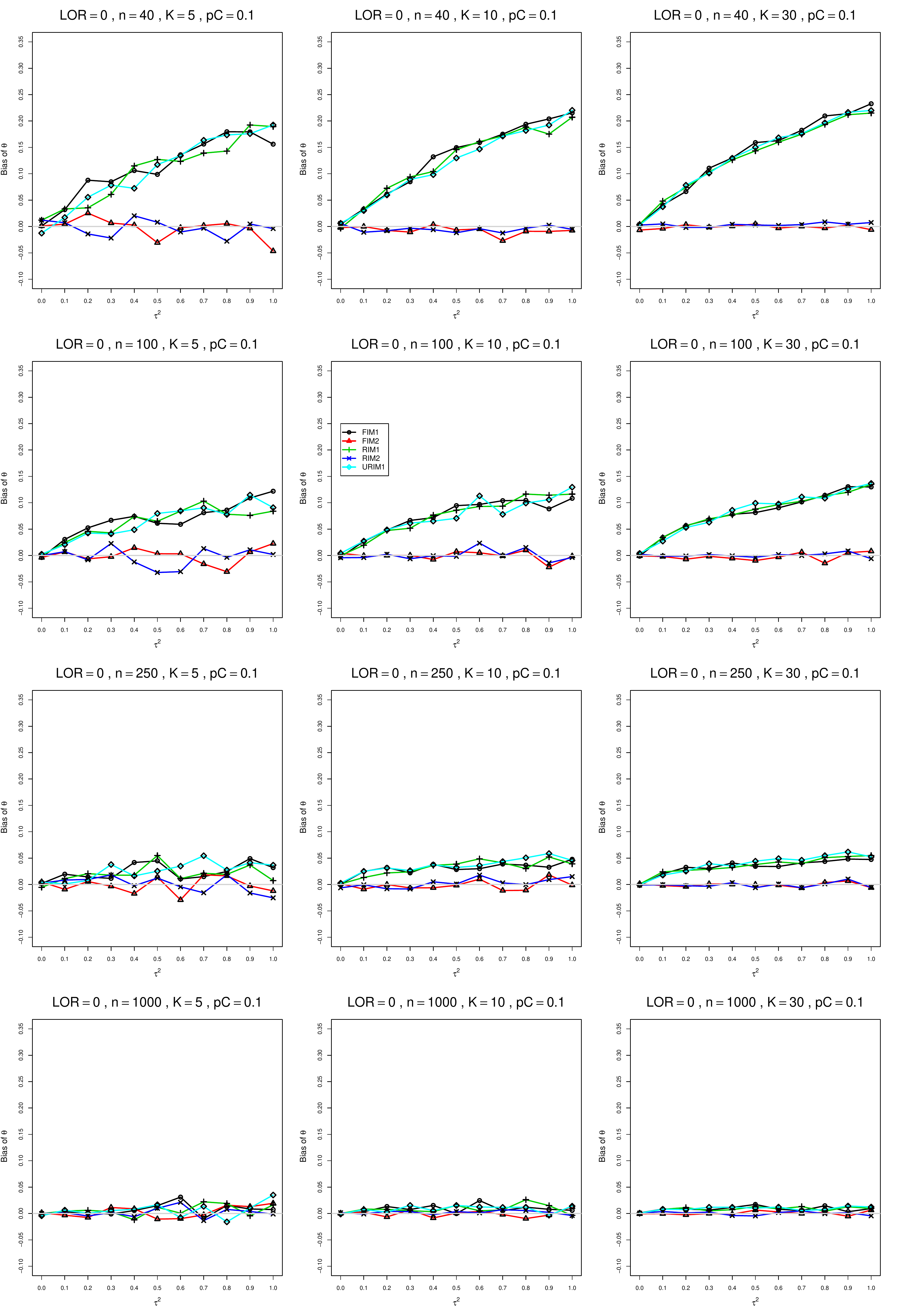}
	\caption{Bias of  overall log-odds ratio $\hat{\theta}_{REML}$ for $\theta=0$, $p_{C}=0.1$, $\sigma^2=0.1$, constant sample sizes $n=40,\;100,\;250,\;1000$.
The data-generation mechanisms are FIM1 ($\circ$), FIM2 ($\triangle$), RIM1 (+), RIM2 ($\times$), and URIM1 ($\diamond$).
		\label{PlotBiasThetamu0andpC01LOR_REMLsigma01}}
\end{figure}
\begin{figure}[t]
	\centering
	\includegraphics[scale=0.33]{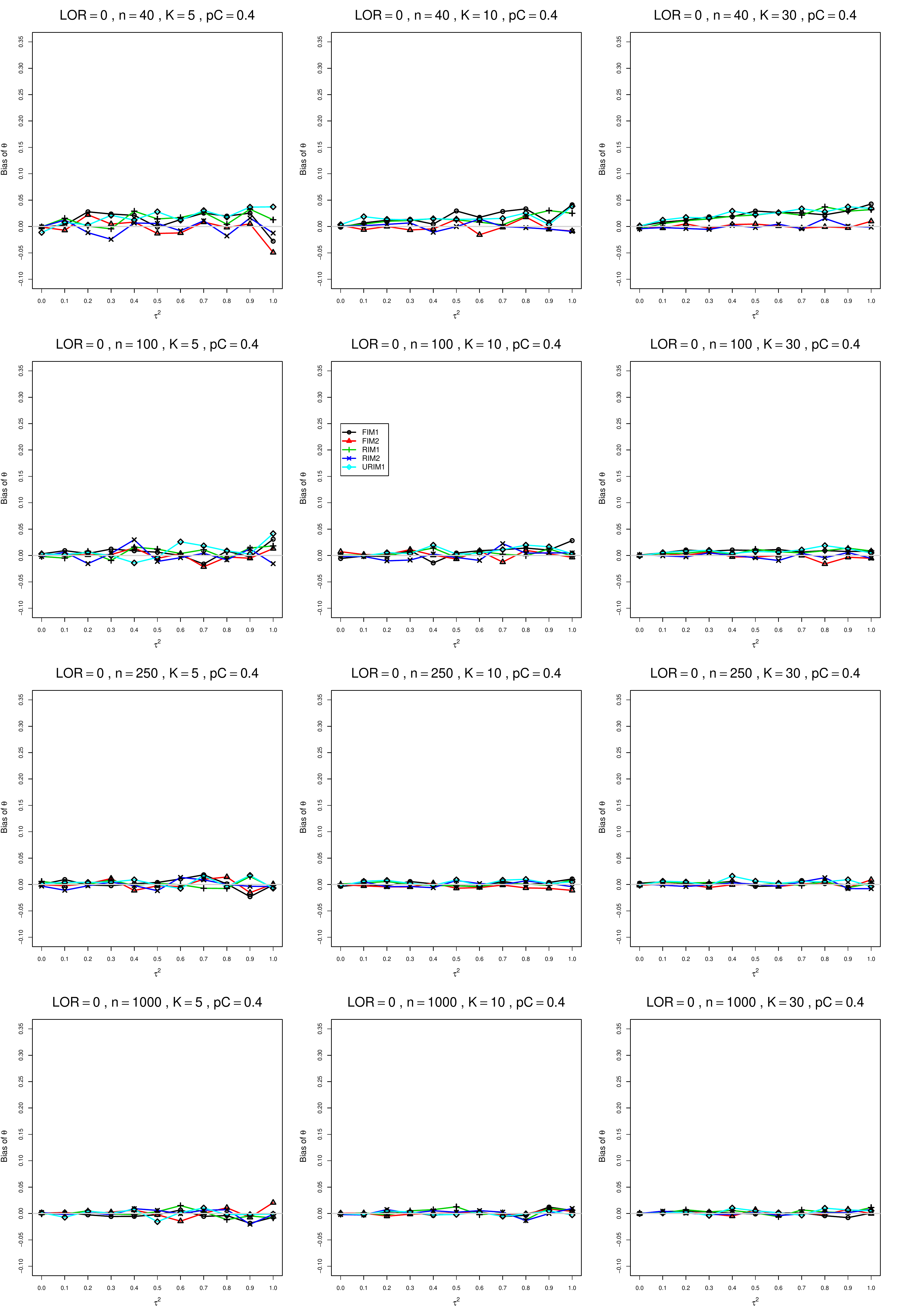}
	\caption{Bias of  overall log-odds ratio $\hat{\theta}_{REML}$ for $\theta=0$, $p_{C}=0.4$, $\sigma^2=0.1$, constant sample sizes $n=40,\;100,\;250,\;1000$.
The data-generation mechanisms are FIM1 ($\circ$), FIM2 ($\triangle$), RIM1 (+), RIM2 ($\times$), and URIM1 ($\diamond$).
		\label{PlotBiasThetamu0andpC04LOR_REMLsigma01}}
\end{figure}
\begin{figure}[t]
	\centering
	\includegraphics[scale=0.33]{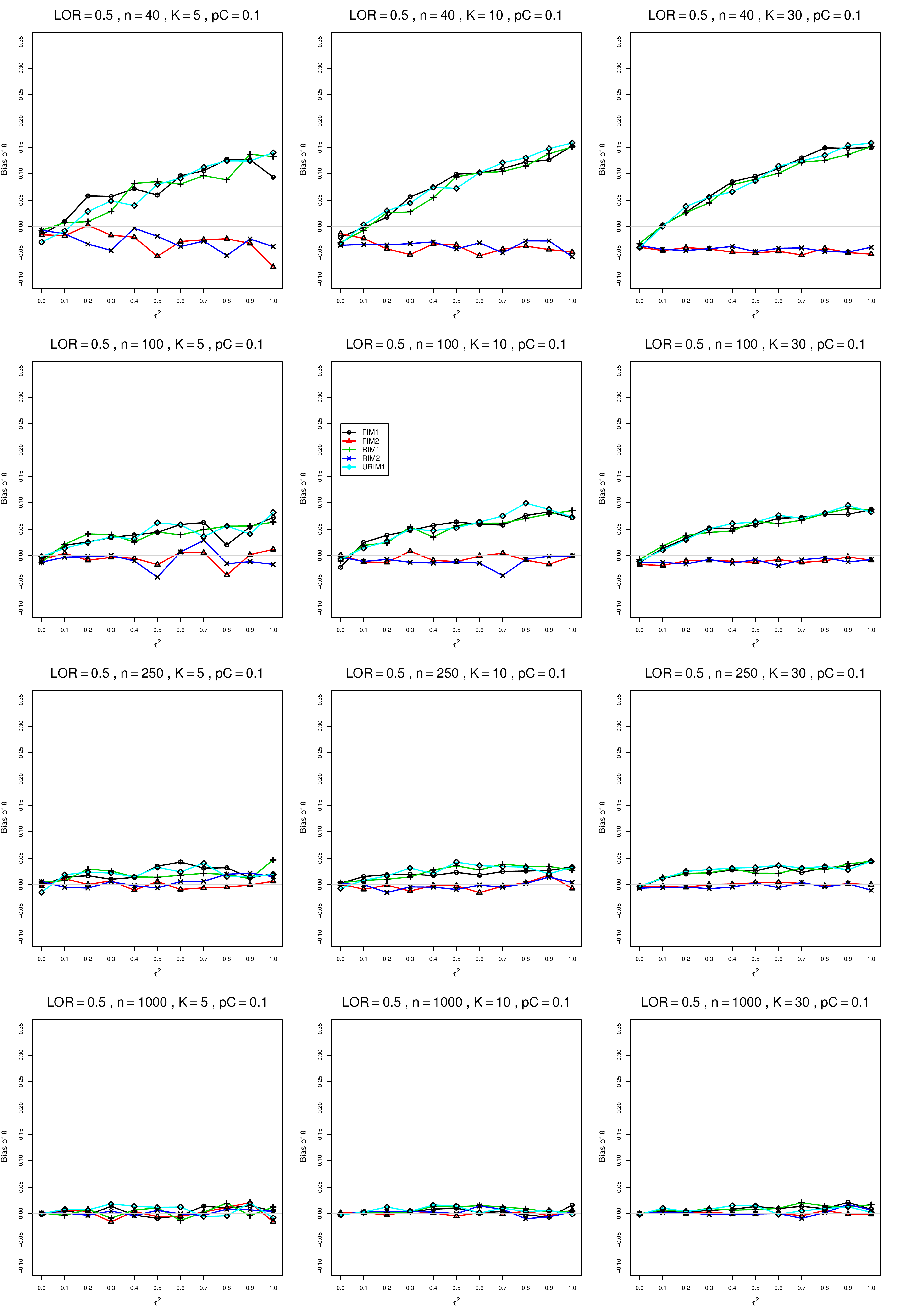}
	\caption{Bias of  overall log-odds ratio $\hat{\theta}_{REML}$ for $\theta=0.5$, $p_{C}=0.1$, $\sigma^2=0.1$, constant sample sizes $n=40,\;100,\;250,\;1000$.
The data-generation mechanisms are FIM1 ($\circ$), FIM2 ($\triangle$), RIM1 (+), RIM2 ($\times$), and URIM1 ($\diamond$).
		\label{PlotBiasThetamu05andpC01LOR_REMLsigma01}}
\end{figure}
\begin{figure}[t]
	\centering
	\includegraphics[scale=0.33]{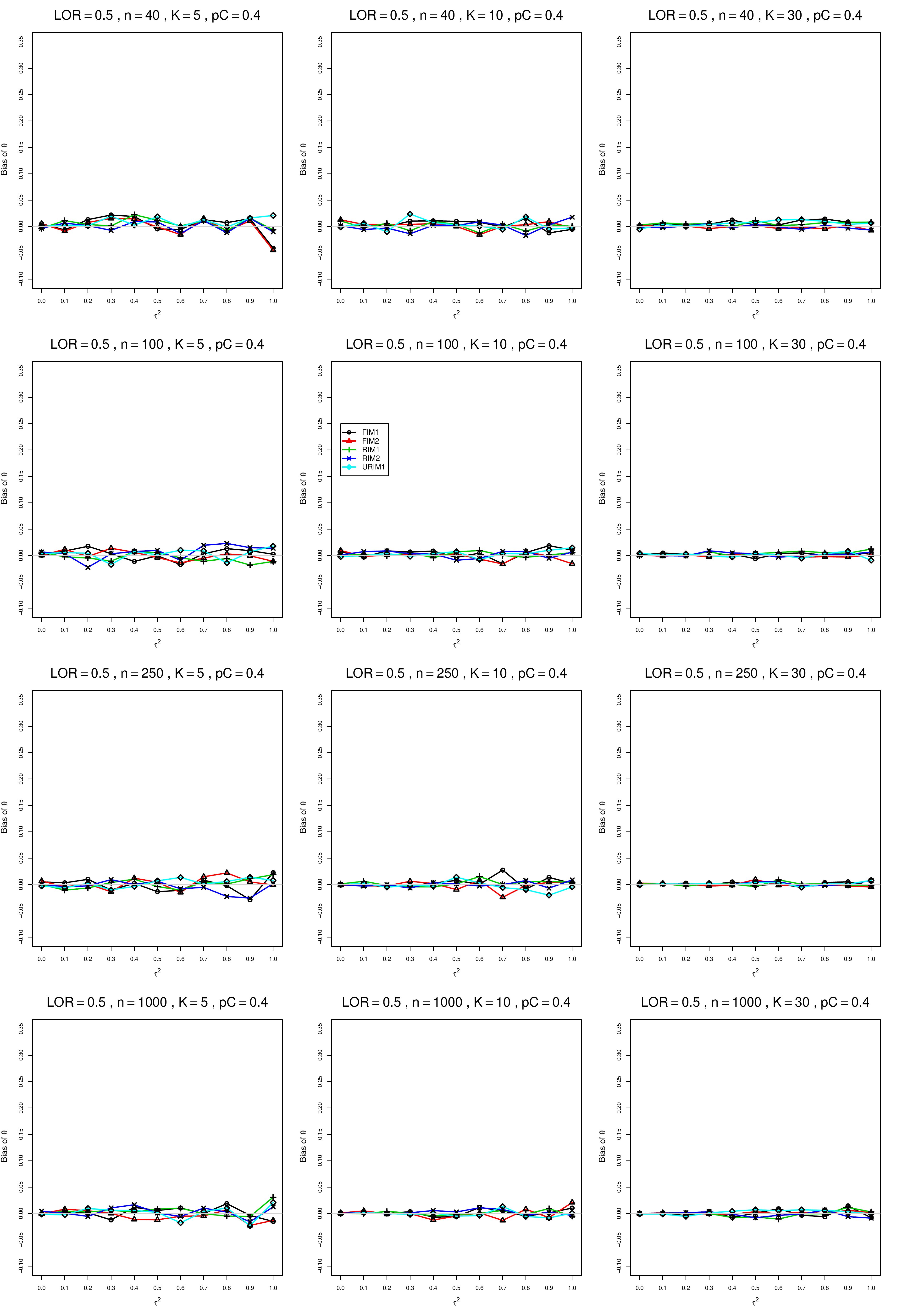}
	\caption{Bias of  overall log-odds ratio $\hat{\theta}_{REML}$ for $\theta=0.5$, $p_{C}=0.4$, $\sigma^2=0.1$, constant sample sizes $n=40,\;100,\;250,\;1000$.
The data-generation mechanisms are FIM1 ($\circ$), FIM2 ($\triangle$), RIM1 (+), RIM2 ($\times$), and URIM1 ($\diamond$).
		\label{PlotBiasThetamu05andpC04LOR_REMLsigma01}}
\end{figure}
\begin{figure}[t]
	\centering
	\includegraphics[scale=0.33]{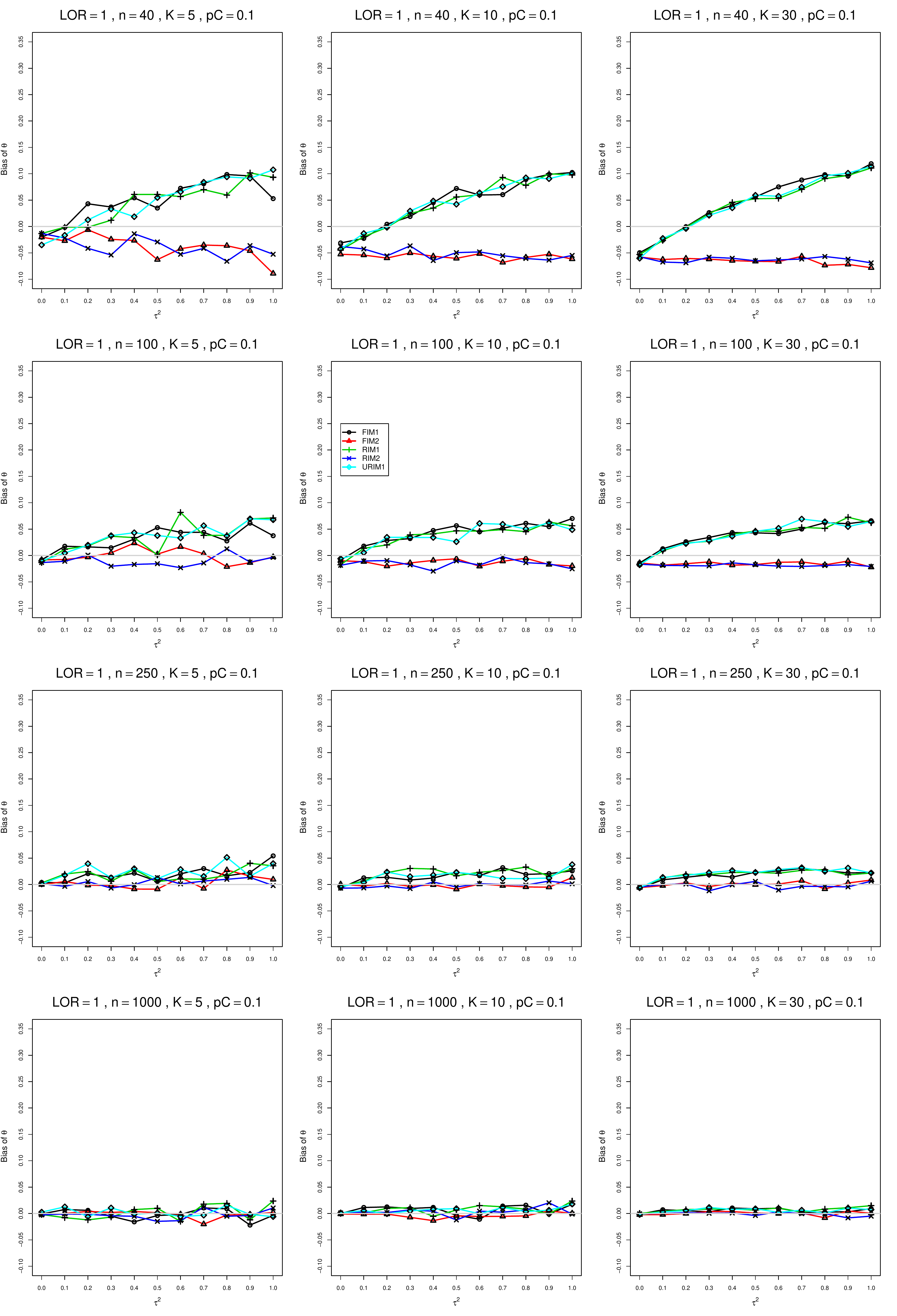}
	\caption{Bias of  overall log-odds ratio $\hat{\theta}_{REML}$ for $\theta=1$, $p_{C}=0.1$, $\sigma^2=0.1$, constant sample sizes $n=40,\;100,\;250,\;1000$.
The data-generation mechanisms are FIM1 ($\circ$), FIM2 ($\triangle$), RIM1 (+), RIM2 ($\times$), and URIM1 ($\diamond$).
		\label{PlotBiasThetamu1andpC01LOR_REMLsigma01}}
\end{figure}
\begin{figure}[t]
	\centering
	\includegraphics[scale=0.33]{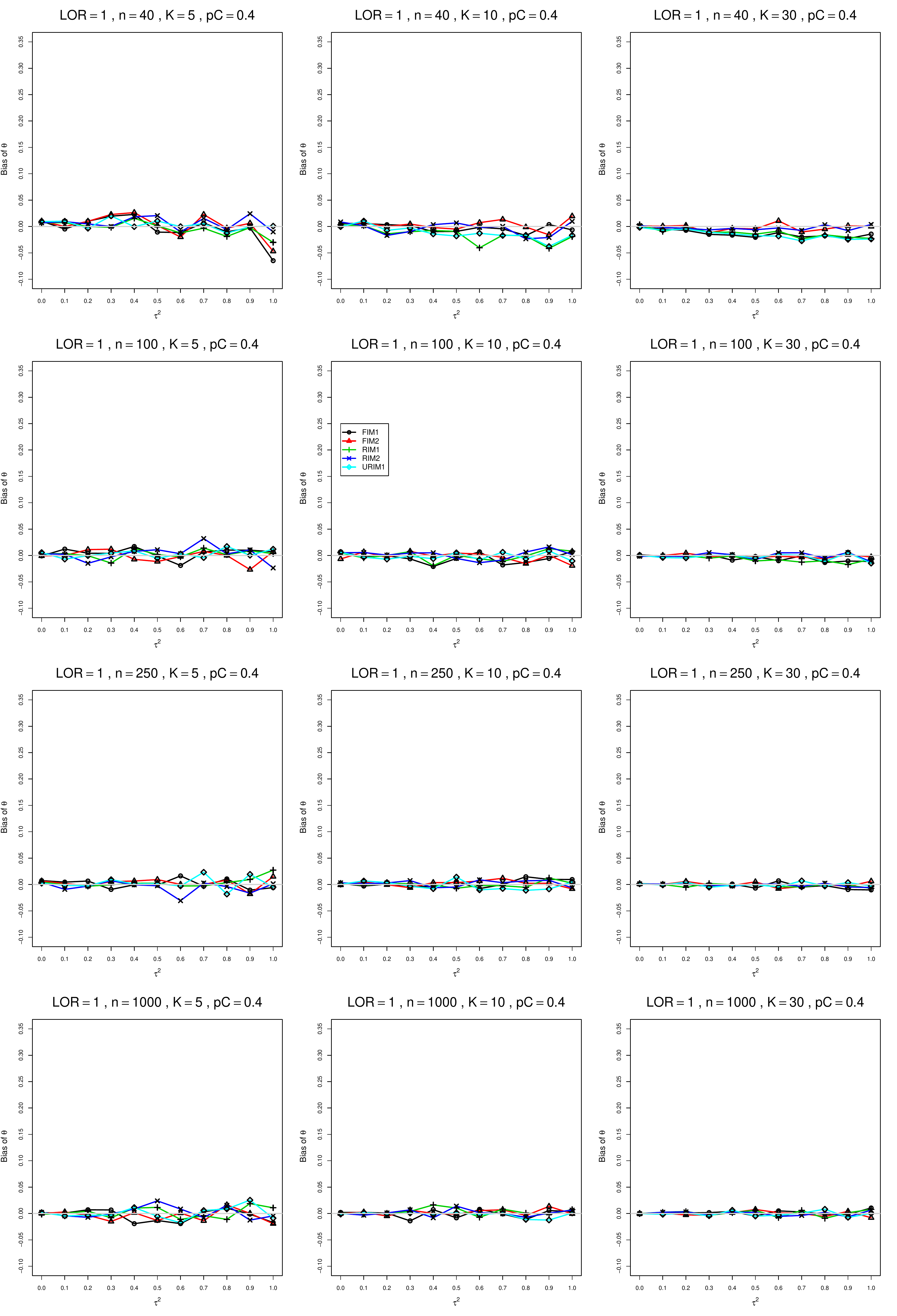}
	\caption{Bias of  overall log-odds ratio $\hat{\theta}_{REML}$ for $\theta=1$, $p_{C}=0.4$, $\sigma^2=0.1$, constant sample sizes $n=40,\;100,\;250,\;1000$.
The data-generation mechanisms are FIM1 ($\circ$), FIM2 ($\triangle$), RIM1 (+), RIM2 ($\times$), and URIM1 ($\diamond$).
		\label{PlotBiasThetamu1andpC04LOR_REMLsigma01}}
\end{figure}
\begin{figure}[t]
	\centering
	\includegraphics[scale=0.33]{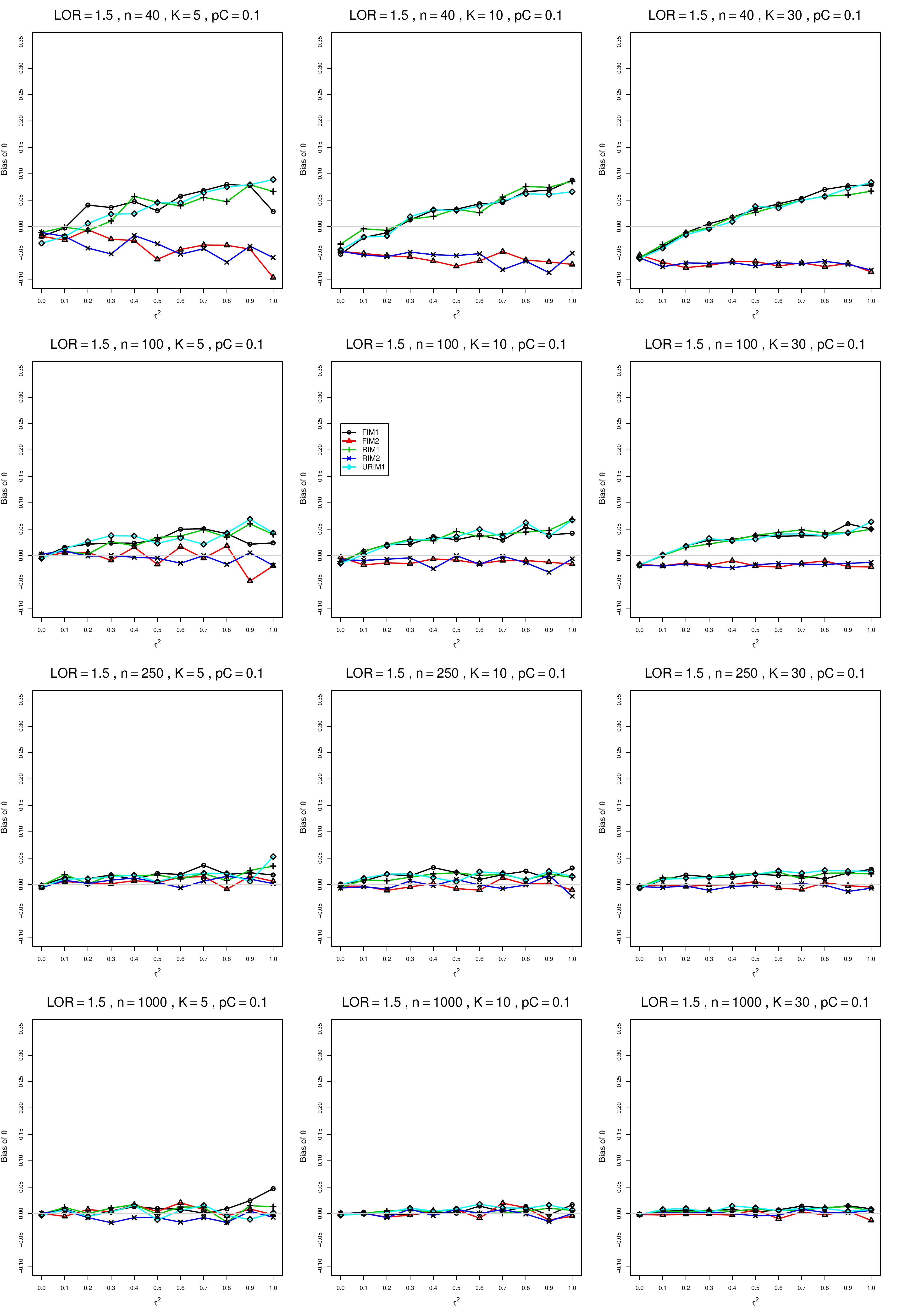}
	\caption{Bias of  overall log-odds ratio $\hat{\theta}_{REML}$ for $\theta=1.5$, $p_{C}=0.1$, $\sigma^2=0.1$, constant sample sizes $n=40,\;100,\;250,\;1000$.
The data-generation mechanisms are FIM1 ($\circ$), FIM2 ($\triangle$), RIM1 (+), RIM2 ($\times$), and URIM1 ($\diamond$).
		\label{PlotBiasThetamu15andpC01LOR_REMLsigma01}}
\end{figure}
\begin{figure}[t]
	\centering
	\includegraphics[scale=0.33]{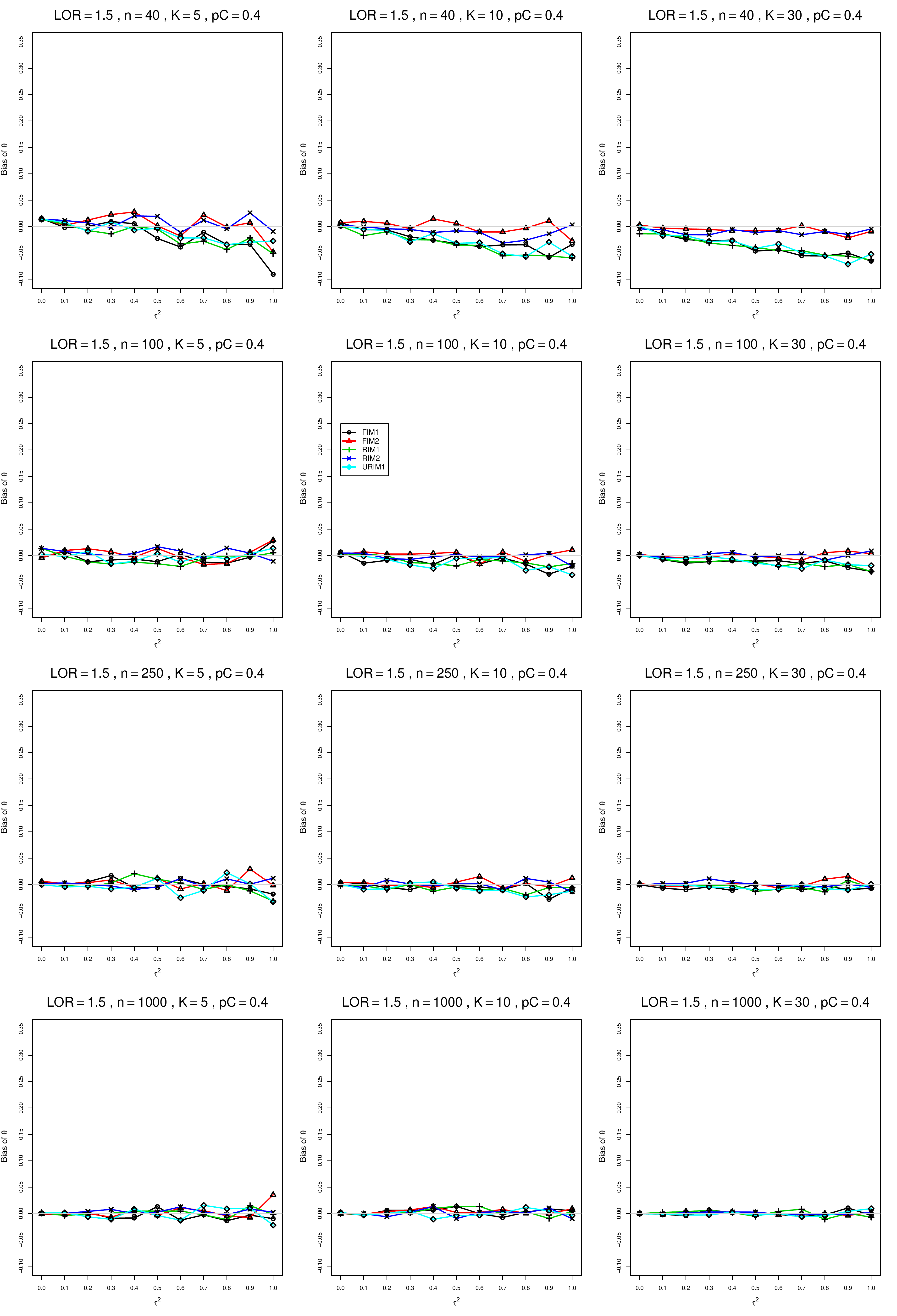}
	\caption{Bias of  overall log-odds ratio $\hat{\theta}_{REML}$ for $\theta=1.5$, $p_{C}=0.4$, $\sigma^2=0.1$, constant sample sizes $n=40,\;100,\;250,\;1000$.
The data-generation mechanisms are FIM1 ($\circ$), FIM2 ($\triangle$), RIM1 (+), RIM2 ($\times$), and URIM1 ($\diamond$).
		\label{PlotBiasThetamu15andpC04LOR_REMLsigma01}}
\end{figure}
\begin{figure}[t]
	\centering
	\includegraphics[scale=0.33]{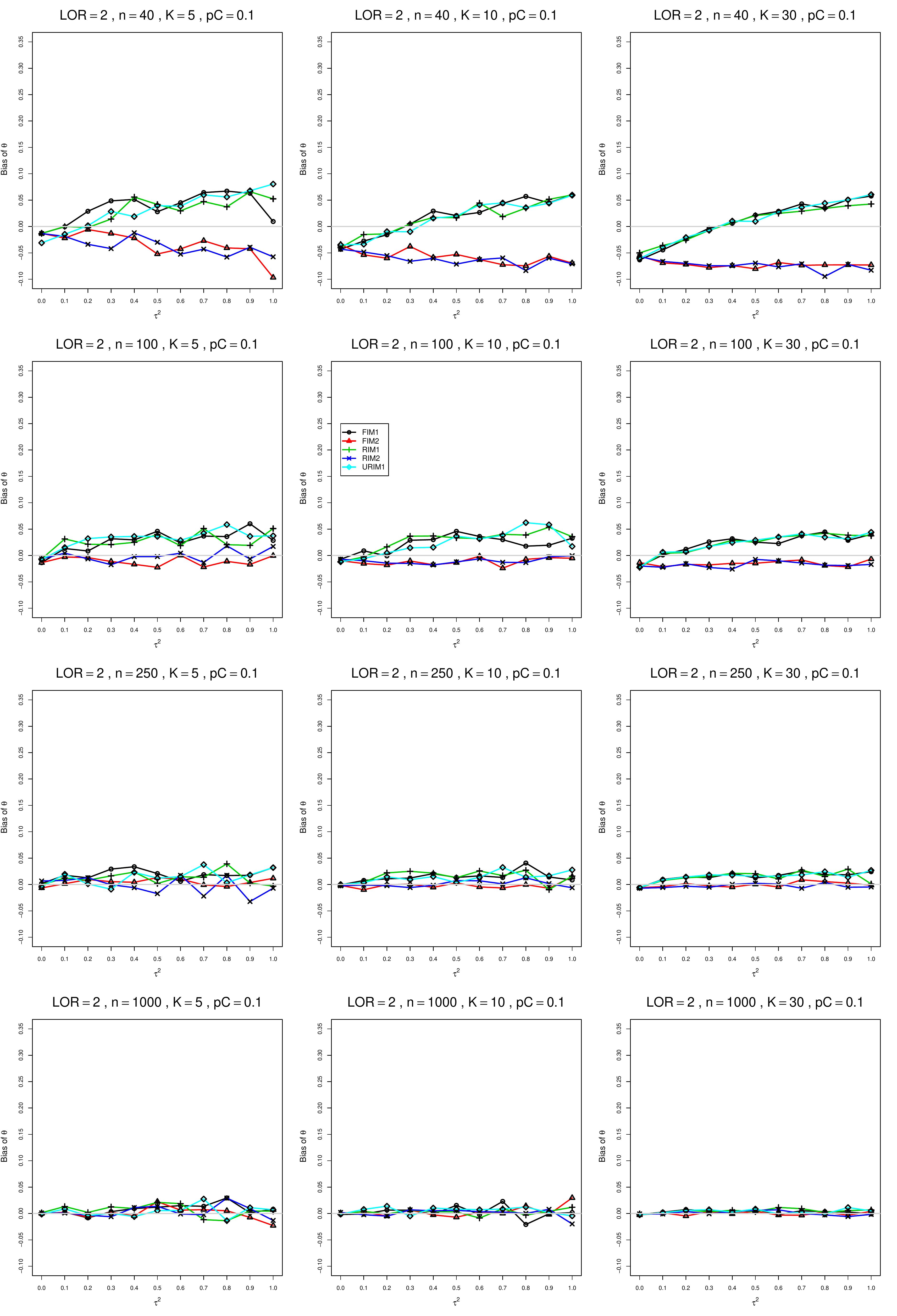}
	\caption{Bias of  overall log-odds ratio $\hat{\theta}_{REML}$ for $\theta=2$, $p_{C}=0.1$, $\sigma^2=0.1$, constant sample sizes $n=40,\;100,\;250,\;1000$.
The data-generation mechanisms are FIM1 ($\circ$), FIM2 ($\triangle$), RIM1 (+), RIM2 ($\times$), and URIM1 ($\diamond$).
		\label{PlotBiasThetamu2andpC01LOR_REMLsigma01}}
\end{figure}
\begin{figure}[t]
	\centering
	\includegraphics[scale=0.33]{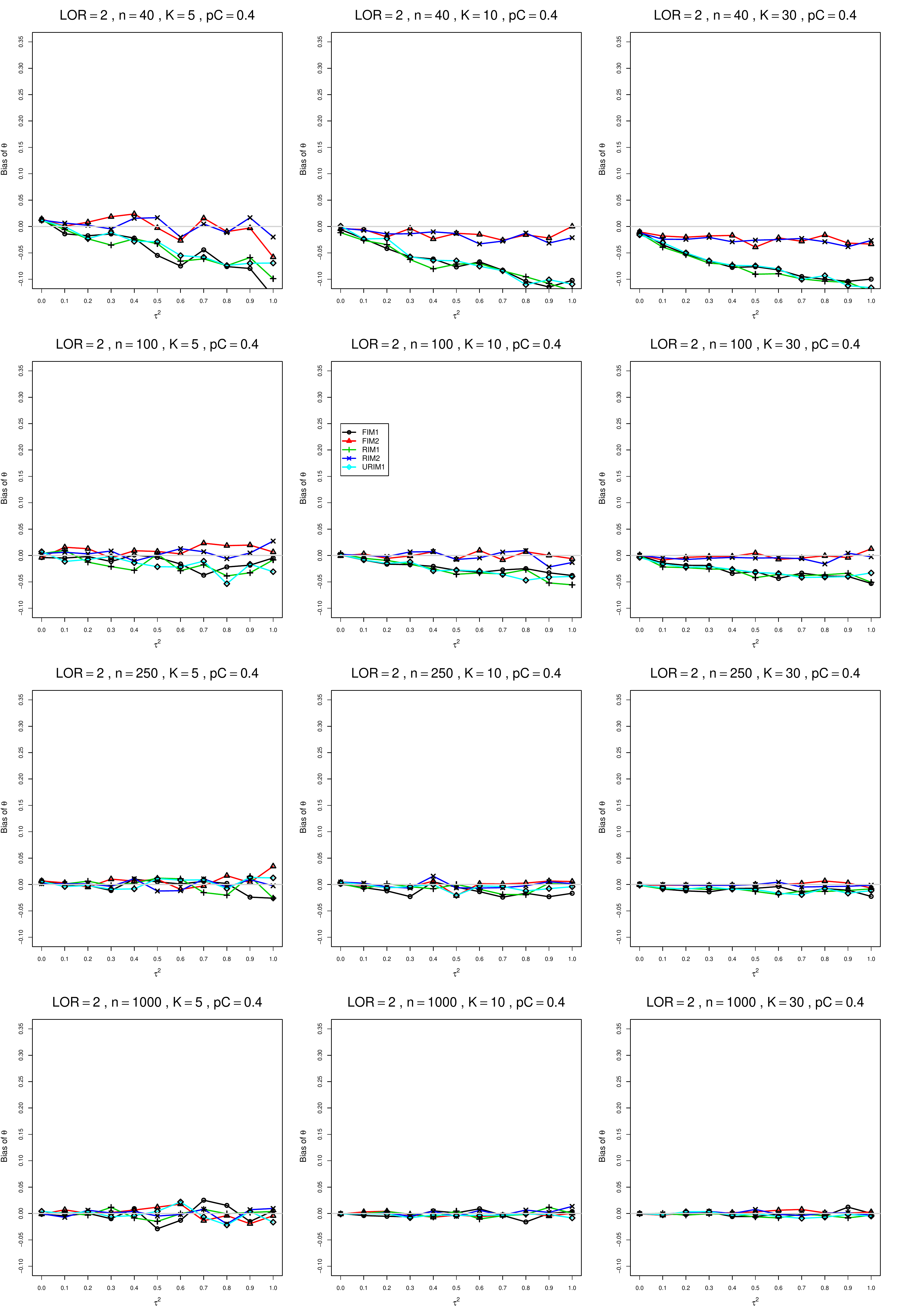}
	\caption{Bias of  overall log-odds ratio $\hat{\theta}_{REML}$ for $\theta=2$, $p_{C}=0.4$, $\sigma^2=0.1$, constant sample sizes $n=40,\;100,\;250,\;1000$.
The data-generation mechanisms are FIM1 ($\circ$), FIM2 ($\triangle$), RIM1 (+), RIM2 ($\times$), and URIM1 ($\diamond$).
		\label{PlotBiasThetamu2andpC04LOR_REMLsigma01}}
\end{figure}
\begin{figure}[t]
	\centering
	\includegraphics[scale=0.33]{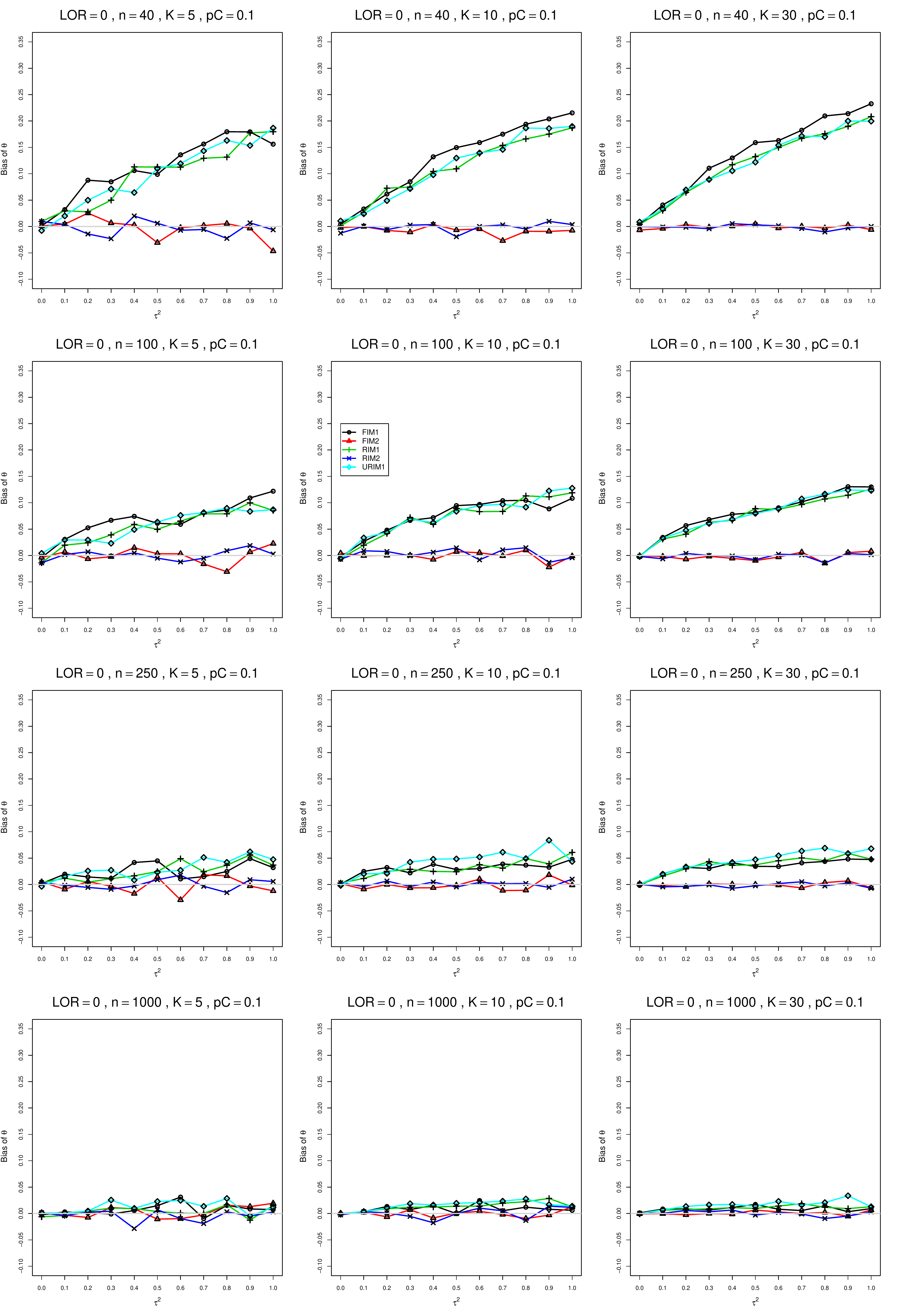}
	\caption{Bias of  overall log-odds ratio $\hat{\theta}_{REML}$ for $\theta=0$, $p_{C}=0.1$, $\sigma^2=0.4$, constant sample sizes $n=40,\;100,\;250,\;1000$.
The data-generation mechanisms are FIM1 ($\circ$), FIM2 ($\triangle$), RIM1 (+), RIM2 ($\times$), and URIM1 ($\diamond$).
		\label{PlotBiasThetamu0andpC01LOR_REMLsigma04}}
\end{figure}
\begin{figure}[t]
	\centering
	\includegraphics[scale=0.33]{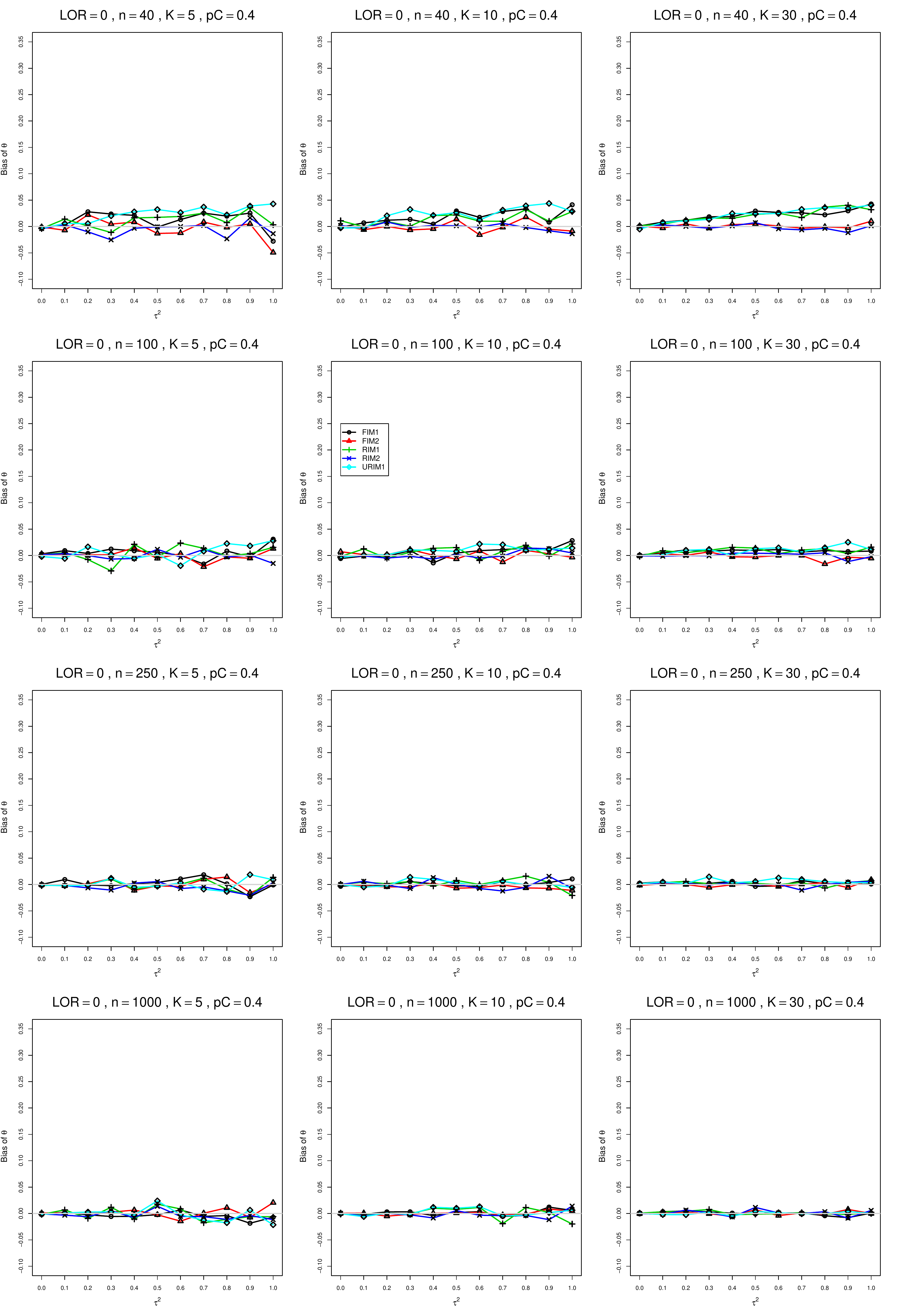}
	\caption{Bias of  overall log-odds ratio $\hat{\theta}_{REML}$ for $\theta=0$, $p_{C}=0.4$, $\sigma^2=0.4$, constant sample sizes $n=40,\;100,\;250,\;1000$.
The data-generation mechanisms are FIM1 ($\circ$), FIM2 ($\triangle$), RIM1 (+), RIM2 ($\times$), and URIM1 ($\diamond$).
		\label{PlotBiasThetamu0andpC04LOR_REMLsigma04}}
\end{figure}
\begin{figure}[t]
	\centering
	\includegraphics[scale=0.33]{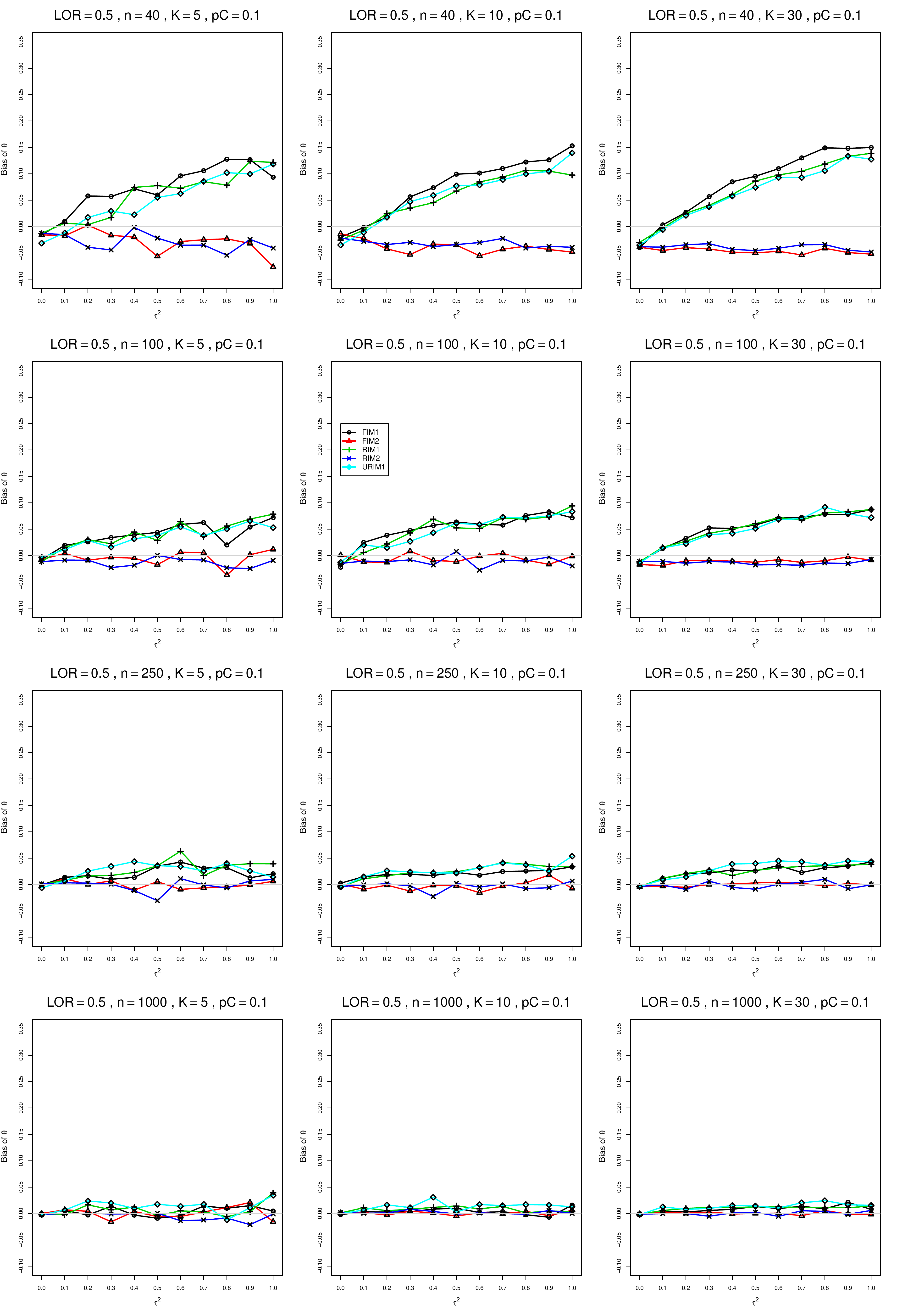}
	\caption{Bias of  overall log-odds ratio $\hat{\theta}_{REML}$ for $\theta=0.5$, $p_{C}=0.1$, $\sigma^2=0.4$, constant sample sizes $n=40,\;100,\;250,\;1000$.
The data-generation mechanisms are FIM1 ($\circ$), FIM2 ($\triangle$), RIM1 (+), RIM2 ($\times$), and URIM1 ($\diamond$).
		\label{PlotBiasThetamu05andpC01LOR_REMLsigma04}}
\end{figure}
\begin{figure}[t]
	\centering
	\includegraphics[scale=0.33]{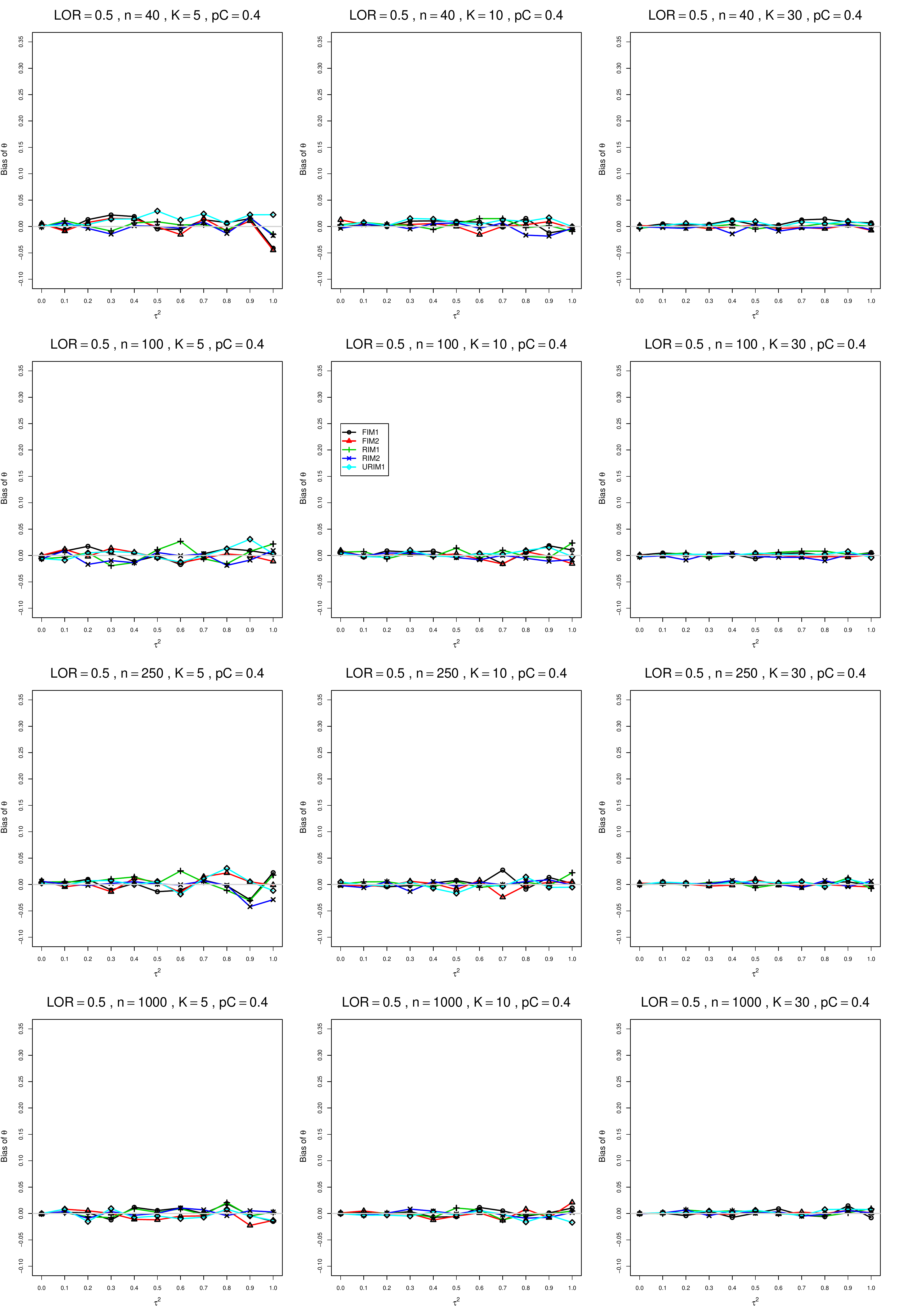}
	\caption{Bias of  overall log-odds ratio $\hat{\theta}_{REML}$ for $\theta=0.5$, $p_{C}=0.4$, $\sigma^2=0.4$, constant sample sizes $n=40,\;100,\;250,\;1000$.
The data-generation mechanisms are FIM1 ($\circ$), FIM2 ($\triangle$), RIM1 (+), RIM2 ($\times$), and URIM1 ($\diamond$).
		\label{PlotBiasThetamu05andpC04LOR_REMLsigma04}}
\end{figure}
\begin{figure}[t]
	\centering
	\includegraphics[scale=0.33]{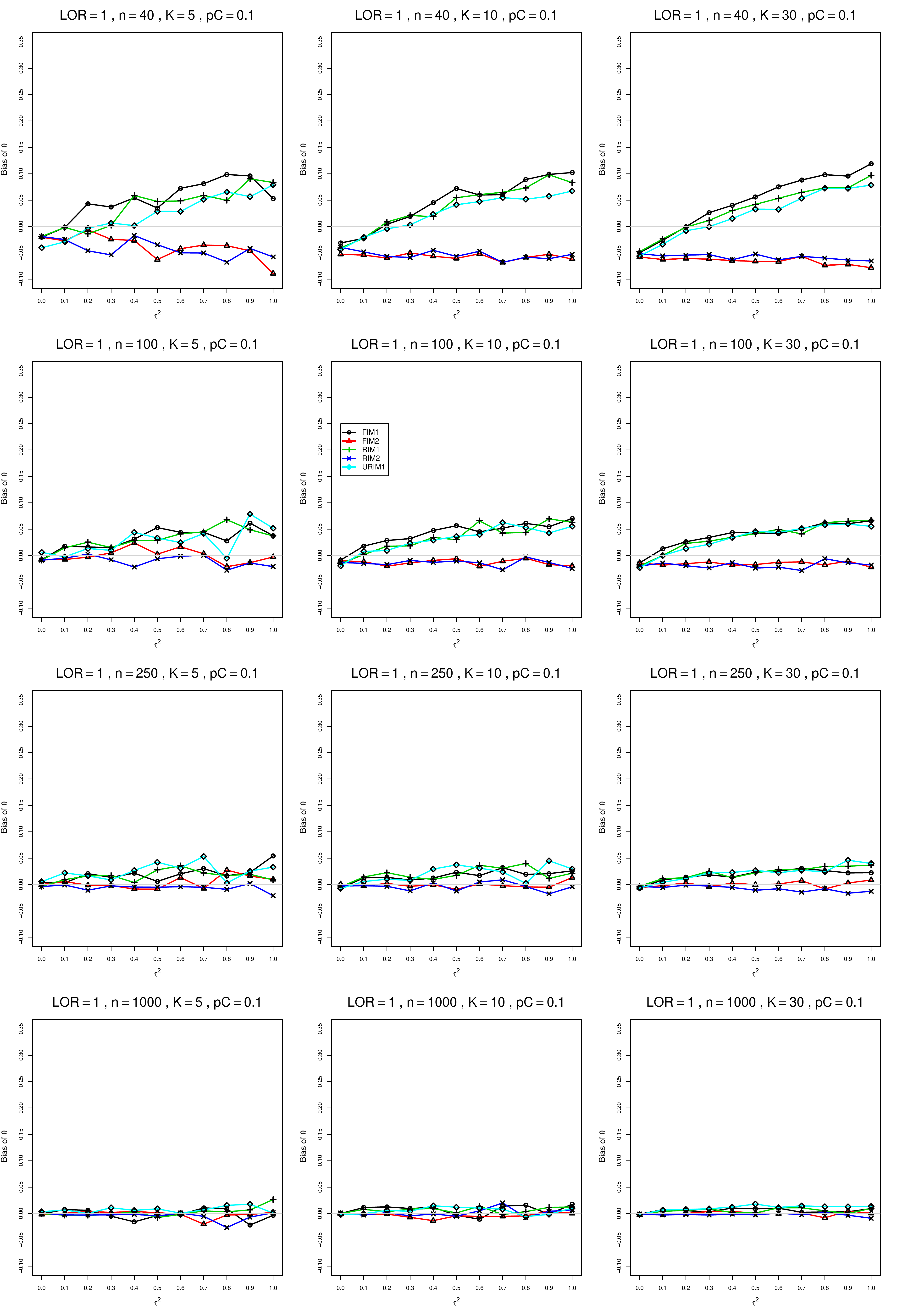}
	\caption{Bias of  overall log-odds ratio $\hat{\theta}_{REML}$ for $\theta=1$, $p_{C}=0.1$, $\sigma^2=0.4$, constant sample sizes $n=40,\;100,\;250,\;1000$.
The data-generation mechanisms are FIM1 ($\circ$), FIM2 ($\triangle$), RIM1 (+), RIM2 ($\times$), and URIM1 ($\diamond$).
		\label{PlotBiasThetamu1andpC01LOR_REMLsigma04}}
\end{figure}
\begin{figure}[t]
	\centering
	\includegraphics[scale=0.33]{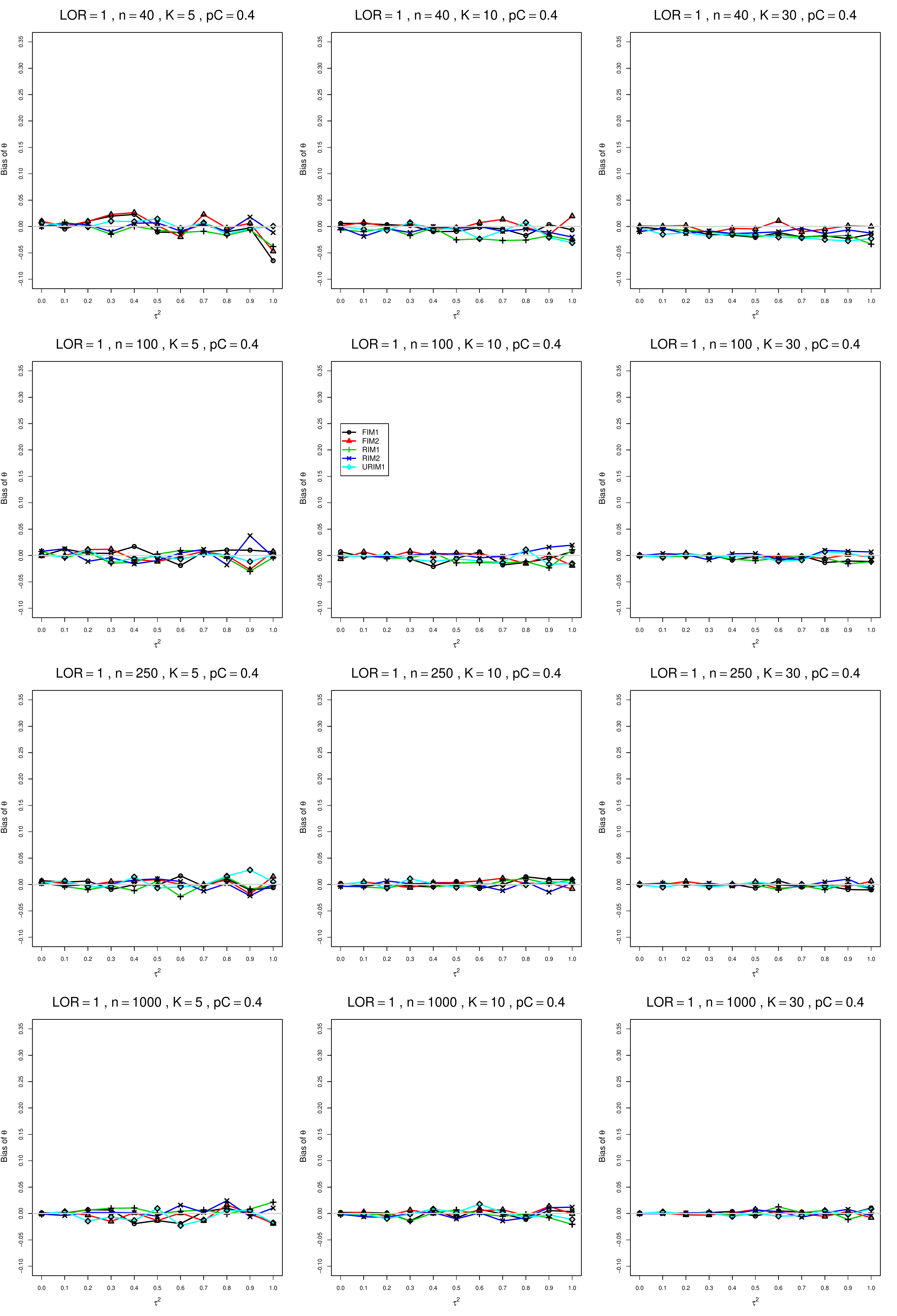}
	\caption{Bias of  overall log-odds ratio $\hat{\theta}_{REML}$ for $\theta=1$, $p_{C}=0.4$, $\sigma^2=0.4$, constant sample sizes $n=40,\;100,\;250,\;1000$.
The data-generation mechanisms are FIM1 ($\circ$), FIM2 ($\triangle$), RIM1 (+), RIM2 ($\times$), and URIM1 ($\diamond$).
		\label{PlotBiasThetamu1andpC04LOR_REMLsigma04}}
\end{figure}
\begin{figure}[t]
	\centering
	\includegraphics[scale=0.33]{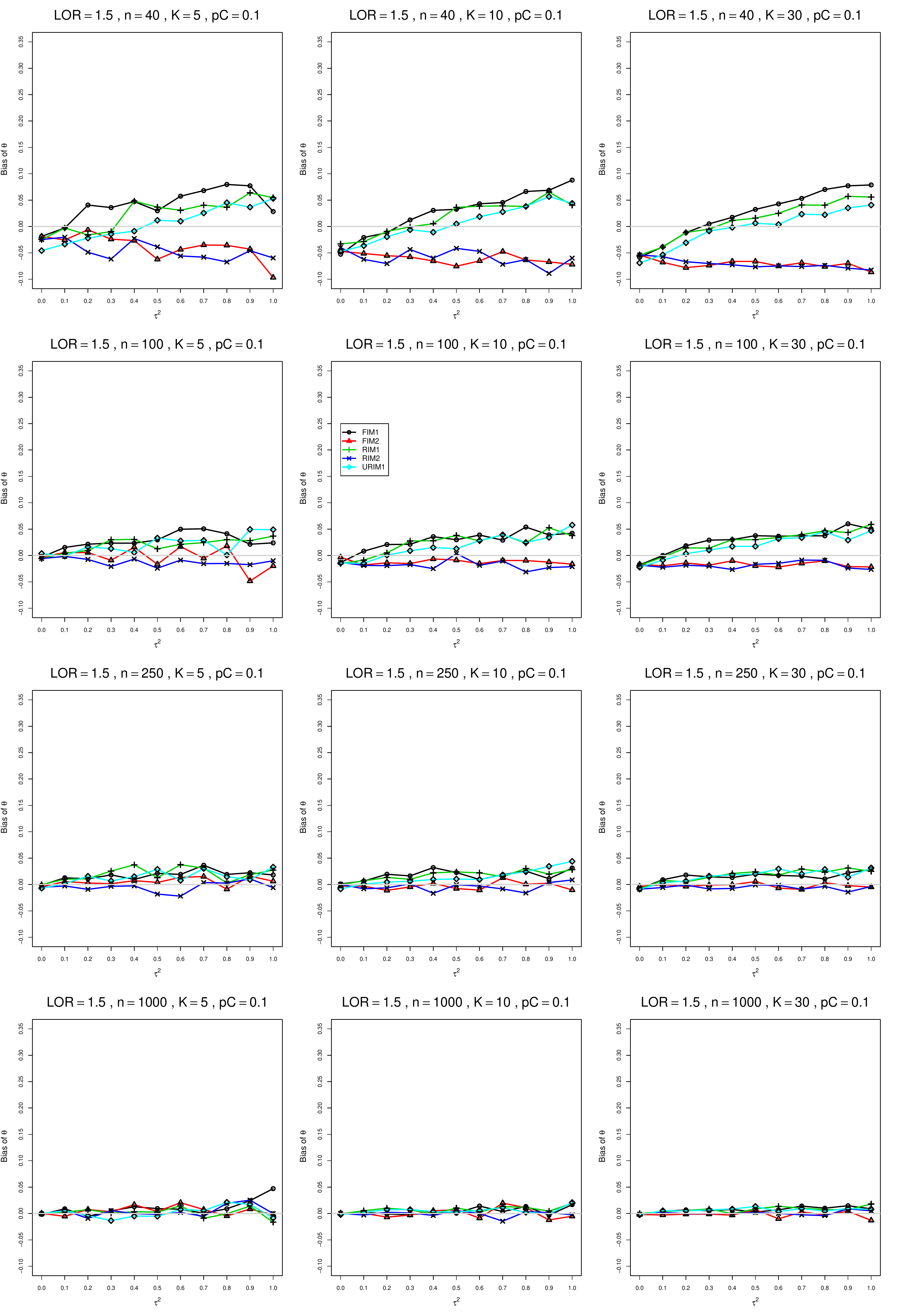}
	\caption{Bias of  overall log-odds ratio $\hat{\theta}_{REML}$ for $\theta=1.5$, $p_{C}=0.1$, $\sigma^2=0.4$, constant sample sizes $n=40,\;100,\;250,\;1000$.
The data-generation mechanisms are FIM1 ($\circ$), FIM2 ($\triangle$), RIM1 (+), RIM2 ($\times$), and URIM1 ($\diamond$).
		\label{PlotBiasThetamu15andpC01LOR_REMLsigma04}}
\end{figure}
\begin{figure}[t]
	\centering
	\includegraphics[scale=0.33]{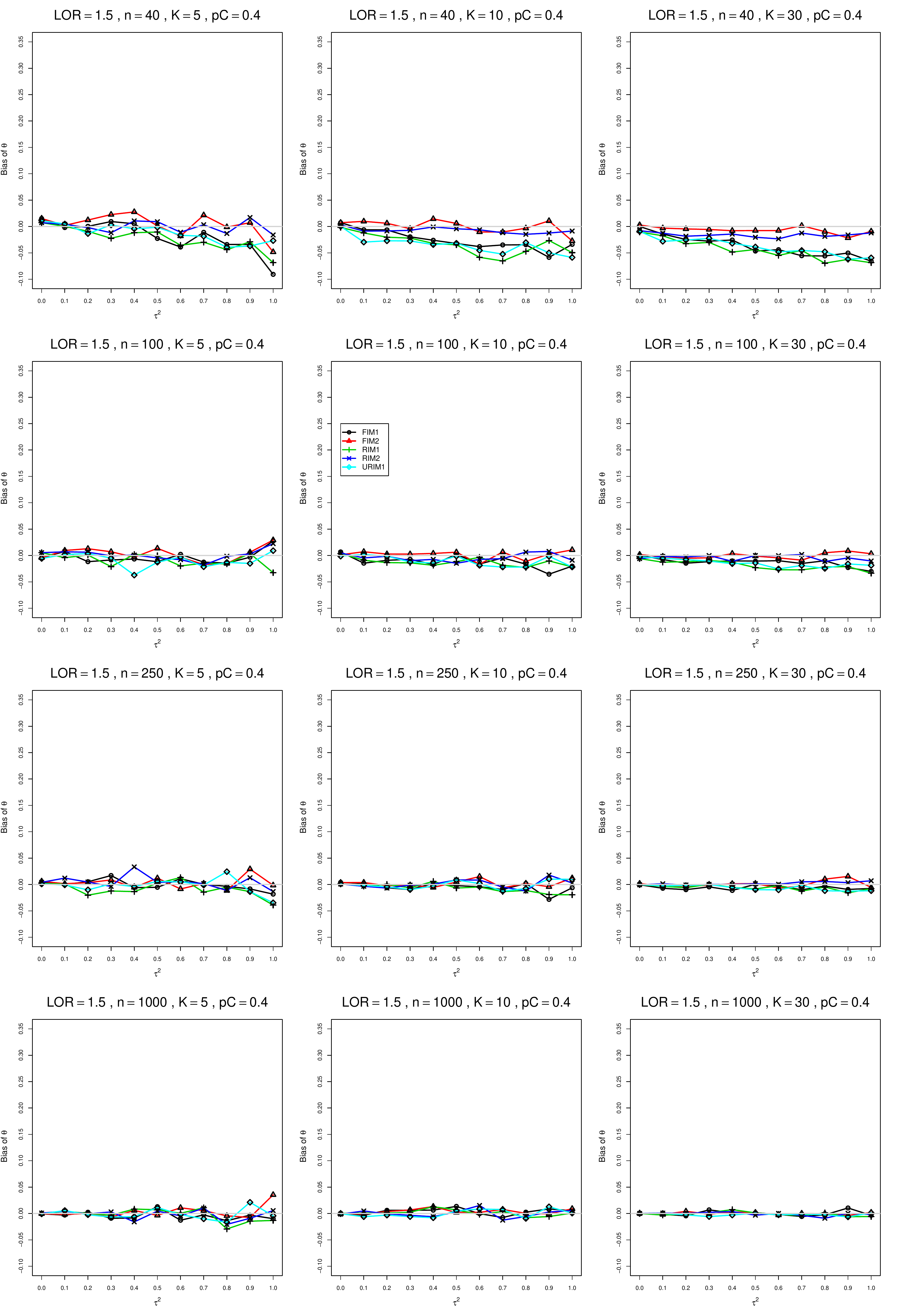}
	\caption{Bias of  overall log-odds ratio $\hat{\theta}_{REML}$ for $\theta=1.5$, $p_{C}=0.4$, $\sigma^2=0.4$, constant sample sizes $n=40,\;100,\;250,\;1000$.
The data-generation mechanisms are FIM1 ($\circ$), FIM2 ($\triangle$), RIM1 (+), RIM2 ($\times$), and URIM1 ($\diamond$).
		\label{PlotBiasThetamu15andpC04LOR_REMLsigma04}}
\end{figure}
\begin{figure}[t]
	\centering
	\includegraphics[scale=0.33]{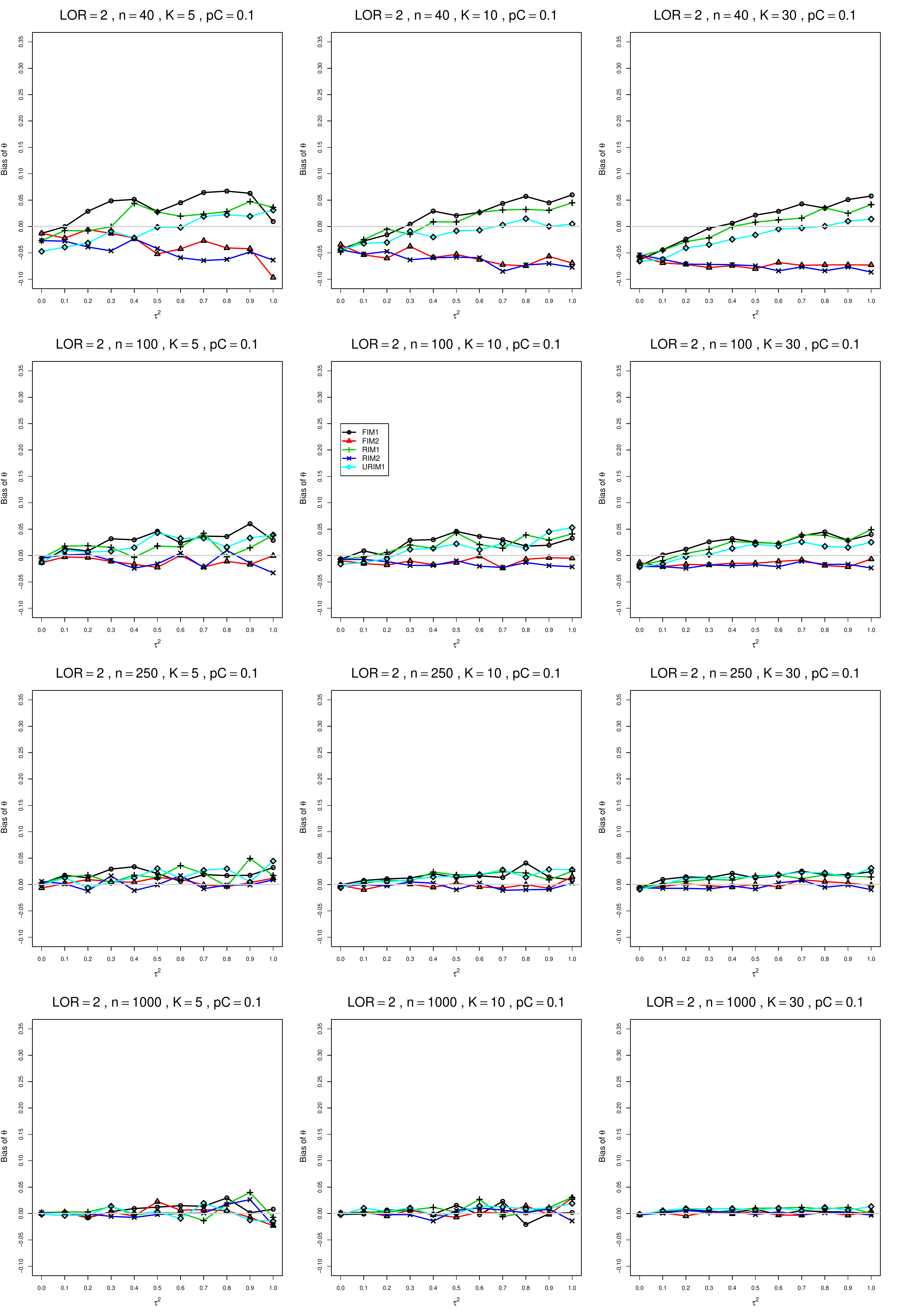}
	\caption{Bias of  overall log-odds ratio $\hat{\theta}_{REML}$ for $\theta=2$, $p_{C}=0.1$, $\sigma^2=0.4$, constant sample sizes $n=40,\;100,\;250,\;1000$.
The data-generation mechanisms are FIM1 ($\circ$), FIM2 ($\triangle$), RIM1 (+), RIM2 ($\times$), and URIM1 ($\diamond$).
		\label{PlotBiasThetamu2andpC01LOR_REMLsigma04}}
\end{figure}
\begin{figure}[t]
	\centering
	\includegraphics[scale=0.33]{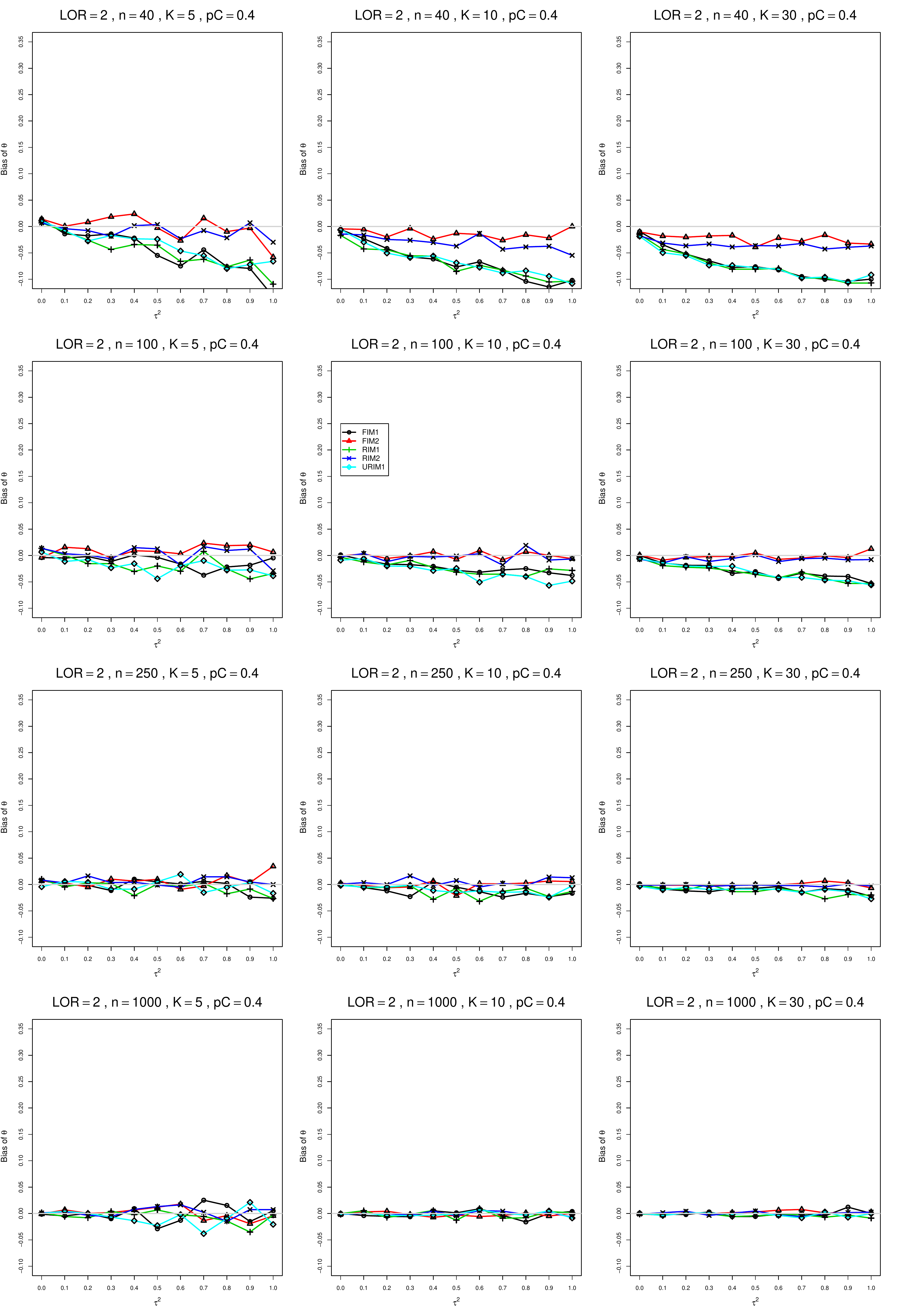}
	\caption{Bias of  overall log-odds ratio $\hat{\theta}_{REML}$ for $\theta=2$, $p_{C}=0.4$, $\sigma^2=0.4$, constant sample sizes $n=40,\;100,\;250,\;1000$.
The data-generation mechanisms are FIM1 ($\circ$), FIM2 ($\triangle$), RIM1 (+), RIM2 ($\times$), and URIM1 ($\diamond$).
		\label{PlotBiasThetamu2andpC04LOR_REMLsigma04}}
\end{figure}

\clearpage
\subsection*{A2.3 Bias of $\hat{\theta}_{MP}$}
\renewcommand{\thefigure}{A2.3.\arabic{figure}}
\setcounter{figure}{0}

\begin{figure}[t]
	\centering
	\includegraphics[scale=0.33]{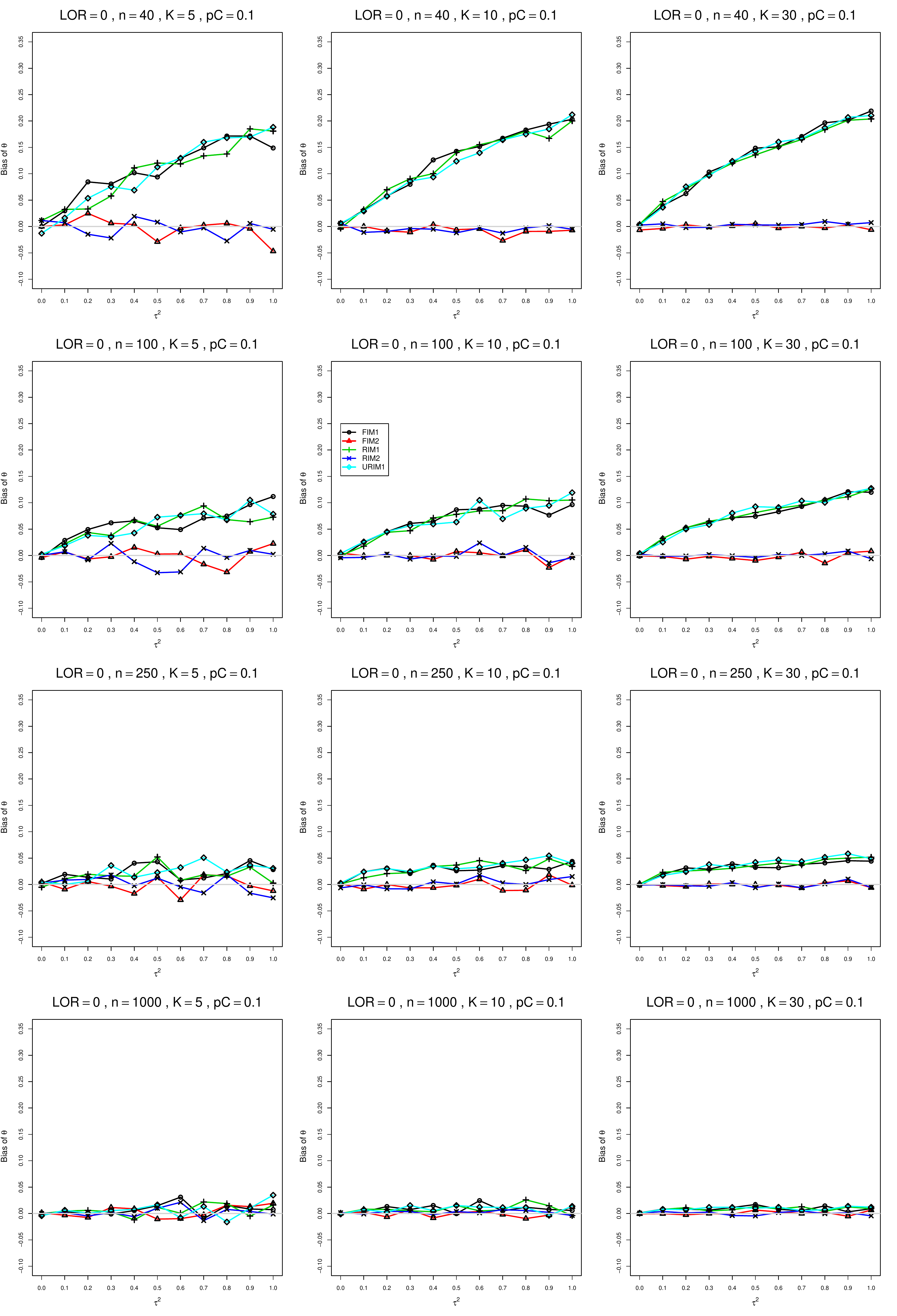}
	\caption{Bias of  overall log-odds ratio $\hat{\theta}_{MP}$ for $\theta=0$, $p_{C}=0.1$, $\sigma^2=0.1$, constant sample sizes $n=40,\;100,\;250,\;1000$.
The data-generation mechanisms are FIM1 ($\circ$), FIM2 ($\triangle$), RIM1 (+), RIM2 ($\times$), and URIM1 ($\diamond$).
		\label{PlotBiasThetamu0andpC01LOR_MPsigma01}}
\end{figure}
\begin{figure}[t]
	\centering
	\includegraphics[scale=0.33]{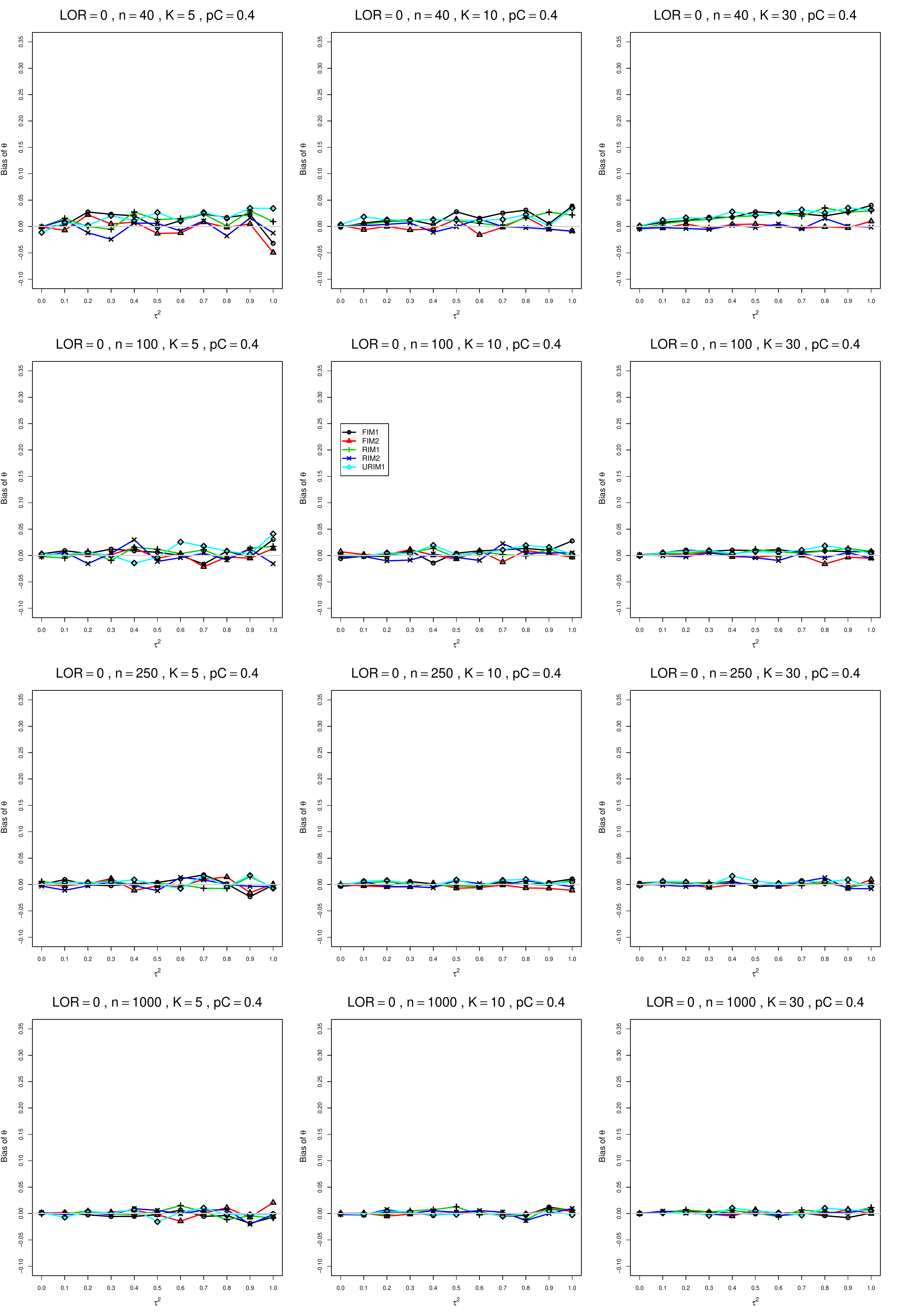}
	\caption{Bias of  overall log-odds ratio $\hat{\theta}_{MP}$ for $\theta=0$, $p_{C}=0.4$, $\sigma^2=0.1$, constant sample sizes $n=40,\;100,\;250,\;1000$.
The data-generation mechanisms are FIM1 ($\circ$), FIM2 ($\triangle$), RIM1 (+), RIM2 ($\times$), and URIM1 ($\diamond$).
		\label{PlotBiasThetamu0andpC04LOR_MPsigma01}}
\end{figure}
\begin{figure}[t]
	\centering
	\includegraphics[scale=0.33]{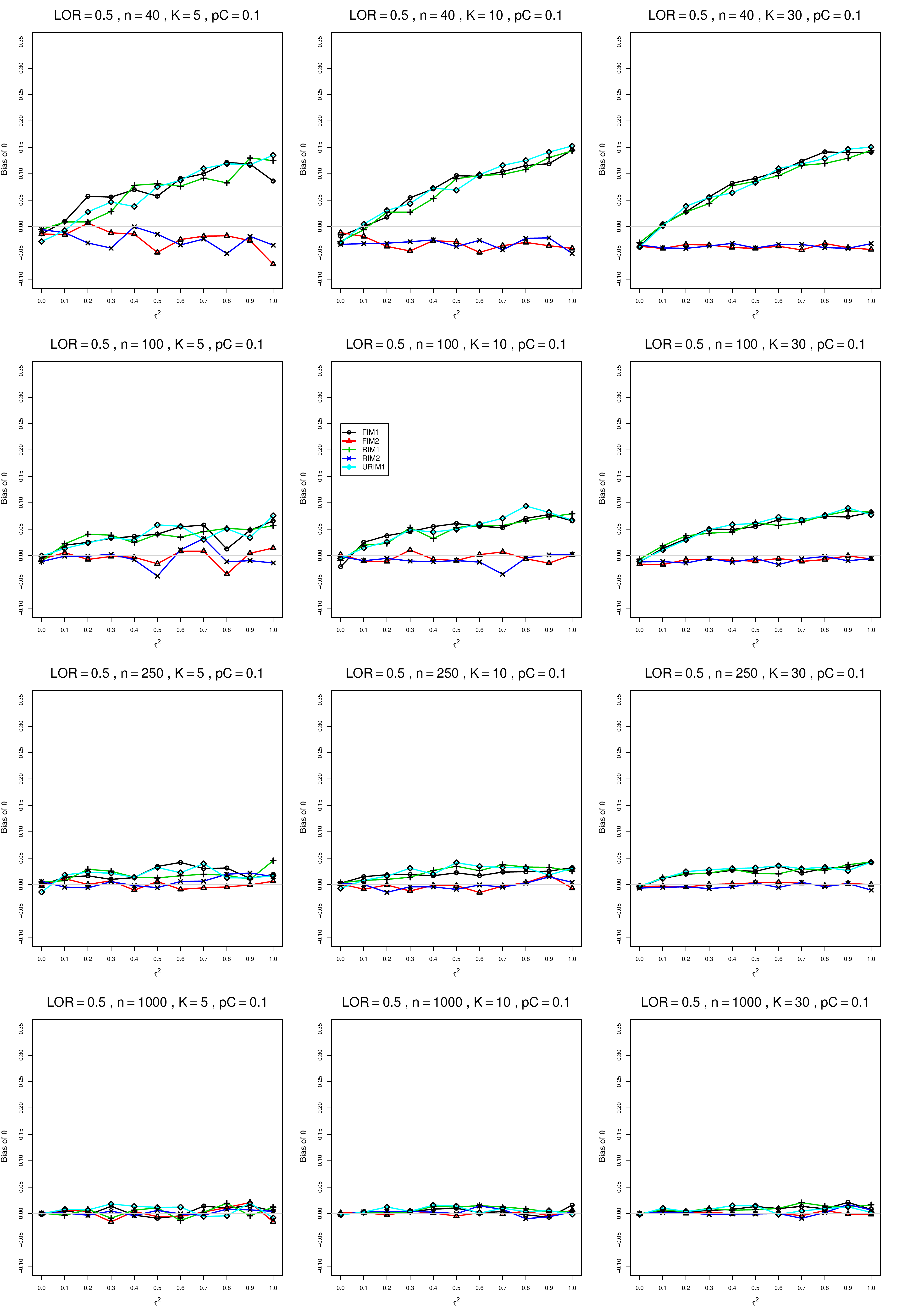}
	\caption{Bias of  overall log-odds ratio $\hat{\theta}_{MP}$ for $\theta=0.5$, $p_{C}=0.1$, $\sigma^2=0.1$, constant sample sizes $n=40,\;100,\;250,\;1000$.
The data-generation mechanisms are FIM1 ($\circ$), FIM2 ($\triangle$), RIM1 (+), RIM2 ($\times$), and URIM1 ($\diamond$).
		\label{PlotBiasThetamu05andpC01LOR_MPsigma01}}
\end{figure}
\begin{figure}[t]
	\centering
	\includegraphics[scale=0.33]{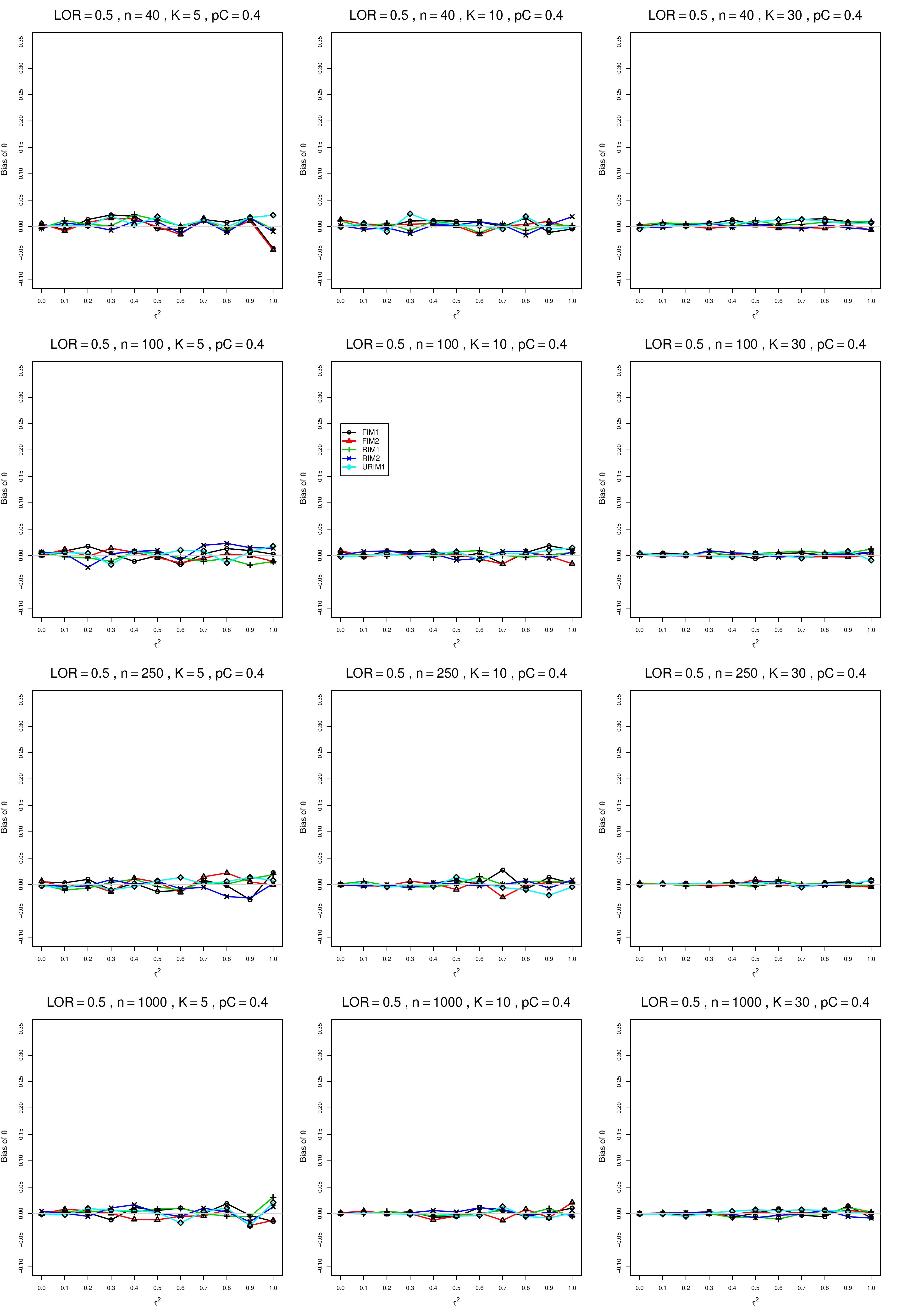}
	\caption{Bias of  overall log-odds ratio $\hat{\theta}_{MP}$ for $\theta=0.5$, $p_{C}=0.4$, $\sigma^2=0.1$, constant sample sizes $n=40,\;100,\;250,\;1000$.
The data-generation mechanisms are FIM1 ($\circ$), FIM2 ($\triangle$), RIM1 (+), RIM2 ($\times$), and URIM1 ($\diamond$).
		\label{PlotBiasThetamu05andpC04LOR_MPsigma01}}
\end{figure}
\begin{figure}[t]
	\centering
	\includegraphics[scale=0.33]{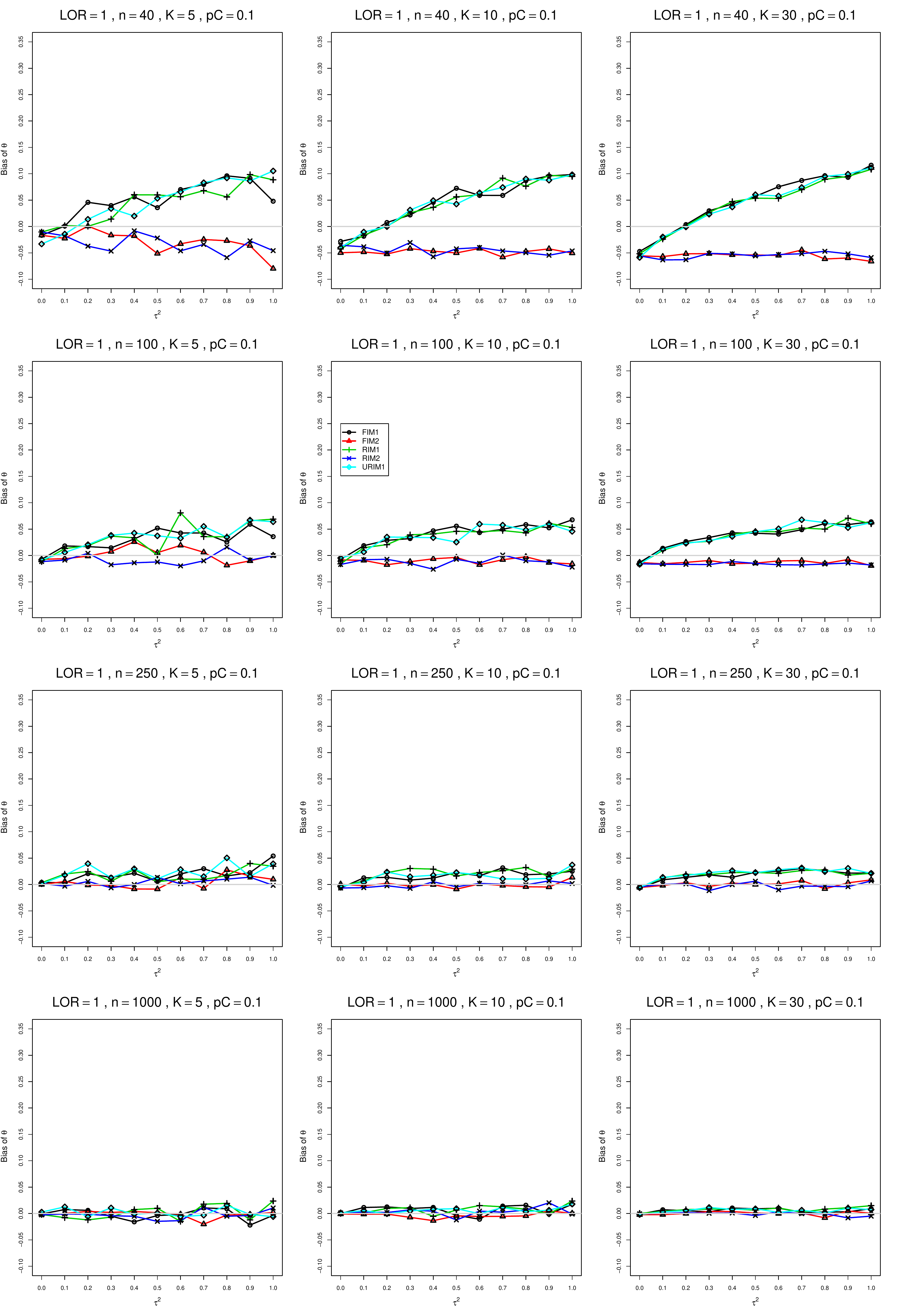}
	\caption{Bias of  overall log-odds ratio $\hat{\theta}_{MP}$ for $\theta=1$, $p_{C}=0.1$, $\sigma^2=0.1$, constant sample sizes $n=40,\;100,\;250,\;1000$.
The data-generation mechanisms are FIM1 ($\circ$), FIM2 ($\triangle$), RIM1 (+), RIM2 ($\times$), and URIM1 ($\diamond$).
		\label{PlotBiasThetamu1andpC01LOR_MPsigma01}}
\end{figure}
\begin{figure}[t]
	\centering
	\includegraphics[scale=0.33]{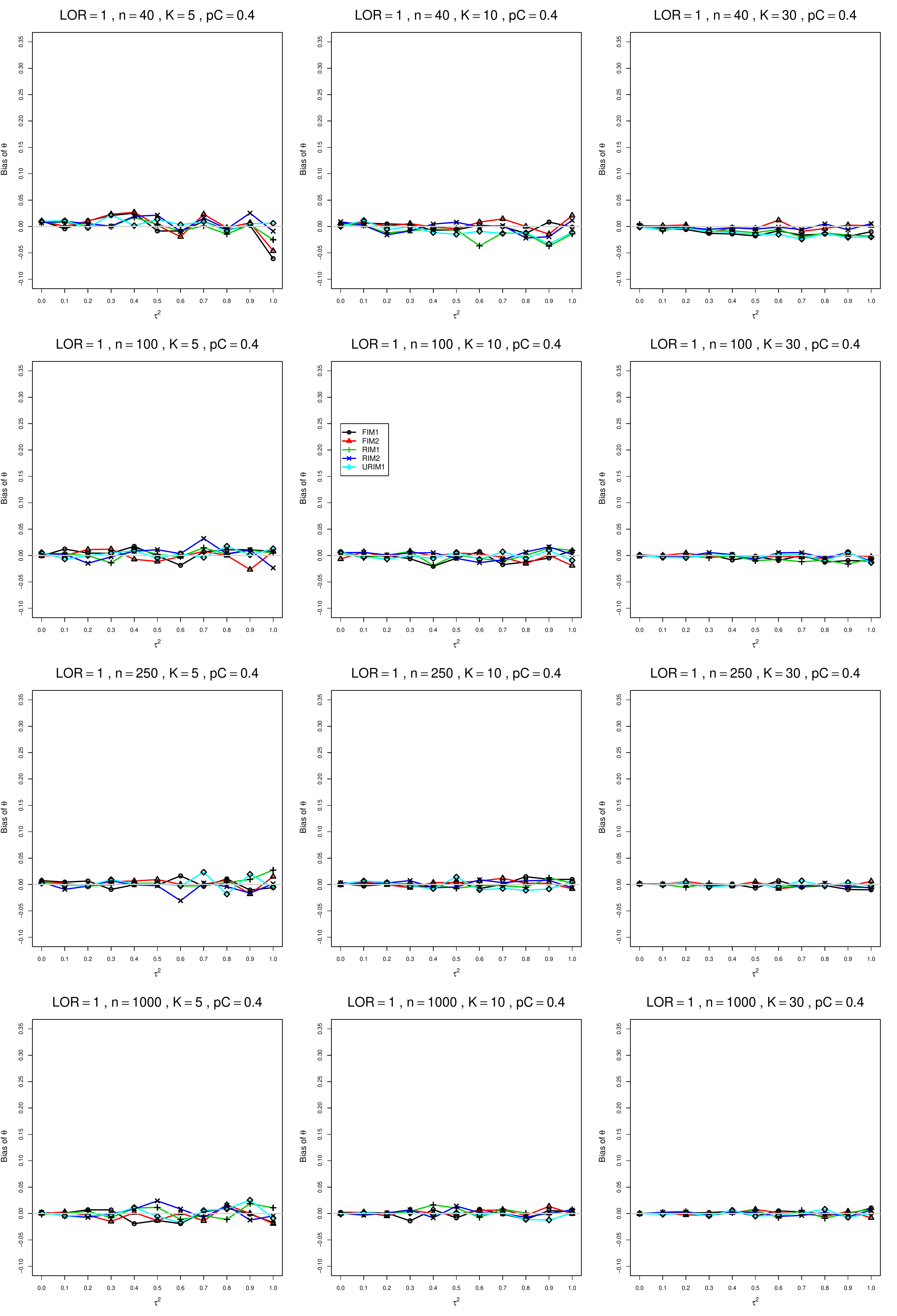}
	\caption{Bias of  overall log-odds ratio $\hat{\theta}_{MP}$ for $\theta=1$, $p_{C}=0.4$, $\sigma^2=0.1$, constant sample sizes $n=40,\;100,\;250,\;1000$.
The data-generation mechanisms are FIM1 ($\circ$), FIM2 ($\triangle$), RIM1 (+), RIM2 ($\times$), and URIM1 ($\diamond$).
		\label{PlotBiasThetamu1andpC04LOR_MPsigma01}}
\end{figure}
\begin{figure}[t]
	\centering
	\includegraphics[scale=0.33]{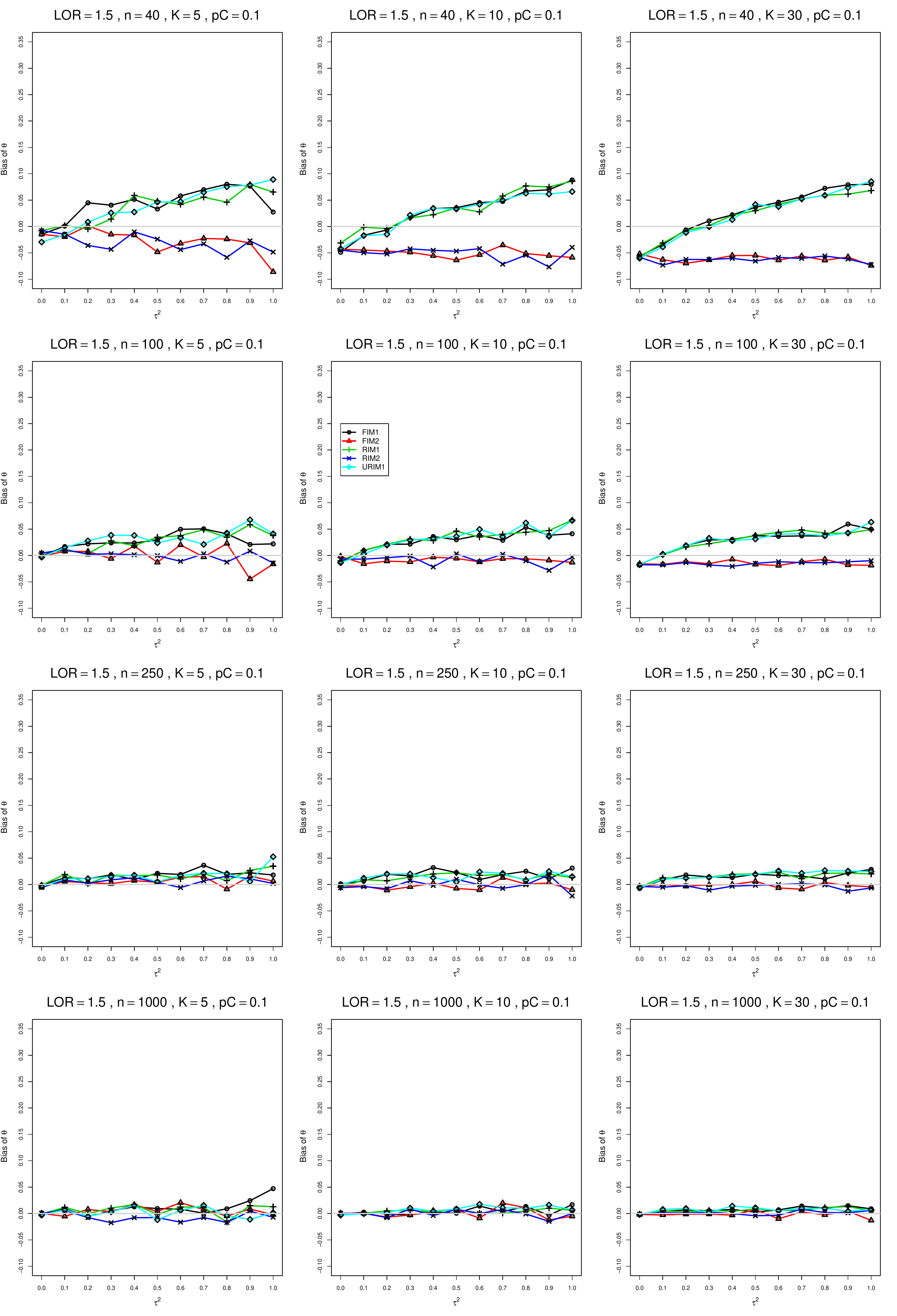}
	\caption{Bias of  overall log-odds ratio $\hat{\theta}_{MP}$ for $\theta=1.5$, $p_{C}=0.1$, $\sigma^2=0.1$, constant sample sizes $n=40,\;100,\;250,\;1000$.
The data-generation mechanisms are FIM1 ($\circ$), FIM2 ($\triangle$), RIM1 (+), RIM2 ($\times$), and URIM1 ($\diamond$).
		\label{PlotBiasThetamu15andpC01LOR_MPsigma01}}
\end{figure}
\begin{figure}[t]
	\centering
	\includegraphics[scale=0.33]{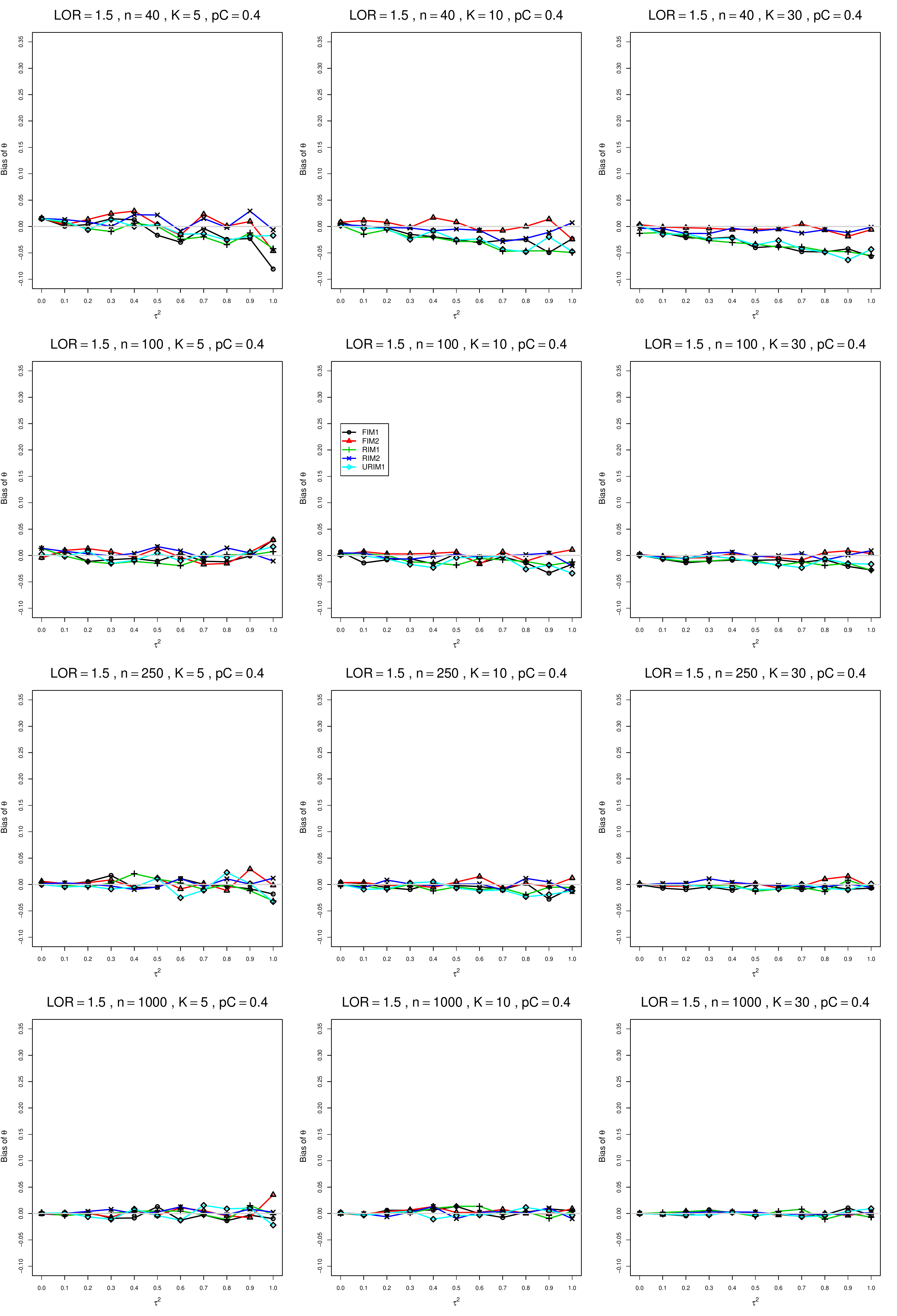}
	\caption{Bias of  overall log-odds ratio $\hat{\theta}_{MP}$ for $\theta=1.5$, $p_{C}=0.4$, $\sigma^2=0.1$, constant sample sizes $n=40,\;100,\;250,\;1000$.
The data-generation mechanisms are FIM1 ($\circ$), FIM2 ($\triangle$), RIM1 (+), RIM2 ($\times$), and URIM1 ($\diamond$).
		\label{PlotBiasThetamu15andpC04LOR_MPsigma01}}
\end{figure}
\begin{figure}[t]
	\centering
	\includegraphics[scale=0.33]{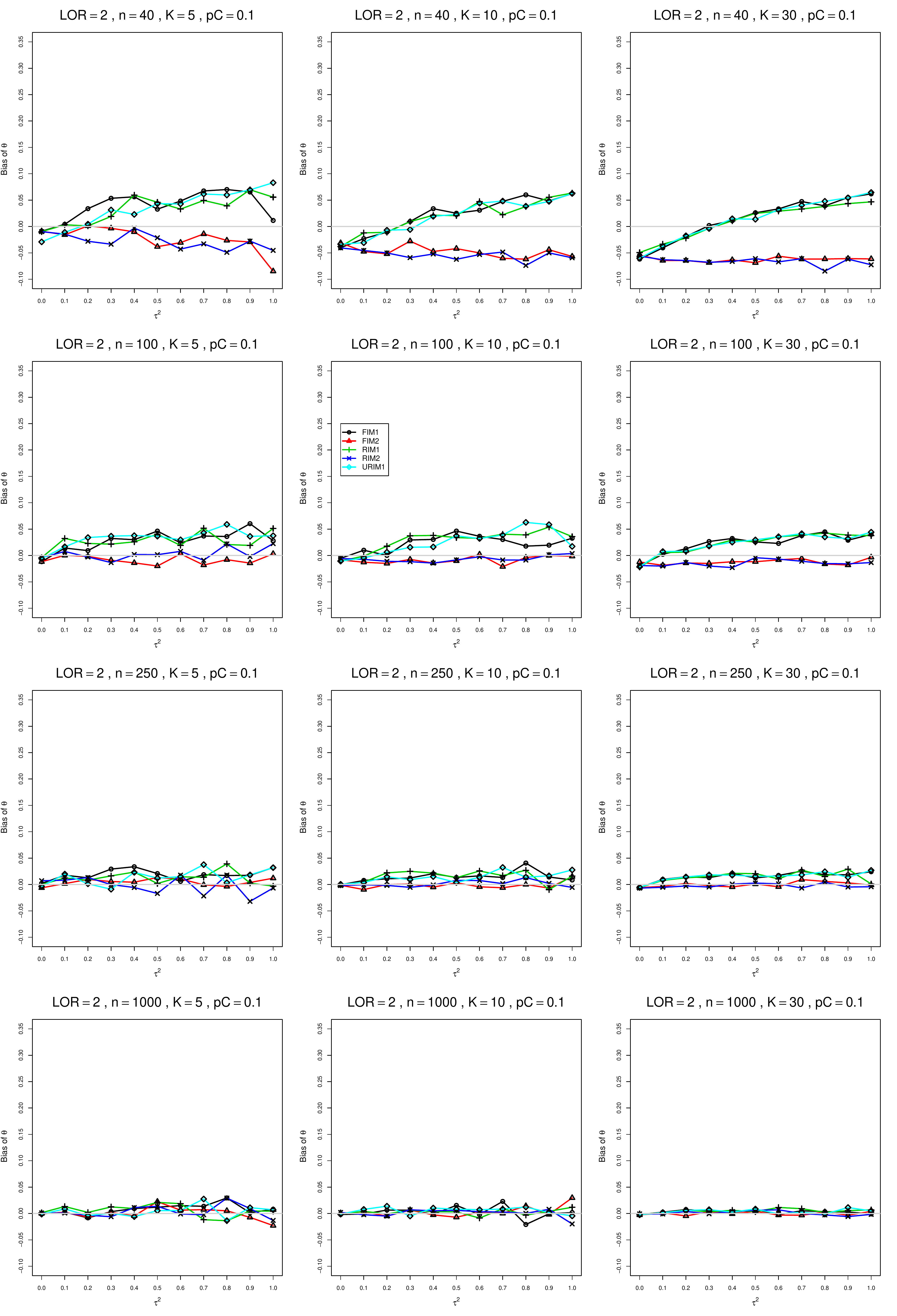}
	\caption{Bias of  overall log-odds ratio $\hat{\theta}_{MP}$ for $\theta=2$, $p_{C}=0.1$, $\sigma^2=0.1$, constant sample sizes $n=40,\;100,\;250,\;1000$.
The data-generation mechanisms are FIM1 ($\circ$), FIM2 ($\triangle$), RIM1 (+), RIM2 ($\times$), and URIM1 ($\diamond$).
		\label{PlotBiasThetamu2andpC01LOR_MPsigma01}}
\end{figure}
\begin{figure}[t]
	\centering
	\includegraphics[scale=0.33]{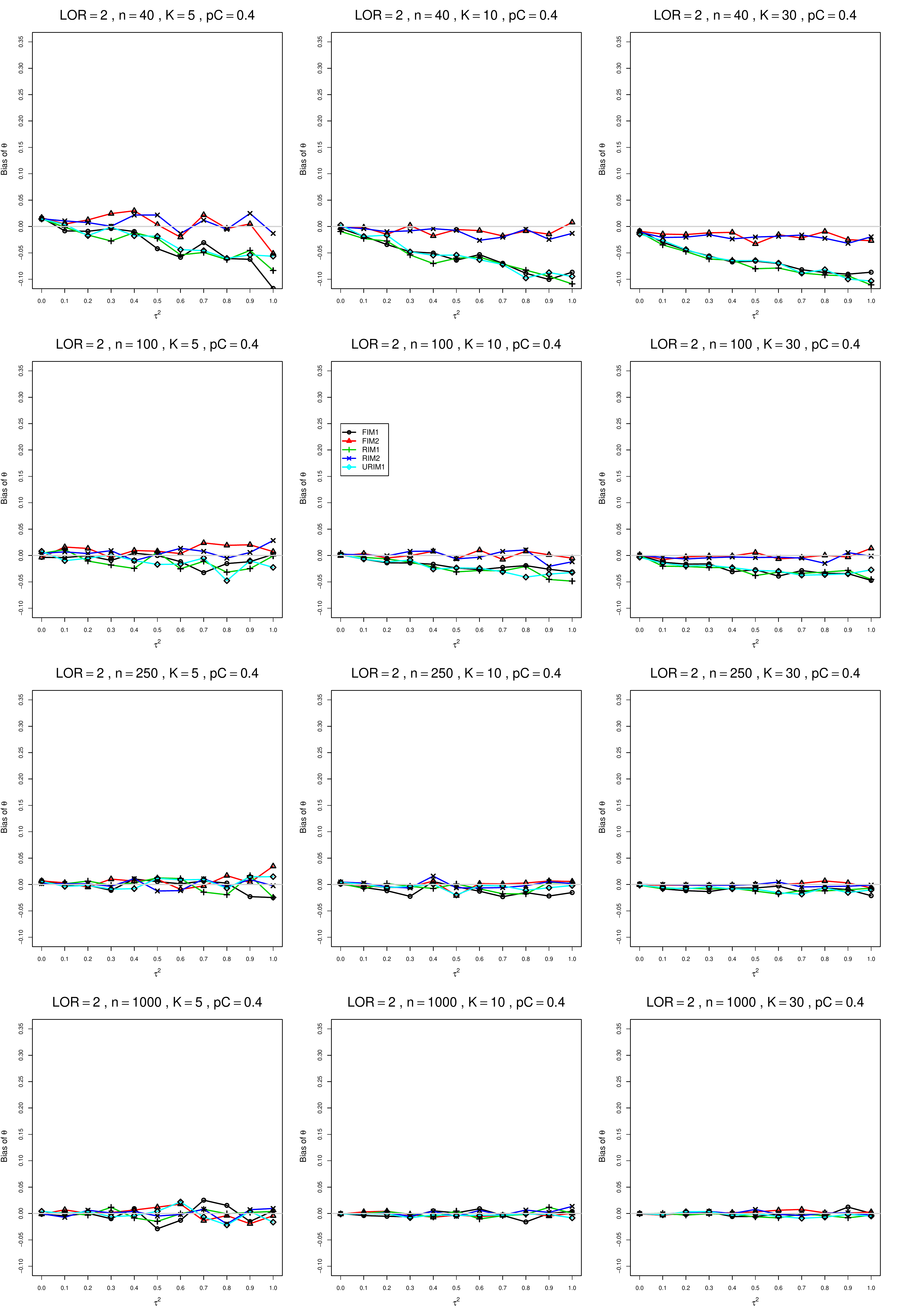}
	\caption{Bias of  overall log-odds ratio $\hat{\theta}_{MP}$ for $\theta=2$, $p_{C}=0.4$, $\sigma^2=0.1$, constant sample sizes $n=40,\;100,\;250,\;1000$.
The data-generation mechanisms are FIM1 ($\circ$), FIM2 ($\triangle$), RIM1 (+), RIM2 ($\times$), and URIM1 ($\diamond$).
		\label{PlotBiasThetamu2andpC04LOR_MPsigma01}}
\end{figure}
\begin{figure}[t]
	\centering
	\includegraphics[scale=0.33]{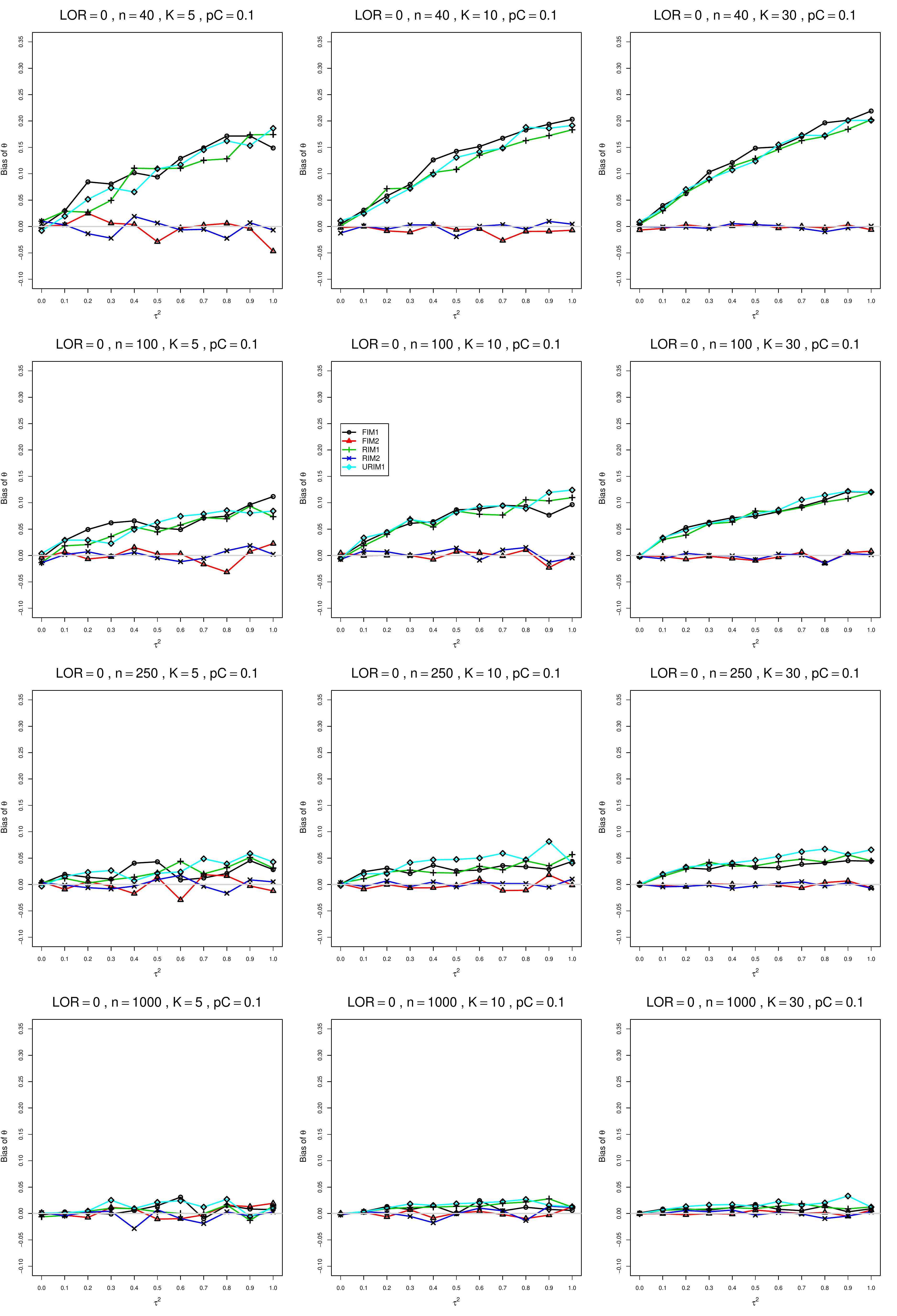}
	\caption{Bias of  overall log-odds ratio $\hat{\theta}_{MP}$ for $\theta=0$, $p_{C}=0.1$, $\sigma^2=0.4$, constant sample sizes $n=40,\;100,\;250,\;1000$.
The data-generation mechanisms are FIM1 ($\circ$), FIM2 ($\triangle$), RIM1 (+), RIM2 ($\times$), and URIM1 ($\diamond$).
		\label{PlotBiasThetamu0andpC01LOR_MPsigma04}}
\end{figure}
\begin{figure}[t]
	\centering
	\includegraphics[scale=0.33]{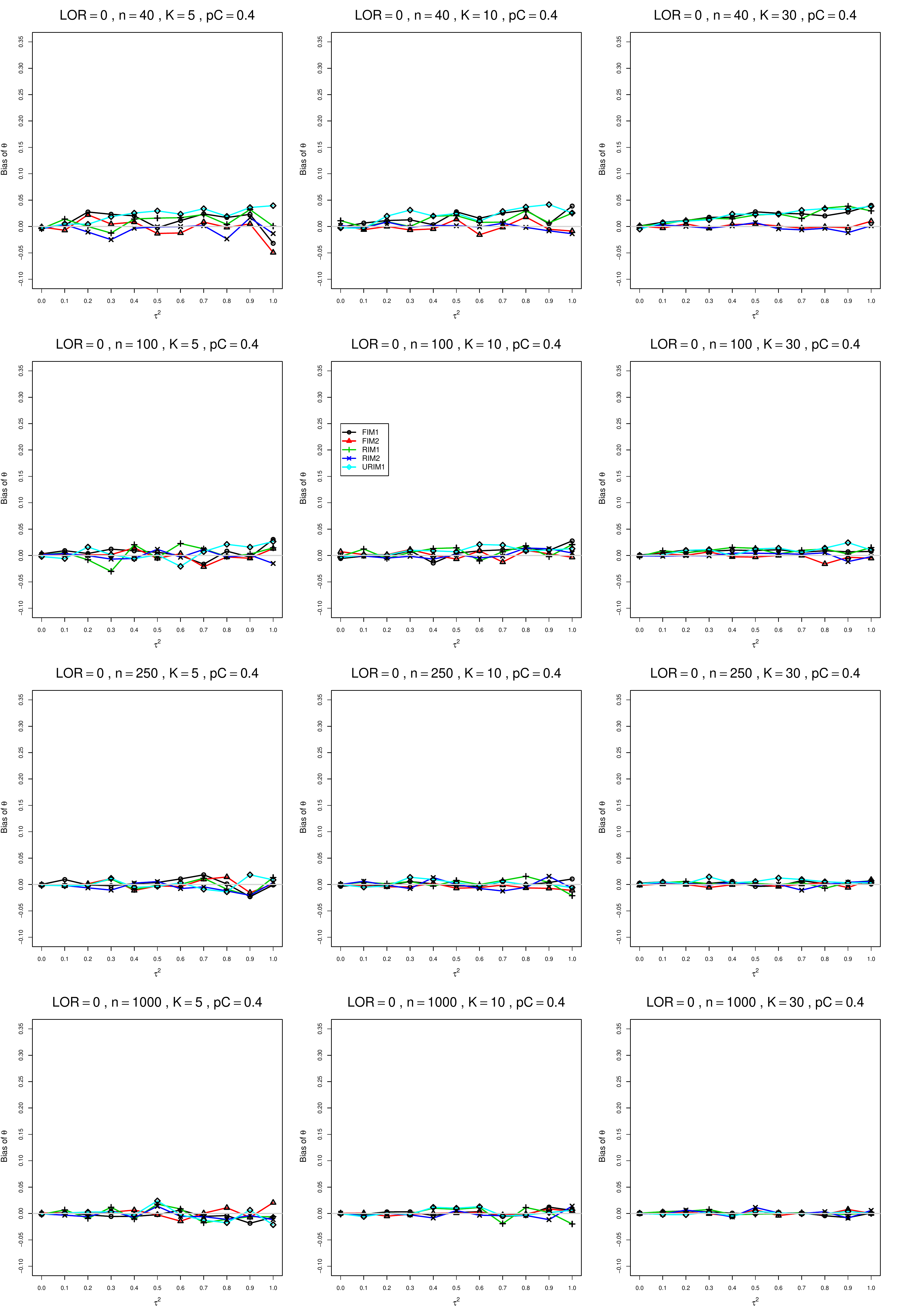}
	\caption{Bias of  overall log-odds ratio $\hat{\theta}_{MP}$ for $\theta=0$, $p_{C}=0.4$, $\sigma^2=0.4$, constant sample sizes $n=40,\;100,\;250,\;1000$.
The data-generation mechanisms are FIM1 ($\circ$), FIM2 ($\triangle$), RIM1 (+), RIM2 ($\times$), and URIM1 ($\diamond$).
		\label{PlotBiasThetamu0andpC04LOR_MPsigma04}}
\end{figure}
\begin{figure}[t]
	\centering
	\includegraphics[scale=0.33]{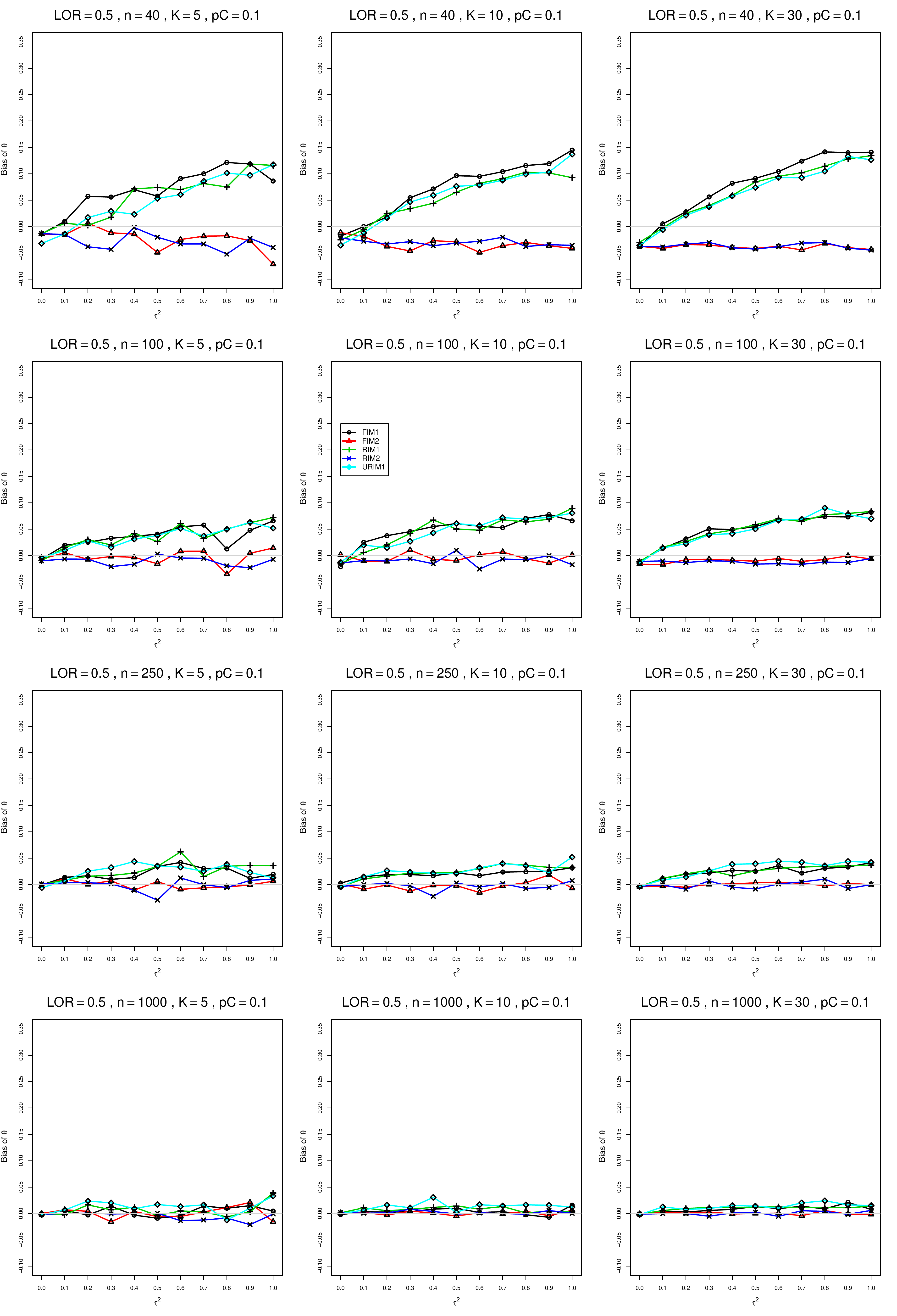}
	\caption{Bias of  overall log-odds ratio $\hat{\theta}_{MP}$ for $\theta=0.5$, $p_{C}=0.1$, $\sigma^2=0.4$, constant sample sizes $n=40,\;100,\;250,\;1000$.
The data-generation mechanisms are FIM1 ($\circ$), FIM2 ($\triangle$), RIM1 (+), RIM2 ($\times$), and URIM1 ($\diamond$).
		\label{PlotBiasThetamu05andpC01LOR_MPsigma04}}
\end{figure}
\begin{figure}[t]
	\centering
	\includegraphics[scale=0.33]{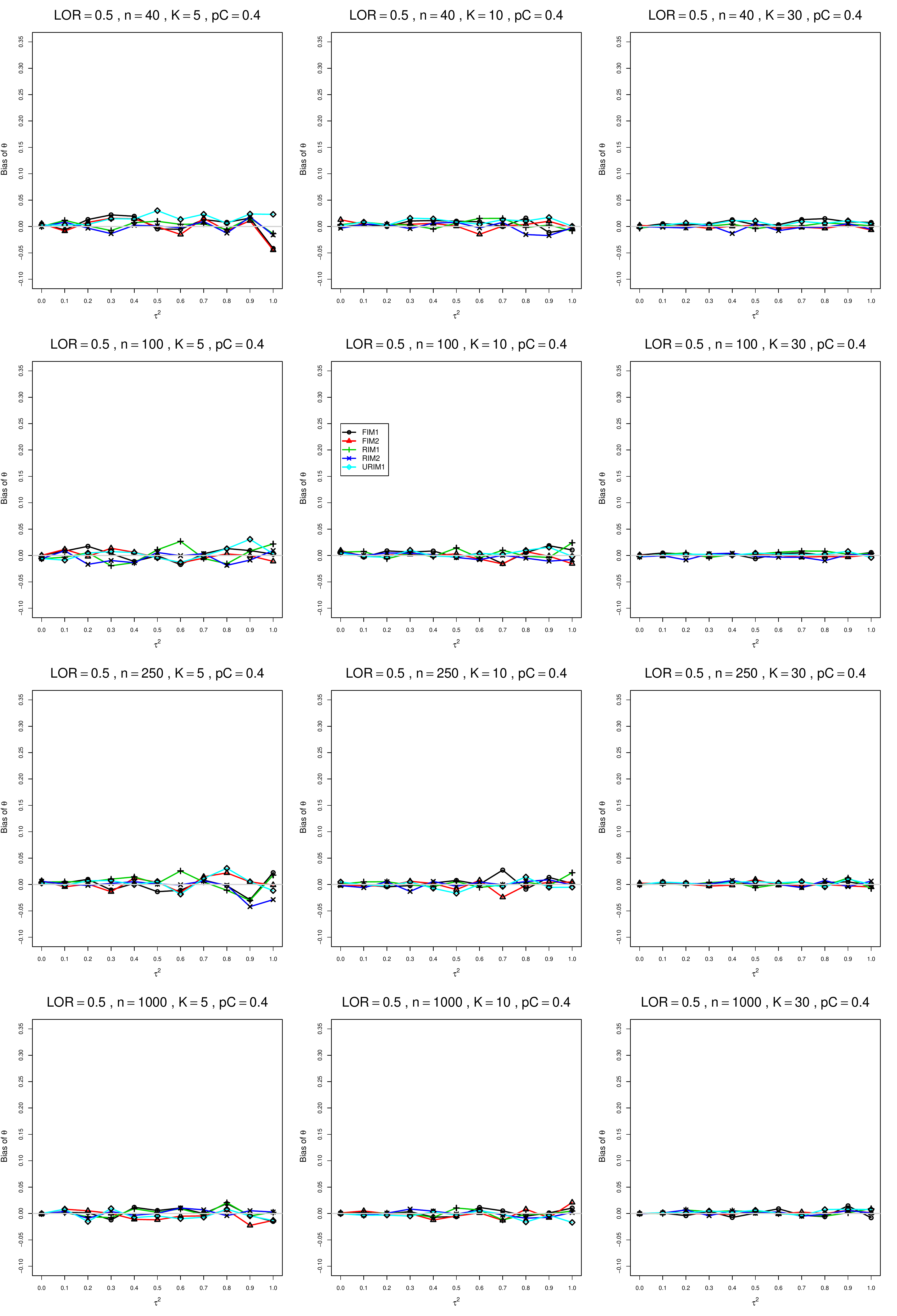}
	\caption{Bias of  overall log-odds ratio $\hat{\theta}_{MP}$ for $\theta=0.5$, $p_{C}=0.4$, $\sigma^2=0.4$, constant sample sizes $n=40,\;100,\;250,\;1000$.
The data-generation mechanisms are FIM1 ($\circ$), FIM2 ($\triangle$), RIM1 (+), RIM2 ($\times$), and URIM1 ($\diamond$).
		\label{PlotBiasThetamu05andpC04LOR_MPsigma04}}
\end{figure}
\begin{figure}[t]
	\centering
	\includegraphics[scale=0.33]{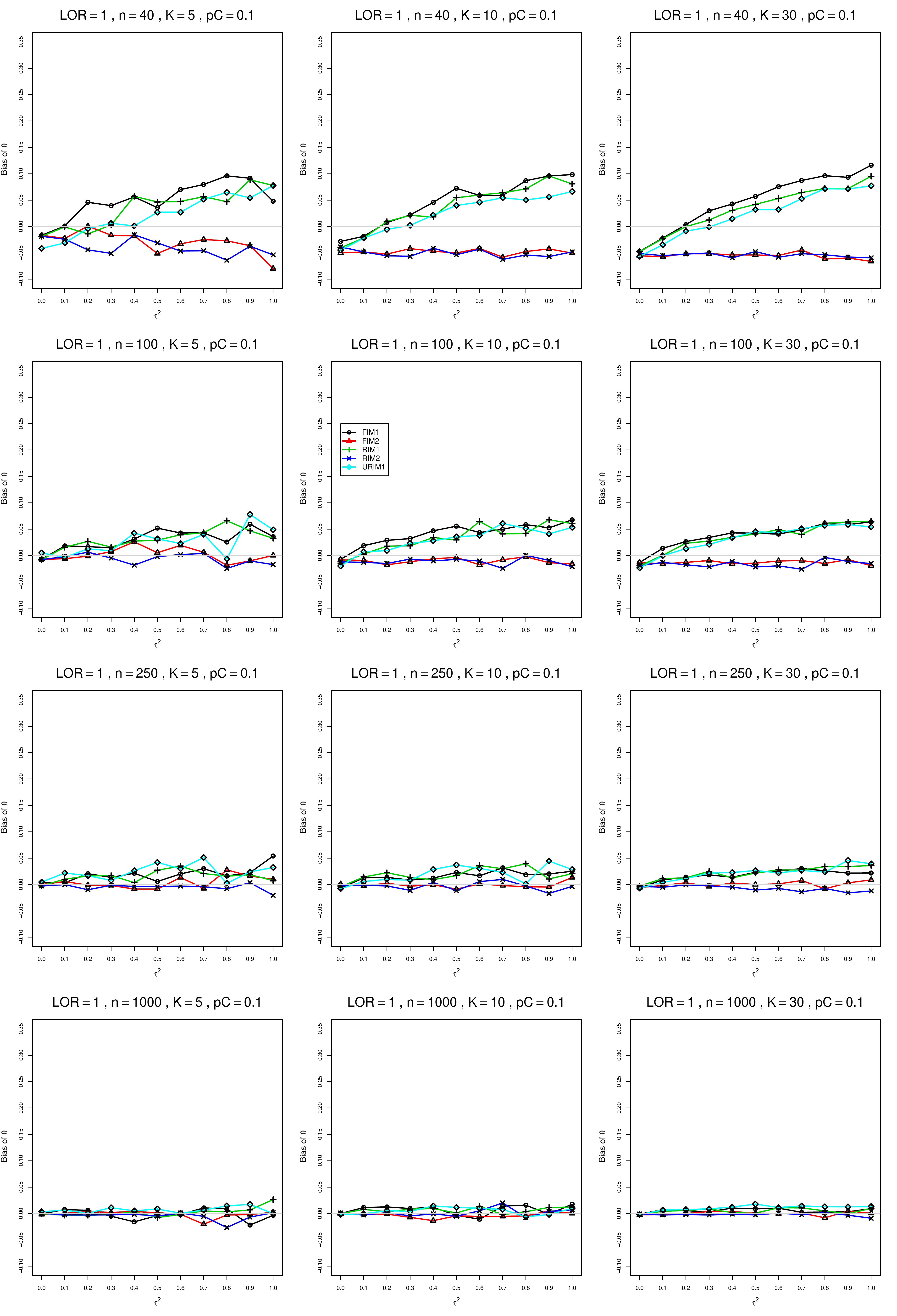}
	\caption{Bias of  overall log-odds ratio $\hat{\theta}_{MP}$ for $\theta=1$, $p_{C}=0.1$, $\sigma^2=0.4$, constant sample sizes $n=40,\;100,\;250,\;1000$.
The data-generation mechanisms are FIM1 ($\circ$), FIM2 ($\triangle$), RIM1 (+), RIM2 ($\times$), and URIM1 ($\diamond$).
		\label{PlotBiasThetamu1andpC01LOR_MPsigma04}}
\end{figure}
\begin{figure}[t]
	\centering
	\includegraphics[scale=0.33]{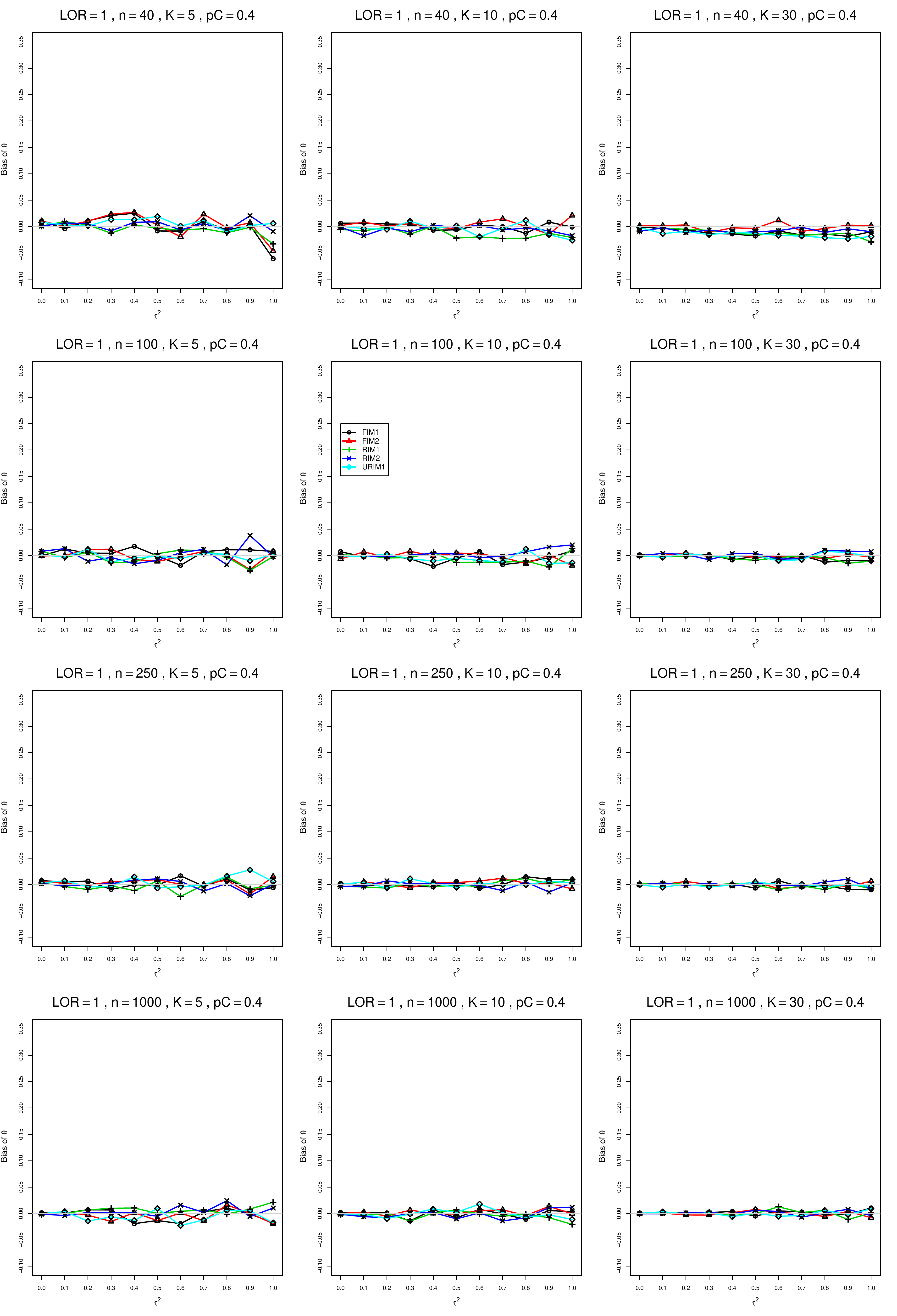}
	\caption{Bias of  overall log-odds ratio $\hat{\theta}_{MP}$ for $\theta=1$, $p_{C}=0.4$, $\sigma^2=0.4$, constant sample sizes $n=40,\;100,\;250,\;1000$.
The data-generation mechanisms are FIM1 ($\circ$), FIM2 ($\triangle$), RIM1 (+), RIM2 ($\times$), and URIM1 ($\diamond$).
		\label{PlotBiasThetamu1andpC04LOR_MPsigma04}}
\end{figure}
\begin{figure}[t]
	\centering
	\includegraphics[scale=0.33]{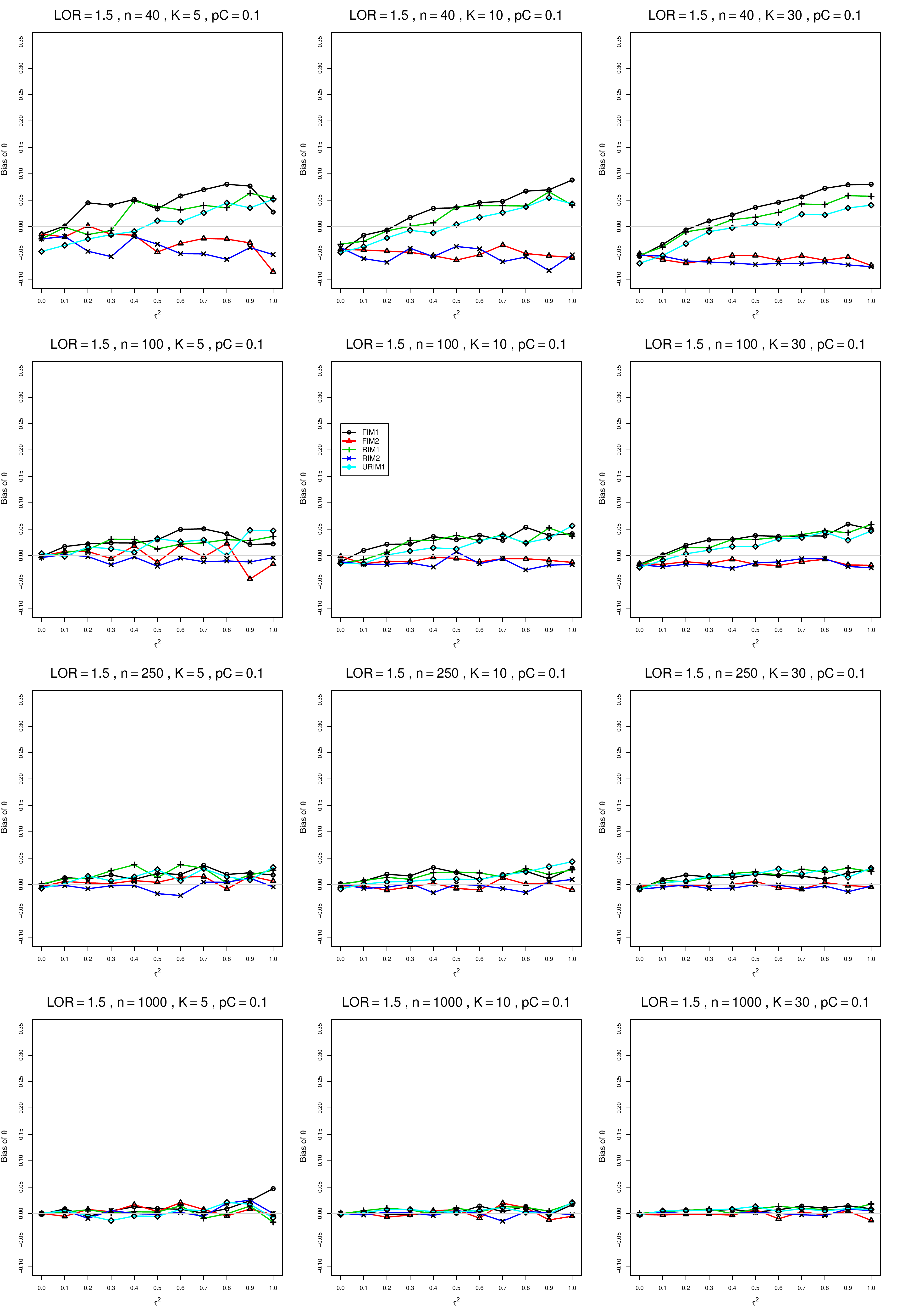}
	\caption{Bias of  overall log-odds ratio $\hat{\theta}_{MP}$ for $\theta=1.5$, $p_{C}=0.1$, $\sigma^2=0.4$, constant sample sizes $n=40,\;100,\;250,\;1000$.
The data-generation mechanisms are FIM1 ($\circ$), FIM2 ($\triangle$), RIM1 (+), RIM2 ($\times$), and URIM1 ($\diamond$).
		\label{PlotBiasThetamu15andpC01LOR_MPsigma04}}
\end{figure}
\begin{figure}[t]
	\centering
	\includegraphics[scale=0.33]{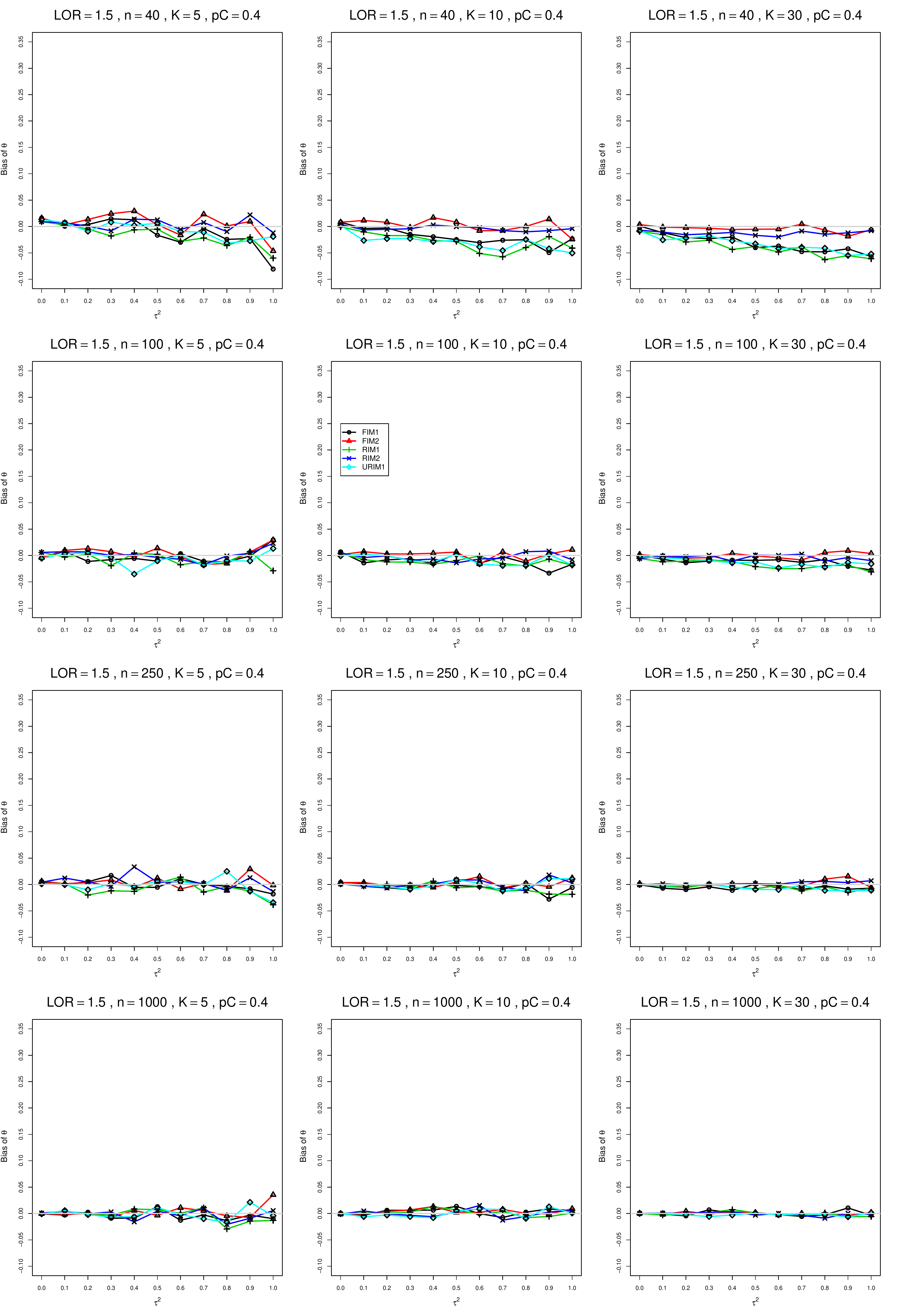}
	\caption{Bias of  overall log-odds ratio $\hat{\theta}_{MP}$ for $\theta=1.5$, $p_{C}=0.4$, $\sigma^2=0.4$, constant sample sizes $n=40,\;100,\;250,\;1000$.
The data-generation mechanisms are FIM1 ($\circ$), FIM2 ($\triangle$), RIM1 (+), RIM2 ($\times$), and URIM1 ($\diamond$).
		\label{PlotBiasThetamu15andpC04LOR_MPsigma04}}
\end{figure}
\begin{figure}[t]
	\centering
	\includegraphics[scale=0.33]{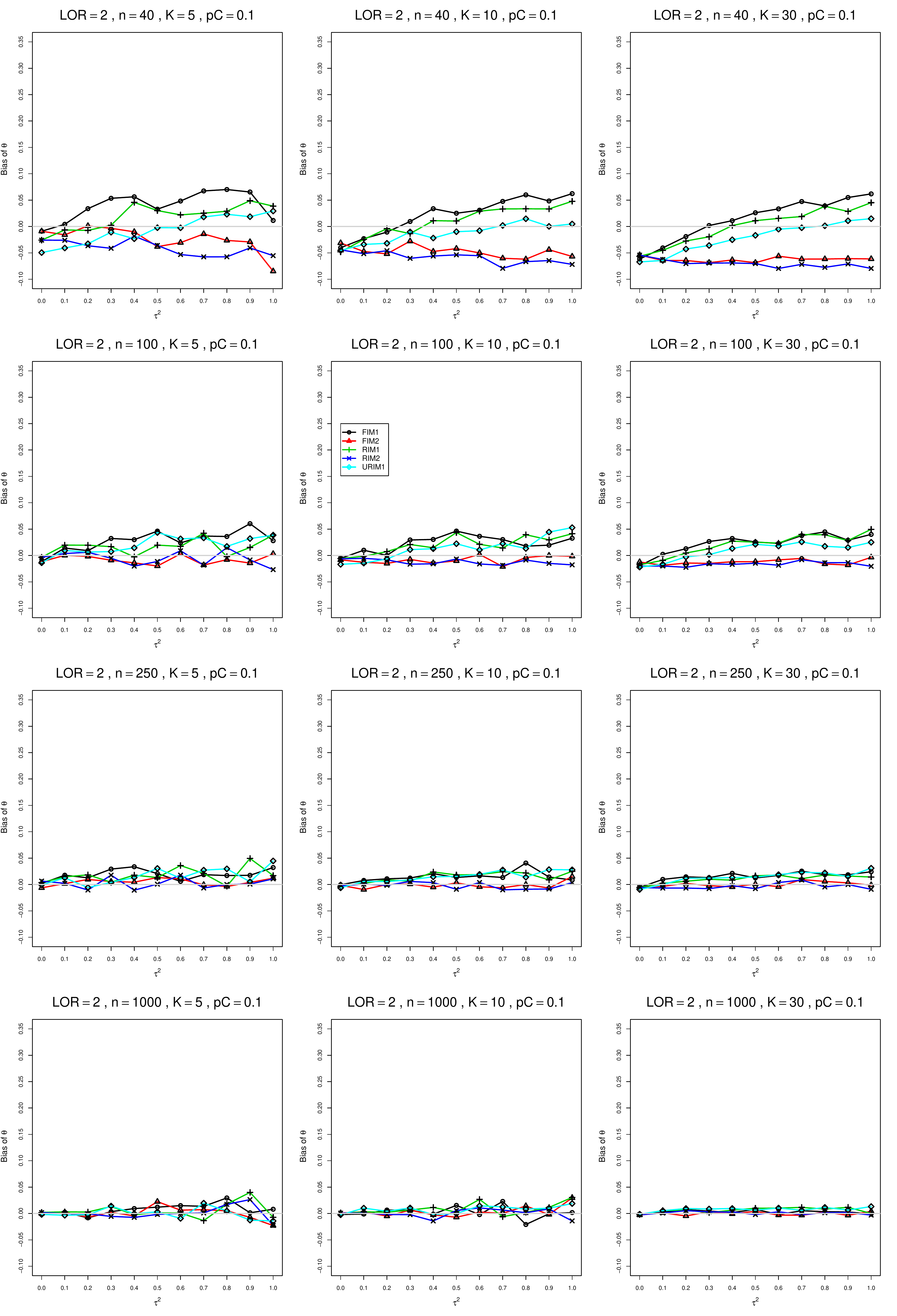}
	\caption{Bias of  overall log-odds ratio $\hat{\theta}_{MP}$ for $\theta=2$, $p_{C}=0.1$, $\sigma^2=0.4$, constant sample sizes $n=40,\;100,\;250,\;1000$.
The data-generation mechanisms are FIM1 ($\circ$), FIM2 ($\triangle$), RIM1 (+), RIM2 ($\times$), and URIM1 ($\diamond$).
		\label{PlotBiasThetamu2andpC01LOR_MPsigma04}}
\end{figure}
\begin{figure}[t]
	\centering
	\includegraphics[scale=0.33]{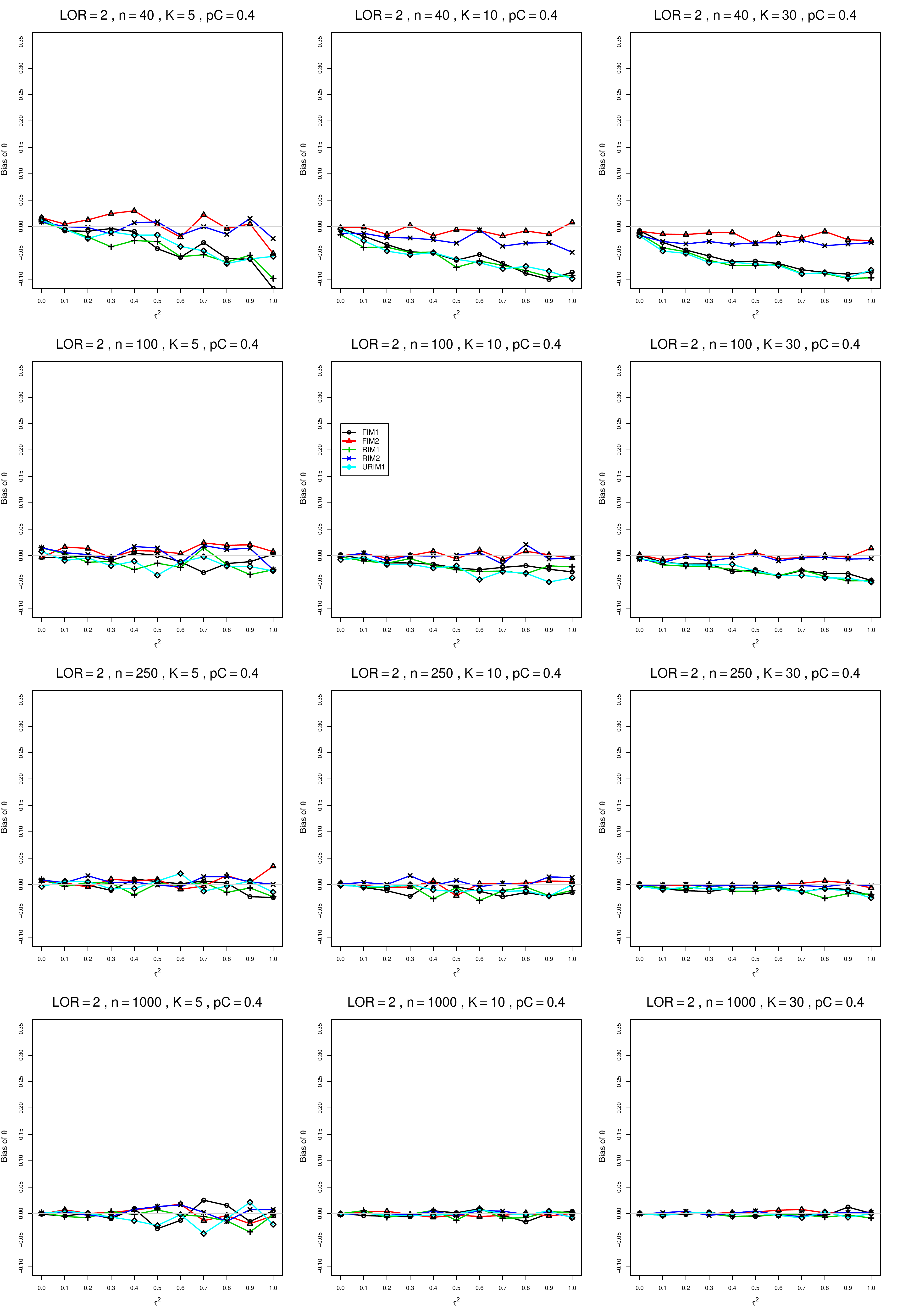}
	\caption{Bias of  overall log-odds ratio $\hat{\theta}_{MP}$ for $\theta=2$, $p_{C}=0.4$, $\sigma^2=0.4$, constant sample sizes $n=40,\;100,\;250,\;1000$.
The data-generation mechanisms are FIM1 ($\circ$), FIM2 ($\triangle$), RIM1 (+), RIM2 ($\times$), and URIM1 ($\diamond$).
		\label{PlotBiasThetamu2andpC04LOR_MPsigma04}}
\end{figure}

\clearpage
\subsection*{A2.4 Bias of $\hat{\theta}_{KD}$}
\renewcommand{\thefigure}{A2.4.\arabic{figure}}
\setcounter{figure}{0}

\begin{figure}[t]
	\centering
	\includegraphics[scale=0.33]{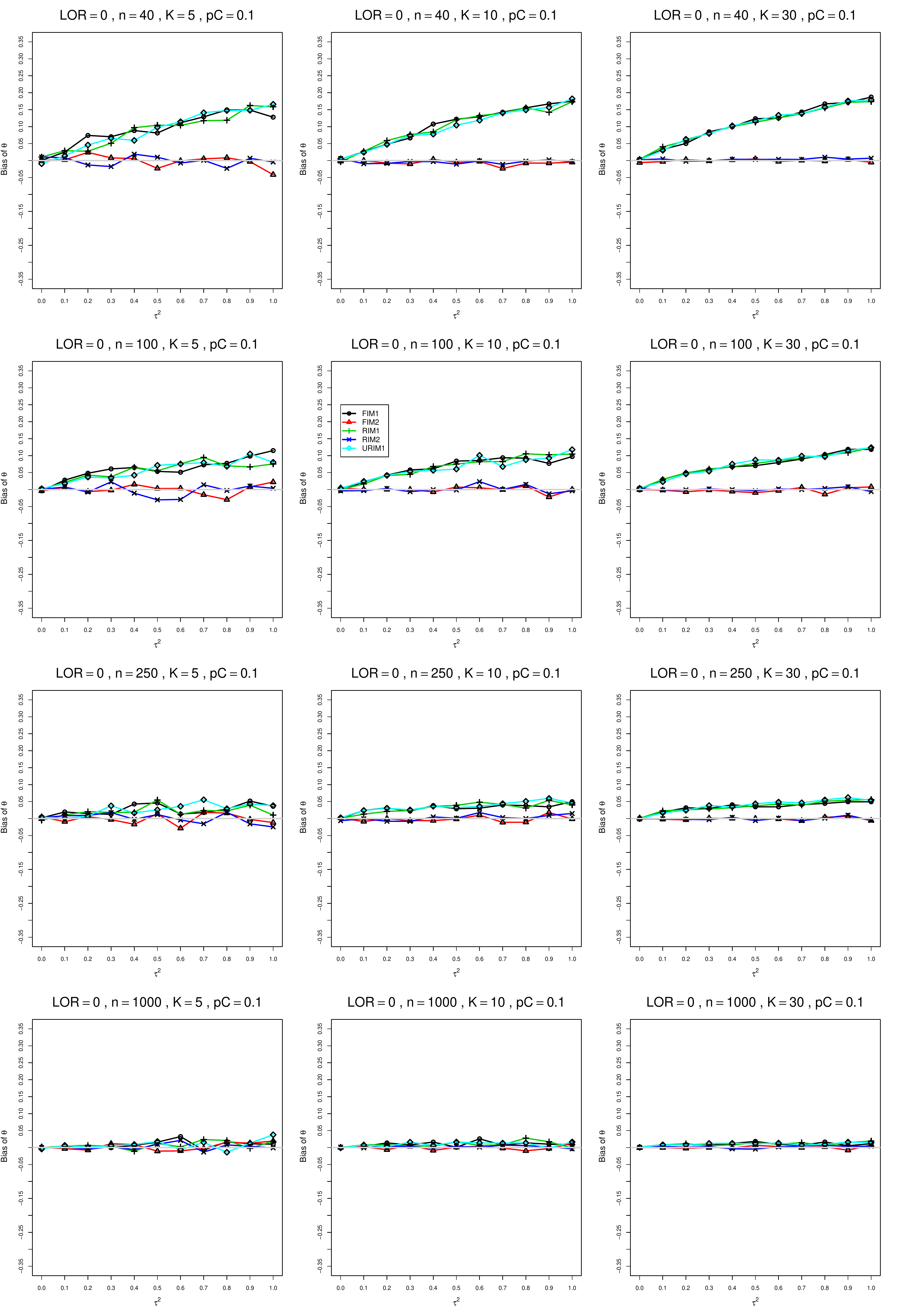}
	\caption{Bias of  overall log-odds ratio $\hat{\theta}_{KD}$ for $\theta=0$, $p_{C}=0.1$, $\sigma^2=0.1$, constant sample sizes $n=40,\;100,\;250,\;1000$.
The data-generation mechanisms are FIM1 ($\circ$), FIM2 ($\triangle$), RIM1 (+), RIM2 ($\times$), and URIM1 ($\diamond$).
		\label{PlotBiasThetamu0andpC01LOR_KDsigma01}}
\end{figure}
\begin{figure}[t]
	\centering
	\includegraphics[scale=0.33]{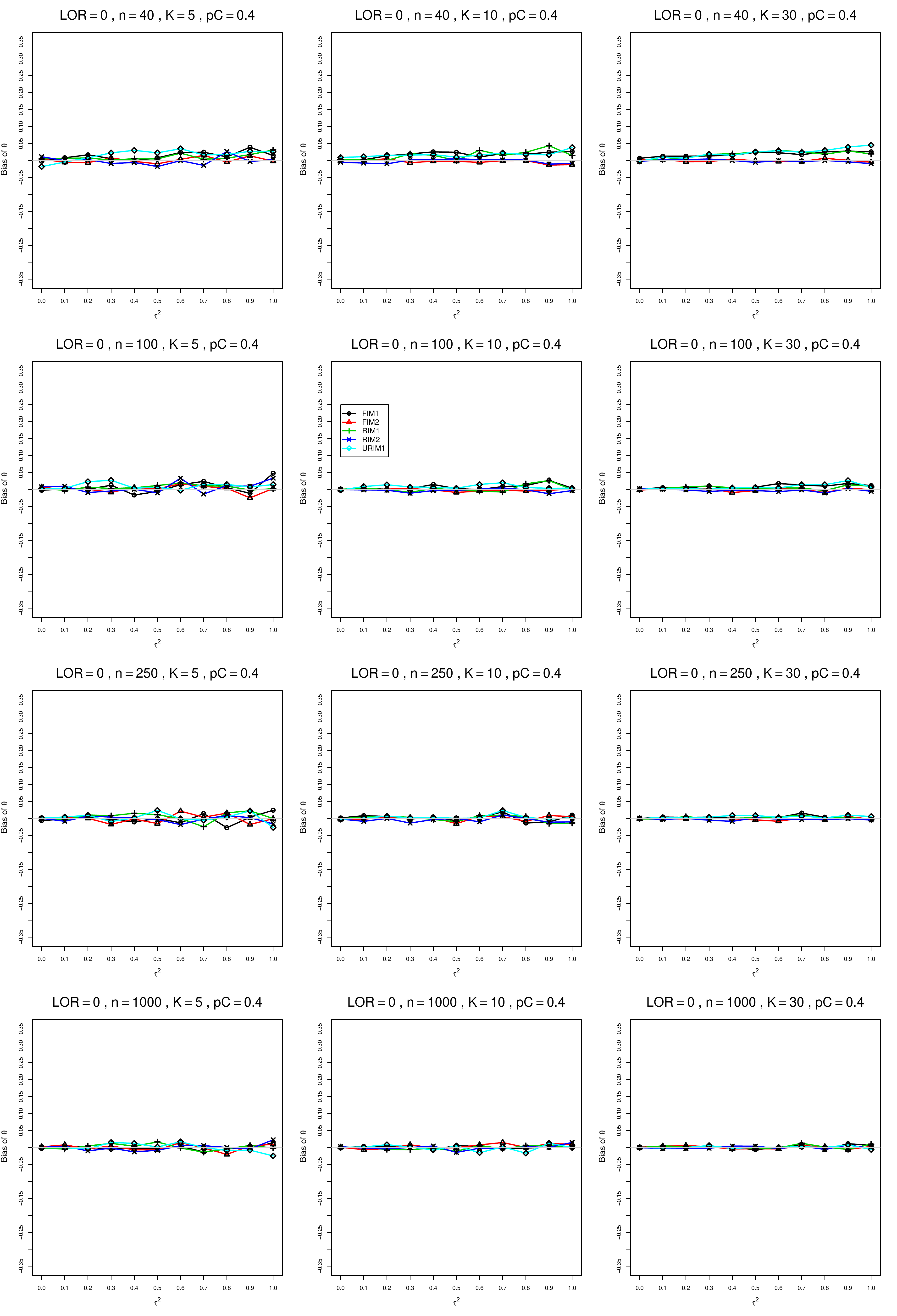}
	\caption{Bias of  overall log-odds ratio $\hat{\theta}_{KD}$ for $\theta=0$, $p_{C}=0.4$, $\sigma^2=0.1$, constant sample sizes $n=40,\;100,\;250,\;1000$.
The data-generation mechanisms are FIM1 ($\circ$), FIM2 ($\triangle$), RIM1 (+), RIM2 ($\times$), and URIM1 ($\diamond$).
		\label{PlotBiasThetamu0andpC04LOR_KDsigma01}}
\end{figure}
\begin{figure}[t]
	\centering
	\includegraphics[scale=0.33]{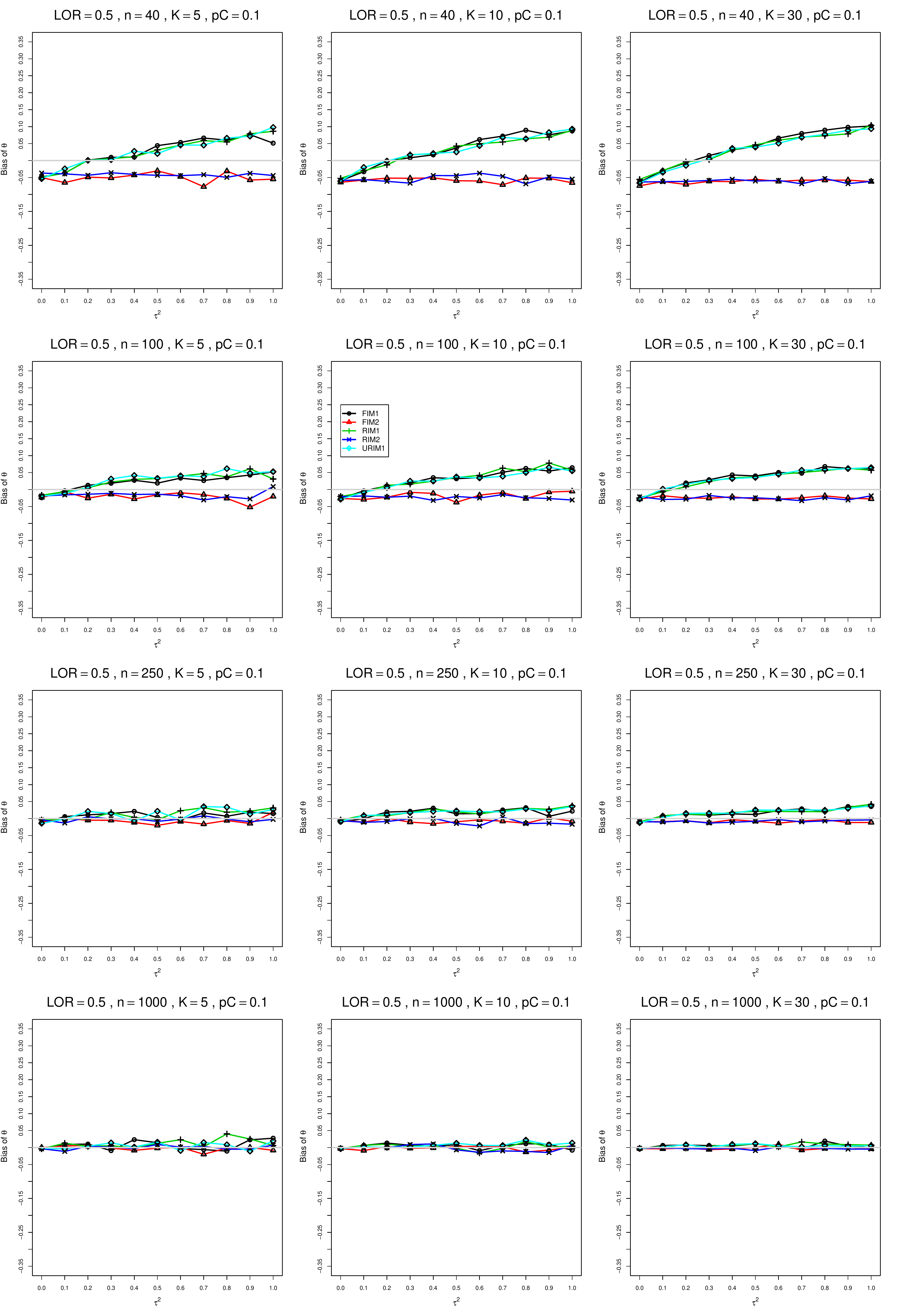}
	\caption{Bias of  overall log-odds ratio $\hat{\theta}_{KD}$ for $\theta=0.5$, $p_{C}=0.1$, $\sigma^2=0.1$, constant sample sizes $n=40,\;100,\;250,\;1000$.
The data-generation mechanisms are FIM1 ($\circ$), FIM2 ($\triangle$), RIM1 (+), RIM2 ($\times$), and URIM1 ($\diamond$).
		\label{PlotBiasThetamu05andpC01LOR_KDsigma01}}
\end{figure}
\begin{figure}[t]
	\centering
	\includegraphics[scale=0.33]{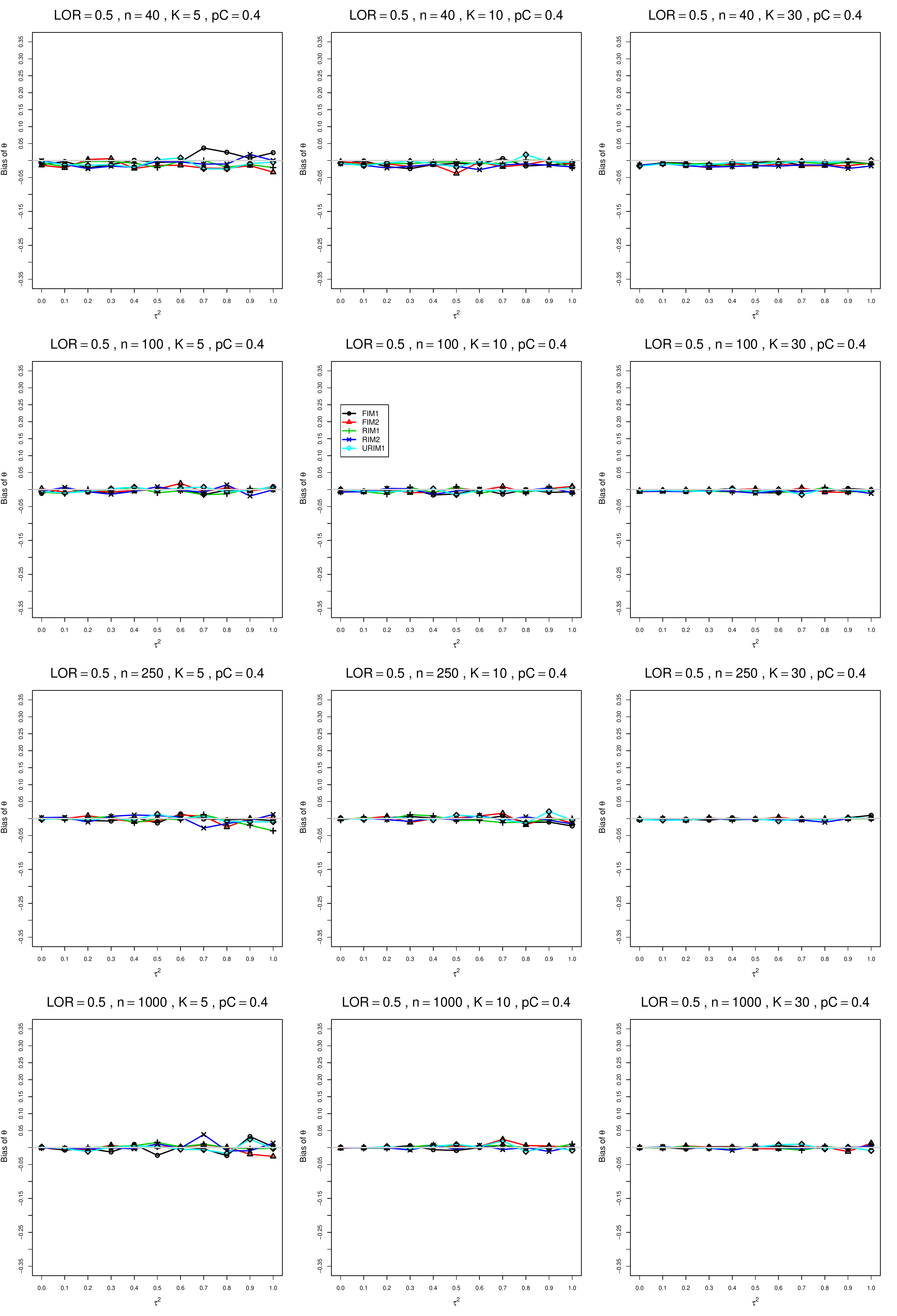}
	\caption{Bias of  overall log-odds ratio $\hat{\theta}_{KD}$ for $\theta=0.5$, $p_{C}=0.4$, $\sigma^2=0.1$, constant sample sizes $n=40,\;100,\;250,\;1000$.
The data-generation mechanisms are FIM1 ($\circ$), FIM2 ($\triangle$), RIM1 (+), RIM2 ($\times$), and URIM1 ($\diamond$).
		\label{PlotBiasThetamu05andpC04LOR_KDsigma01}}
\end{figure}
\begin{figure}[t]
	\centering
	\includegraphics[scale=0.33]{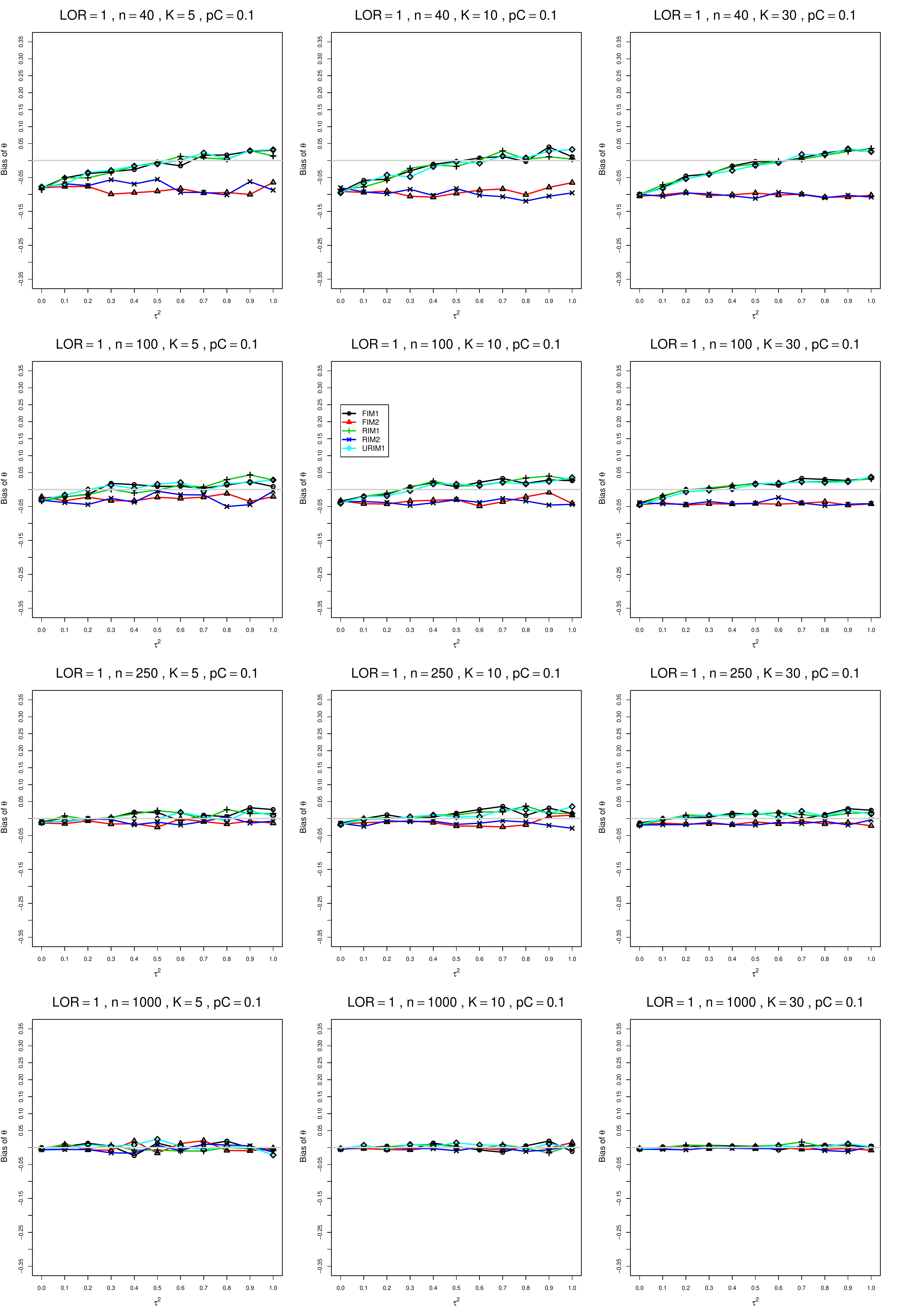}
	\caption{Bias of  overall log-odds ratio $\hat{\theta}_{KD}$ for $\theta=1$, $p_{C}=0.1$, $\sigma^2=0.1$, constant sample sizes $n=40,\;100,\;250,\;1000$.
The data-generation mechanisms are FIM1 ($\circ$), FIM2 ($\triangle$), RIM1 (+), RIM2 ($\times$), and URIM1 ($\diamond$).
		\label{PlotBiasThetamu1andpC01LOR_KDsigma01}}
\end{figure}
\begin{figure}[t]
	\centering
	\includegraphics[scale=0.33]{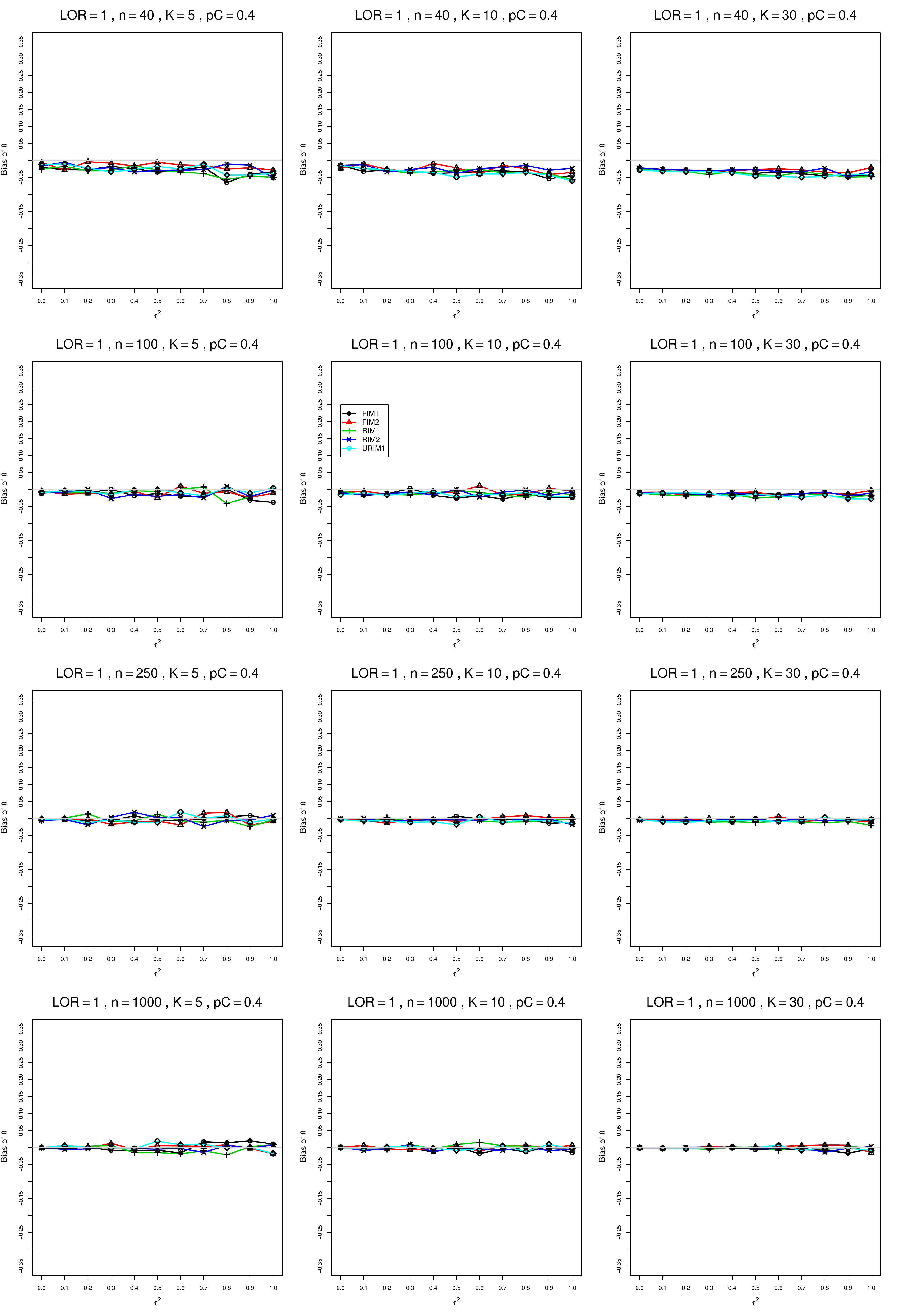}
	\caption{Bias of  overall log-odds ratio $\hat{\theta}_{KD}$ for $\theta=1$, $p_{C}=0.4$, $\sigma^2=0.1$, constant sample sizes $n=40,\;100,\;250,\;1000$.
The data-generation mechanisms are FIM1 ($\circ$), FIM2 ($\triangle$), RIM1 (+), RIM2 ($\times$), and URIM1 ($\diamond$).
		\label{PlotBiasThetamu1andpC04LOR_KDsigma01}}
\end{figure}
\begin{figure}[t]
	\centering
	\includegraphics[scale=0.33]{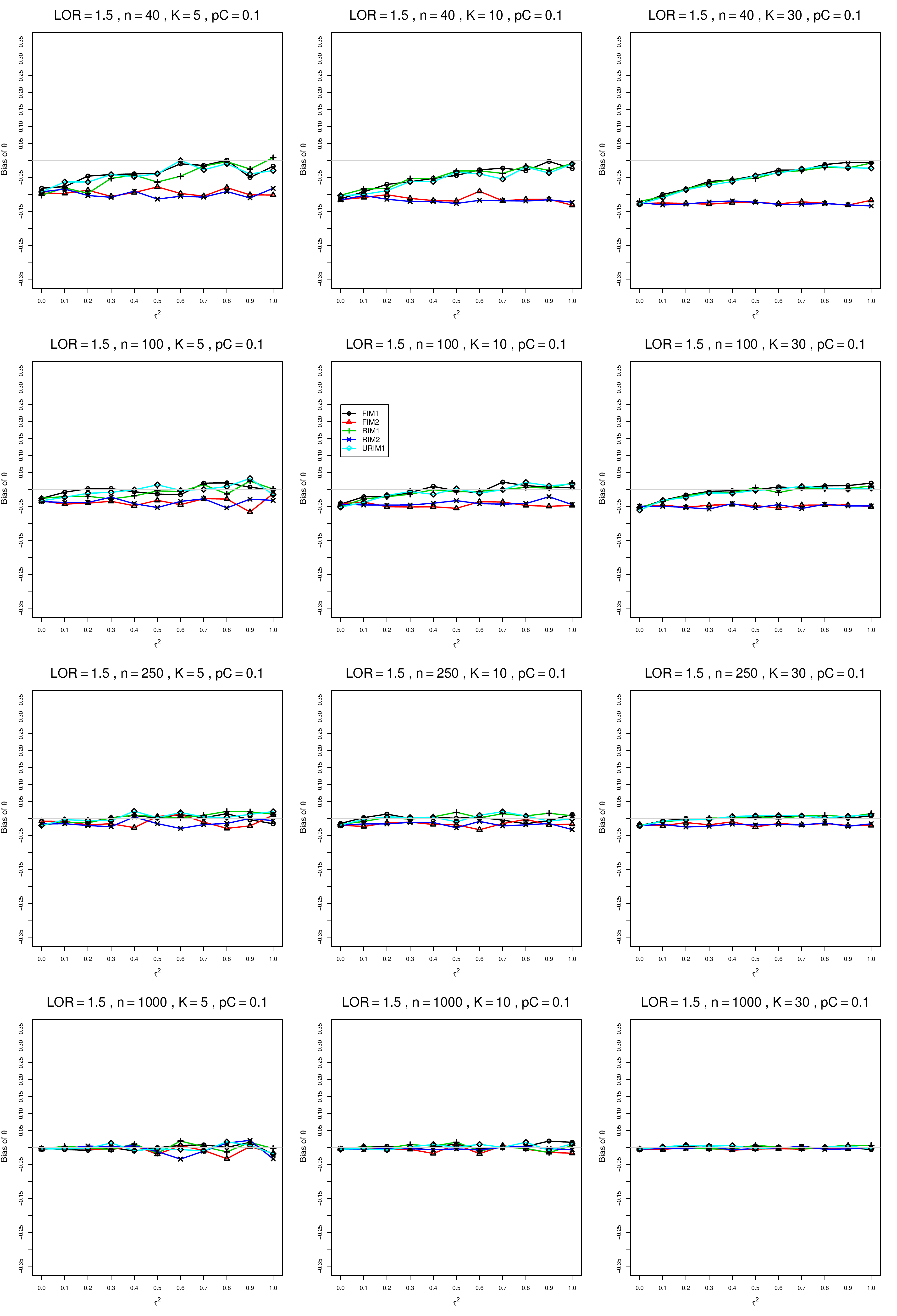}
	\caption{Bias of  overall log-odds ratio $\hat{\theta}_{KD}$ for $\theta=1.5$, $p_{C}=0.1$, $\sigma^2=0.1$, constant sample sizes $n=40,\;100,\;250,\;1000$.
The data-generation mechanisms are FIM1 ($\circ$), FIM2 ($\triangle$), RIM1 (+), RIM2 ($\times$), and URIM1 ($\diamond$).
		\label{PlotBiasThetamu15andpC01LOR_KDsigma01}}
\end{figure}
\begin{figure}[t]
	\centering
	\includegraphics[scale=0.33]{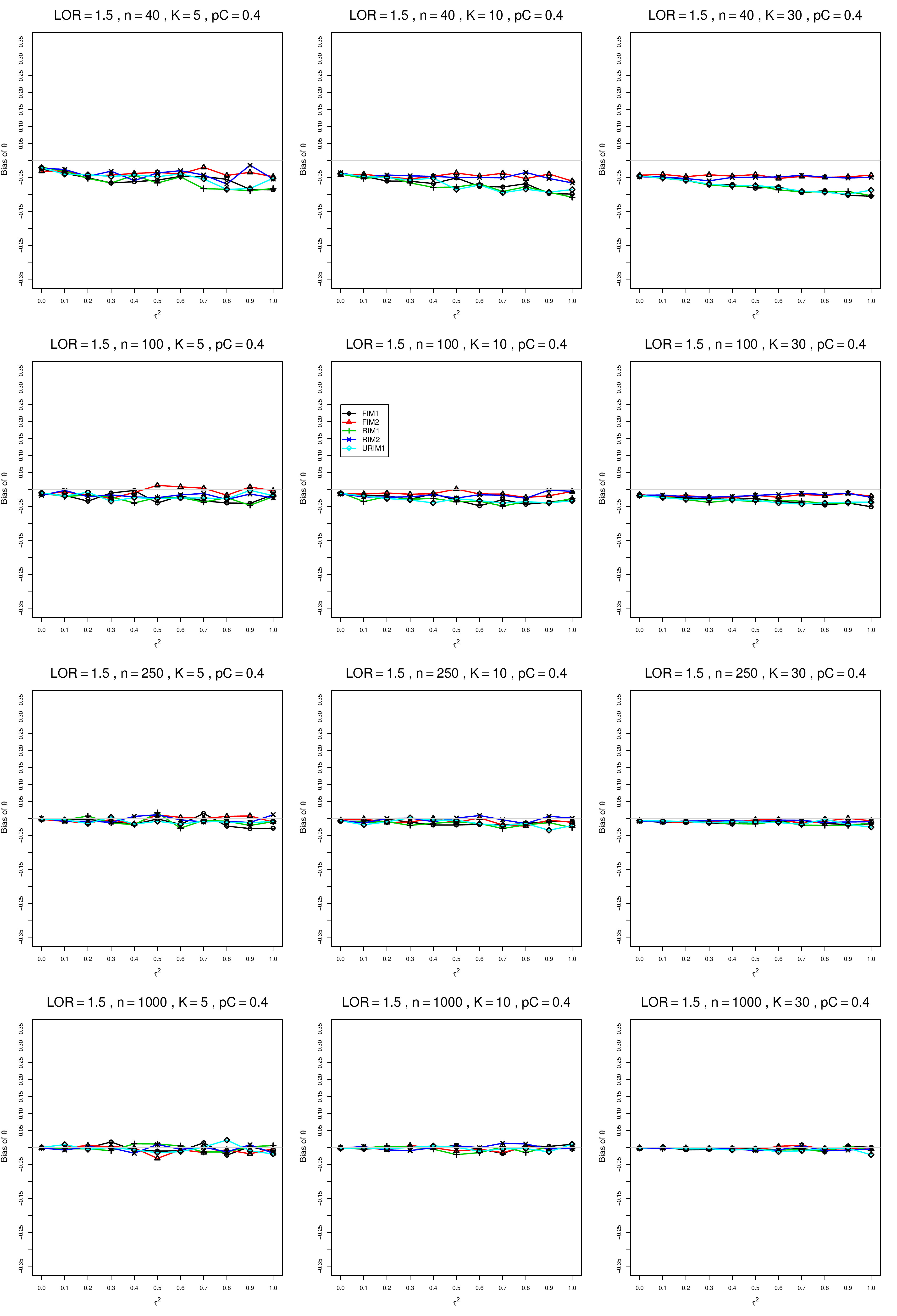}
	\caption{Bias of  overall log-odds ratio $\hat{\theta}_{KD}$ for $\theta=1.5$, $p_{C}=0.4$, $\sigma^2=0.1$, constant sample sizes $n=40,\;100,\;250,\;1000$.
The data-generation mechanisms are FIM1 ($\circ$), FIM2 ($\triangle$), RIM1 (+), RIM2 ($\times$), and URIM1 ($\diamond$).
		\label{PlotBiasThetamu15andpC04LOR_KDsigma01}}
\end{figure}
\begin{figure}[t]
	\centering
	\includegraphics[scale=0.33]{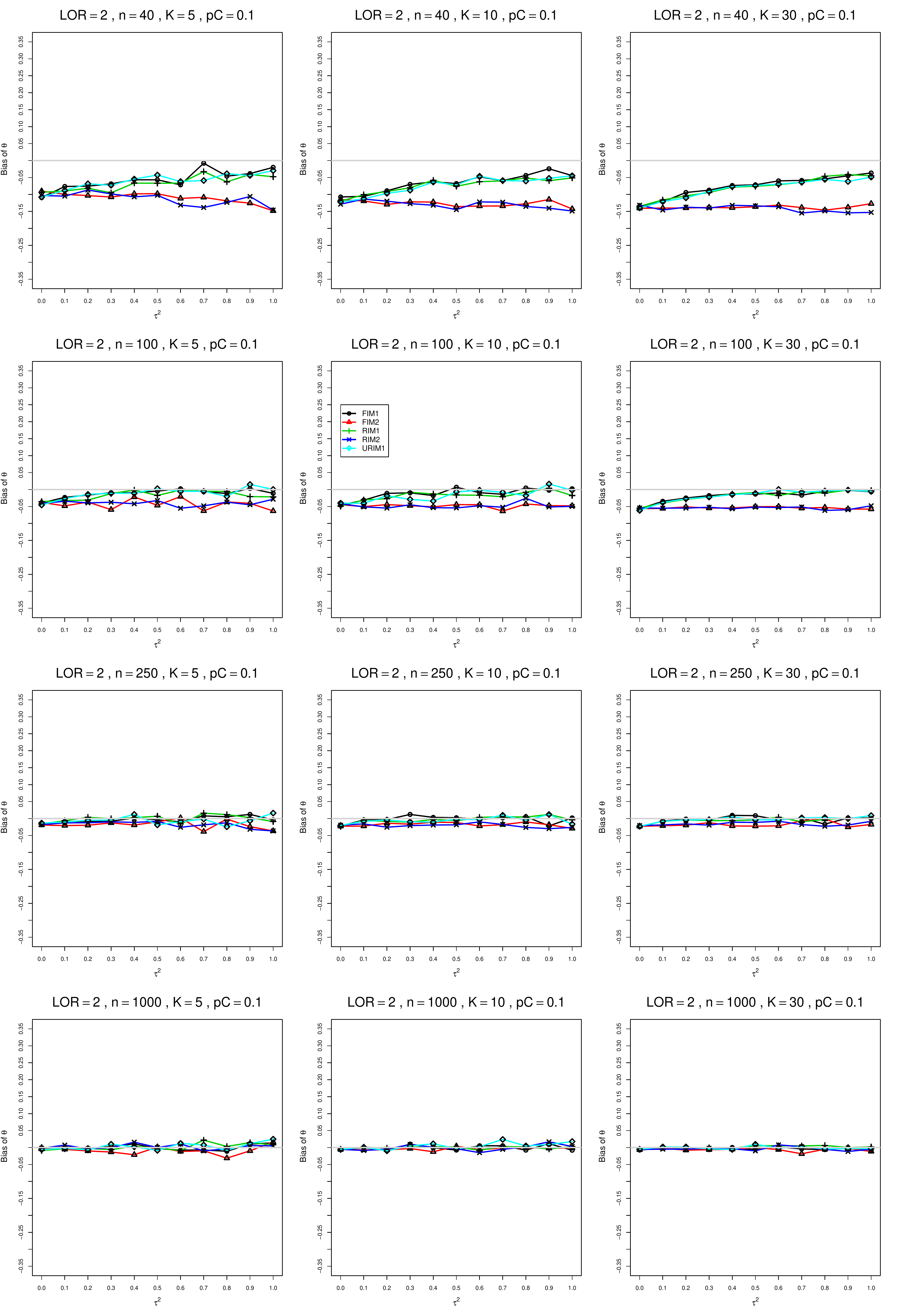}
	\caption{Bias of  overall log-odds ratio $\hat{\theta}_{KD}$ for $\theta=2$, $p_{C}=0.1$, $\sigma^2=0.1$, constant sample sizes $n=40,\;100,\;250,\;1000$.
The data-generation mechanisms are FIM1 ($\circ$), FIM2 ($\triangle$), RIM1 (+), RIM2 ($\times$), and URIM1 ($\diamond$).
		\label{PlotBiasThetamu2andpC01LOR_KDsigma01}}
\end{figure}
\begin{figure}[t]
	\centering
	\includegraphics[scale=0.33]{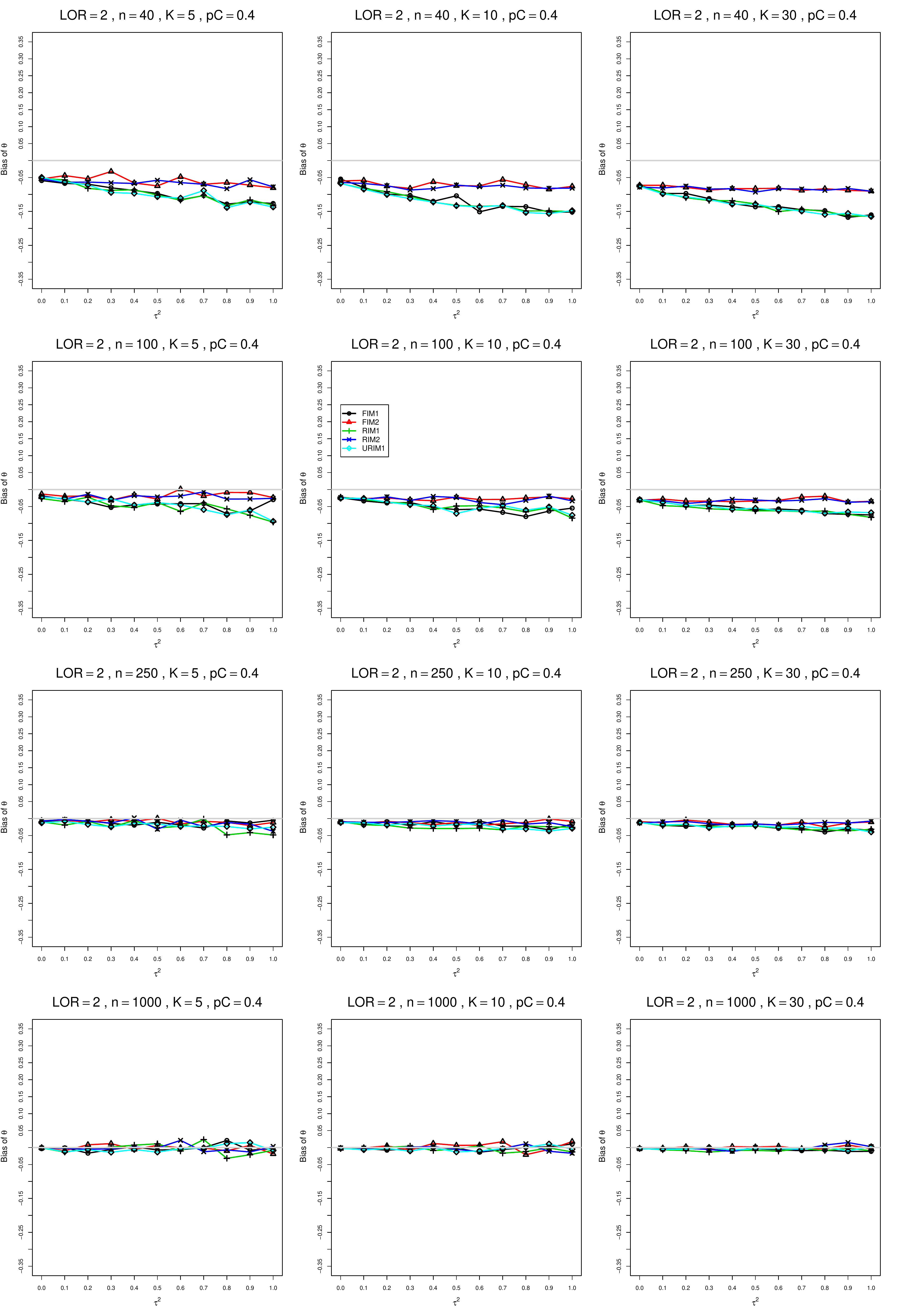}
	\caption{Bias of  overall log-odds ratio $\hat{\theta}_{KD}$ for $\theta=2$, $p_{C}=0.4$, $\sigma^2=0.1$, constant sample sizes $n=40,\;100,\;250,\;1000$.
The data-generation mechanisms are FIM1 ($\circ$), FIM2 ($\triangle$), RIM1 (+), RIM2 ($\times$), and URIM1 ($\diamond$).
		\label{PlotBiasThetamu2andpC04LOR_KDsigma01}}
\end{figure}
\begin{figure}[t]
	\centering
	\includegraphics[scale=0.33]{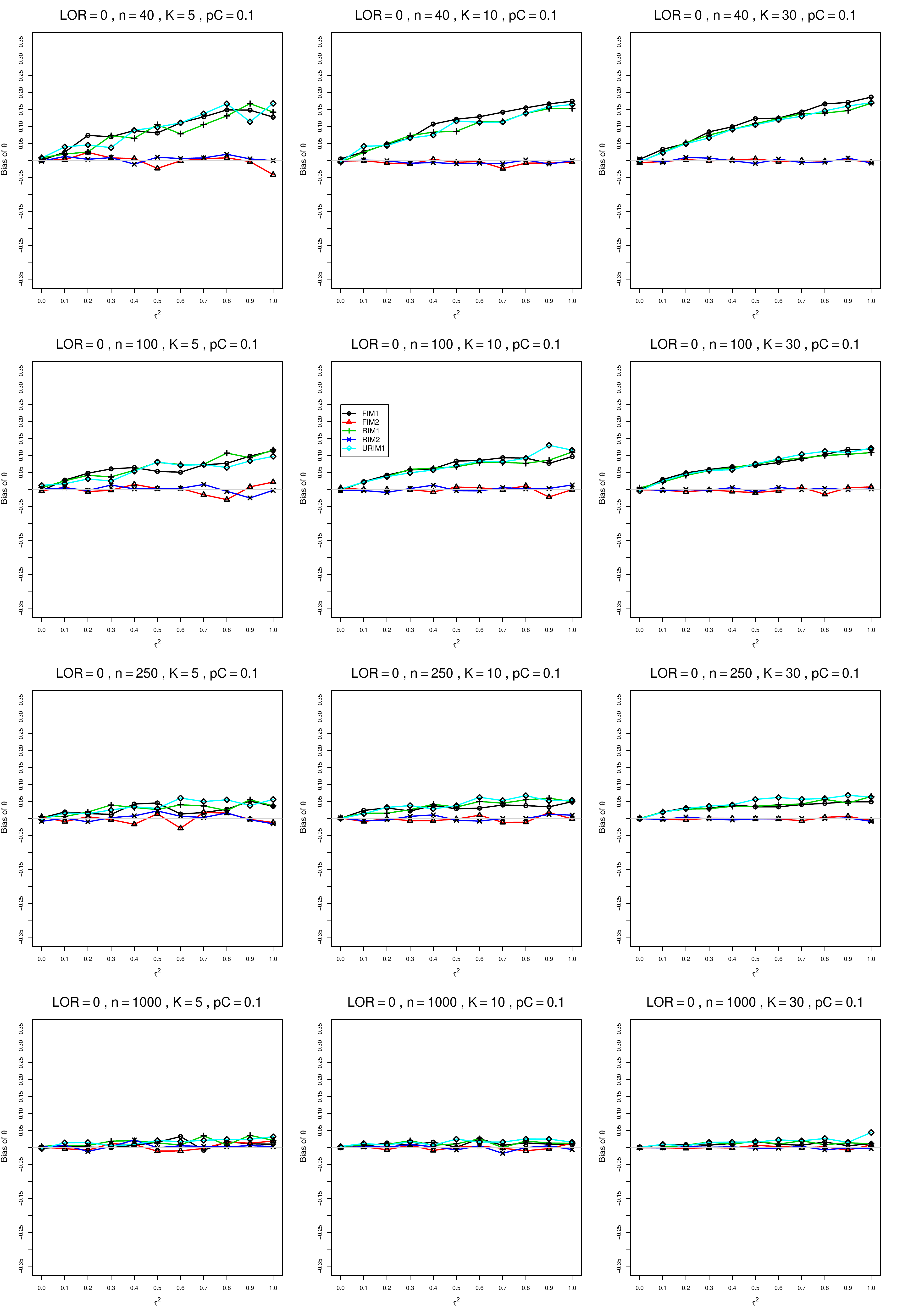}
	\caption{Bias of  overall log-odds ratio $\hat{\theta}_{KD}$ for $\theta=0$, $p_{C}=0.1$, $\sigma^2=0.4$, constant sample sizes $n=40,\;100,\;250,\;1000$.
The data-generation mechanisms are FIM1 ($\circ$), FIM2 ($\triangle$), RIM1 (+), RIM2 ($\times$), and URIM1 ($\diamond$).
		\label{PlotBiasThetamu0andpC01LOR_KDsigma04}}
\end{figure}
\begin{figure}[t]
	\centering
	\includegraphics[scale=0.33]{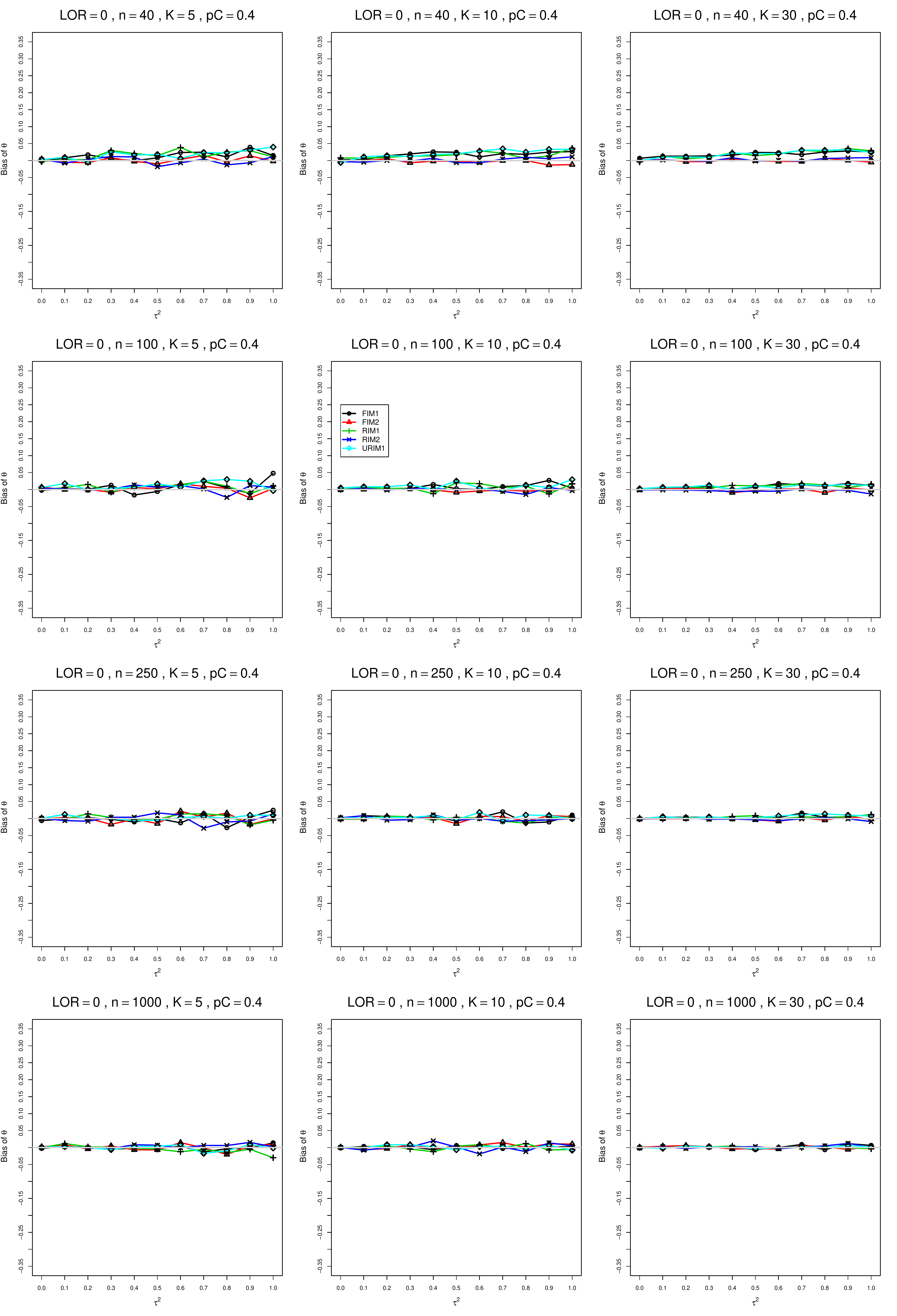}
	\caption{Bias of  overall log-odds ratio $\hat{\theta}_{KD}$ for $\theta=0$, $p_{C}=0.4$, $\sigma^2=0.4$, constant sample sizes $n=40,\;100,\;250,\;1000$.
The data-generation mechanisms are FIM1 ($\circ$), FIM2 ($\triangle$), RIM1 (+), RIM2 ($\times$), and URIM1 ($\diamond$).
		\label{PlotBiasThetamu0andpC04LOR_KDsigma04}}
\end{figure}
\begin{figure}[t]
	\centering
	\includegraphics[scale=0.33]{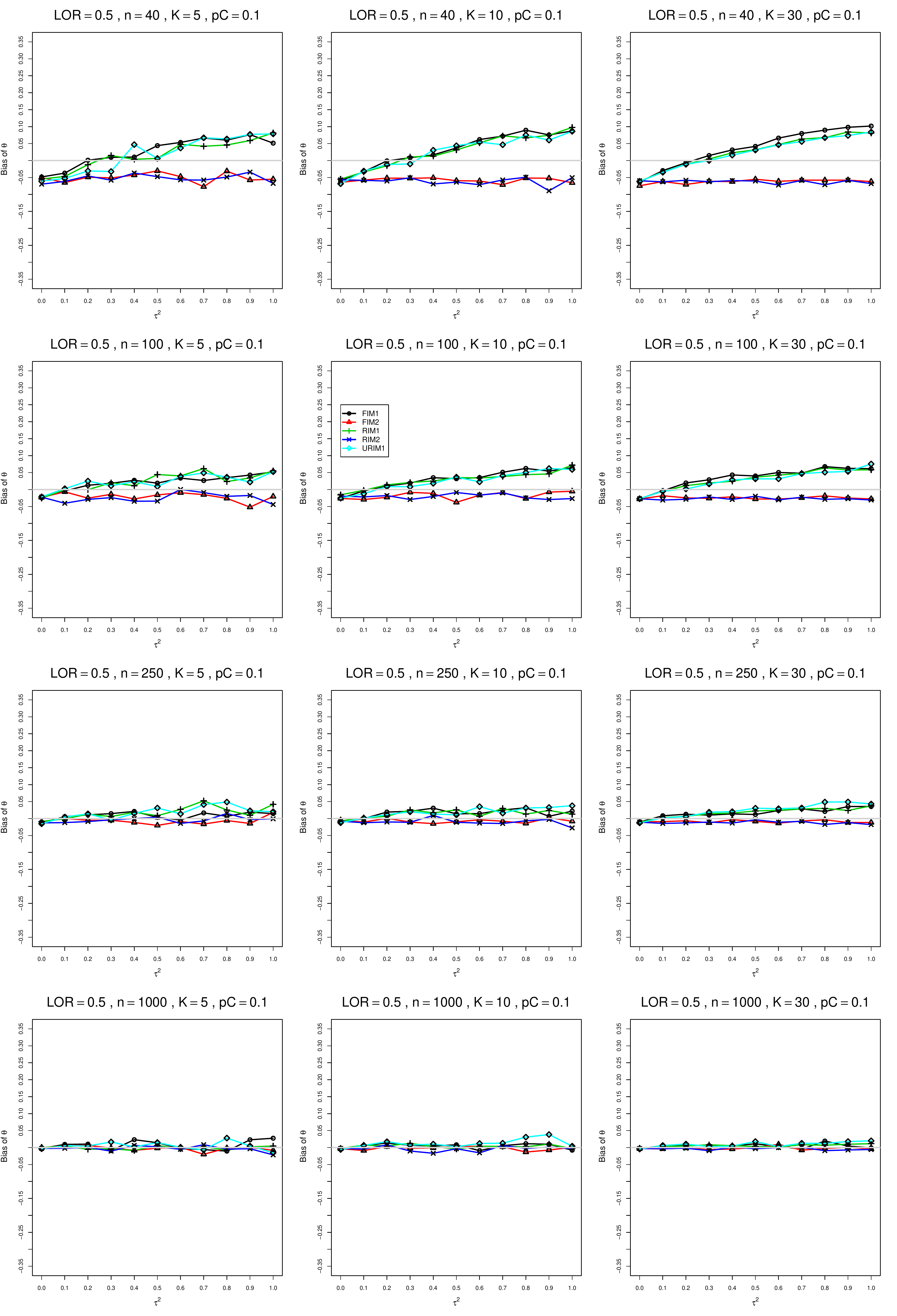}
	\caption{Bias of  overall log-odds ratio $\hat{\theta}_{KD}$ for $\theta=0.5$, $p_{C}=0.1$, $\sigma^2=0.4$, constant sample sizes $n=40,\;100,\;250,\;1000$.
The data-generation mechanisms are FIM1 ($\circ$), FIM2 ($\triangle$), RIM1 (+), RIM2 ($\times$), and URIM1 ($\diamond$).
		\label{PlotBiasThetamu05andpC01LOR_KDsigma04}}
\end{figure}
\begin{figure}[t]
	\centering
	\includegraphics[scale=0.33]{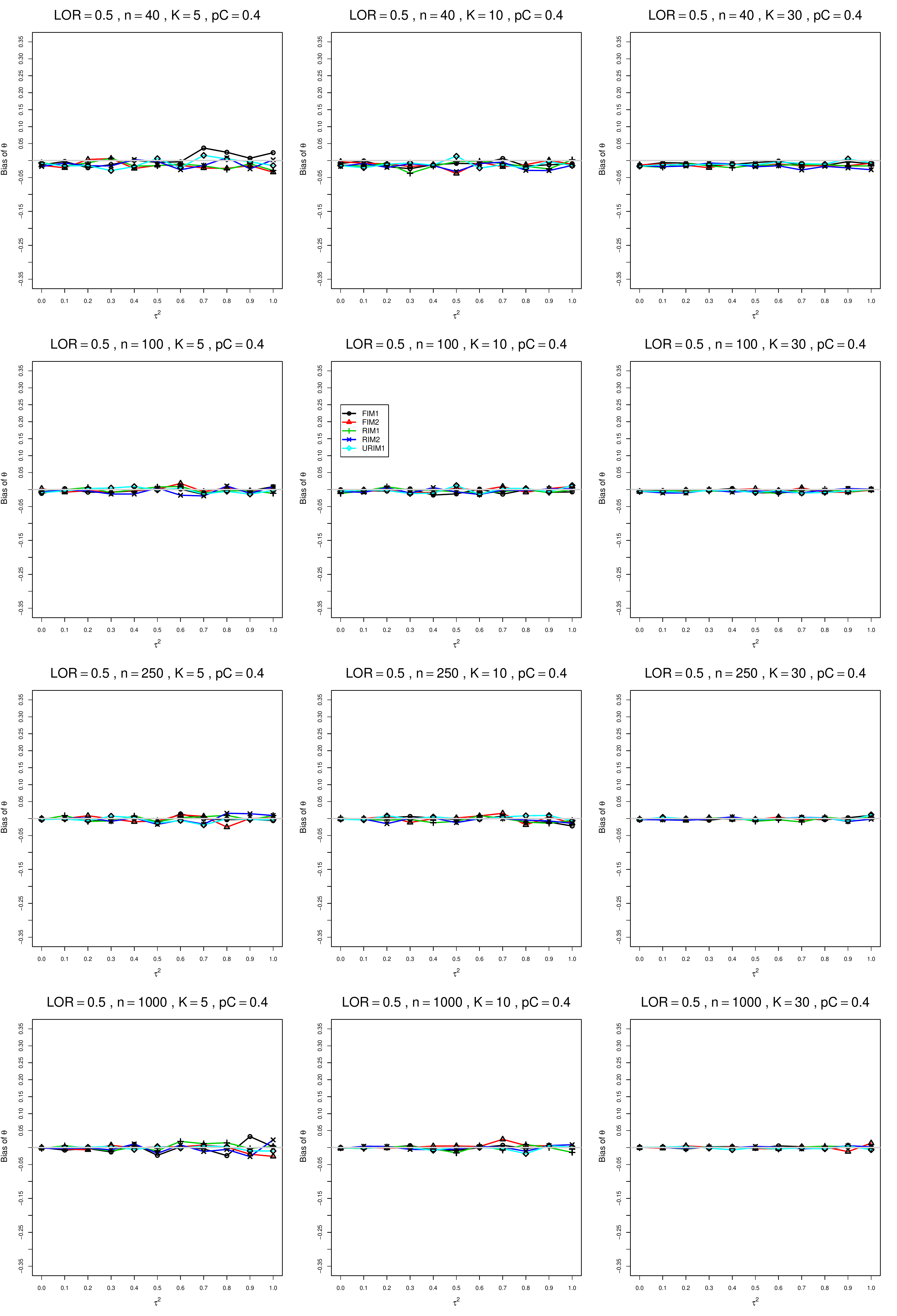}
	\caption{Bias of  overall log-odds ratio $\hat{\theta}_{KD}$ for $\theta=0.5$, $p_{C}=0.4$, $\sigma^2=0.4$, constant sample sizes $n=40,\;100,\;250,\;1000$.
The data-generation mechanisms are FIM1 ($\circ$), FIM2 ($\triangle$), RIM1 (+), RIM2 ($\times$), and URIM1 ($\diamond$).
		\label{PlotBiasThetamu05andpC04LOR_KDsigma04}}
\end{figure}
\begin{figure}[t]
	\centering
	\includegraphics[scale=0.33]{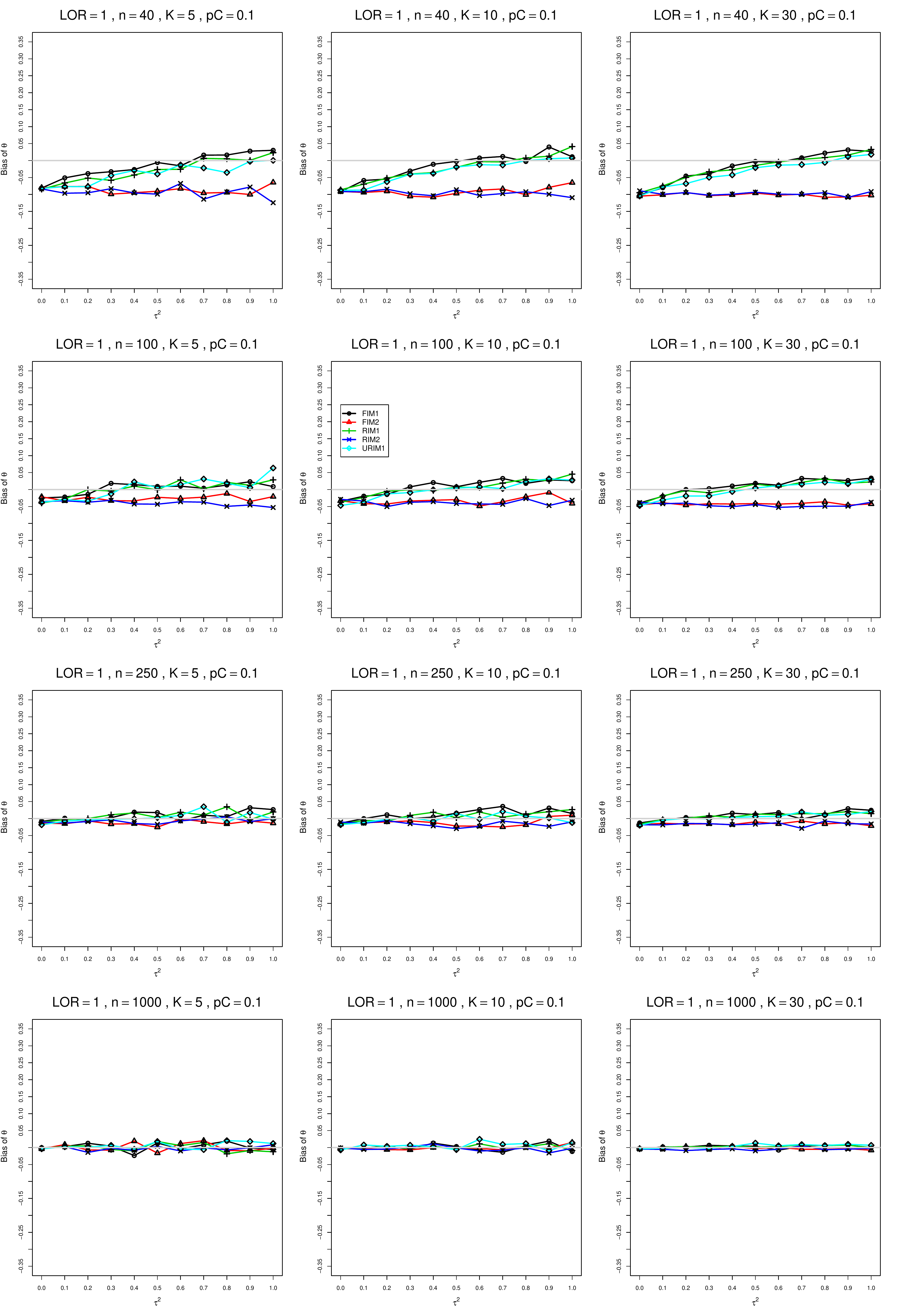}
	\caption{Bias of  overall log-odds ratio $\hat{\theta}_{KD}$ for $\theta=1$, $p_{C}=0.1$, $\sigma^2=0.4$, constant sample sizes $n=40,\;100,\;250,\;1000$.
The data-generation mechanisms are FIM1 ($\circ$), FIM2 ($\triangle$), RIM1 (+), RIM2 ($\times$), and URIM1 ($\diamond$).
		\label{PlotBiasThetamu1andpC01LOR_KDsigma04}}
\end{figure}
\begin{figure}[t]
	\centering
	\includegraphics[scale=0.33]{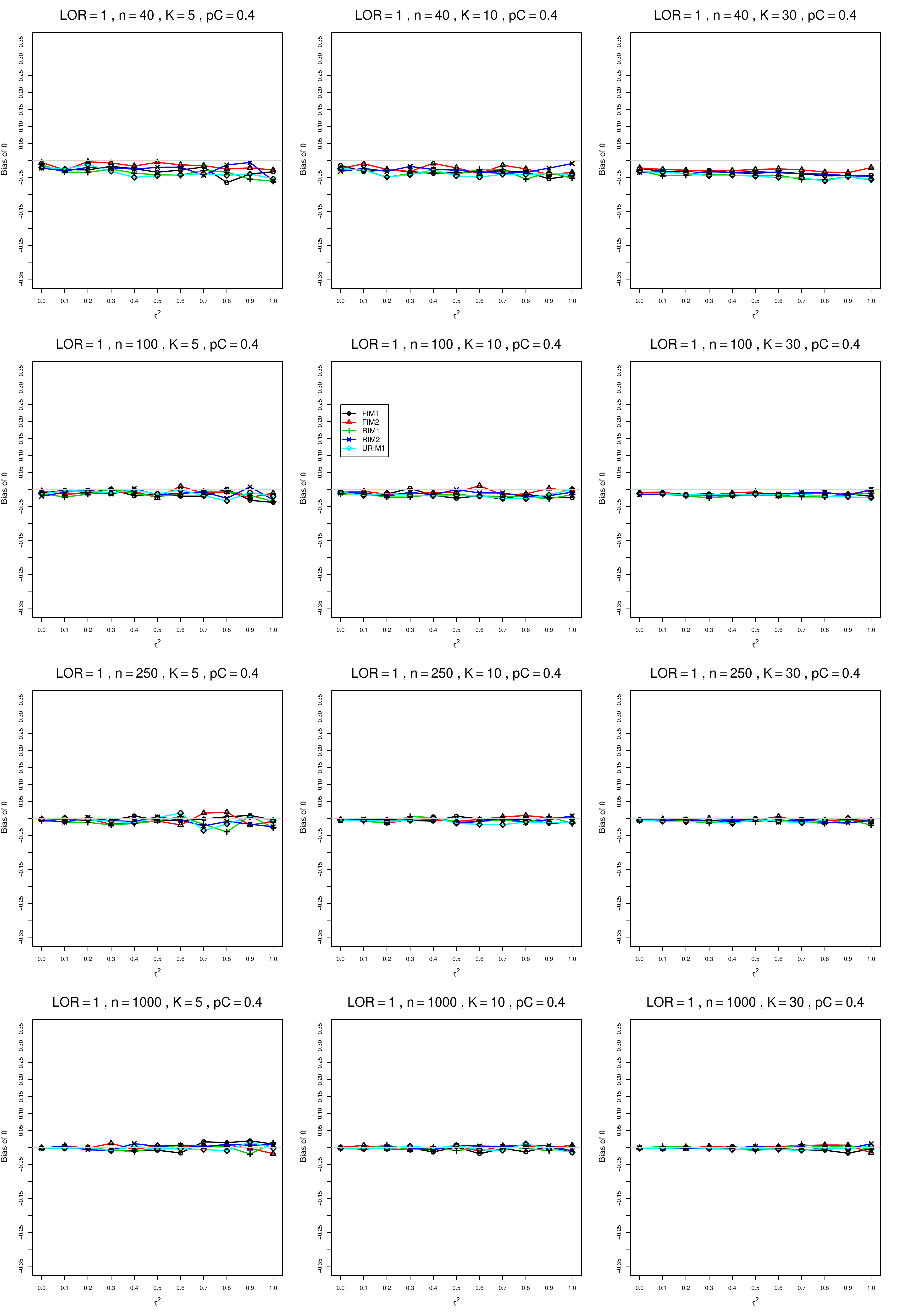}
	\caption{Bias of  overall log-odds ratio $\hat{\theta}_{KD}$ for $\theta=1$, $p_{C}=0.4$, $\sigma^2=0.4$, constant sample sizes $n=40,\;100,\;250,\;1000$.
The data-generation mechanisms are FIM1 ($\circ$), FIM2 ($\triangle$), RIM1 (+), RIM2 ($\times$), and URIM1 ($\diamond$).
		\label{PlotBiasThetamu1andpC04LOR_KDsigma04}}
\end{figure}
\begin{figure}[t]
	\centering
	\includegraphics[scale=0.33]{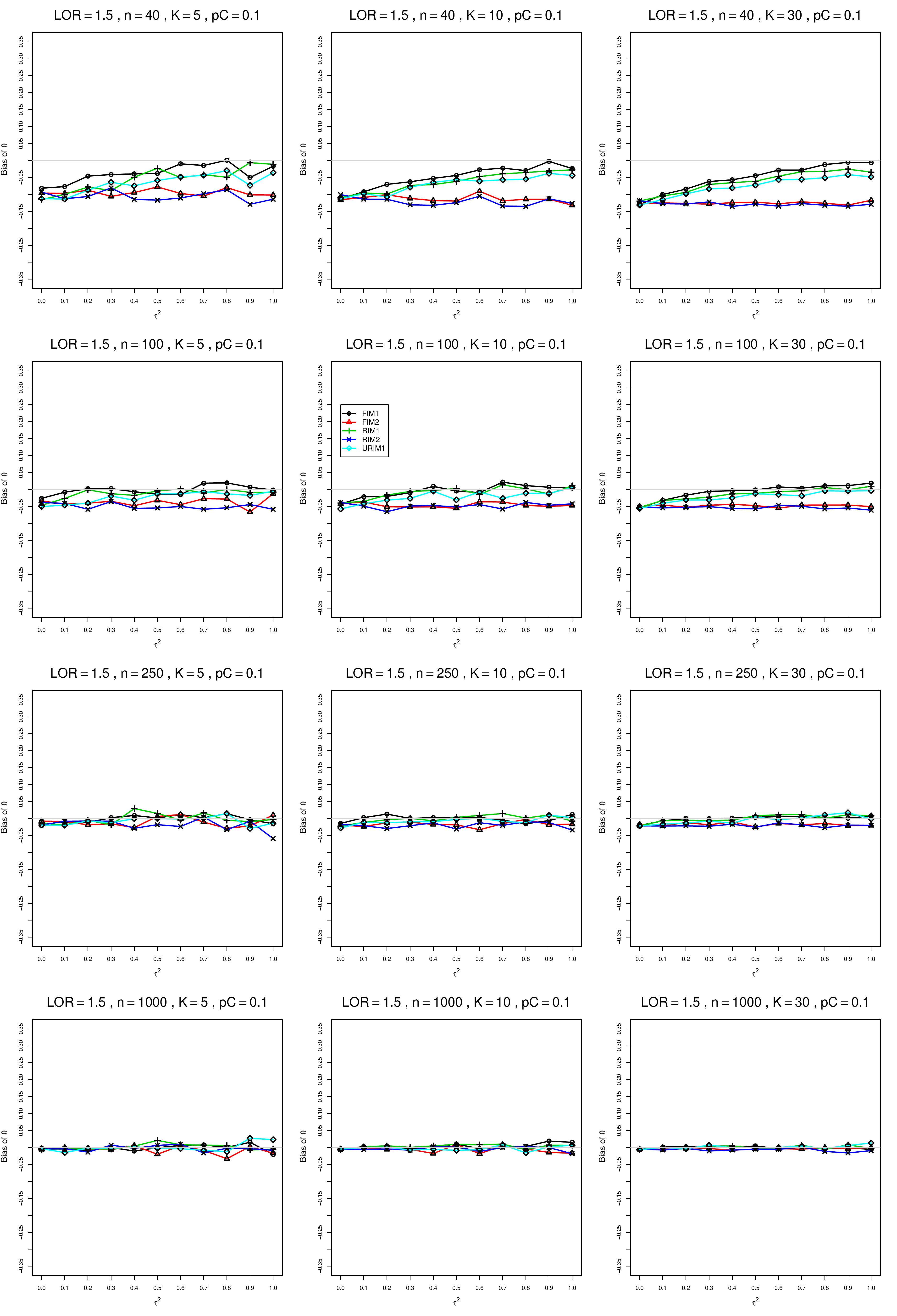}
	\caption{Bias of  overall log-odds ratio $\hat{\theta}_{KD}$ for $\theta=1.5$, $p_{C}=0.1$, $\sigma^2=0.4$, constant sample sizes $n=40,\;100,\;250,\;1000$.
The data-generation mechanisms are FIM1 ($\circ$), FIM2 ($\triangle$), RIM1 (+), RIM2 ($\times$), and URIM1 ($\diamond$).
		\label{PlotBiasThetamu15andpC01LOR_KDsigma04}}
\end{figure}
\begin{figure}[t]
	\centering
	\includegraphics[scale=0.33]{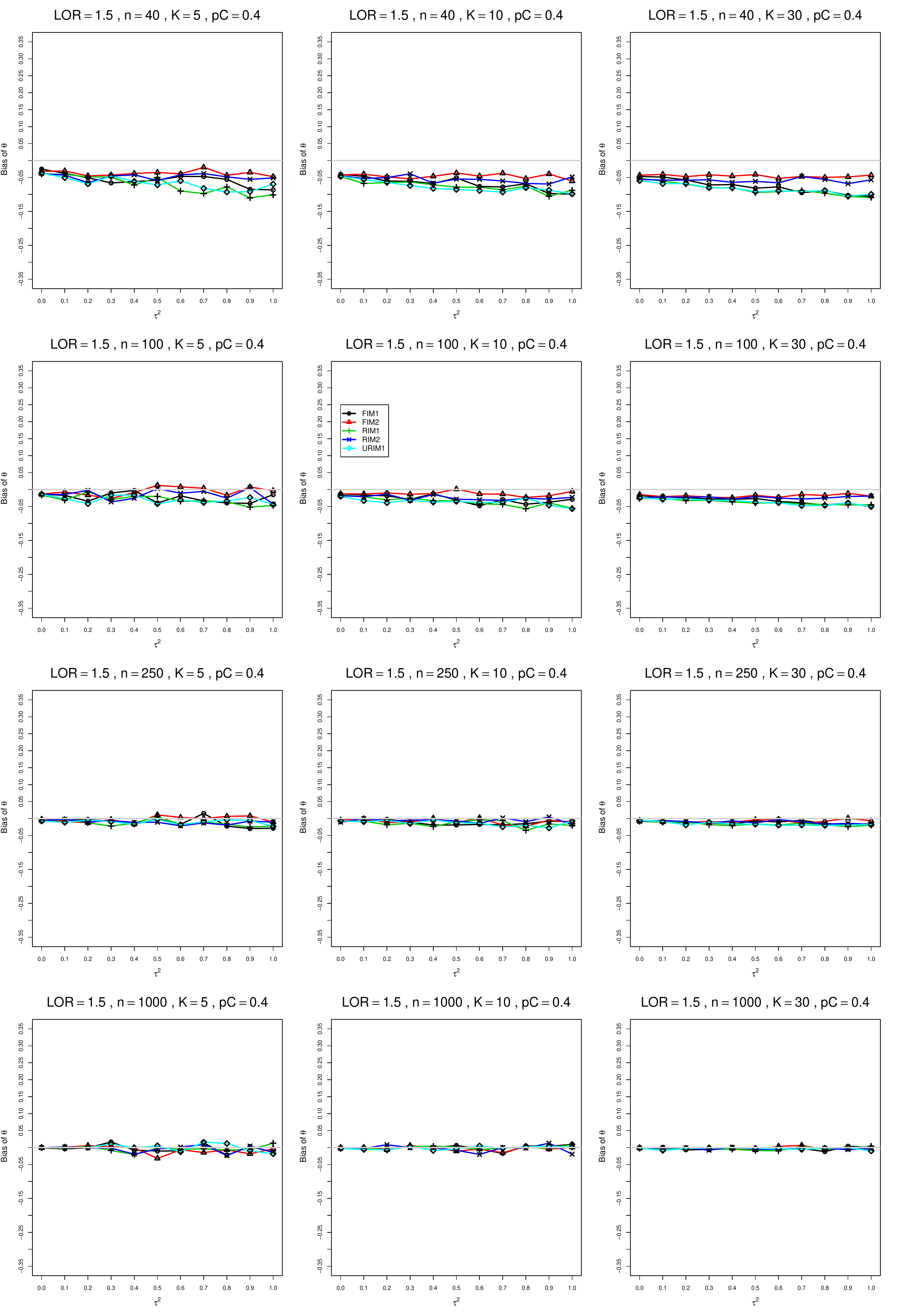}
	\caption{Bias of  overall log-odds ratio $\hat{\theta}_{KD}$ for $\theta=1.5$, $p_{C}=0.4$, $\sigma^2=0.4$, constant sample sizes $n=40,\;100,\;250,\;1000$.
The data-generation mechanisms are FIM1 ($\circ$), FIM2 ($\triangle$), RIM1 (+), RIM2 ($\times$), and URIM1 ($\diamond$).
		\label{PlotBiasThetamu15andpC04LOR_KDsigma04}}
\end{figure}
\begin{figure}[t]
	\centering
	\includegraphics[scale=0.33]{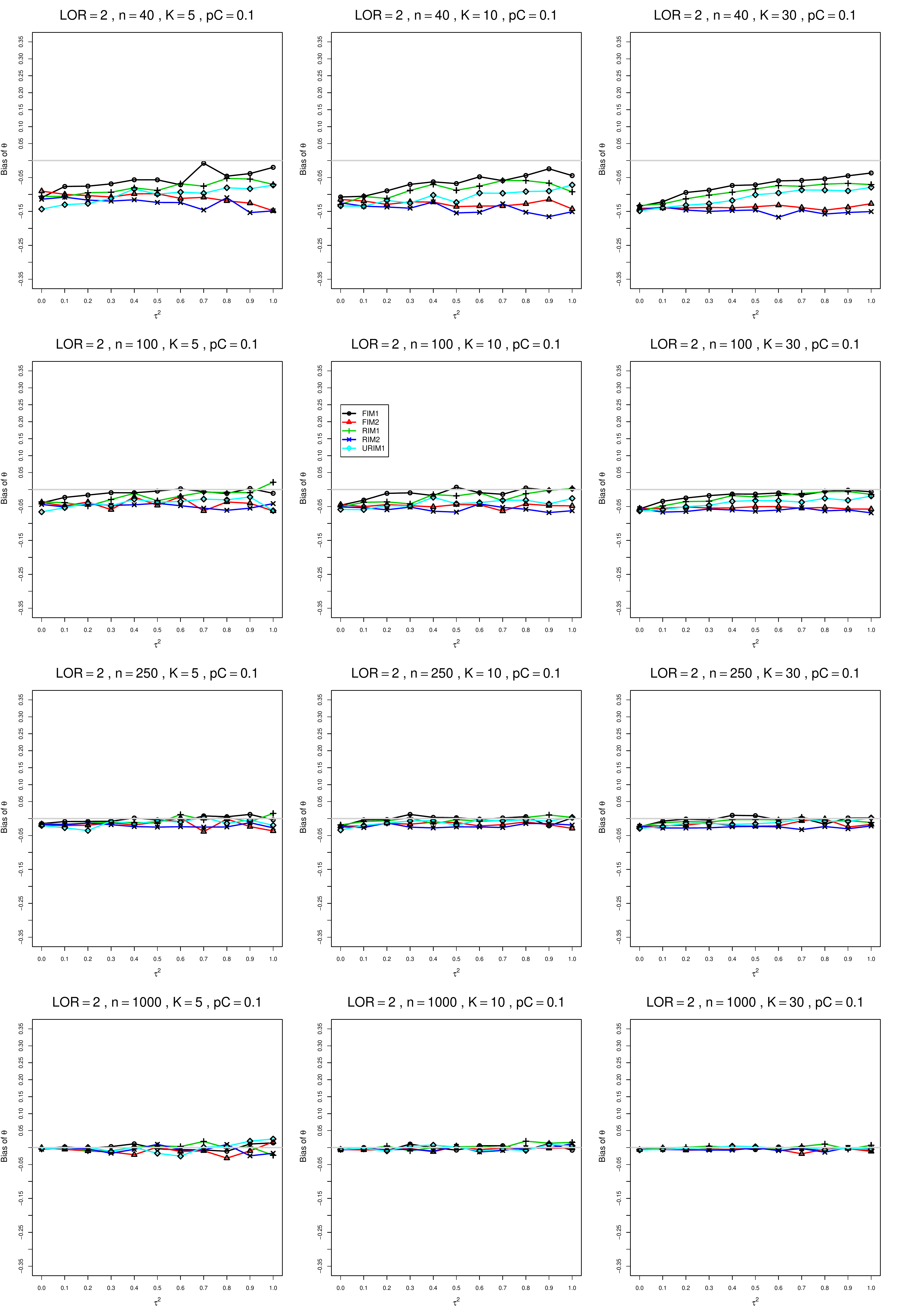}
	\caption{Bias of  overall log-odds ratio $\hat{\theta}_{KD}$ for $\theta=2$, $p_{C}=0.1$, $\sigma^2=0.4$, constant sample sizes $n=40,\;100,\;250,\;1000$.
The data-generation mechanisms are FIM1 ($\circ$), FIM2 ($\triangle$), RIM1 (+), RIM2 ($\times$), and URIM1 ($\diamond$).
		\label{PlotBiasThetamu2andpC01LOR_KDsigma04}}
\end{figure}
\begin{figure}[t]
	\centering
	\includegraphics[scale=0.33]{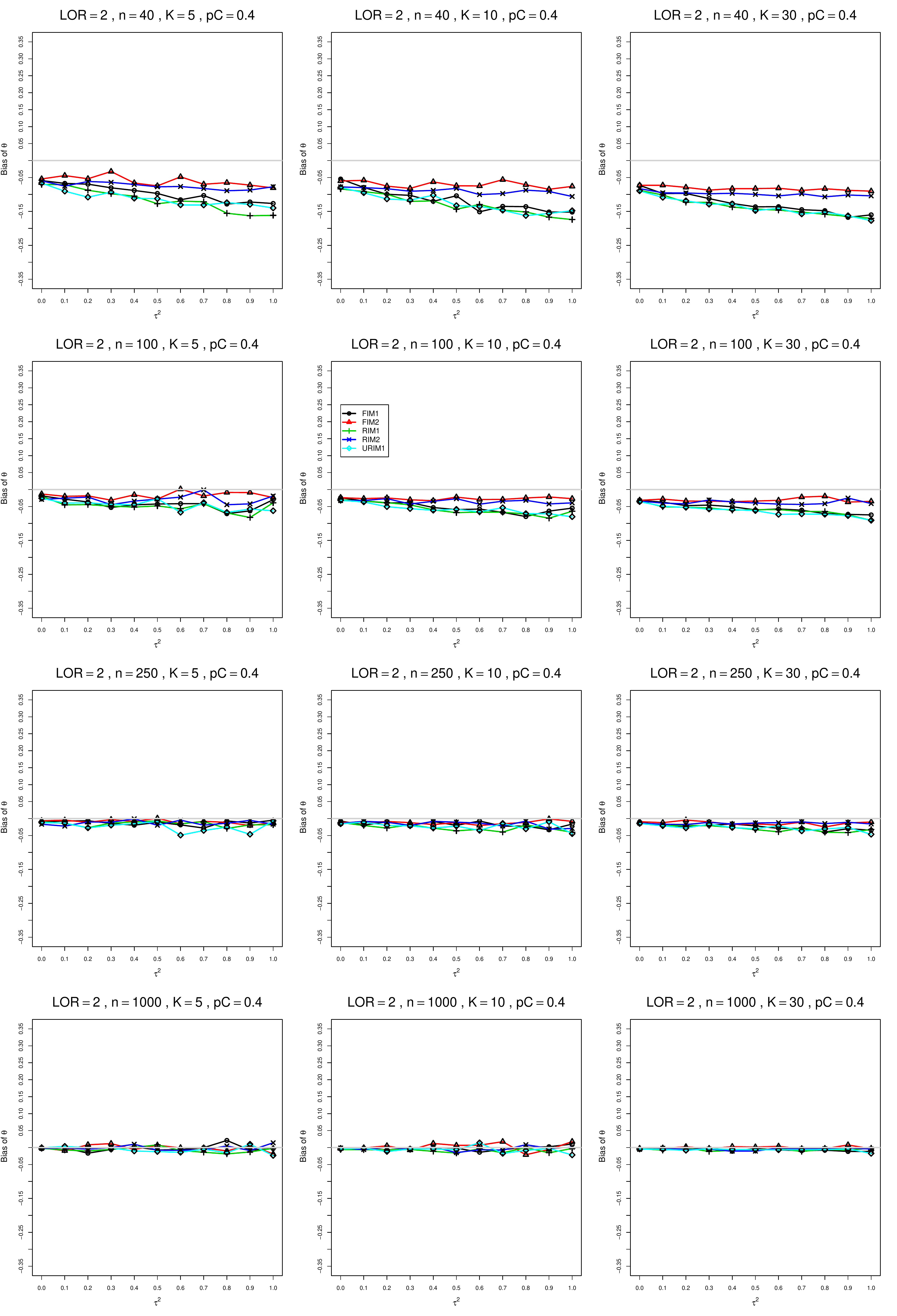}
	\caption{Bias of  overall log-odds ratio $\hat{\theta}_{KD}$ for $\theta=2$, $p_{C}=0.4$, $\sigma^2=0.4$, constant sample sizes $n=40,\;100,\;250,\;1000$.
The data-generation mechanisms are FIM1 ($\circ$), FIM2 ($\triangle$), RIM1 (+), RIM2 ($\times$), and URIM1 ($\diamond$).
		\label{PlotBiasThetamu2andpC04LOR_KDsigma04}}
\end{figure}

\clearpage
\subsection*{A2.5 Bias of $\hat{\theta}_{FIM2}$}
\renewcommand{\thefigure}{A2.5.\arabic{figure}}
\setcounter{figure}{0}

\begin{figure}[t]
	\centering
	\includegraphics[scale=0.33]{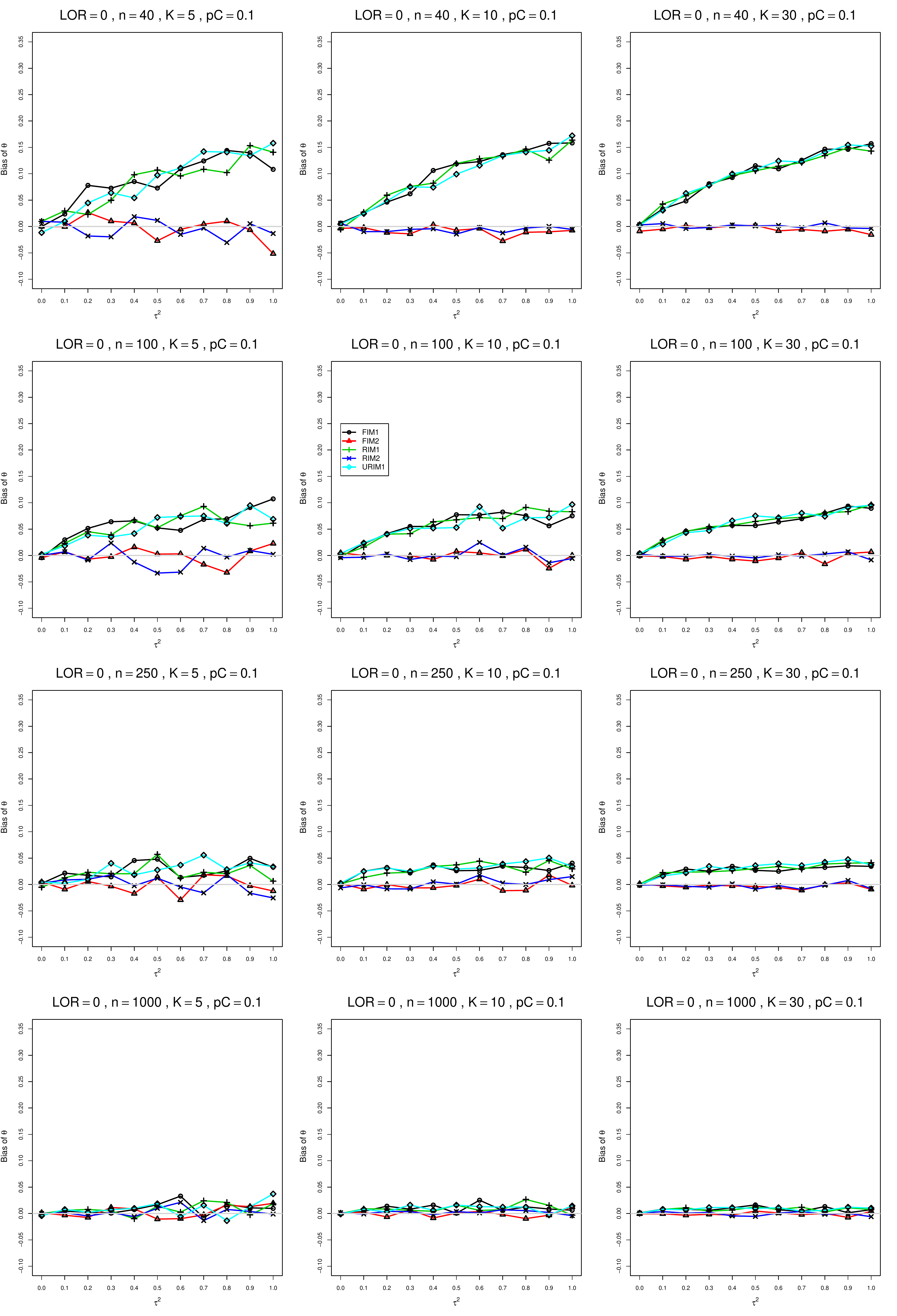}
	\caption{Bias of  overall log-odds ratio $\hat{\theta}_{FIM2}$ for $\theta=0$, $p_{C}=0.1$, $\sigma^2=0.1$, constant sample sizes $n=40,\;100,\;250,\;1000$.
The data-generation mechanisms are FIM1 ($\circ$), FIM2 ($\triangle$), RIM1 (+), RIM2 ($\times$), and URIM1 ($\diamond$).
		\label{PlotBiasThetamu0andpC01LOR_UMFSsigma01}}
\end{figure}
\begin{figure}[t]
	\centering
	\includegraphics[scale=0.33]{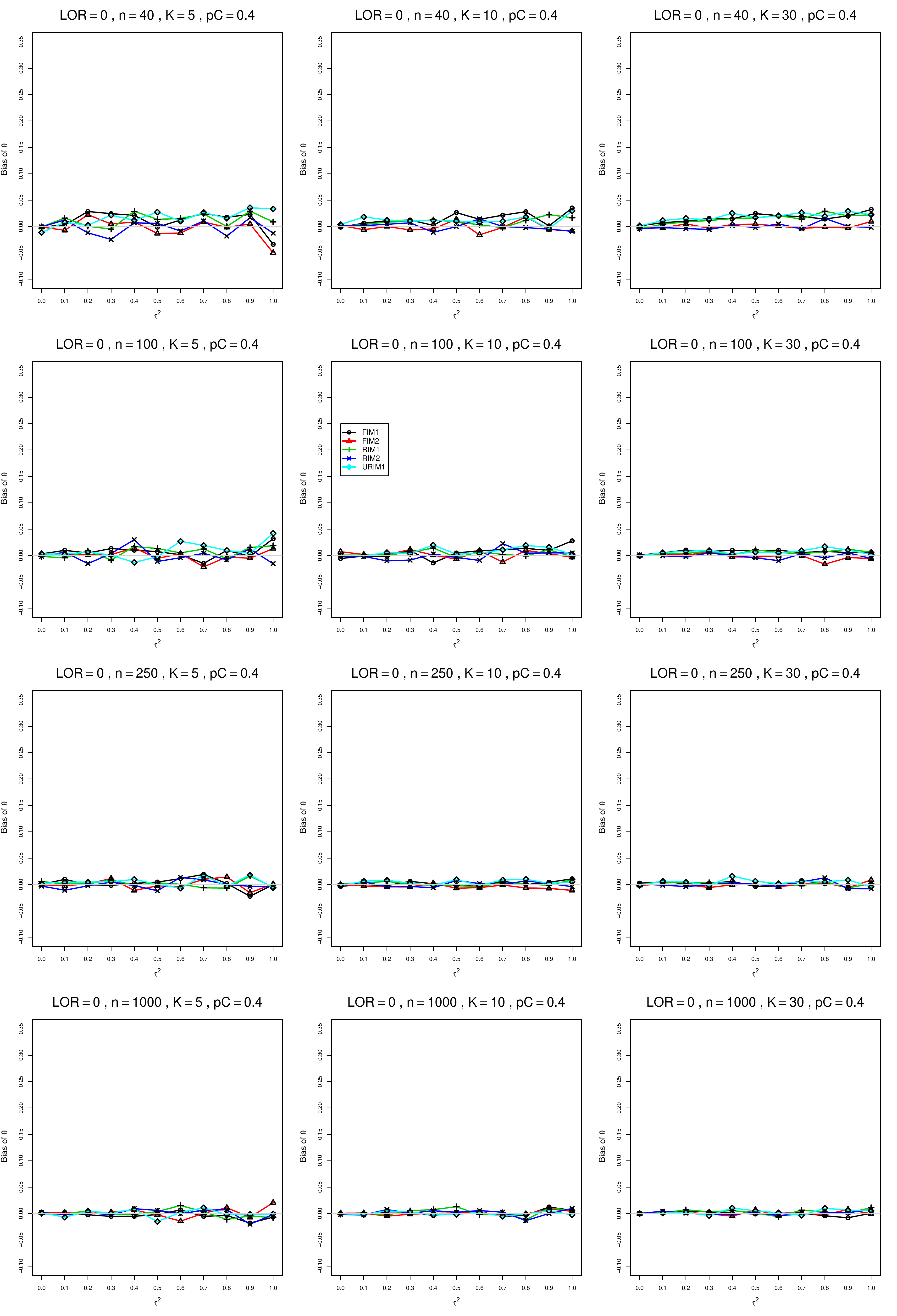}
	\caption{Bias of  overall log-odds ratio $\hat{\theta}_{FIM2}$ for $\theta=0$, $p_{C}=0.4$, $\sigma^2=0.1$, constant sample sizes $n=40,\;100,\;250,\;1000$.
The data-generation mechanisms are FIM1 ($\circ$), FIM2 ($\triangle$), RIM1 (+), RIM2 ($\times$), and URIM1 ($\diamond$).
		\label{PlotBiasThetamu0andpC04LOR_UMFSsigma01}}
\end{figure}
\begin{figure}[t]
	\centering
	\includegraphics[scale=0.33]{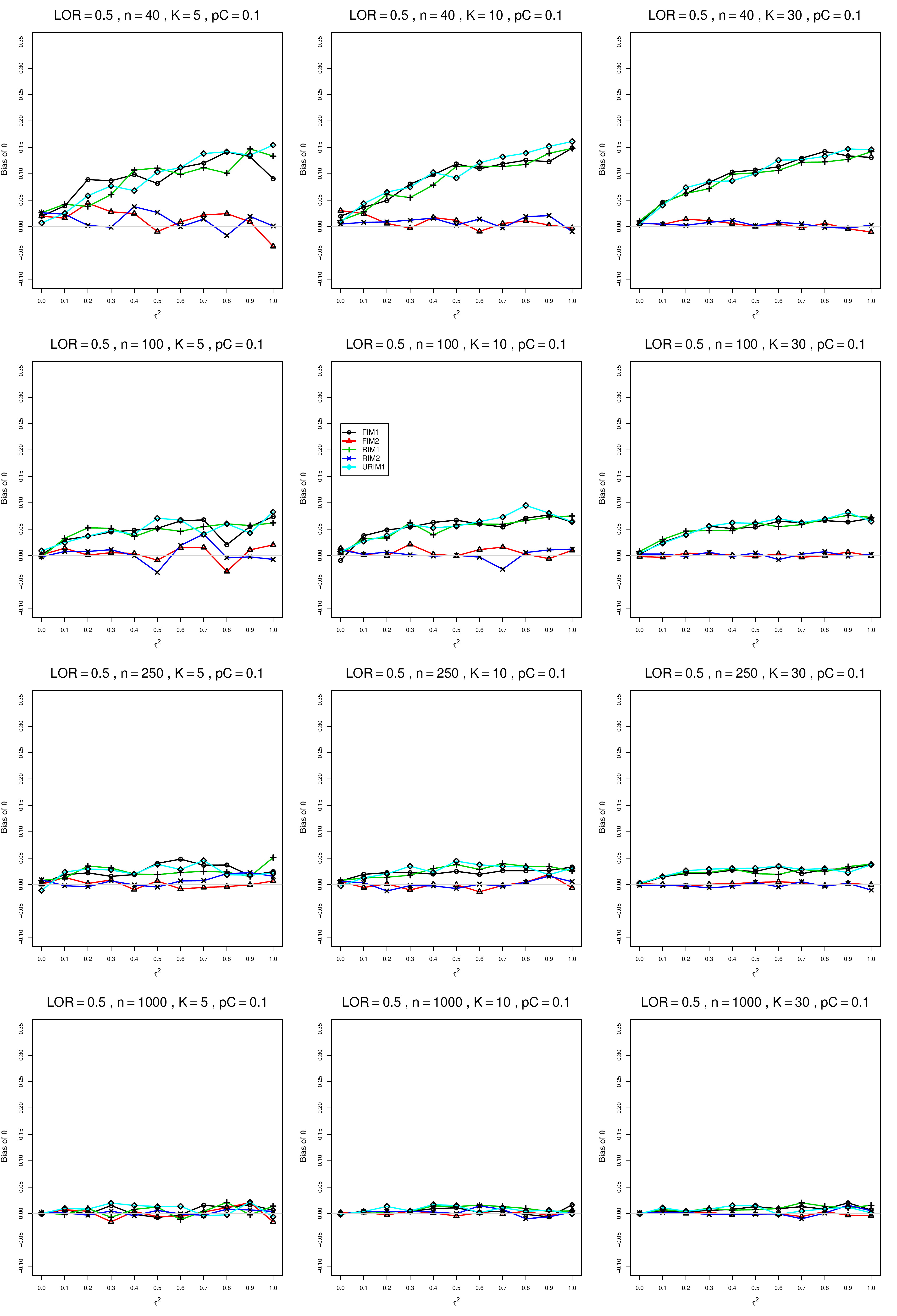}
	\caption{Bias of  overall log-odds ratio $\hat{\theta}_{FIM2}$ for $\theta=0.5$, $p_{C}=0.1$, $\sigma^2=0.1$, constant sample sizes $n=40,\;100,\;250,\;1000$.
The data-generation mechanisms are FIM1 ($\circ$), FIM2 ($\triangle$), RIM1 (+), RIM2 ($\times$), and URIM1 ($\diamond$).
		\label{PlotBiasThetamu05andpC01LOR_UMFSsigma01}}
\end{figure}
\begin{figure}[t]
	\centering
	\includegraphics[scale=0.33]{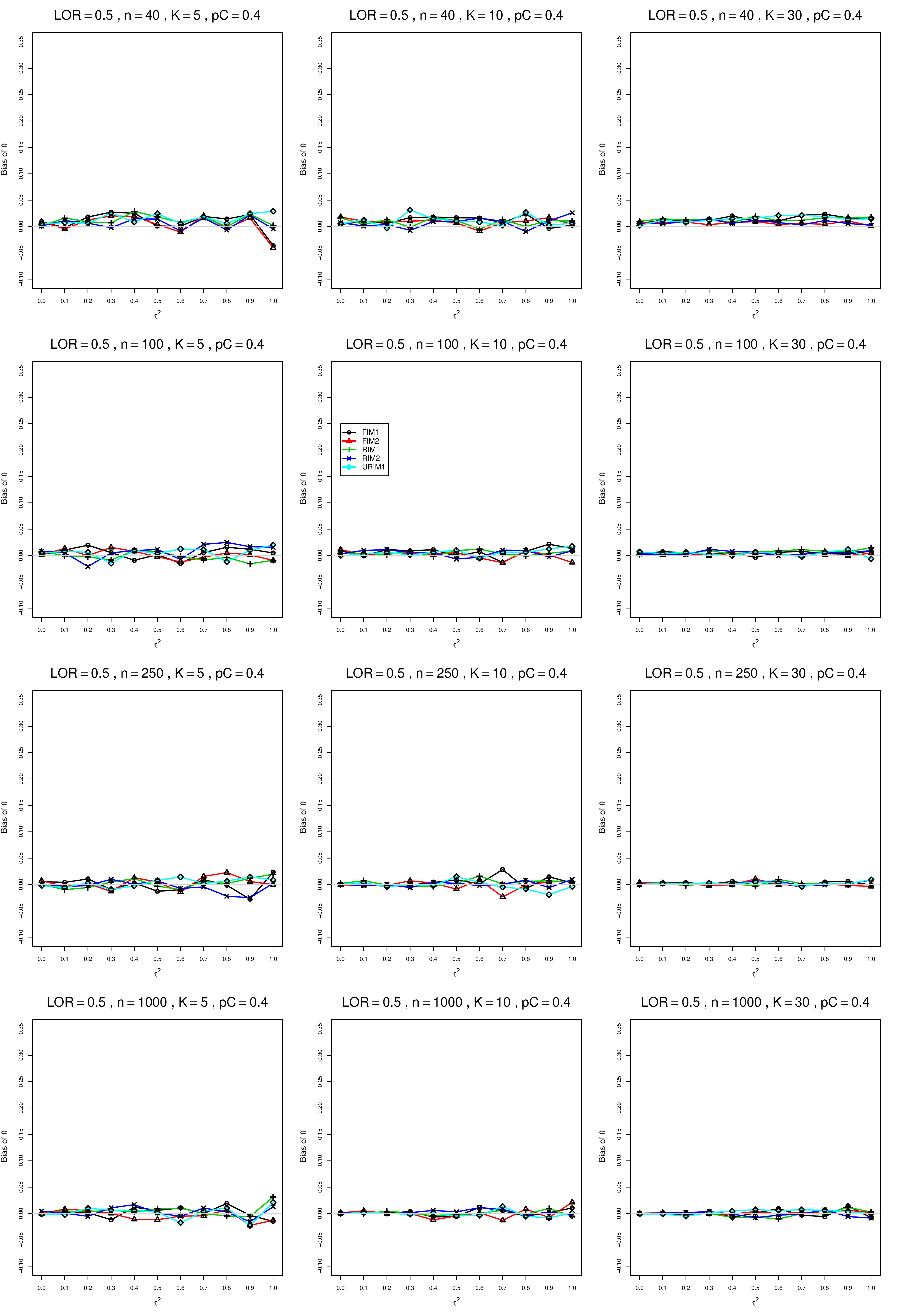}
	\caption{Bias of  overall log-odds ratio $\hat{\theta}_{FIM2}$ for $\theta=0.5$, $p_{C}=0.4$, $\sigma^2=0.1$, constant sample sizes $n=40,\;100,\;250,\;1000$.
The data-generation mechanisms are FIM1 ($\circ$), FIM2 ($\triangle$), RIM1 (+), RIM2 ($\times$), and URIM1 ($\diamond$).
		\label{PlotBiasThetamu05andpC04LOR_UMFSsigma01}}
\end{figure}
\begin{figure}[t]
	\centering
	\includegraphics[scale=0.33]{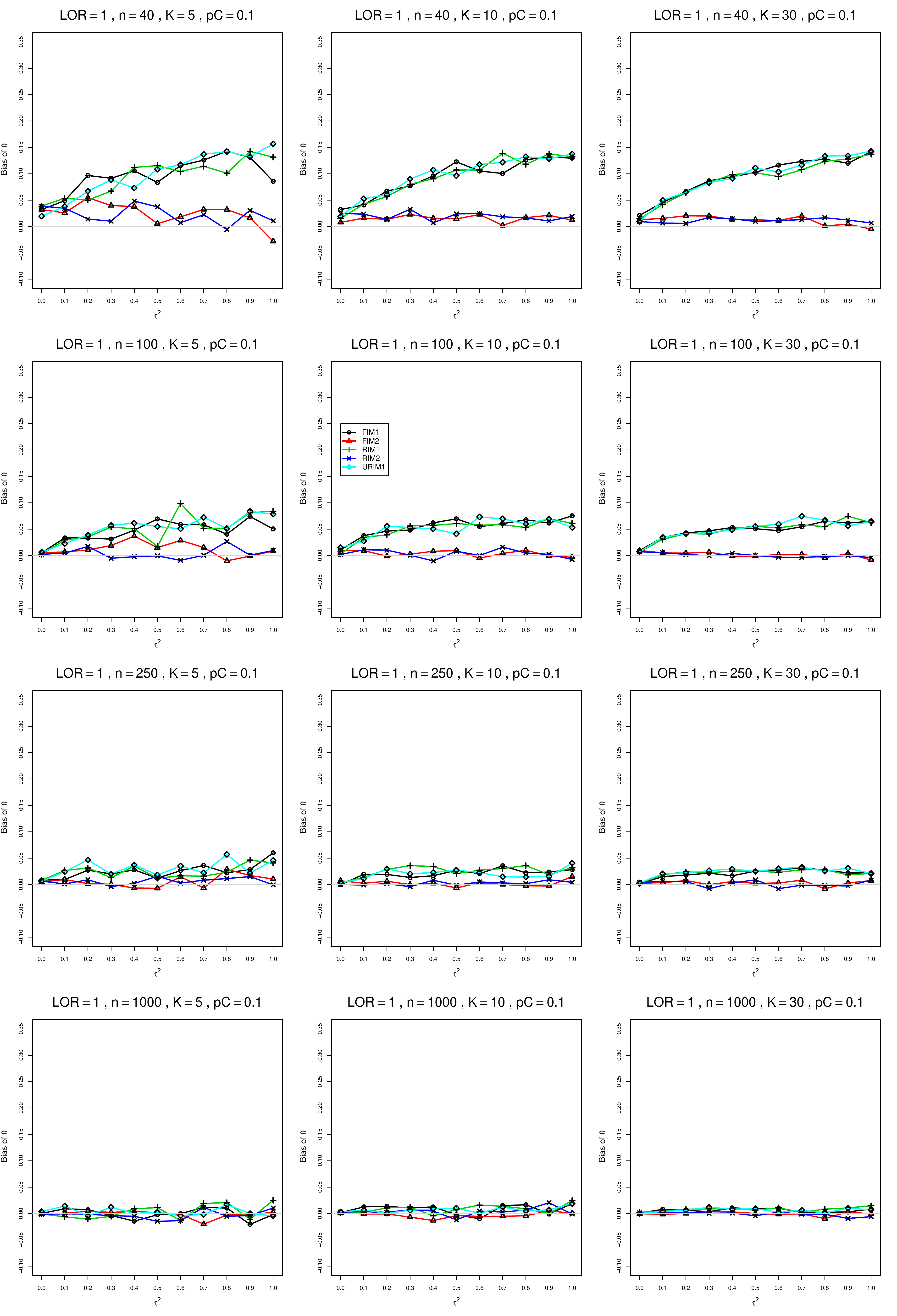}
	\caption{Bias of  overall log-odds ratio $\hat{\theta}_{FIM2}$ for $\theta=1$, $p_{C}=0.1$, $\sigma^2=0.1$, constant sample sizes $n=40,\;100,\;250,\;1000$.
The data-generation mechanisms are FIM1 ($\circ$), FIM2 ($\triangle$), RIM1 (+), RIM2 ($\times$), and URIM1 ($\diamond$).
		\label{PlotBiasThetamu1andpC01LOR_UMFSsigma01}}
\end{figure}
\begin{figure}[t]
	\centering
	\includegraphics[scale=0.33]{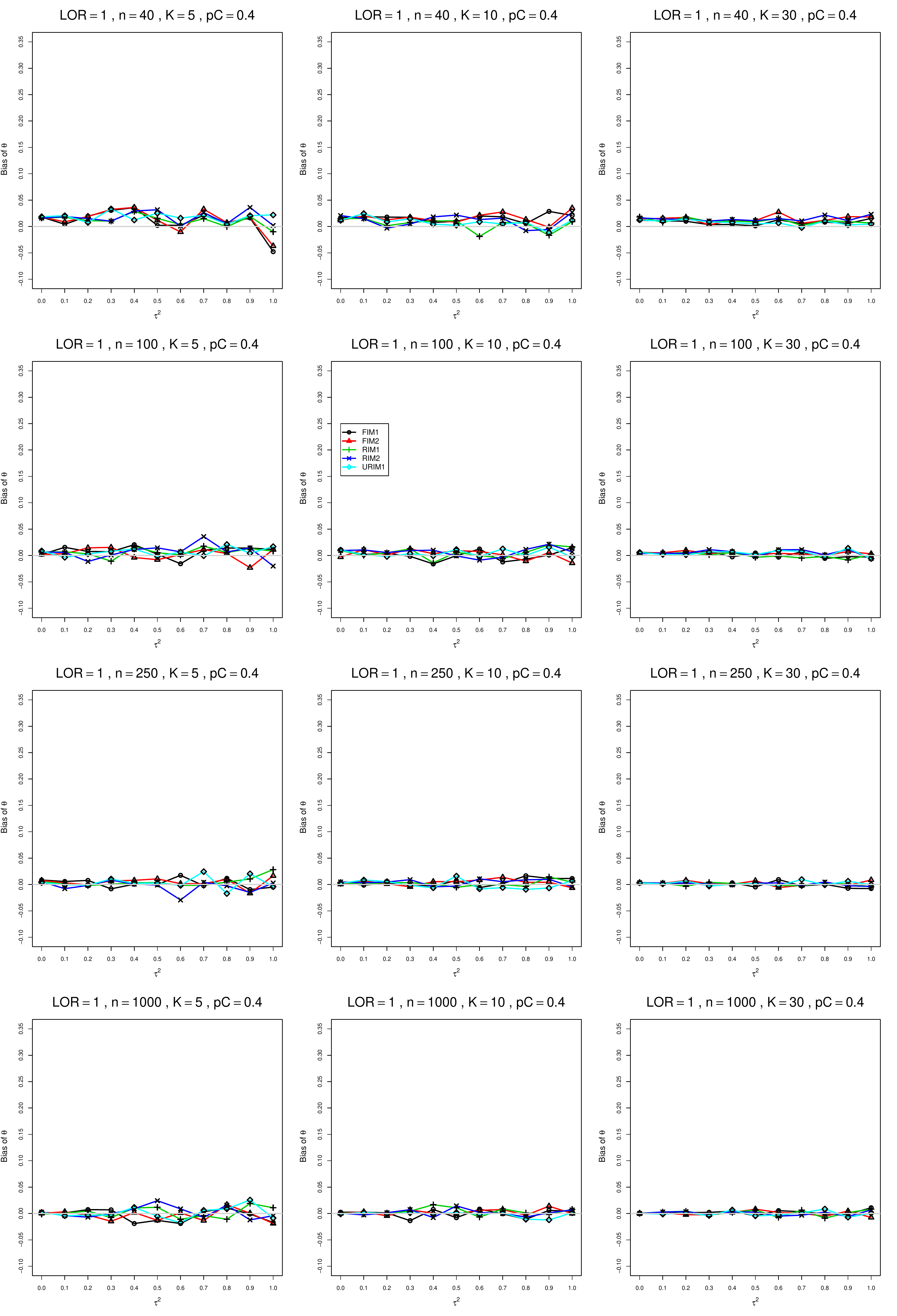}
	\caption{Bias of  overall log-odds ratio $\hat{\theta}_{FIM2}$ for $\theta=1$, $p_{C}=0.4$, $\sigma^2=0.1$, constant sample sizes $n=40,\;100,\;250,\;1000$.
The data-generation mechanisms are FIM1 ($\circ$), FIM2 ($\triangle$), RIM1 (+), RIM2 ($\times$), and URIM1 ($\diamond$).
		\label{PlotBiasThetamu1andpC04LOR_UMFSsigma01}}
\end{figure}
\begin{figure}[t]
	\centering
	\includegraphics[scale=0.33]{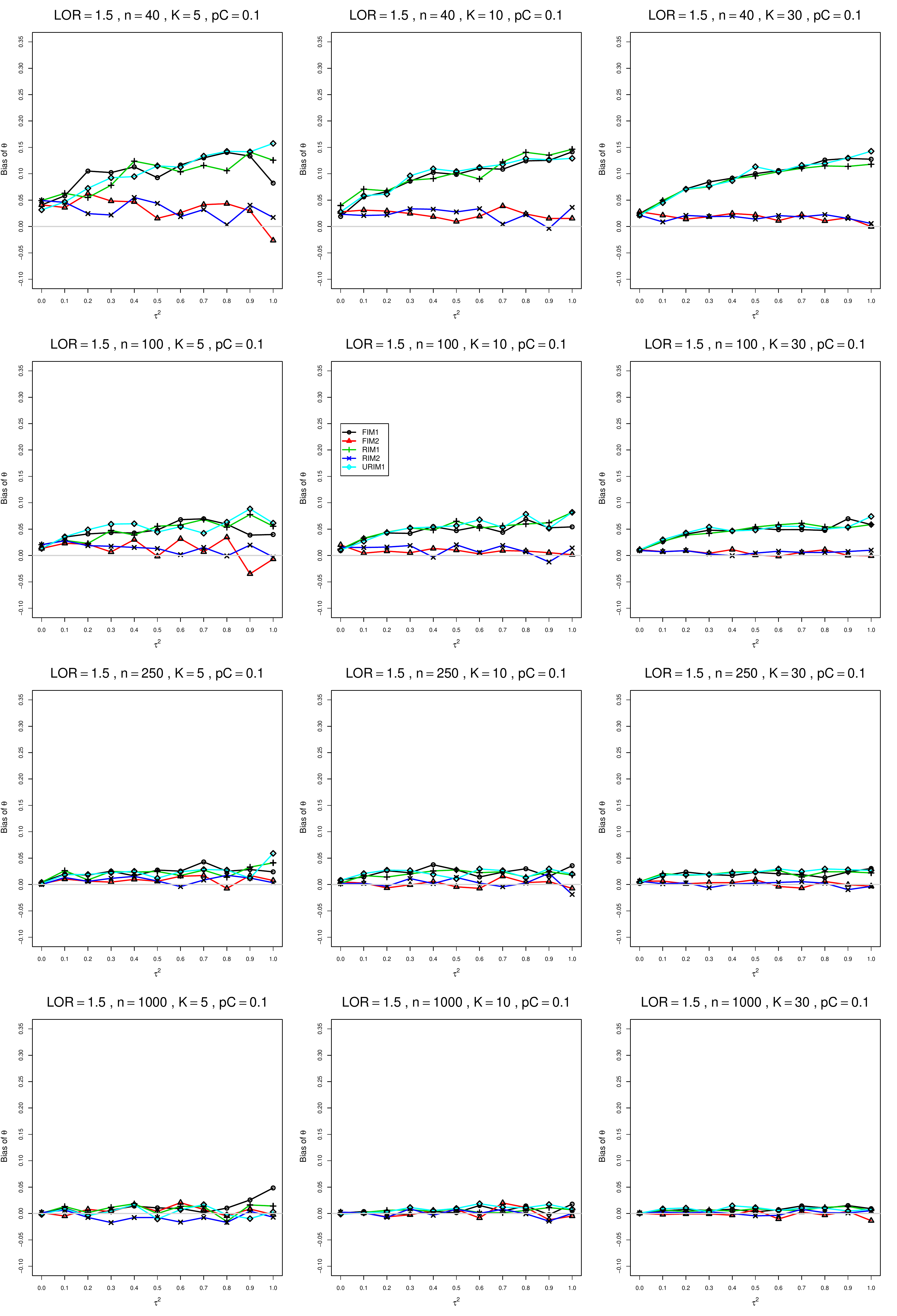}
	\caption{Bias of  overall log-odds ratio $\hat{\theta}_{FIM2}$ for $\theta=1.5$, $p_{C}=0.1$, $\sigma^2=0.1$, constant sample sizes $n=40,\;100,\;250,\;1000$.
The data-generation mechanisms are FIM1 ($\circ$), FIM2 ($\triangle$), RIM1 (+), RIM2 ($\times$), and URIM1 ($\diamond$).
		\label{PlotBiasThetamu15andpC01LOR_UMFSsigma01}}
\end{figure}
\begin{figure}[t]
	\centering
	\includegraphics[scale=0.33]{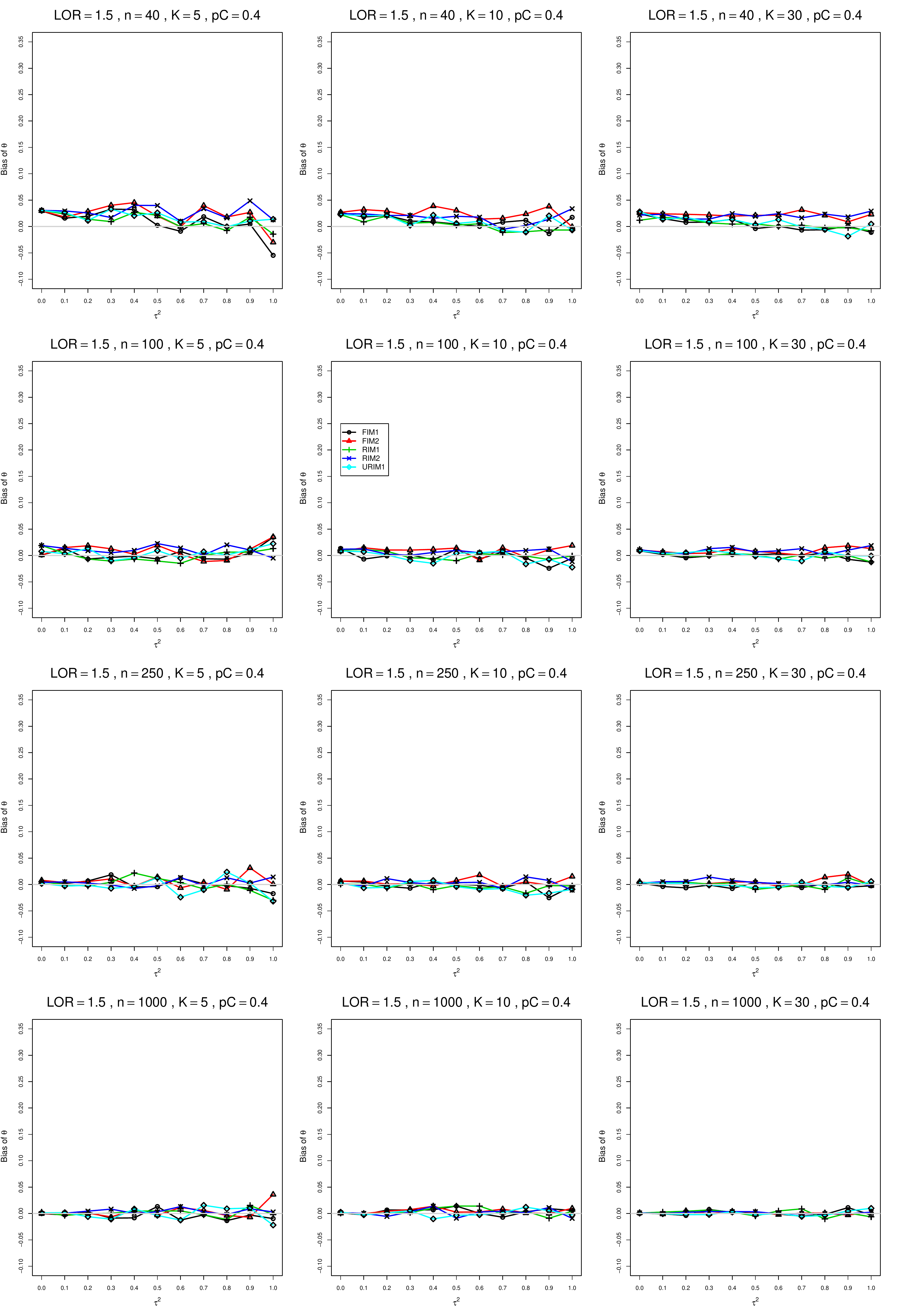}
	\caption{Bias of  overall log-odds ratio $\hat{\theta}_{FIM2}$ for $\theta=1.5$, $p_{C}=0.4$, $\sigma^2=0.1$, constant sample sizes $n=40,\;100,\;250,\;1000$.
The data-generation mechanisms are FIM1 ($\circ$), FIM2 ($\triangle$), RIM1 (+), RIM2 ($\times$), and URIM1 ($\diamond$).
		\label{PlotBiasThetamu15andpC04LOR_UMFSsigma01}}
\end{figure}
\begin{figure}[t]
	\centering
	\includegraphics[scale=0.33]{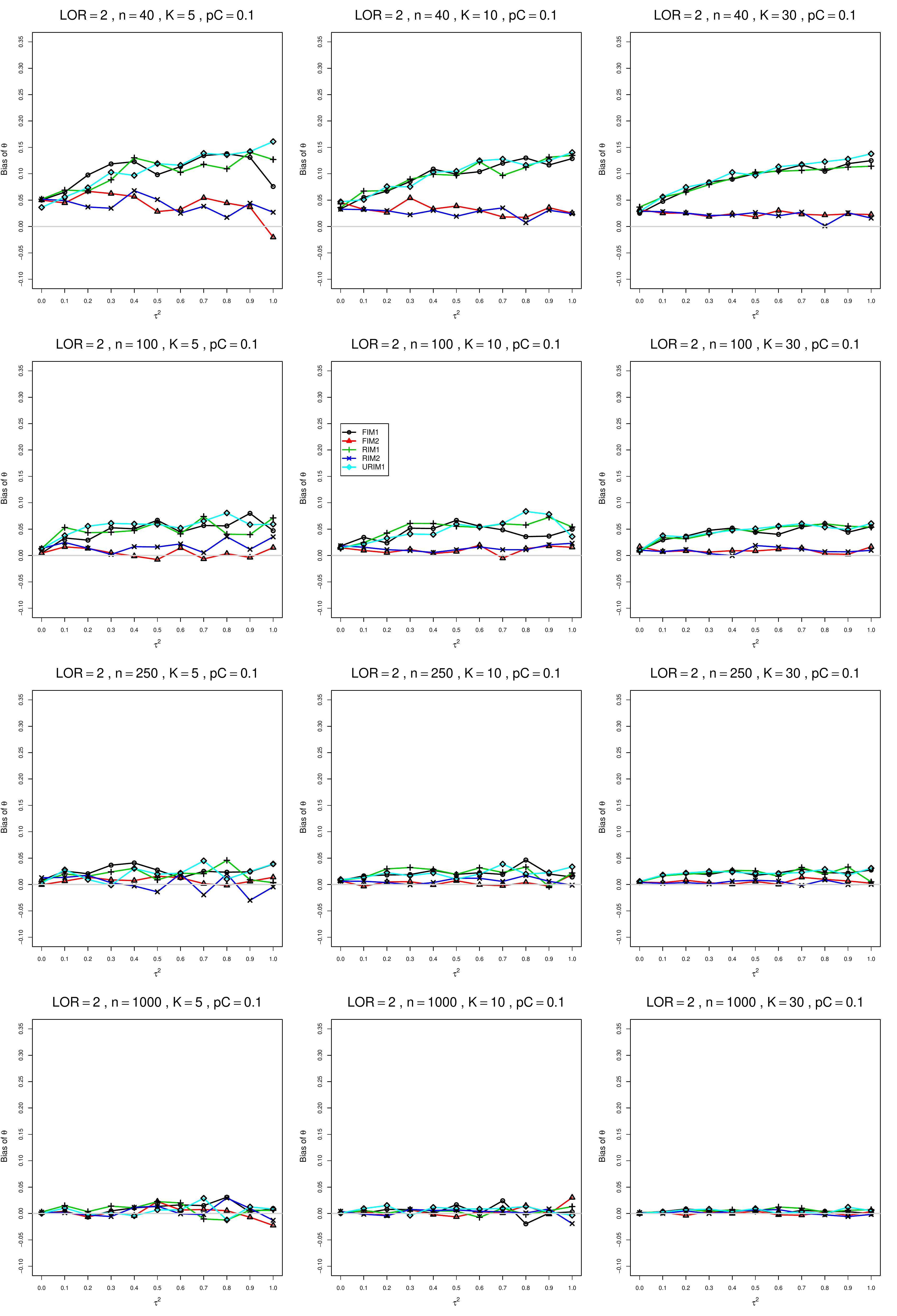}
	\caption{Bias of  overall log-odds ratio $\hat{\theta}_{FIM2}$ for $\theta=2$, $p_{C}=0.1$, $\sigma^2=0.1$, constant sample sizes $n=40,\;100,\;250,\;1000$.
The data-generation mechanisms are FIM1 ($\circ$), FIM2 ($\triangle$), RIM1 (+), RIM2 ($\times$), and URIM1 ($\diamond$).
		\label{PlotBiasThetamu2andpC01LOR_UMFSsigma01}}
\end{figure}
\begin{figure}[t]
	\centering
	\includegraphics[scale=0.33]{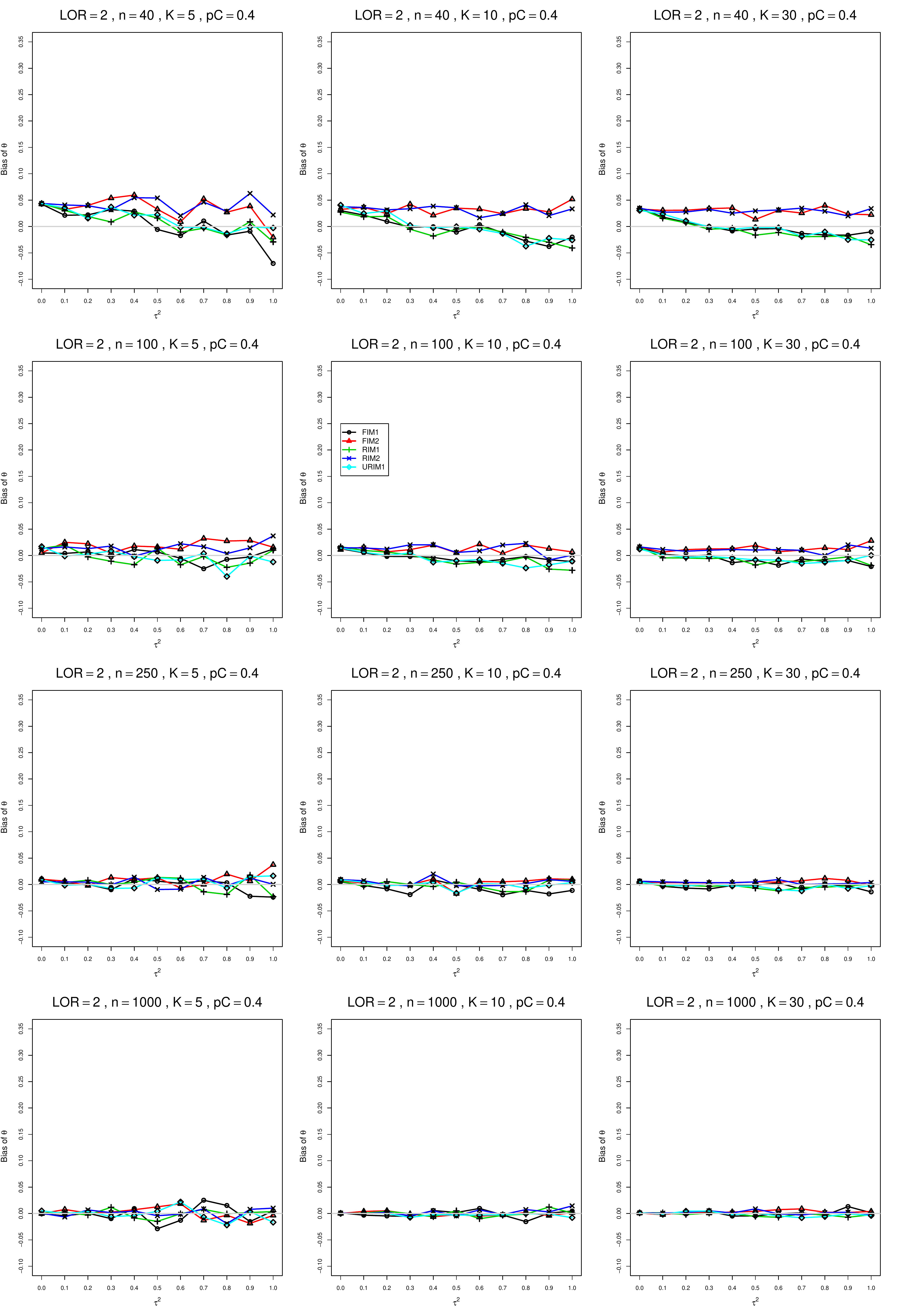}
	\caption{Bias of  overall log-odds ratio $\hat{\theta}_{FIM2}$ for $\theta=2$, $p_{C}=0.4$, $\sigma^2=0.1$, constant sample sizes $n=40,\;100,\;250,\;1000$.
The data-generation mechanisms are FIM1 ($\circ$), FIM2 ($\triangle$), RIM1 (+), RIM2 ($\times$), and URIM1 ($\diamond$).
		\label{PlotBiasThetamu2andpC04LOR_UMFSsigma01}}
\end{figure}
\begin{figure}[t]
	\centering
	\includegraphics[scale=0.33]{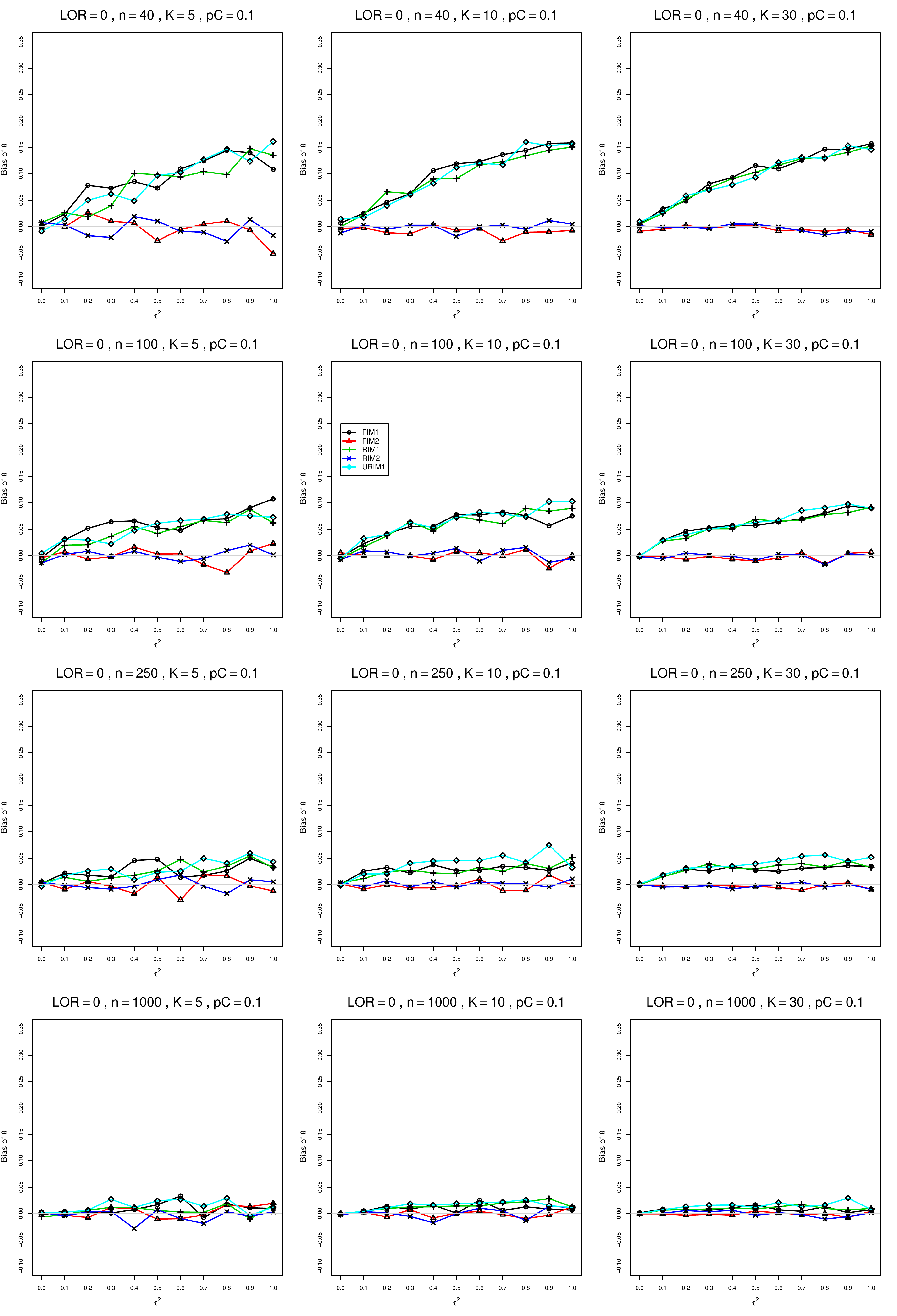}
	\caption{Bias of  overall log-odds ratio $\hat{\theta}_{FIM2}$ for $\theta=0$, $p_{C}=0.1$, $\sigma^2=0.4$, constant sample sizes $n=40,\;100,\;250,\;1000$.
The data-generation mechanisms are FIM1 ($\circ$), FIM2 ($\triangle$), RIM1 (+), RIM2 ($\times$), and URIM1 ($\diamond$).
		\label{PlotBiasThetamu0andpC01LOR_UMFSsigma04}}
\end{figure}
\begin{figure}[t]
	\centering
	\includegraphics[scale=0.33]{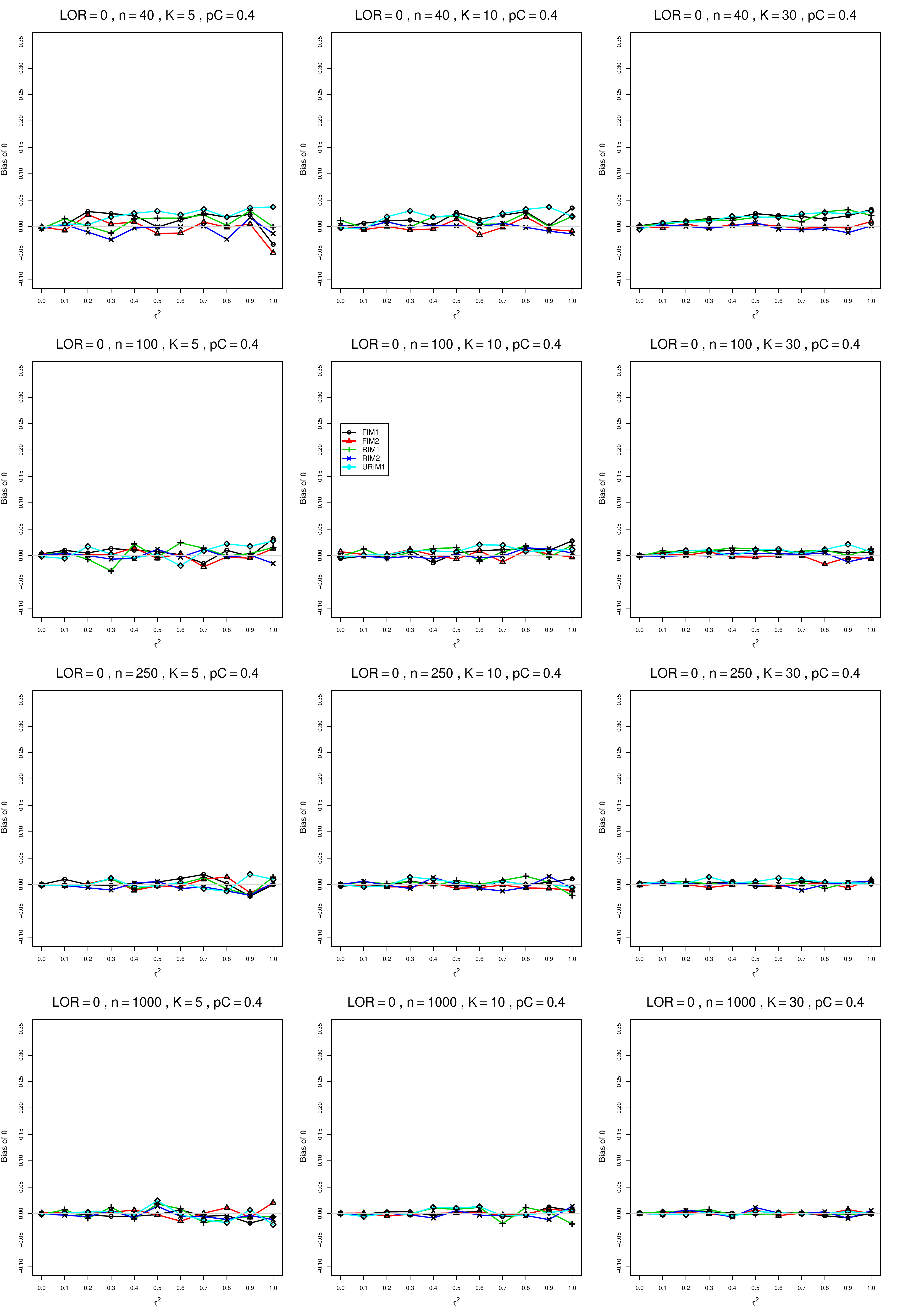}
	\caption{Bias of  overall log-odds ratio $\hat{\theta}_{FIM2}$ for $\theta=0$, $p_{C}=0.4$, $\sigma^2=0.4$, constant sample sizes $n=40,\;100,\;250,\;1000$.
The data-generation mechanisms are FIM1 ($\circ$), FIM2 ($\triangle$), RIM1 (+), RIM2 ($\times$), and URIM1 ($\diamond$).
		\label{PlotBiasThetamu0andpC04LOR_UMFSsigma04}}
\end{figure}
\begin{figure}[t]
	\centering
	\includegraphics[scale=0.33]{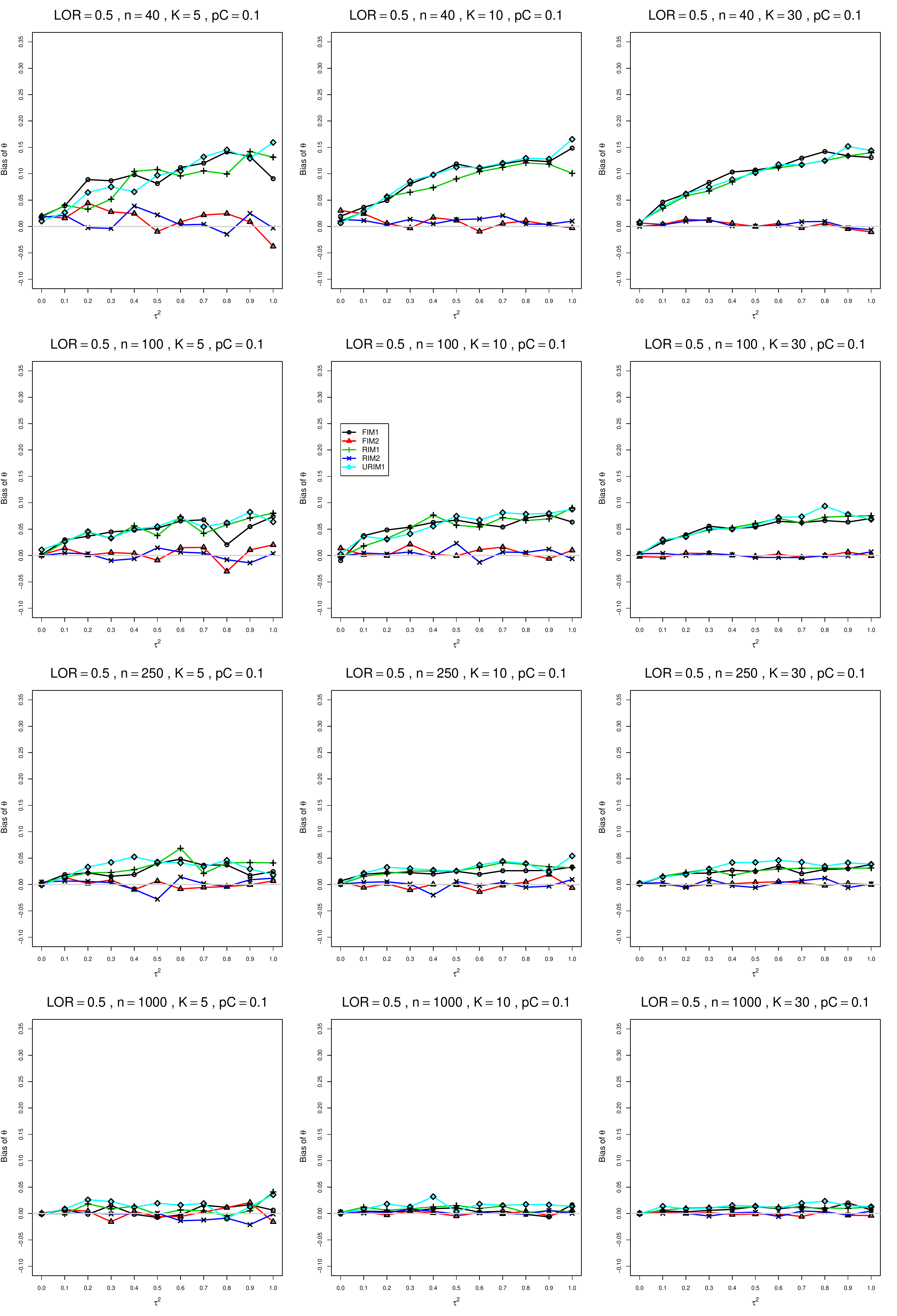}
	\caption{Bias of  overall log-odds ratio $\hat{\theta}_{FIM2}$ for $\theta=0.5$, $p_{C}=0.1$, $\sigma^2=0.4$, constant sample sizes $n=40,\;100,\;250,\;1000$.
The data-generation mechanisms are FIM1 ($\circ$), FIM2 ($\triangle$), RIM1 (+), RIM2 ($\times$), and URIM1 ($\diamond$).
		\label{PlotBiasThetamu05andpC01LOR_UMFSsigma04}}
\end{figure}
\begin{figure}[t]
	\centering
	\includegraphics[scale=0.33]{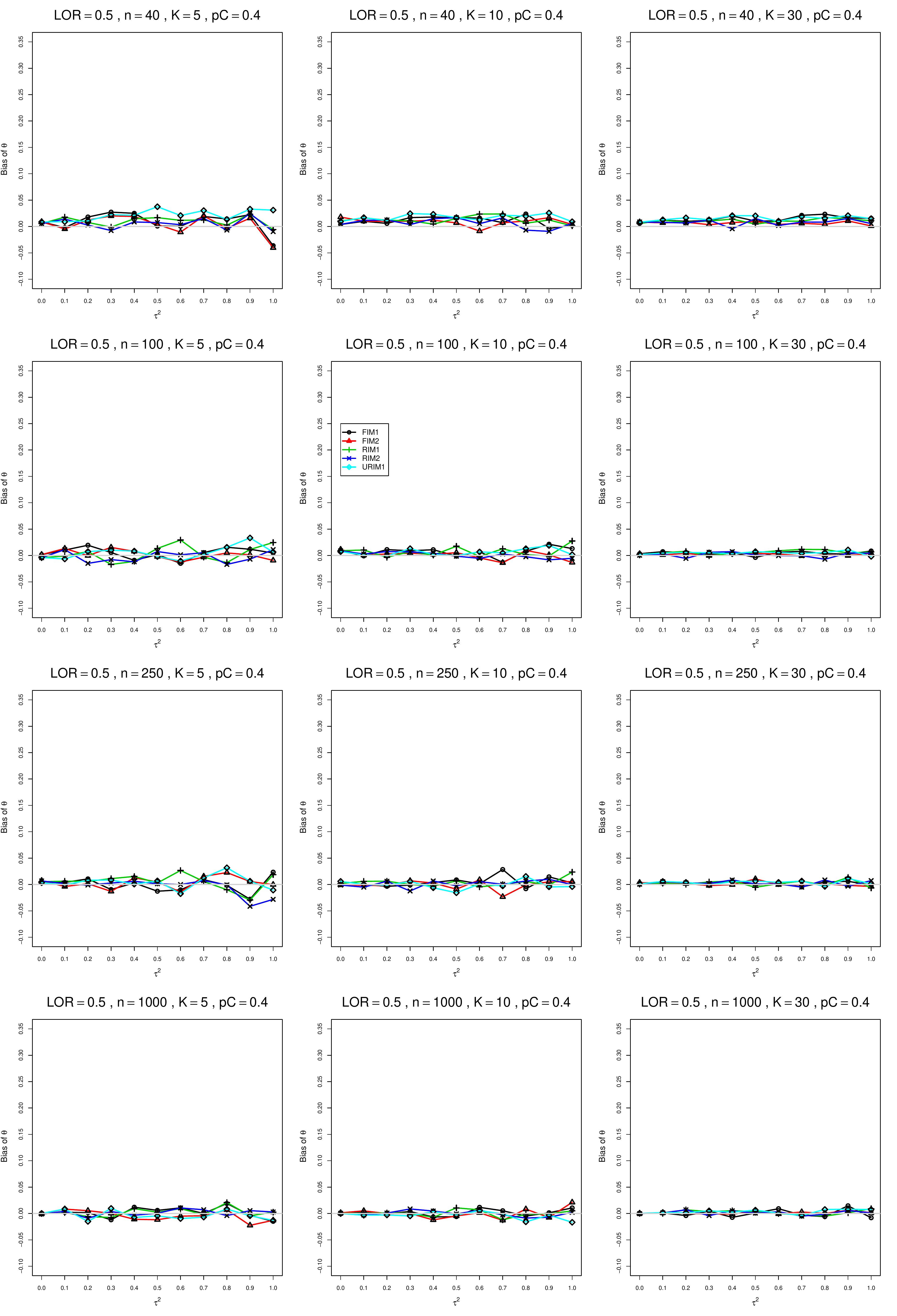}
	\caption{Bias of  overall log-odds ratio $\hat{\theta}_{FIM2}$ for $\theta=0.5$, $p_{C}=0.4$, $\sigma^2=0.4$, constant sample sizes $n=40,\;100,\;250,\;1000$.
The data-generation mechanisms are FIM1 ($\circ$), FIM2 ($\triangle$), RIM1 (+), RIM2 ($\times$), and URIM1 ($\diamond$).
		\label{PlotBiasThetamu05andpC04LOR_UMFSsigma04}}
\end{figure}
\begin{figure}[t]
	\centering
	\includegraphics[scale=0.33]{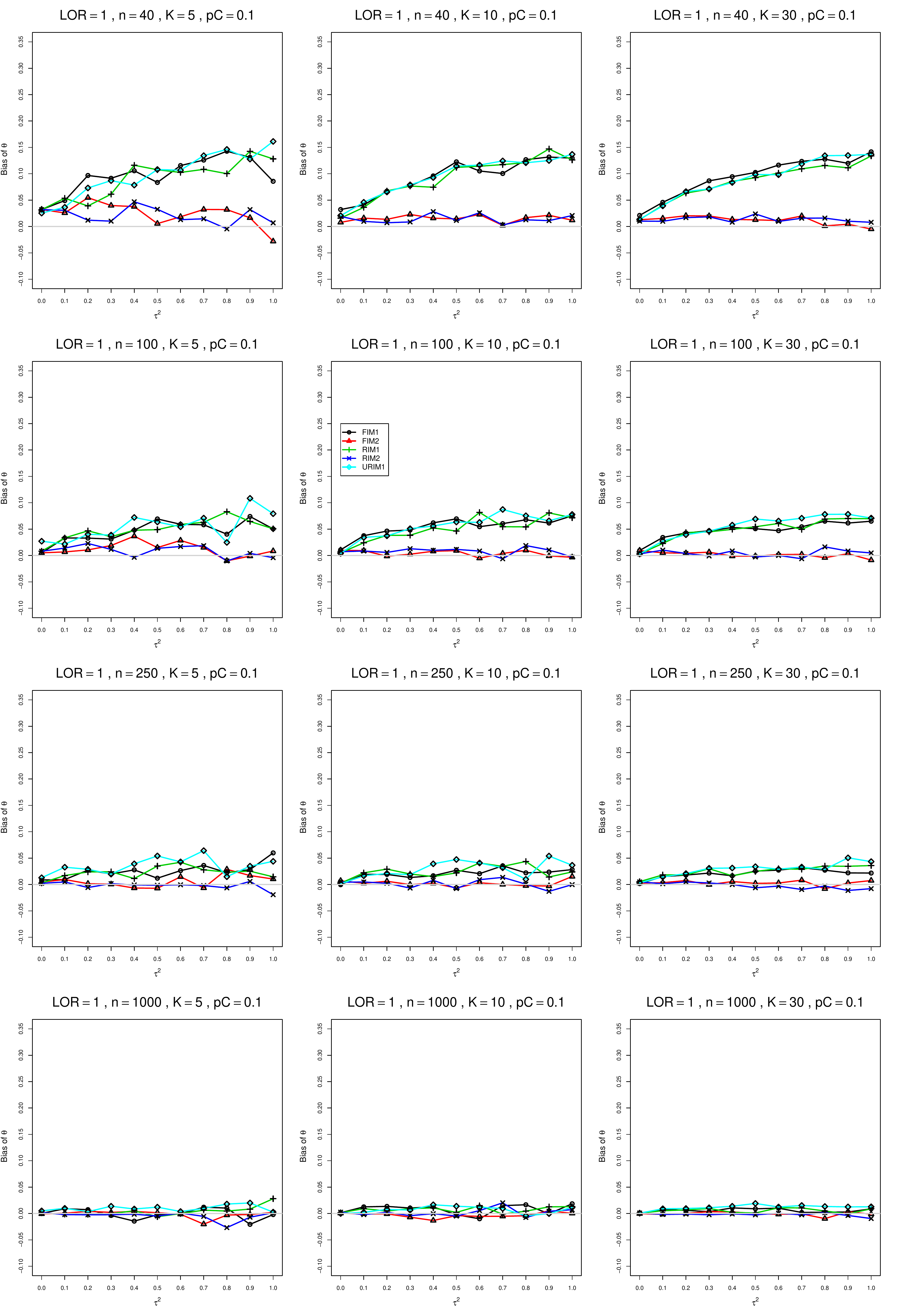}
	\caption{Bias of  overall log-odds ratio $\hat{\theta}_{FIM2}$ for $\theta=1$, $p_{C}=0.1$, $\sigma^2=0.4$, constant sample sizes $n=40,\;100,\;250,\;1000$.
The data-generation mechanisms are FIM1 ($\circ$), FIM2 ($\triangle$), RIM1 (+), RIM2 ($\times$), and URIM1 ($\diamond$).
		\label{PlotBiasThetamu1andpC01LOR_UMFSsigma04}}
\end{figure}
\begin{figure}[t]
	\centering
	\includegraphics[scale=0.33]{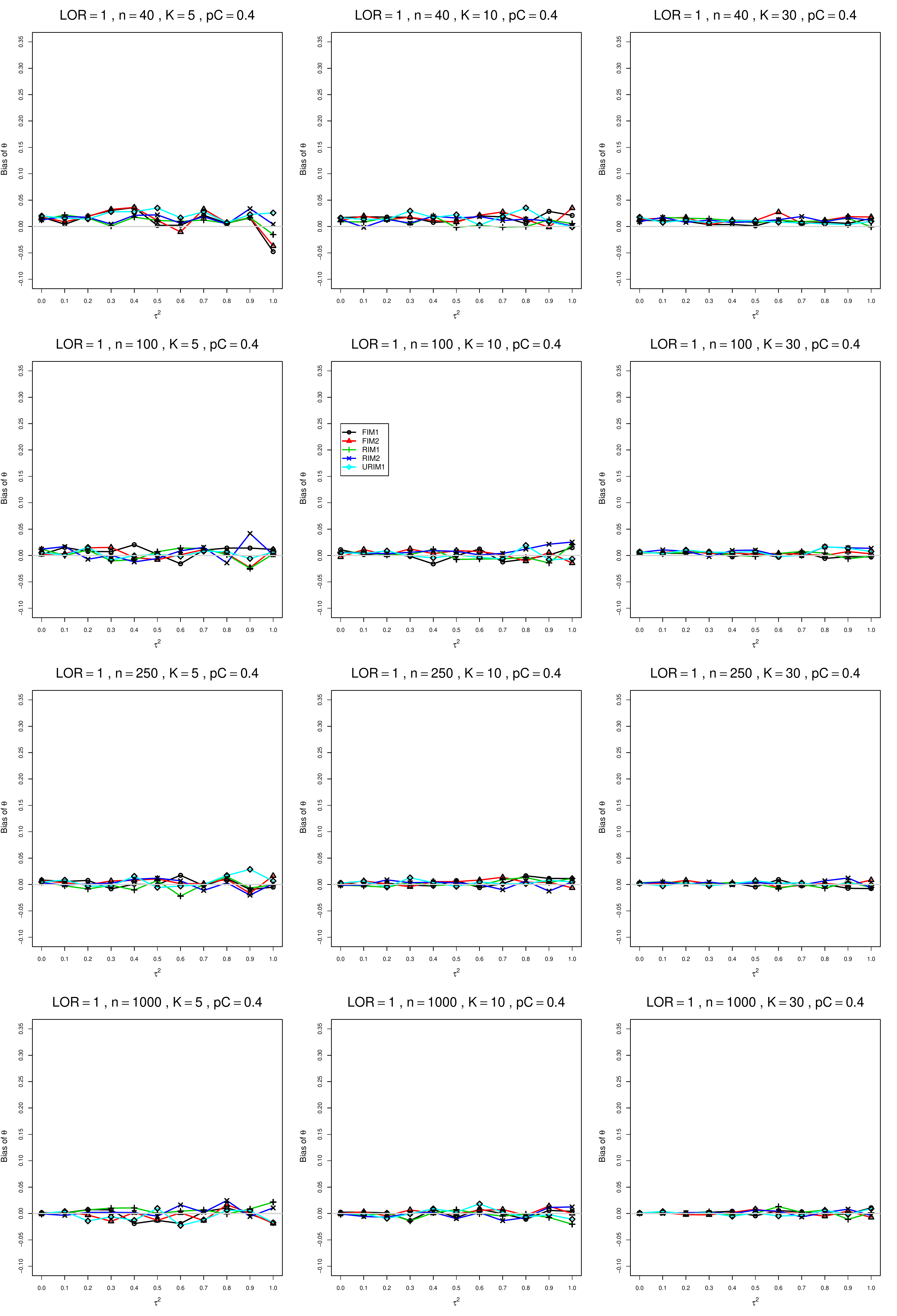}
	\caption{Bias of  overall log-odds ratio $\hat{\theta}_{FIM2}$ for $\theta=1$, $p_{C}=0.4$, $\sigma^2=0.4$, constant sample sizes $n=40,\;100,\;250,\;1000$.
The data-generation mechanisms are FIM1 ($\circ$), FIM2 ($\triangle$), RIM1 (+), RIM2 ($\times$), and URIM1 ($\diamond$).
		\label{PlotBiasThetamu1andpC04LOR_UMFSsigma04}}
\end{figure}
\begin{figure}[t]
	\centering
	\includegraphics[scale=0.33]{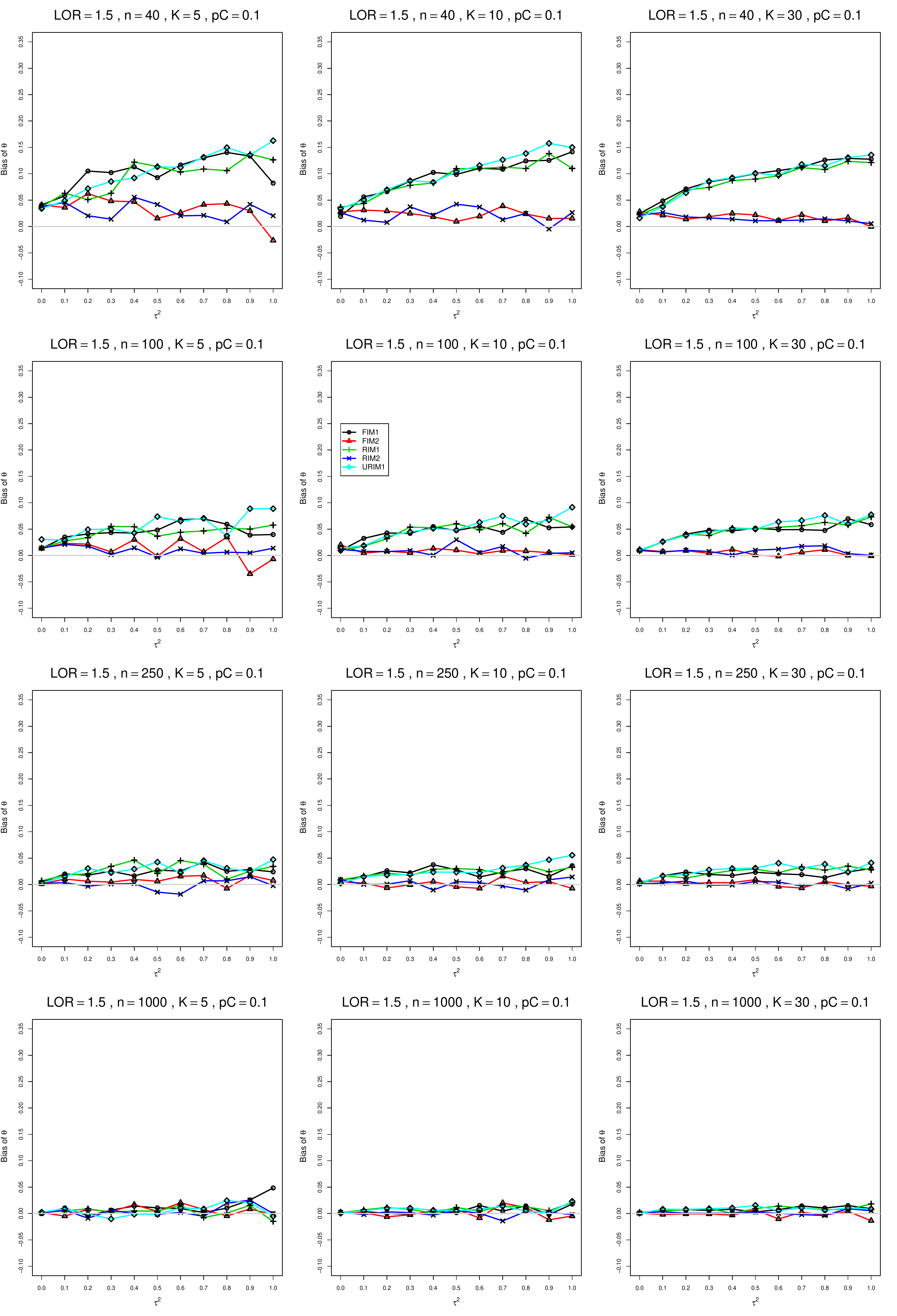}
	\caption{Bias of  overall log-odds ratio $\hat{\theta}_{FIM2}$ for $\theta=1.5$, $p_{C}=0.1$, $\sigma^2=0.4$, constant sample sizes $n=40,\;100,\;250,\;1000$.
The data-generation mechanisms are FIM1 ($\circ$), FIM2 ($\triangle$), RIM1 (+), RIM2 ($\times$), and URIM1 ($\diamond$).
		\label{PlotBiasThetamu15andpC01LOR_UMFSsigma04}}
\end{figure}
\begin{figure}[t]
	\centering
	\includegraphics[scale=0.33]{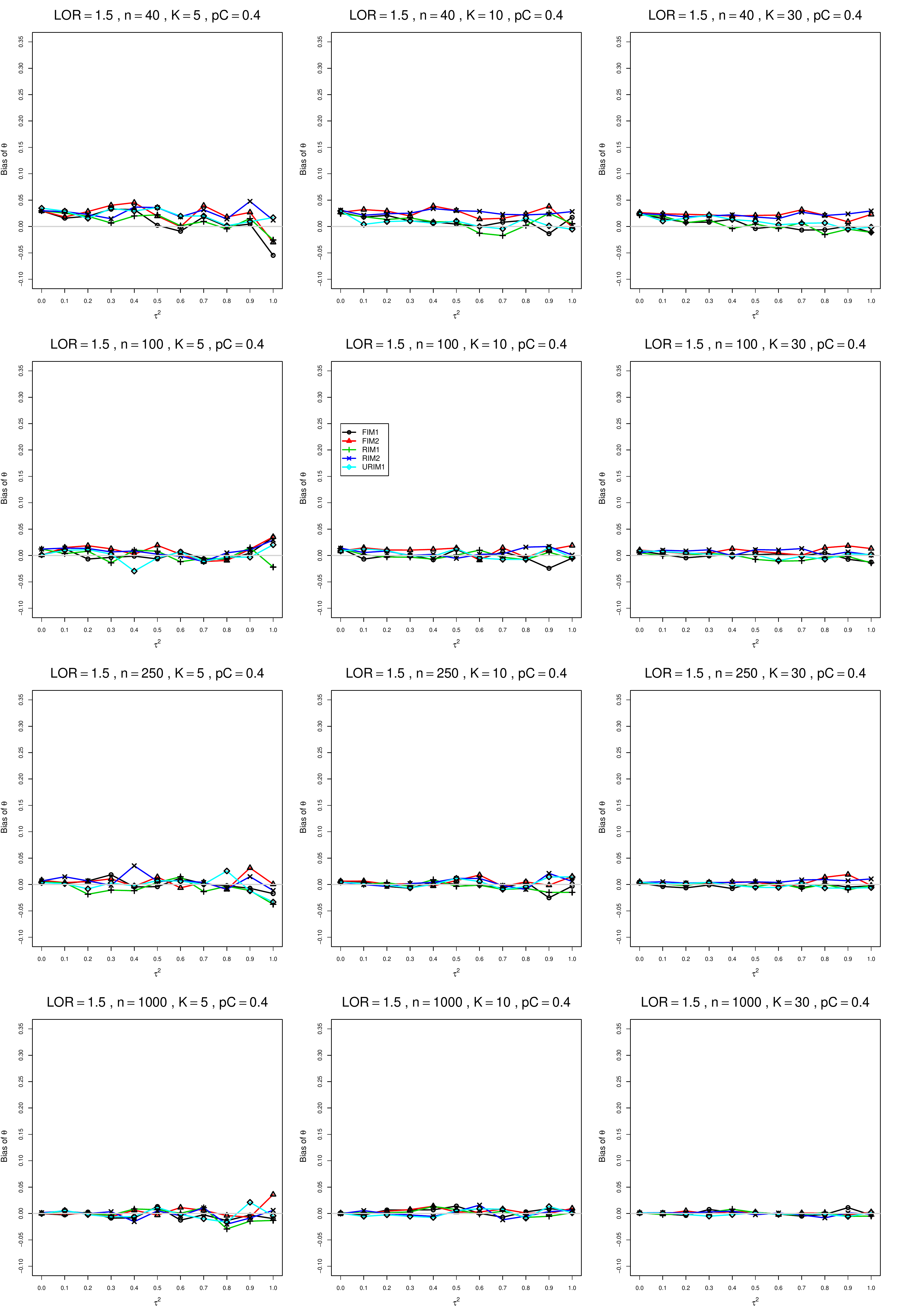}
	\caption{Bias of  overall log-odds ratio $\hat{\theta}_{FIM2}$ for $\theta=1.5$, $p_{C}=0.4$, $\sigma^2=0.4$, constant sample sizes $n=40,\;100,\;250,\;1000$.
The data-generation mechanisms are FIM1 ($\circ$), FIM2 ($\triangle$), RIM1 (+), RIM2 ($\times$), and URIM1 ($\diamond$).
		\label{PlotBiasThetamu15andpC04LOR_UMFSsigma04}}
\end{figure}
\begin{figure}[t]
	\centering
	\includegraphics[scale=0.33]{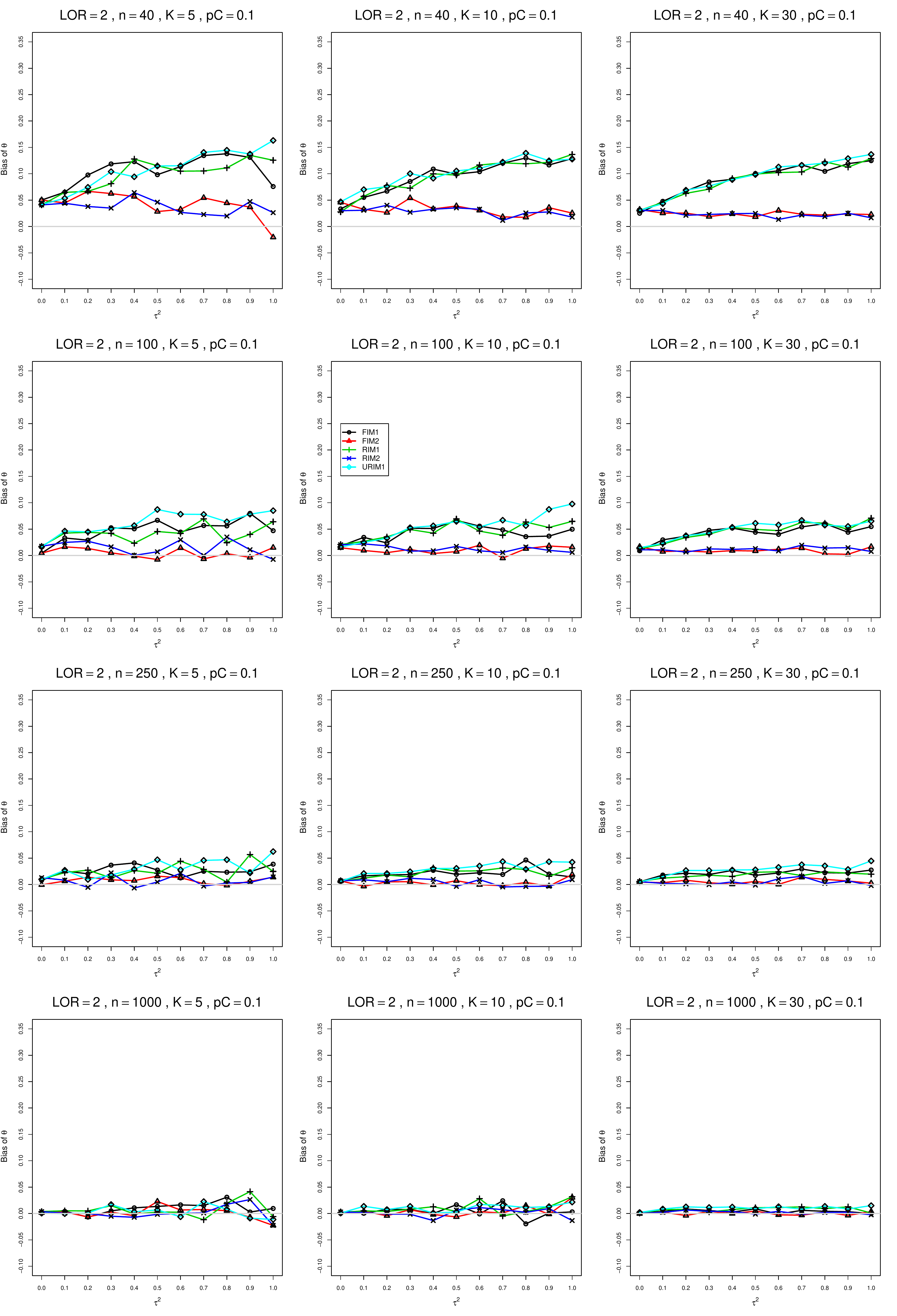}
	\caption{Bias of  overall log-odds ratio $\hat{\theta}_{FIM2}$ for $\theta=2$, $p_{C}=0.1$, $\sigma^2=0.4$, constant sample sizes $n=40,\;100,\;250,\;1000$.
The data-generation mechanisms are FIM1 ($\circ$), FIM2 ($\triangle$), RIM1 (+), RIM2 ($\times$), and URIM1 ($\diamond$).
		\label{PlotBiasThetamu2andpC01LOR_UMFSsigma04}}
\end{figure}
\begin{figure}[t]
	\centering
	\includegraphics[scale=0.33]{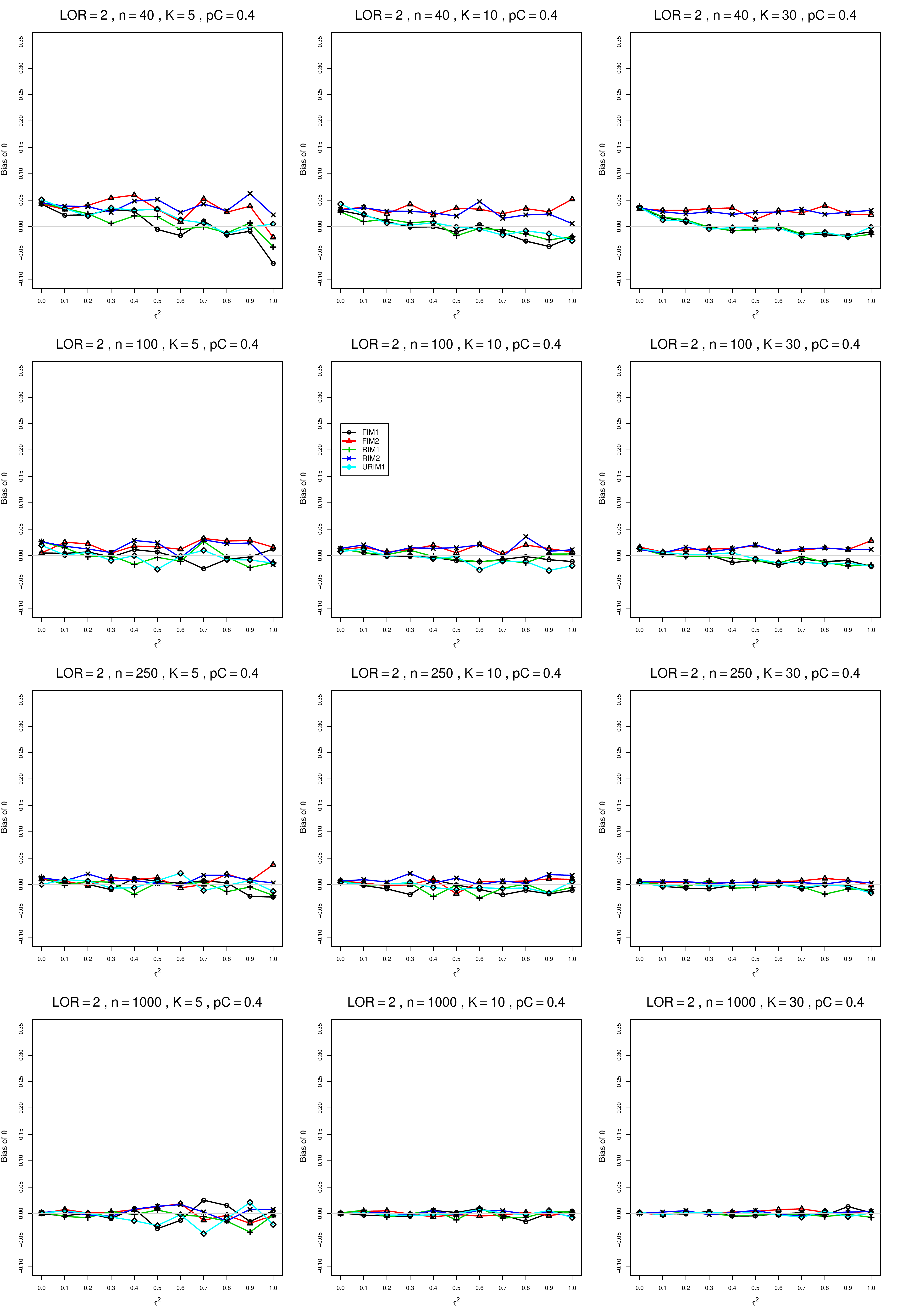}
	\caption{Bias of  overall log-odds ratio $\hat{\theta}_{FIM2}$ for $\theta=2$, $p_{C}=0.4$, $\sigma^2=0.4$, constant sample sizes $n=40,\;100,\;250,\;1000$.
The data-generation mechanisms are FIM1 ($\circ$), FIM2 ($\triangle$), RIM1 (+), RIM2 ($\times$), and URIM1 ($\diamond$).
		\label{PlotBiasThetamu2andpC04LOR_UMFSsigma04}}
\end{figure}

\clearpage
\subsection*{A2.6 Bias of $\hat{\theta}_{RIM2}$}
\renewcommand{\thefigure}{A2.6.\arabic{figure}}
\setcounter{figure}{0}

\begin{figure}[t]
	\centering
	\includegraphics[scale=0.33]{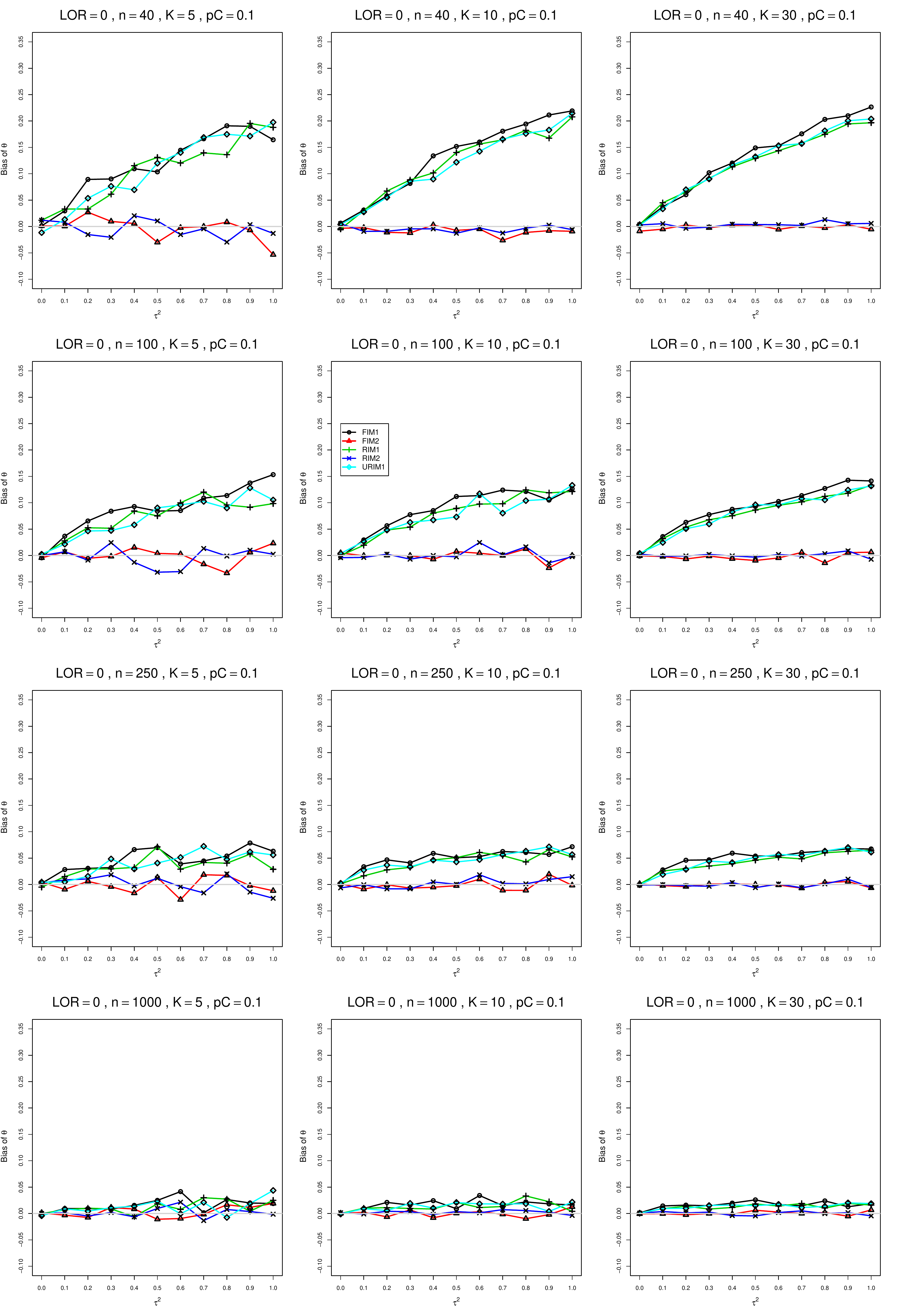}
	\caption{Bias of  overall log-odds ratio $\hat{\theta}_{RIM2}$ for $\theta=0$, $p_{C}=0.1$, $\sigma^2=0.1$, constant sample sizes $n=40,\;100,\;250,\;1000$.
The data-generation mechanisms are FIM1 ($\circ$), FIM2 ($\triangle$), RIM1 (+), RIM2 ($\times$), and URIM1 ($\diamond$).
		\label{PlotBiasThetamu0andpC01LOR_UMRSsigma01}}
\end{figure}
\begin{figure}[t]
	\centering
	\includegraphics[scale=0.33]{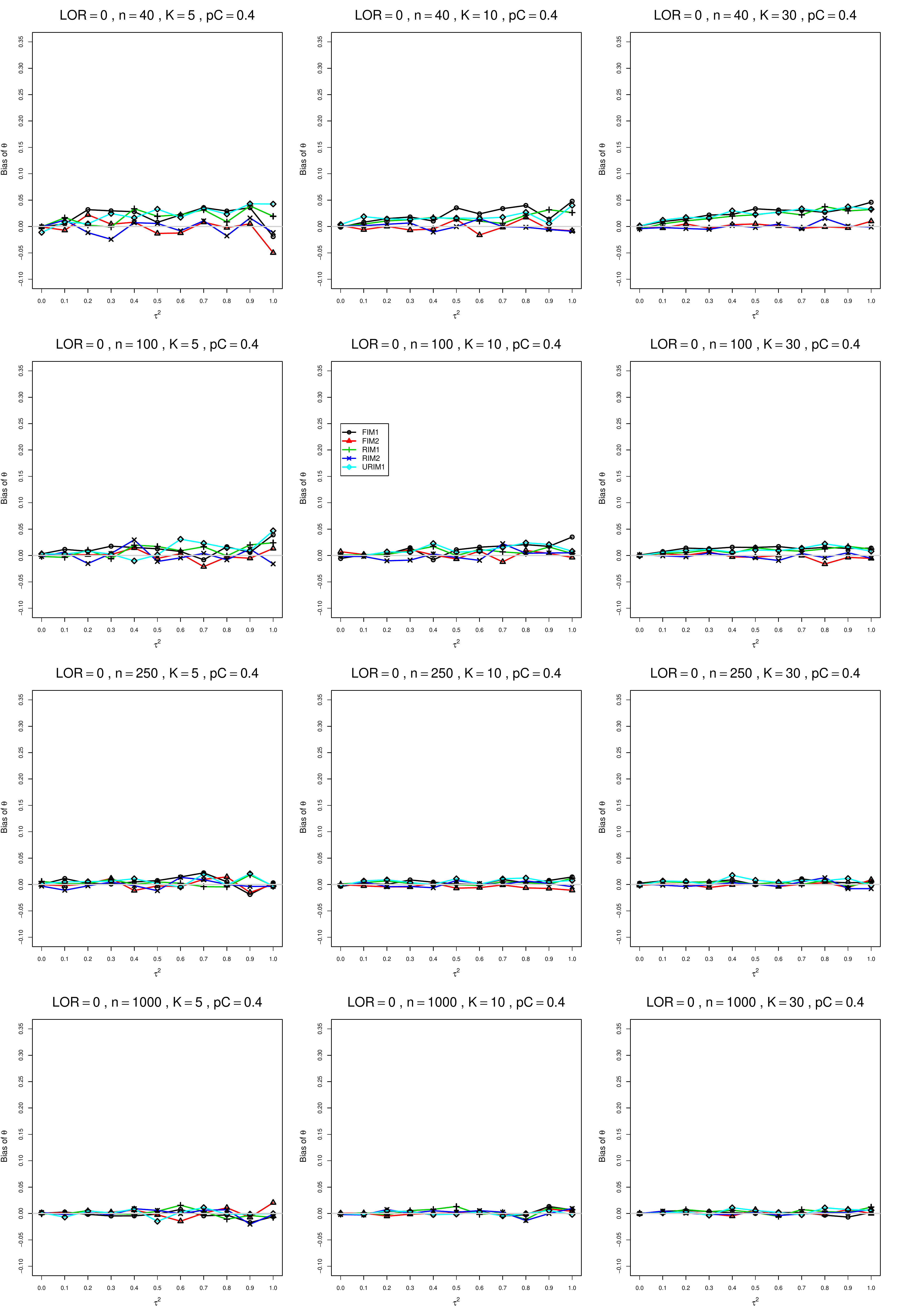}
	\caption{Bias of  overall log-odds ratio $\hat{\theta}_{RIM2}$ for $\theta=0$, $p_{C}=0.4$, $\sigma^2=0.1$, constant sample sizes $n=40,\;100,\;250,\;1000$.
The data-generation mechanisms are FIM1 ($\circ$), FIM2 ($\triangle$), RIM1 (+), RIM2 ($\times$), and URIM1 ($\diamond$).
		\label{PlotBiasThetamu0andpC04LOR_UMRSsigma01}}
\end{figure}
\begin{figure}[t]
	\centering
	\includegraphics[scale=0.33]{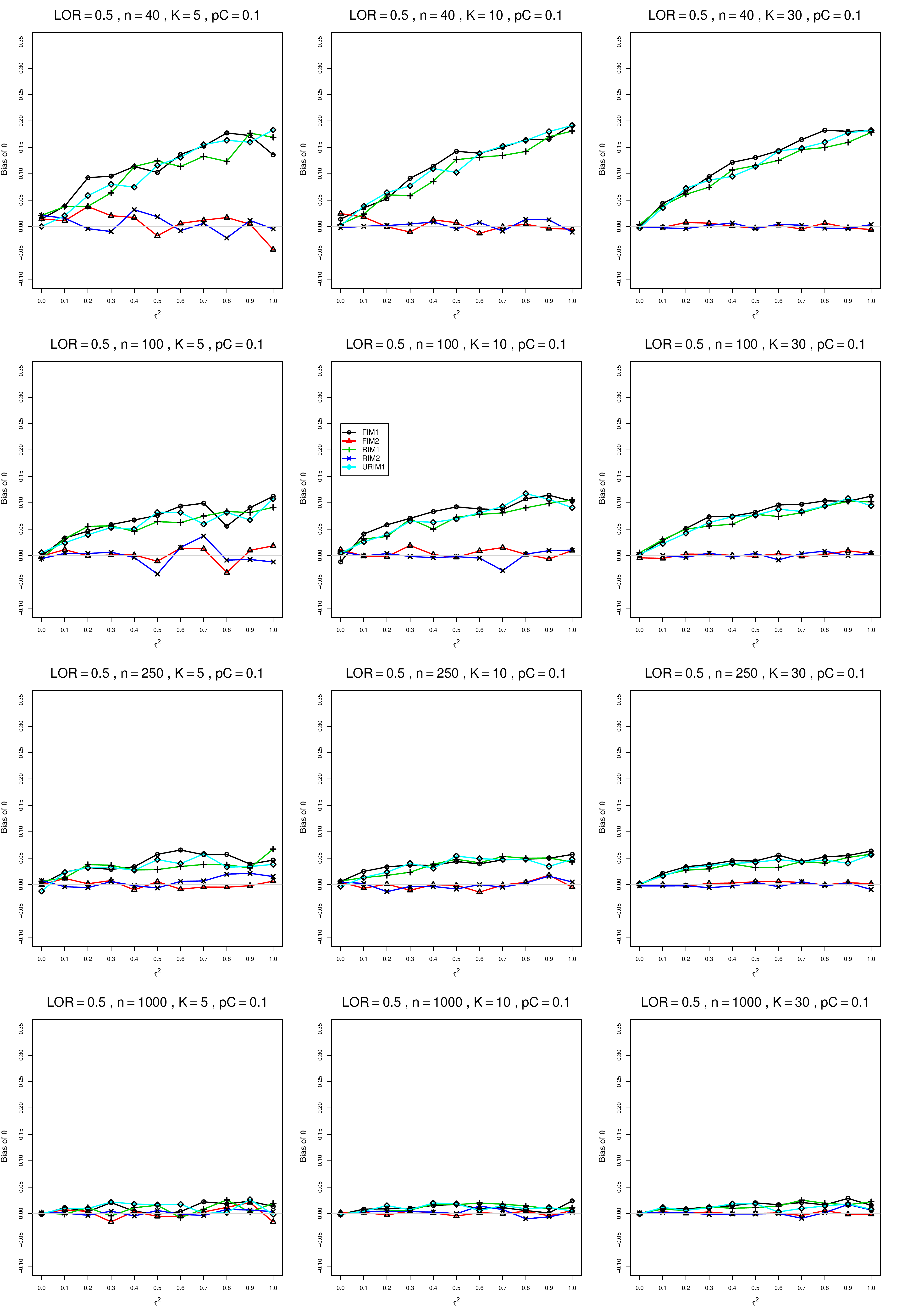}
	\caption{Bias of  overall log-odds ratio $\hat{\theta}_{RIM2}$ for $\theta=0.5$, $p_{C}=0.1$, $\sigma^2=0.1$, constant sample sizes $n=40,\;100,\;250,\;1000$.
The data-generation mechanisms are FIM1 ($\circ$), FIM2 ($\triangle$), RIM1 (+), RIM2 ($\times$), and URIM1 ($\diamond$).
		\label{PlotBiasThetamu05andpC01LOR_UMRSsigma01}}
\end{figure}
\begin{figure}[t]
	\centering
	\includegraphics[scale=0.33]{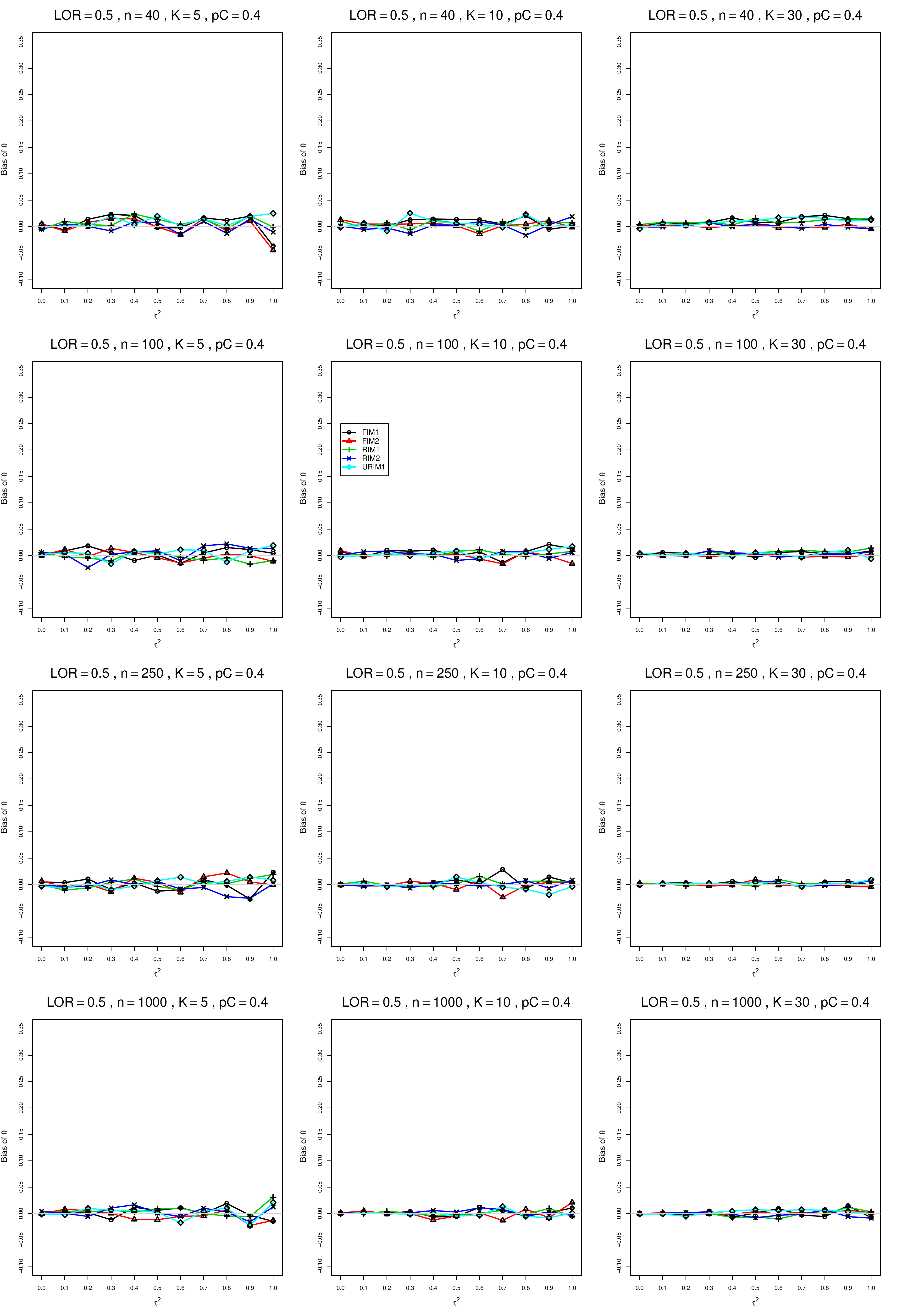}
	\caption{Bias of  overall log-odds ratio $\hat{\theta}_{RIM2}$ for $\theta=0.5$, $p_{C}=0.4$, $\sigma^2=0.1$, constant sample sizes $n=40,\;100,\;250,\;1000$.
The data-generation mechanisms are FIM1 ($\circ$), FIM2 ($\triangle$), RIM1 (+), RIM2 ($\times$), and URIM1 ($\diamond$).
		\label{PlotBiasThetamu05andpC04LOR_UMRSsigma01}}
\end{figure}
\begin{figure}[t]
	\centering
	\includegraphics[scale=0.33]{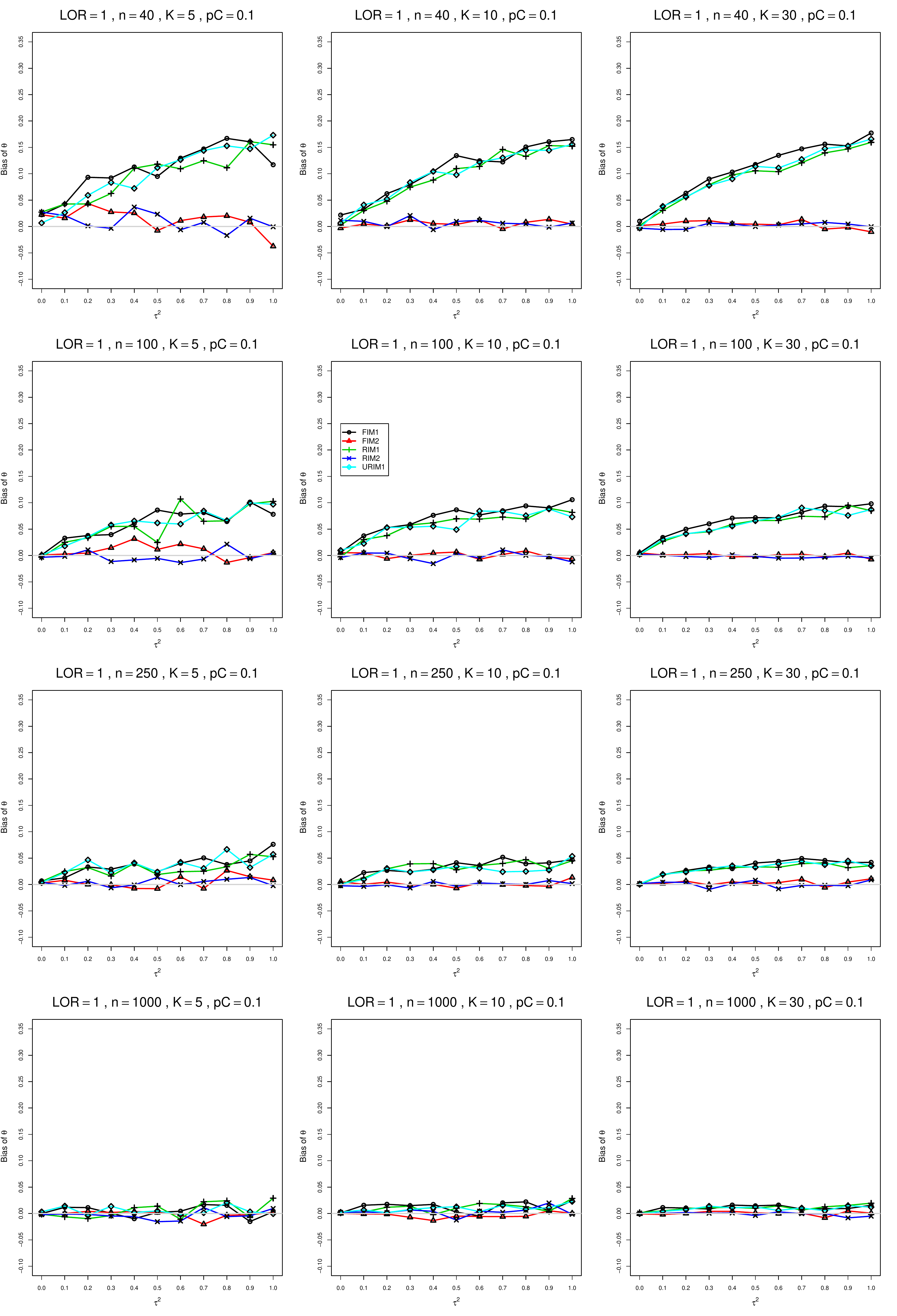}
	\caption{Bias of  overall log-odds ratio $\hat{\theta}_{RIM2}$ for $\theta=1$, $p_{C}=0.1$, $\sigma^2=0.1$, constant sample sizes $n=40,\;100,\;250,\;1000$.
The data-generation mechanisms are FIM1 ($\circ$), FIM2 ($\triangle$), RIM1 (+), RIM2 ($\times$), and URIM1 ($\diamond$).
		\label{PlotBiasThetamu1andpC01LOR_UMRSsigma01}}
\end{figure}
\begin{figure}[t]
	\centering
	\includegraphics[scale=0.33]{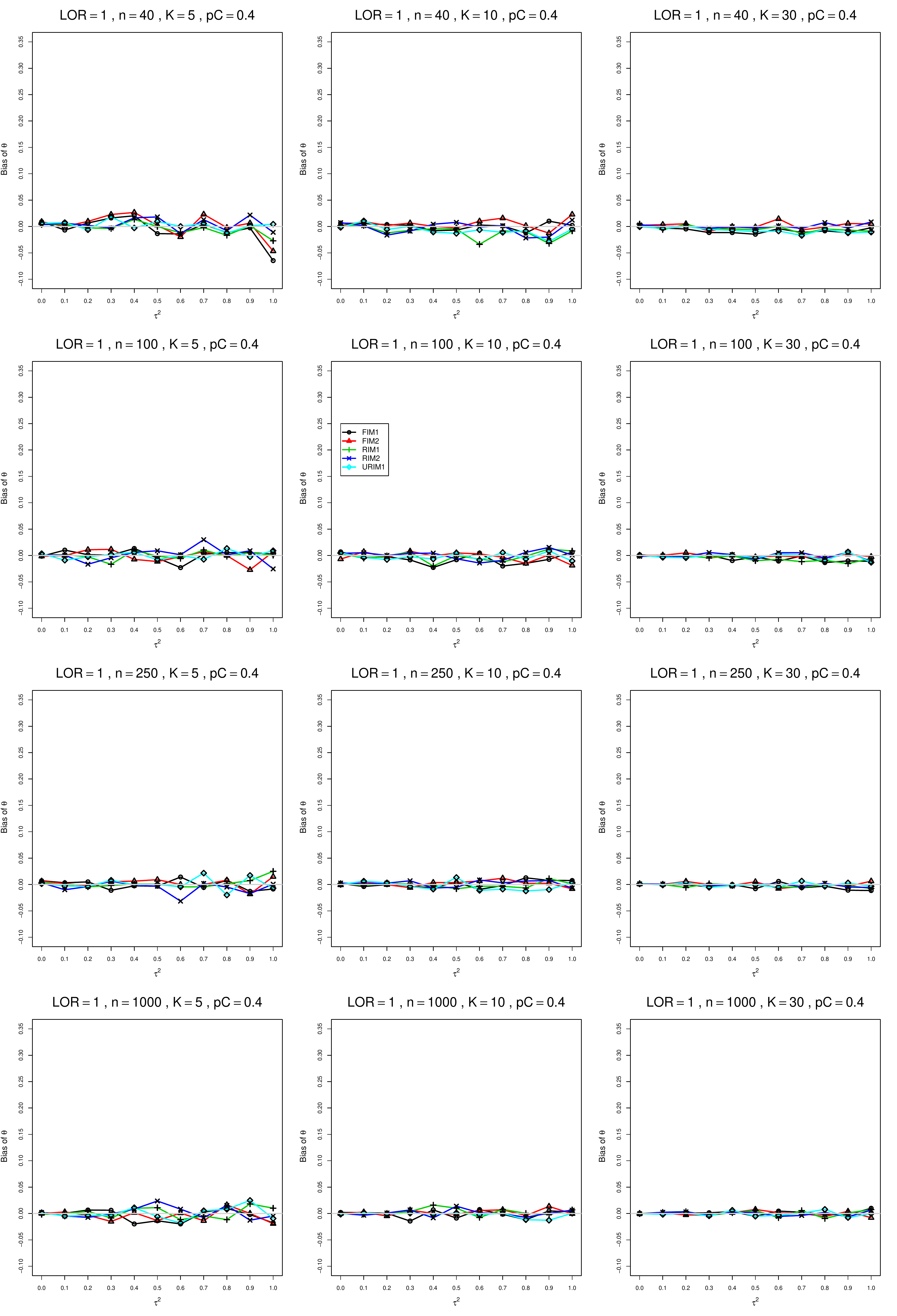}
	\caption{Bias of  overall log-odds ratio $\hat{\theta}_{RIM2}$ for $\theta=1$, $p_{C}=0.4$, $\sigma^2=0.1$, constant sample sizes $n=40,\;100,\;250,\;1000$.
The data-generation mechanisms are FIM1 ($\circ$), FIM2 ($\triangle$), RIM1 (+), RIM2 ($\times$), and URIM1 ($\diamond$).
		\label{PlotBiasThetamu1andpC04LOR_UMRSsigma01}}
\end{figure}
\begin{figure}[t]
	\centering
	\includegraphics[scale=0.33]{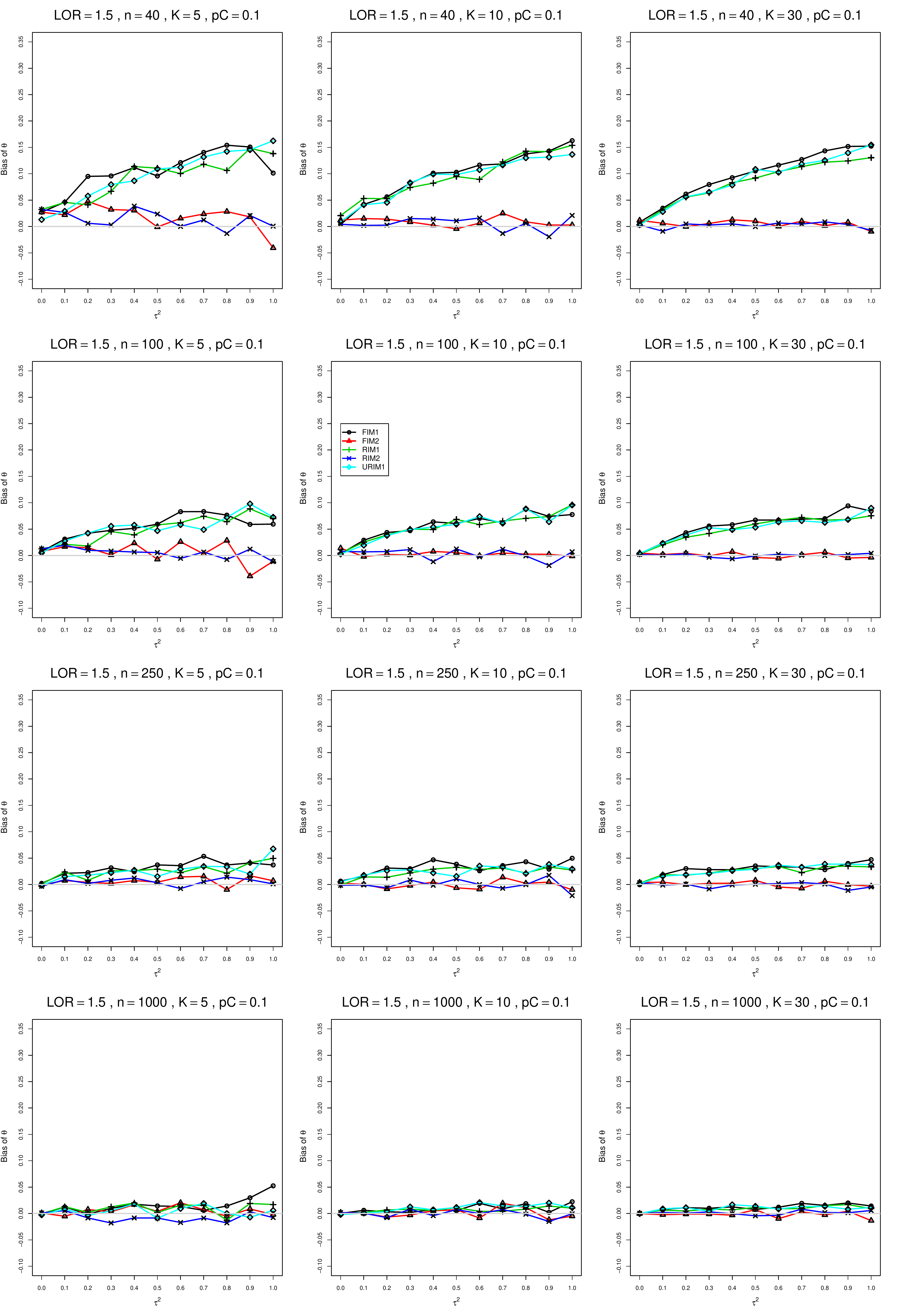}
	\caption{Bias of  overall log-odds ratio $\hat{\theta}_{RIM2}$ for $\theta=1.5$, $p_{C}=0.1$, $\sigma^2=0.1$, constant sample sizes $n=40,\;100,\;250,\;1000$.
The data-generation mechanisms are FIM1 ($\circ$), FIM2 ($\triangle$), RIM1 (+), RIM2 ($\times$), and URIM1 ($\diamond$).
		\label{PlotBiasThetamu15andpC01LOR_UMRSsigma01}}
\end{figure}
\begin{figure}[t]
	\centering
	\includegraphics[scale=0.33]{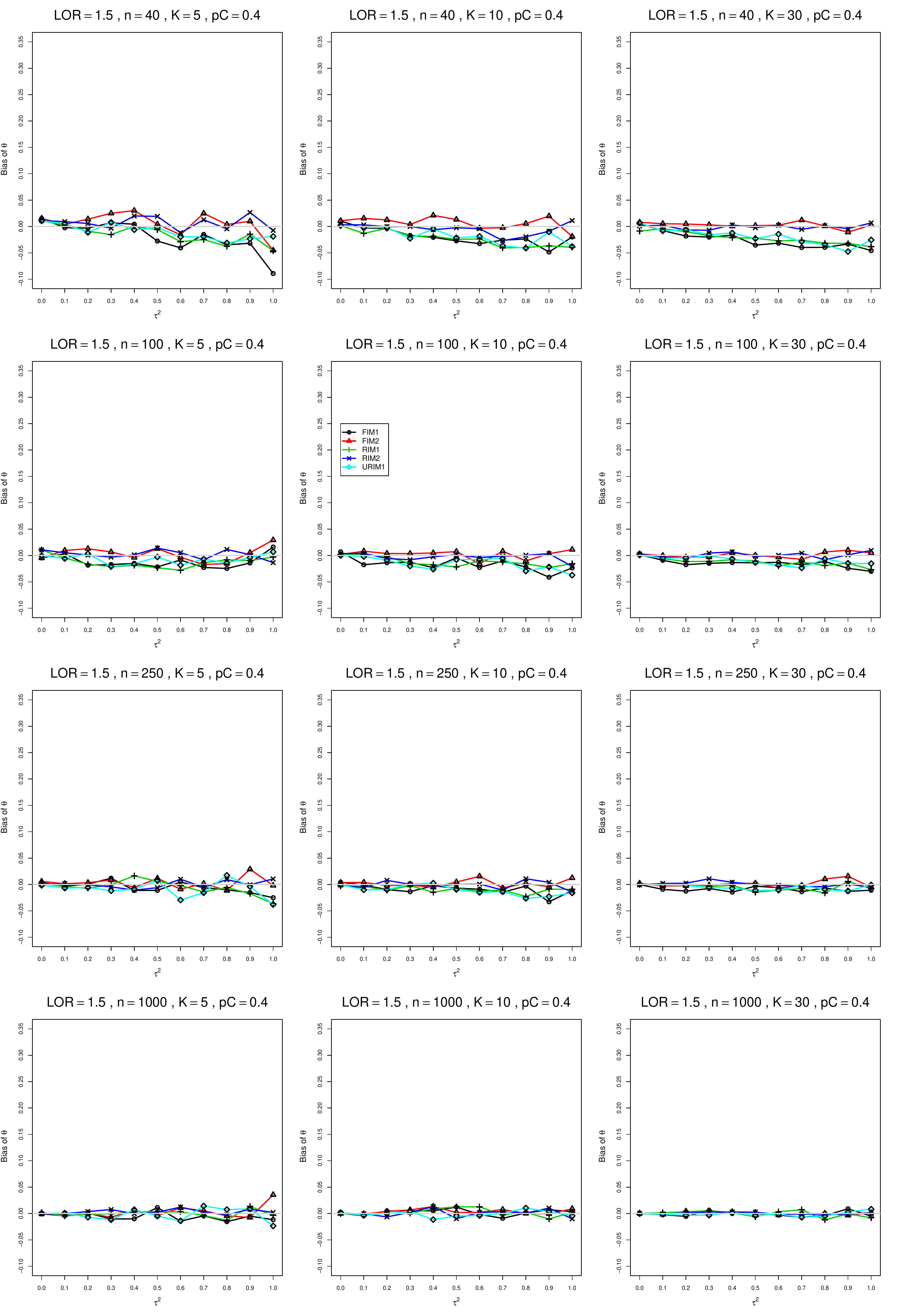}
	\caption{Bias of  overall log-odds ratio $\hat{\theta}_{RIM2}$ for $\theta=1.5$, $p_{C}=0.4$, $\sigma^2=0.1$, constant sample sizes $n=40,\;100,\;250,\;1000$.
The data-generation mechanisms are FIM1 ($\circ$), FIM2 ($\triangle$), RIM1 (+), RIM2 ($\times$), and URIM1 ($\diamond$).
		\label{PlotBiasThetamu15andpC04LOR_UMRSsigma01}}
\end{figure}
\begin{figure}[t]
	\centering
	\includegraphics[scale=0.33]{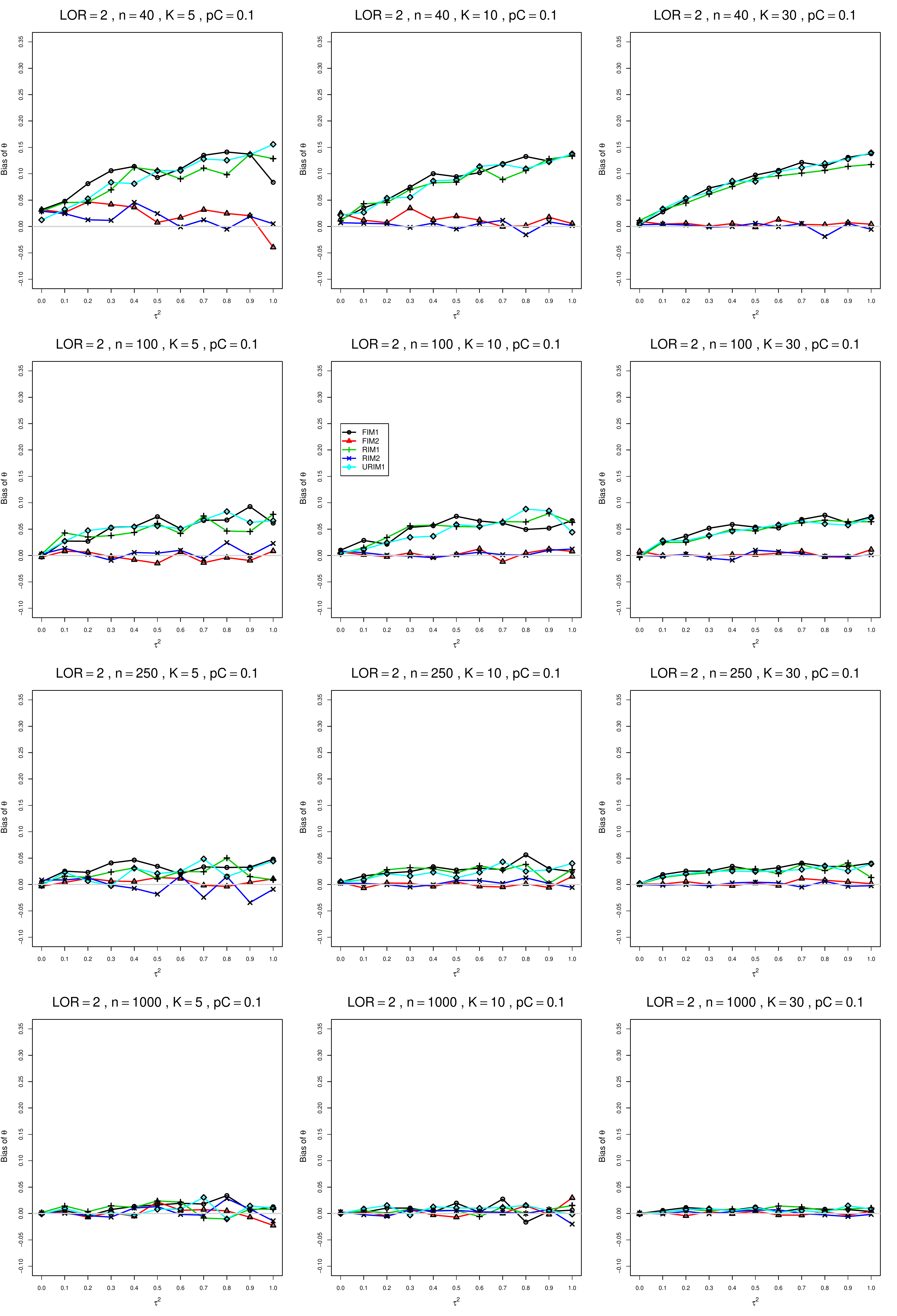}
	\caption{Bias of  overall log-odds ratio $\hat{\theta}_{RIM2}$ for $\theta=2$, $p_{C}=0.1$, $\sigma^2=0.1$, constant sample sizes $n=40,\;100,\;250,\;1000$.
The data-generation mechanisms are FIM1 ($\circ$), FIM2 ($\triangle$), RIM1 (+), RIM2 ($\times$), and URIM1 ($\diamond$).
		\label{PlotBiasThetamu2andpC01LOR_UMRSsigma01}}
\end{figure}
\begin{figure}[t]
	\centering
	\includegraphics[scale=0.33]{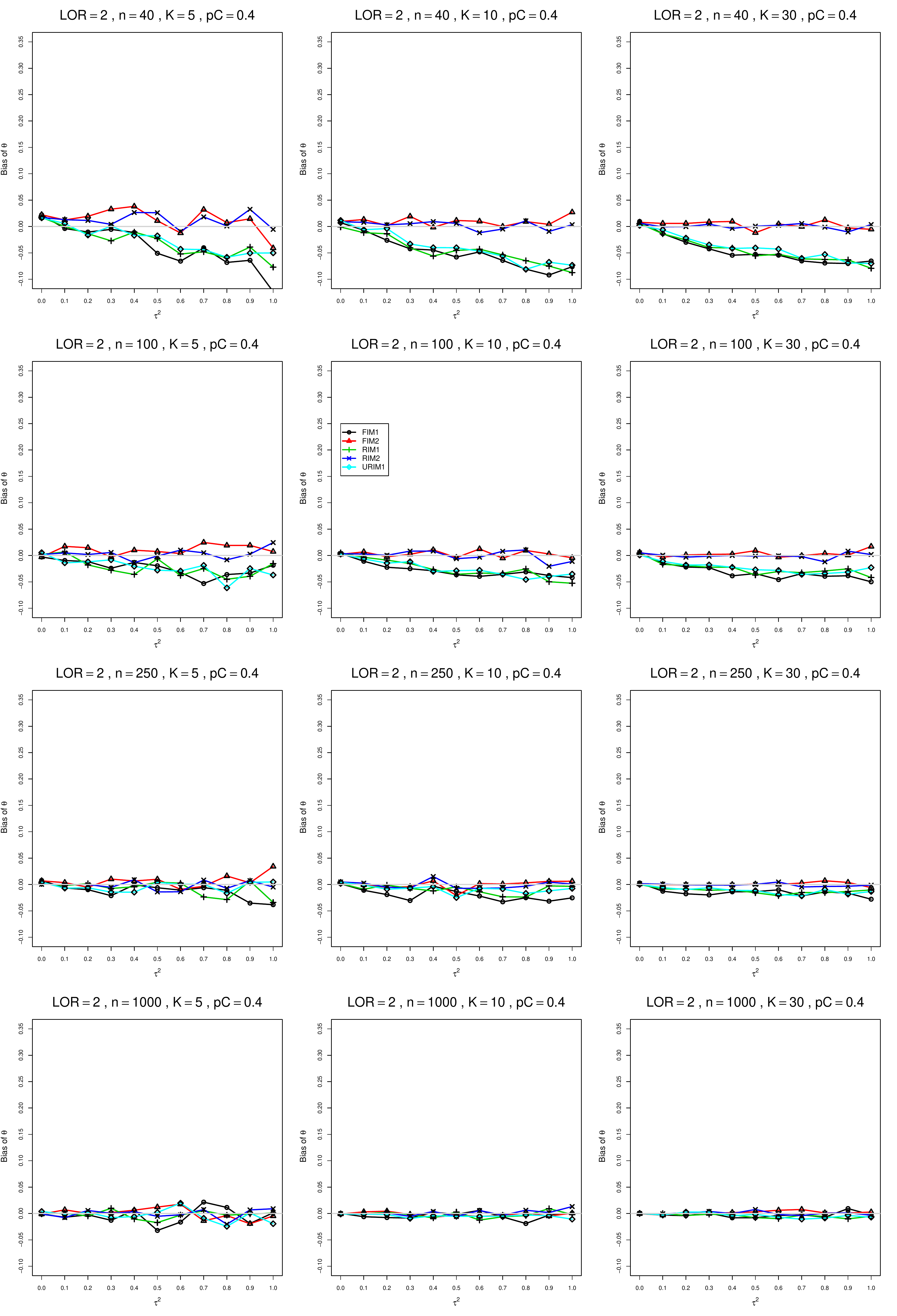}
	\caption{Bias of  overall log-odds ratio $\hat{\theta}_{RIM2}$ for $\theta=2$, $p_{C}=0.4$, $\sigma^2=0.1$, constant sample sizes $n=40,\;100,\;250,\;1000$.
The data-generation mechanisms are FIM1 ($\circ$), FIM2 ($\triangle$), RIM1 (+), RIM2 ($\times$), and URIM1 ($\diamond$).
		\label{PlotBiasThetamu2andpC04LOR_UMRSsigma01}}
\end{figure}
\begin{figure}[t]
	\centering
	\includegraphics[scale=0.33]{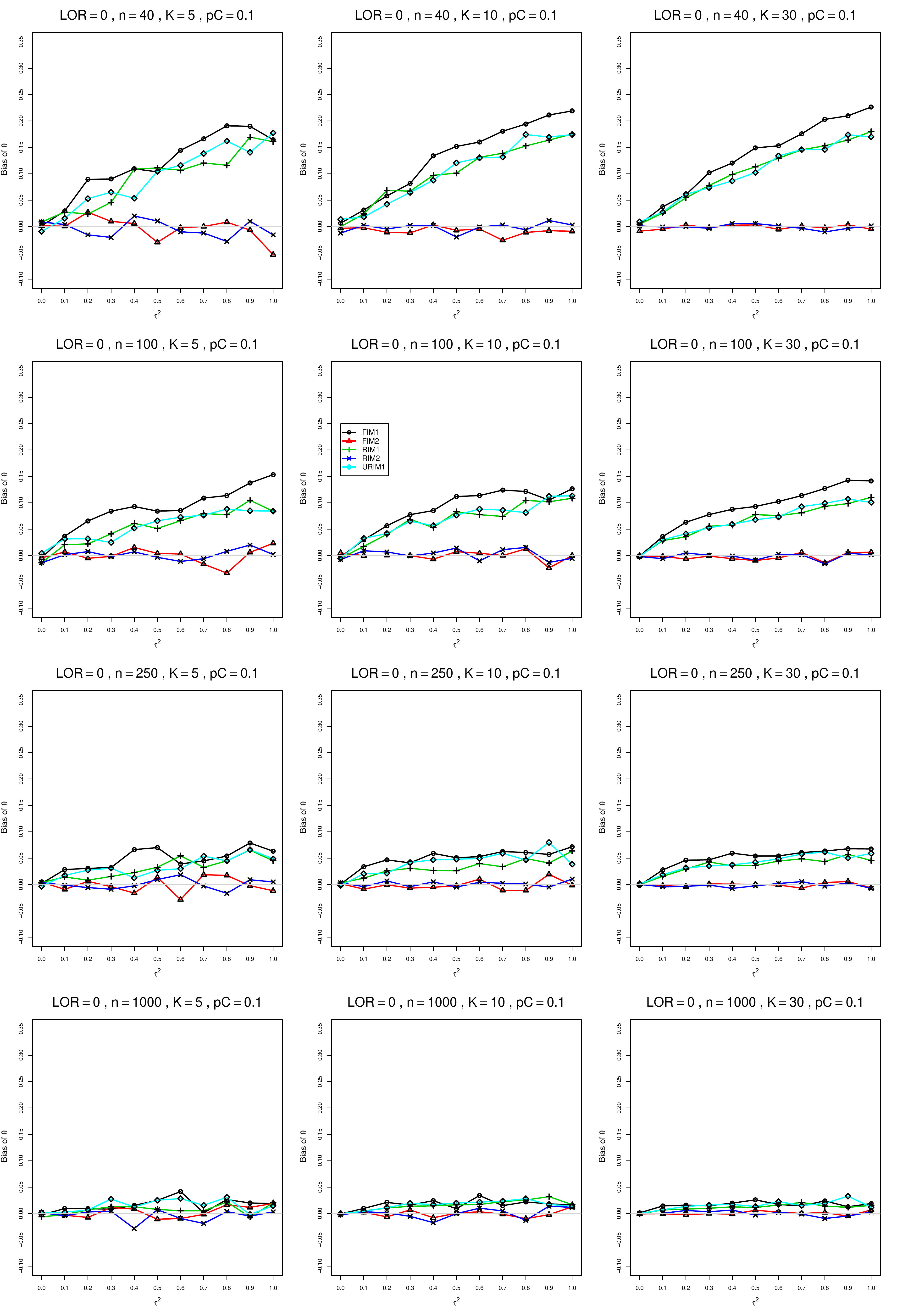}
	\caption{Bias of  overall log-odds ratio $\hat{\theta}_{RIM2}$ for $\theta=0$, $p_{C}=0.1$, $\sigma^2=0.4$, constant sample sizes $n=40,\;100,\;250,\;1000$.
The data-generation mechanisms are FIM1 ($\circ$), FIM2 ($\triangle$), RIM1 (+), RIM2 ($\times$), and URIM1 ($\diamond$).
		\label{PlotBiasThetamu0andpC01LOR_UMRSsigma04}}
\end{figure}
\begin{figure}[t]
	\centering
	\includegraphics[scale=0.33]{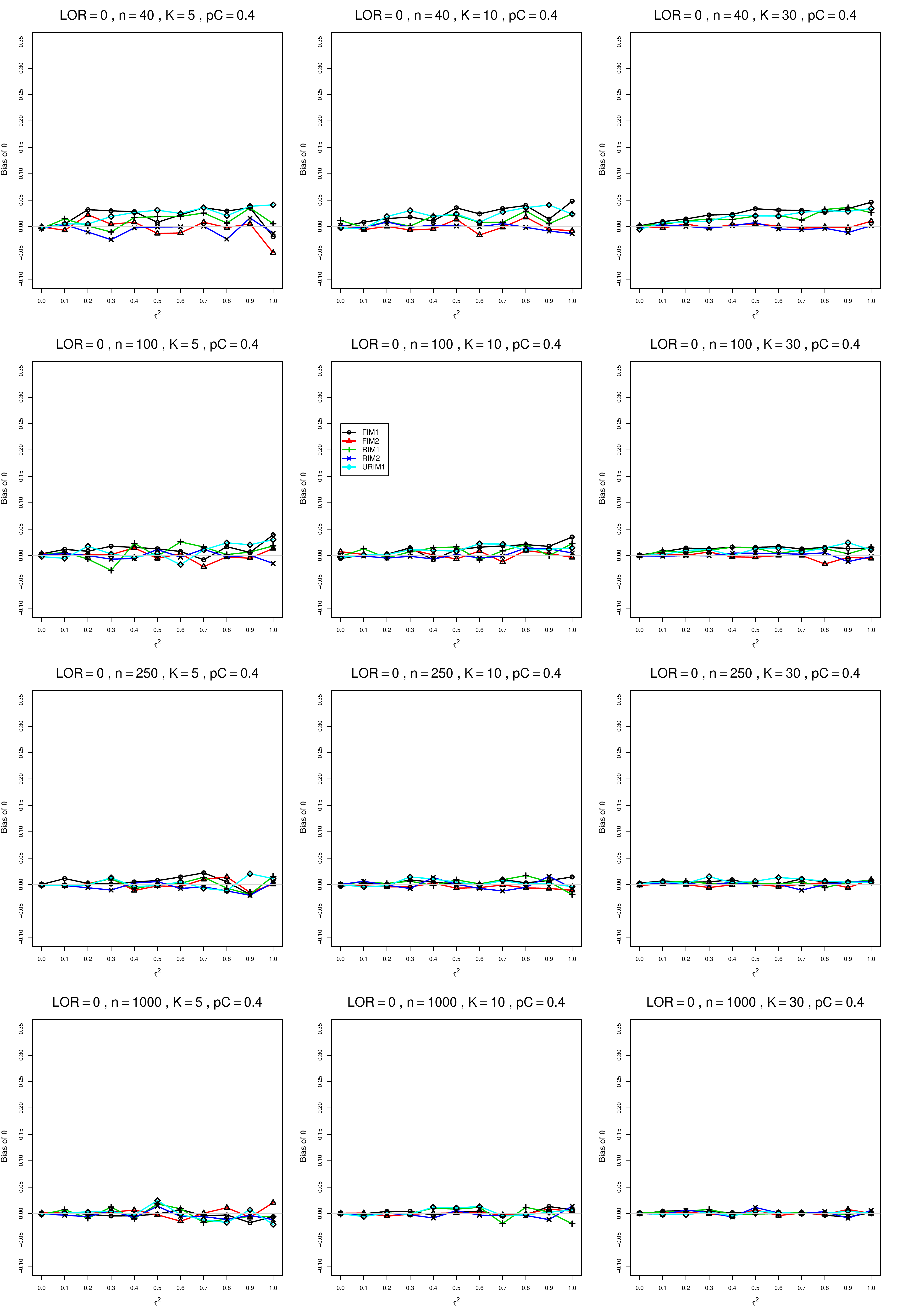}
	\caption{Bias of  overall log-odds ratio $\hat{\theta}_{RIM2}$ for $\theta=0$, $p_{C}=0.4$, $\sigma^2=0.4$, constant sample sizes $n=40,\;100,\;250,\;1000$.
The data-generation mechanisms are FIM1 ($\circ$), FIM2 ($\triangle$), RIM1 (+), RIM2 ($\times$), and URIM1 ($\diamond$).
		\label{PlotBiasThetamu0andpC04LOR_UMRSsigma04}}
\end{figure}
\begin{figure}[t]
	\centering
	\includegraphics[scale=0.33]{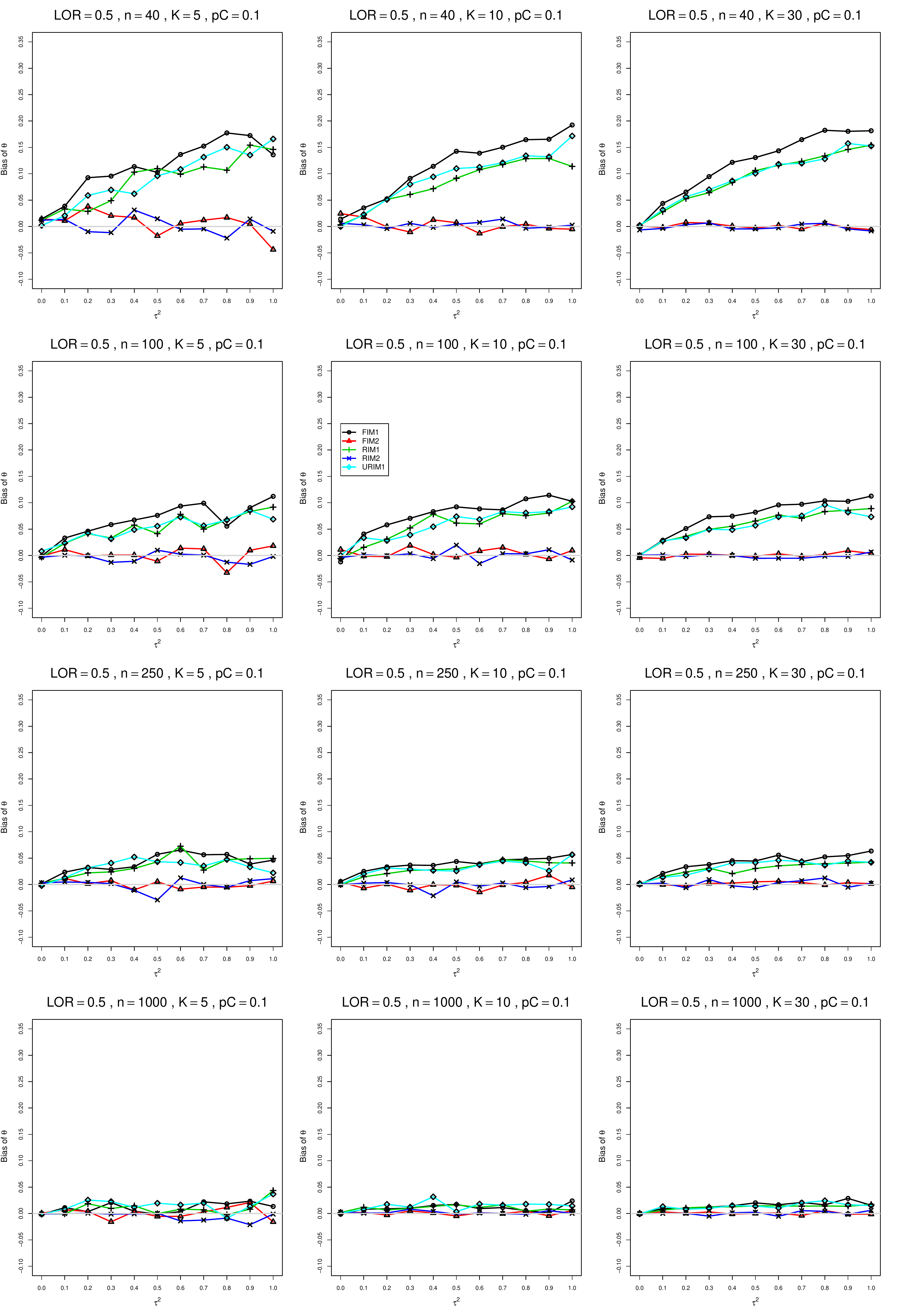}
	\caption{Bias of  overall log-odds ratio $\hat{\theta}_{RIM2}$ for $\theta=0.5$, $p_{C}=0.1$, $\sigma^2=0.4$, constant sample sizes $n=40,\;100,\;250,\;1000$.
The data-generation mechanisms are FIM1 ($\circ$), FIM2 ($\triangle$), RIM1 (+), RIM2 ($\times$), and URIM1 ($\diamond$).
		\label{PlotBiasThetamu05andpC01LOR_UMRSsigma04}}
\end{figure}
\begin{figure}[t]
	\centering
	\includegraphics[scale=0.33]{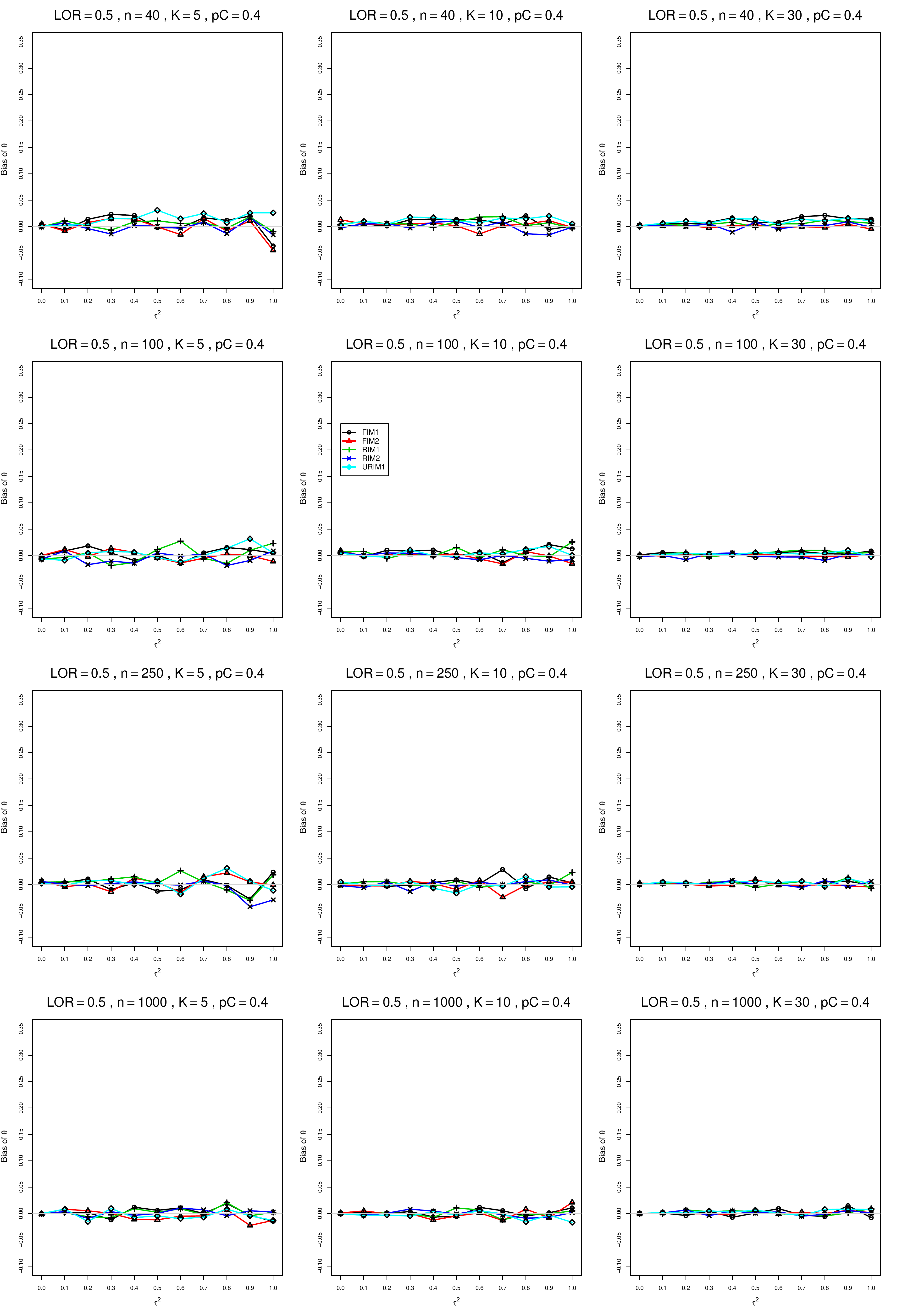}
	\caption{Bias of  overall log-odds ratio $\hat{\theta}_{RIM2}$ for $\theta=0.5$, $p_{C}=0.4$, $\sigma^2=0.4$, constant sample sizes $n=40,\;100,\;250,\;1000$.
The data-generation mechanisms are FIM1 ($\circ$), FIM2 ($\triangle$), RIM1 (+), RIM2 ($\times$), and URIM1 ($\diamond$).
		\label{PlotBiasThetamu05andpC04LOR_UMRSsigma04}}
\end{figure}
\begin{figure}[t]
	\centering
	\includegraphics[scale=0.33]{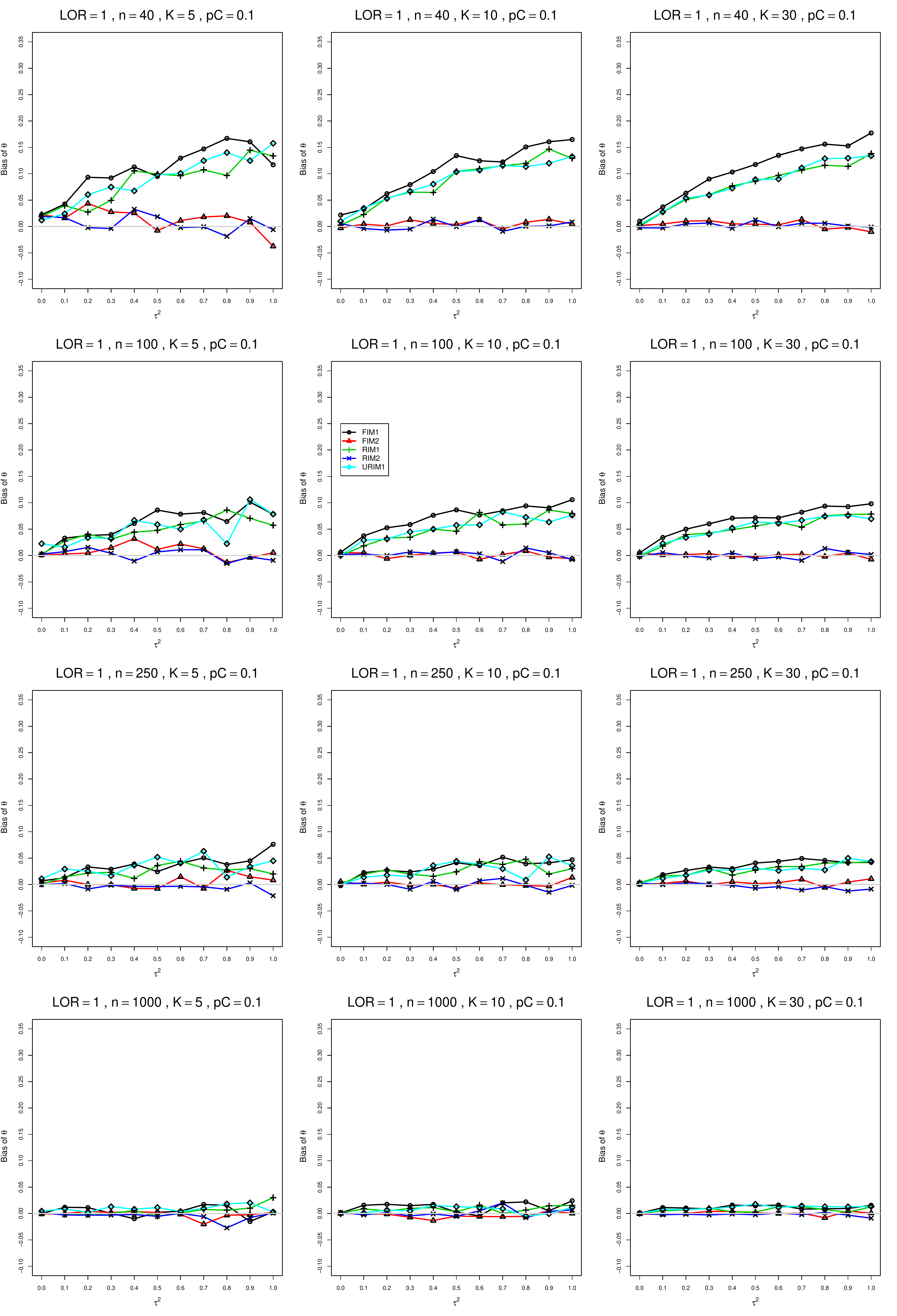}
	\caption{Bias of  overall log-odds ratio $\hat{\theta}_{RIM2}$ for $\theta=1$, $p_{C}=0.1$, $\sigma^2=0.4$, constant sample sizes $n=40,\;100,\;250,\;1000$.
The data-generation mechanisms are FIM1 ($\circ$), FIM2 ($\triangle$), RIM1 (+), RIM2 ($\times$), and URIM1 ($\diamond$).
		\label{PlotBiasThetamu1andpC01LOR_UMRSsigma04}}
\end{figure}
\begin{figure}[t]
	\centering
	\includegraphics[scale=0.33]{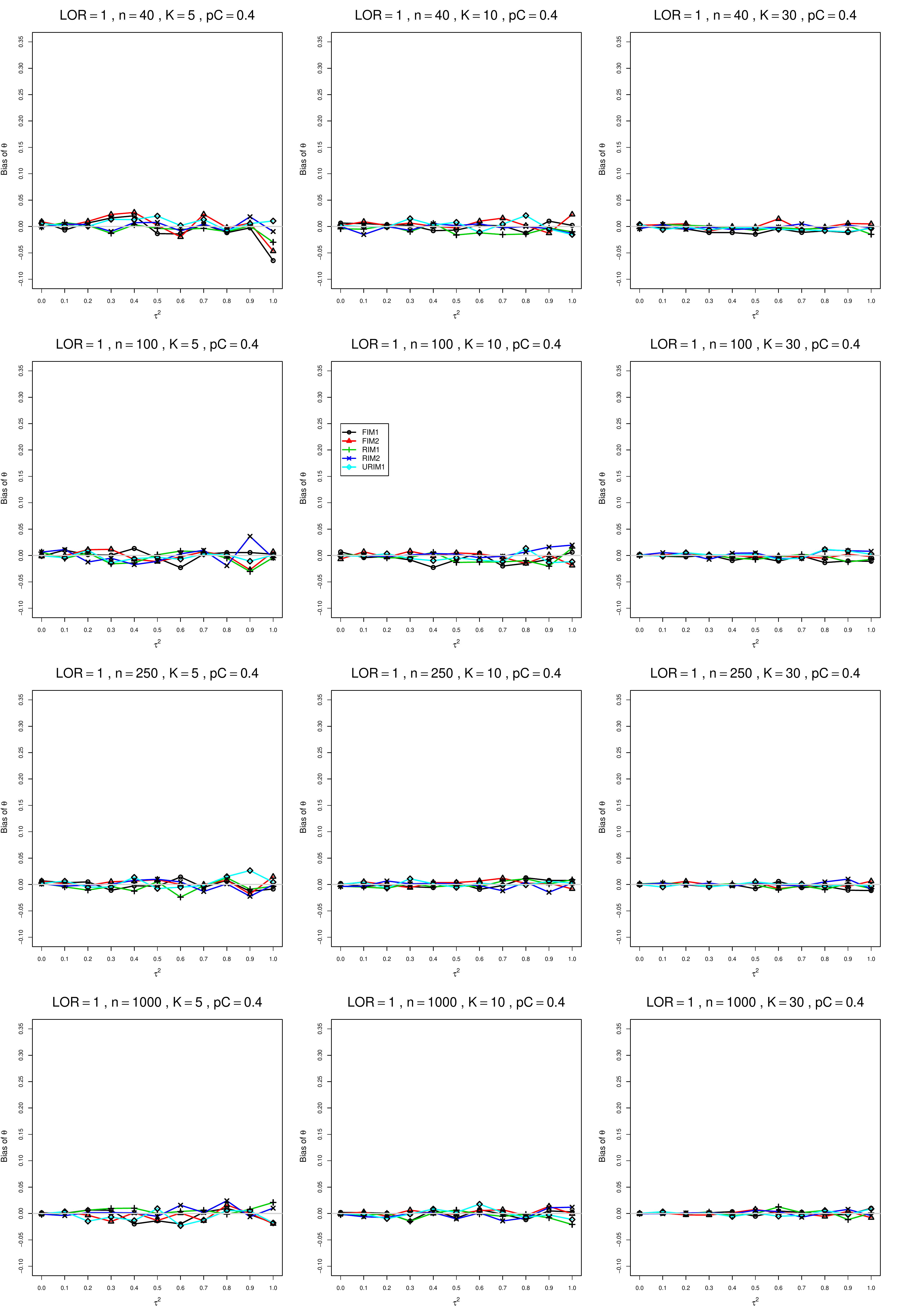}
	\caption{Bias of  overall log-odds ratio $\hat{\theta}_{RIM2}$ for $\theta=1$, $p_{C}=0.4$, $\sigma^2=0.4$, constant sample sizes $n=40,\;100,\;250,\;1000$.
The data-generation mechanisms are FIM1 ($\circ$), FIM2 ($\triangle$), RIM1 (+), RIM2 ($\times$), and URIM1 ($\diamond$).
		\label{PlotBiasThetamu1andpC04LOR_UMRSsigma04}}
\end{figure}
\begin{figure}[t]
	\centering
	\includegraphics[scale=0.33]{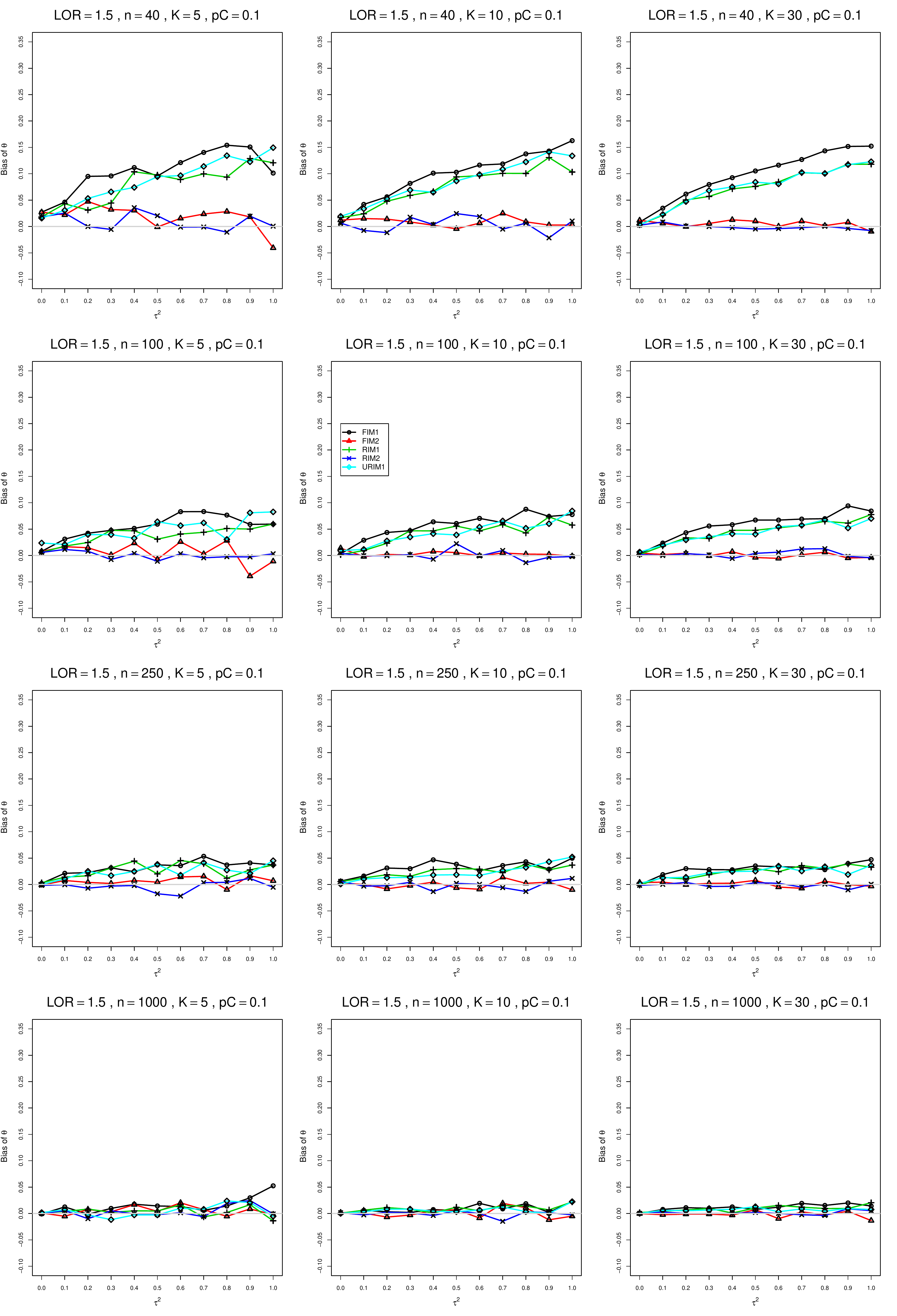}
	\caption{Bias of  overall log-odds ratio $\hat{\theta}_{RIM2}$ for $\theta=1.5$, $p_{C}=0.1$, $\sigma^2=0.4$, constant sample sizes $n=40,\;100,\;250,\;1000$.
The data-generation mechanisms are FIM1 ($\circ$), FIM2 ($\triangle$), RIM1 (+), RIM2 ($\times$), and URIM1 ($\diamond$).
		\label{PlotBiasThetamu15andpC01LOR_UMRSsigma04}}
\end{figure}
\begin{figure}[t]
	\centering
	\includegraphics[scale=0.33]{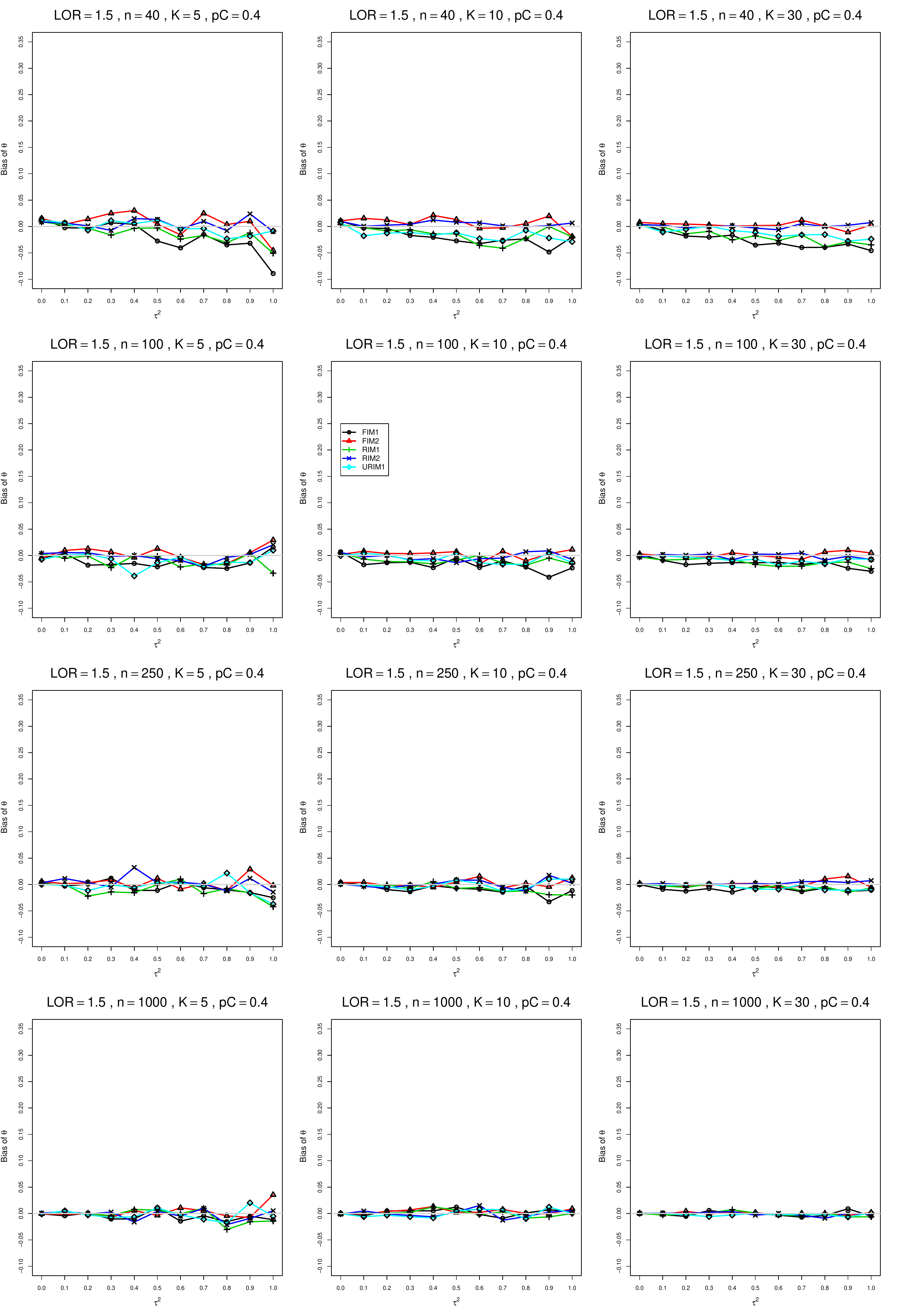}
	\caption{Bias of  overall log-odds ratio $\hat{\theta}_{RIM2}$ for $\theta=1.5$, $p_{C}=0.4$, $\sigma^2=0.4$, constant sample sizes $n=40,\;100,\;250,\;1000$.
The data-generation mechanisms are FIM1 ($\circ$), FIM2 ($\triangle$), RIM1 (+), RIM2 ($\times$), and URIM1 ($\diamond$).
		\label{PlotBiasThetamu15andpC04LOR_UMRSsigma04}}
\end{figure}
\begin{figure}[t]
	\centering
	\includegraphics[scale=0.33]{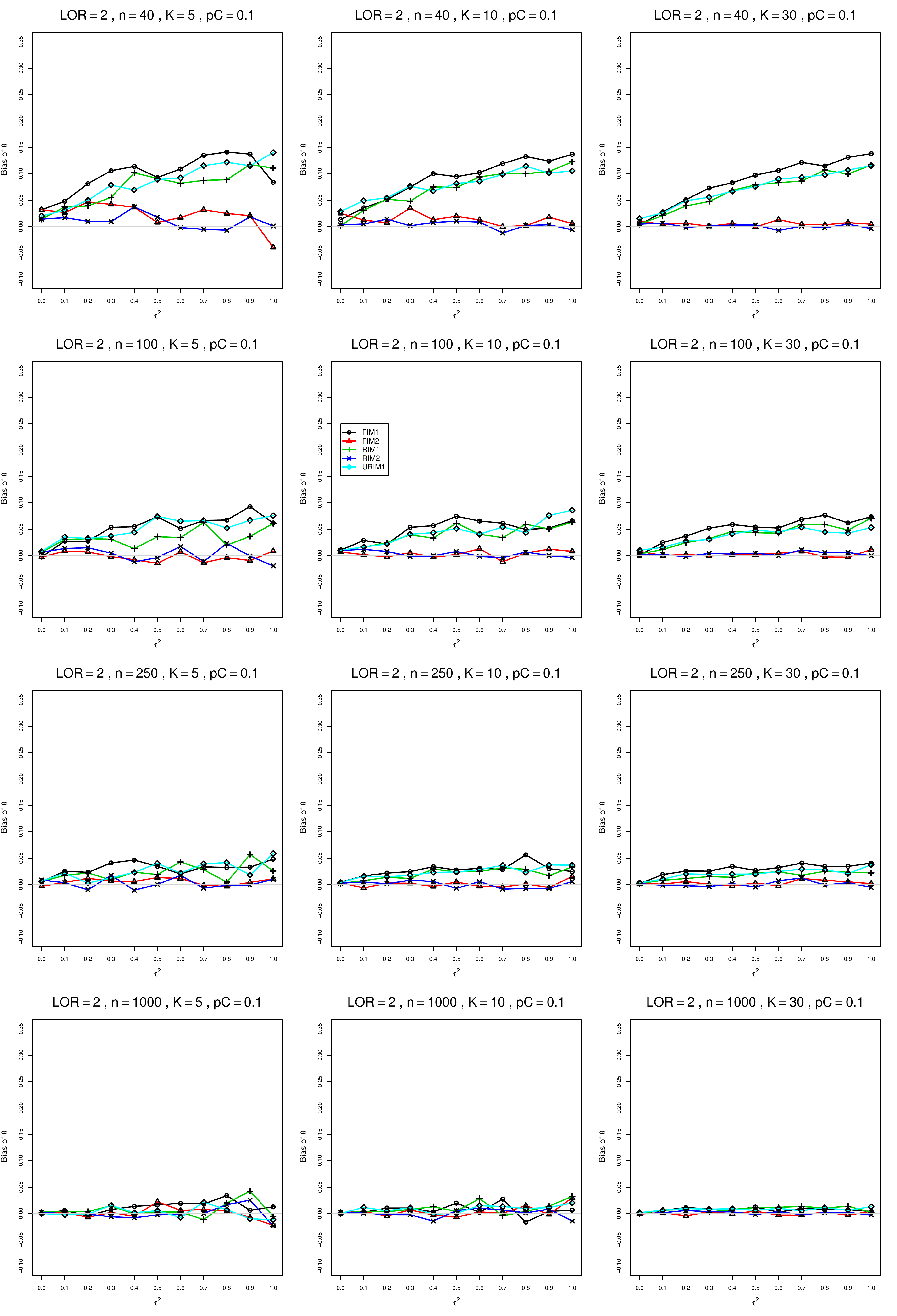}
	\caption{Bias of  overall log-odds ratio $\hat{\theta}_{RIM2}$ for $\theta=2$, $p_{C}=0.1$, $\sigma^2=0.4$, constant sample sizes $n=40,\;100,\;250,\;1000$.
The data-generation mechanisms are FIM1 ($\circ$), FIM2 ($\triangle$), RIM1 (+), RIM2 ($\times$), and URIM1 ($\diamond$).
		\label{PlotBiasThetamu2andpC01LOR_UMRSsigma04}}
\end{figure}
\begin{figure}[t]
	\centering
	\includegraphics[scale=0.33]{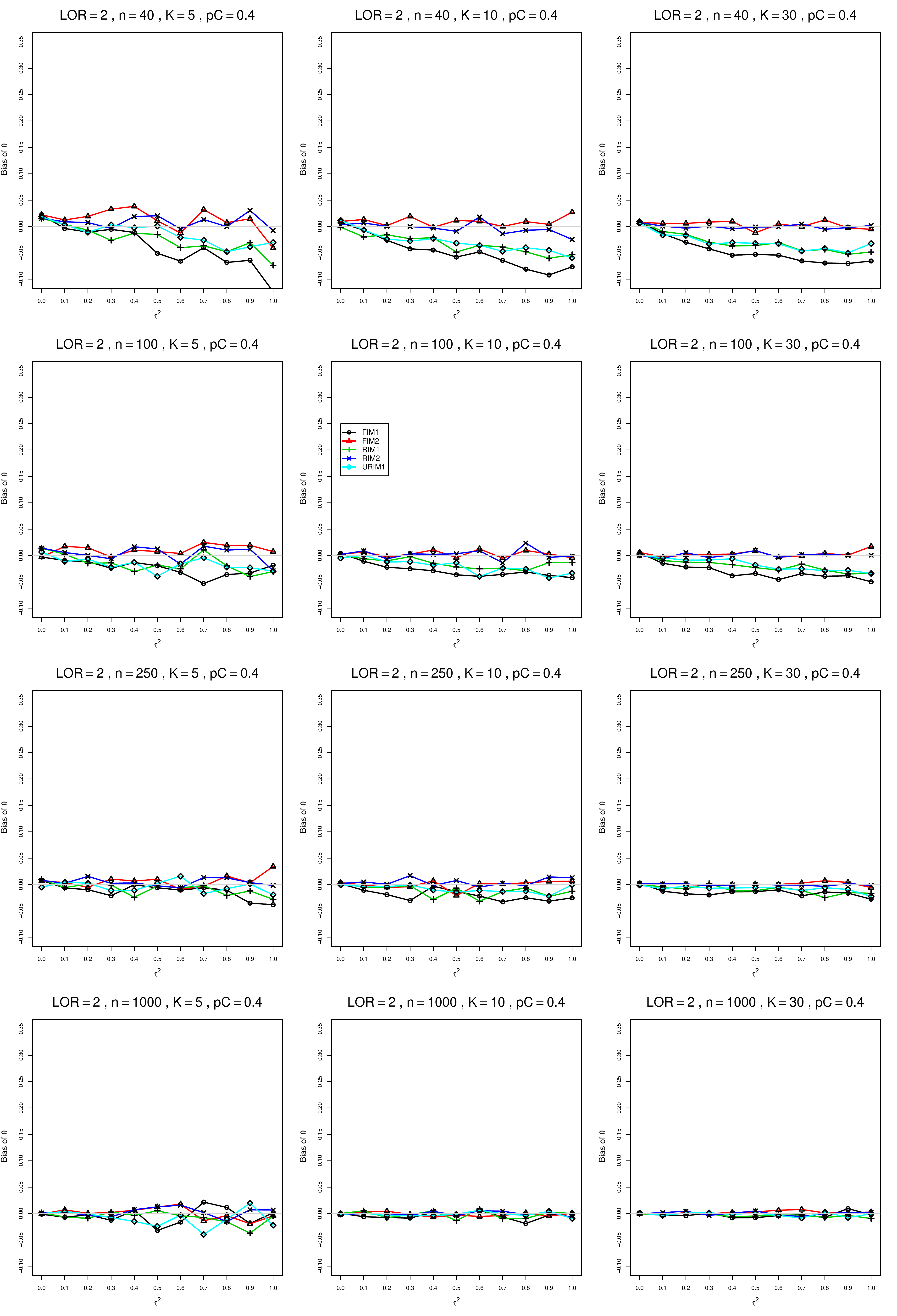}
	\caption{Bias of  overall log-odds ratio $\hat{\theta}_{RIM2}$ for $\theta=2$, $p_{C}=0.4$, $\sigma^2=0.4$, constant sample sizes $n=40,\;100,\;250,\;1000$.
The data-generation mechanisms are FIM1 ($\circ$), FIM2 ($\triangle$), RIM1 (+), RIM2 ($\times$), and URIM1 ($\diamond$).
		\label{PlotBiasThetamu2andpC04LOR_UMRSsigma04}}
\end{figure}

\clearpage
\subsection*{A2.7 Bias of $\hat{\theta}_{SSW}$}
\renewcommand{\thefigure}{A2.7.\arabic{figure}}
\setcounter{figure}{0}

\begin{figure}[t]
	\centering
	\includegraphics[scale=0.33]{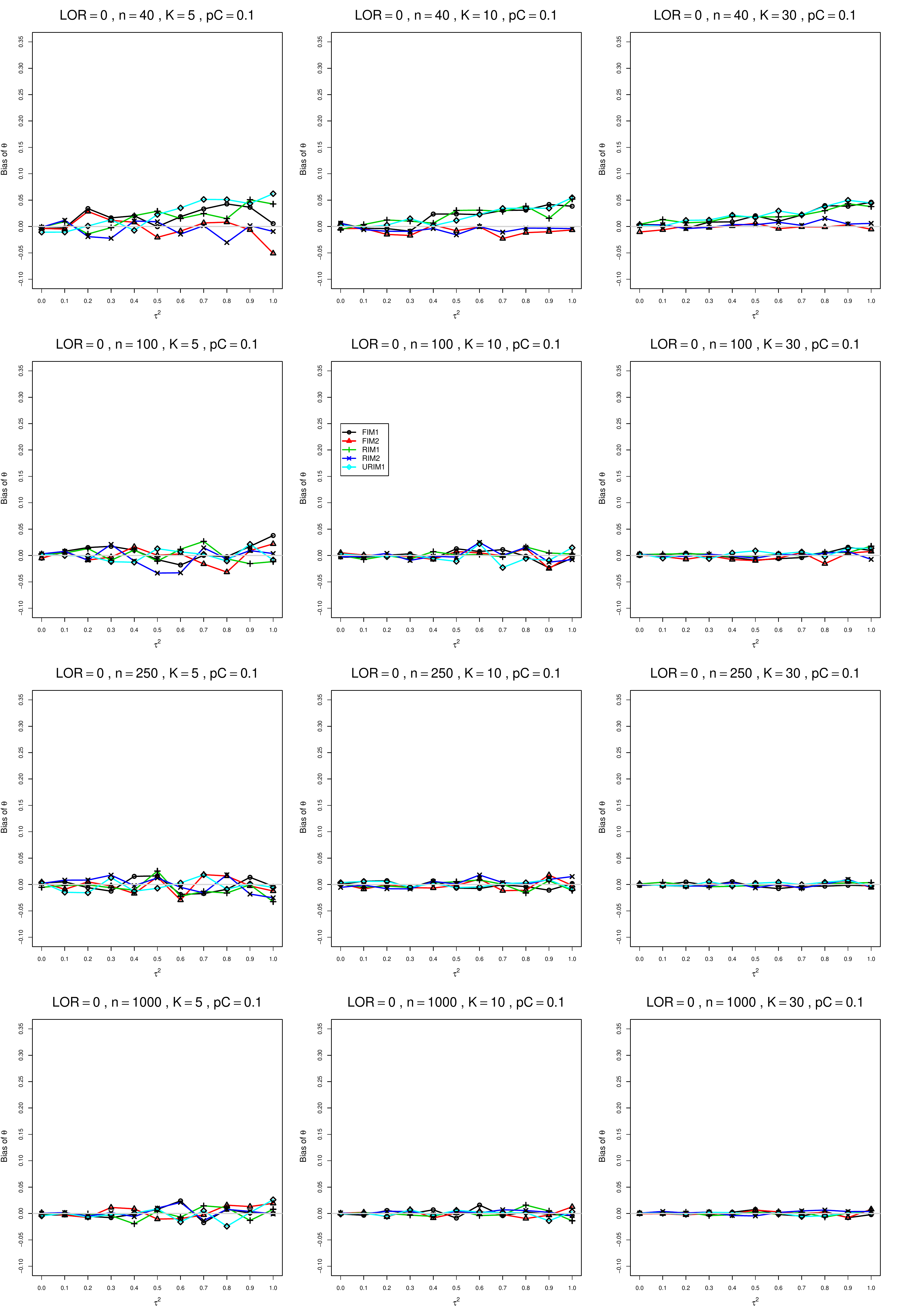}
	\caption{Bias of overall log-odds ratio $\hat{\theta}_{SSW}$ for $\theta=0$, $p_{C}=0.1$, $\sigma^2=0.1$, constant sample sizes $n=40,\;100,\;250,\;1000$.
The data-generation mechanisms are FIM1 ($\circ$), FIM2 ($\triangle$), RIM1 (+), RIM2 ($\times$), and URIM1 ($\diamond$).
		\label{PlotBiasThetamu0andpC01LOR_SSWsigma01}}
\end{figure}
\begin{figure}[t]
	\centering
	\includegraphics[scale=0.33]{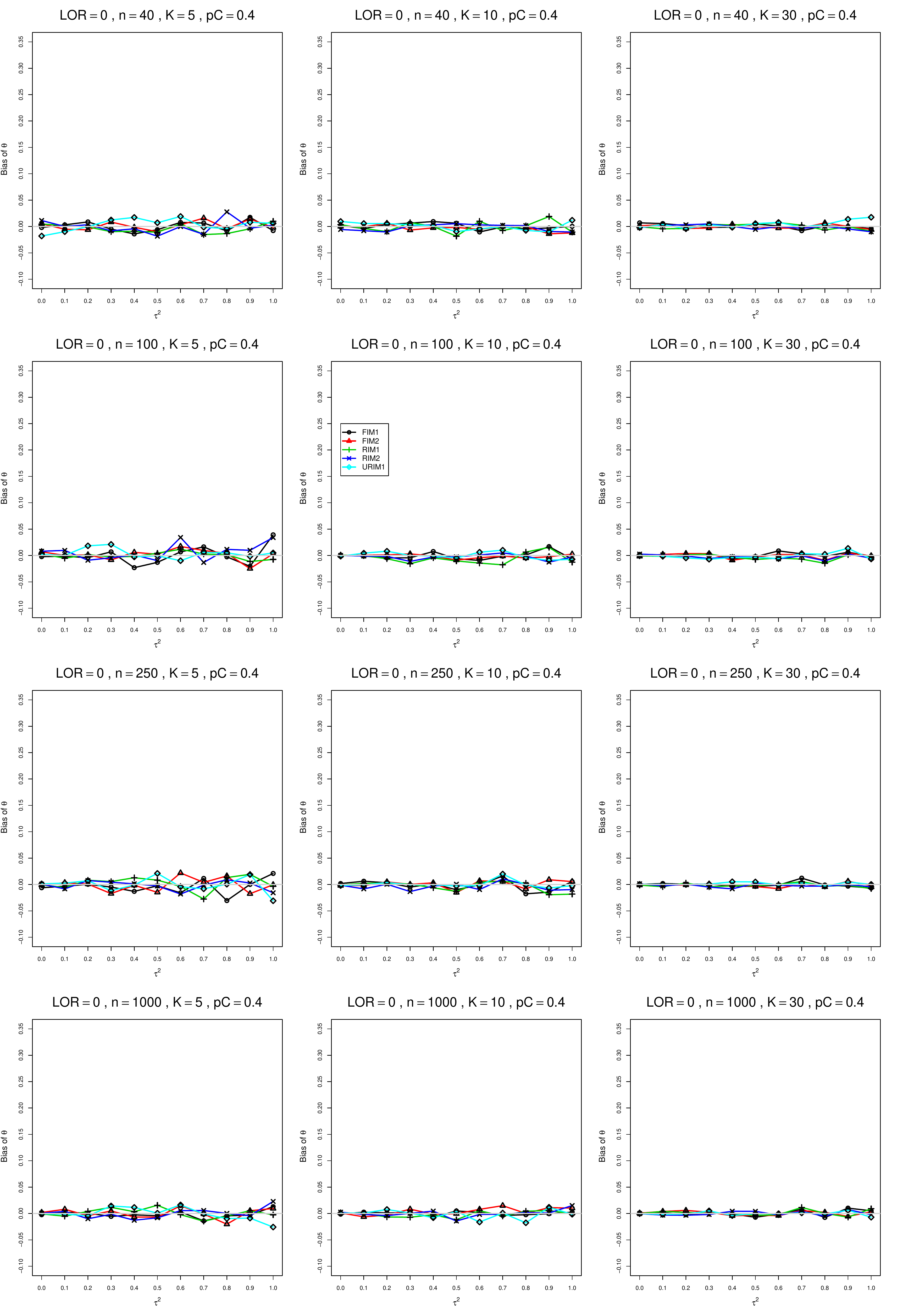}
	\caption{Bias of overall log-odds ratio $\hat{\theta}_{SSW}$ for $\theta=0$, $p_{C}=0.4$, $\sigma^2=0.1$, constant sample sizes $n=40,\;100,\;250,\;1000$.
The data-generation mechanisms are FIM1 ($\circ$), FIM2 ($\triangle$), RIM1 (+), RIM2 ($\times$), and URIM1 ($\diamond$).
		\label{PlotBiasThetamu0andpC04LOR_SSWsigma01}}
\end{figure}
\begin{figure}[t]
	\centering
	\includegraphics[scale=0.33]{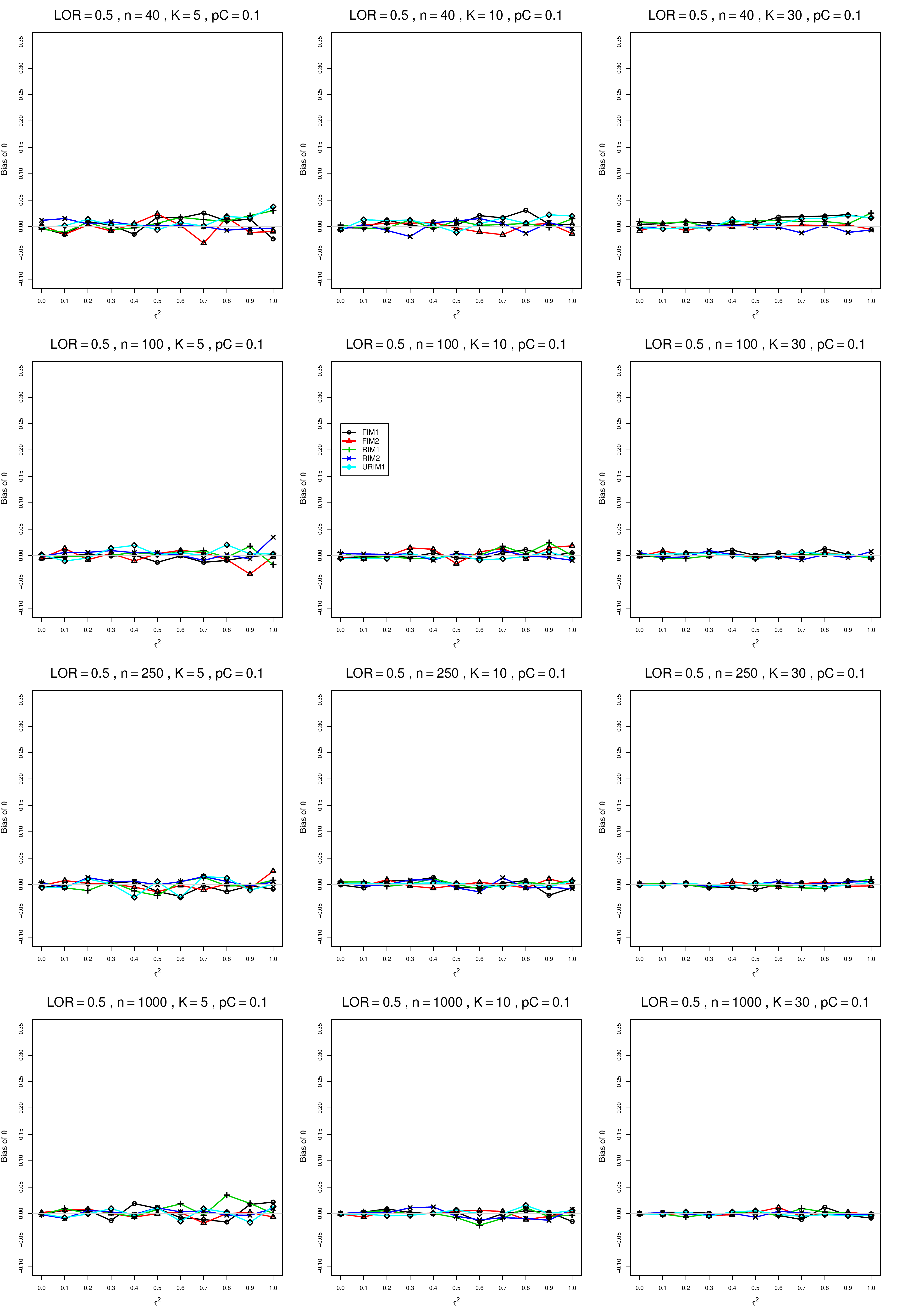}
	\caption{Bias of overall log-odds ratio $\hat{\theta}_{SSW}$ for $\theta=0.5$, $p_{C}=0.1$, $\sigma^2=0.1$, constant sample sizes $n=40,\;100,\;250,\;1000$.
The data-generation mechanisms are FIM1 ($\circ$), FIM2 ($\triangle$), RIM1 (+), RIM2 ($\times$), and URIM1 ($\diamond$).
		\label{PlotBiasThetamu05andpC01LOR_SSWsigma01}}
\end{figure}
\begin{figure}[t]
	\centering
	\includegraphics[scale=0.33]{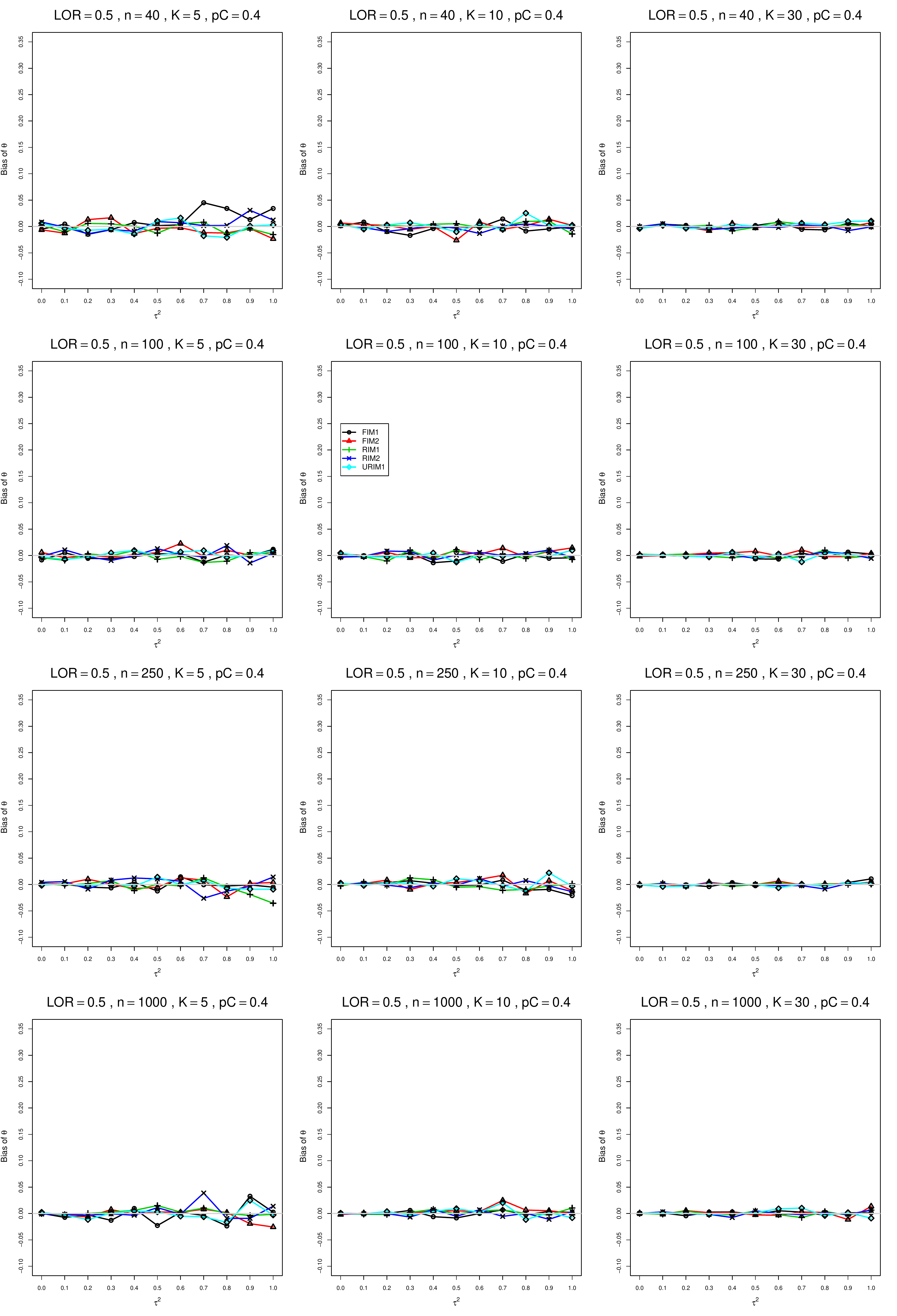}
	\caption{Bias of overall log-odds ratio $\hat{\theta}_{SSW}$ for $\theta=0.5$, $p_{C}=0.4$, $\sigma^2=0.1$, constant sample sizes $n=40,\;100,\;250,\;1000$.
The data-generation mechanisms are FIM1 ($\circ$), FIM2 ($\triangle$), RIM1 (+), RIM2 ($\times$), and URIM1 ($\diamond$).
		\label{PlotBiasThetamu05andpC04LOR_SSWsigma01}}
\end{figure}
\begin{figure}[t]
	\centering
	\includegraphics[scale=0.33]{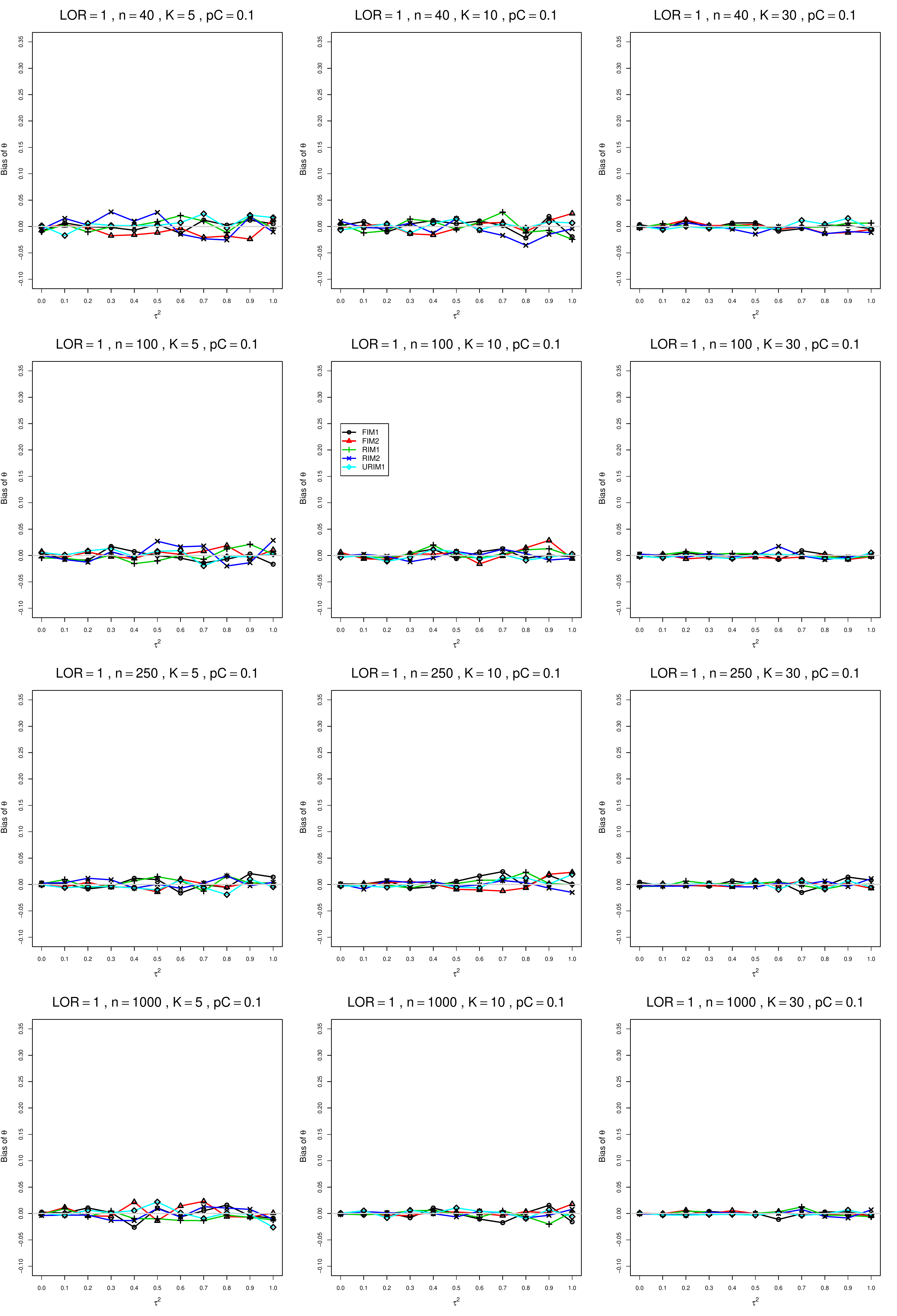}
	\caption{Bias of overall log-odds ratio $\hat{\theta}_{SSW}$ for $\theta=1$, $p_{C}=0.1$, $\sigma^2=0.1$, constant sample sizes $n=40,\;100,\;250,\;1000$.
The data-generation mechanisms are FIM1 ($\circ$), FIM2 ($\triangle$), RIM1 (+), RIM2 ($\times$), and URIM1 ($\diamond$).
		\label{PlotBiasThetamu1andpC01LOR_SSWsigma01}}
\end{figure}
\begin{figure}[t]
	\centering
	\includegraphics[scale=0.33]{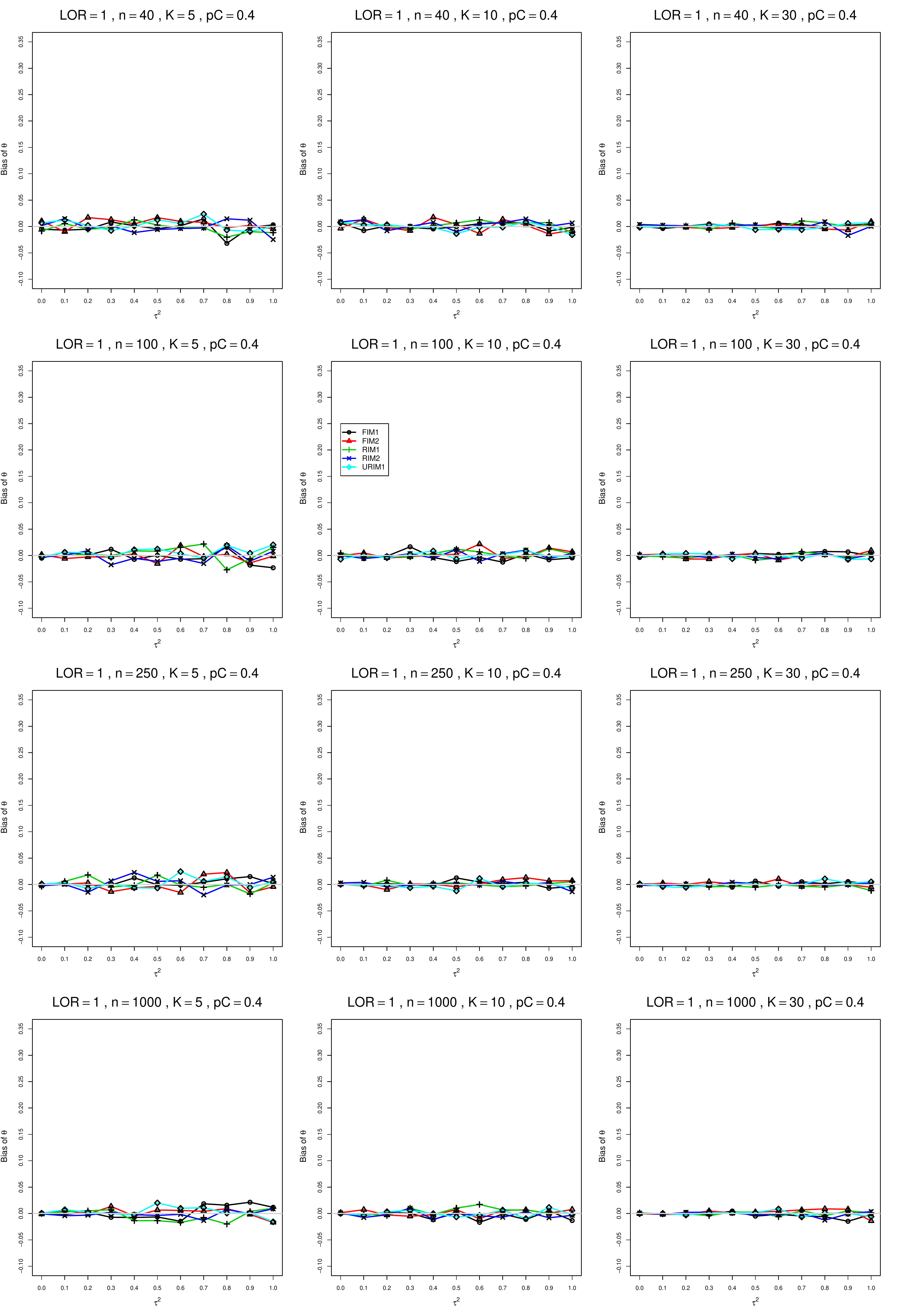}
	\caption{Bias of overall log-odds ratio $\hat{\theta}_{SSW}$ for $\theta=1$, $p_{C}=0.4$, $\sigma^2=0.1$, constant sample sizes $n=40,\;100,\;250,\;1000$.
The data-generation mechanisms are FIM1 ($\circ$), FIM2 ($\triangle$), RIM1 (+), RIM2 ($\times$), and URIM1 ($\diamond$).
		\label{PlotBiasThetamu1andpC04LOR_SSWsigma01}}
\end{figure}
\begin{figure}[t]
	\centering
	\includegraphics[scale=0.33]{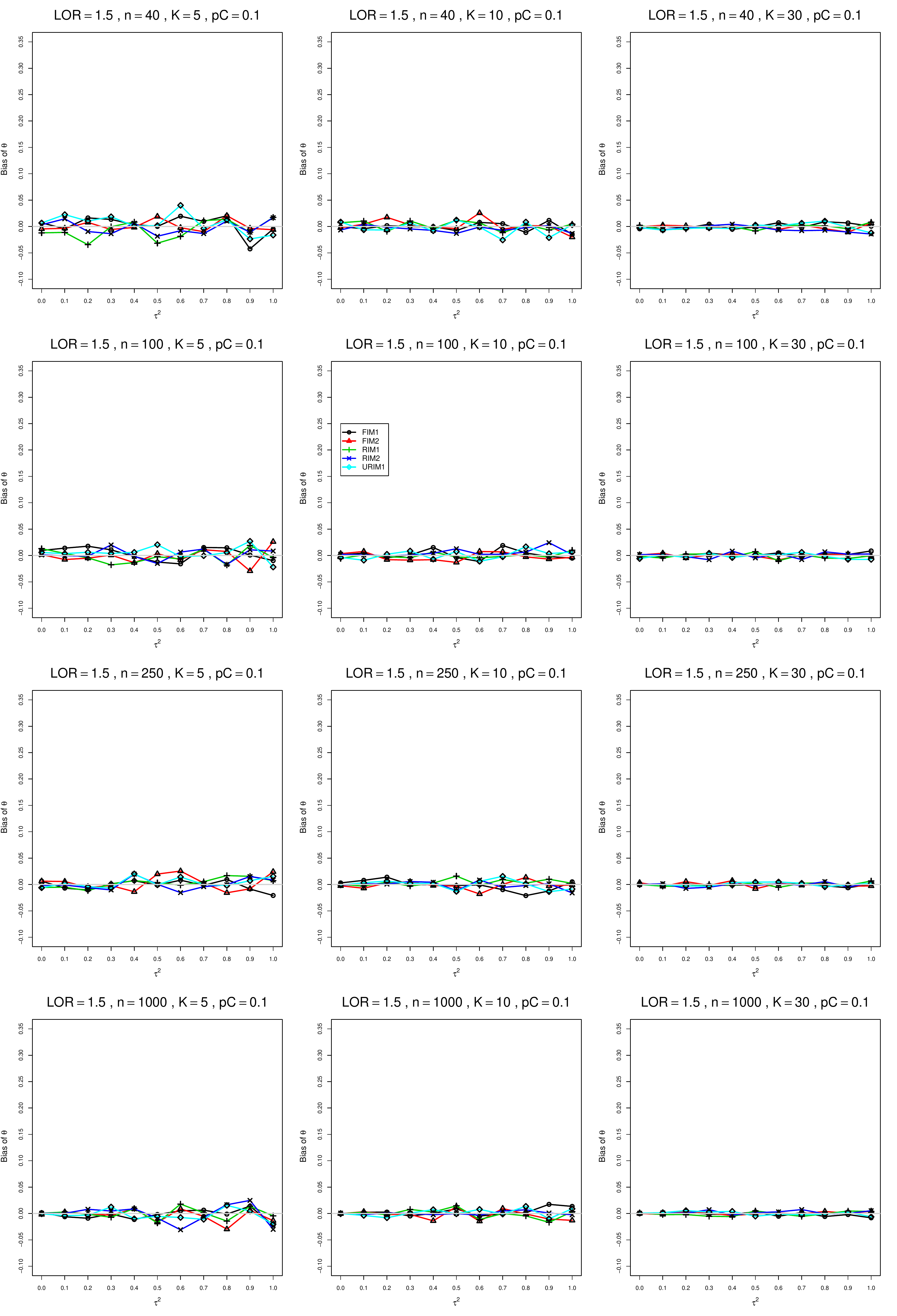}
	\caption{Bias of overall log-odds ratio $\hat{\theta}_{SSW}$ for $\theta=1.5$, $p_{C}=0.1$, $\sigma^2=0.1$, constant sample sizes $n=40,\;100,\;250,\;1000$.
The data-generation mechanisms are FIM1 ($\circ$), FIM2 ($\triangle$), RIM1 (+), RIM2 ($\times$), and URIM1 ($\diamond$).
		\label{PlotBiasThetamu15andpC01LOR_SSWsigma01}}
\end{figure}
\begin{figure}[t]
	\centering
	\includegraphics[scale=0.33]{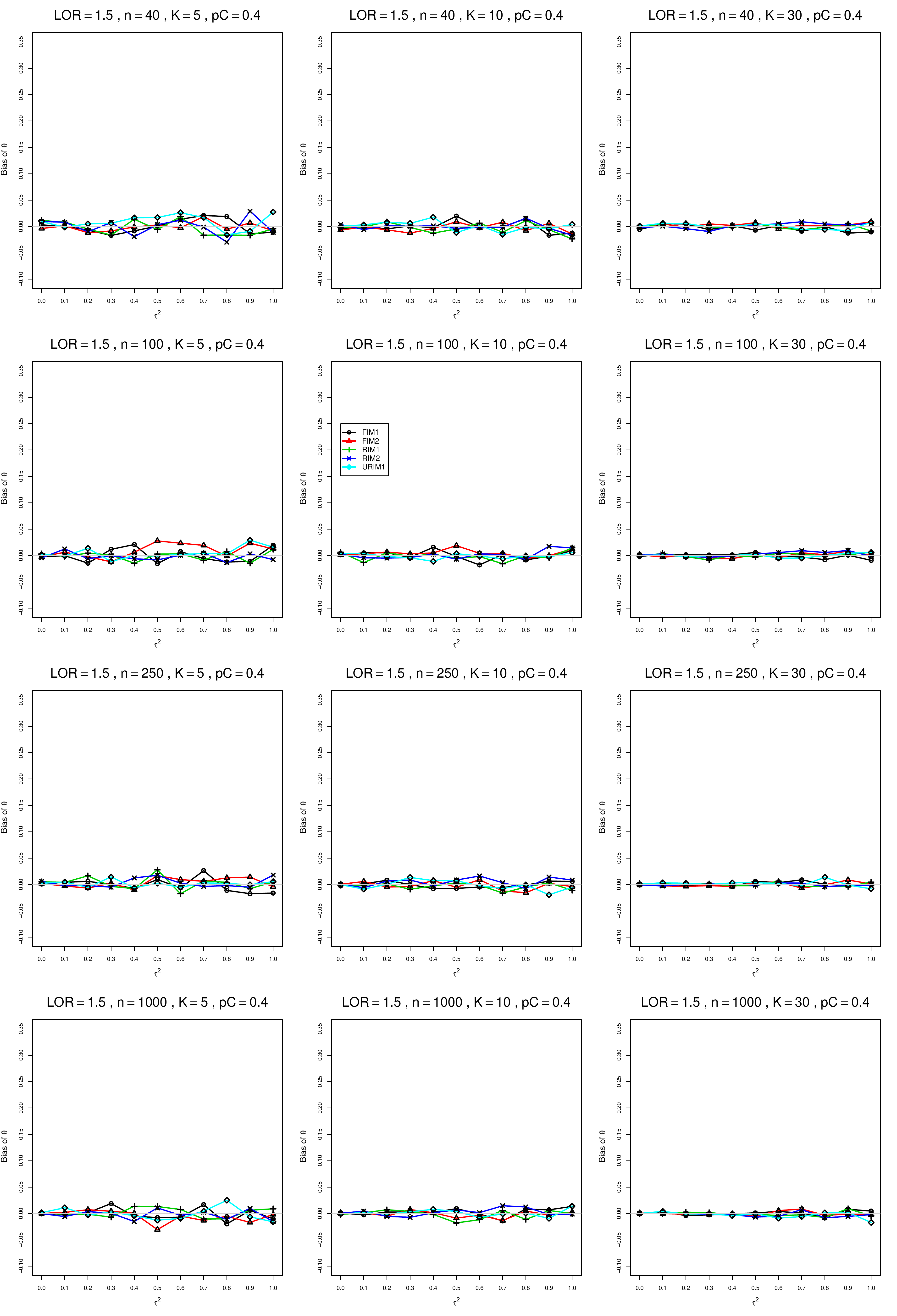}
	\caption{Bias of overall log-odds ratio $\hat{\theta}_{SSW}$ for $\theta=1.5$, $p_{C}=0.4$, $\sigma^2=0.1$, constant sample sizes $n=40,\;100,\;250,\;1000$.
The data-generation mechanisms are FIM1 ($\circ$), FIM2 ($\triangle$), RIM1 (+), RIM2 ($\times$), and URIM1 ($\diamond$).
		\label{PlotBiasThetamu15andpC04LOR_SSWsigma01}}
\end{figure}
\begin{figure}[t]
	\centering
	\includegraphics[scale=0.33]{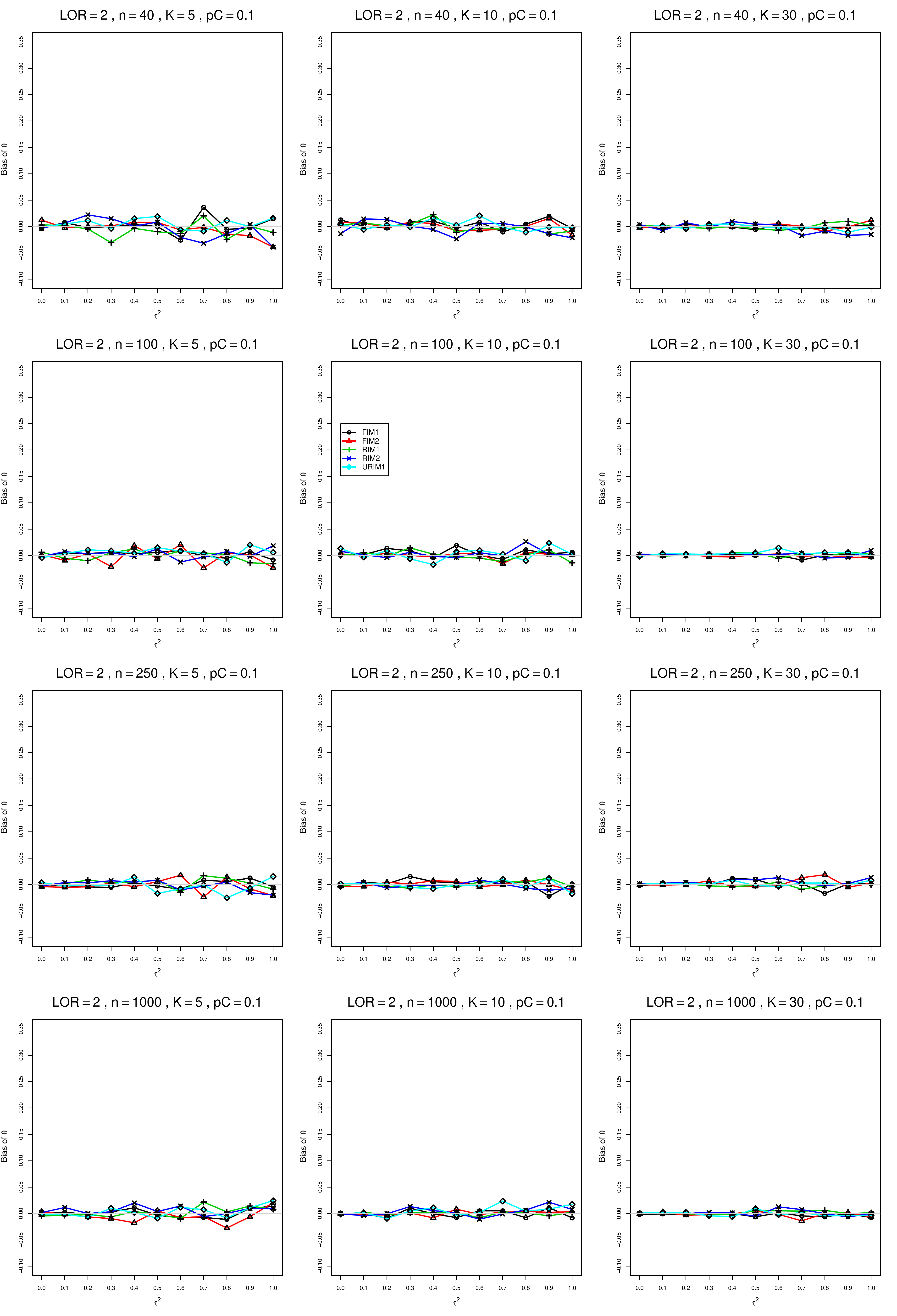}
	\caption{Bias of overall log-odds ratio $\hat{\theta}_{SSW}$ for $\theta=2$, $p_{C}=0.1$, $\sigma^2=0.1$, constant sample sizes $n=40,\;100,\;250,\;1000$.
The data-generation mechanisms are FIM1 ($\circ$), FIM2 ($\triangle$), RIM1 (+), RIM2 ($\times$), and URIM1 ($\diamond$).
		\label{PlotBiasThetamu2andpC01LOR_SSWsigma01}}
\end{figure}
\begin{figure}[t]
	\centering
	\includegraphics[scale=0.33]{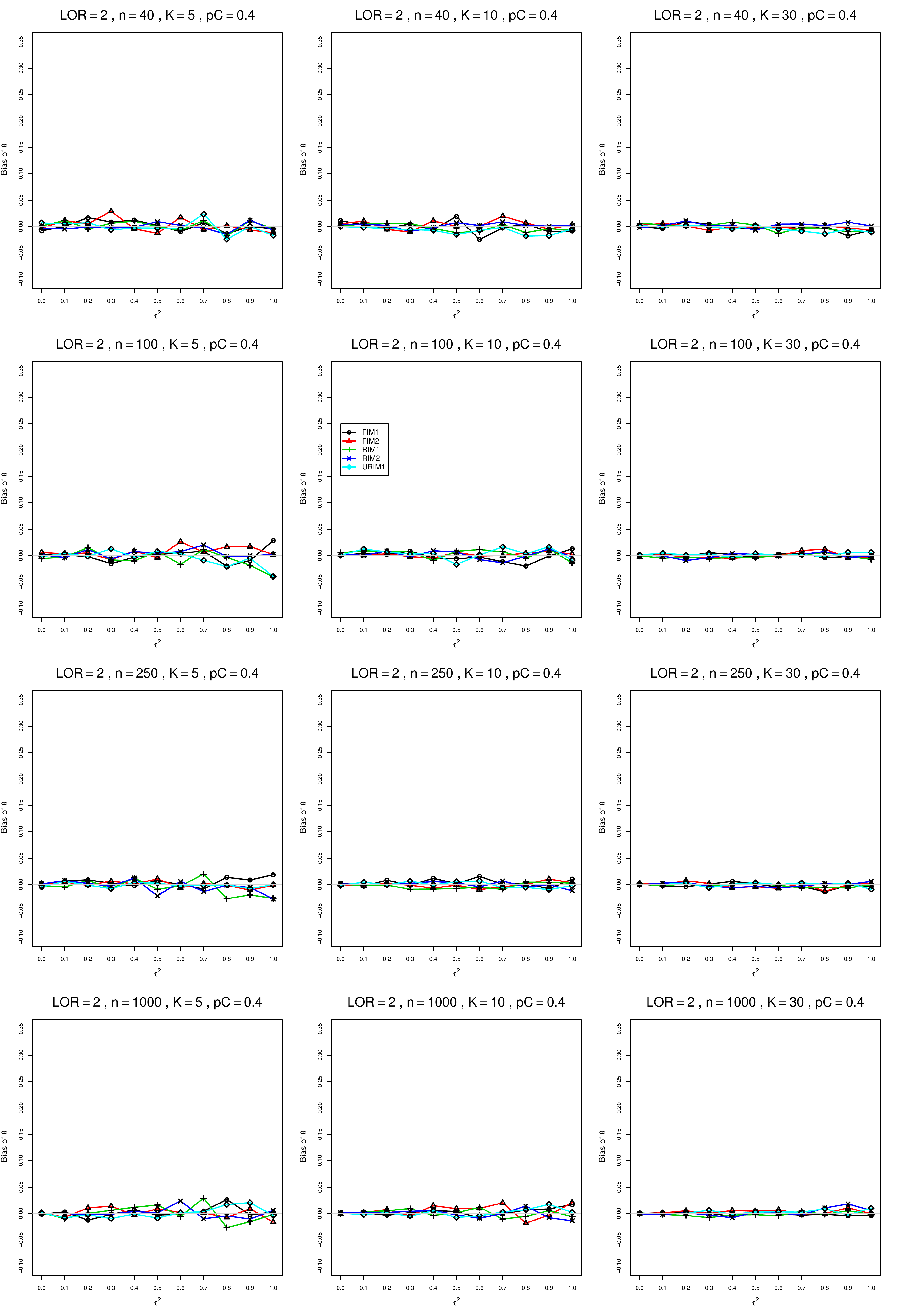}
	\caption{Bias of overall log-odds ratio $\hat{\theta}_{SSW}$ for $\theta=2$, $p_{C}=0.4$, $\sigma^2=0.1$, constant sample sizes $n=40,\;100,\;250,\;1000$.
The data-generation mechanisms are FIM1 ($\circ$), FIM2 ($\triangle$), RIM1 (+), RIM2 ($\times$), and URIM1 ($\diamond$).
		\label{PlotBiasThetamu2andpC04LOR_SSWsigma01}}
\end{figure}
\begin{figure}[t]
	\centering
	\includegraphics[scale=0.33]{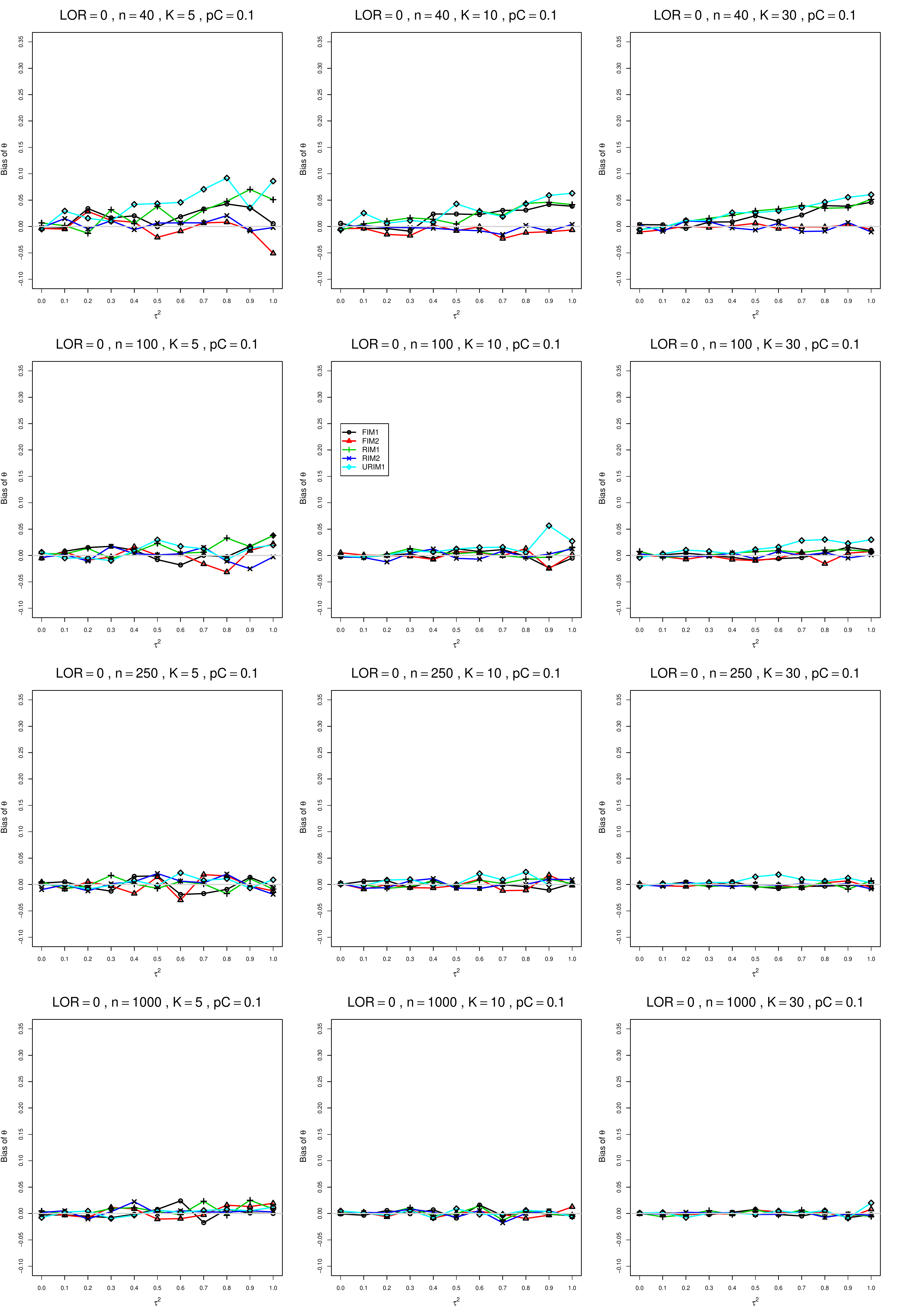}
	\caption{Bias of overall log-odds ratio $\hat{\theta}_{SSW}$ for $\theta=0$, $p_{C}=0.1$, $\sigma^2=0.4$, constant sample sizes $n=40,\;100,\;250,\;1000$.
The data-generation mechanisms are FIM1 ($\circ$), FIM2 ($\triangle$), RIM1 (+), RIM2 ($\times$), and URIM1 ($\diamond$).
		\label{PlotBiasThetamu0andpC01LOR_SSWsigma04}}
\end{figure}
\begin{figure}[t]
	\centering
	\includegraphics[scale=0.33]{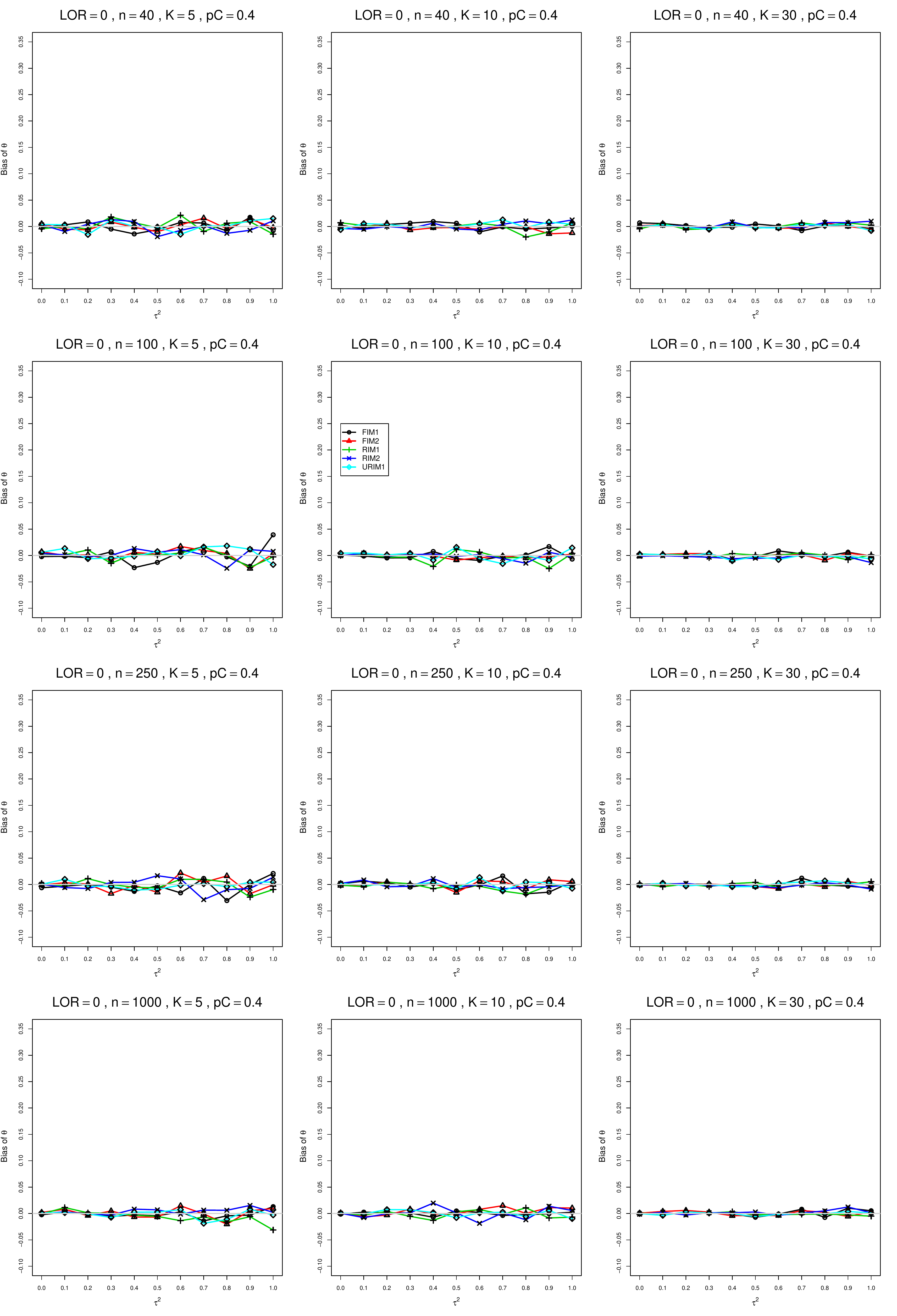}
	\caption{Bias of overall log-odds ratio $\hat{\theta}_{SSW}$ for $\theta=0$, $p_{C}=0.4$, $\sigma^2=0.4$, constant sample sizes $n=40,\;100,\;250,\;1000$.
The data-generation mechanisms are FIM1 ($\circ$), FIM2 ($\triangle$), RIM1 (+), RIM2 ($\times$), and URIM1 ($\diamond$).
		\label{PlotBiasThetamu0andpC04LOR_SSWsigma04}}
\end{figure}
\begin{figure}[t]
	\centering
	\includegraphics[scale=0.33]{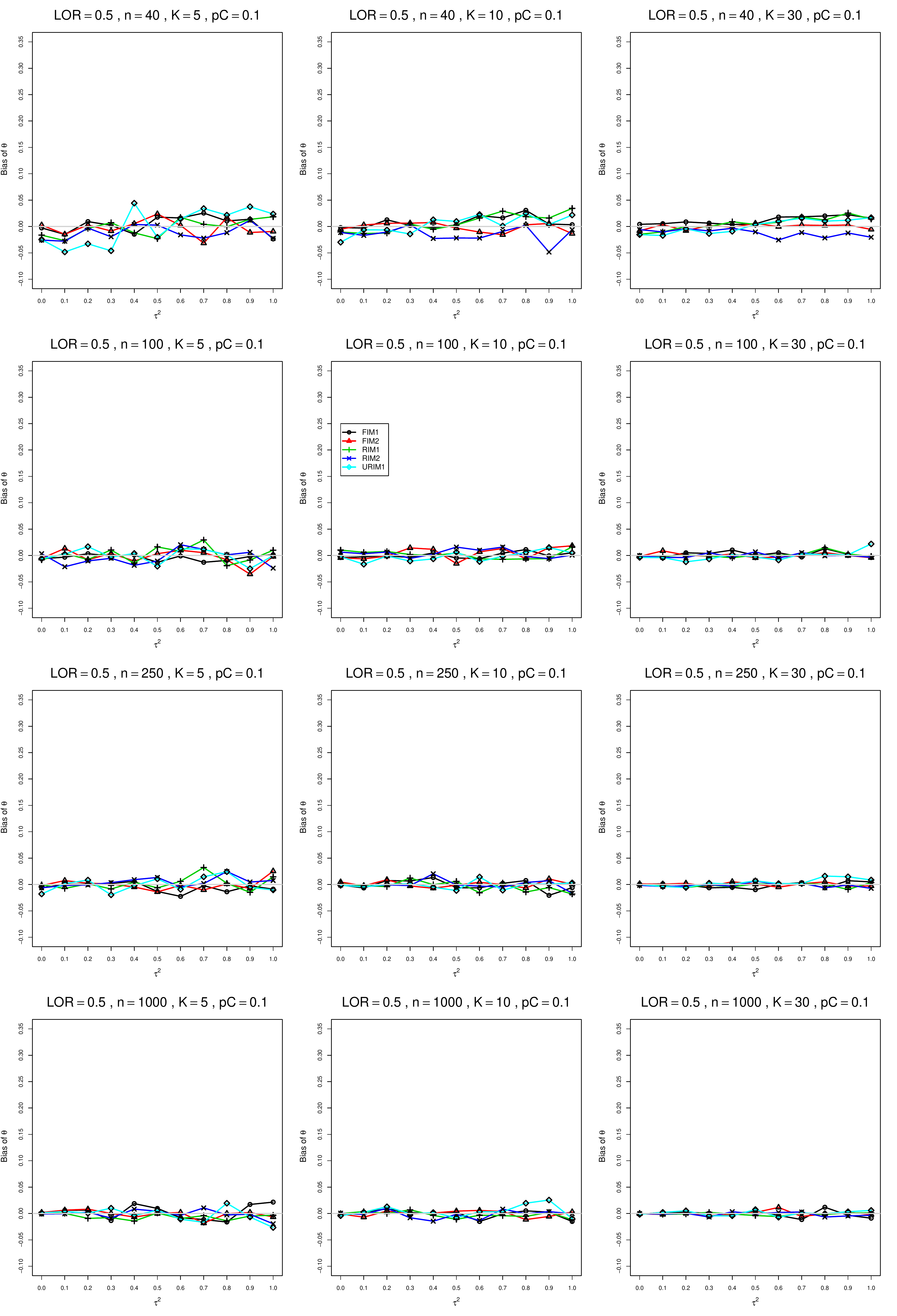}
	\caption{Bias of overall log-odds ratio $\hat{\theta}_{SSW}$ for $\theta=0.5$, $p_{C}=0.1$, $\sigma^2=0.4$, constant sample sizes $n=40,\;100,\;250,\;1000$.
The data-generation mechanisms are FIM1 ($\circ$), FIM2 ($\triangle$), RIM1 (+), RIM2 ($\times$), and URIM1 ($\diamond$).
		\label{PlotBiasThetamu05andpC01LOR_SSWsigma04}}
\end{figure}
\begin{figure}[t]
	\centering
	\includegraphics[scale=0.33]{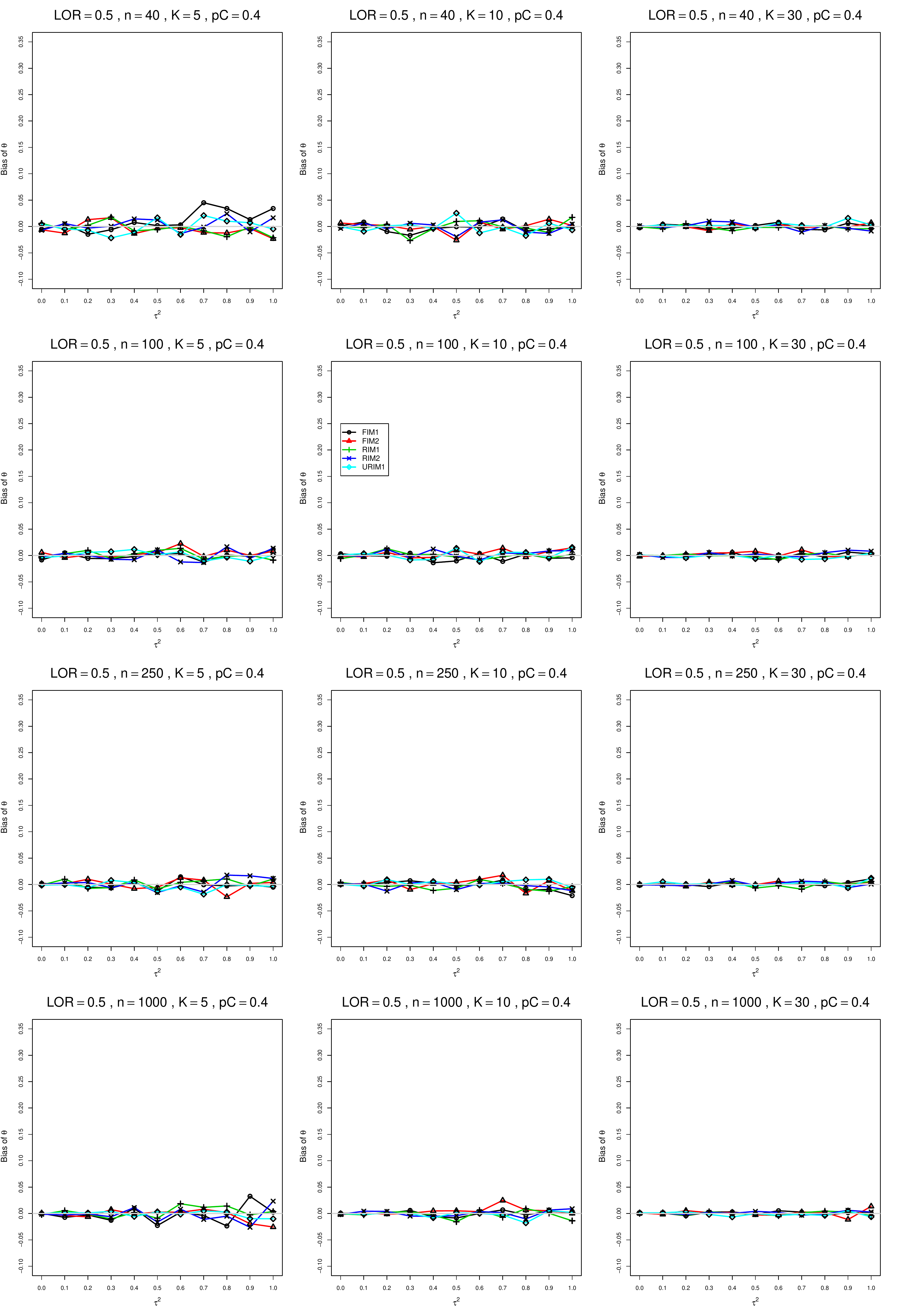}
	\caption{Bias of overall log-odds ratio $\hat{\theta}_{SSW}$ for $\theta=0.5$, $p_{C}=0.4$, $\sigma^2=0.4$, constant sample sizes $n=40,\;100,\;250,\;1000$.
The data-generation mechanisms are FIM1 ($\circ$), FIM2 ($\triangle$), RIM1 (+), RIM2 ($\times$), and URIM1 ($\diamond$).
		\label{PlotBiasThetamu05andpC04LOR_SSWsigma04}}
\end{figure}
\begin{figure}[t]
	\centering
	\includegraphics[scale=0.33]{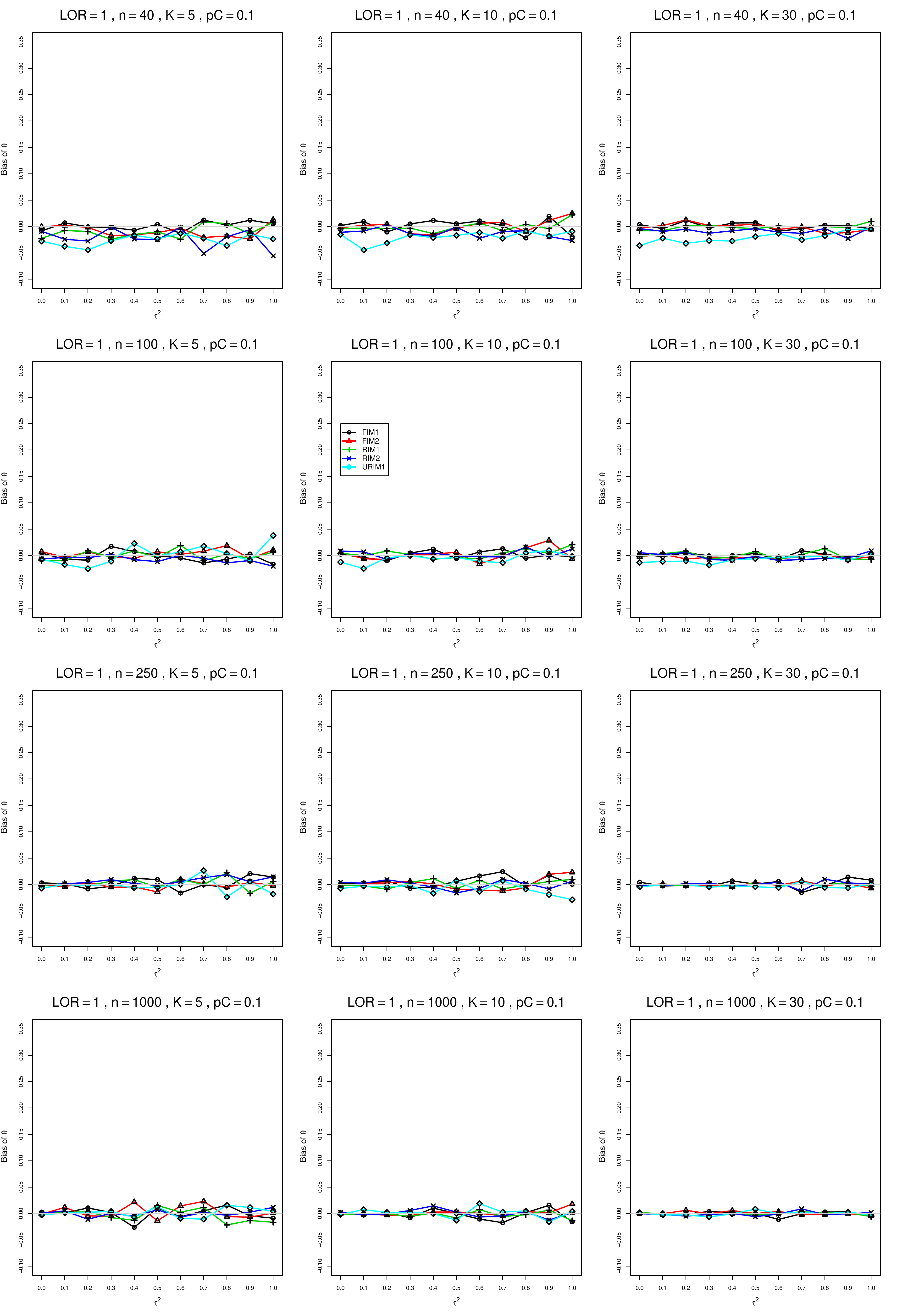}
	\caption{Bias of overall log-odds ratio $\hat{\theta}_{SSW}$ for $\theta=1$, $p_{C}=0.1$, $\sigma^2=0.4$, constant sample sizes $n=40,\;100,\;250,\;1000$.
The data-generation mechanisms are FIM1 ($\circ$), FIM2 ($\triangle$), RIM1 (+), RIM2 ($\times$), and URIM1 ($\diamond$).
		\label{PlotBiasThetamu1andpC01LOR_SSWsigma04}}
\end{figure}
\begin{figure}[t]
	\centering
	\includegraphics[scale=0.33]{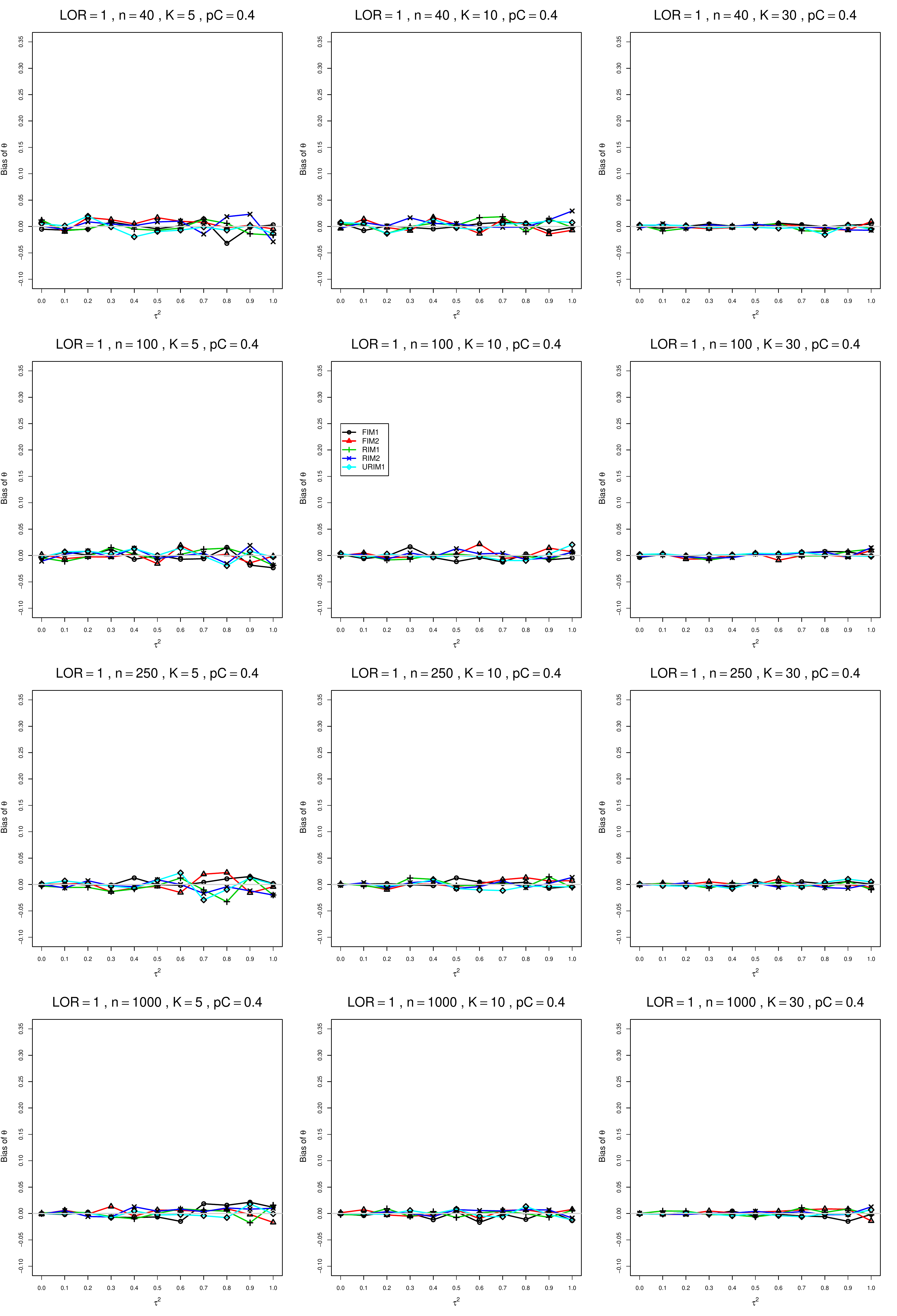}
	\caption{Bias of overall log-odds ratio $\hat{\theta}_{SSW}$ for $\theta=1$, $p_{C}=0.4$, $\sigma^2=0.4$, constant sample sizes $n=40,\;100,\;250,\;1000$.
The data-generation mechanisms are FIM1 ($\circ$), FIM2 ($\triangle$), RIM1 (+), RIM2 ($\times$), and URIM1 ($\diamond$).
		\label{PlotBiasThetamu1andpC04LOR_SSWsigma04}}
\end{figure}
\begin{figure}[t]
	\centering
	\includegraphics[scale=0.33]{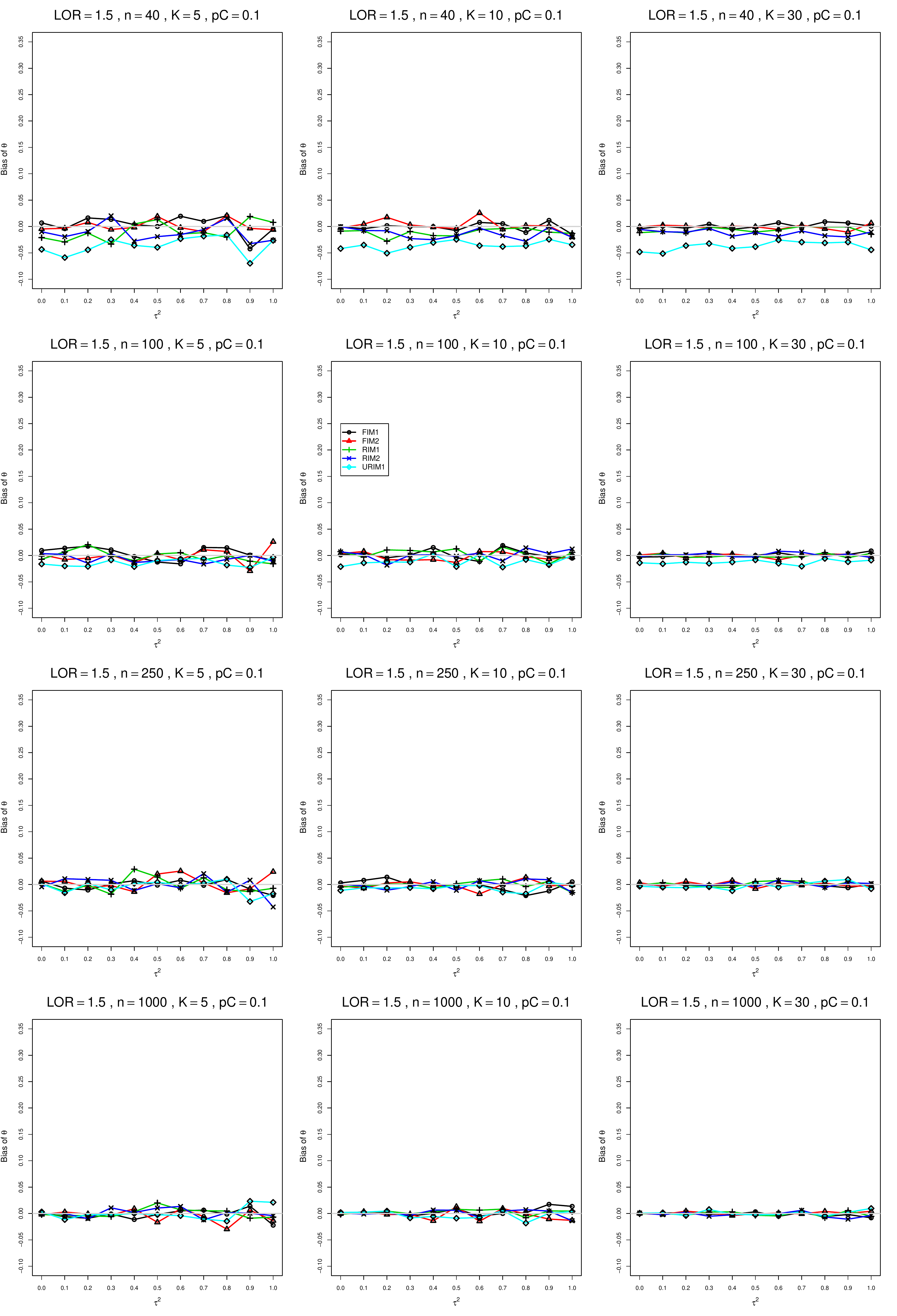}
	\caption{Bias of overall log-odds ratio $\hat{\theta}_{SSW}$ for $\theta=1.5$, $p_{C}=0.1$, $\sigma^2=0.4$, constant sample sizes $n=40,\;100,\;250,\;1000$.
The data-generation mechanisms are FIM1 ($\circ$), FIM2 ($\triangle$), RIM1 (+), RIM2 ($\times$), and URIM1 ($\diamond$).
		\label{PlotBiasThetamu15andpC01LOR_SSWsigma04}}
\end{figure}
\begin{figure}[t]
	\centering
	\includegraphics[scale=0.33]{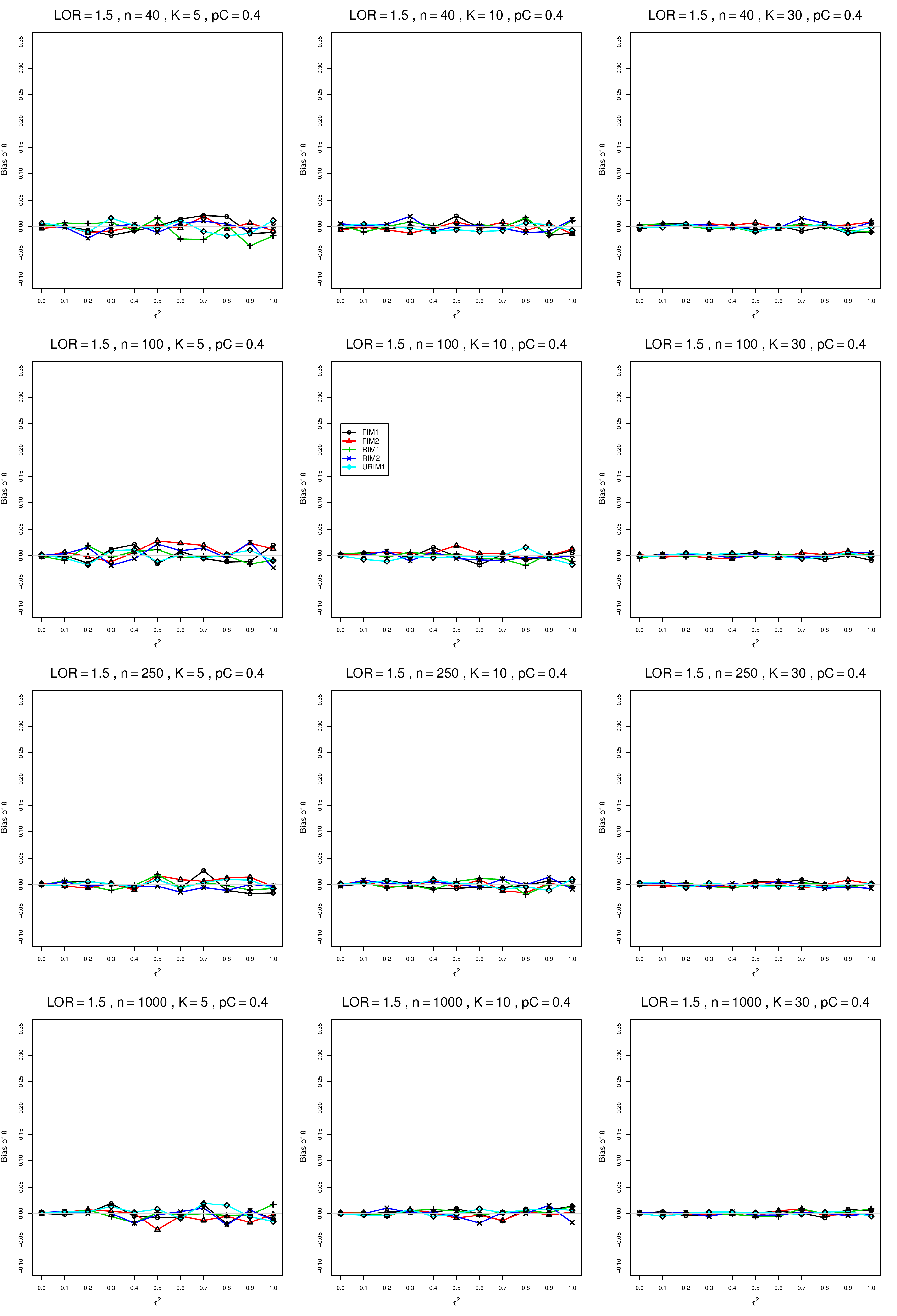}
	\caption{Bias of overall log-odds ratio $\hat{\theta}_{SSW}$ for $\theta=1.5$, $p_{C}=0.4$, $\sigma^2=0.4$, constant sample sizes $n=40,\;100,\;250,\;1000$.
The data-generation mechanisms are FIM1 ($\circ$), FIM2 ($\triangle$), RIM1 (+), RIM2 ($\times$), and URIM1 ($\diamond$).
		\label{PlotBiasThetamu15andpC04LOR_SSWsigma04}}
\end{figure}
\begin{figure}[t]
	\centering
	\includegraphics[scale=0.33]{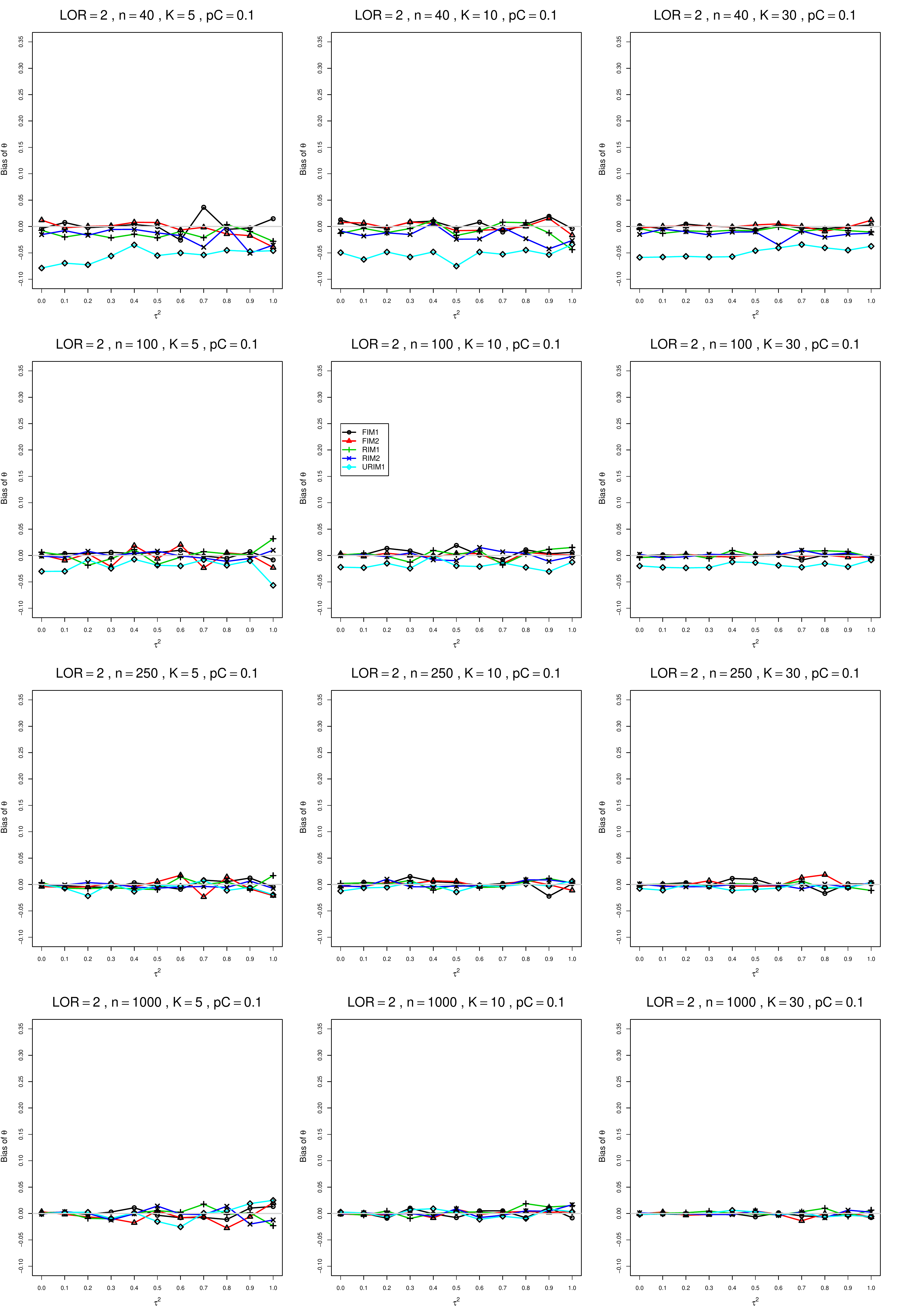}
	\caption{Bias of overall log-odds ratio $\hat{\theta}_{SSW}$ for $\theta=2$, $p_{C}=0.1$, $\sigma^2=0.4$, constant sample sizes $n=40,\;100,\;250,\;1000$.
The data-generation mechanisms are FIM1 ($\circ$), FIM2 ($\triangle$), RIM1 (+), RIM2 ($\times$), and URIM1 ($\diamond$).
		\label{PlotBiasThetamu2andpC01LOR_SSWsigma04}}
\end{figure}
\begin{figure}[t]
	\centering
	\includegraphics[scale=0.33]{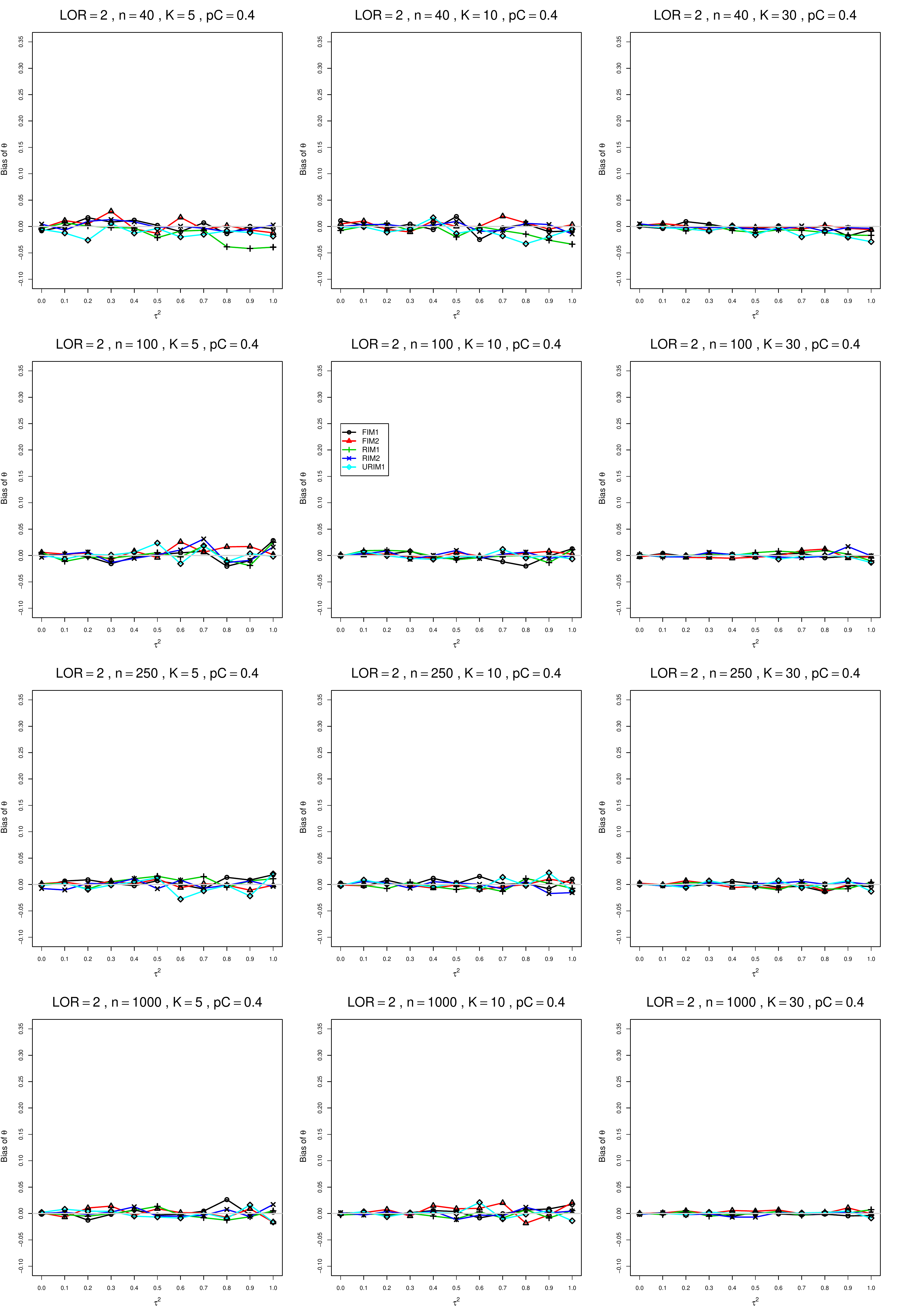}
	\caption{Bias of overall log-odds ratio $\hat{\theta}_{SSW}$ for $\theta=2$, $p_{C}=0.4$, $\sigma^2=0.4$, constant sample sizes $n=40,\;100,\;250,\;1000$.
The data-generation mechanisms are FIM1 ($\circ$), FIM2 ($\triangle$), RIM1 (+), RIM2 ($\times$), and URIM1 ($\diamond$).
		\label{PlotBiasThetamu2andpC04LOR_SSWsigma04}}
\end{figure}

\clearpage
\section*{A3: Coverage  estimators of $\theta$ for log-odds-ratio}
Each panel of a figure corresponds to a value of n (= 40, 100, 250, 1000) and a value of K (= 5, 10, 30) and has $\tau^2$ = 0.0(0.1)1.0 on the horizontal axis. \\
The data-generation mechanisms are
\begin{itemize}
	\item FIM1 - Fixed-intercept model with $c = 0$
	\item FIM2 - Fixed-intercept model with $c = 1/2$
	\item RIM1 - Random-intercept model with $c = 0$
	\item RIM2 - Random-intercept model with $c = 1/2$
	\item URIM1 - Random-intercept model with $c = 0$ and $p_{iC}$ is uniformly distributed on [$p_{iC}-\sigma\sqrt{3}p_{iC}(1-p_{iC})$, $p_{iC}+\sigma\sqrt{3}p_{iC}(1-p_{iC})$]
\end{itemize}
The interval point estimators of $\theta$ are
\begin{itemize}
	\item $\hat{\theta}_{DL}$ - DerSimonian-Laird
	\item $\hat{\theta}_{REML}$ - Restricted maximum-likelihood
	\item $\hat{\theta}_{MP}$ - Mandel-Paule
	\item $\hat{\theta}_{KD}$ - Kulinskaya-Dollinger
	\item $\hat{\theta}_{FIM2}$ - Estimator of $\theta$ in the FIM2 GLMM
	\item $\hat{\theta}_{RIM2}$ - Estimator of $\theta$ in the RIM2 GLMM
	\item $\hat{\theta}_{SSW}$ - Estimator of $\theta$ based on the Hartung-Knapp-Sidik-Jonkman with KD based modification and an interval based on the sample-size-weighted estimator
\end{itemize}

\clearpage
\subsection*{A3.1 Coverage of $\hat{\theta}_{DL}$}
\renewcommand{\thefigure}{A3.1.\arabic{figure}}
\setcounter{figure}{0}

\begin{figure}[t]
	\centering
	\includegraphics[scale=0.33]{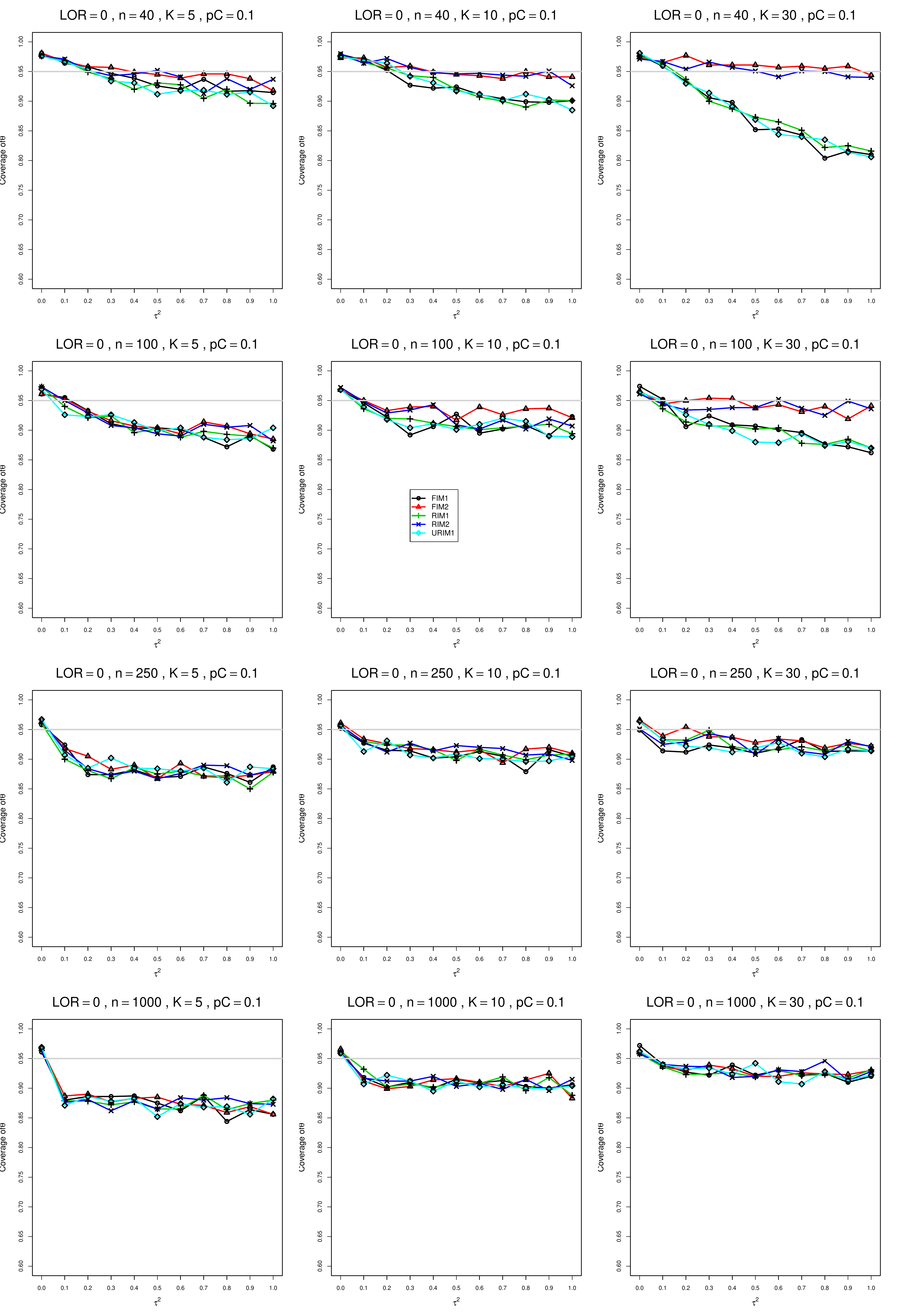}
	\caption{Coverage of the DerSimonian-Laird confidence interval for $\theta=0$, $p_{C}=0.1$, $\sigma^2=0.1$, constant sample sizes $n=40,\;100,\;250,\;1000$.
The data-generation mechanisms are FIM1 ($\circ$), FIM2 ($\triangle$), RIM1 (+), RIM2 ($\times$), and URIM1 ($\diamond$).
		\label{PlotCovThetamu0andpC01LOR_DLsigma01}}
\end{figure}
\begin{figure}[t]
	\centering
	\includegraphics[scale=0.33]{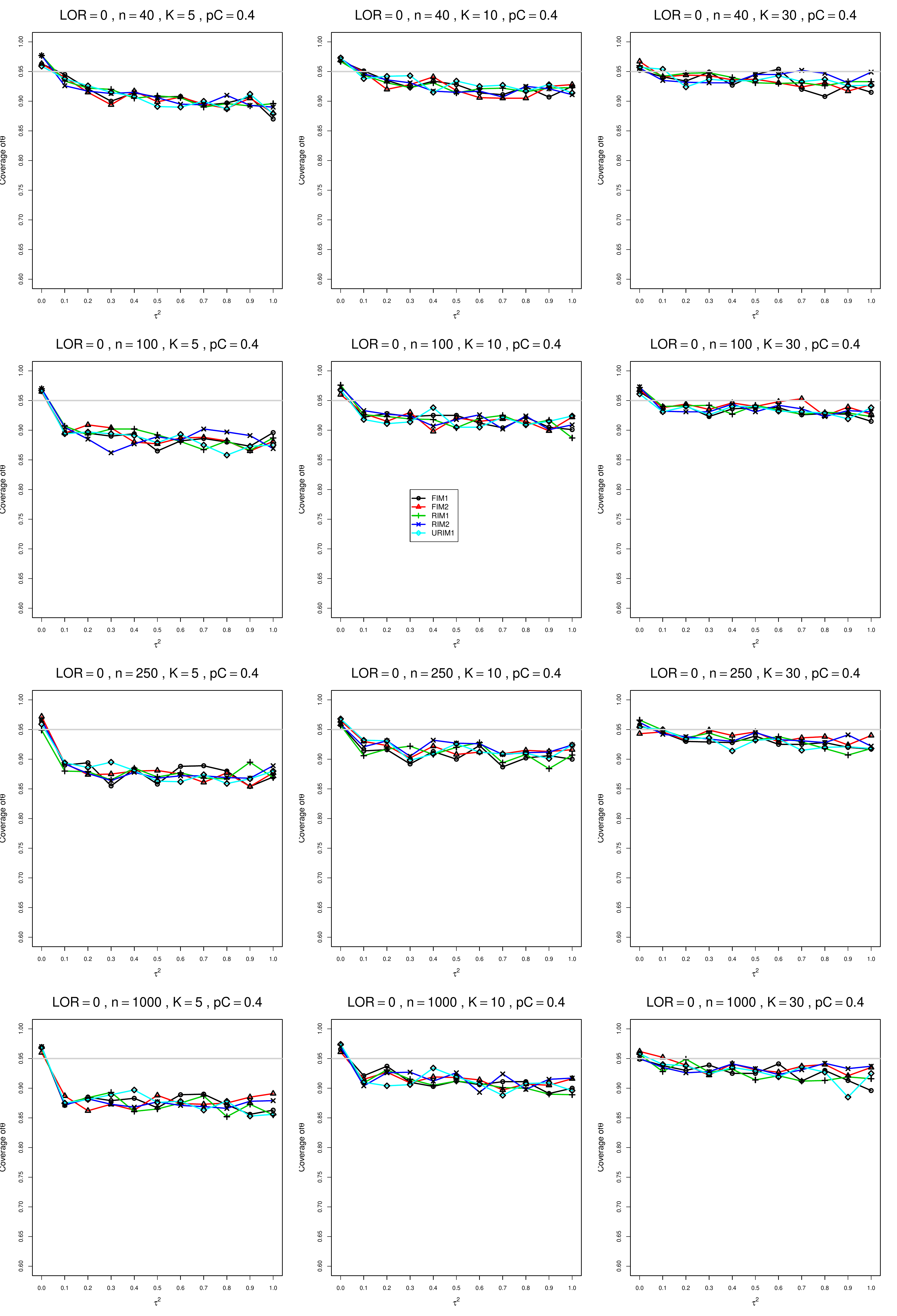}
	\caption{Coverage of the DerSimonian-Laird confidence interval for $\theta=0$, $p_{C}=0.4$, $\sigma^2=0.1$, constant sample sizes $n=40,\;100,\;250,\;1000$.
The data-generation mechanisms are FIM1 ($\circ$), FIM2 ($\triangle$), RIM1 (+), RIM2 ($\times$), and URIM1 ($\diamond$).
		\label{PlotCovThetamu0andpC04LOR_DLsigma01}}
\end{figure}
\begin{figure}[t]
	\centering
	\includegraphics[scale=0.33]{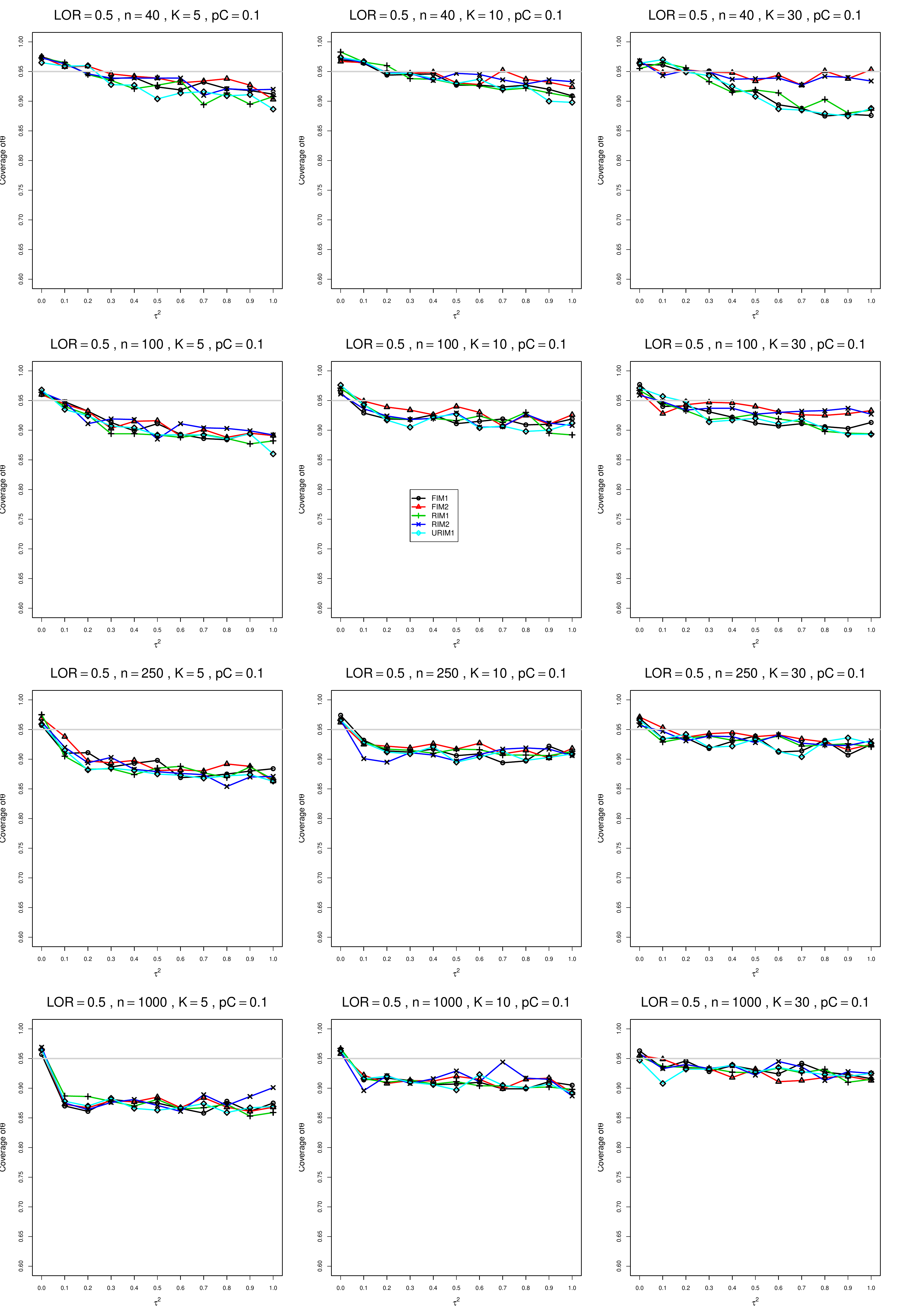}
	\caption{Coverage of the DerSimonian-Laird confidence interval for $\theta=0.5$, $p_{C}=0.1$, $\sigma^2=0.1$, constant sample sizes $n=40,\;100,\;250,\;1000$.
The data-generation mechanisms are FIM1 ($\circ$), FIM2 ($\triangle$), RIM1 (+), RIM2 ($\times$), and URIM1 ($\diamond$).
		\label{PlotCovThetamu05andpC01LOR_DLsigma01}}
\end{figure}
\begin{figure}[t]
	\centering
	\includegraphics[scale=0.33]{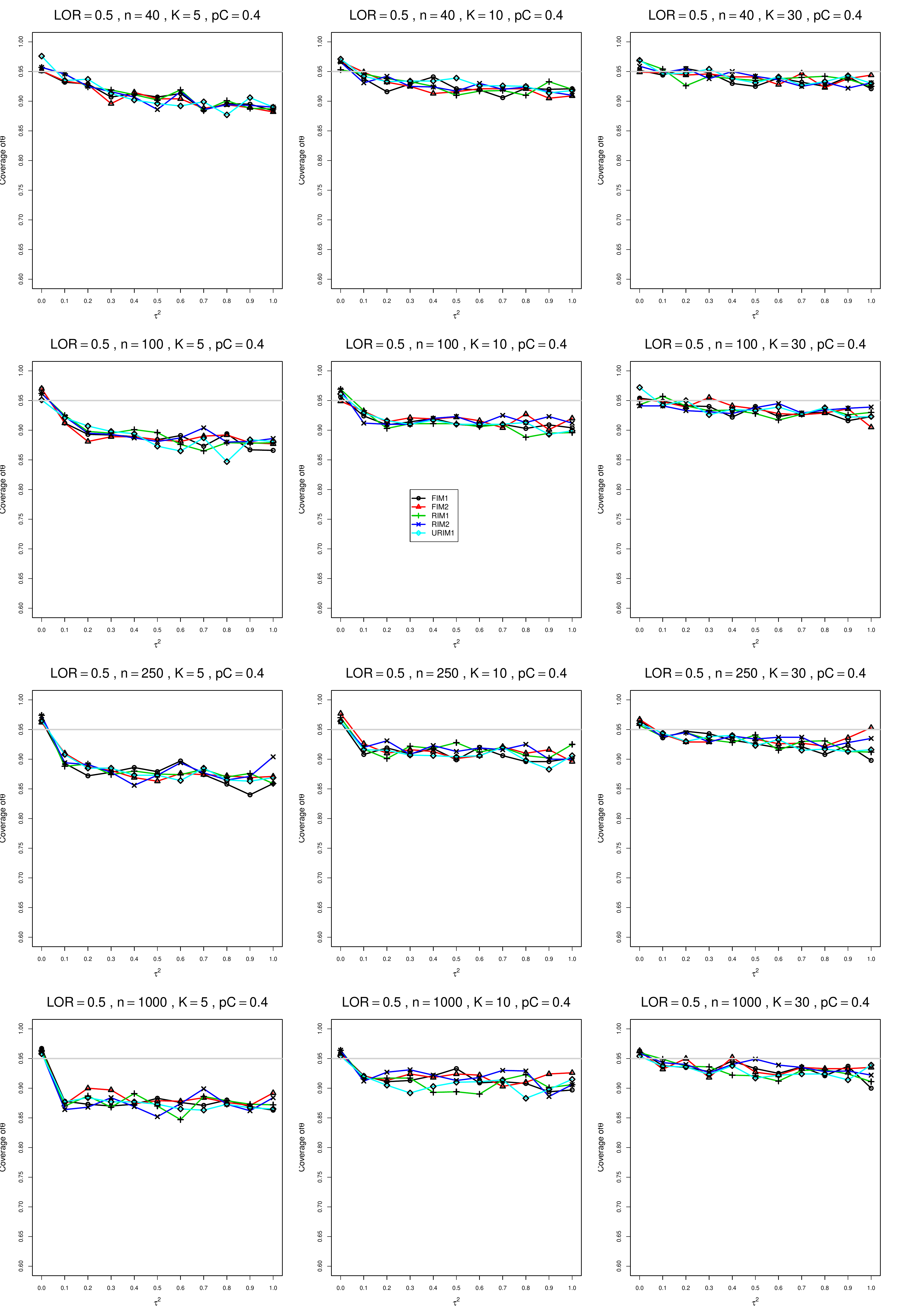}
	\caption{Coverage of the DerSimonian-Laird confidence interval for $\theta=0.5$, $p_{C}=0.4$, $\sigma^2=0.1$, constant sample sizes $n=40,\;100,\;250,\;1000$.
The data-generation mechanisms are FIM1 ($\circ$), FIM2 ($\triangle$), RIM1 (+), RIM2 ($\times$), and URIM1 ($\diamond$).
		\label{PlotCovThetamu05andpC04LOR_DLsigma01}}
\end{figure}
\begin{figure}[t]
	\centering
	\includegraphics[scale=0.33]{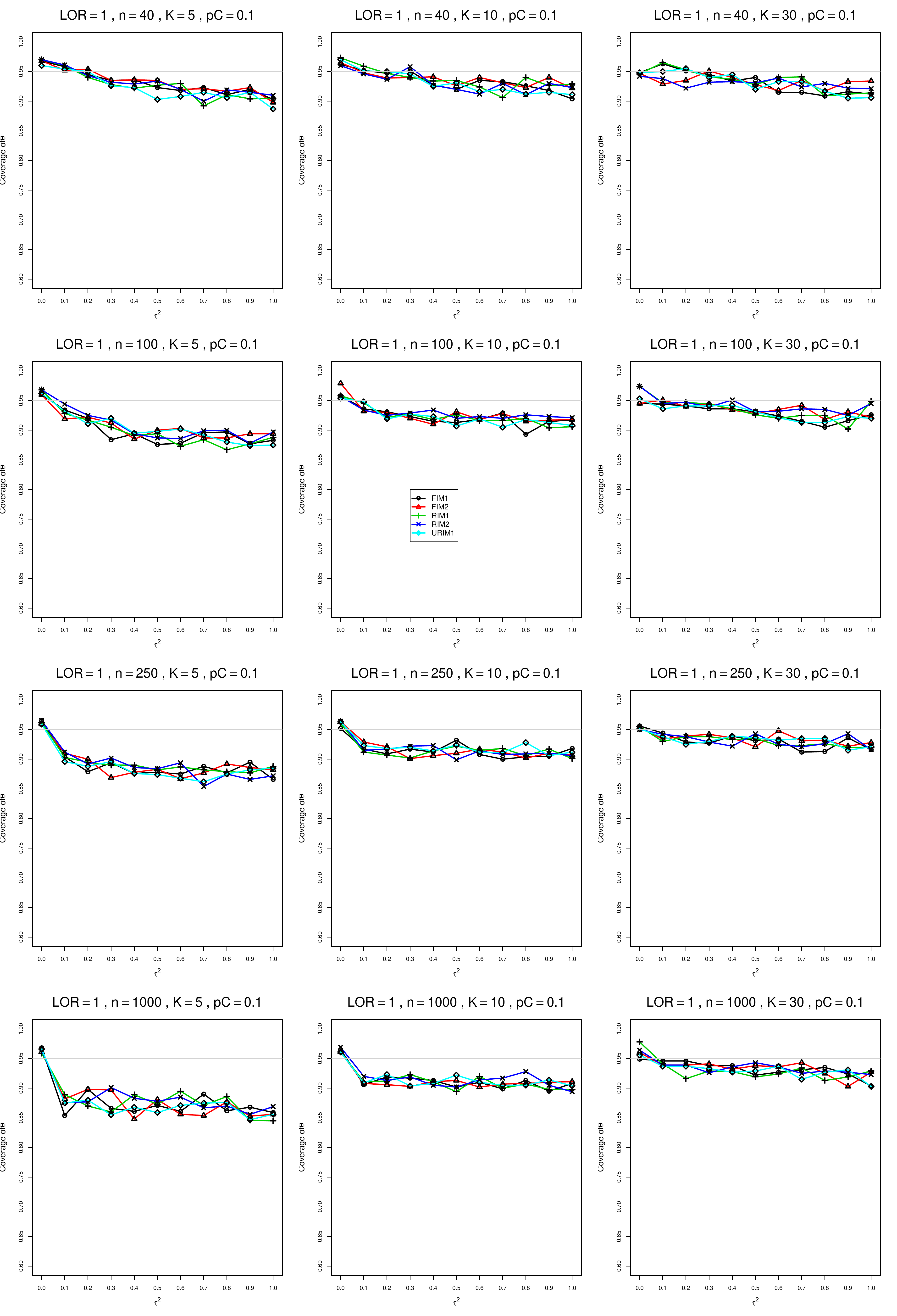}
	\caption{Coverage of the DerSimonian-Laird confidence interval for $\theta=1$, $p_{C}=0.1$, $\sigma^2=0.1$, constant sample sizes $n=40,\;100,\;250,\;1000$.
The data-generation mechanisms are FIM1 ($\circ$), FIM2 ($\triangle$), RIM1 (+), RIM2 ($\times$), and URIM1 ($\diamond$).
		\label{PlotCovThetamu1andpC01LOR_DLsigma01}}
\end{figure}
\begin{figure}[t]
	\centering
	\includegraphics[scale=0.33]{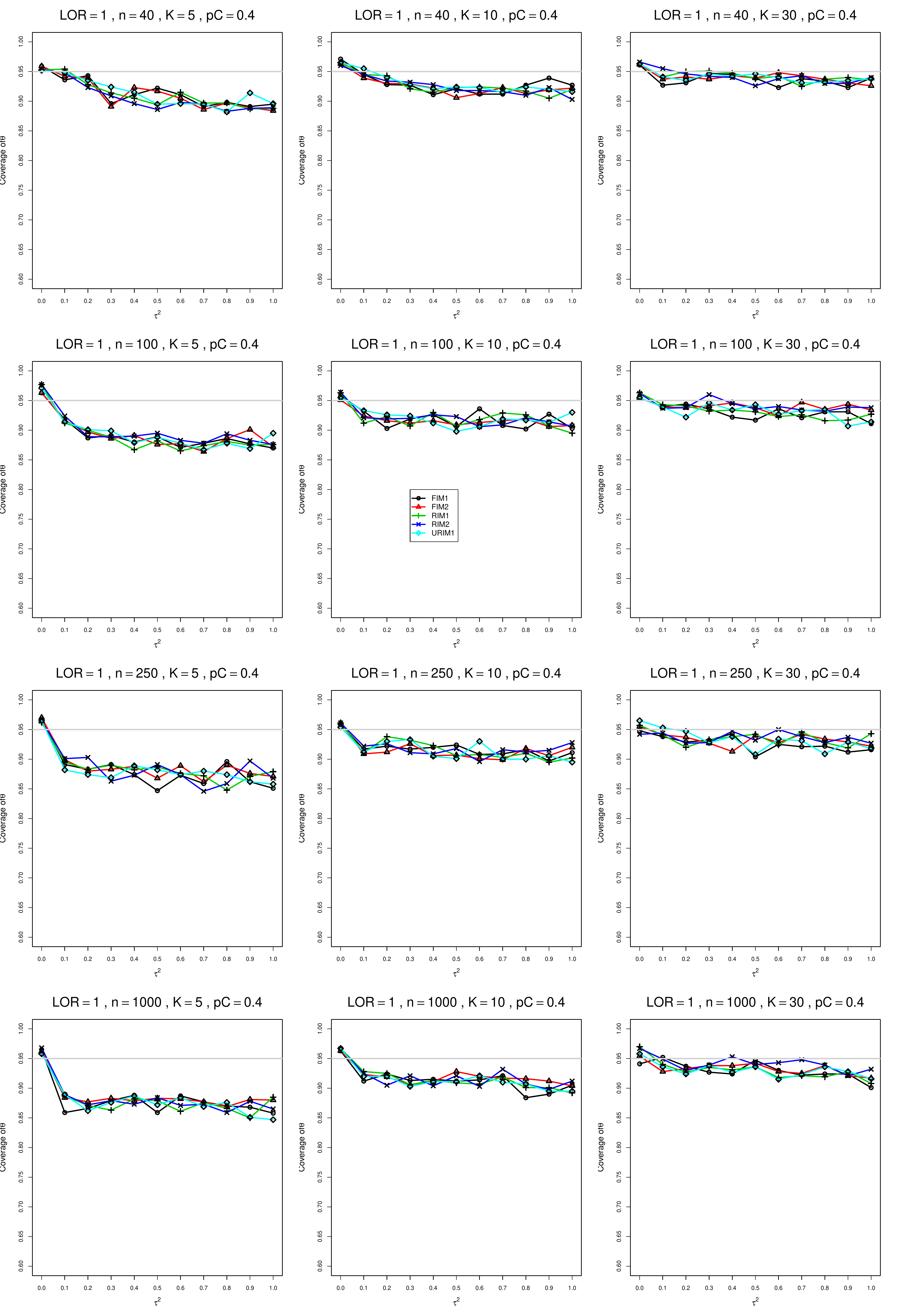}
	\caption{Coverage of the DerSimonian-Laird confidence interval for $\theta=1$, $p_{C}=0.4$, $\sigma^2=0.1$, constant sample sizes $n=40,\;100,\;250,\;1000$.
The data-generation mechanisms are FIM1 ($\circ$), FIM2 ($\triangle$), RIM1 (+), RIM2 ($\times$), and URIM1 ($\diamond$).
		\label{PlotCovThetamu1andpC04LOR_DLsigma01}}
\end{figure}
\begin{figure}[t]
	\centering
	\includegraphics[scale=0.33]{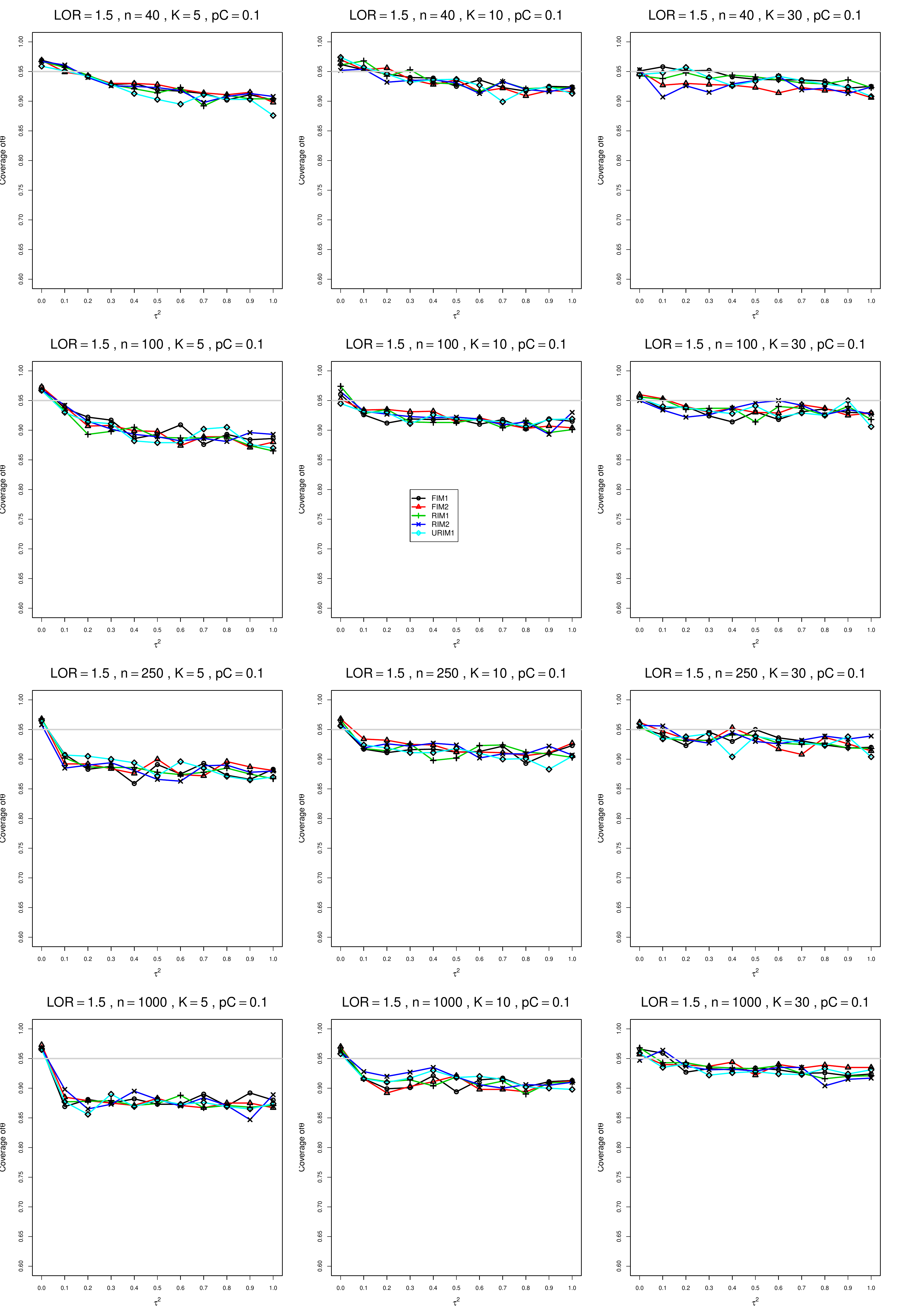}
	\caption{Coverage of the DerSimonian-Laird confidence interval for $\theta=1.5$, $p_{C}=0.1$, $\sigma^2=0.1$, constant sample sizes $n=40,\;100,\;250,\;1000$.
The data-generation mechanisms are FIM1 ($\circ$), FIM2 ($\triangle$), RIM1 (+), RIM2 ($\times$), and URIM1 ($\diamond$).
		\label{PlotCovThetamu15andpC01LOR_DLsigma01}}
\end{figure}
\begin{figure}[t]
	\centering
	\includegraphics[scale=0.33]{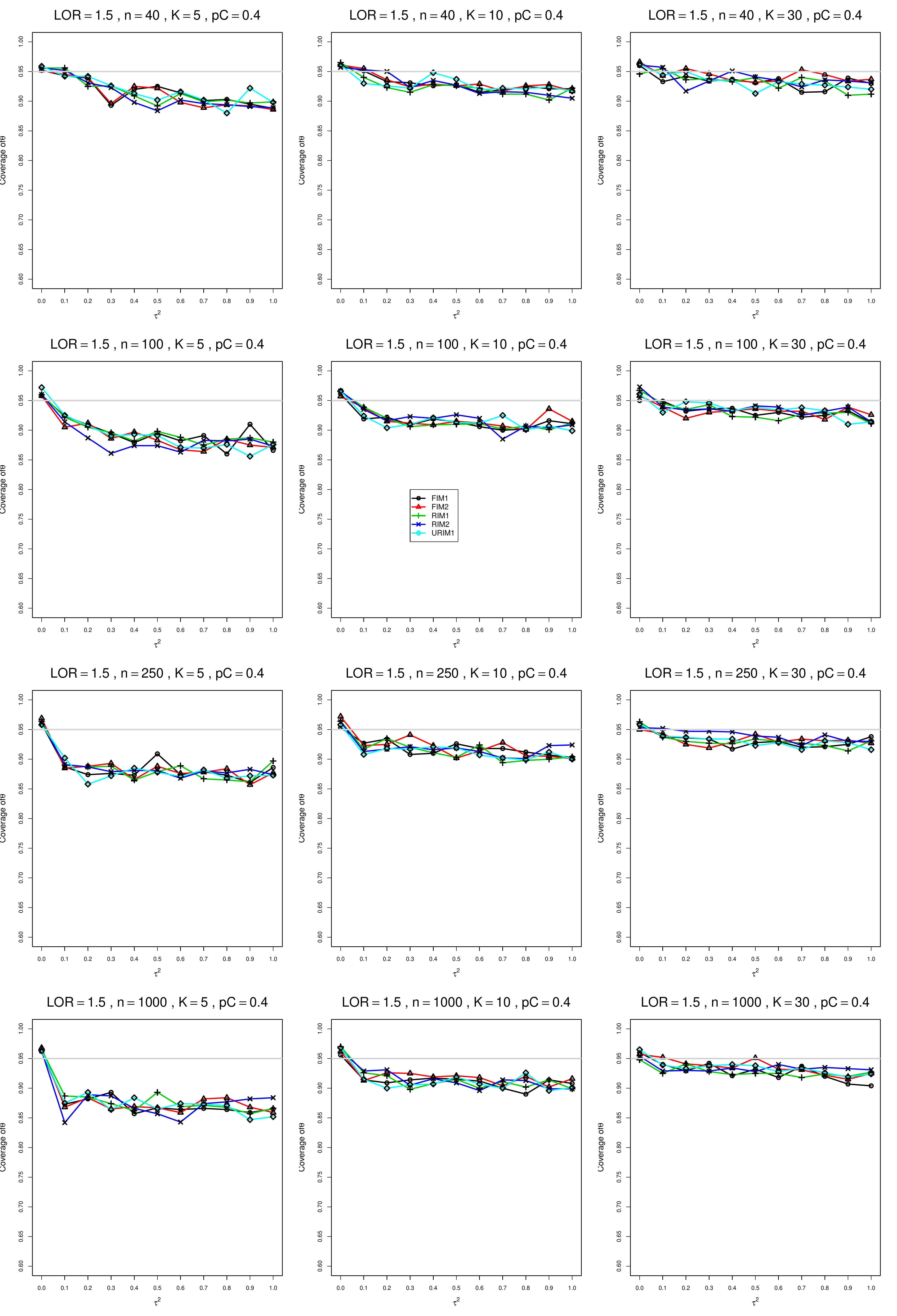}
	\caption{Coverage of the DerSimonian-Laird confidence interval for $\theta=1.5$, $p_{C}=0.4$, $\sigma^2=0.1$, constant sample sizes $n=40,\;100,\;250,\;1000$.
The data-generation mechanisms are FIM1 ($\circ$), FIM2 ($\triangle$), RIM1 (+), RIM2 ($\times$), and URIM1 ($\diamond$).
		\label{PlotCovThetamu15andpC04LOR_DLsigma01}}
\end{figure}
\begin{figure}[t]
	\centering
	\includegraphics[scale=0.33]{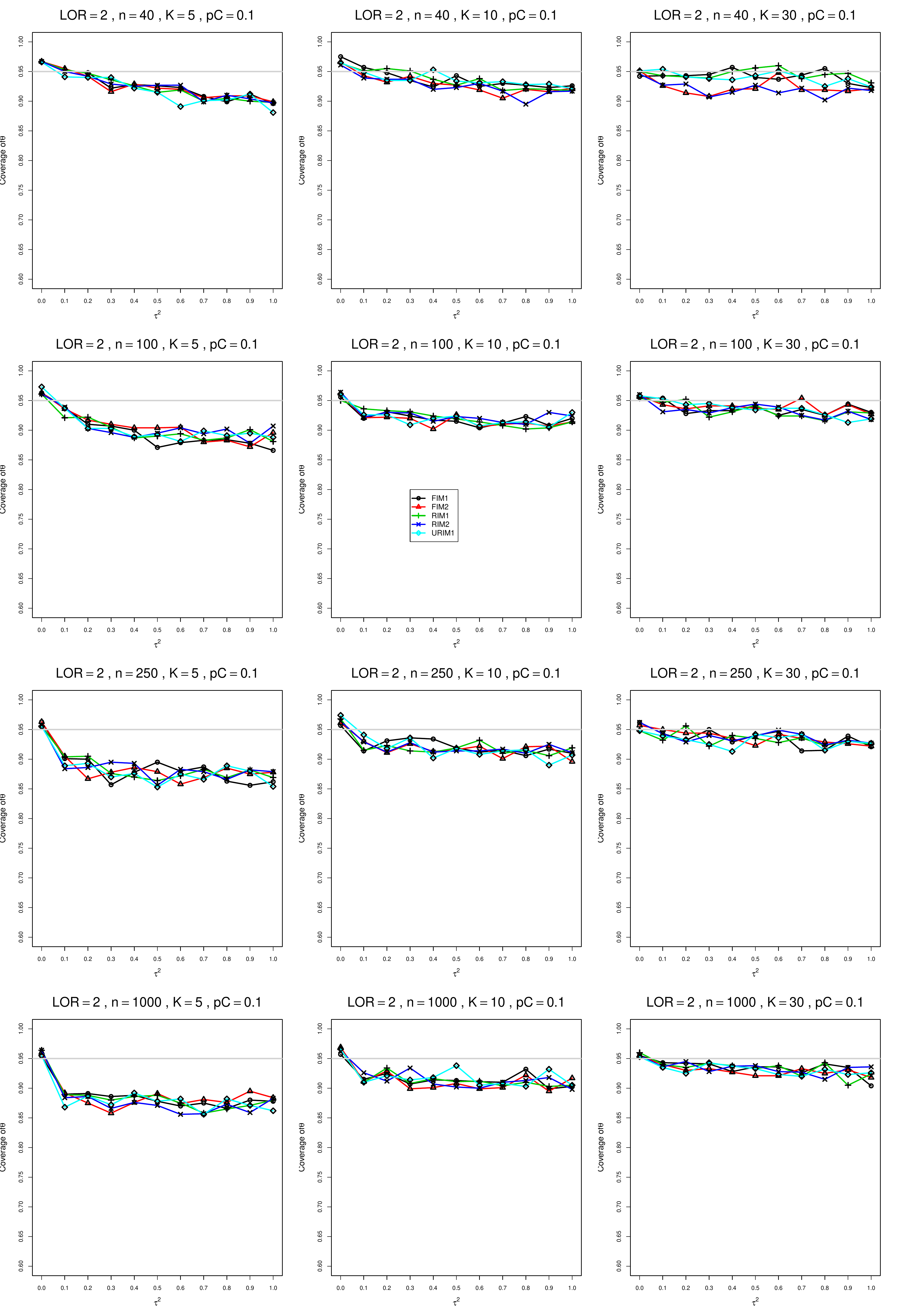}
	\caption{Coverage of the DerSimonian-Laird confidence interval for $\theta=2$, $p_{C}=0.1$, $\sigma^2=0.1$, constant sample sizes $n=40,\;100,\;250,\;1000$.
The data-generation mechanisms are FIM1 ($\circ$), FIM2 ($\triangle$), RIM1 (+), RIM2 ($\times$), and URIM1 ($\diamond$).
		\label{PlotCovThetamu2andpC01LOR_DLsigma01}}
\end{figure}
\begin{figure}[t]
	\centering
	\includegraphics[scale=0.33]{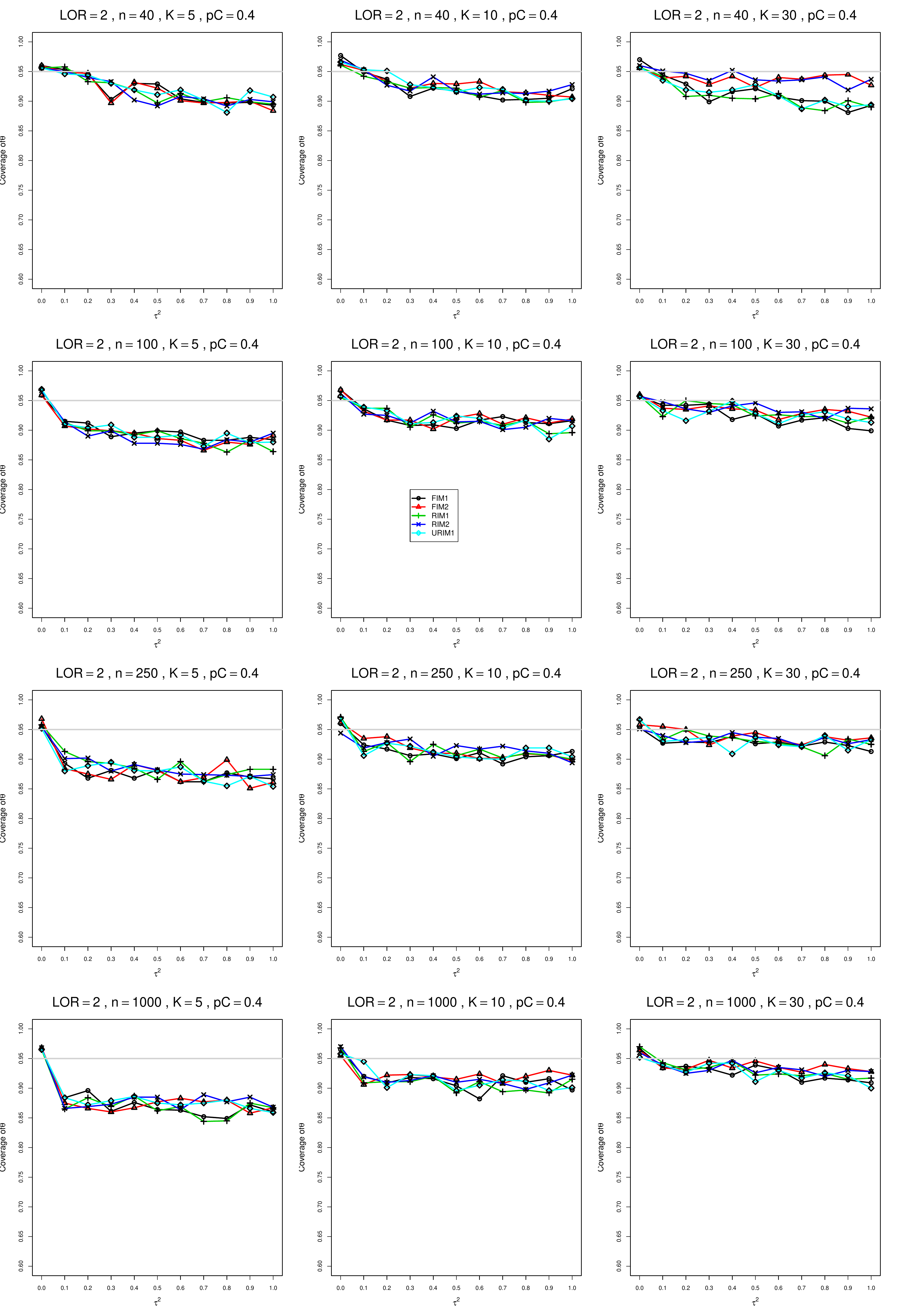}
	\caption{Coverage of the DerSimonian-Laird confidence interval for $\theta=2$, $p_{C}=0.4$, $\sigma^2=0.1$, constant sample sizes $n=40,\;100,\;250,\;1000$.
The data-generation mechanisms are FIM1 ($\circ$), FIM2 ($\triangle$), RIM1 (+), RIM2 ($\times$), and URIM1 ($\diamond$).
		\label{PlotCovThetamu2andpC04LOR_DLsigma01}}
\end{figure}
\begin{figure}[t]
	\centering
	\includegraphics[scale=0.33]{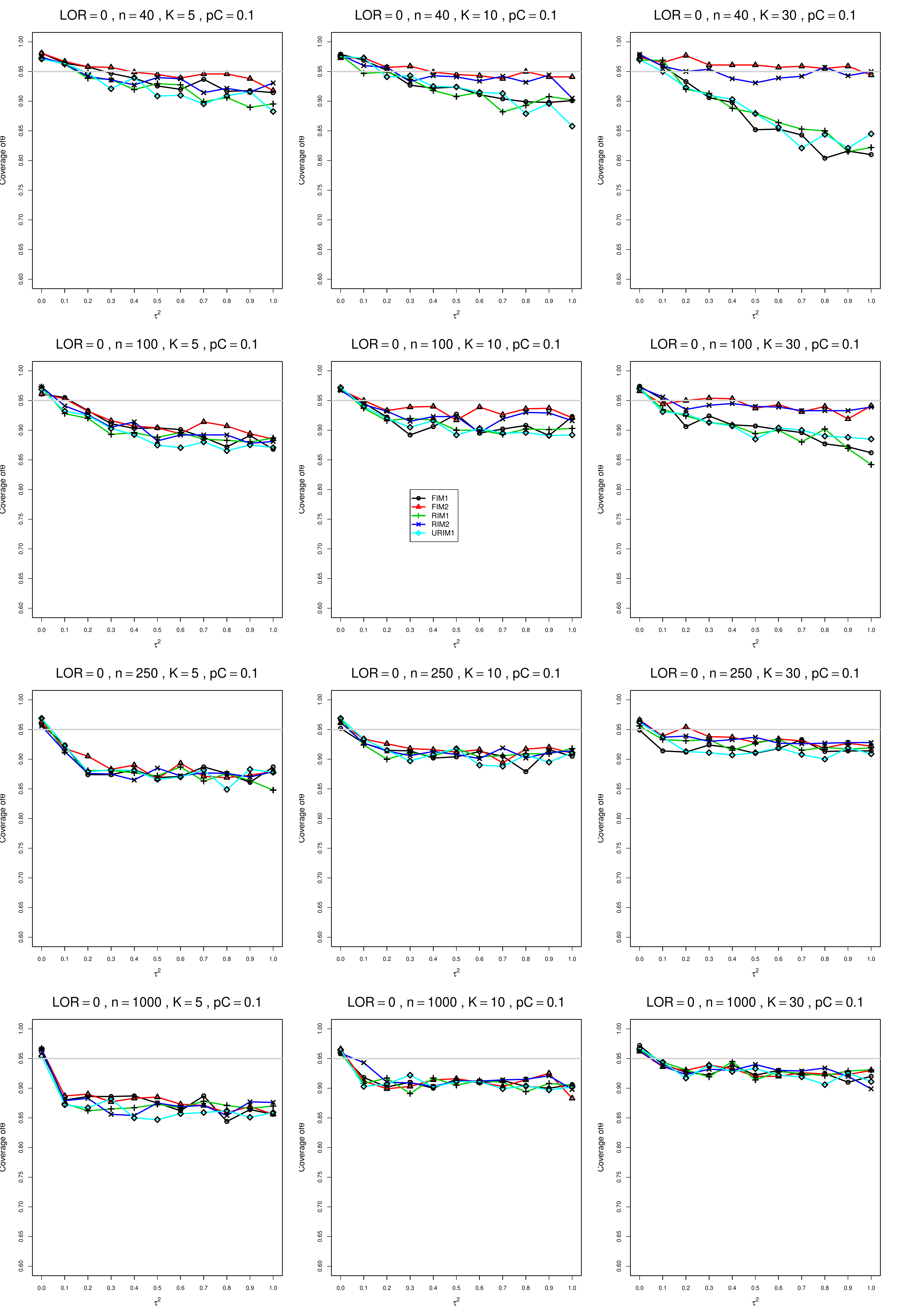}
	\caption{Coverage of the DerSimonian-Laird confidence interval for $\theta=0$, $p_{C}=0.1$, $\sigma^2=0.4$, constant sample sizes $n=40,\;100,\;250,\;1000$.
The data-generation mechanisms are FIM1 ($\circ$), FIM2 ($\triangle$), RIM1 (+), RIM2 ($\times$), and URIM1 ($\diamond$).
		\label{PlotCovThetamu0andpC01LOR_DLsigma04}}
\end{figure}
\begin{figure}[t]
	\centering
	\includegraphics[scale=0.33]{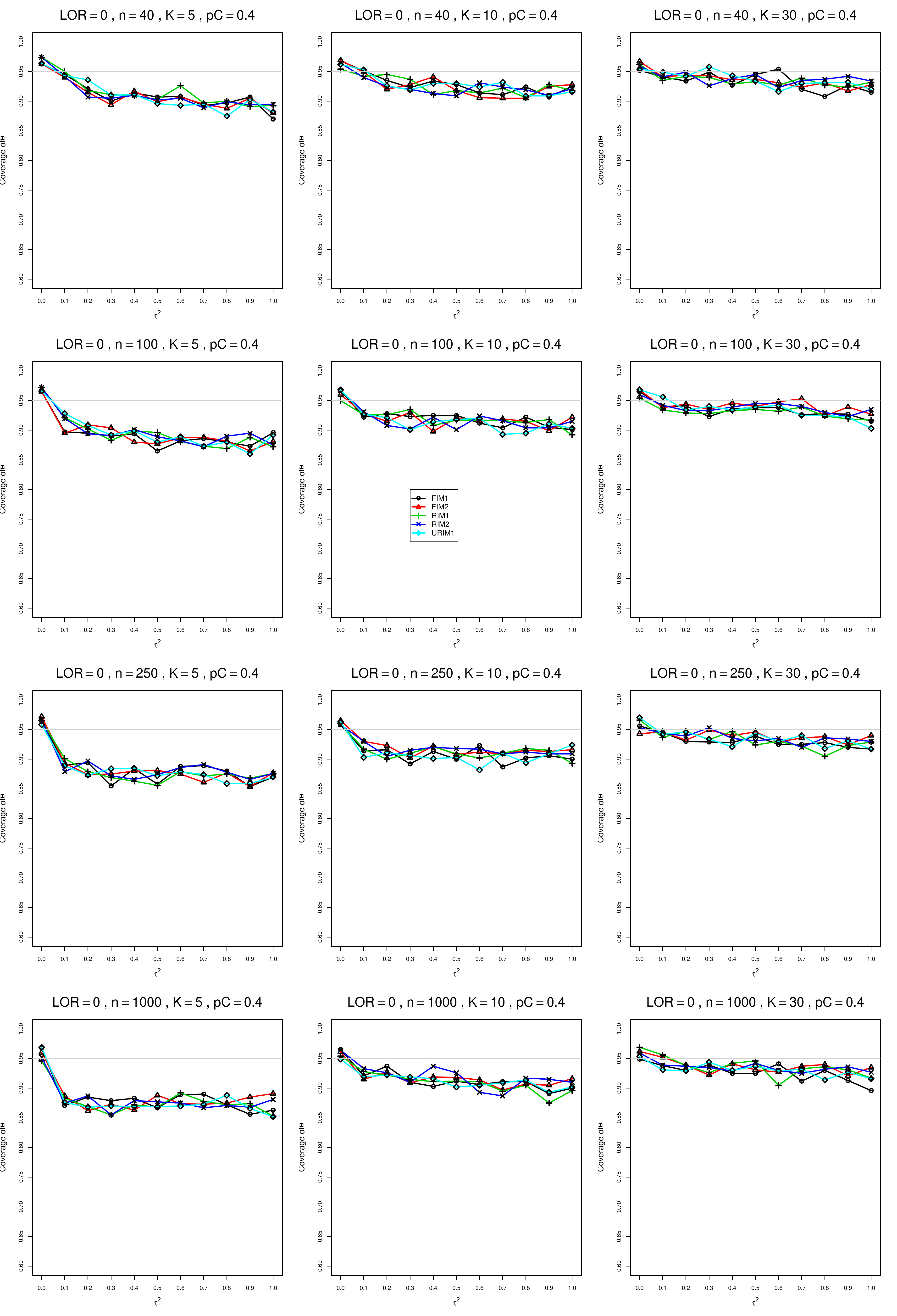}
	\caption{Coverage of the DerSimonian-Laird confidence interval for $\theta=0$, $p_{C}=0.4$, $\sigma^2=0.4$, constant sample sizes $n=40,\;100,\;250,\;1000$.
The data-generation mechanisms are FIM1 ($\circ$), FIM2 ($\triangle$), RIM1 (+), RIM2 ($\times$), and URIM1 ($\diamond$).
		\label{PlotCovThetamu0andpC04LOR_DLsigma04}}
\end{figure}
\begin{figure}[t]
	\centering
	\includegraphics[scale=0.33]{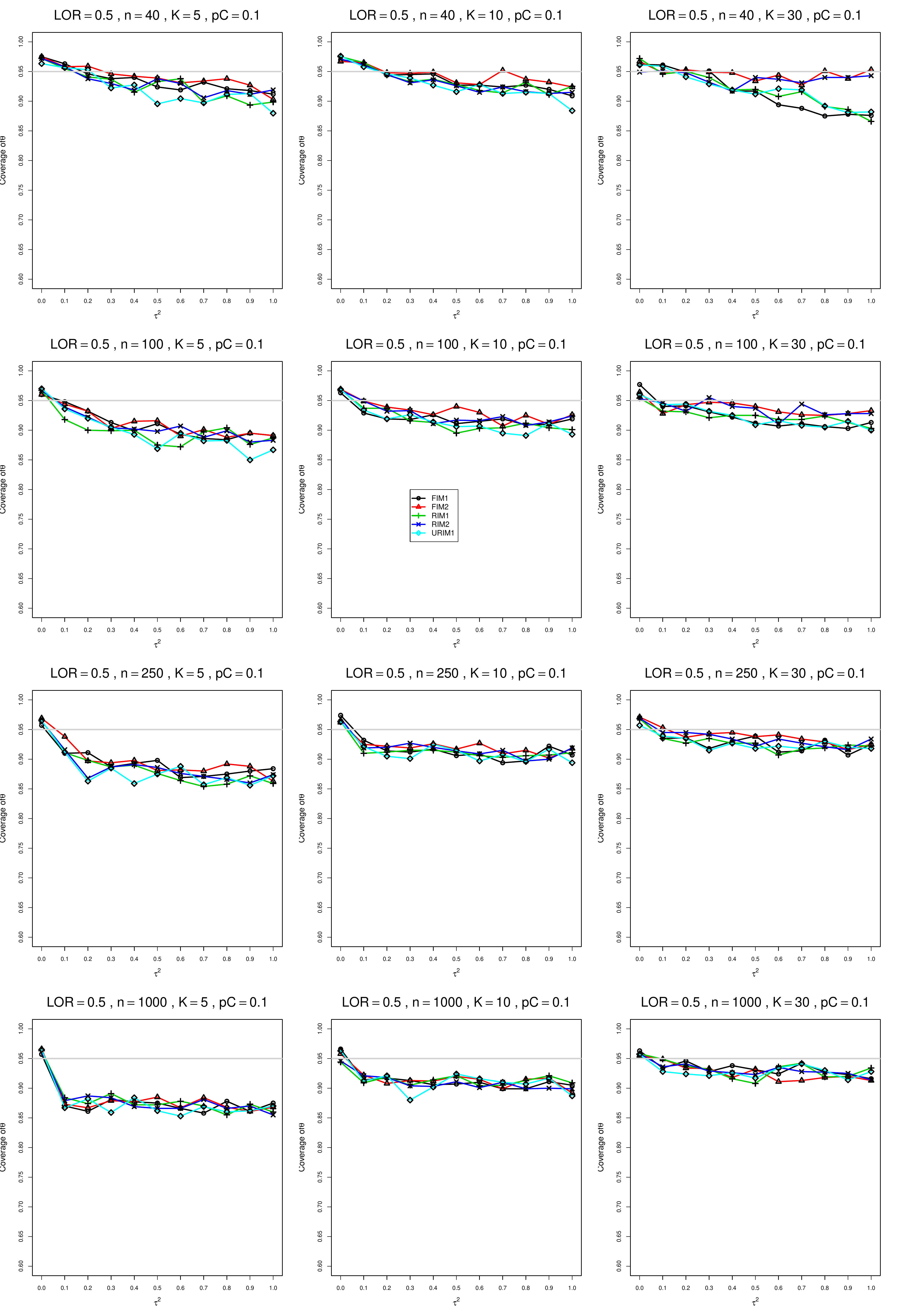}
	\caption{Coverage of the DerSimonian-Laird confidence interval for $\theta=0.5$, $p_{C}=0.1$, $\sigma^2=0.4$, constant sample sizes $n=40,\;100,\;250,\;1000$.
The data-generation mechanisms are FIM1 ($\circ$), FIM2 ($\triangle$), RIM1 (+), RIM2 ($\times$), and URIM1 ($\diamond$).
		\label{PlotCovThetamu05andpC01LOR_DLsigma04}}
\end{figure}
\begin{figure}[t]
	\centering
	\includegraphics[scale=0.33]{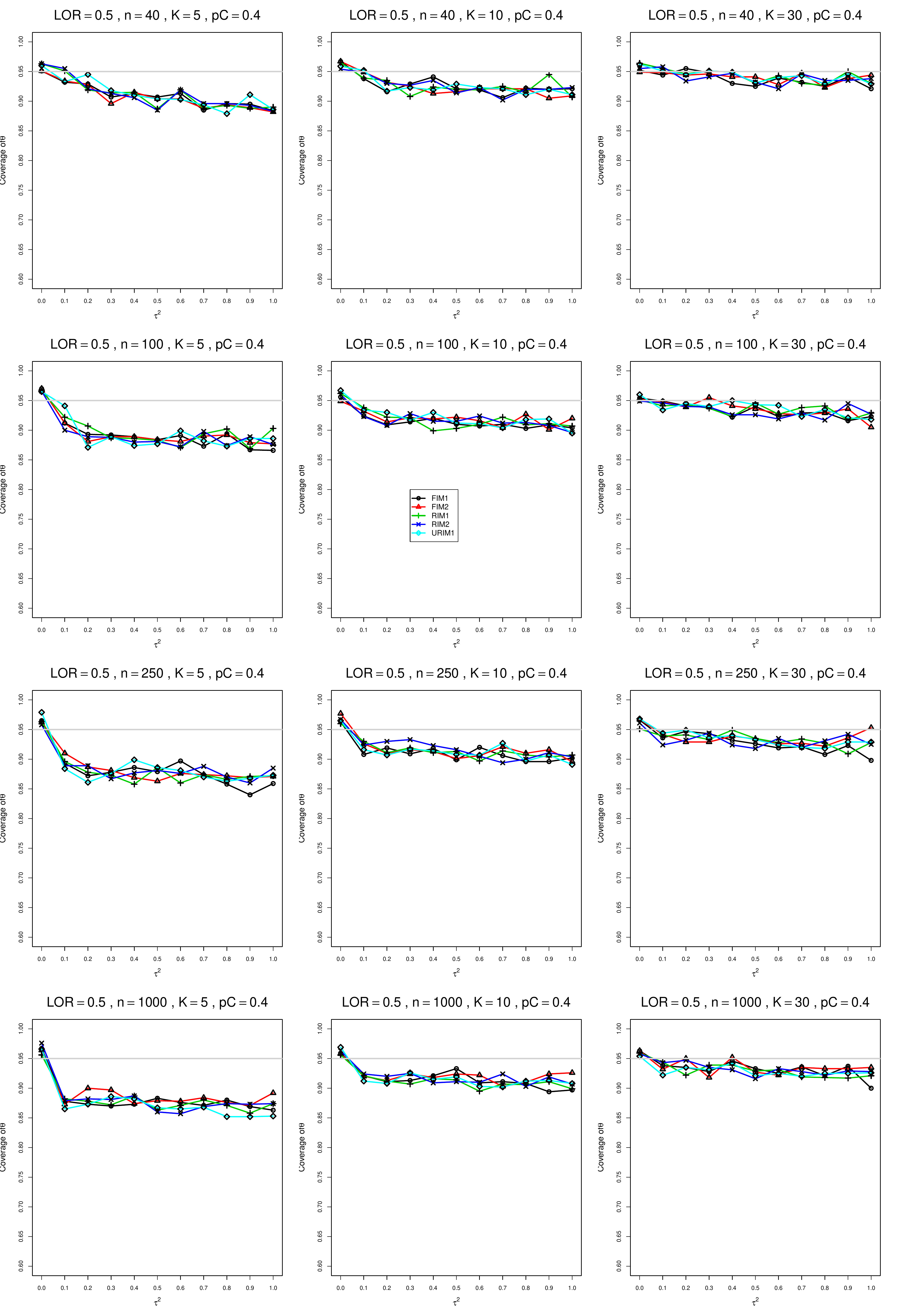}
	\caption{Coverage of the DerSimonian-Laird confidence interval for $\theta=0.5$, $p_{C}=0.4$, $\sigma^2=0.4$, constant sample sizes $n=40,\;100,\;250,\;1000$.
The data-generation mechanisms are FIM1 ($\circ$), FIM2 ($\triangle$), RIM1 (+), RIM2 ($\times$), and URIM1 ($\diamond$).
		\label{PlotCovThetamu05andpC04LOR_DLsigma04}}
\end{figure}
\begin{figure}[t]
	\centering
	\includegraphics[scale=0.33]{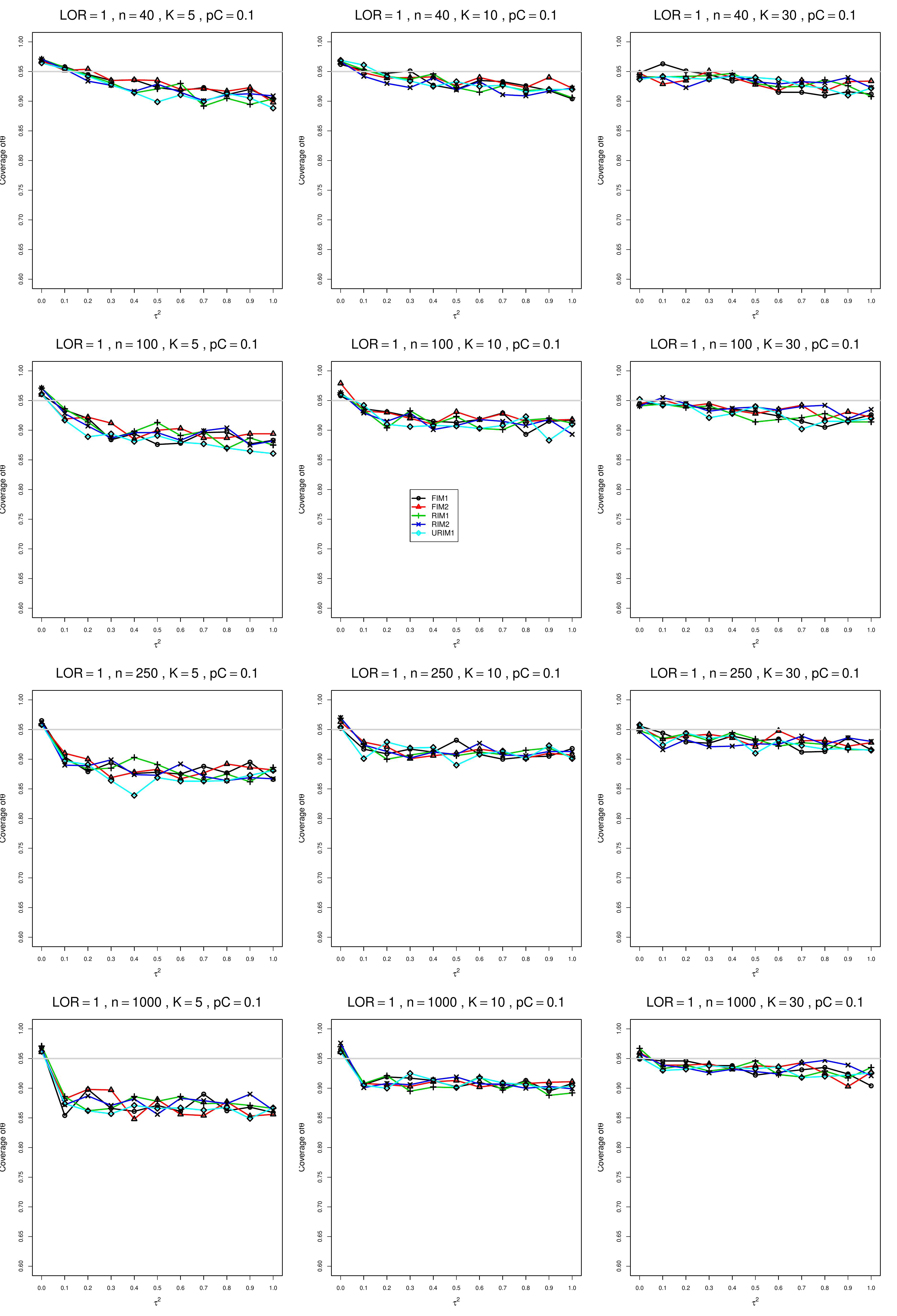}
	\caption{Coverage of the DerSimonian-Laird confidence interval for $\theta=1$, $p_{C}=0.1$, $\sigma^2=0.4$, constant sample sizes $n=40,\;100,\;250,\;1000$.
The data-generation mechanisms are FIM1 ($\circ$), FIM2 ($\triangle$), RIM1 (+), RIM2 ($\times$), and URIM1 ($\diamond$).
		\label{PlotCovThetamu1andpC01LOR_DLsigma04}}
\end{figure}
\begin{figure}[t]
	\centering
	\includegraphics[scale=0.33]{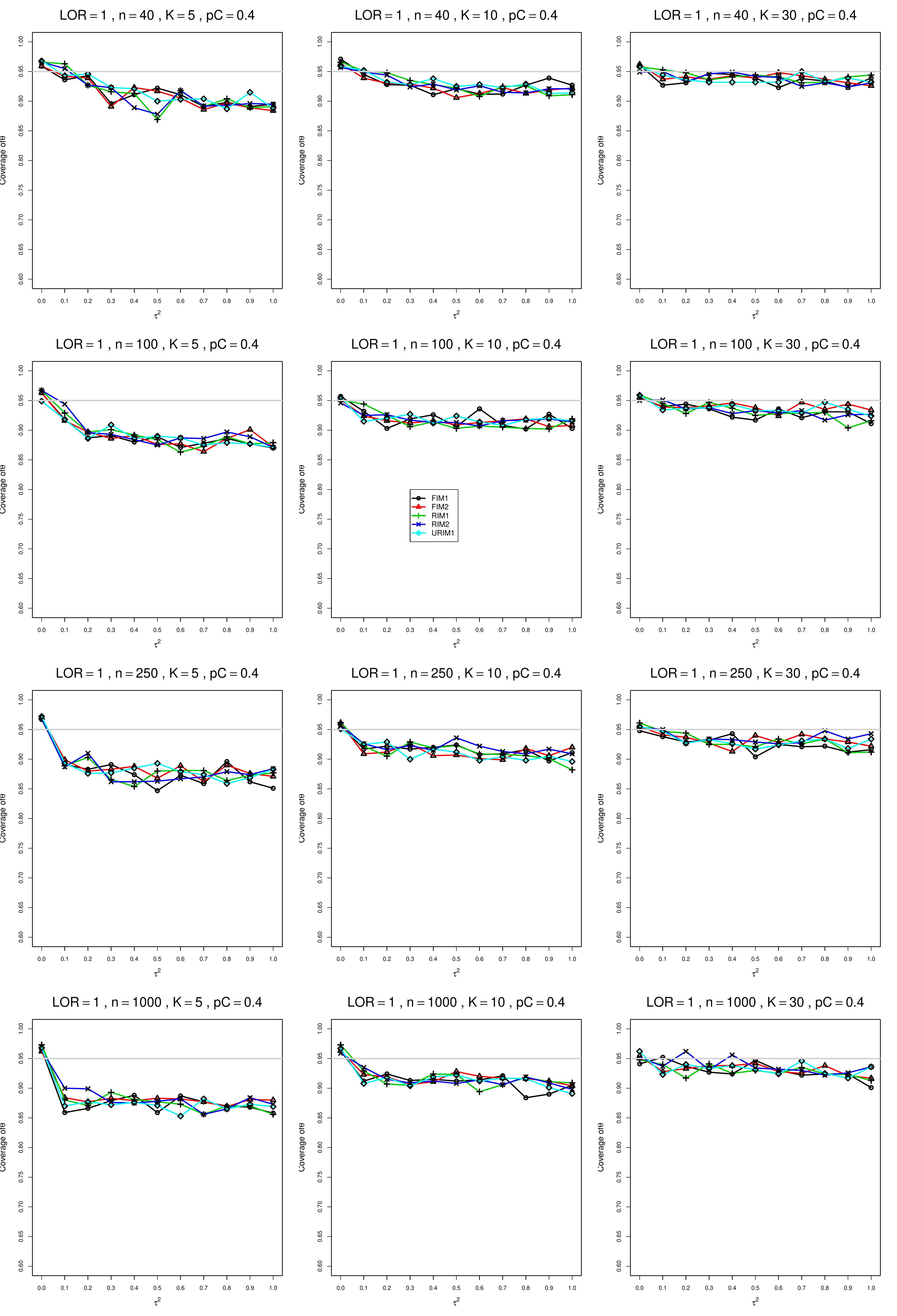}
	\caption{Coverage of the DerSimonian-Laird confidence interval for $\theta=1$, $p_{C}=0.4$, $\sigma^2=0.4$, constant sample sizes $n=40,\;100,\;250,\;1000$.
The data-generation mechanisms are FIM1 ($\circ$), FIM2 ($\triangle$), RIM1 (+), RIM2 ($\times$), and URIM1 ($\diamond$).
		\label{PlotCovThetamu1andpC04LOR_DLsigma04}}
\end{figure}
\begin{figure}[t]
	\centering
	\includegraphics[scale=0.33]{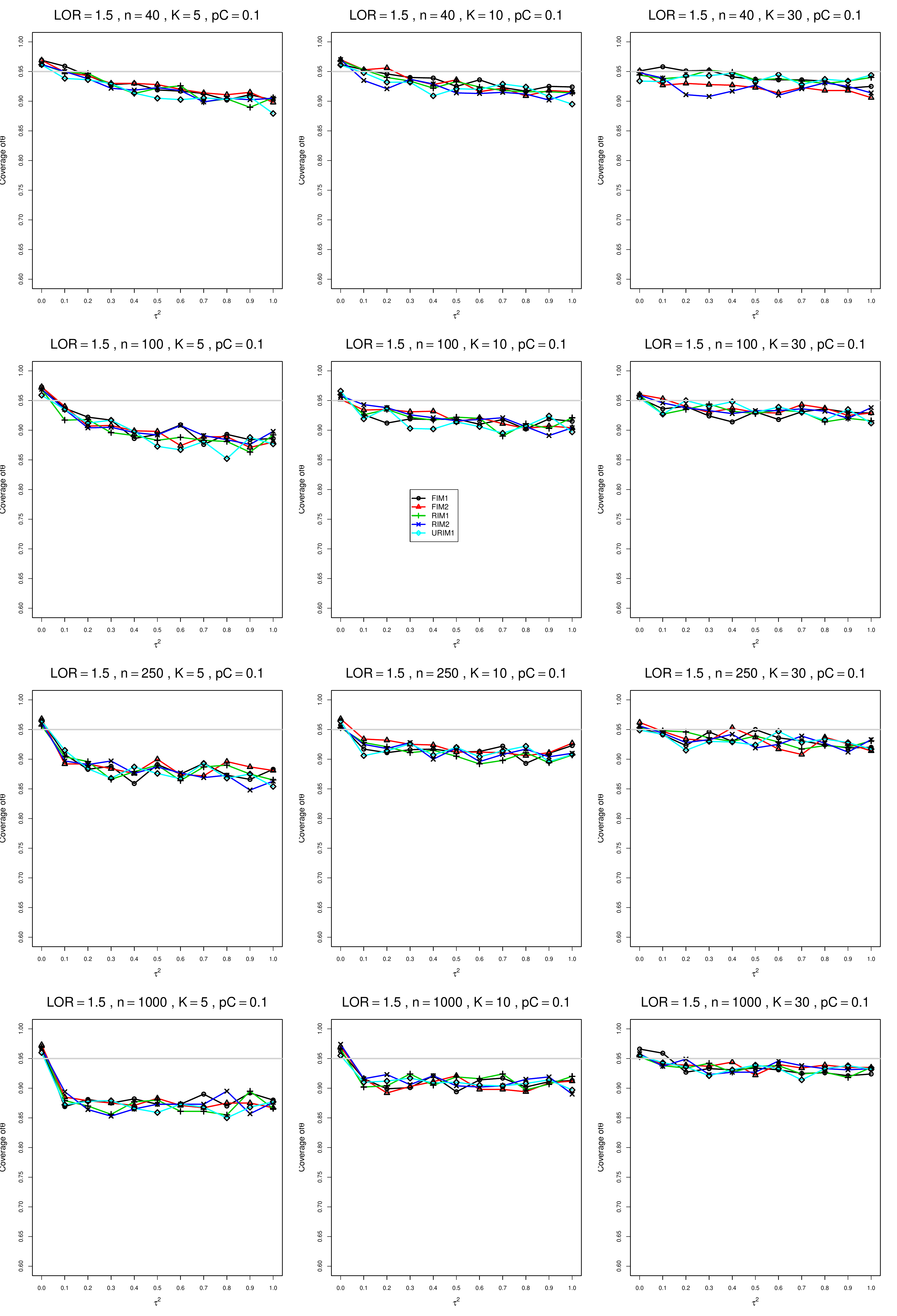}
	\caption{Coverage of the DerSimonian-Laird confidence interval for $\theta=1.5$, $p_{C}=0.1$, $\sigma^2=0.4$, constant sample sizes $n=40,\;100,\;250,\;1000$.
The data-generation mechanisms are FIM1 ($\circ$), FIM2 ($\triangle$), RIM1 (+), RIM2 ($\times$), and URIM1 ($\diamond$).
		\label{PlotCovThetamu15andpC01LOR_DLsigma04}}
\end{figure}
\begin{figure}[t]
	\centering
	\includegraphics[scale=0.33]{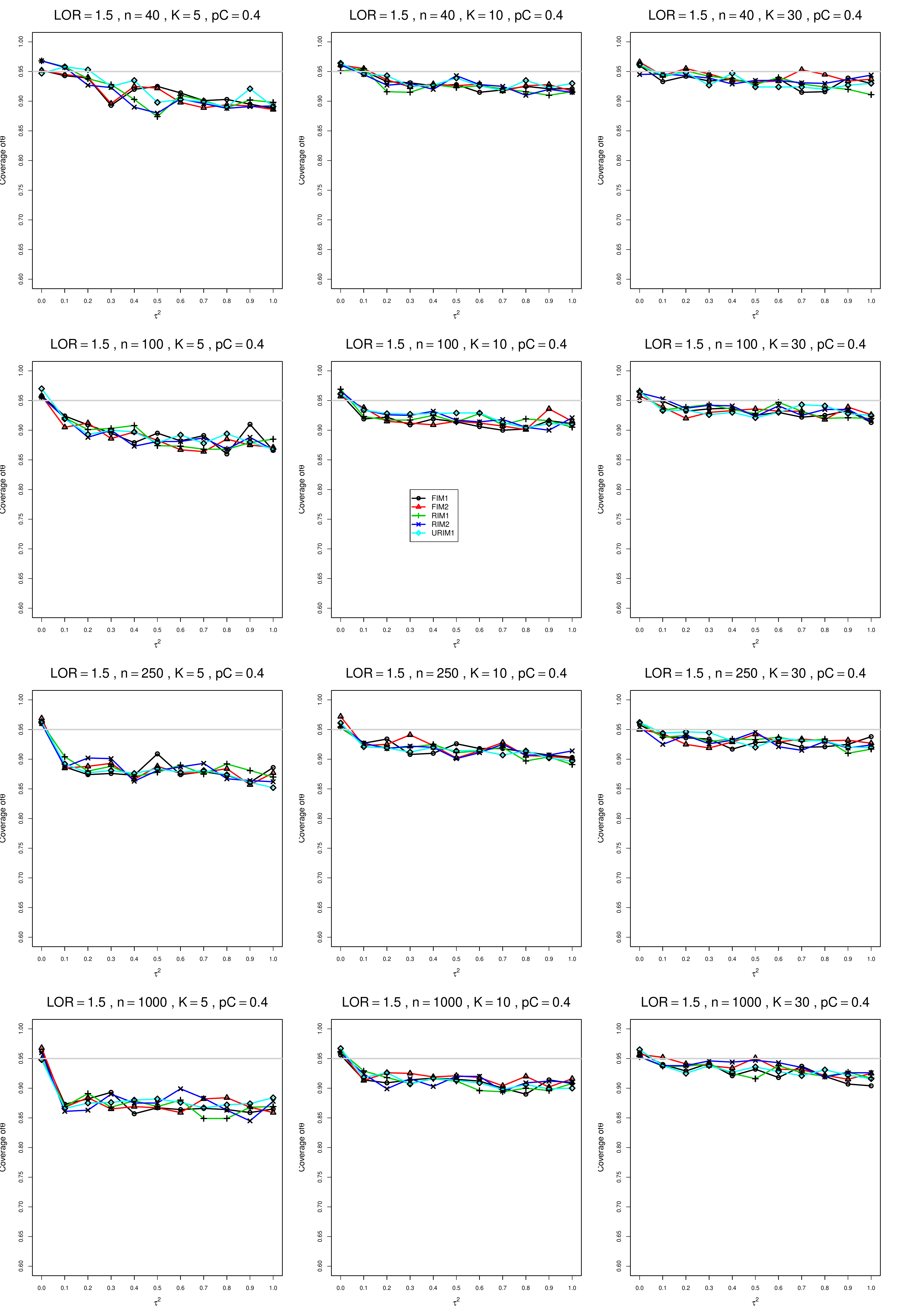}
	\caption{Coverage of the DerSimonian-Laird confidence interval for $\theta=1.5$, $p_{C}=0.4$, $\sigma^2=0.4$, constant sample sizes $n=40,\;100,\;250,\;1000$.
The data-generation mechanisms are FIM1 ($\circ$), FIM2 ($\triangle$), RIM1 (+), RIM2 ($\times$), and URIM1 ($\diamond$).
		\label{PlotCovThetamu15andpC04LOR_DLsigma04}}
\end{figure}
\begin{figure}[t]
	\centering
	\includegraphics[scale=0.33]{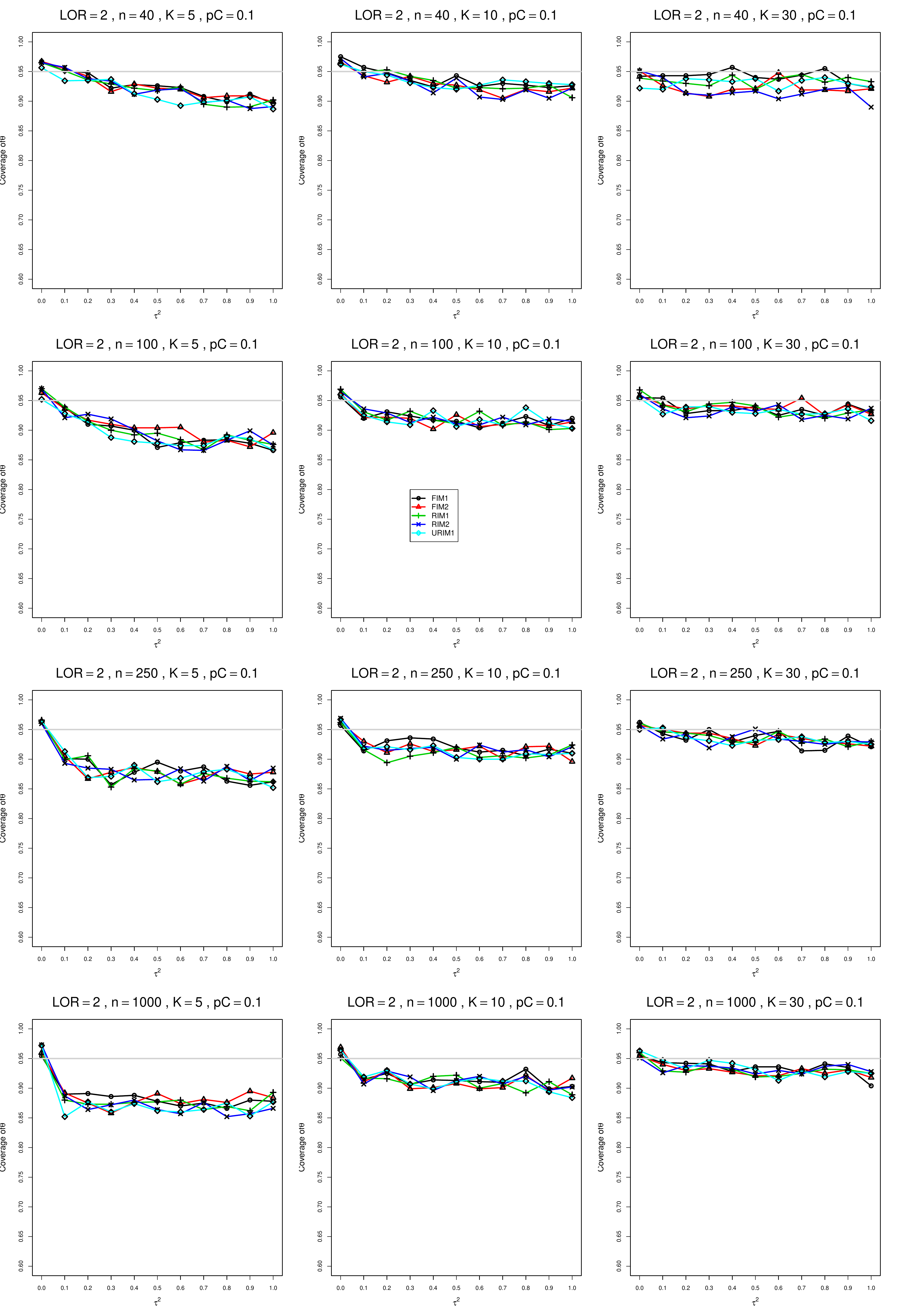}
	\caption{Coverage of the DerSimonian-Laird confidence interval for $\theta=2$, $p_{C}=0.1$, $\sigma^2=0.4$, constant sample sizes $n=40,\;100,\;250,\;1000$.
The data-generation mechanisms are FIM1 ($\circ$), FIM2 ($\triangle$), RIM1 (+), RIM2 ($\times$), and URIM1 ($\diamond$).
		\label{PlotCovThetamu2andpC01LOR_DLsigma04}}
\end{figure}
\begin{figure}[t]
	\centering
	\includegraphics[scale=0.33]{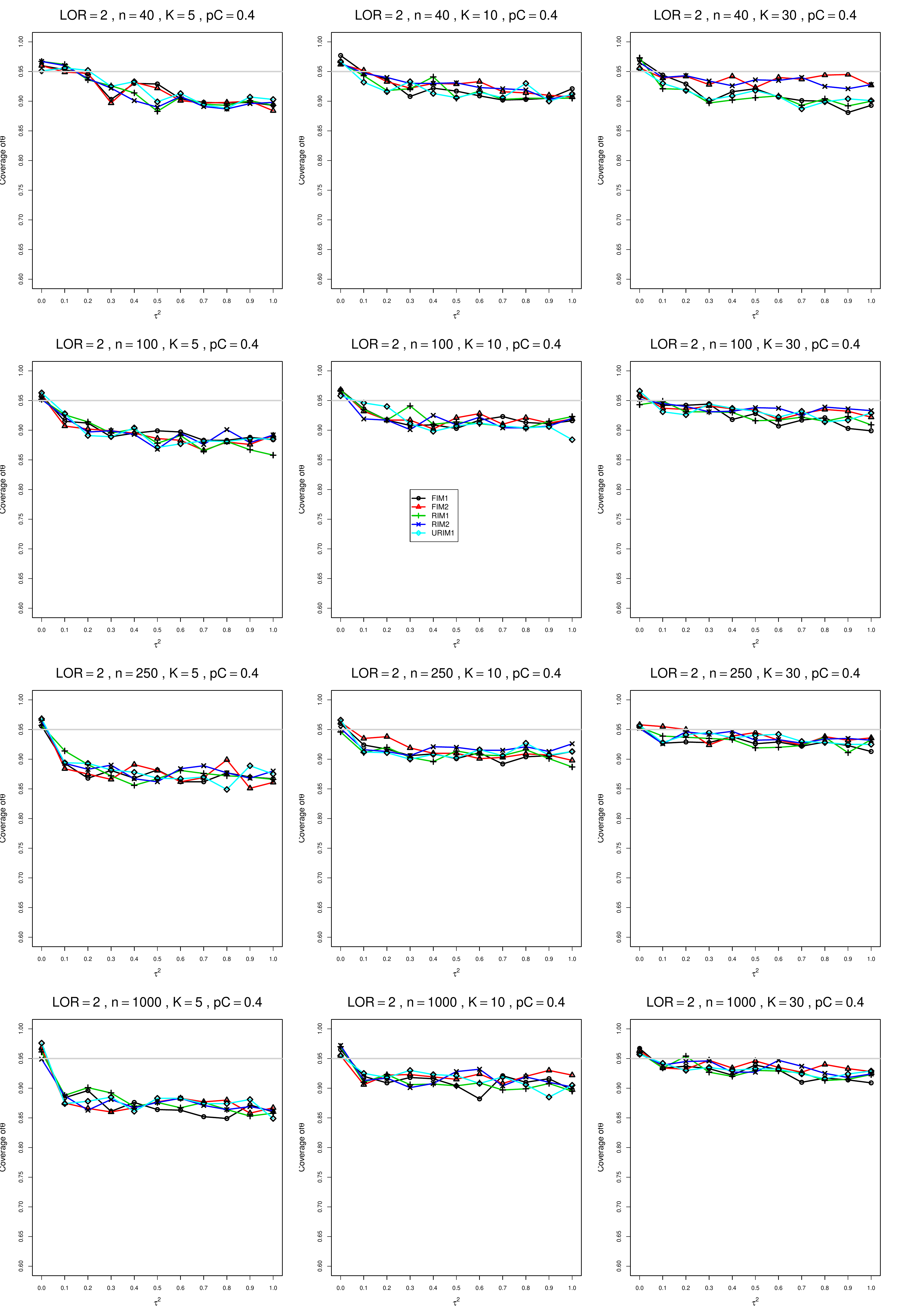}
	\caption{Coverage of the DerSimonian-Laird confidence interval for $\theta=2$, $p_{C}=0.4$, $\sigma^2=0.4$, constant sample sizes $n=40,\;100,\;250,\;1000$.
The data-generation mechanisms are FIM1 ($\circ$), FIM2 ($\triangle$), RIM1 (+), RIM2 ($\times$), and URIM1 ($\diamond$).
		\label{PlotCovThetamu2andpC04LOR_DLsigma04}}
\end{figure}

\clearpage
\subsection*{A3.2 Coverage of $\hat{\theta}_{REML}$}
\renewcommand{\thefigure}{A3.2.\arabic{figure}}
\setcounter{figure}{0}

\begin{figure}[t]
	\centering
	\includegraphics[scale=0.33]{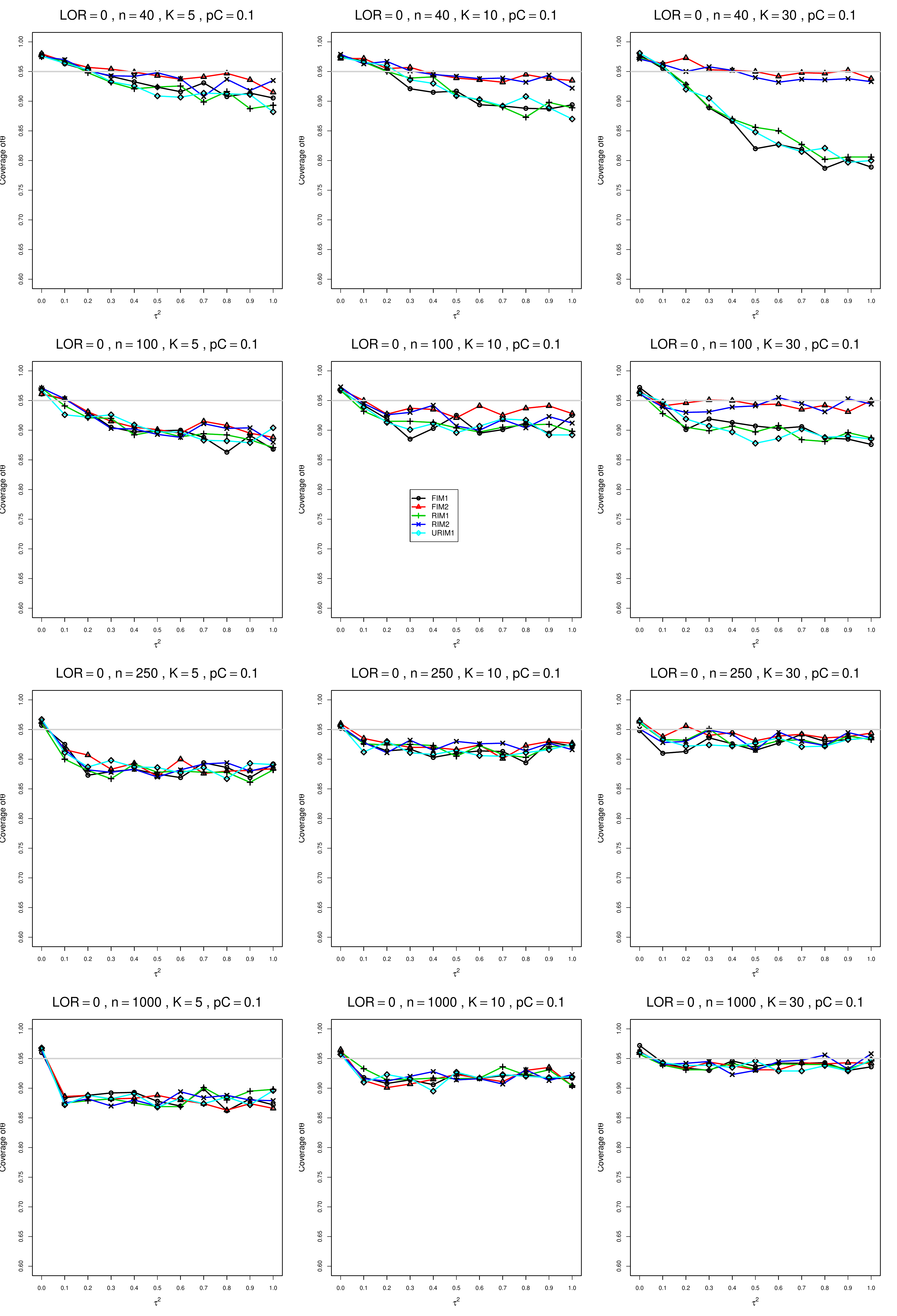}
	\caption{Coverage of the Restricted Maximum Likelihood confidence interval for $\theta=0$, $p_{C}=0.1$, $\sigma^2=0.1$, constant sample sizes $n=40,\;100,\;250,\;1000$.
The data-generation mechanisms are FIM1 ($\circ$), FIM2 ($\triangle$), RIM1 (+), RIM2 ($\times$), and URIM1 ($\diamond$).
		\label{PlotCovThetamu0andpC01LOR_REMLsigma01}}
\end{figure}
\begin{figure}[t]
	\centering
	\includegraphics[scale=0.33]{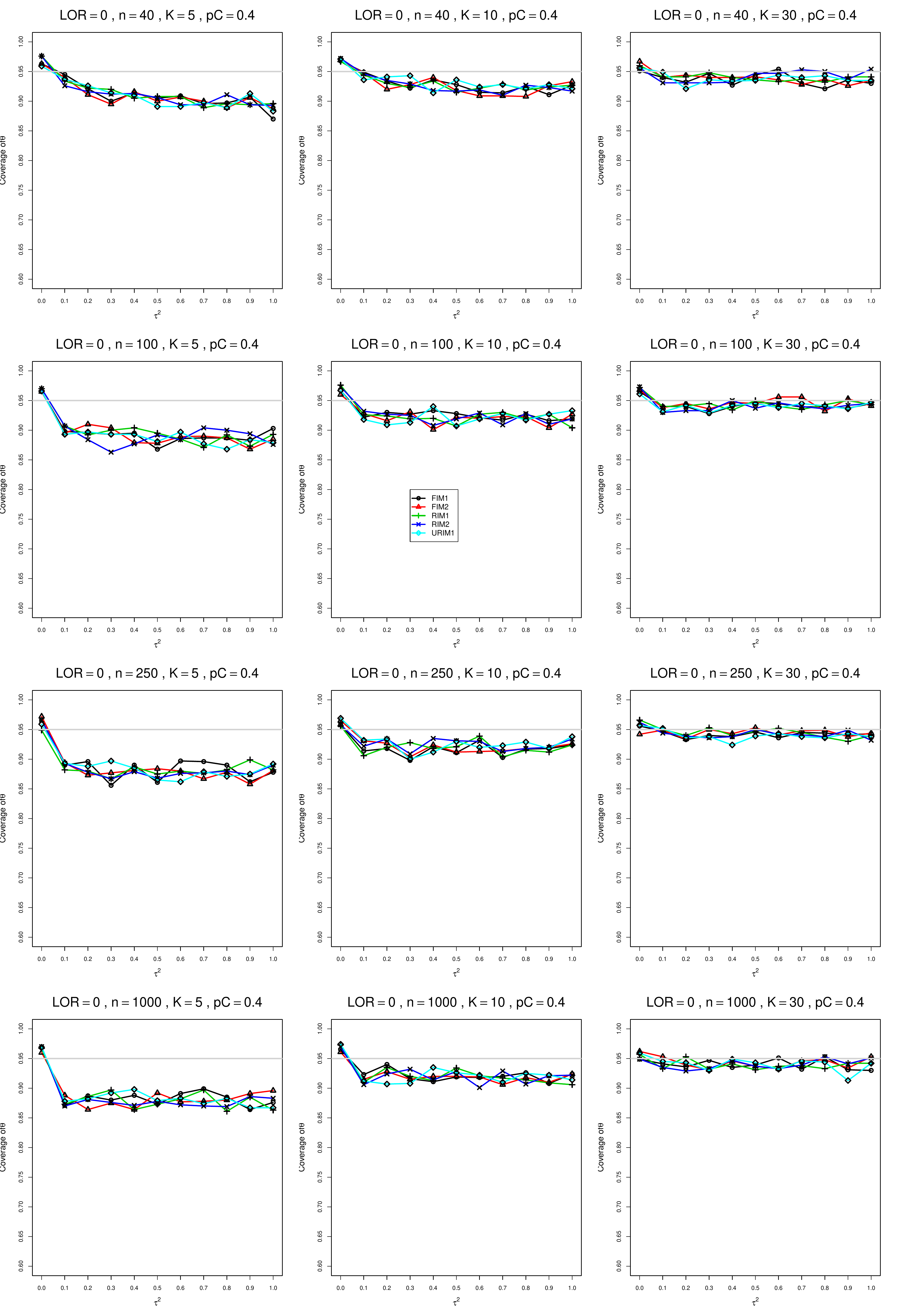}
	\caption{Coverage of the Restricted Maximum Likelihood confidence interval for $\theta=0$, $p_{C}=0.4$, $\sigma^2=0.1$, constant sample sizes $n=40,\;100,\;250,\;1000$.
The data-generation mechanisms are FIM1 ($\circ$), FIM2 ($\triangle$), RIM1 (+), RIM2 ($\times$), and URIM1 ($\diamond$).
		\label{PlotCovThetamu0andpC04LOR_REMLsigma01}}
\end{figure}
\begin{figure}[t]
	\centering
	\includegraphics[scale=0.33]{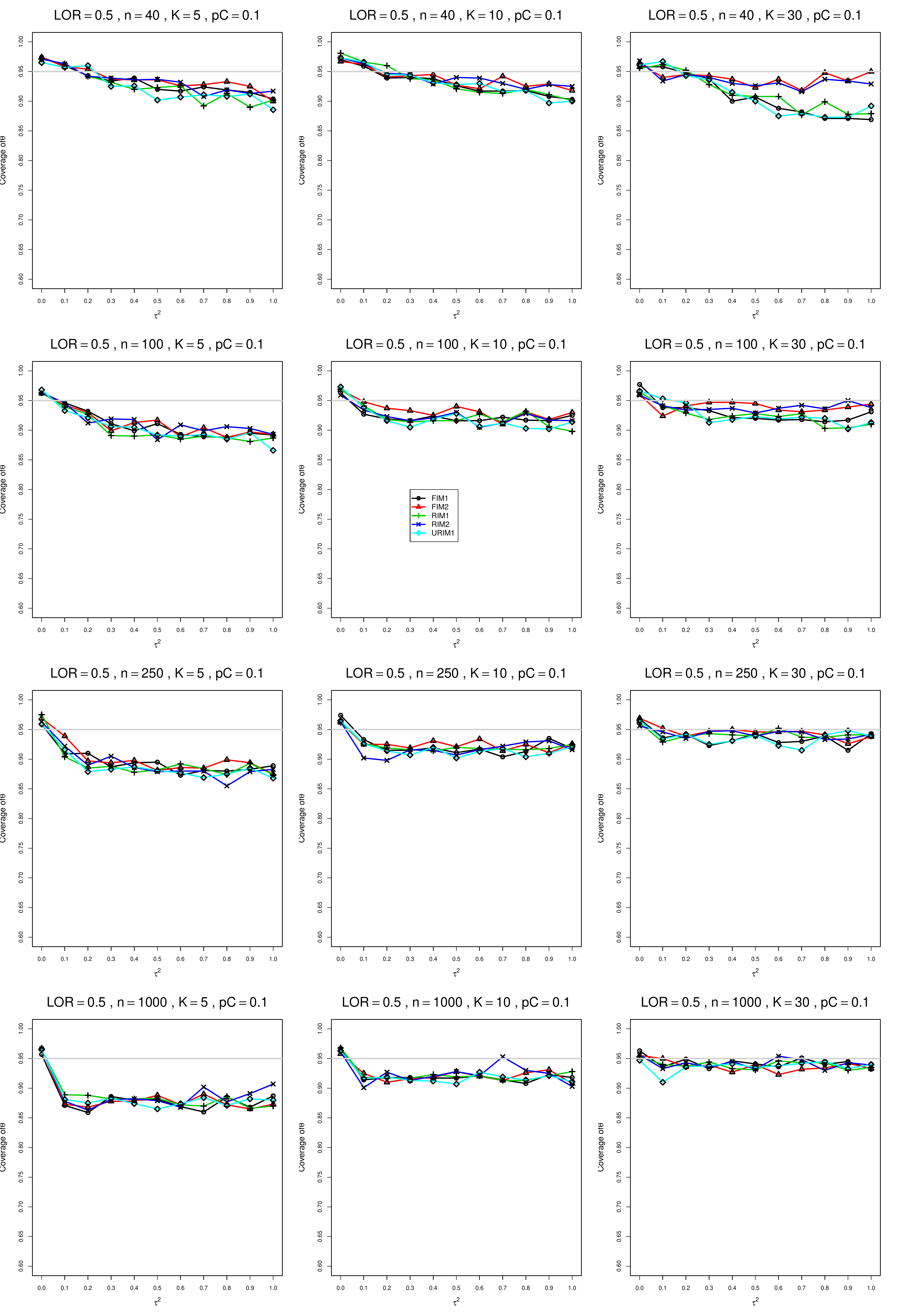}
	\caption{Coverage of the Restricted Maximum Likelihood confidence interval for $\theta=0.5$, $p_{C}=0.1$, $\sigma^2=0.1$, constant sample sizes $n=40,\;100,\;250,\;1000$.
The data-generation mechanisms are FIM1 ($\circ$), FIM2 ($\triangle$), RIM1 (+), RIM2 ($\times$), and URIM1 ($\diamond$).
		\label{PlotCovThetamu05andpC01LOR_REMLsigma01}}
\end{figure}
\begin{figure}[t]
	\centering
	\includegraphics[scale=0.33]{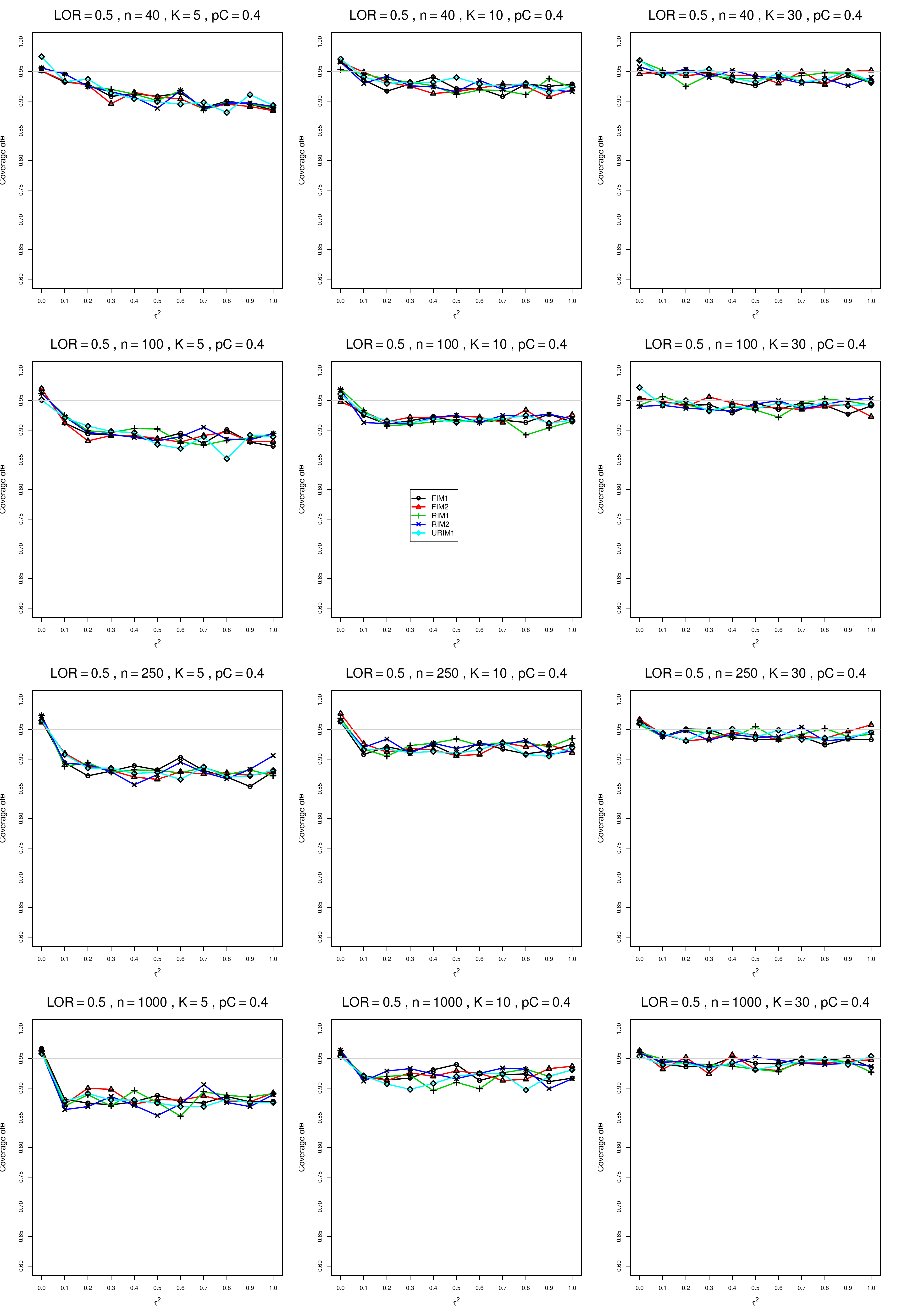}
	\caption{Coverage of the Restricted Maximum Likelihood confidence interval for $\theta=0.5$, $p_{C}=0.4$, $\sigma^2=0.1$, constant sample sizes $n=40,\;100,\;250,\;1000$.
The data-generation mechanisms are FIM1 ($\circ$), FIM2 ($\triangle$), RIM1 (+), RIM2 ($\times$), and URIM1 ($\diamond$).
		\label{PlotCovThetamu05andpC04LOR_REMLsigma01}}
\end{figure}
\begin{figure}[t]
	\centering
	\includegraphics[scale=0.33]{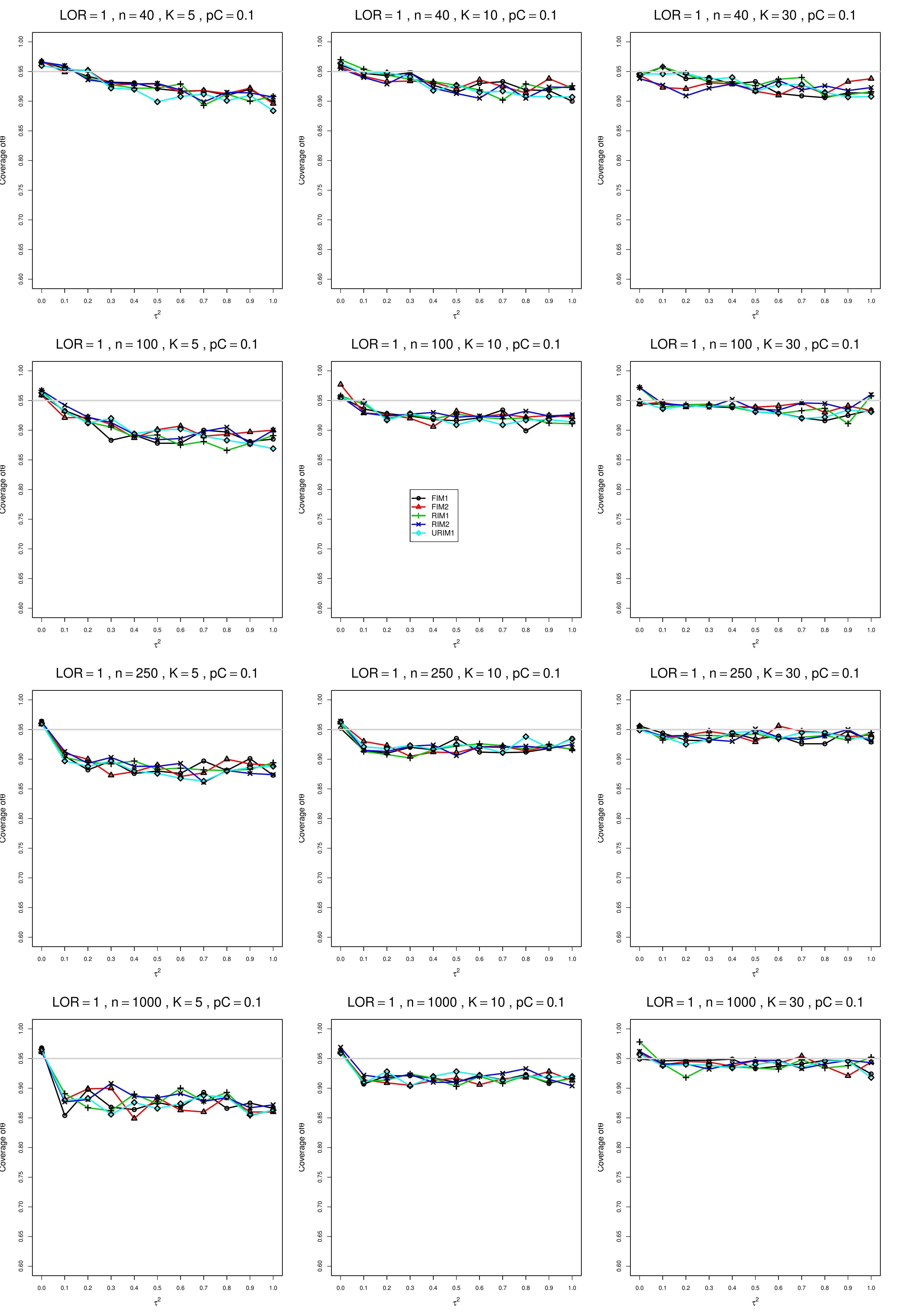}
	\caption{Coverage of the Restricted Maximum Likelihood confidence interval for $\theta=1$, $p_{C}=0.1$, $\sigma^2=0.1$, constant sample sizes $n=40,\;100,\;250,\;1000$.
The data-generation mechanisms are FIM1 ($\circ$), FIM2 ($\triangle$), RIM1 (+), RIM2 ($\times$), and URIM1 ($\diamond$).
		\label{PlotCovThetamu1andpC01LOR_REMLsigma01}}
\end{figure}
\begin{figure}[t]
	\centering
	\includegraphics[scale=0.33]{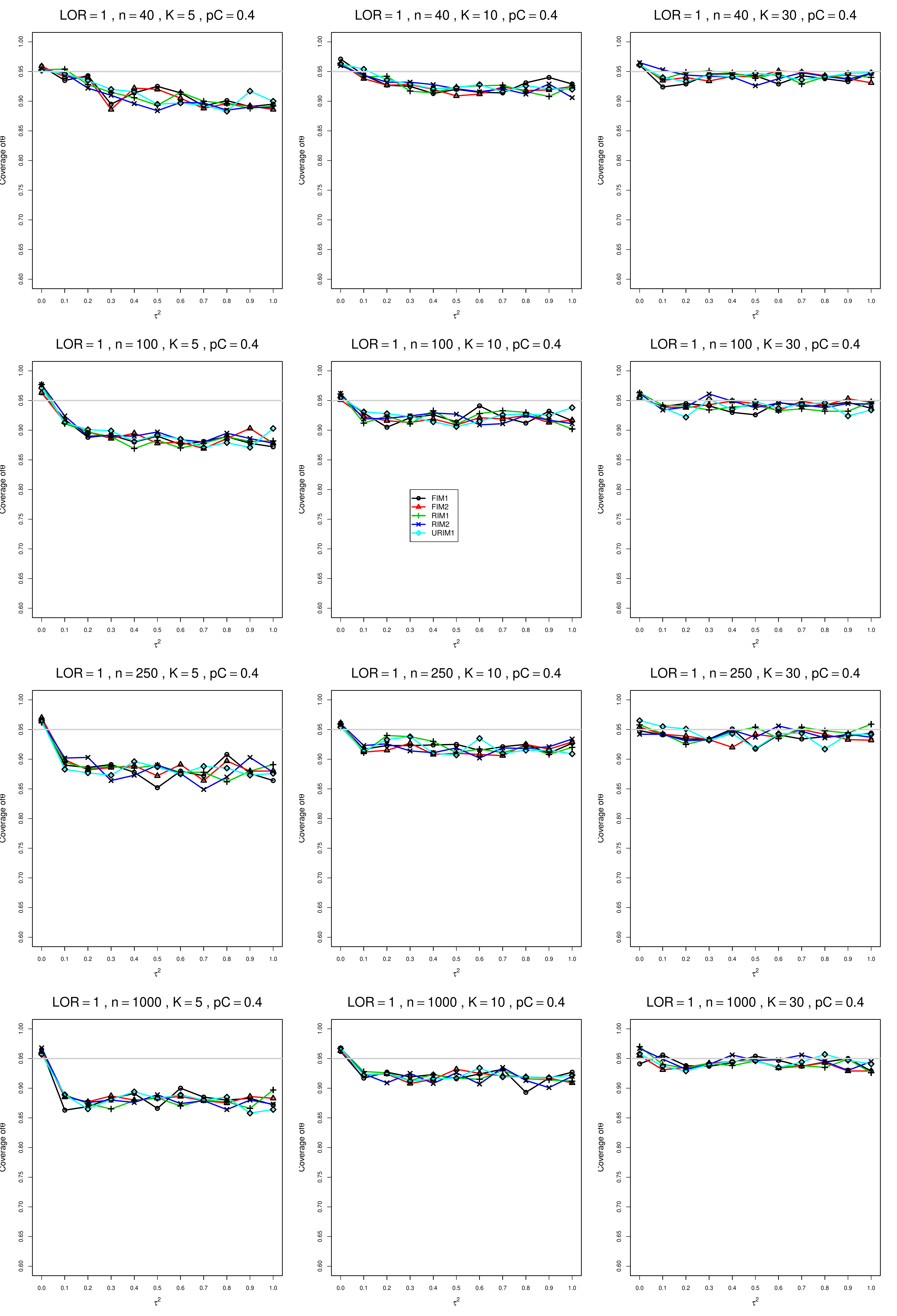}
	\caption{Coverage of the Restricted Maximum Likelihood confidence interval for $\theta=1$, $p_{C}=0.4$, $\sigma^2=0.1$, constant sample sizes $n=40,\;100,\;250,\;1000$.
The data-generation mechanisms are FIM1 ($\circ$), FIM2 ($\triangle$), RIM1 (+), RIM2 ($\times$), and URIM1 ($\diamond$).
		\label{PlotCovThetamu1andpC04LOR_REMLsigma01}}
\end{figure}
\begin{figure}[t]
	\centering
	\includegraphics[scale=0.33]{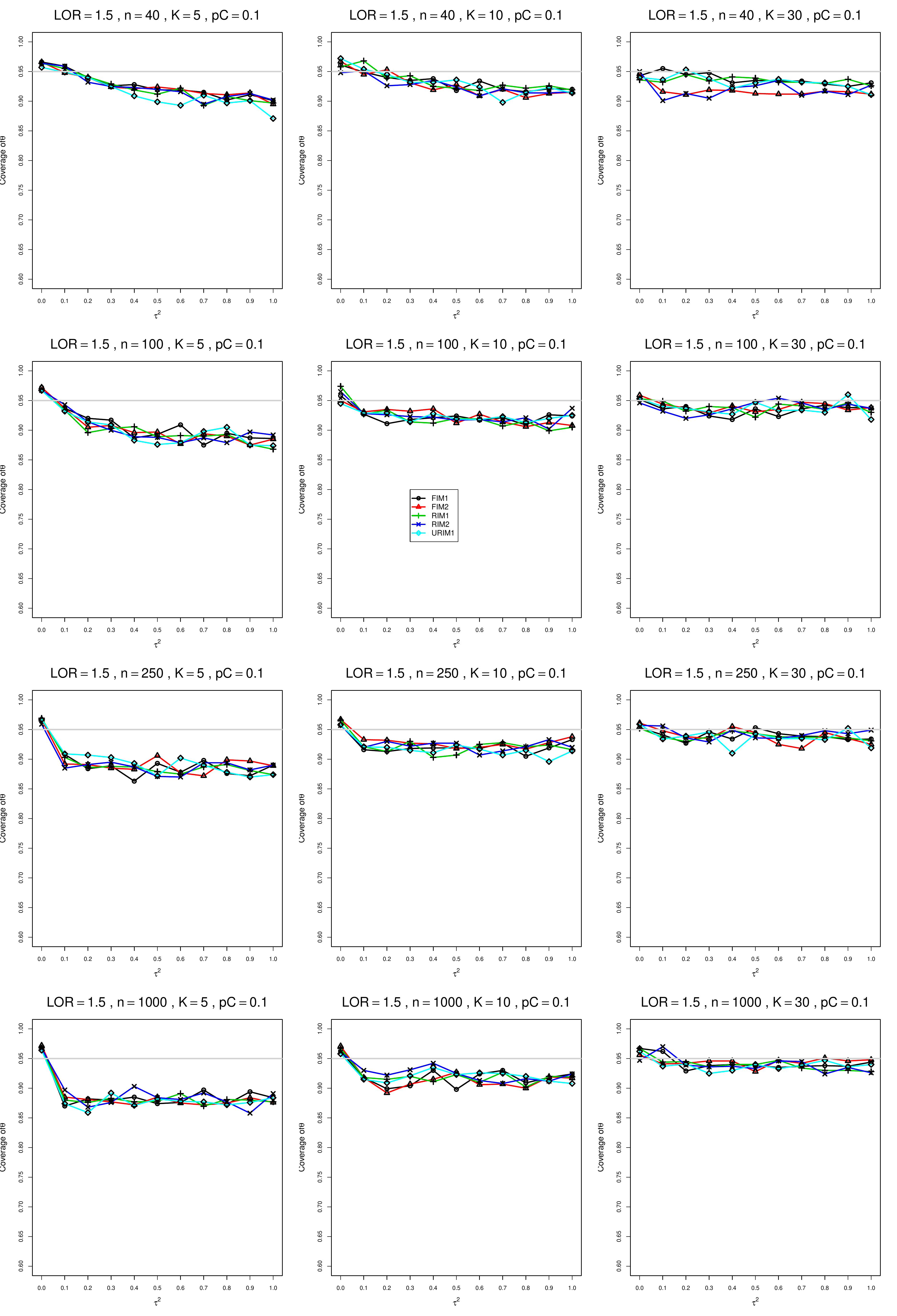}
	\caption{Coverage of the Restricted Maximum Likelihood confidence interval for $\theta=1.5$, $p_{C}=0.1$, $\sigma^2=0.1$, constant sample sizes $n=40,\;100,\;250,\;1000$.
The data-generation mechanisms are FIM1 ($\circ$), FIM2 ($\triangle$), RIM1 (+), RIM2 ($\times$), and URIM1 ($\diamond$).
		\label{PlotCovThetamu15andpC01LOR_REMLsigma01}}
\end{figure}
\begin{figure}[t]
	\centering
	\includegraphics[scale=0.33]{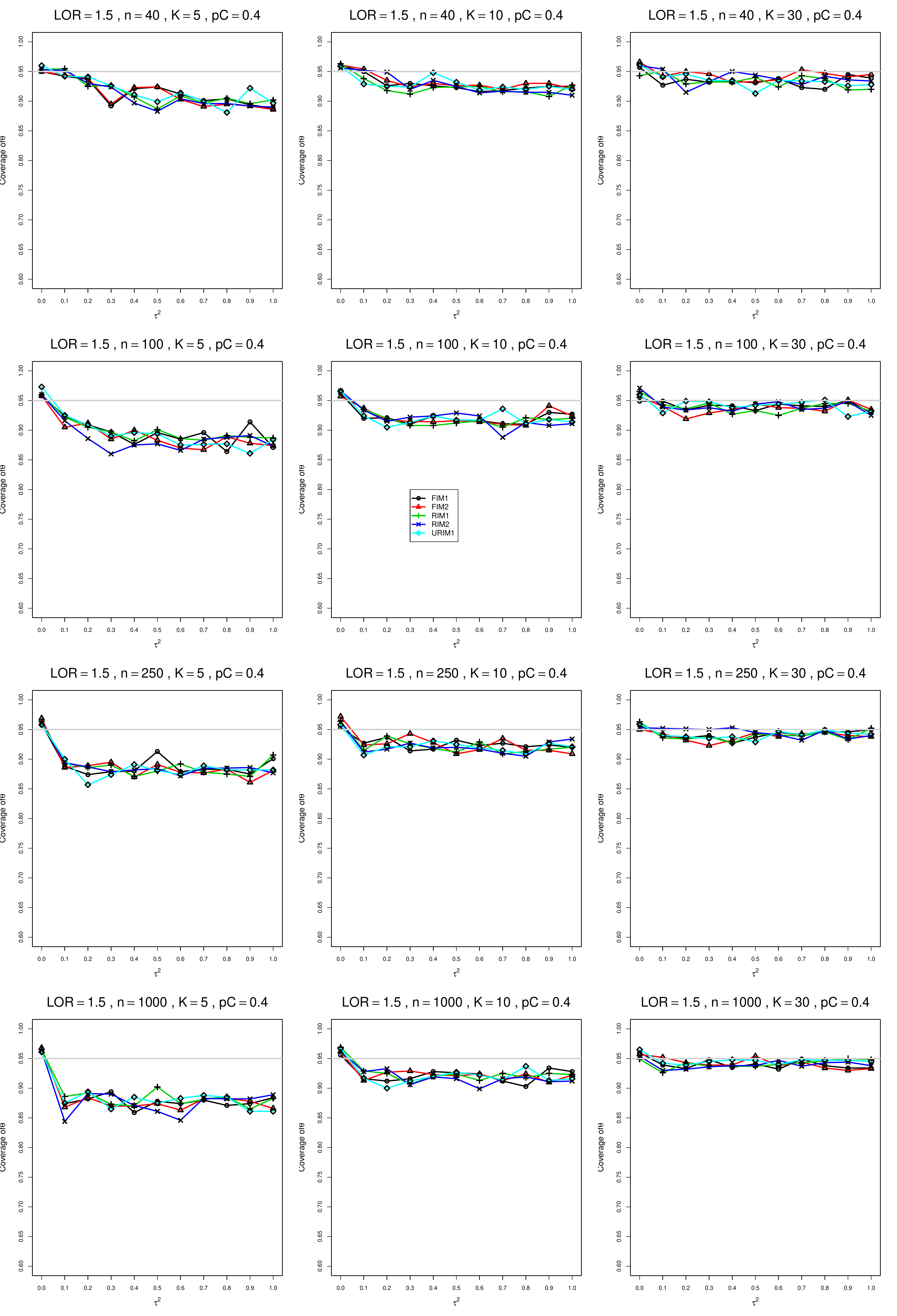}
	\caption{Coverage of the Restricted Maximum Likelihood confidence interval for $\theta=1.5$, $p_{C}=0.4$, $\sigma^2=0.1$, constant sample sizes $n=40,\;100,\;250,\;1000$.
The data-generation mechanisms are FIM1 ($\circ$), FIM2 ($\triangle$), RIM1 (+), RIM2 ($\times$), and URIM1 ($\diamond$).
		\label{PlotCovThetamu15andpC04LOR_REMLsigma01}}
\end{figure}
\begin{figure}[t]
	\centering
	\includegraphics[scale=0.33]{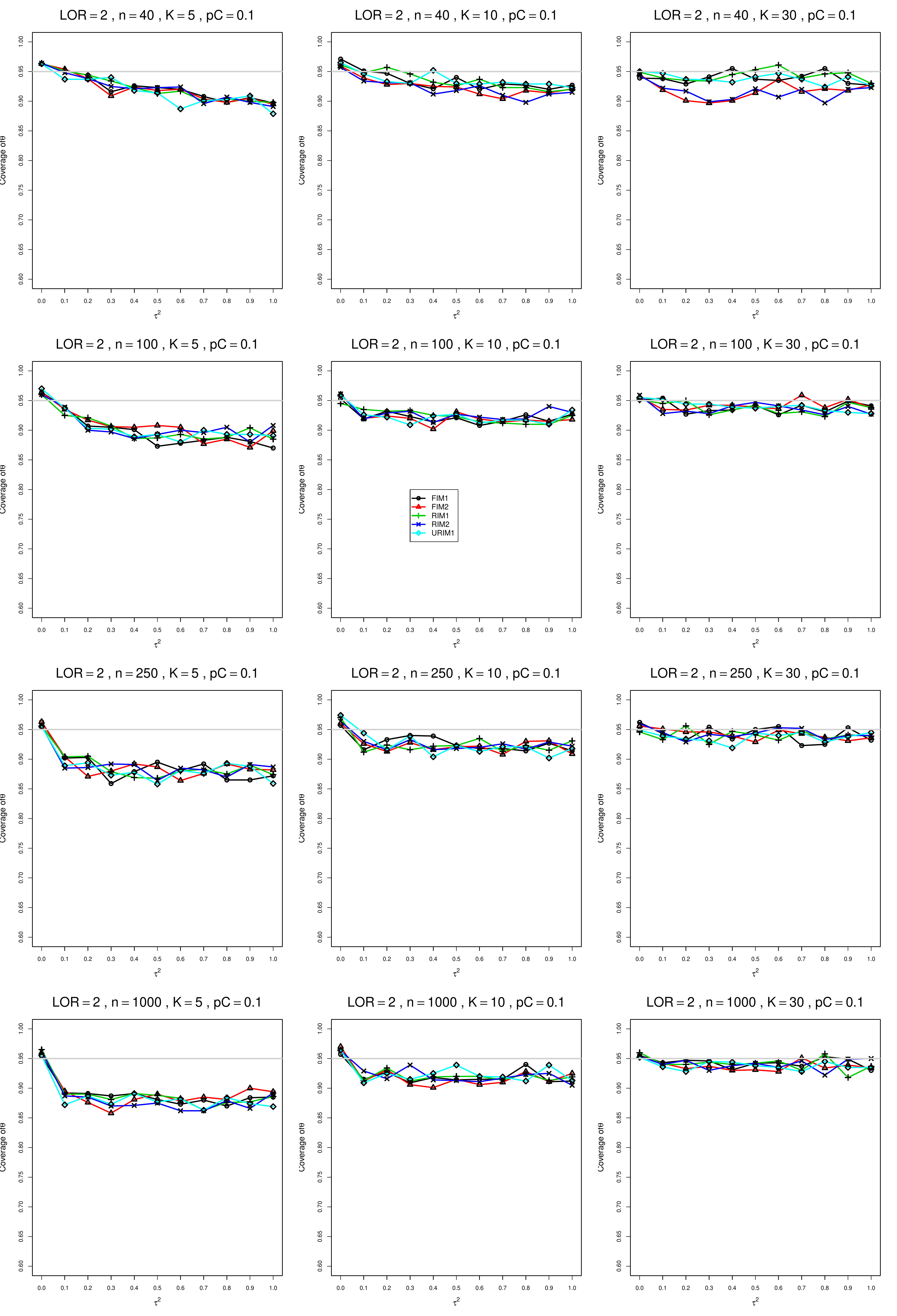}
	\caption{Coverage of the Restricted Maximum Likelihood confidence interval for $\theta=2$, $p_{C}=0.1$, $\sigma^2=0.1$, constant sample sizes $n=40,\;100,\;250,\;1000$.
The data-generation mechanisms are FIM1 ($\circ$), FIM2 ($\triangle$), RIM1 (+), RIM2 ($\times$), and URIM1 ($\diamond$).
		\label{PlotCovThetamu2andpC01LOR_REMLsigma01}}
\end{figure}
\begin{figure}[t]
	\centering
	\includegraphics[scale=0.33]{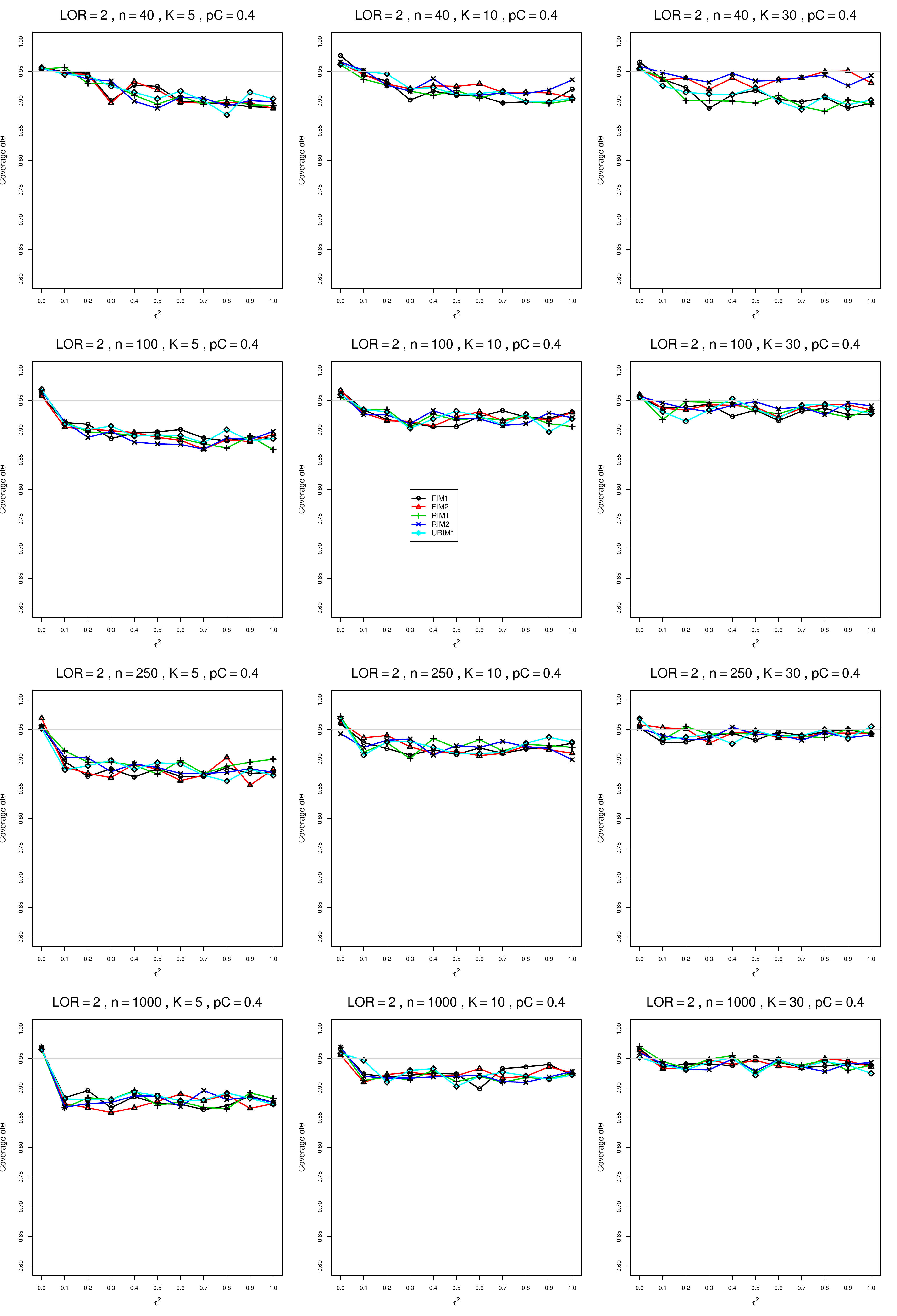}
	\caption{Coverage of the Restricted Maximum Likelihood confidence interval for $\theta=2$, $p_{C}=0.4$, $\sigma^2=0.1$, constant sample sizes $n=40,\;100,\;250,\;1000$.
The data-generation mechanisms are FIM1 ($\circ$), FIM2 ($\triangle$), RIM1 (+), RIM2 ($\times$), and URIM1 ($\diamond$).
		\label{PlotCovThetamu2andpC04LOR_REMLsigma01}}
\end{figure}
\begin{figure}[t]
	\centering
	\includegraphics[scale=0.33]{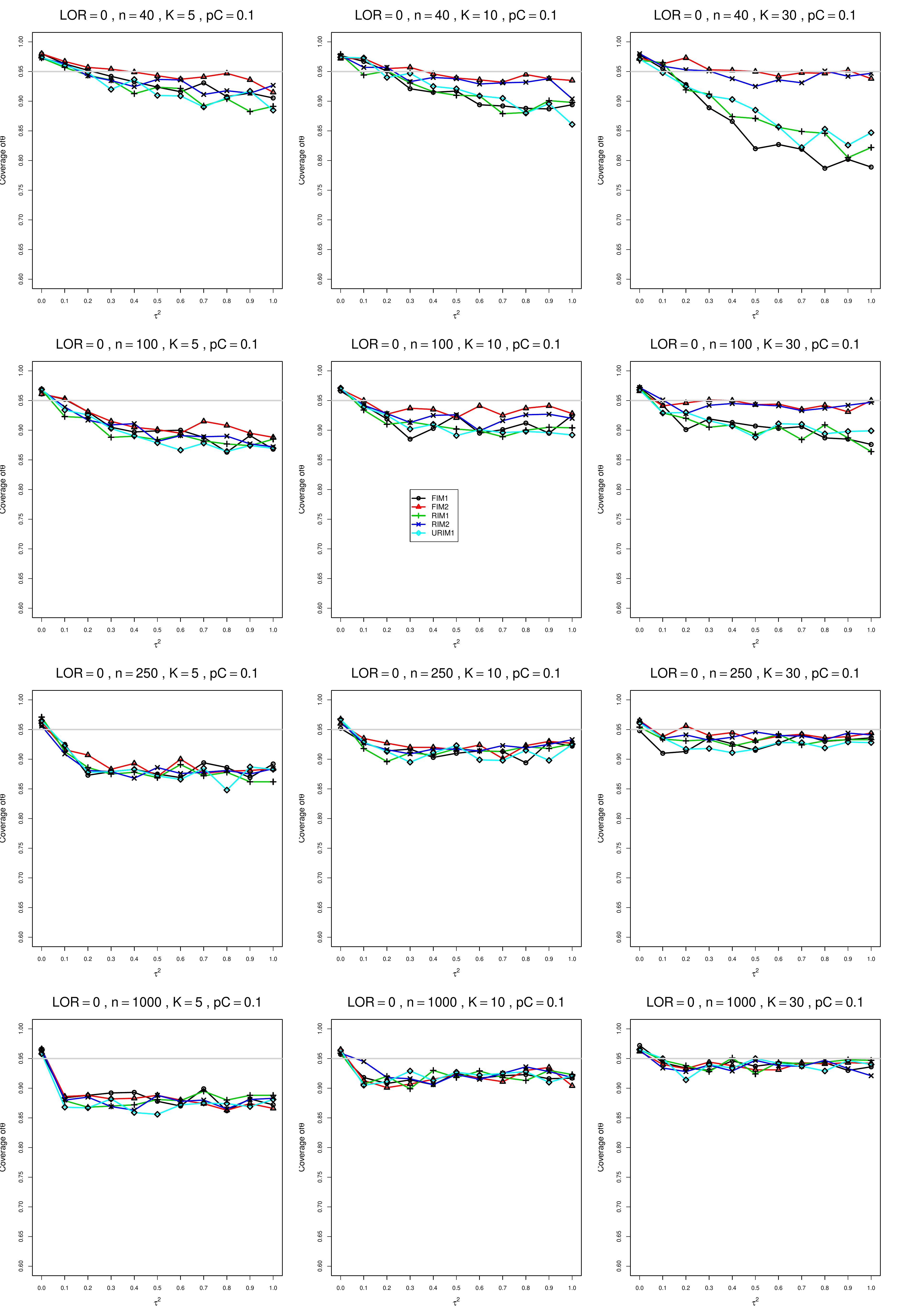}
	\caption{Coverage of the Restricted Maximum Likelihood confidence interval for $\theta=0$, $p_{C}=0.1$, $\sigma^2=0.4$, constant sample sizes $n=40,\;100,\;250,\;1000$.
The data-generation mechanisms are FIM1 ($\circ$), FIM2 ($\triangle$), RIM1 (+), RIM2 ($\times$), and URIM1 ($\diamond$).
		\label{PlotCovThetamu0andpC01LOR_REMLsigma04}}
\end{figure}
\begin{figure}[t]
	\centering
	\includegraphics[scale=0.33]{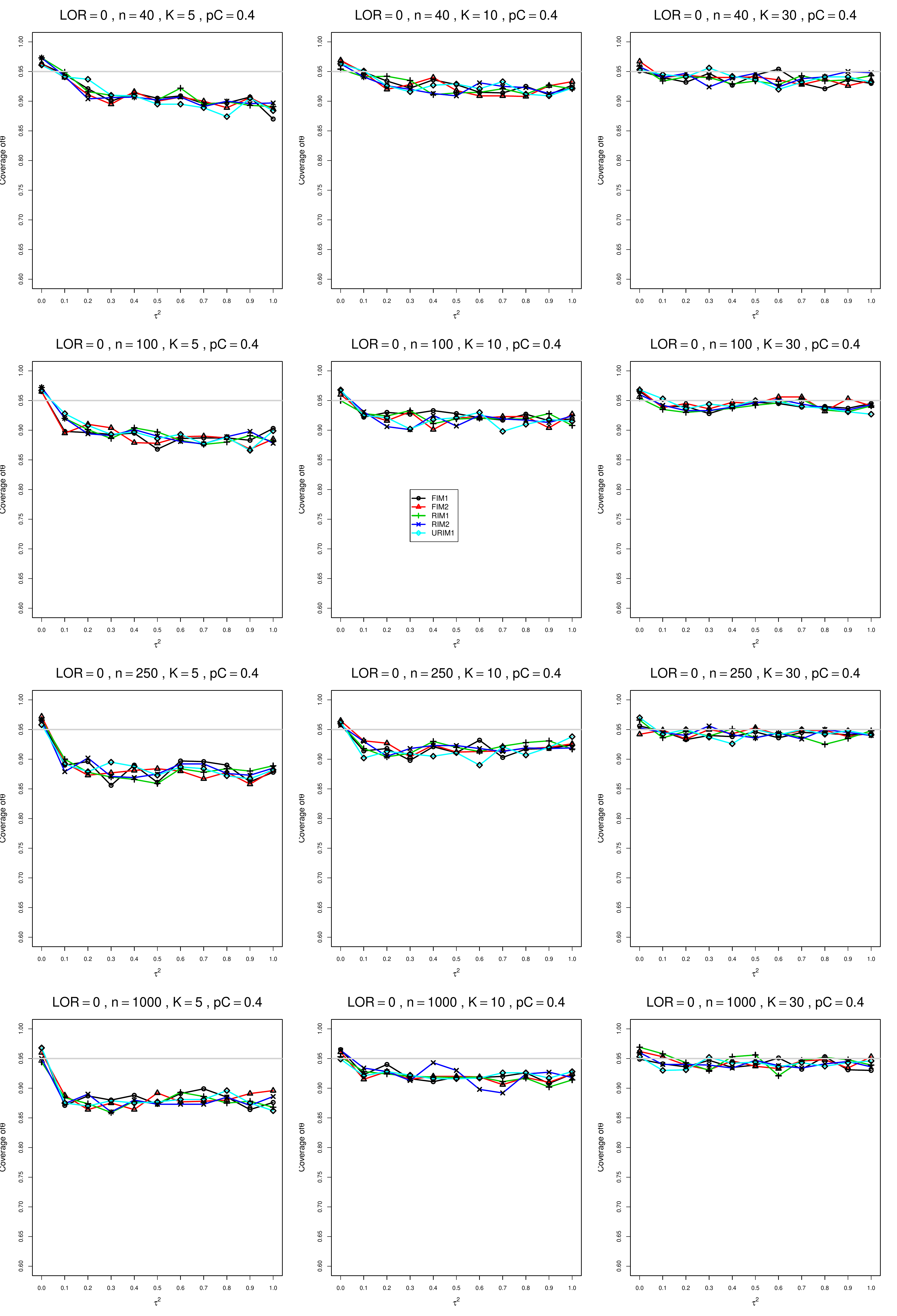}
	\caption{Coverage of the Restricted Maximum Likelihood confidence interval for $\theta=0$, $p_{C}=0.4$, $\sigma^2=0.4$, constant sample sizes $n=40,\;100,\;250,\;1000$.
The data-generation mechanisms are FIM1 ($\circ$), FIM2 ($\triangle$), RIM1 (+), RIM2 ($\times$), and URIM1 ($\diamond$).
		\label{PlotCovThetamu0andpC04LOR_REMLsigma04}}
\end{figure}
\begin{figure}[t]
	\centering
	\includegraphics[scale=0.33]{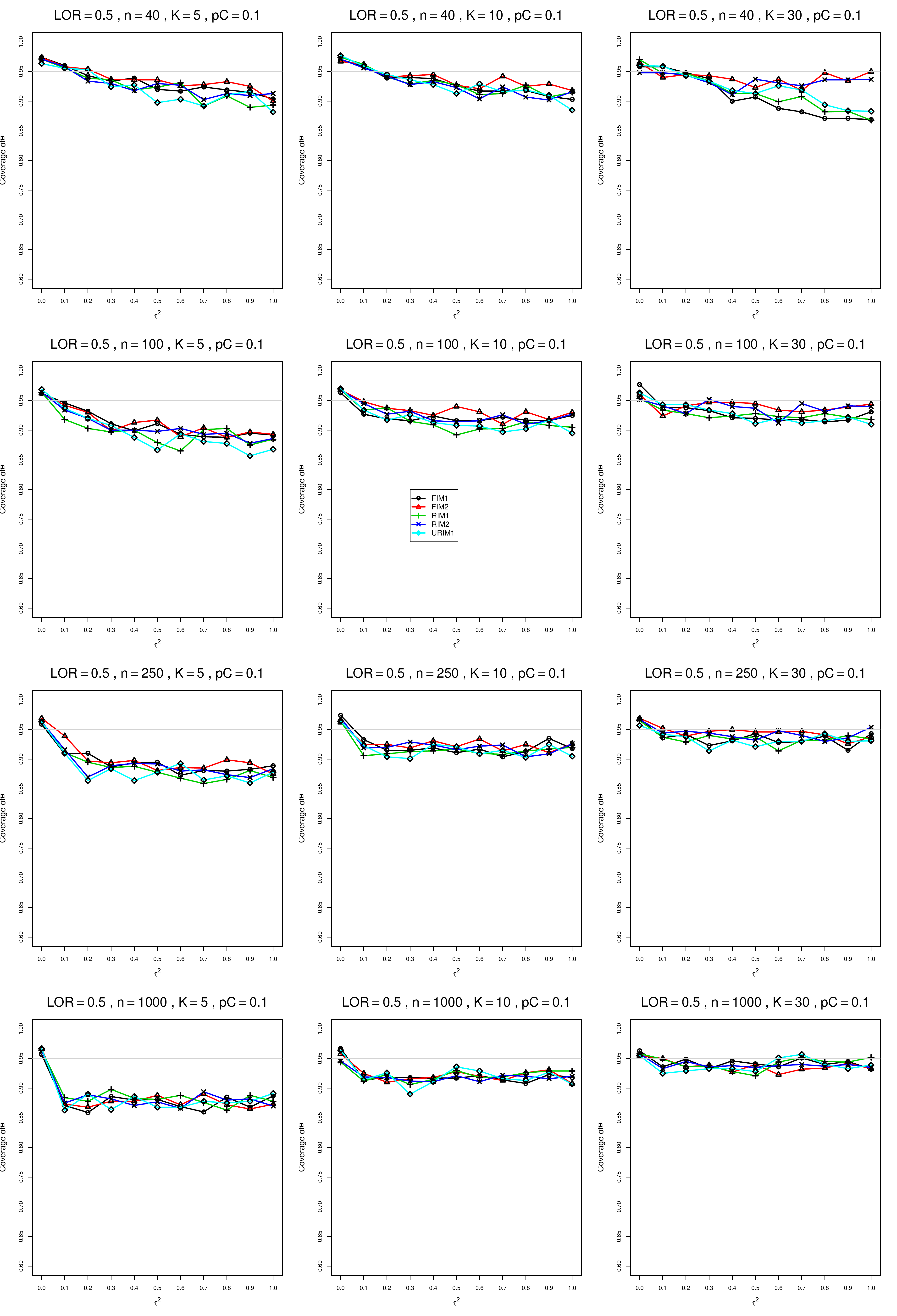}
	\caption{Coverage of the Restricted Maximum Likelihood confidence interval for $\theta=0.5$, $p_{C}=0.1$, $\sigma^2=0.4$, constant sample sizes $n=40,\;100,\;250,\;1000$.
The data-generation mechanisms are FIM1 ($\circ$), FIM2 ($\triangle$), RIM1 (+), RIM2 ($\times$), and URIM1 ($\diamond$).
		\label{PlotCovThetamu05andpC01LOR_REMLsigma04}}
\end{figure}
\begin{figure}[t]
	\centering
	\includegraphics[scale=0.33]{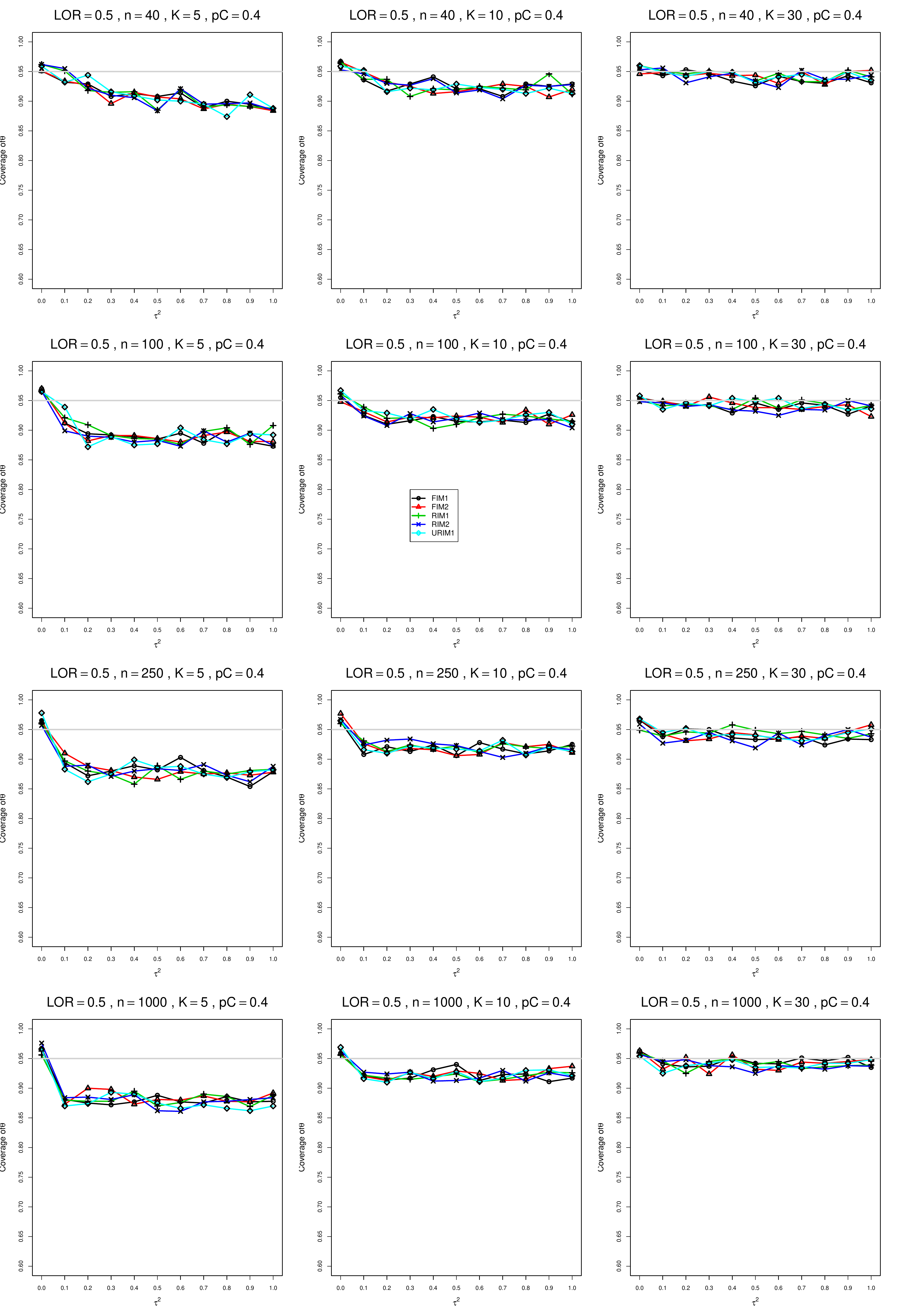}
	\caption{Coverage of the Restricted Maximum Likelihood confidence interval for $\theta=0.5$, $p_{C}=0.4$, $\sigma^2=0.4$, constant sample sizes $n=40,\;100,\;250,\;1000$.
The data-generation mechanisms are FIM1 ($\circ$), FIM2 ($\triangle$), RIM1 (+), RIM2 ($\times$), and URIM1 ($\diamond$).
		\label{PlotCovThetamu05andpC04LOR_REMLsigma04}}
\end{figure}
\begin{figure}[t]
	\centering
	\includegraphics[scale=0.33]{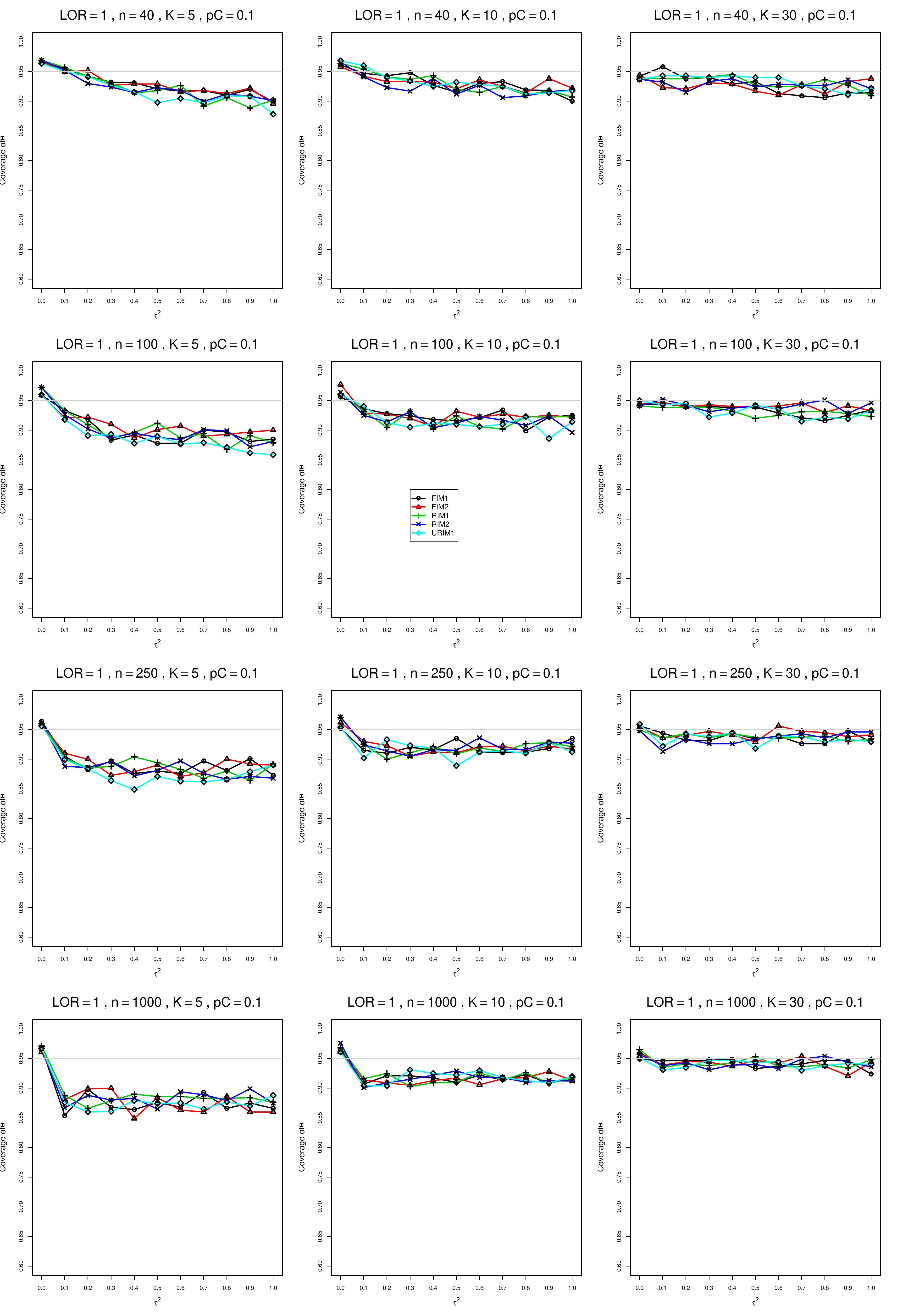}
	\caption{Coverage of the Restricted Maximum Likelihood confidence interval for $\theta=1$, $p_{C}=0.1$, $\sigma^2=0.4$, constant sample sizes $n=40,\;100,\;250,\;1000$.
The data-generation mechanisms are FIM1 ($\circ$), FIM2 ($\triangle$), RIM1 (+), RIM2 ($\times$), and URIM1 ($\diamond$).
		\label{PlotCovThetamu1andpC01LOR_REMLsigma04}}
\end{figure}
\begin{figure}[t]
	\centering
	\includegraphics[scale=0.33]{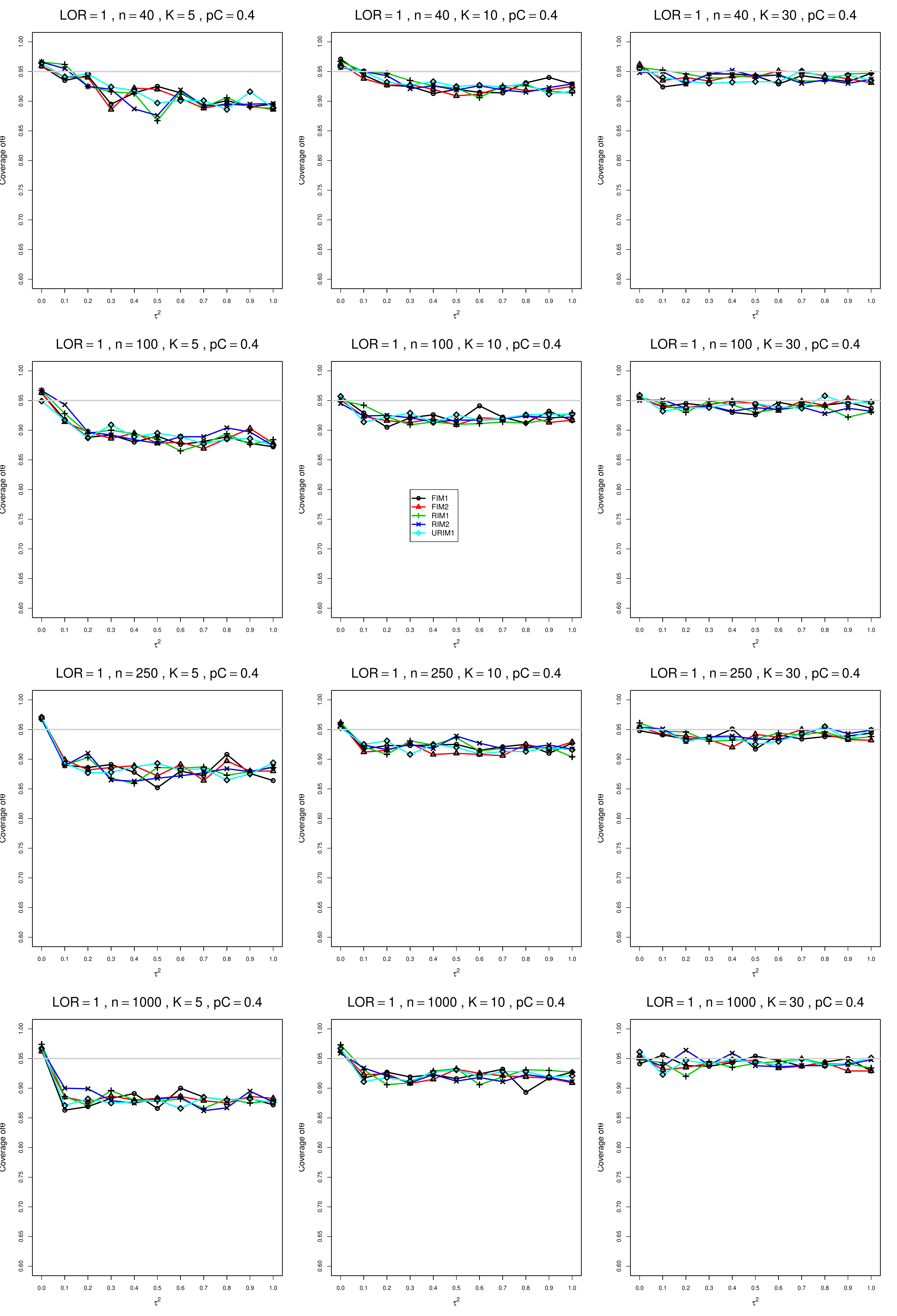}
	\caption{Coverage of the Restricted Maximum Likelihood confidence interval for $\theta=1$, $p_{C}=0.4$, $\sigma^2=0.4$, constant sample sizes $n=40,\;100,\;250,\;1000$.
The data-generation mechanisms are FIM1 ($\circ$), FIM2 ($\triangle$), RIM1 (+), RIM2 ($\times$), and URIM1 ($\diamond$).
		\label{PlotCovThetamu1andpC04LOR_REMLsigma04}}
\end{figure}
\begin{figure}[t]
	\centering
	\includegraphics[scale=0.33]{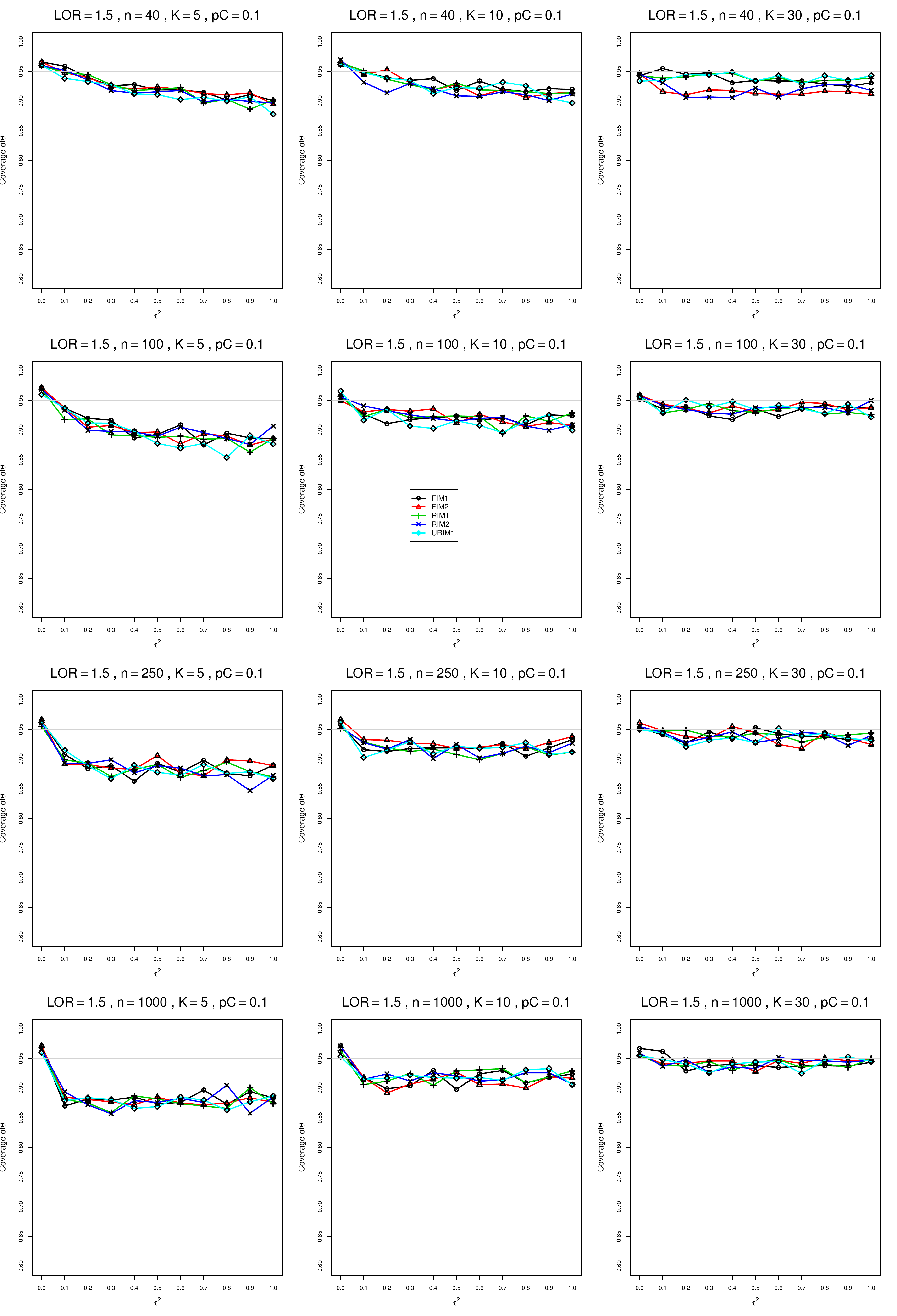}
	\caption{Coverage of the Restricted Maximum Likelihood confidence interval for $\theta=1.5$, $p_{C}=0.1$, $\sigma^2=0.4$, constant sample sizes $n=40,\;100,\;250,\;1000$.
The data-generation mechanisms are FIM1 ($\circ$), FIM2 ($\triangle$), RIM1 (+), RIM2 ($\times$), and URIM1 ($\diamond$).
		\label{PlotCovThetamu15andpC01LOR_REMLsigma04}}
\end{figure}
\begin{figure}[t]
	\centering
	\includegraphics[scale=0.33]{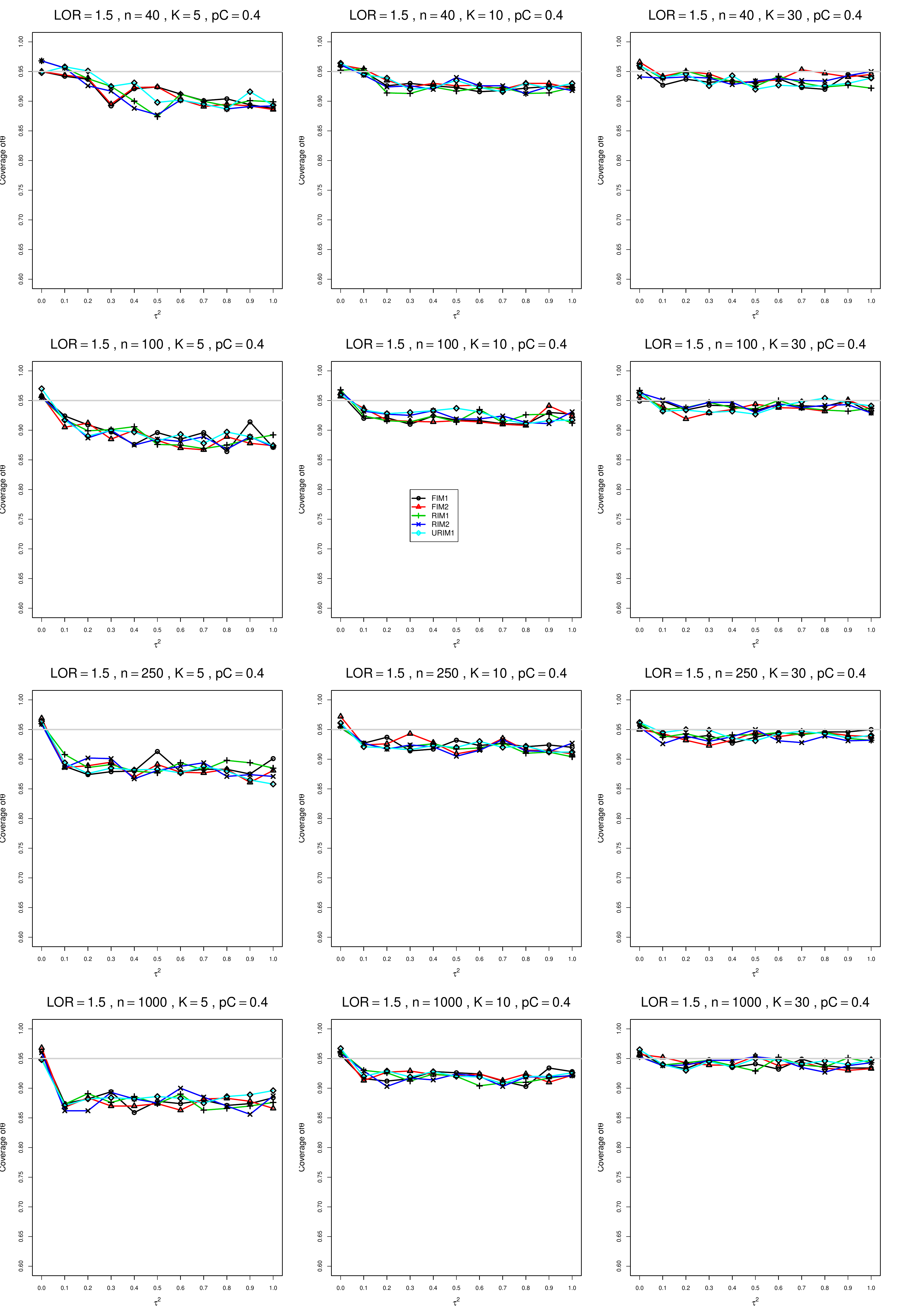}
	\caption{Coverage of the Restricted Maximum Likelihood confidence interval for $\theta=1.5$, $p_{C}=0.4$, $\sigma^2=0.4$, constant sample sizes $n=40,\;100,\;250,\;1000$.
The data-generation mechanisms are FIM1 ($\circ$), FIM2 ($\triangle$), RIM1 (+), RIM2 ($\times$), and URIM1 ($\diamond$).
		\label{PlotCovThetamu15andpC04LOR_REMLsigma04}}
\end{figure}
\begin{figure}[t]
	\centering
	\includegraphics[scale=0.33]{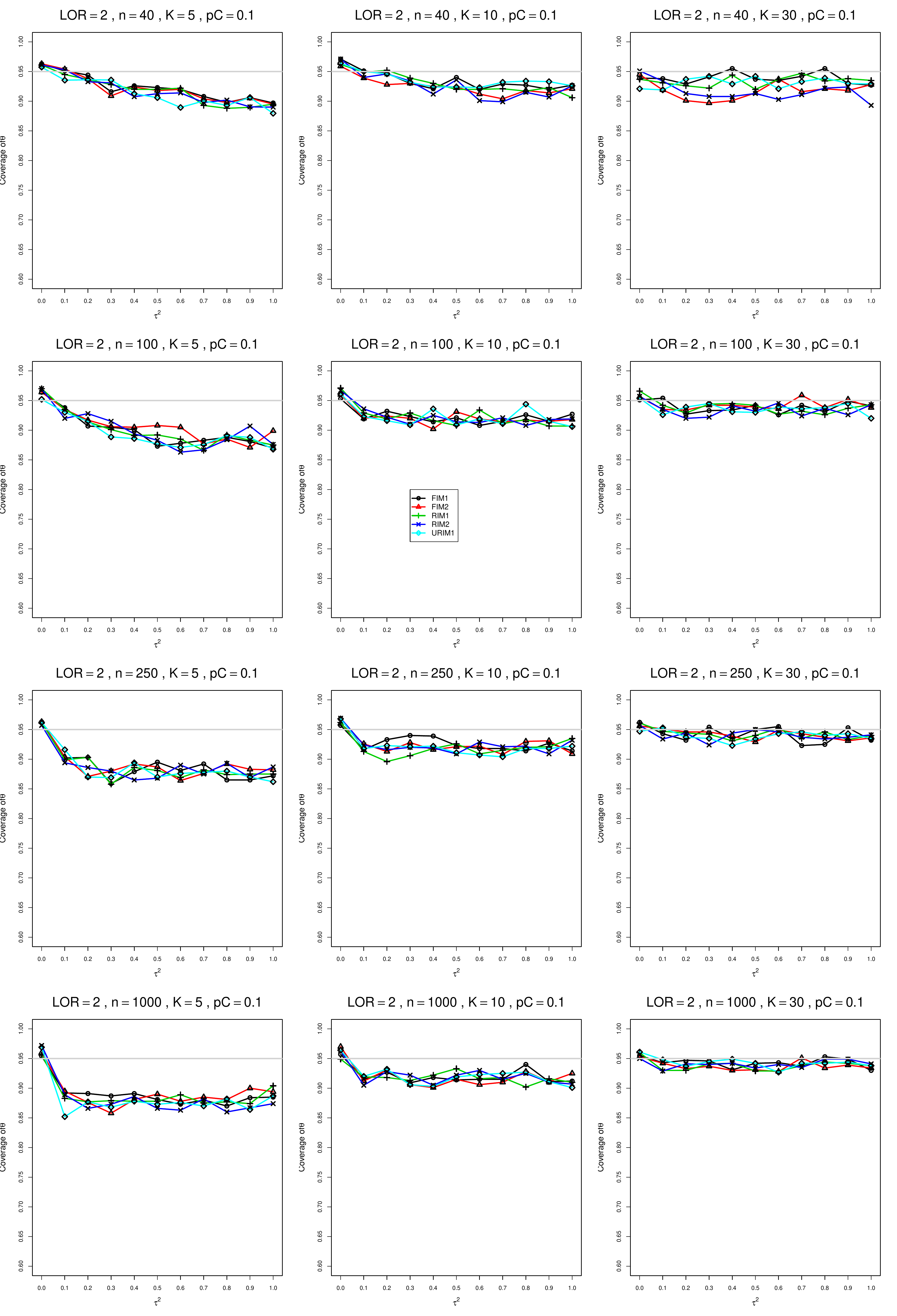}
	\caption{Coverage of the Restricted Maximum Likelihood confidence interval for $\theta=2$, $p_{C}=0.1$, $\sigma^2=0.4$, constant sample sizes $n=40,\;100,\;250,\;1000$.
The data-generation mechanisms are FIM1 ($\circ$), FIM2 ($\triangle$), RIM1 (+), RIM2 ($\times$), and URIM1 ($\diamond$).
		\label{PlotCovThetamu2andpC01LOR_REMLsigma04}}
\end{figure}
\begin{figure}[t]
	\centering
	\includegraphics[scale=0.33]{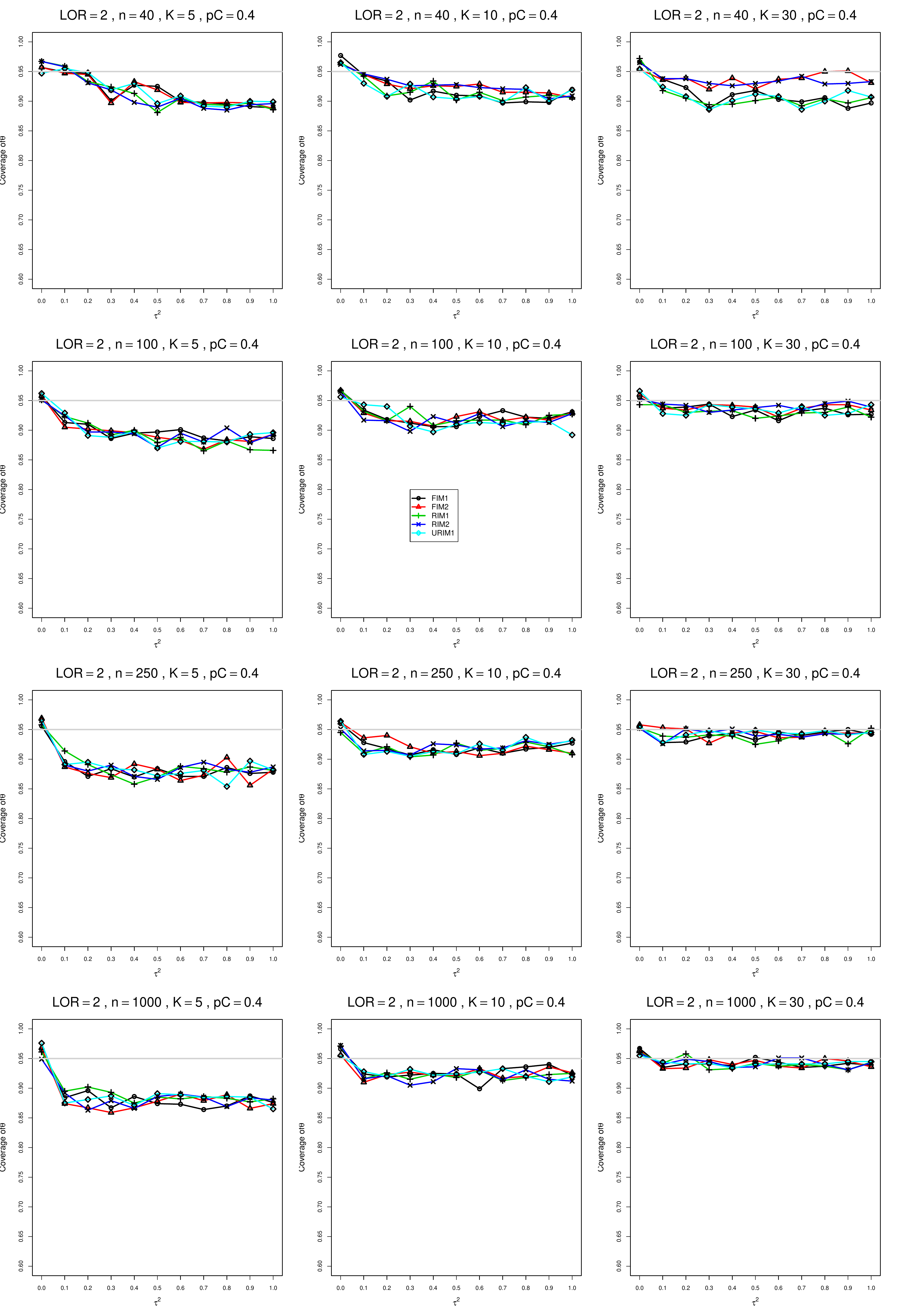}
	\caption{Coverage of the Restricted Maximum Likelihood confidence interval for $\theta=2$, $p_{C}=0.4$, $\sigma^2=0.4$, constant sample sizes $n=40,\;100,\;250,\;1000$.
The data-generation mechanisms are FIM1 ($\circ$), FIM2 ($\triangle$), RIM1 (+), RIM2 ($\times$), and URIM1 ($\diamond$).
		\label{PlotCovThetamu2andpC04LOR_REMLsigma04}}
\end{figure}

\clearpage
\subsection*{A3.3 Coverage of $\hat{\theta}_{MP}$}
\renewcommand{\thefigure}{A3.3.\arabic{figure}}
\setcounter{figure}{0}

\begin{figure}[t]
	\centering
	\includegraphics[scale=0.33]{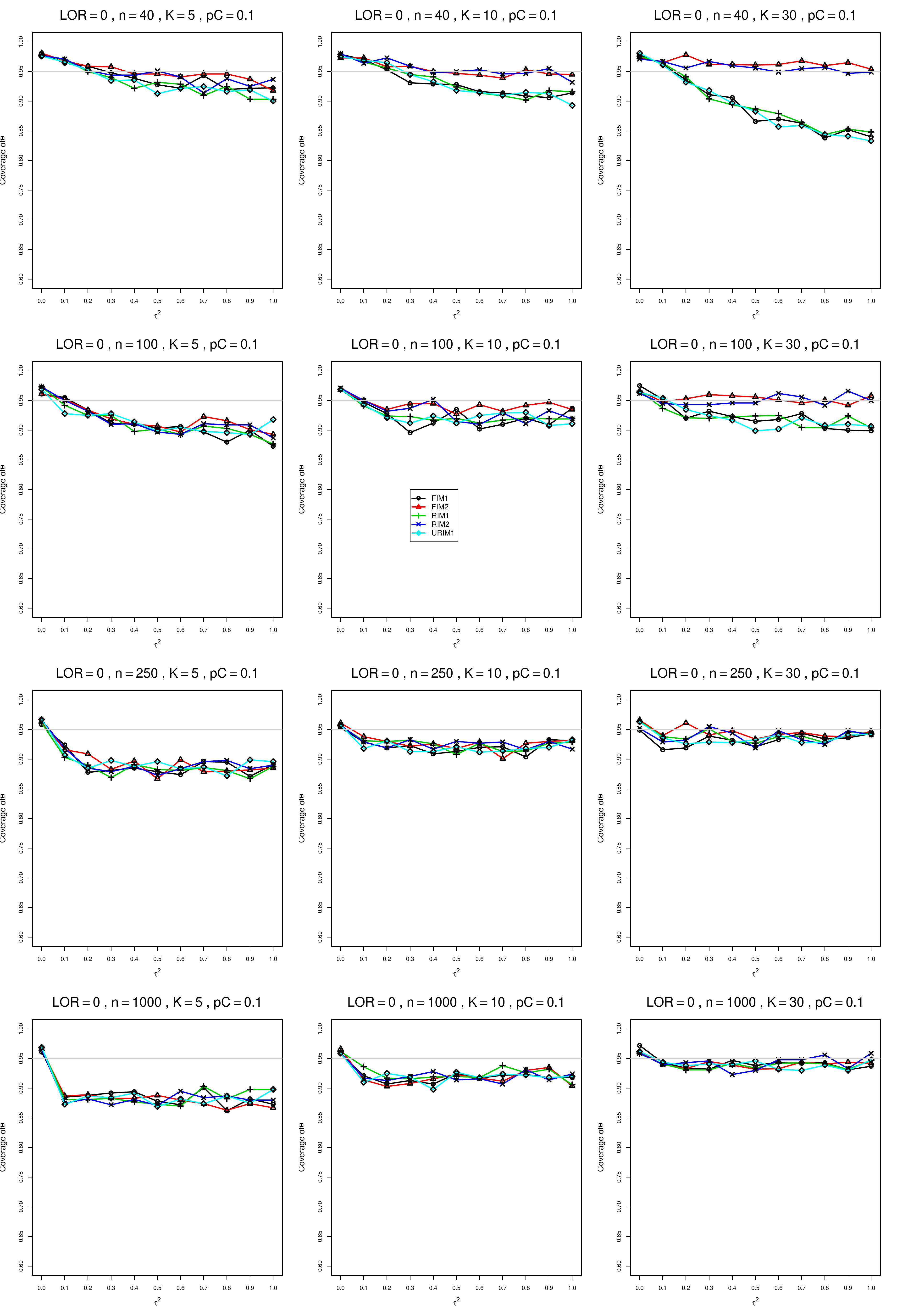}
	\caption{Coverage of the Mandel-Paule confidence interval for $\theta=0$, $p_{C}=0.1$, $\sigma^2=0.1$, constant sample sizes $n=40,\;100,\;250,\;1000$.
The data-generation mechanisms are FIM1 ($\circ$), FIM2 ($\triangle$), RIM1 (+), RIM2 ($\times$), and URIM1 ($\diamond$).
		\label{PlotCovThetamu0andpC01LOR_MPsigma01}}
\end{figure}
\begin{figure}[t]
	\centering
	\includegraphics[scale=0.33]{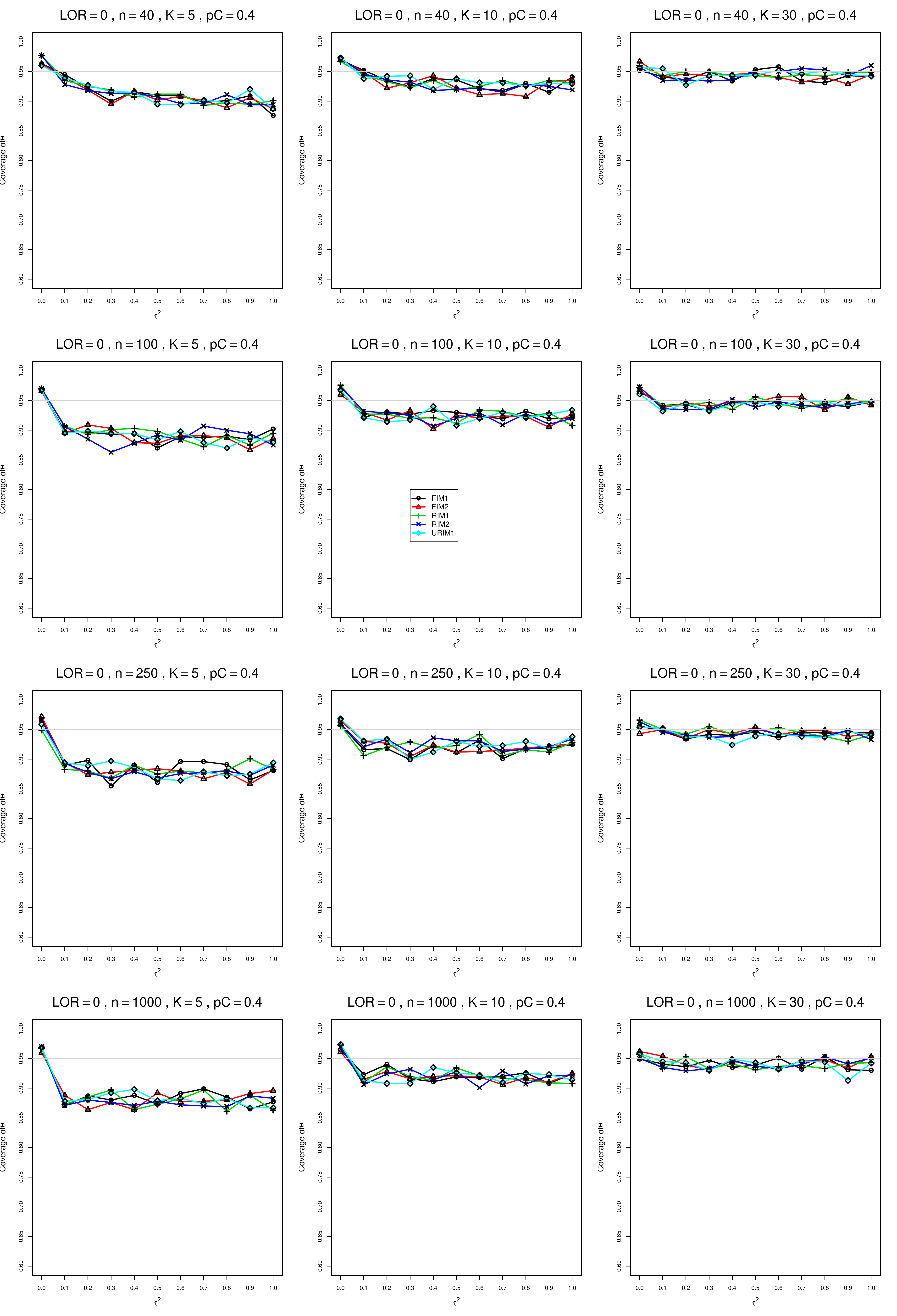}
	\caption{Coverage of the Mandel-Paule confidence interval for $\theta=0$, $p_{C}=0.4$, $\sigma^2=0.1$, constant sample sizes $n=40,\;100,\;250,\;1000$.
The data-generation mechanisms are FIM1 ($\circ$), FIM2 ($\triangle$), RIM1 (+), RIM2 ($\times$), and URIM1 ($\diamond$).
		\label{PlotCovThetamu0andpC04LOR_MPsigma01}}
\end{figure}
\begin{figure}[t]
	\centering
	\includegraphics[scale=0.33]{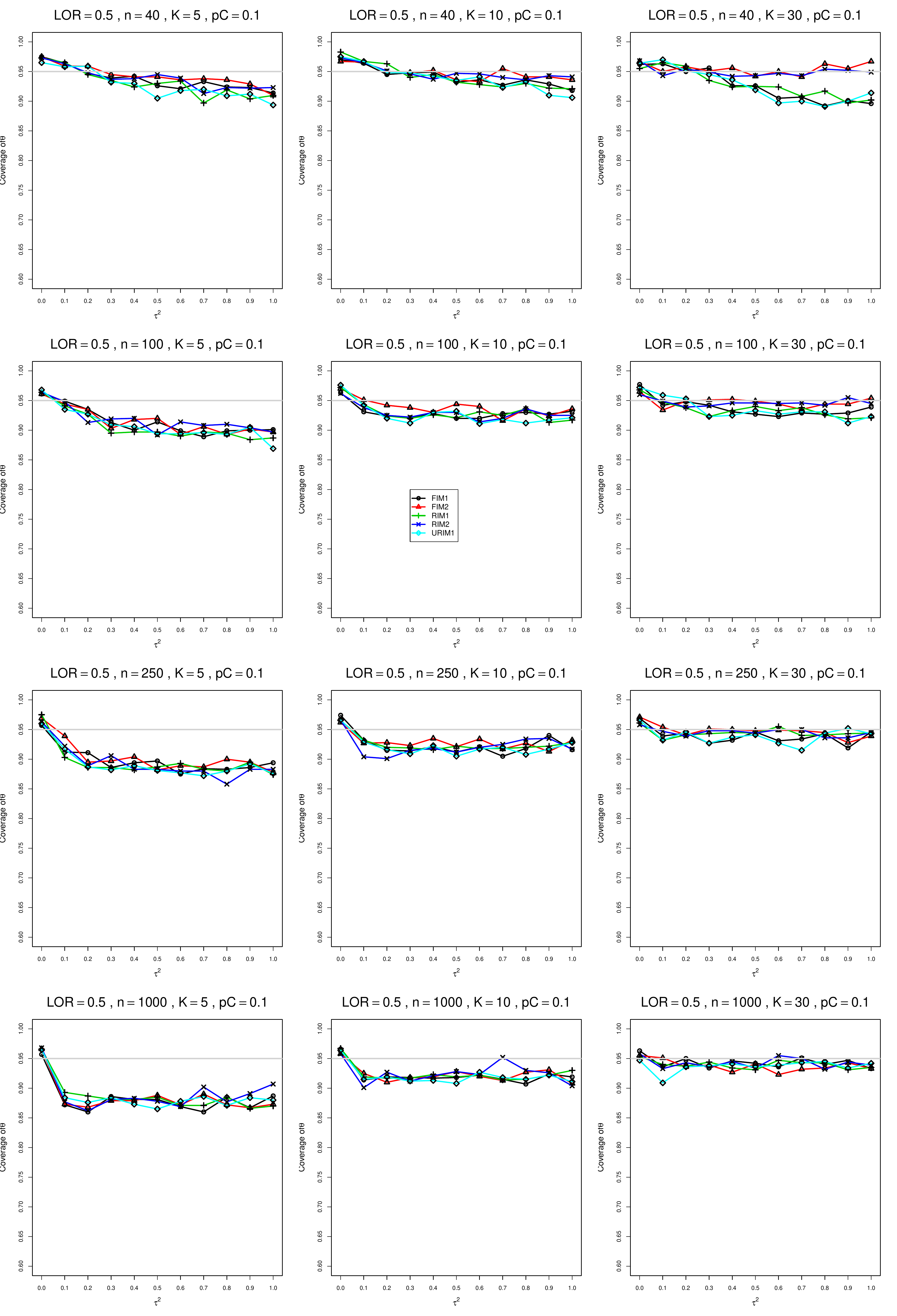}
	\caption{Coverage of the Mandel-Paule confidence interval for $\theta=0.5$, $p_{C}=0.1$, $\sigma^2=0.1$, constant sample sizes $n=40,\;100,\;250,\;1000$.
The data-generation mechanisms are FIM1 ($\circ$), FIM2 ($\triangle$), RIM1 (+), RIM2 ($\times$), and URIM1 ($\diamond$).
		\label{PlotCovThetamu05andpC01LOR_MPsigma01}}
\end{figure}
\begin{figure}[t]
	\centering
	\includegraphics[scale=0.33]{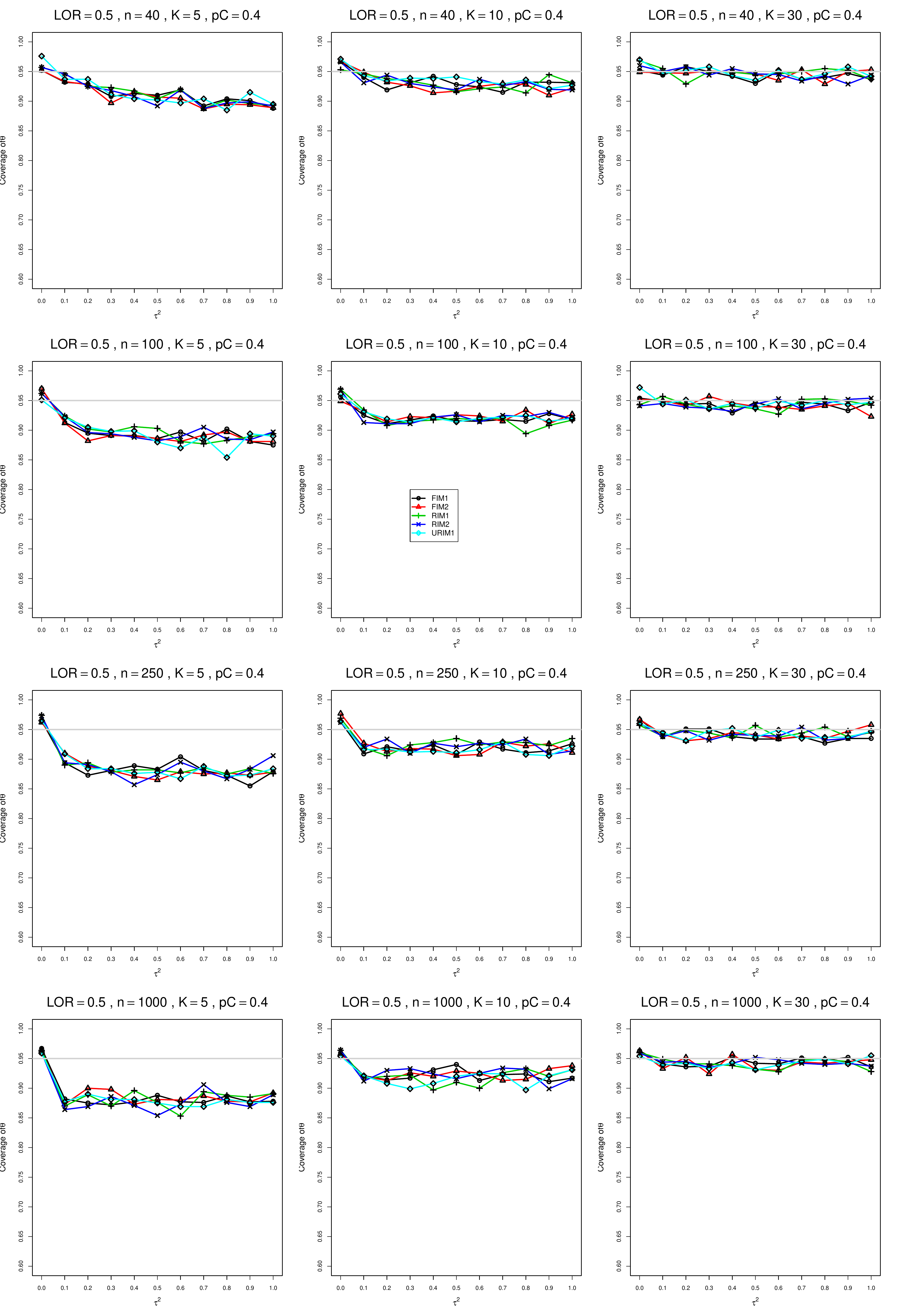}
	\caption{Coverage of the Mandel-Paule confidence interval for $\theta=0.5$, $p_{C}=0.4$, $\sigma^2=0.1$, constant sample sizes $n=40,\;100,\;250,\;1000$.
The data-generation mechanisms are FIM1 ($\circ$), FIM2 ($\triangle$), RIM1 (+), RIM2 ($\times$), and URIM1 ($\diamond$).
		\label{PlotCovThetamu05andpC04LOR_MPsigma01}}
\end{figure}
\begin{figure}[t]
	\centering
	\includegraphics[scale=0.33]{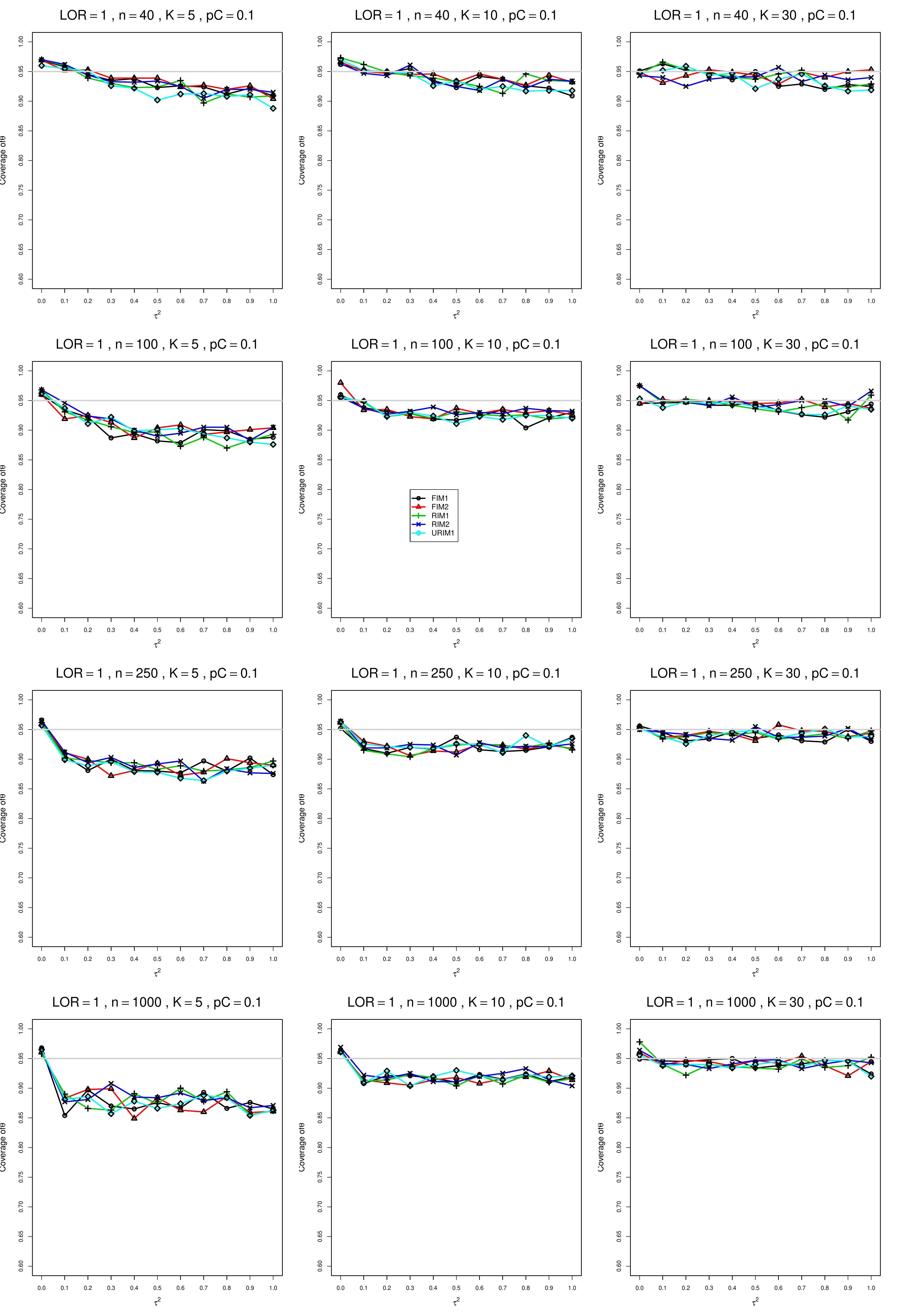}
	\caption{Coverage of the Mandel-Paule confidence interval for $\theta=1$, $p_{C}=0.1$, $\sigma^2=0.1$, constant sample sizes $n=40,\;100,\;250,\;1000$.
The data-generation mechanisms are FIM1 ($\circ$), FIM2 ($\triangle$), RIM1 (+), RIM2 ($\times$), and URIM1 ($\diamond$).
		\label{PlotCovThetamu1andpC01LOR_MPsigma01}}
\end{figure}
\begin{figure}[t]
	\centering
	\includegraphics[scale=0.33]{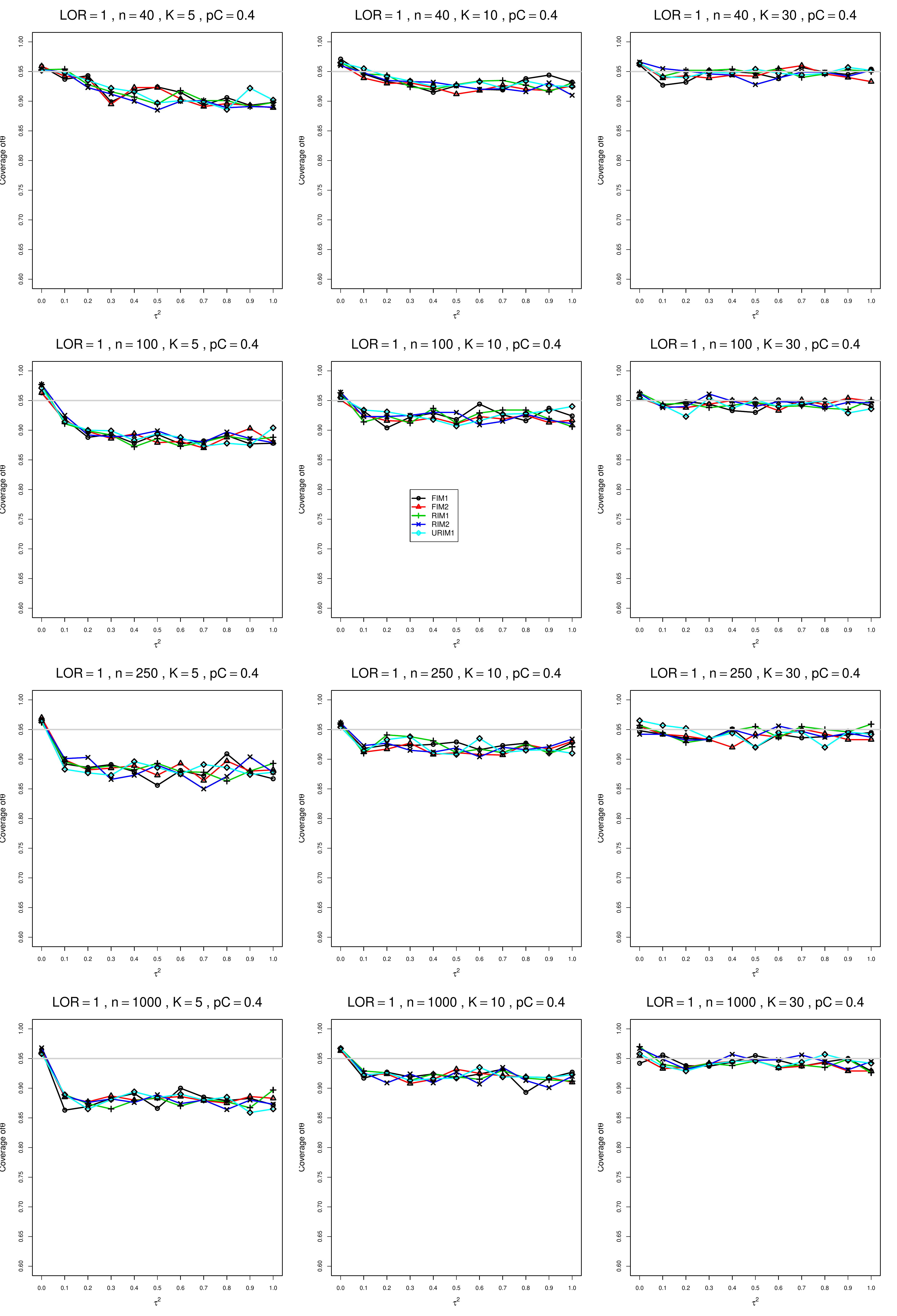}
	\caption{Coverage of the Mandel-Paule confidence interval for $\theta=1$, $p_{C}=0.4$, $\sigma^2=0.1$, constant sample sizes $n=40,\;100,\;250,\;1000$.
The data-generation mechanisms are FIM1 ($\circ$), FIM2 ($\triangle$), RIM1 (+), RIM2 ($\times$), and URIM1 ($\diamond$).
		\label{PlotCovThetamu1andpC04LOR_MPsigma01}}
\end{figure}
\begin{figure}[t]
	\centering
	\includegraphics[scale=0.33]{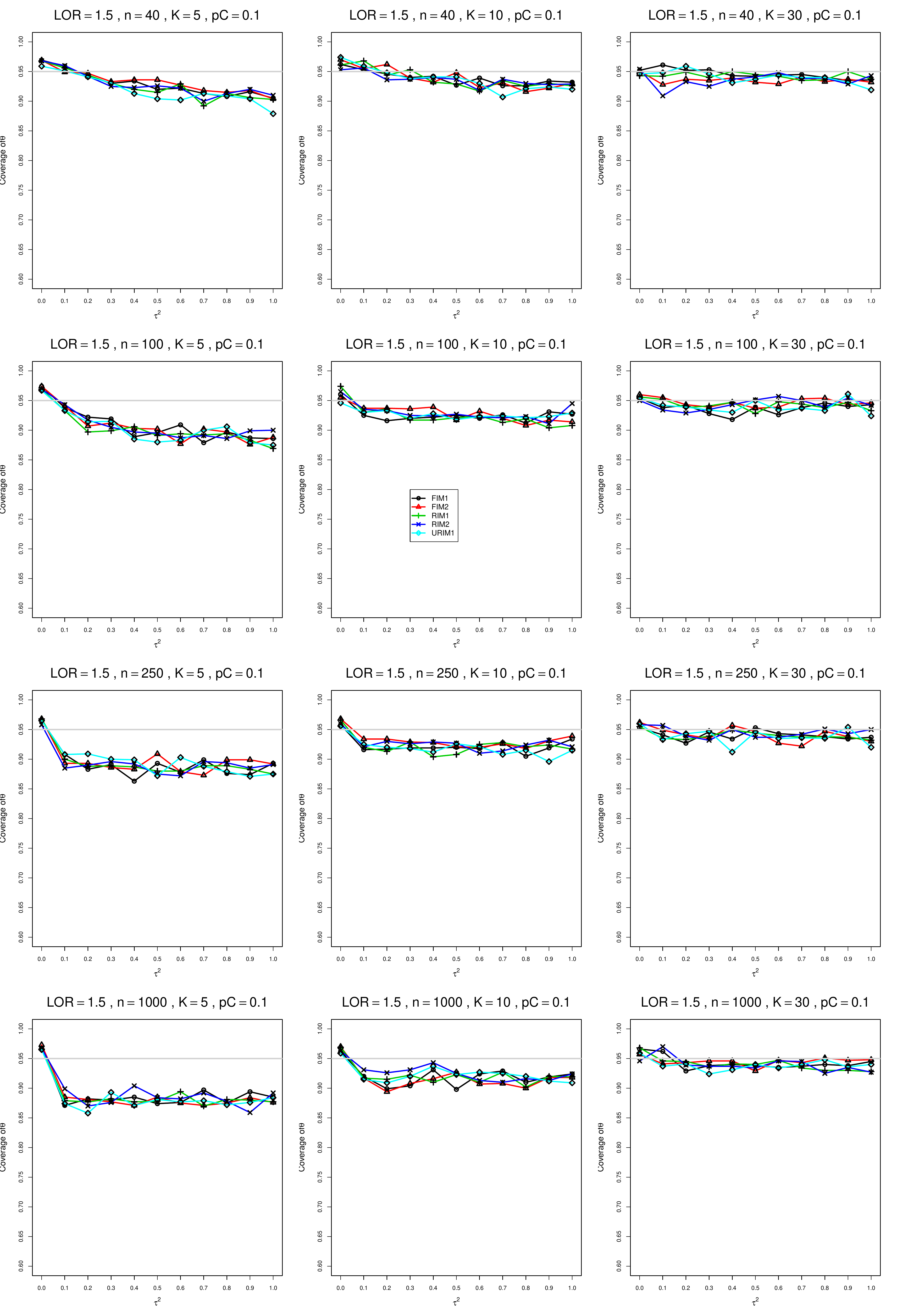}
	\caption{Coverage of the Mandel-Paule confidence interval for $\theta=1.5$, $p_{C}=0.1$, $\sigma^2=0.1$, constant sample sizes $n=40,\;100,\;250,\;1000$.
The data-generation mechanisms are FIM1 ($\circ$), FIM2 ($\triangle$), RIM1 (+), RIM2 ($\times$), and URIM1 ($\diamond$).
		\label{PlotCovThetamu15andpC01LOR_MPsigma01}}
\end{figure}
\begin{figure}[t]
	\centering
	\includegraphics[scale=0.33]{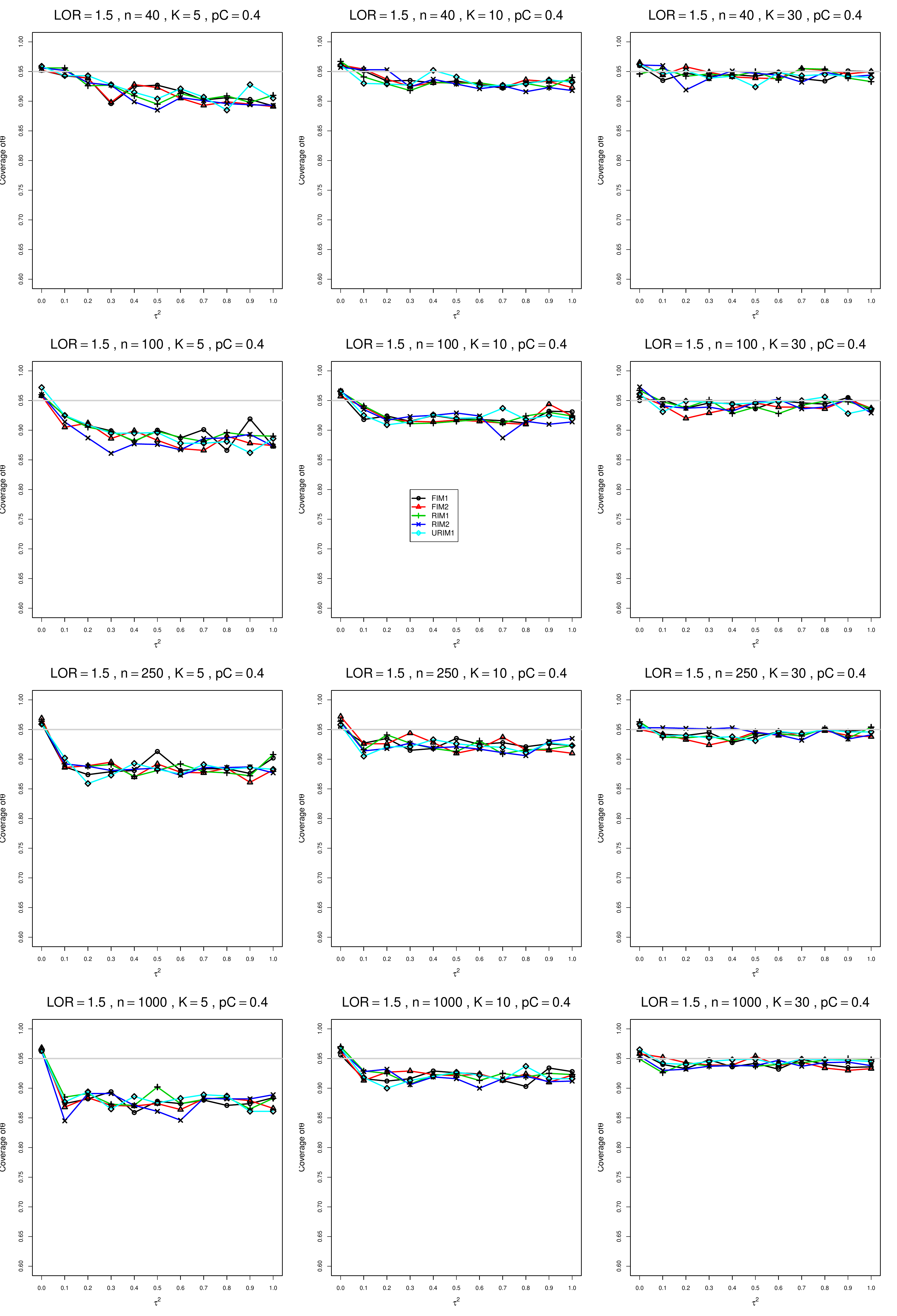}
	\caption{Coverage of the Mandel-Paule confidence interval for $\theta=1.5$, $p_{C}=0.4$, $\sigma^2=0.1$, constant sample sizes $n=40,\;100,\;250,\;1000$.
The data-generation mechanisms are FIM1 ($\circ$), FIM2 ($\triangle$), RIM1 (+), RIM2 ($\times$), and URIM1 ($\diamond$).
		\label{PlotCovThetamu15andpC04LOR_MPsigma01}}
\end{figure}
\begin{figure}[t]
	\centering
	\includegraphics[scale=0.33]{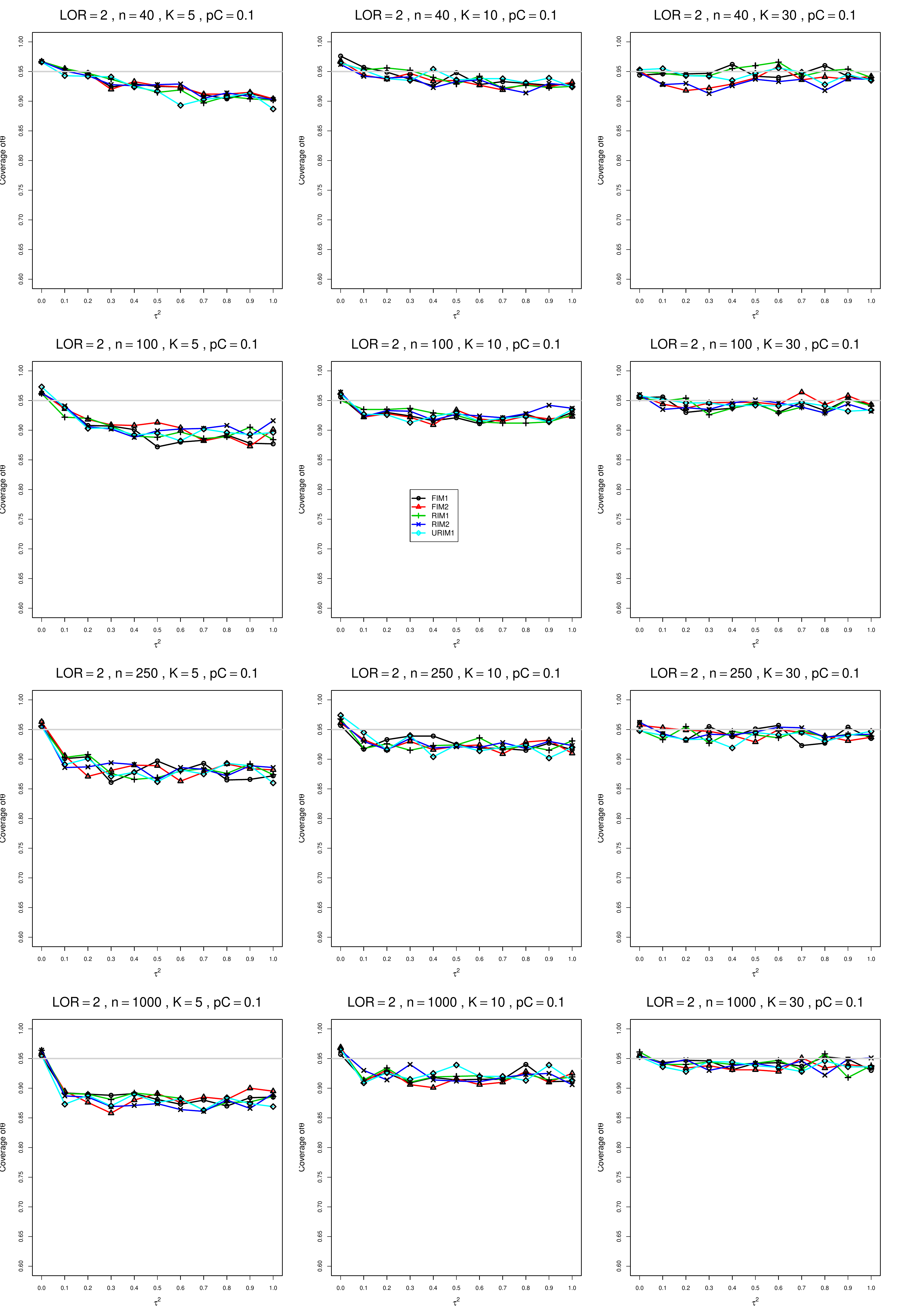}
	\caption{Coverage of the Mandel-Paule confidence interval for $\theta=2$, $p_{C}=0.1$, $\sigma^2=0.1$, constant sample sizes $n=40,\;100,\;250,\;1000$.
The data-generation mechanisms are FIM1 ($\circ$), FIM2 ($\triangle$), RIM1 (+), RIM2 ($\times$), and URIM1 ($\diamond$).
		\label{PlotCovThetamu2andpC01LOR_MPsigma01}}
\end{figure}
\begin{figure}[t]
	\centering
	\includegraphics[scale=0.33]{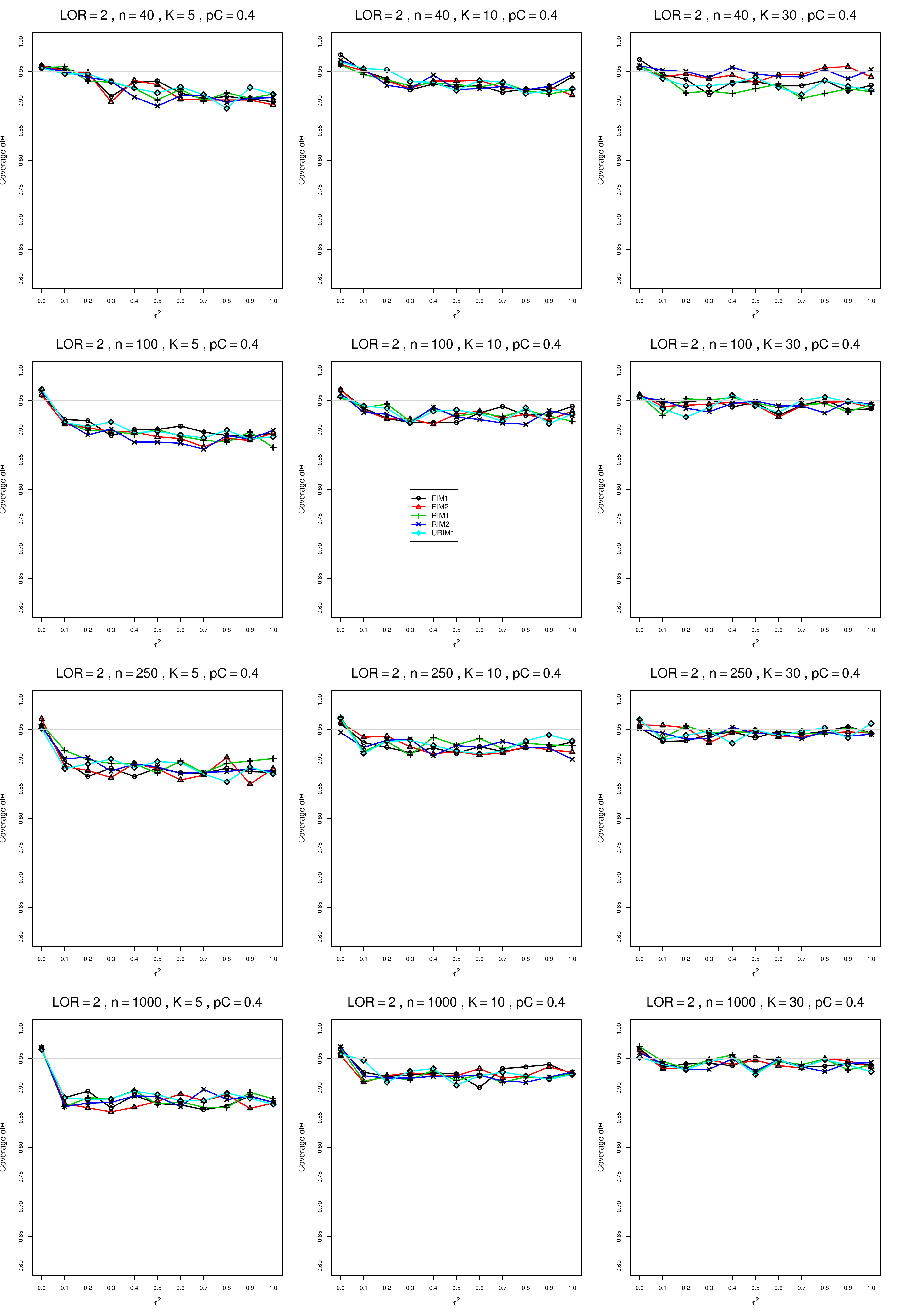}
	\caption{Coverage of the Mandel-Paule confidence interval for $\theta=2$, $p_{C}=0.4$, $\sigma^2=0.1$, constant sample sizes $n=40,\;100,\;250,\;1000$.
The data-generation mechanisms are FIM1 ($\circ$), FIM2 ($\triangle$), RIM1 (+), RIM2 ($\times$), and URIM1 ($\diamond$).
		\label{PlotCovThetamu2andpC04LOR_MPsigma01}}
\end{figure}
\begin{figure}[t]
	\centering
	\includegraphics[scale=0.33]{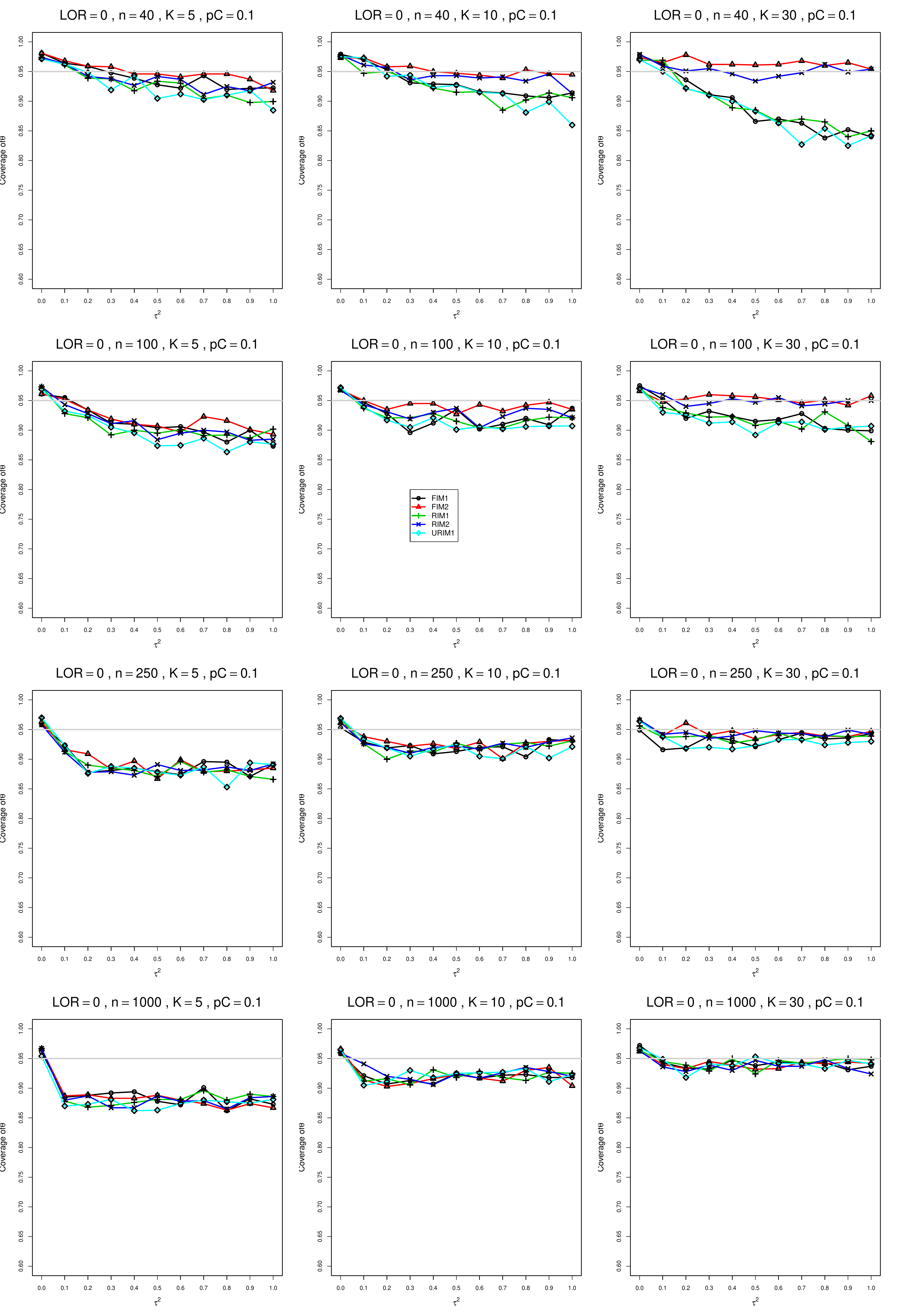}
	\caption{Coverage of the Mandel-Paule confidence interval for $\theta=0$, $p_{C}=0.1$, $\sigma^2=0.4$, constant sample sizes $n=40,\;100,\;250,\;1000$.
The data-generation mechanisms are FIM1 ($\circ$), FIM2 ($\triangle$), RIM1 (+), RIM2 ($\times$), and URIM1 ($\diamond$).
		\label{PlotCovThetamu0andpC01LOR_MPsigma04}}
\end{figure}
\begin{figure}[t]
	\centering
	\includegraphics[scale=0.33]{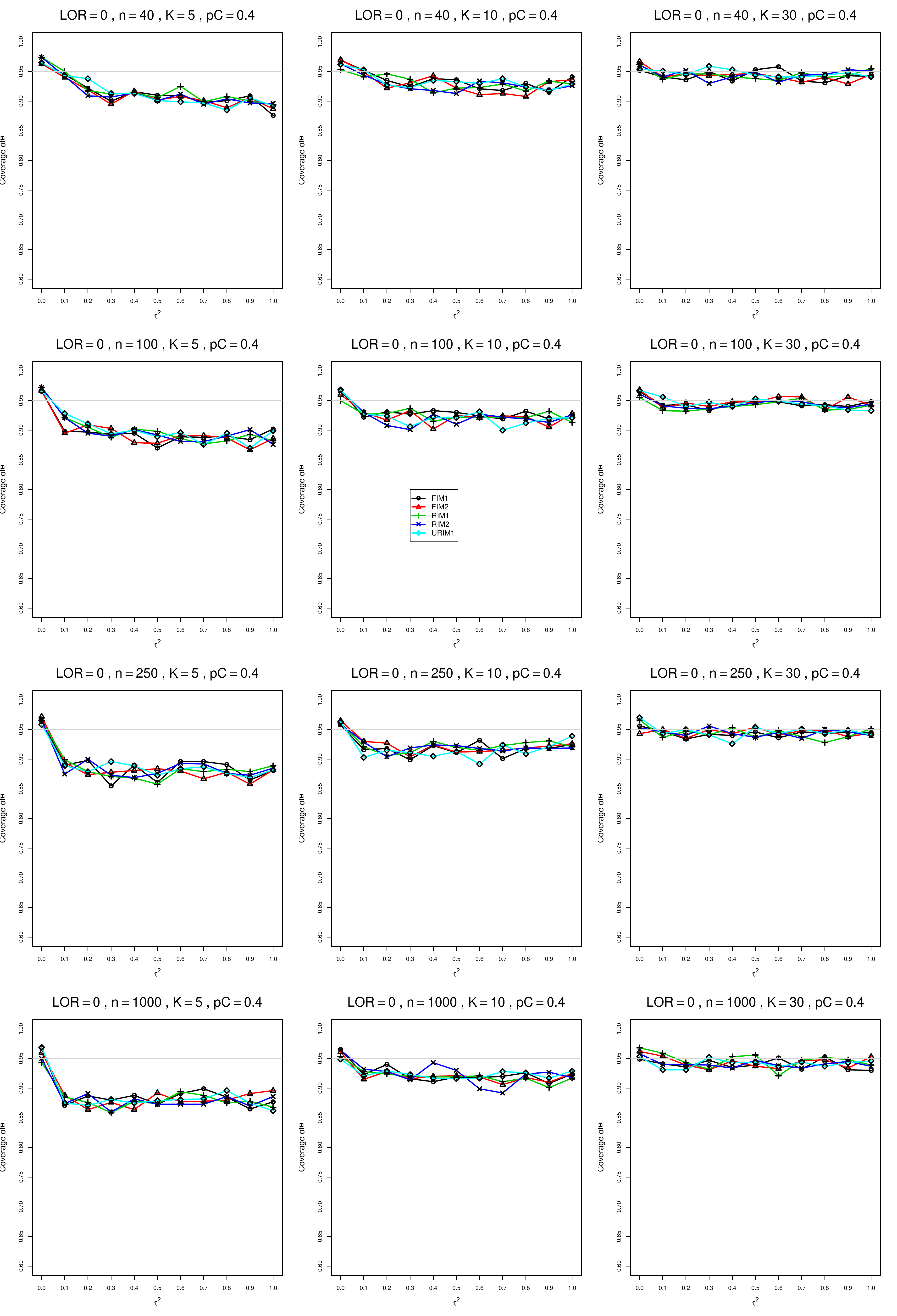}
	\caption{Coverage of the Mandel-Paule confidence interval for $\theta=0$, $p_{C}=0.4$, $\sigma^2=0.4$, constant sample sizes $n=40,\;100,\;250,\;1000$.
The data-generation mechanisms are FIM1 ($\circ$), FIM2 ($\triangle$), RIM1 (+), RIM2 ($\times$), and URIM1 ($\diamond$).
		\label{PlotCovThetamu0andpC04LOR_MPsigma04}}
\end{figure}
\begin{figure}[t]
	\centering
	\includegraphics[scale=0.33]{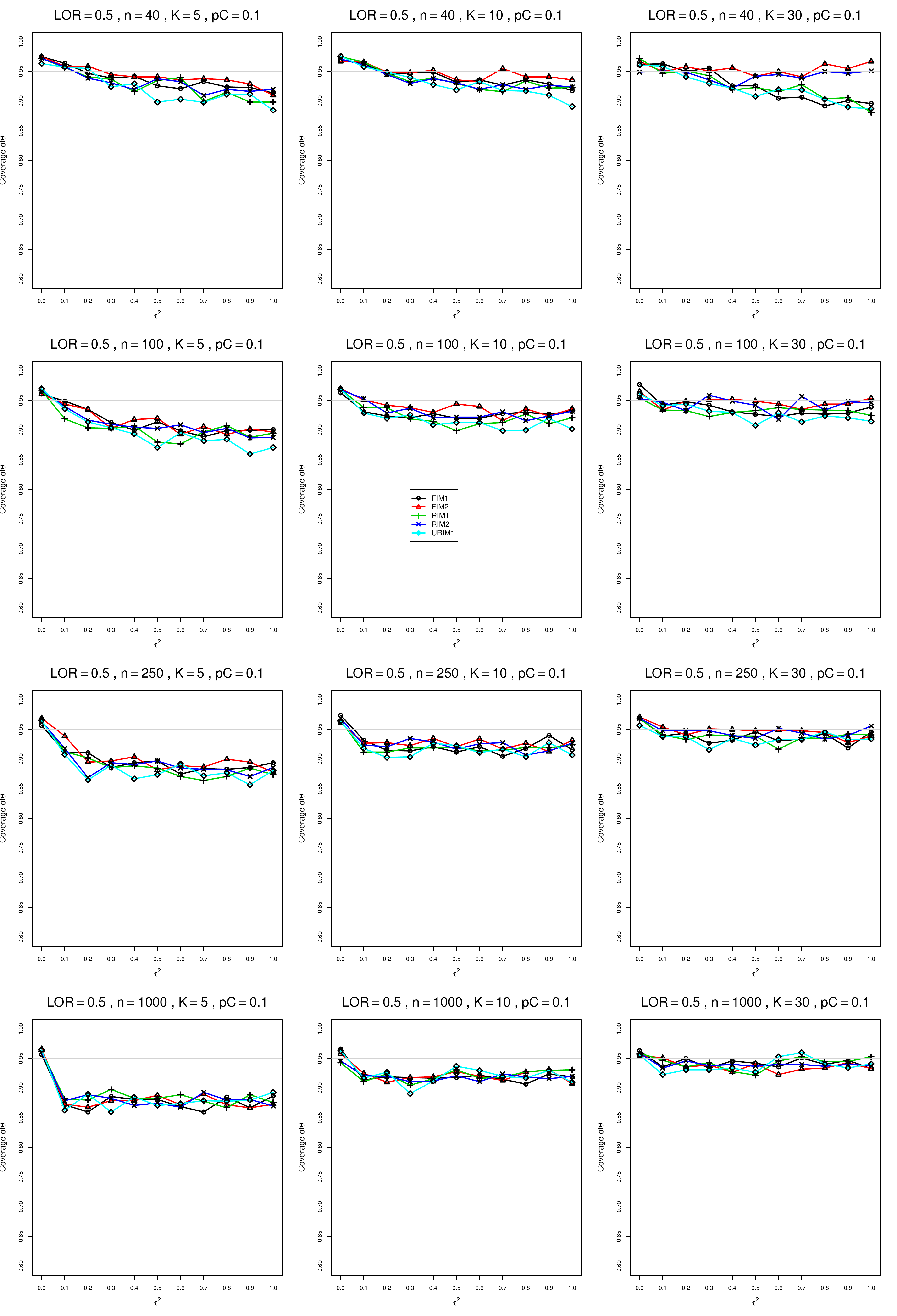}
	\caption{Coverage of the Mandel-Paule confidence interval for $\theta=0.5$, $p_{C}=0.1$, $\sigma^2=0.4$, constant sample sizes $n=40,\;100,\;250,\;1000$.
The data-generation mechanisms are FIM1 ($\circ$), FIM2 ($\triangle$), RIM1 (+), RIM2 ($\times$), and URIM1 ($\diamond$).
		\label{PlotCovThetamu05andpC01LOR_MPsigma04}}
\end{figure}
\begin{figure}[t]
	\centering
	\includegraphics[scale=0.33]{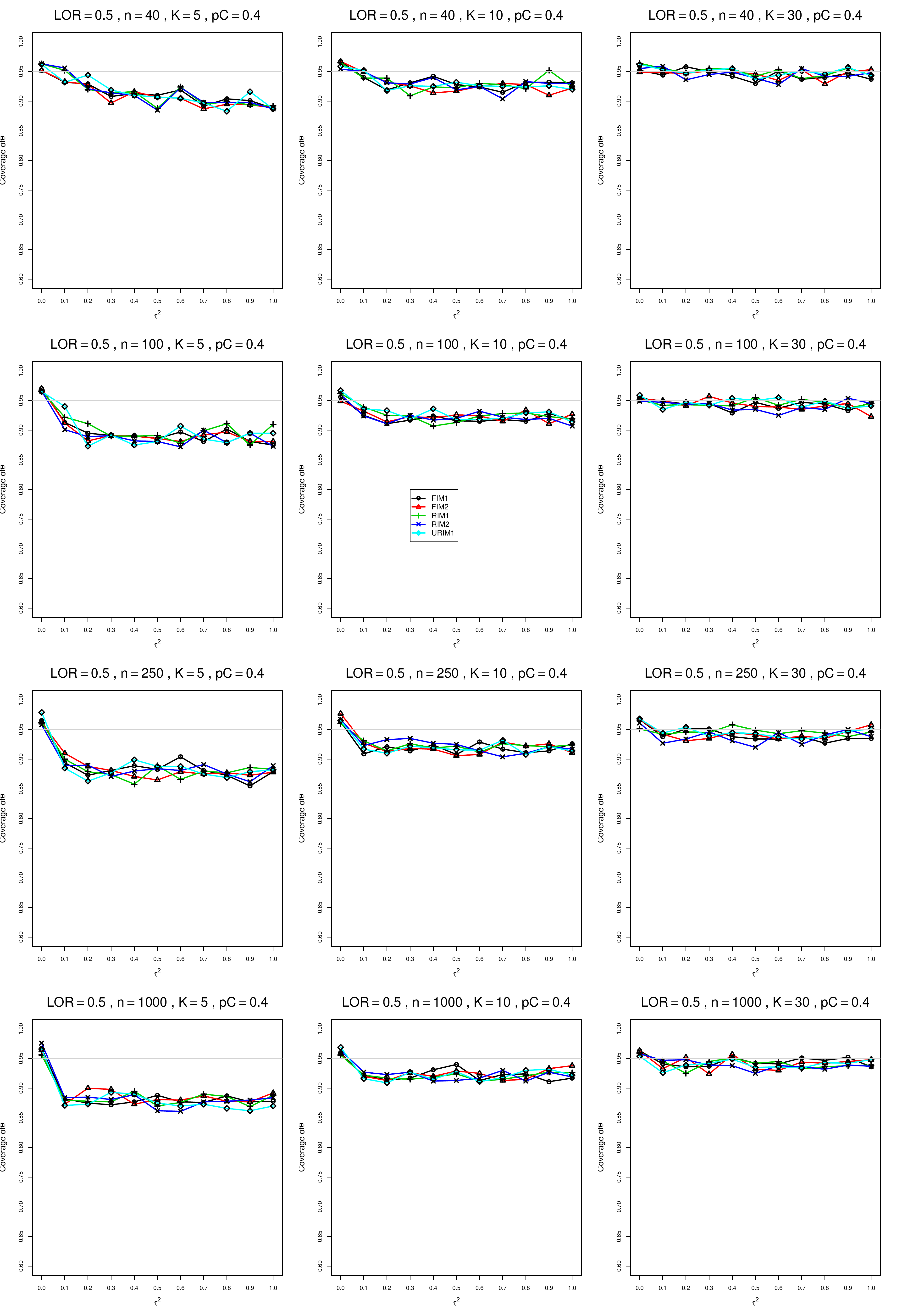}
	\caption{Coverage of the Mandel-Paule confidence interval for $\theta=0.5$, $p_{C}=0.4$, $\sigma^2=0.4$, constant sample sizes $n=40,\;100,\;250,\;1000$.
The data-generation mechanisms are FIM1 ($\circ$), FIM2 ($\triangle$), RIM1 (+), RIM2 ($\times$), and URIM1 ($\diamond$).
		\label{PlotCovThetamu05andpC04LOR_MPsigma04}}
\end{figure}
\begin{figure}[t]
	\centering
	\includegraphics[scale=0.33]{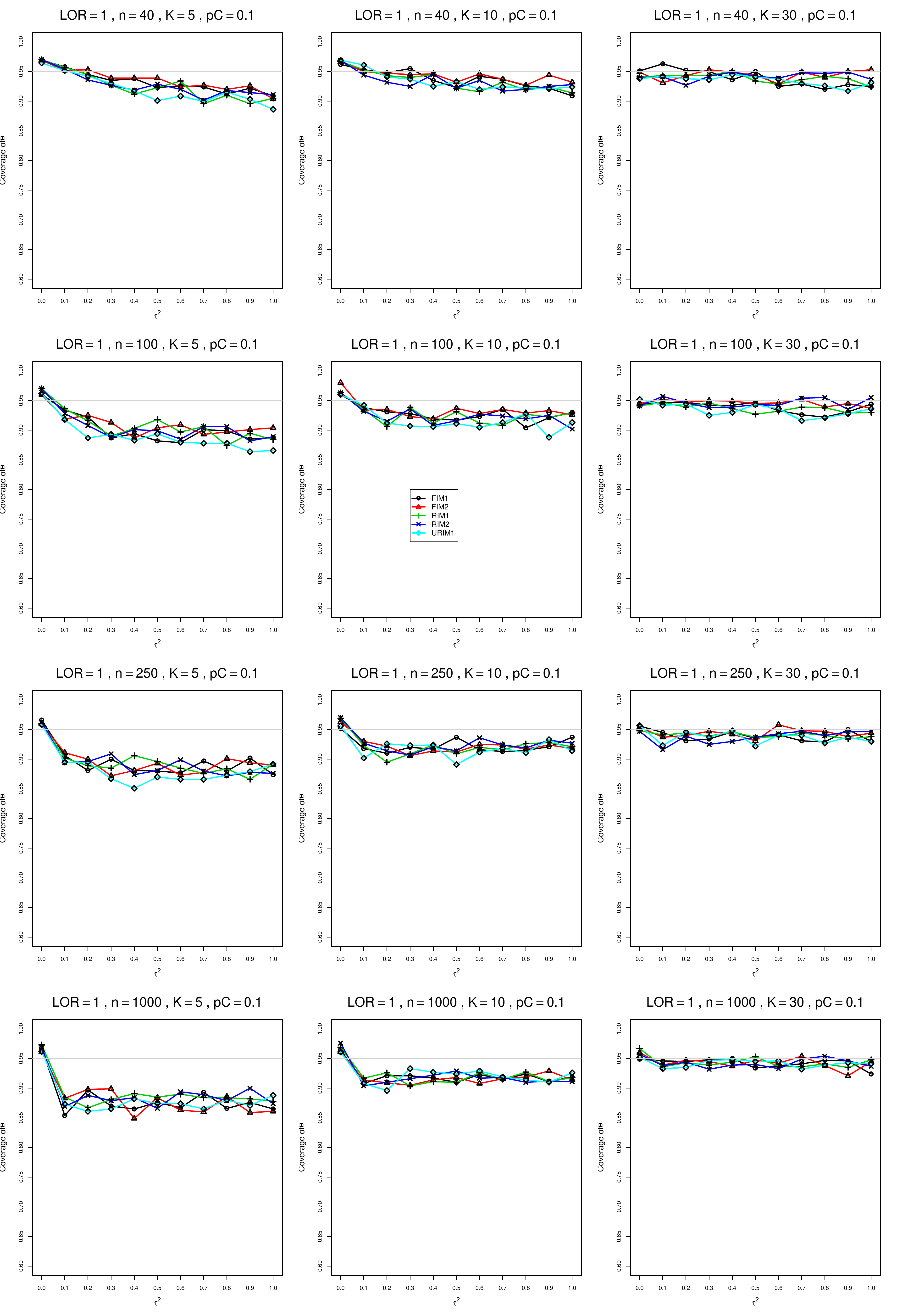}
	\caption{Coverage of the Mandel-Paule confidence interval for $\theta=1$, $p_{C}=0.1$, $\sigma^2=0.4$, constant sample sizes $n=40,\;100,\;250,\;1000$.
The data-generation mechanisms are FIM1 ($\circ$), FIM2 ($\triangle$), RIM1 (+), RIM2 ($\times$), and URIM1 ($\diamond$).
		\label{PlotCovThetamu1andpC01LOR_MPsigma04}}
\end{figure}
\begin{figure}[t]
	\centering
	\includegraphics[scale=0.33]{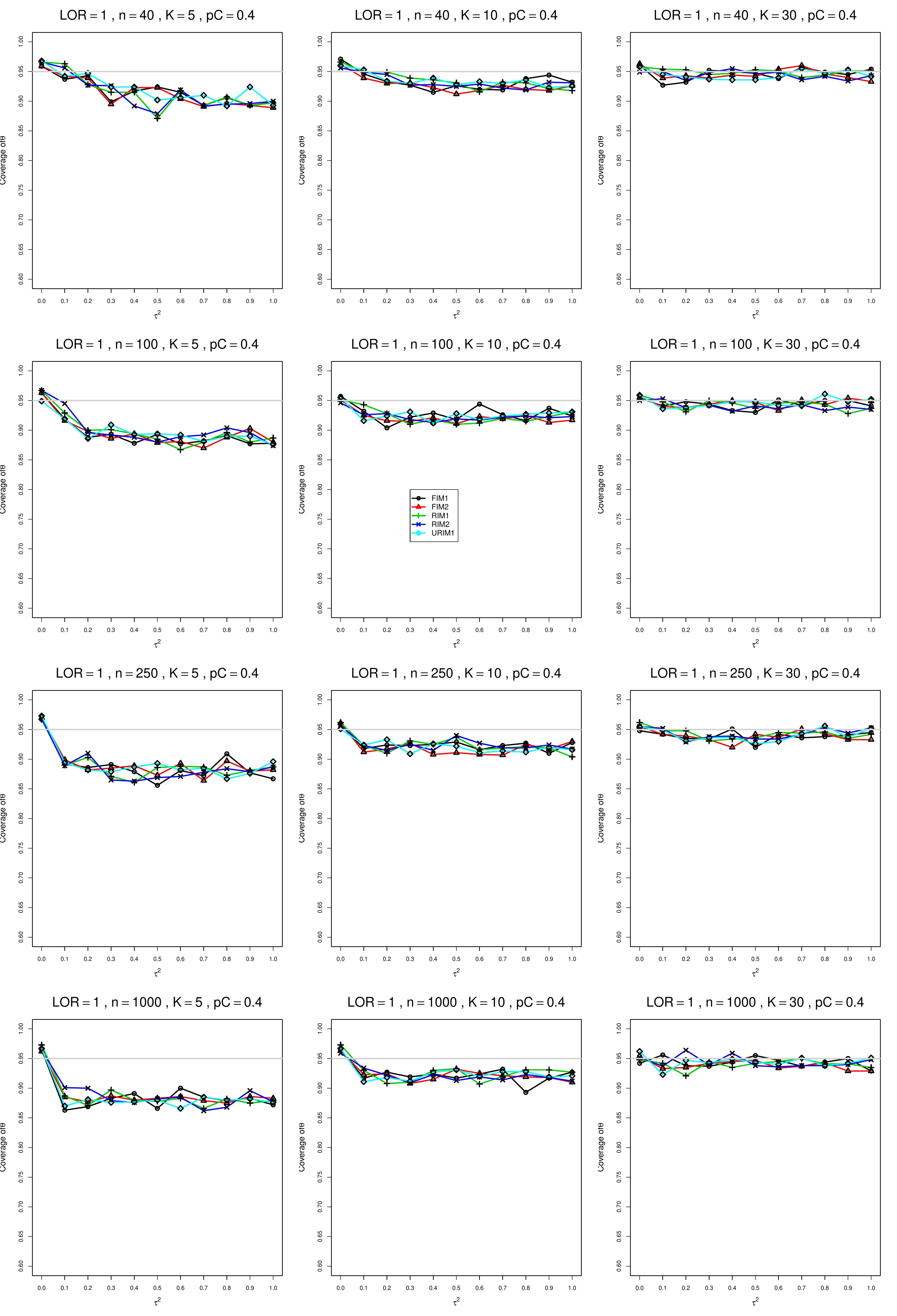}
	\caption{Coverage of the Mandel-Paule confidence interval for $\theta=1$, $p_{C}=0.4$, $\sigma^2=0.4$, constant sample sizes $n=40,\;100,\;250,\;1000$.
The data-generation mechanisms are FIM1 ($\circ$), FIM2 ($\triangle$), RIM1 (+), RIM2 ($\times$), and URIM1 ($\diamond$).
		\label{PlotCovThetamu1andpC04LOR_MPsigma04}}
\end{figure}
\begin{figure}[t]
	\centering
	\includegraphics[scale=0.33]{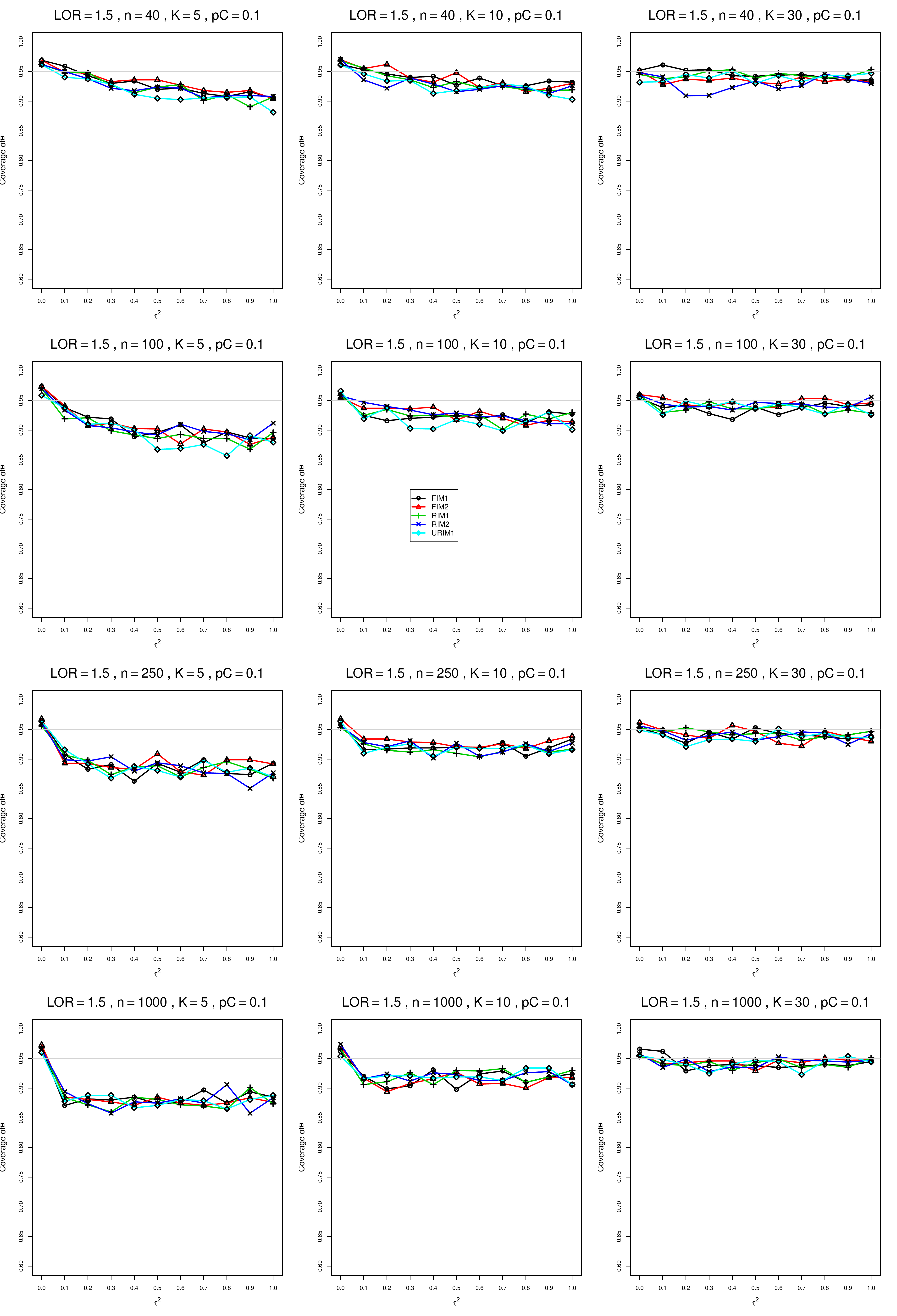}
	\caption{Coverage of the Mandel-Paule confidence interval for $\theta=1.5$, $p_{C}=0.1$, $\sigma^2=0.4$, constant sample sizes $n=40,\;100,\;250,\;1000$.
The data-generation mechanisms are FIM1 ($\circ$), FIM2 ($\triangle$), RIM1 (+), RIM2 ($\times$), and URIM1 ($\diamond$).
		\label{PlotCovThetamu15andpC01LOR_MPsigma04}}
\end{figure}
\begin{figure}[t]
	\centering
	\includegraphics[scale=0.33]{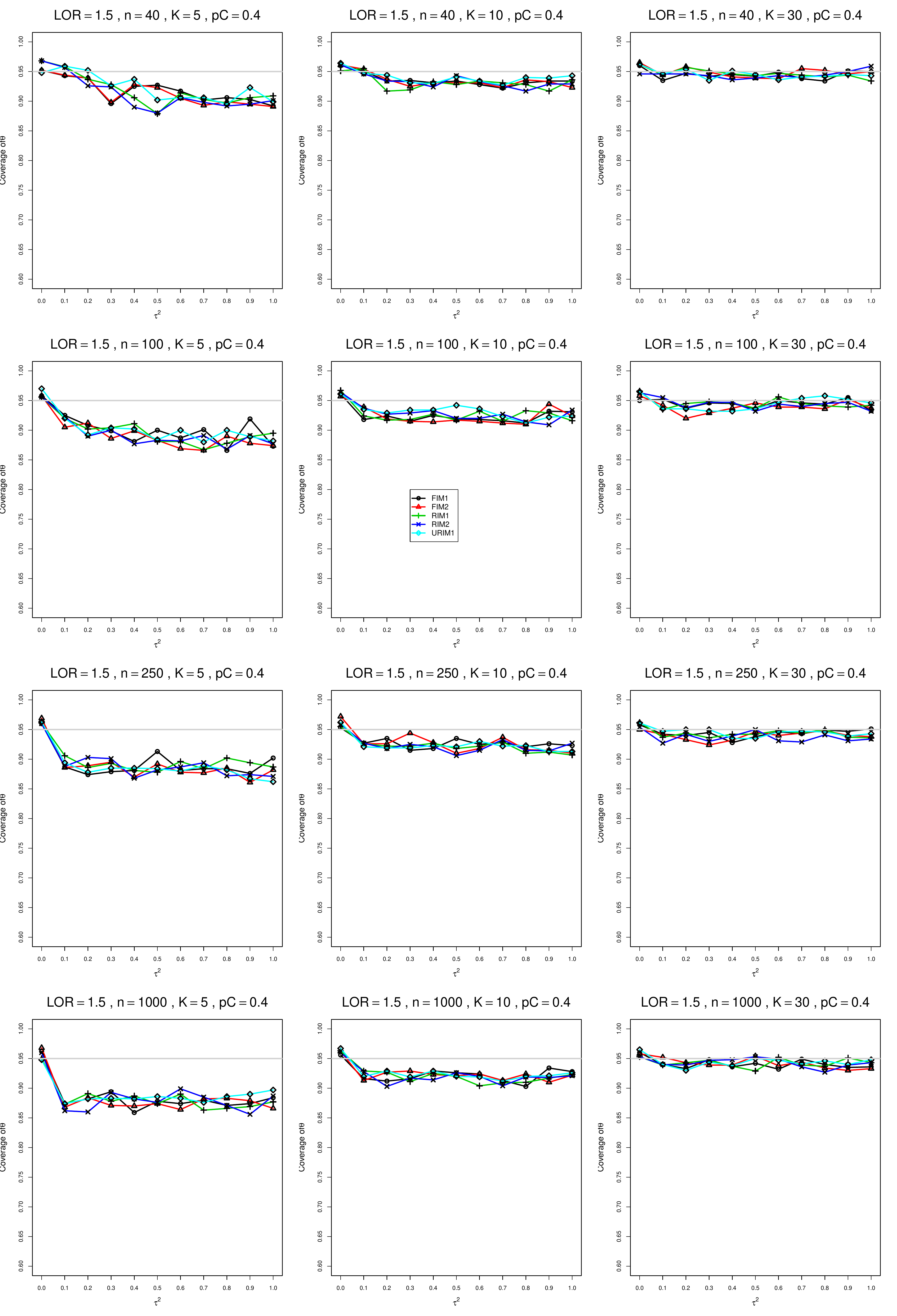}
	\caption{Coverage of the Mandel-Paule confidence interval for $\theta=1.5$, $p_{C}=0.4$, $\sigma^2=0.4$, constant sample sizes $n=40,\;100,\;250,\;1000$.
The data-generation mechanisms are FIM1 ($\circ$), FIM2 ($\triangle$), RIM1 (+), RIM2 ($\times$), and URIM1 ($\diamond$).
		\label{PlotCovThetamu15andpC04LOR_MPsigma04}}
\end{figure}
\begin{figure}[t]
	\centering
	\includegraphics[scale=0.33]{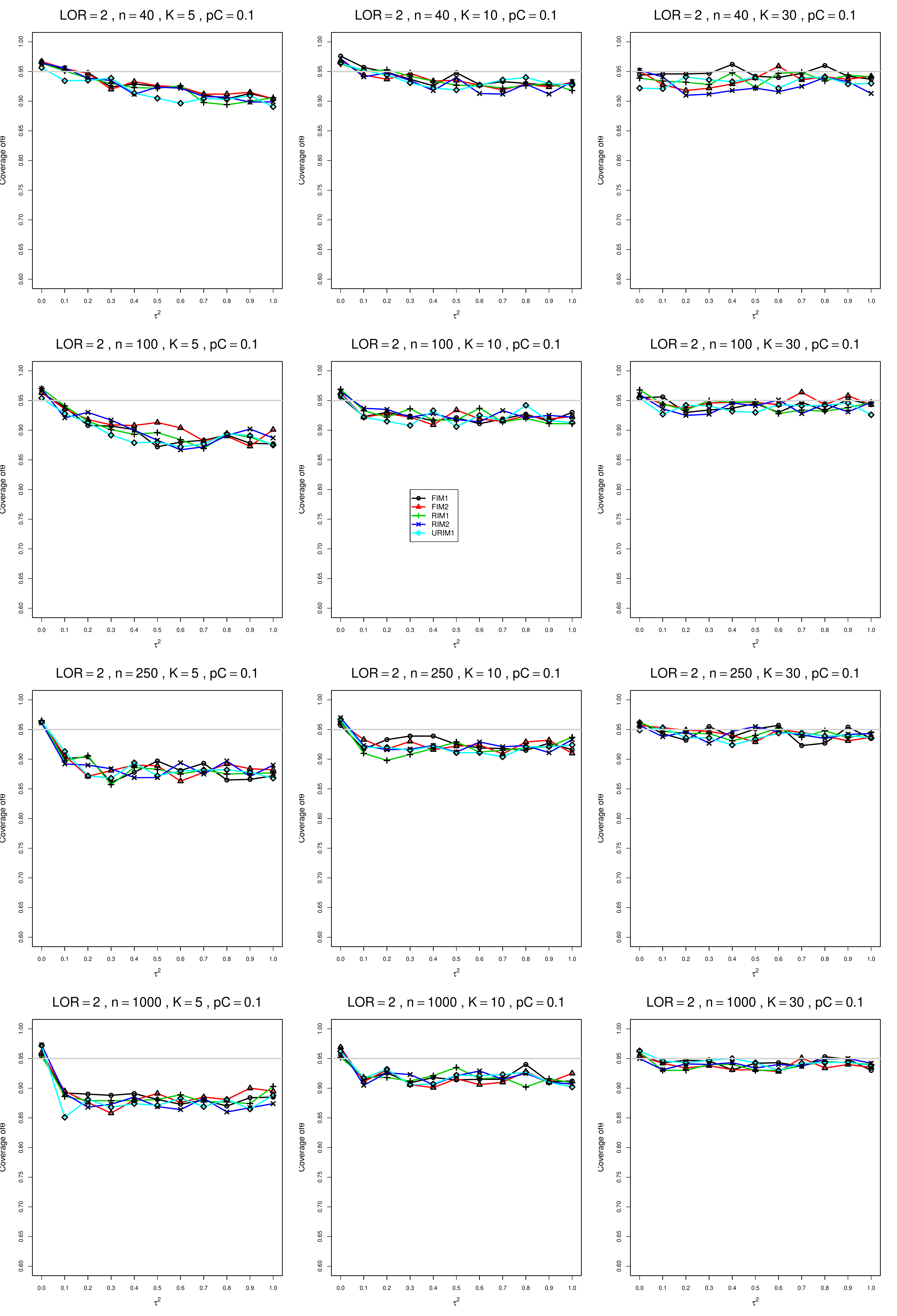}
	\caption{Coverage of the Mandel-Paule confidence interval for $\theta=2$, $p_{C}=0.1$, $\sigma^2=0.4$, constant sample sizes $n=40,\;100,\;250,\;1000$.
The data-generation mechanisms are FIM1 ($\circ$), FIM2 ($\triangle$), RIM1 (+), RIM2 ($\times$), and URIM1 ($\diamond$).
		\label{PlotCovThetamu2andpC01LOR_MPsigma04}}
\end{figure}
\begin{figure}[t]
	\centering
	\includegraphics[scale=0.33]{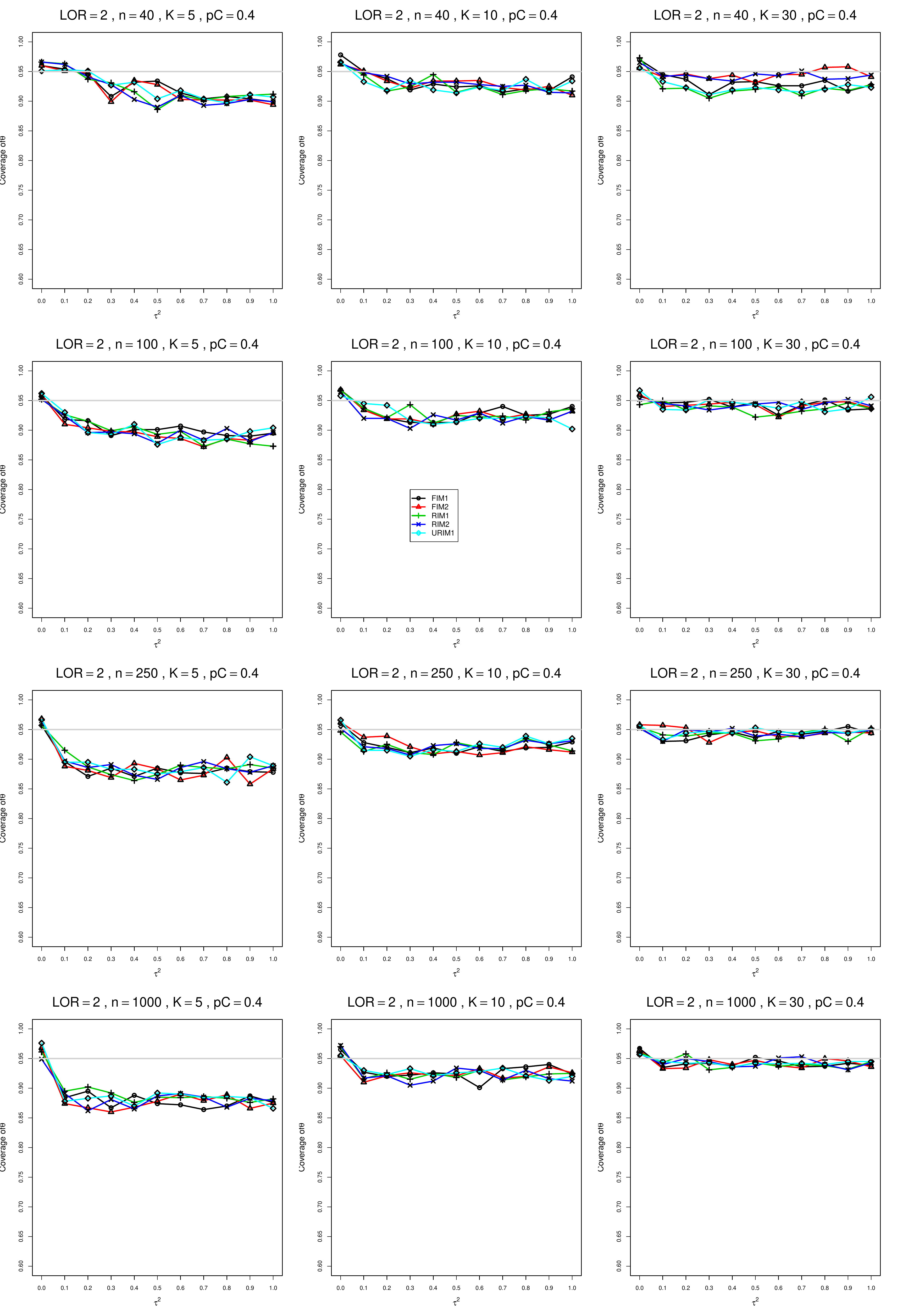}
	\caption{Coverage of the Mandel-Paule confidence interval for $\theta=2$, $p_{C}=0.4$, $\sigma^2=0.4$, constant sample sizes $n=40,\;100,\;250,\;1000$.
The data-generation mechanisms are FIM1 ($\circ$), FIM2 ($\triangle$), RIM1 (+), RIM2 ($\times$), and URIM1 ($\diamond$).
		\label{PlotCovThetamu2andpC04LOR_MPsigma04}}
\end{figure}

\clearpage
\subsection*{A3.4 Coverage of $\hat{\theta}_{KD}$}
\renewcommand{\thefigure}{A3.4.\arabic{figure}}
\setcounter{figure}{0}

\begin{figure}[t]
	\centering
	\includegraphics[scale=0.33]{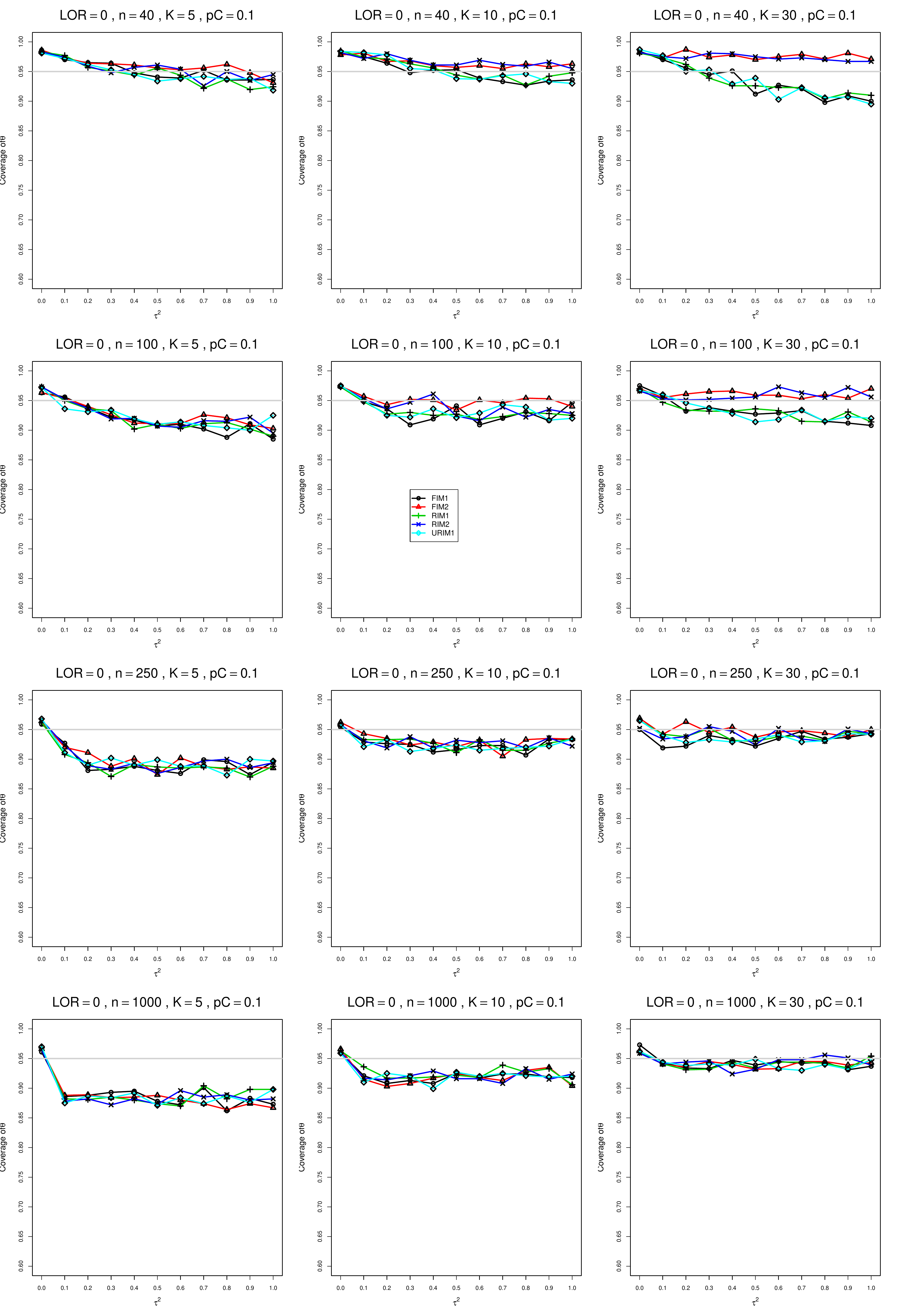}
	\caption{Coverage of the Kulinskaya-Dollinger confidence interval for $\theta=0$, $p_{C}=0.1$, $\sigma^2=0.1$, constant sample sizes $n=40,\;100,\;250,\;1000$.
The data-generation mechanisms are FIM1 ($\circ$), FIM2 ($\triangle$), RIM1 (+), RIM2 ($\times$), and URIM1 ($\diamond$).
		\label{PlotCovThetamu0andpC01LOR_KDsigma01}}
\end{figure}
\begin{figure}[t]
	\centering
	\includegraphics[scale=0.33]{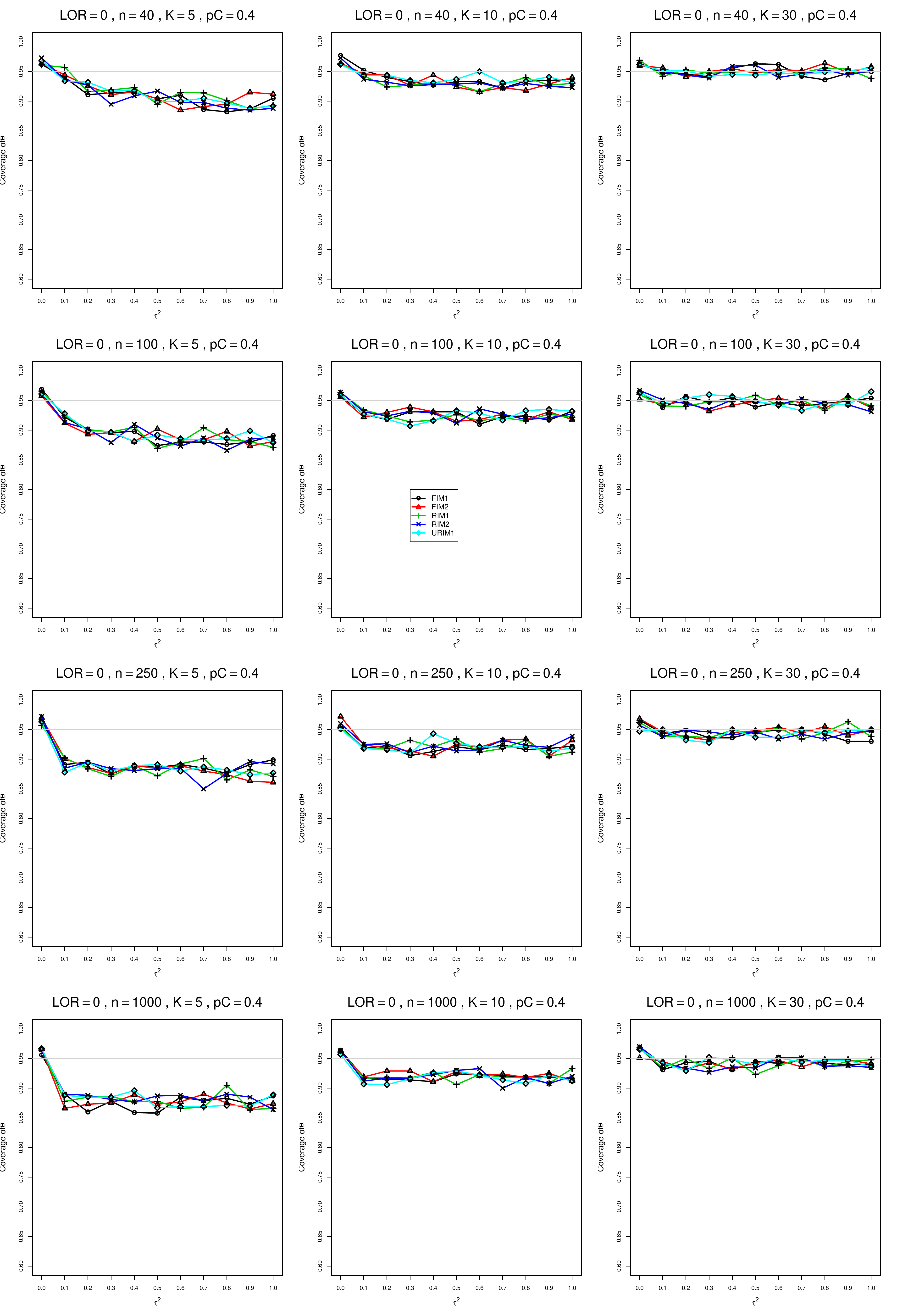}
	\caption{Coverage of the Kulinskaya-Dollinger confidence interval for $\theta=0$, $p_{C}=0.4$, $\sigma^2=0.1$, constant sample sizes $n=40,\;100,\;250,\;1000$.
The data-generation mechanisms are FIM1 ($\circ$), FIM2 ($\triangle$), RIM1 (+), RIM2 ($\times$), and URIM1 ($\diamond$).
		\label{PlotCovThetamu0andpC04LOR_KDsigma01}}
\end{figure}
\begin{figure}[t]
	\centering
	\includegraphics[scale=0.33]{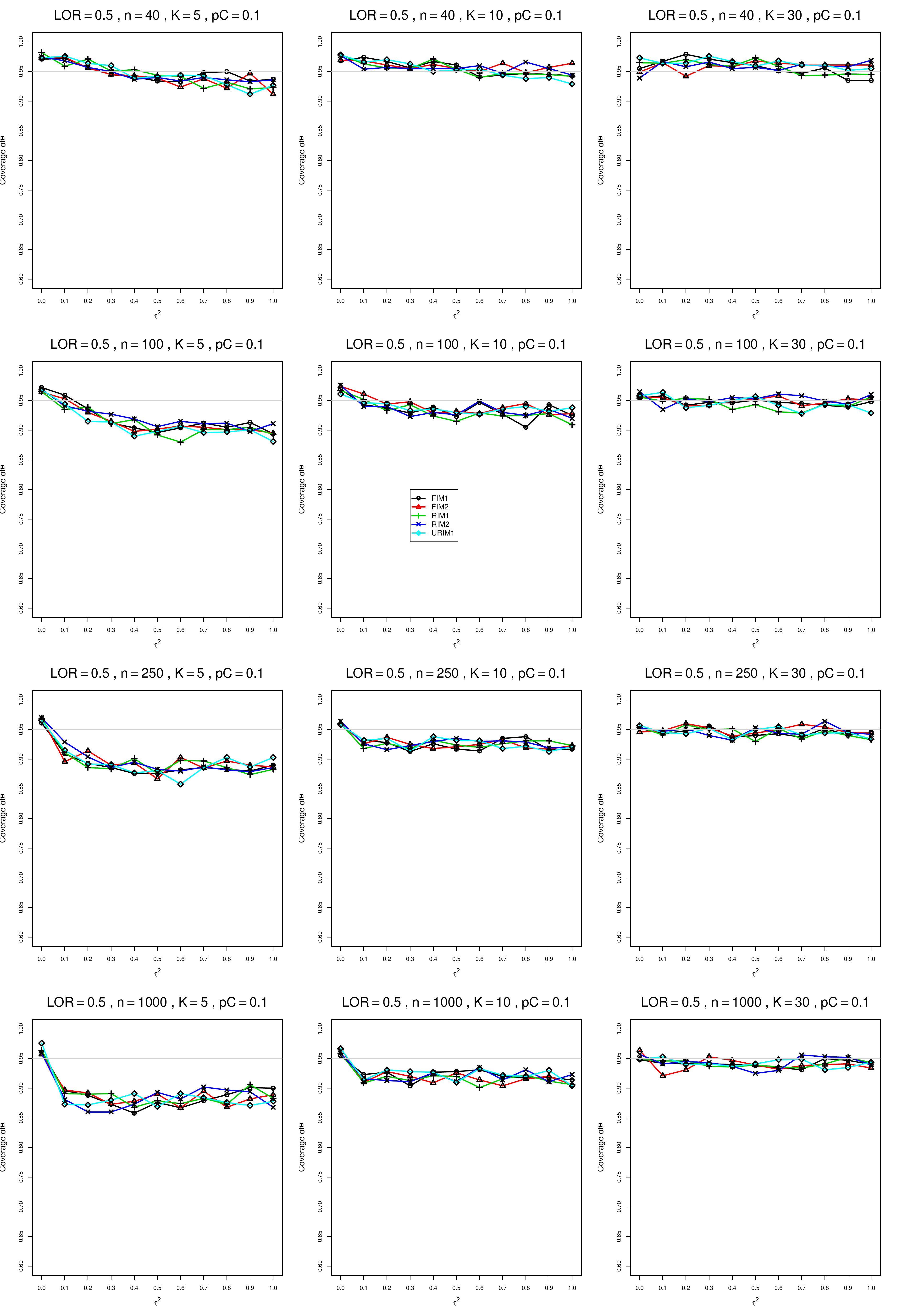}
	\caption{Coverage of the Kulinskaya-Dollinger confidence interval for $\theta=0.5$, $p_{C}=0.1$, $\sigma^2=0.1$, constant sample sizes $n=40,\;100,\;250,\;1000$.
The data-generation mechanisms are FIM1 ($\circ$), FIM2 ($\triangle$), RIM1 (+), RIM2 ($\times$), and URIM1 ($\diamond$).
		\label{PlotCovThetamu05andpC01LOR_KDsigma01}}
\end{figure}
\begin{figure}[t]
	\centering
	\includegraphics[scale=0.33]{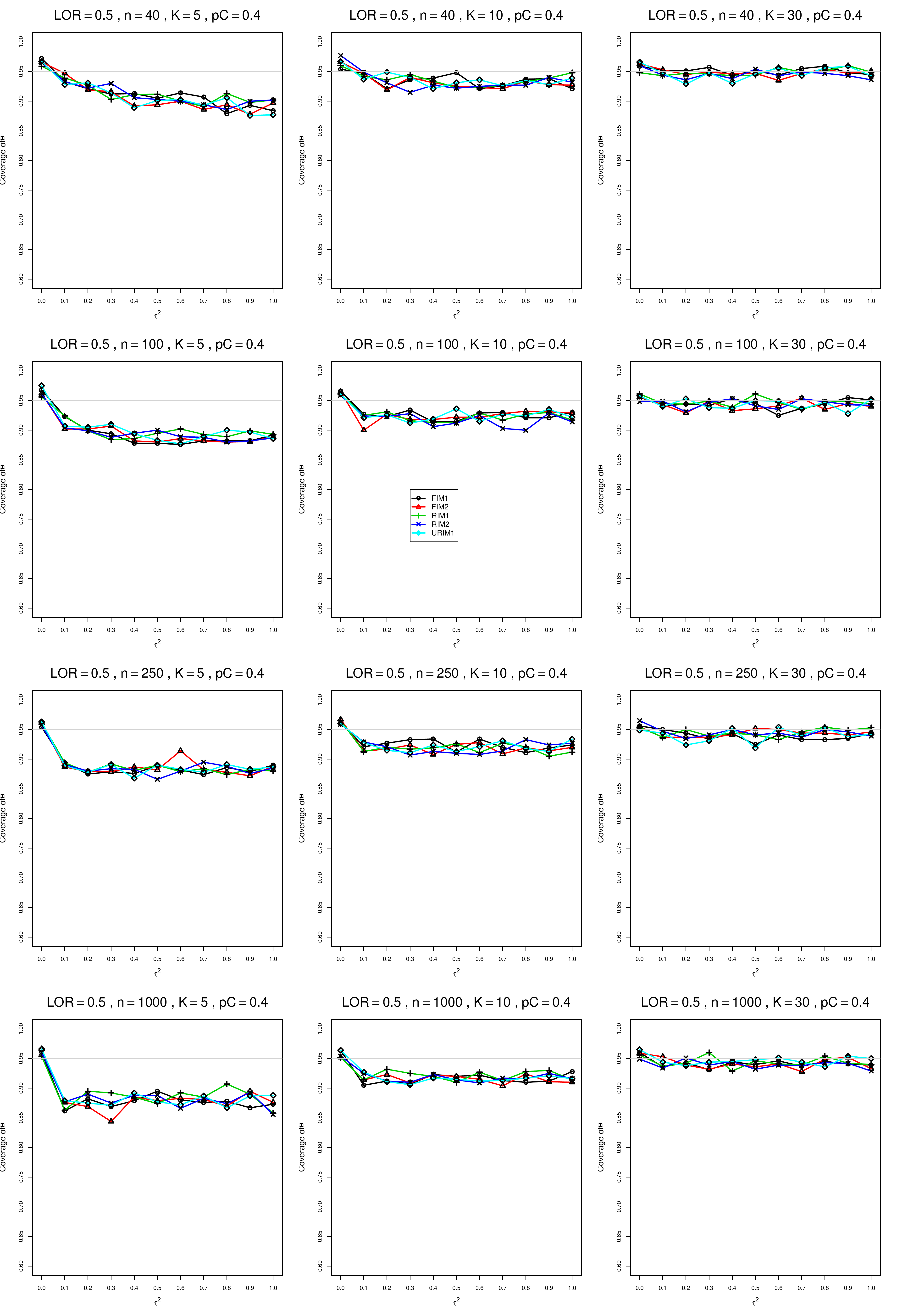}
	\caption{Coverage of the Kulinskaya-Dollinger confidence interval for $\theta=0.5$, $p_{C}=0.4$, $\sigma^2=0.1$, constant sample sizes $n=40,\;100,\;250,\;1000$.
The data-generation mechanisms are FIM1 ($\circ$), FIM2 ($\triangle$), RIM1 (+), RIM2 ($\times$), and URIM1 ($\diamond$).
		\label{PlotCovThetamu05andpC04LOR_KDsigma01}}
\end{figure}
\begin{figure}[t]
	\centering
	\includegraphics[scale=0.33]{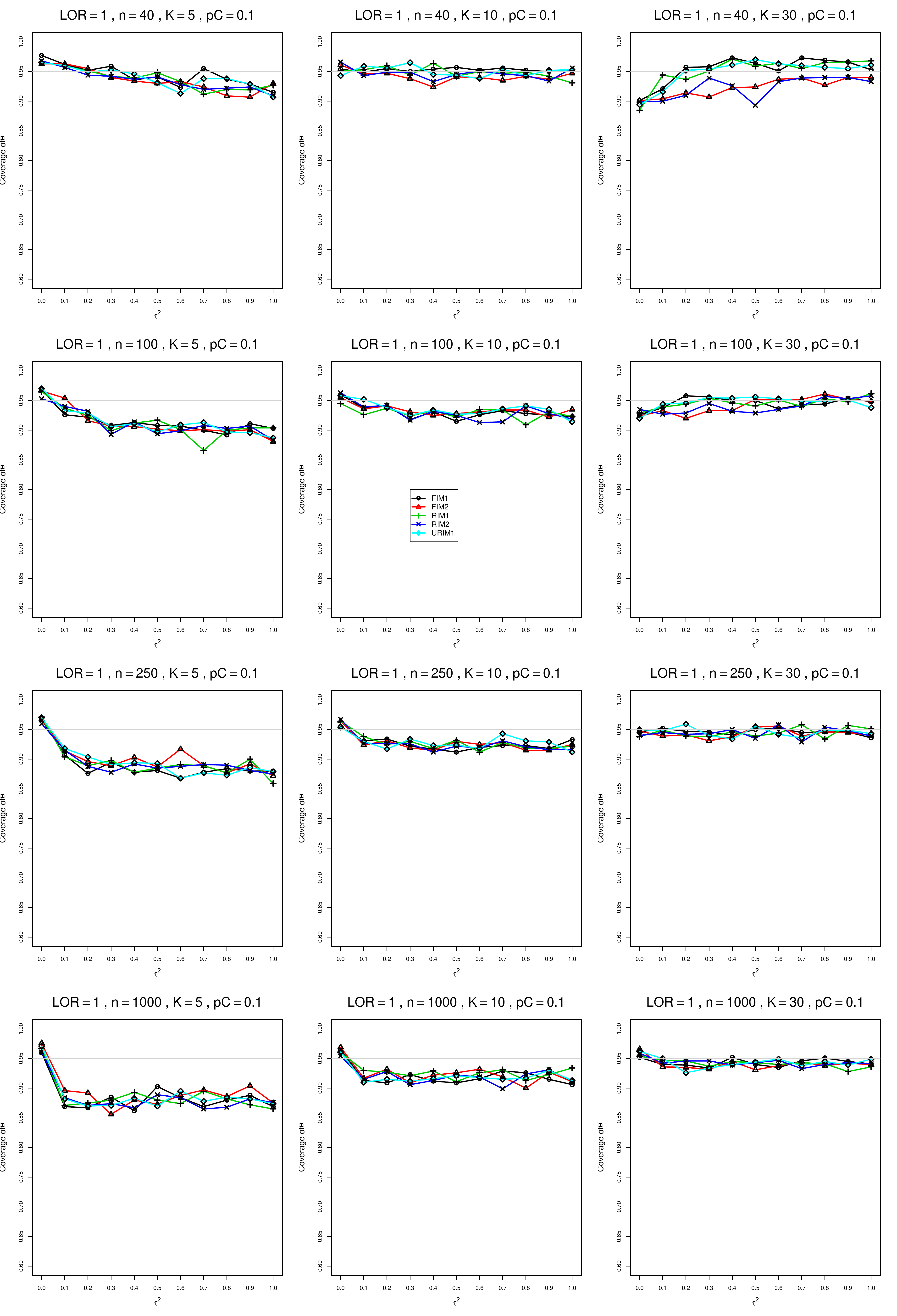}
	\caption{Coverage of the Kulinskaya-Dollinger confidence interval for $\theta=1$, $p_{C}=0.1$, $\sigma^2=0.1$, constant sample sizes $n=40,\;100,\;250,\;1000$.
The data-generation mechanisms are FIM1 ($\circ$), FIM2 ($\triangle$), RIM1 (+), RIM2 ($\times$), and URIM1 ($\diamond$).
		\label{PlotCovThetamu1andpC01LOR_KDsigma01}}
\end{figure}
\begin{figure}[t]
	\centering
	\includegraphics[scale=0.33]{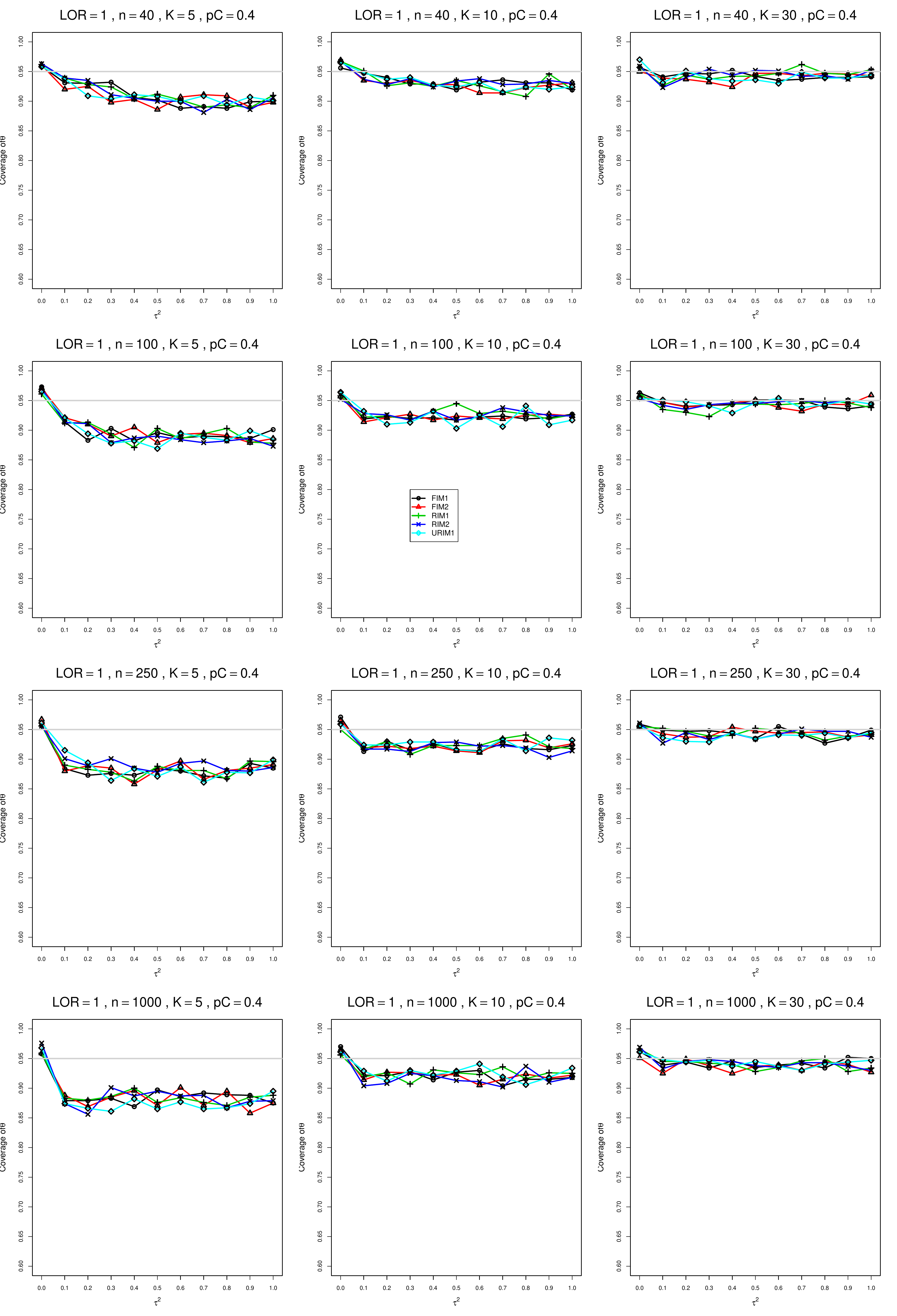}
	\caption{Coverage of the Kulinskaya-Dollinger confidence interval for $\theta=1$, $p_{C}=0.4$, $\sigma^2=0.1$, constant sample sizes $n=40,\;100,\;250,\;1000$.
The data-generation mechanisms are FIM1 ($\circ$), FIM2 ($\triangle$), RIM1 (+), RIM2 ($\times$), and URIM1 ($\diamond$).
		\label{PlotCovThetamu1andpC04LOR_KDsigma01}}
\end{figure}
\begin{figure}[t]
	\centering
	\includegraphics[scale=0.33]{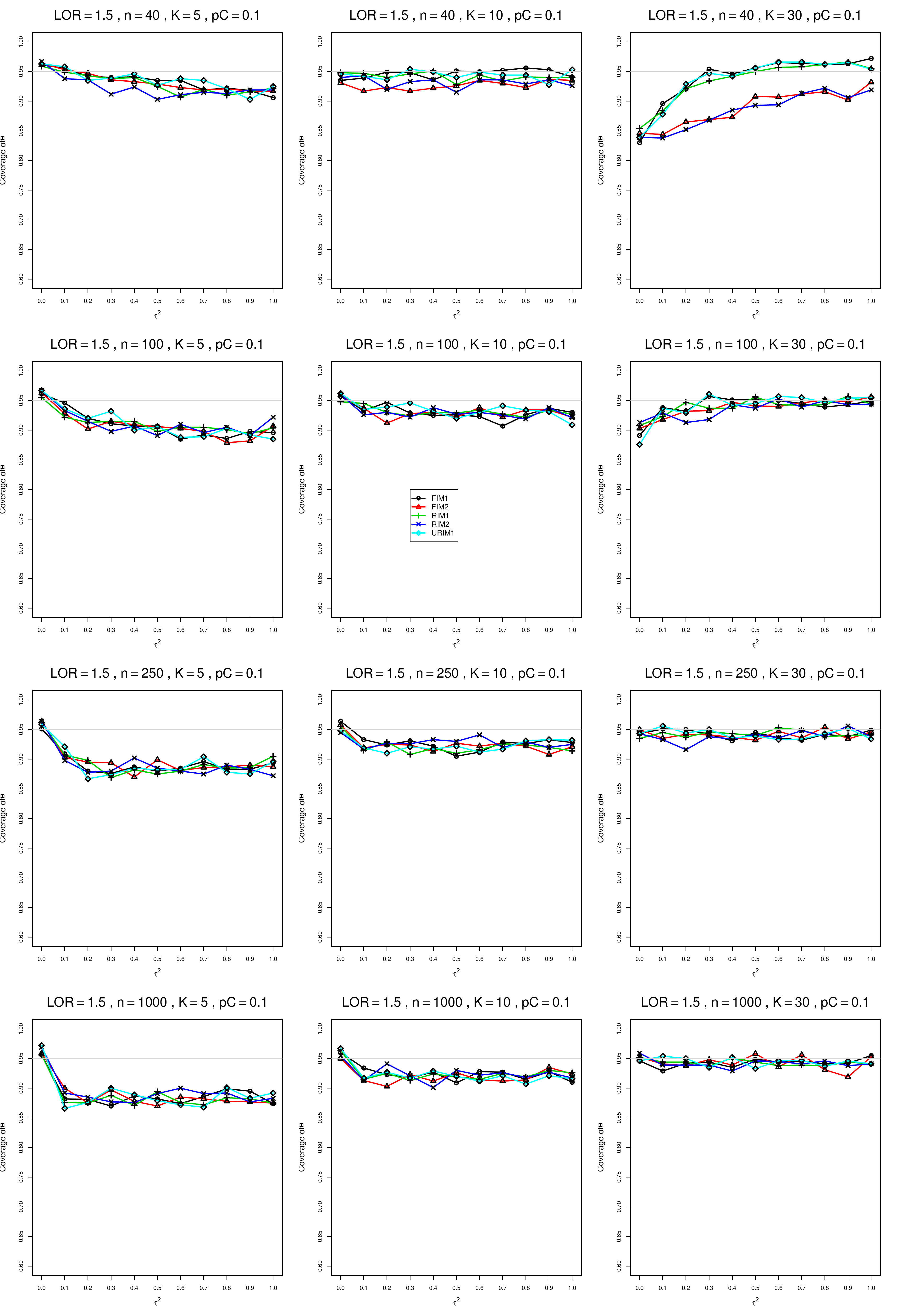}
	\caption{Coverage of the Kulinskaya-Dollinger confidence interval for $\theta=1.5$, $p_{C}=0.1$, $\sigma^2=0.1$, constant sample sizes $n=40,\;100,\;250,\;1000$.
The data-generation mechanisms are FIM1 ($\circ$), FIM2 ($\triangle$), RIM1 (+), RIM2 ($\times$), and URIM1 ($\diamond$).
		\label{PlotCovThetamu15andpC01LOR_KDsigma01}}
\end{figure}
\begin{figure}[t]
	\centering
	\includegraphics[scale=0.33]{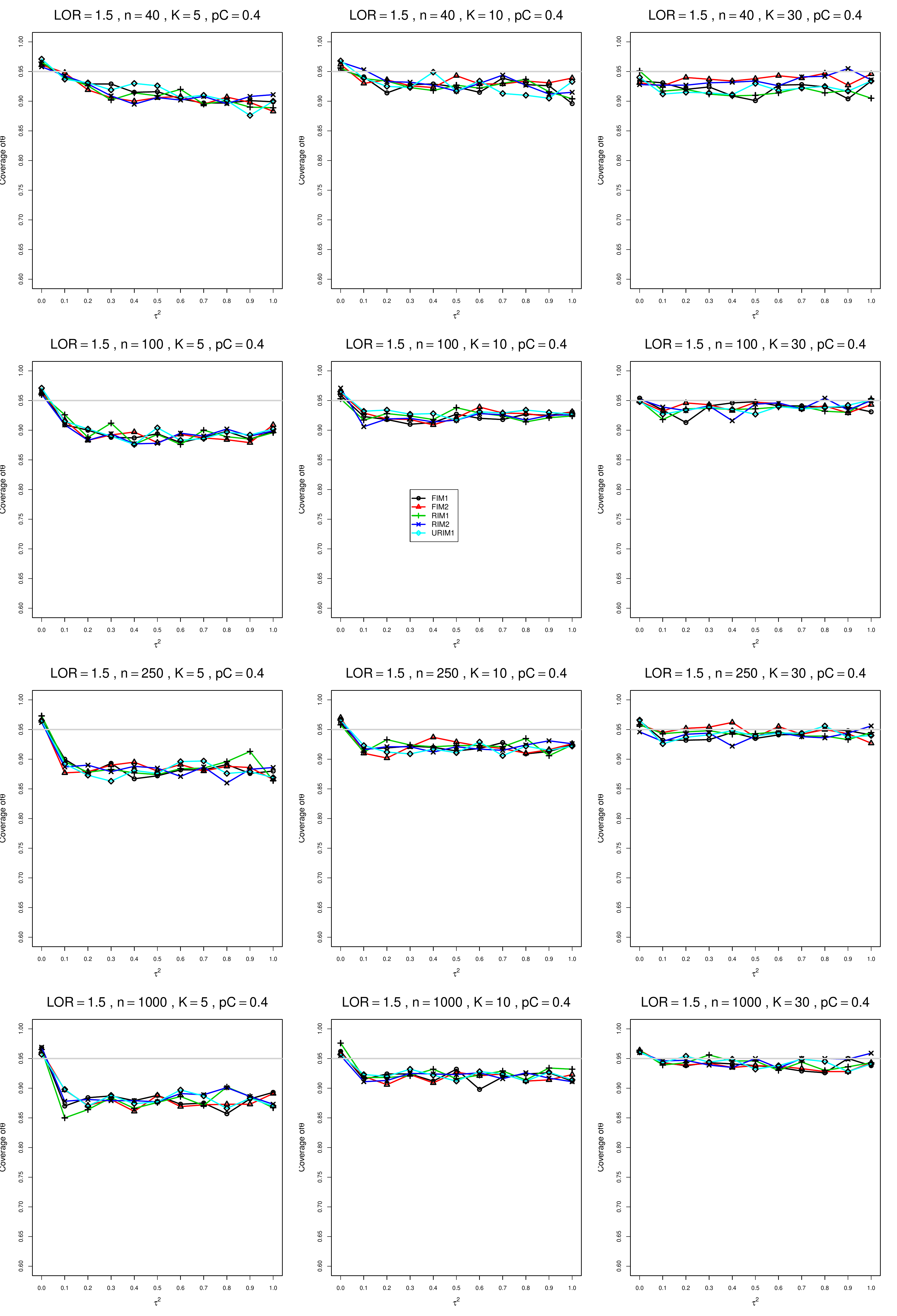}
	\caption{Coverage of the Kulinskaya-Dollinger confidence interval for $\theta=1.5$, $p_{C}=0.4$, $\sigma^2=0.1$, constant sample sizes $n=40,\;100,\;250,\;1000$.
The data-generation mechanisms are FIM1 ($\circ$), FIM2 ($\triangle$), RIM1 (+), RIM2 ($\times$), and URIM1 ($\diamond$).
		\label{PlotCovThetamu15andpC04LOR_KDsigma01}}
\end{figure}
\begin{figure}[t]
	\centering
	\includegraphics[scale=0.33]{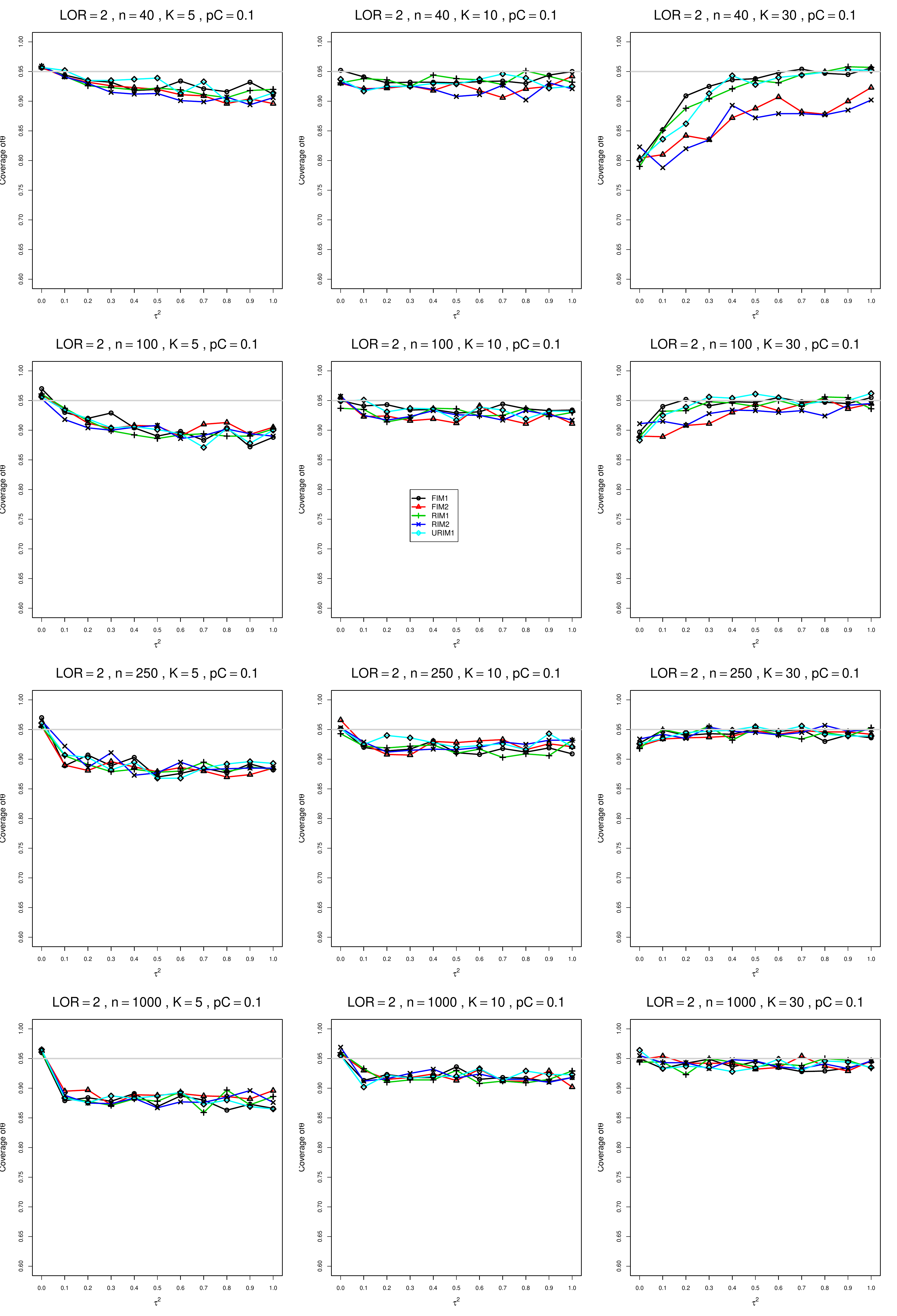}
	\caption{Coverage of the Kulinskaya-Dollinger confidence interval for $\theta=2$, $p_{C}=0.1$, $\sigma^2=0.1$, constant sample sizes $n=40,\;100,\;250,\;1000$.
The data-generation mechanisms are FIM1 ($\circ$), FIM2 ($\triangle$), RIM1 (+), RIM2 ($\times$), and URIM1 ($\diamond$).
		\label{PlotCovThetamu2andpC01LOR_KDsigma01}}
\end{figure}
\begin{figure}[t]
	\centering
	\includegraphics[scale=0.33]{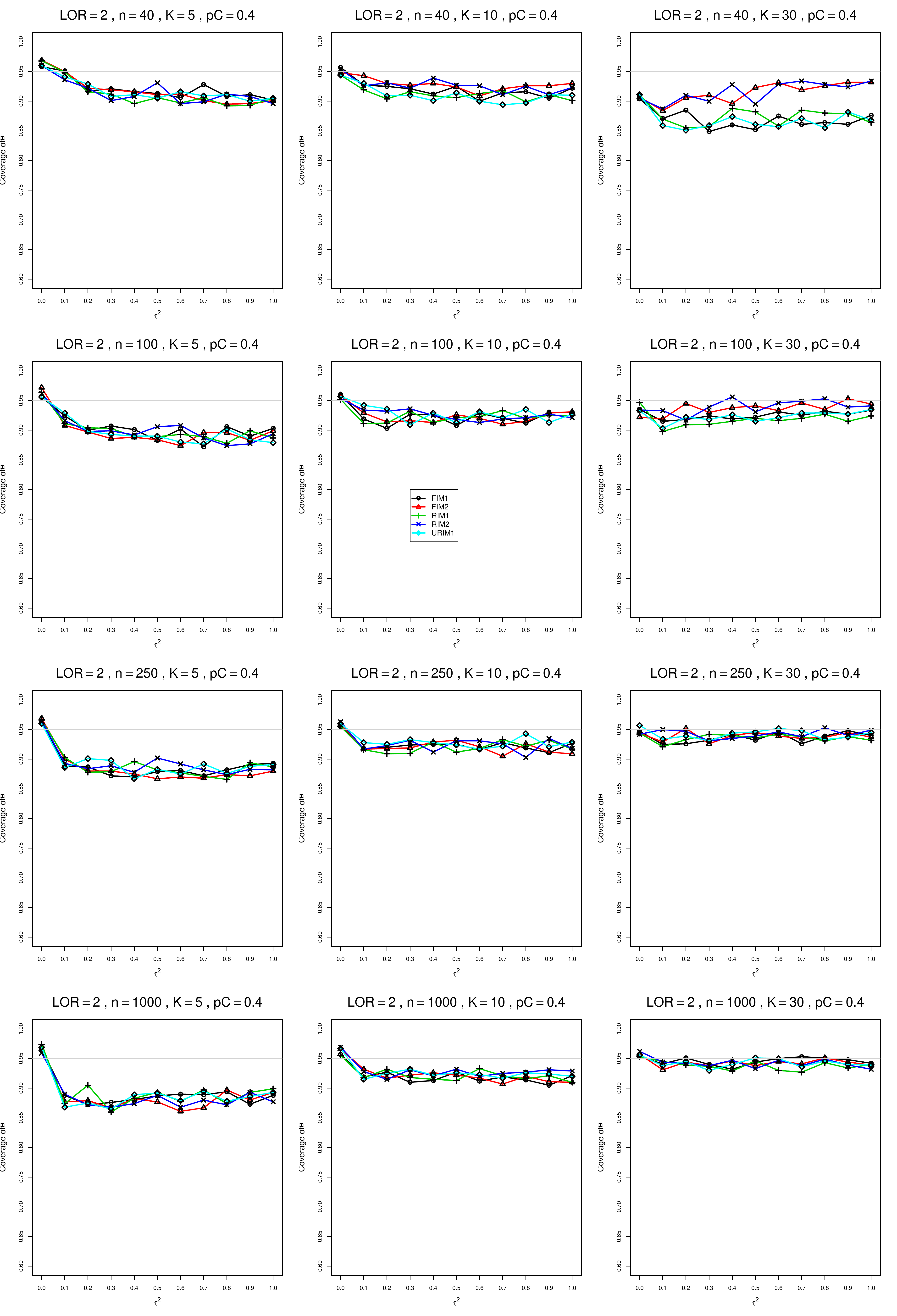}
	\caption{Coverage of the Kulinskaya-Dollinger confidence interval for $\theta=2$, $p_{C}=0.4$, $\sigma^2=0.1$, constant sample sizes $n=40,\;100,\;250,\;1000$.
The data-generation mechanisms are FIM1 ($\circ$), FIM2 ($\triangle$), RIM1 (+), RIM2 ($\times$), and URIM1 ($\diamond$).
		\label{PlotCovThetamu2andpC04LOR_KDsigma01}}
\end{figure}
\begin{figure}[t]
	\centering
	\includegraphics[scale=0.33]{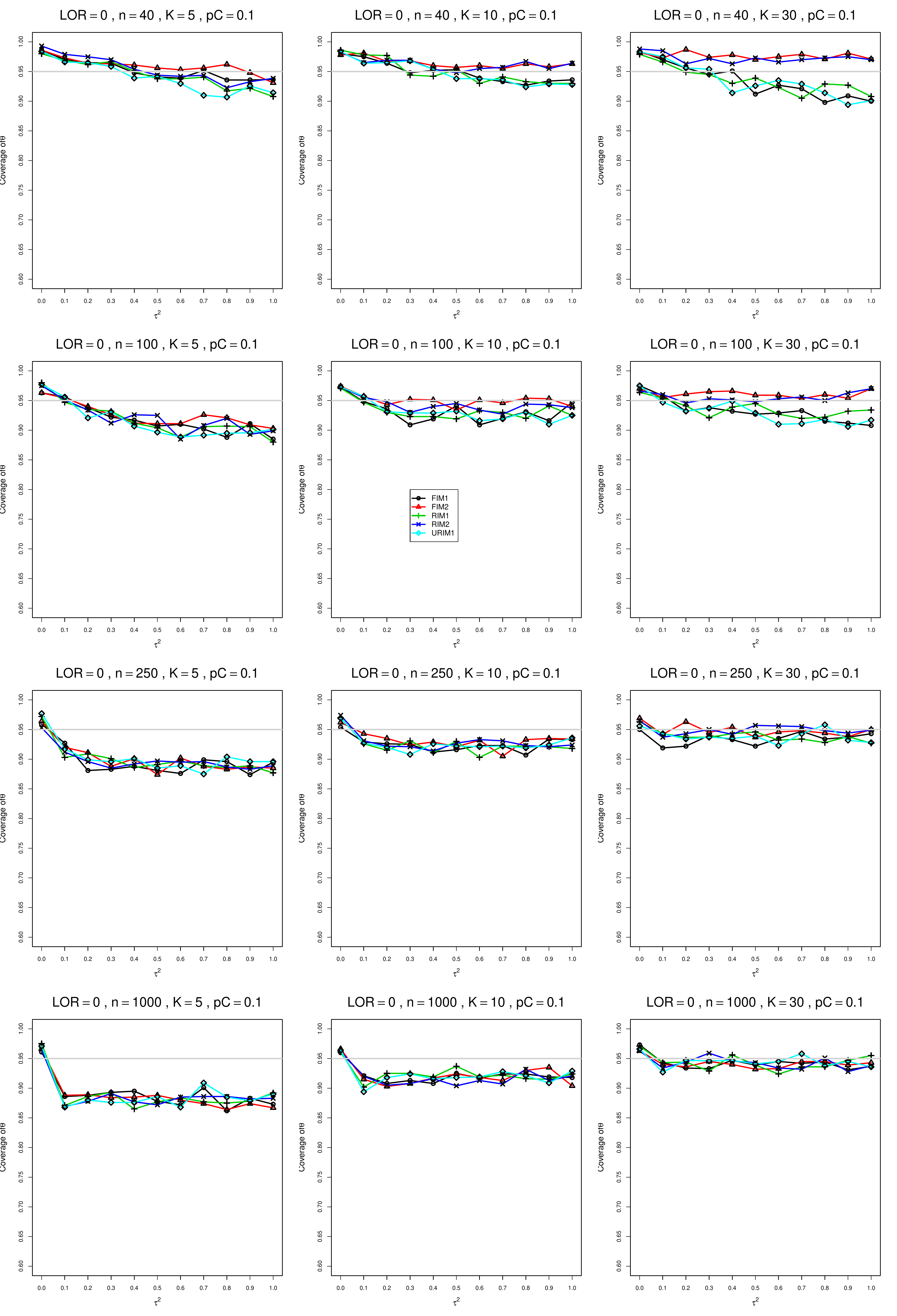}
	\caption{Coverage of the Kulinskaya-Dollinger confidence interval for $\theta=0$, $p_{C}=0.1$, $\sigma^2=0.4$, constant sample sizes $n=40,\;100,\;250,\;1000$.
The data-generation mechanisms are FIM1 ($\circ$), FIM2 ($\triangle$), RIM1 (+), RIM2 ($\times$), and URIM1 ($\diamond$).
		\label{PlotCovThetamu0andpC01LOR_KDsigma04}}
\end{figure}
\begin{figure}[t]
	\centering
	\includegraphics[scale=0.33]{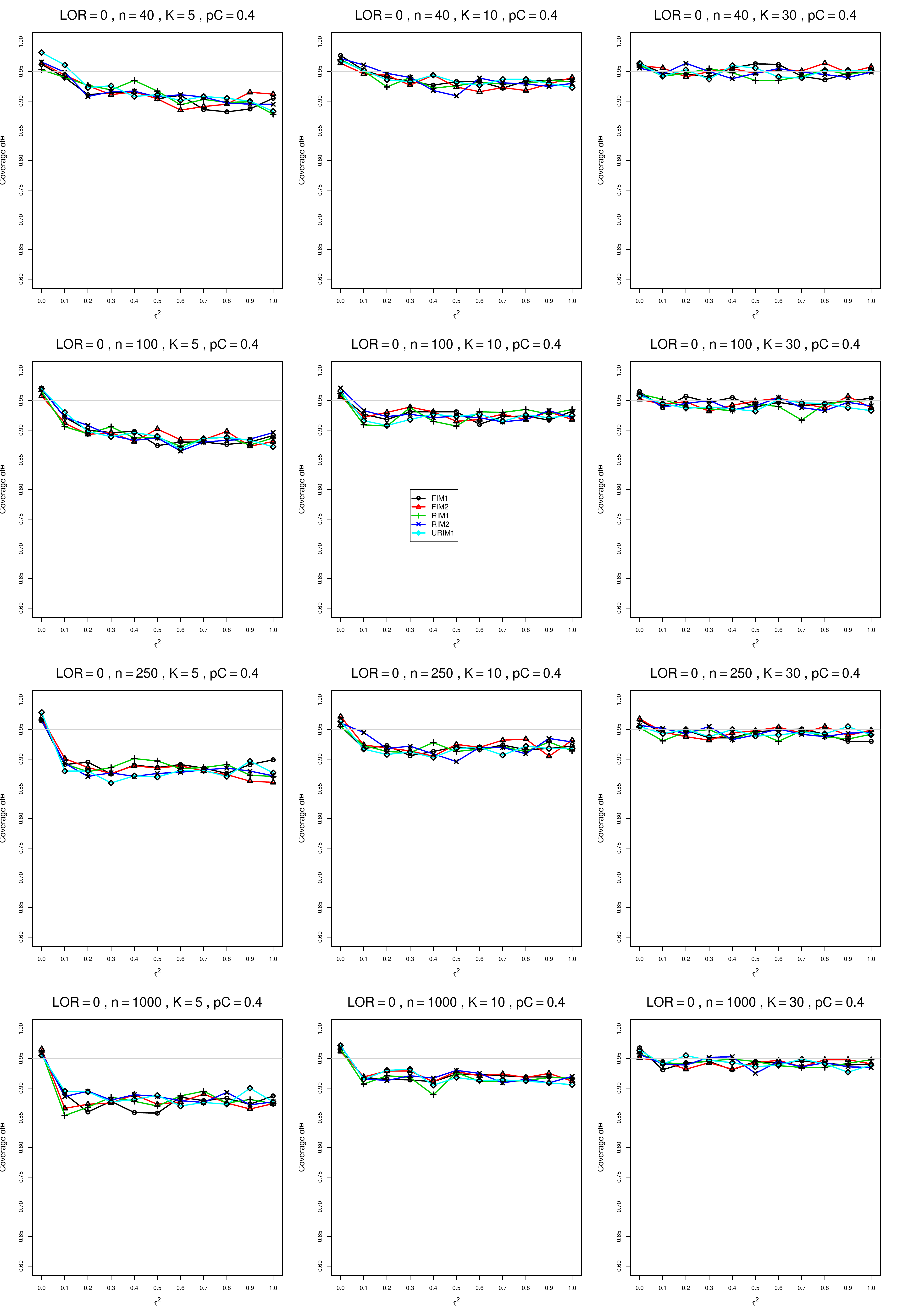}
	\caption{Coverage of the Kulinskaya-Dollinger confidence interval for $\theta=0$, $p_{C}=0.4$, $\sigma^2=0.4$, constant sample sizes $n=40,\;100,\;250,\;1000$.
The data-generation mechanisms are FIM1 ($\circ$), FIM2 ($\triangle$), RIM1 (+), RIM2 ($\times$), and URIM1 ($\diamond$).
		\label{PlotCovThetamu0andpC04LOR_KDsigma04}}
\end{figure}
\begin{figure}[t]
	\centering
	\includegraphics[scale=0.33]{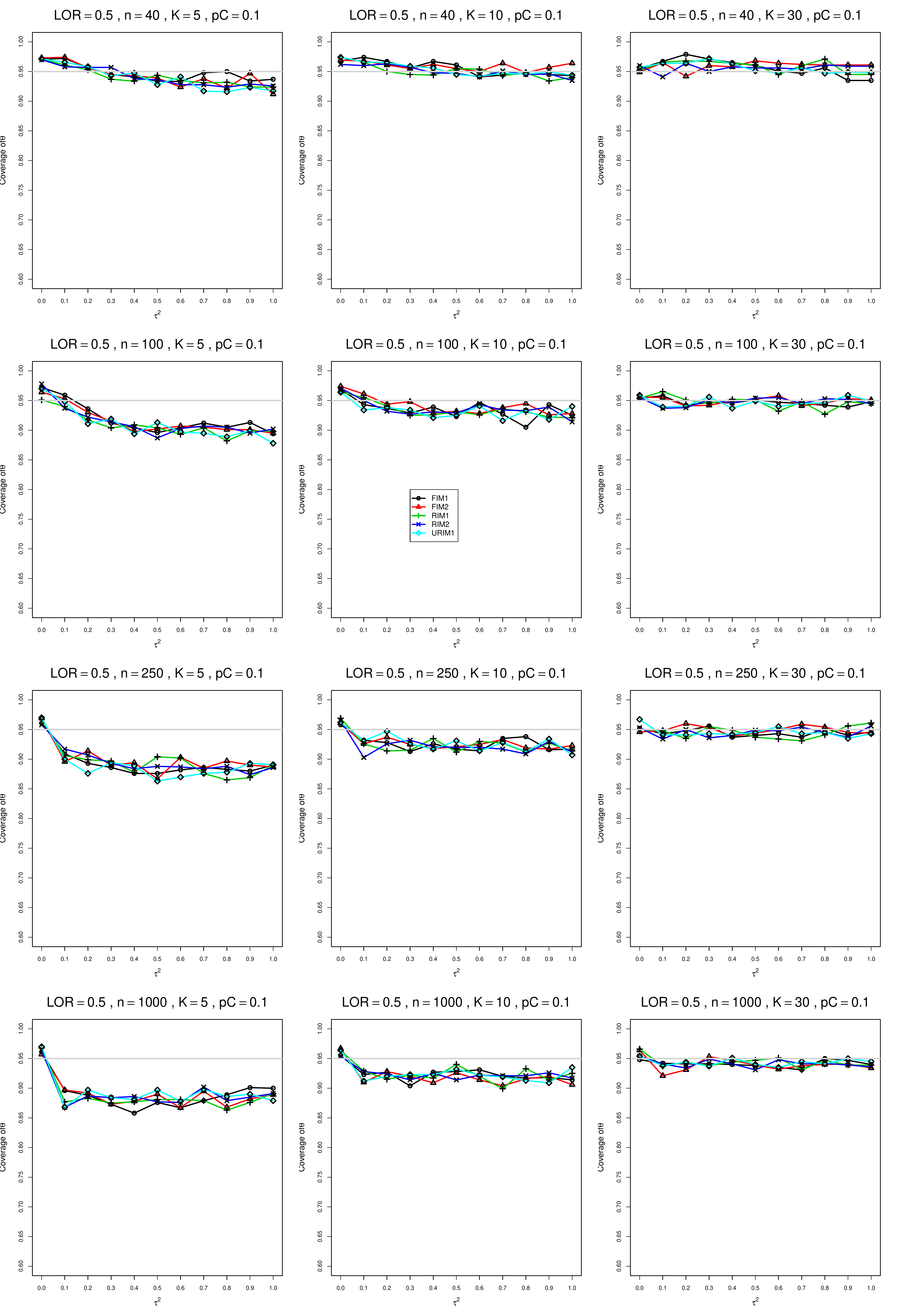}
	\caption{Coverage of the Kulinskaya-Dollinger confidence interval for $\theta=0.5$, $p_{C}=0.1$, $\sigma^2=0.4$, constant sample sizes $n=40,\;100,\;250,\;1000$.
The data-generation mechanisms are FIM1 ($\circ$), FIM2 ($\triangle$), RIM1 (+), RIM2 ($\times$), and URIM1 ($\diamond$).
		\label{PlotCovThetamu05andpC01LOR_KDsigma04}}
\end{figure}
\begin{figure}[t]
	\centering
	\includegraphics[scale=0.33]{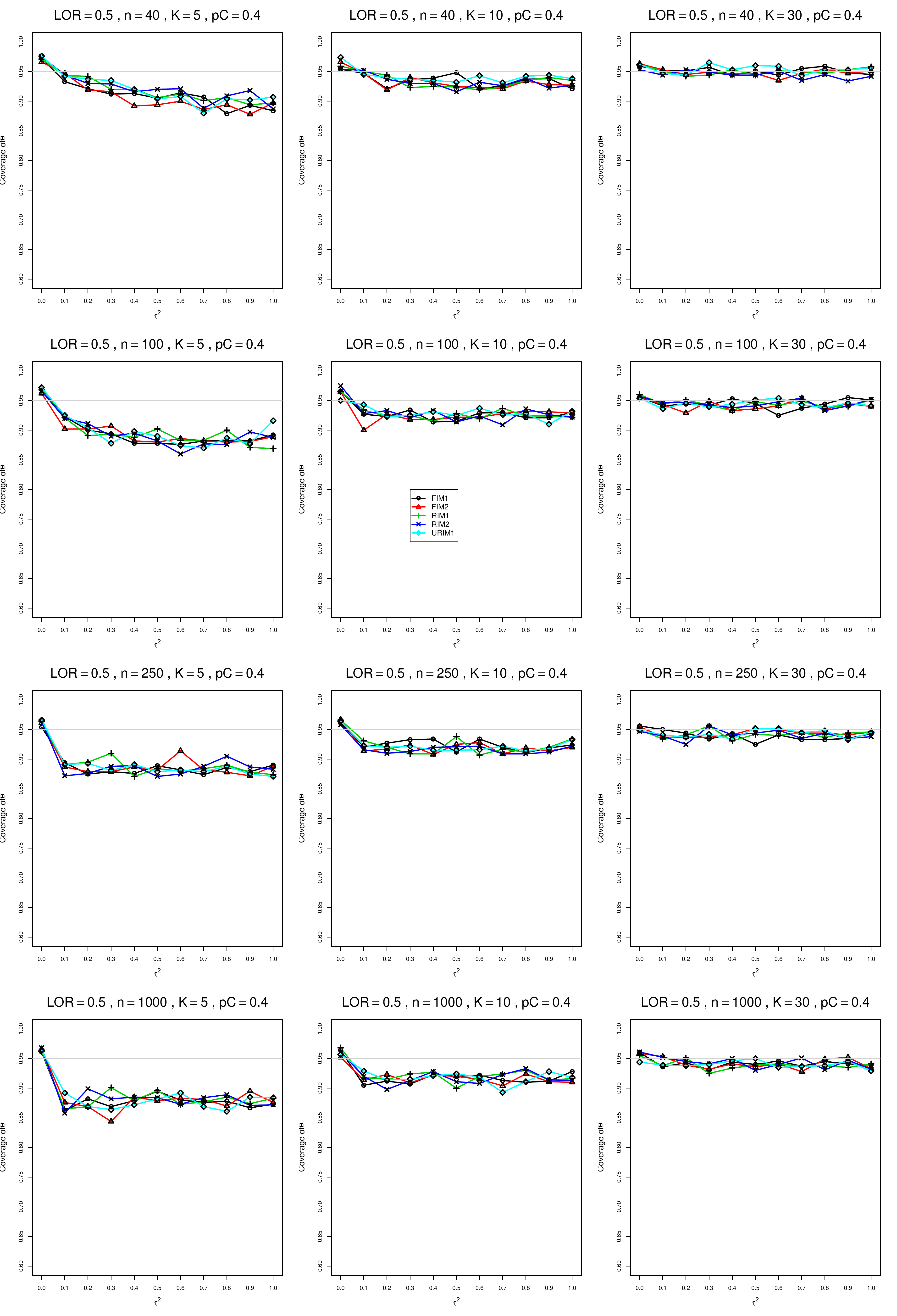}
	\caption{Coverage of the Kulinskaya-Dollinger confidence interval for $\theta=0.5$, $p_{C}=0.4$, $\sigma^2=0.4$, constant sample sizes $n=40,\;100,\;250,\;1000$.
The data-generation mechanisms are FIM1 ($\circ$), FIM2 ($\triangle$), RIM1 (+), RIM2 ($\times$), and URIM1 ($\diamond$).
		\label{PlotCovThetamu05andpC04LOR_KDsigma04}}
\end{figure}
\begin{figure}[t]
	\centering
	\includegraphics[scale=0.33]{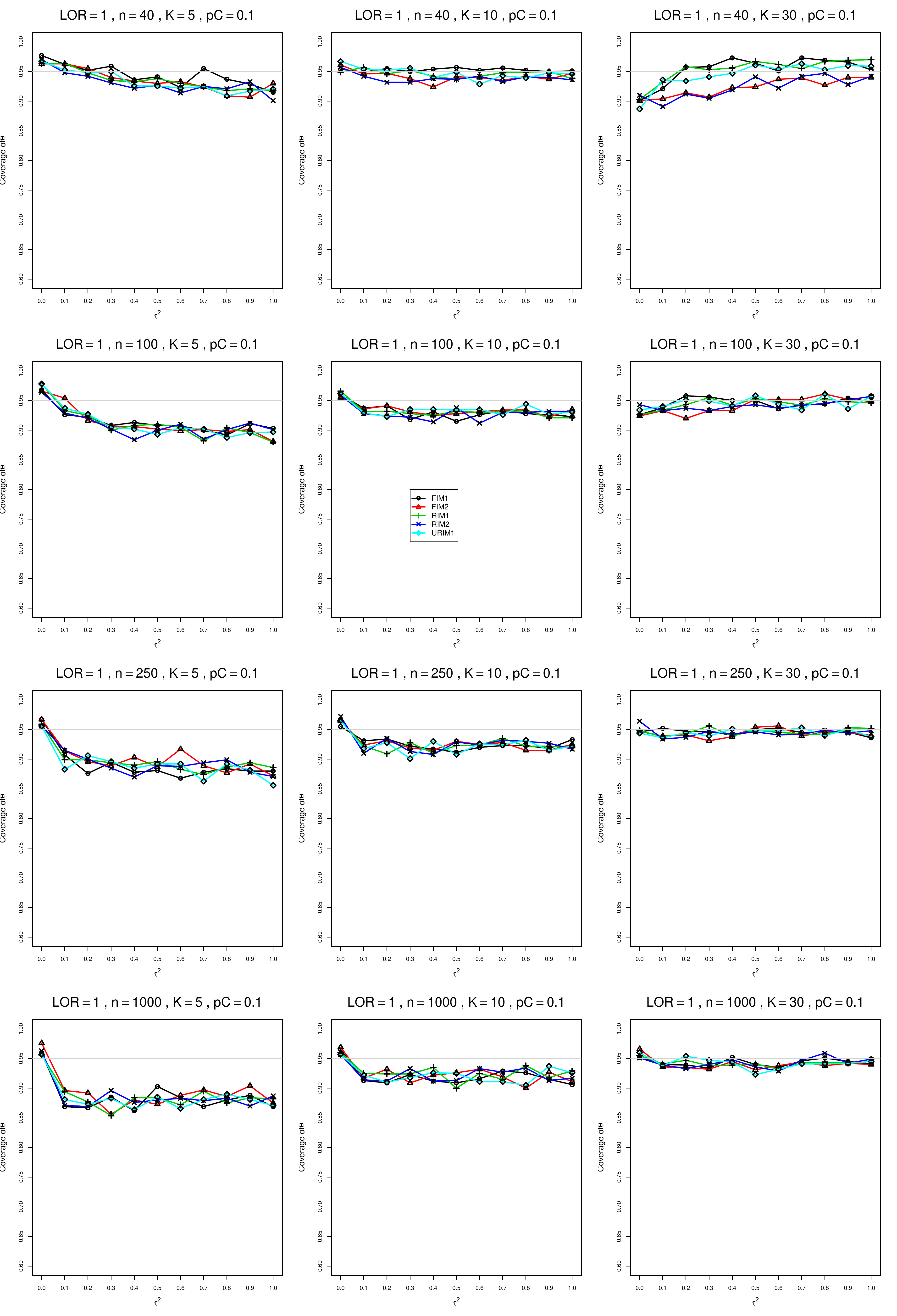}
	\caption{Coverage of the Kulinskaya-Dollinger confidence interval for $\theta=1$, $p_{C}=0.1$, $\sigma^2=0.4$, constant sample sizes $n=40,\;100,\;250,\;1000$.
The data-generation mechanisms are FIM1 ($\circ$), FIM2 ($\triangle$), RIM1 (+), RIM2 ($\times$), and URIM1 ($\diamond$).
		\label{PlotCovThetamu1andpC01LOR_KDsigma04}}
\end{figure}
\begin{figure}[t]
	\centering
	\includegraphics[scale=0.33]{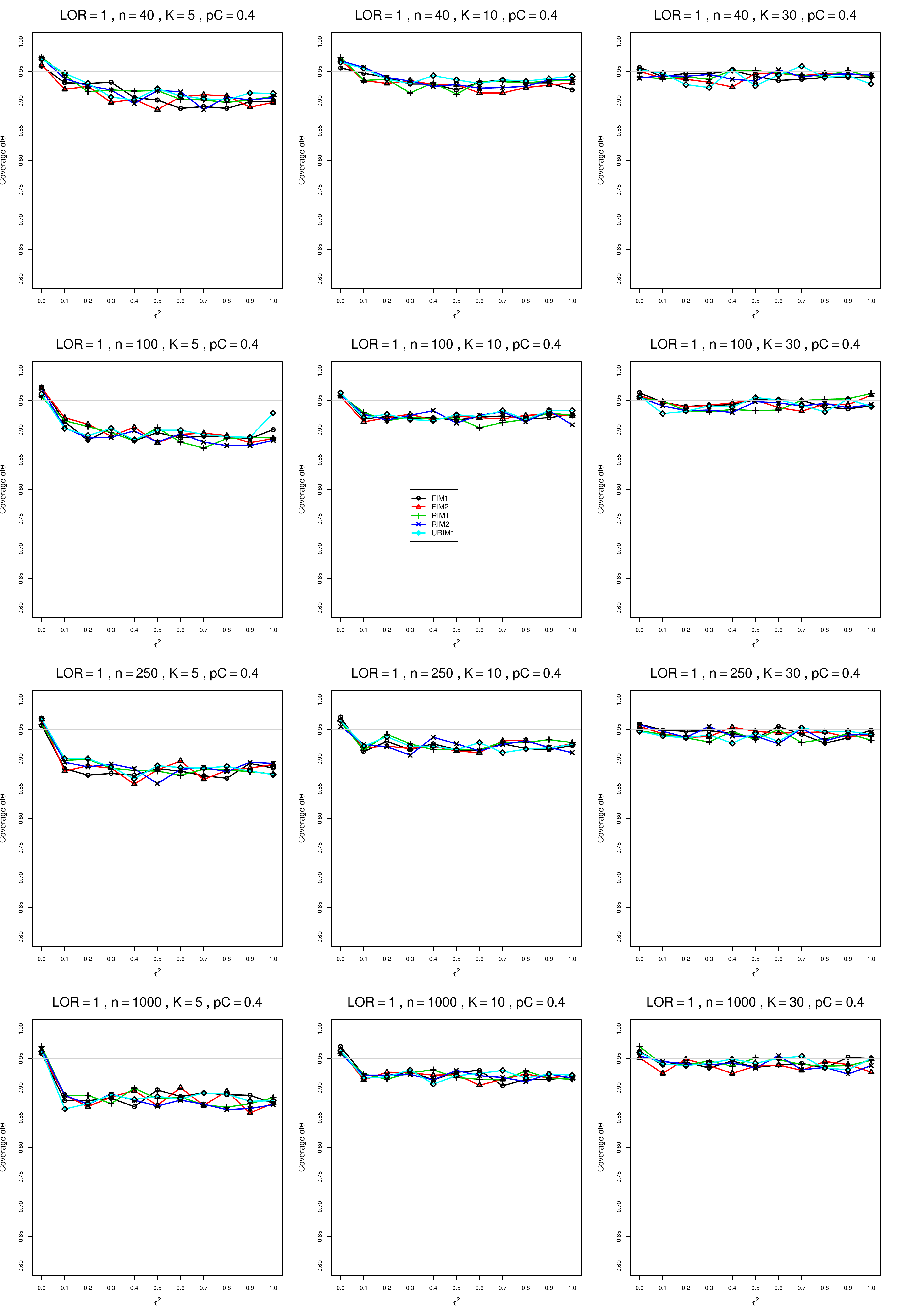}
	\caption{Coverage of the Kulinskaya-Dollinger confidence interval for $\theta=1$, $p_{C}=0.4$, $\sigma^2=0.4$, constant sample sizes $n=40,\;100,\;250,\;1000$.
The data-generation mechanisms are FIM1 ($\circ$), FIM2 ($\triangle$), RIM1 (+), RIM2 ($\times$), and URIM1 ($\diamond$).
		\label{PlotCovThetamu1andpC04LOR_KDsigma04}}
\end{figure}
\begin{figure}[t]
	\centering
	\includegraphics[scale=0.33]{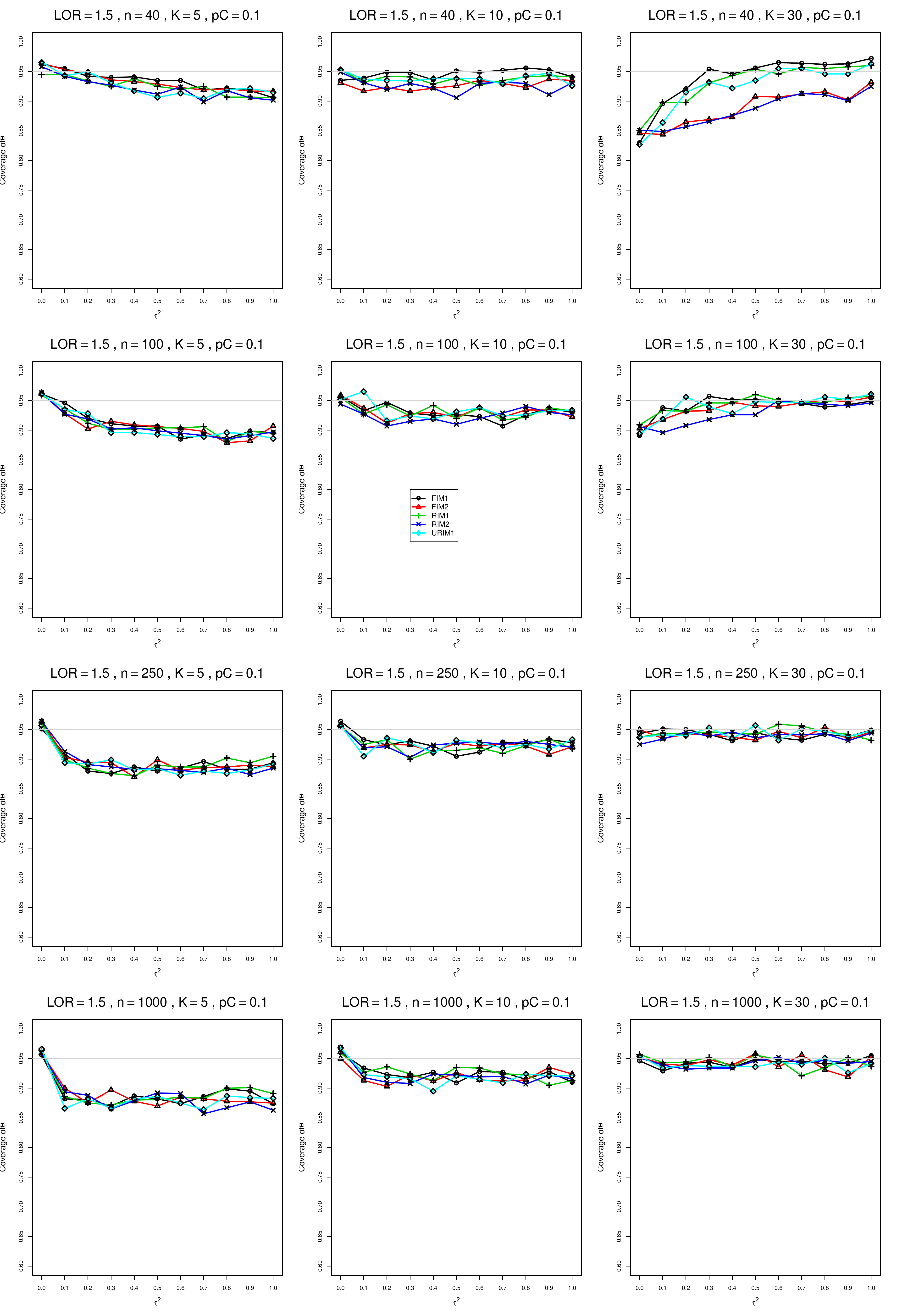}
	\caption{Coverage of the Kulinskaya-Dollinger confidence interval for $\theta=1.5$, $p_{C}=0.1$, $\sigma^2=0.4$, constant sample sizes $n=40,\;100,\;250,\;1000$.
The data-generation mechanisms are FIM1 ($\circ$), FIM2 ($\triangle$), RIM1 (+), RIM2 ($\times$), and URIM1 ($\diamond$).
		\label{PlotCovThetamu15andpC01LOR_KDsigma04}}
\end{figure}
\begin{figure}[t]
	\centering
	\includegraphics[scale=0.33]{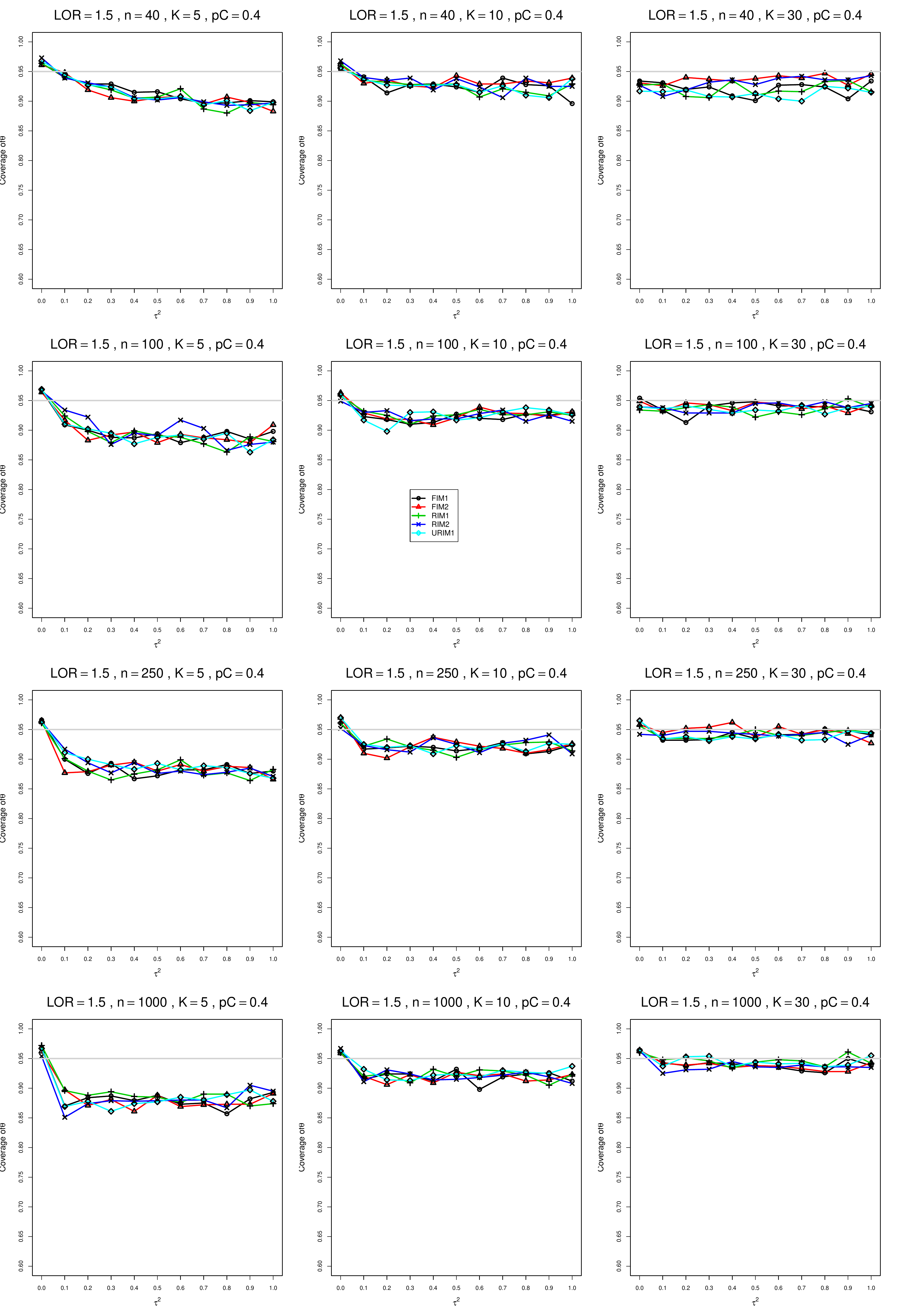}
	\caption{Coverage of the Kulinskaya-Dollinger confidence interval for $\theta=1.5$, $p_{C}=0.4$, $\sigma^2=0.4$, constant sample sizes $n=40,\;100,\;250,\;1000$.
The data-generation mechanisms are FIM1 ($\circ$), FIM2 ($\triangle$), RIM1 (+), RIM2 ($\times$), and URIM1 ($\diamond$).
		\label{PlotCovThetamu15andpC04LOR_KDsigma04}}
\end{figure}
\begin{figure}[t]
	\centering
	\includegraphics[scale=0.33]{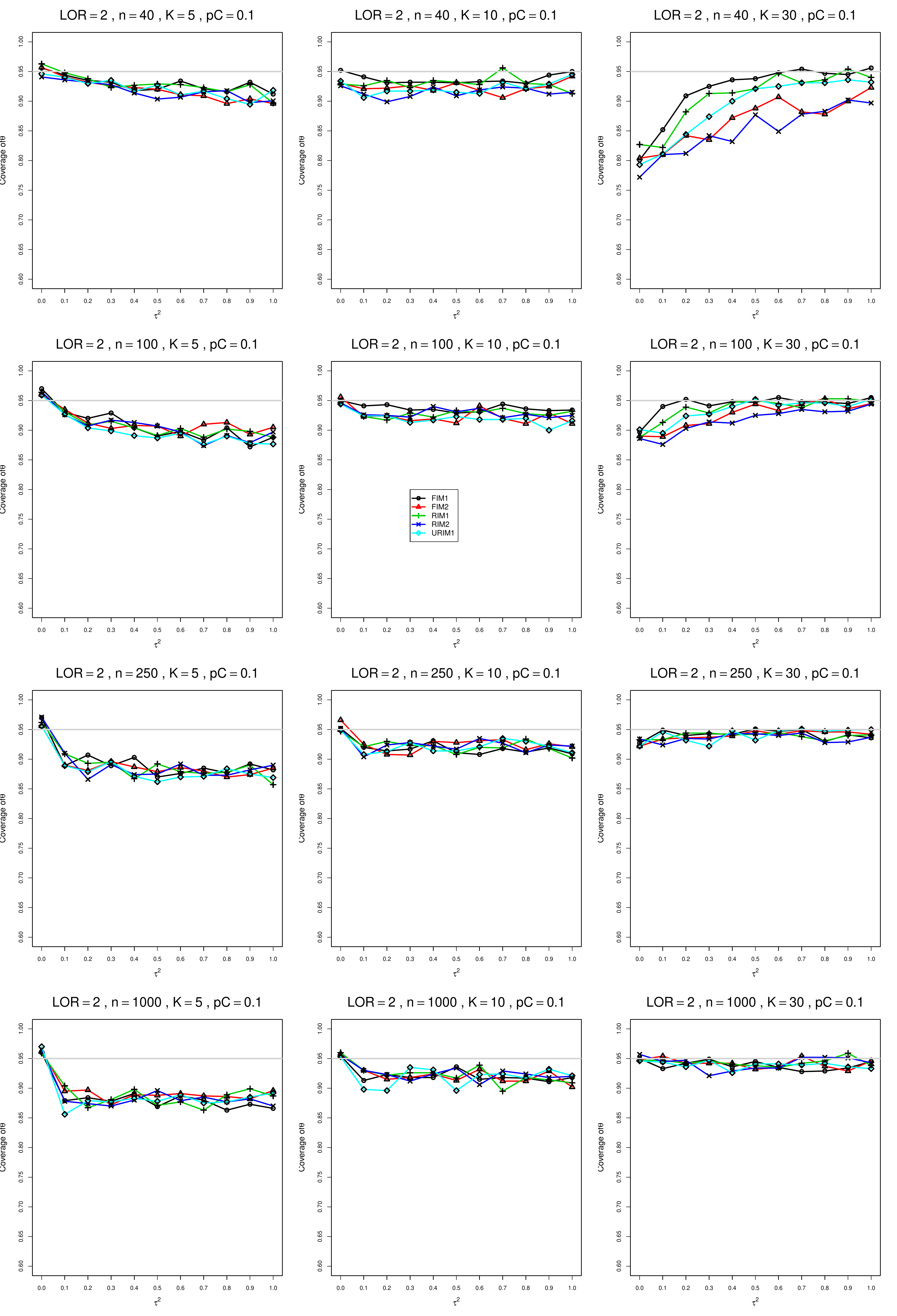}
	\caption{Coverage of the Kulinskaya-Dollinger confidence interval for $\theta=2$, $p_{C}=0.1$, $\sigma^2=0.4$, constant sample sizes $n=40,\;100,\;250,\;1000$.
The data-generation mechanisms are FIM1 ($\circ$), FIM2 ($\triangle$), RIM1 (+), RIM2 ($\times$), and URIM1 ($\diamond$).
		\label{PlotCovThetamu2andpC01LOR_KDsigma04}}
\end{figure}
\begin{figure}[t]
	\centering
	\includegraphics[scale=0.33]{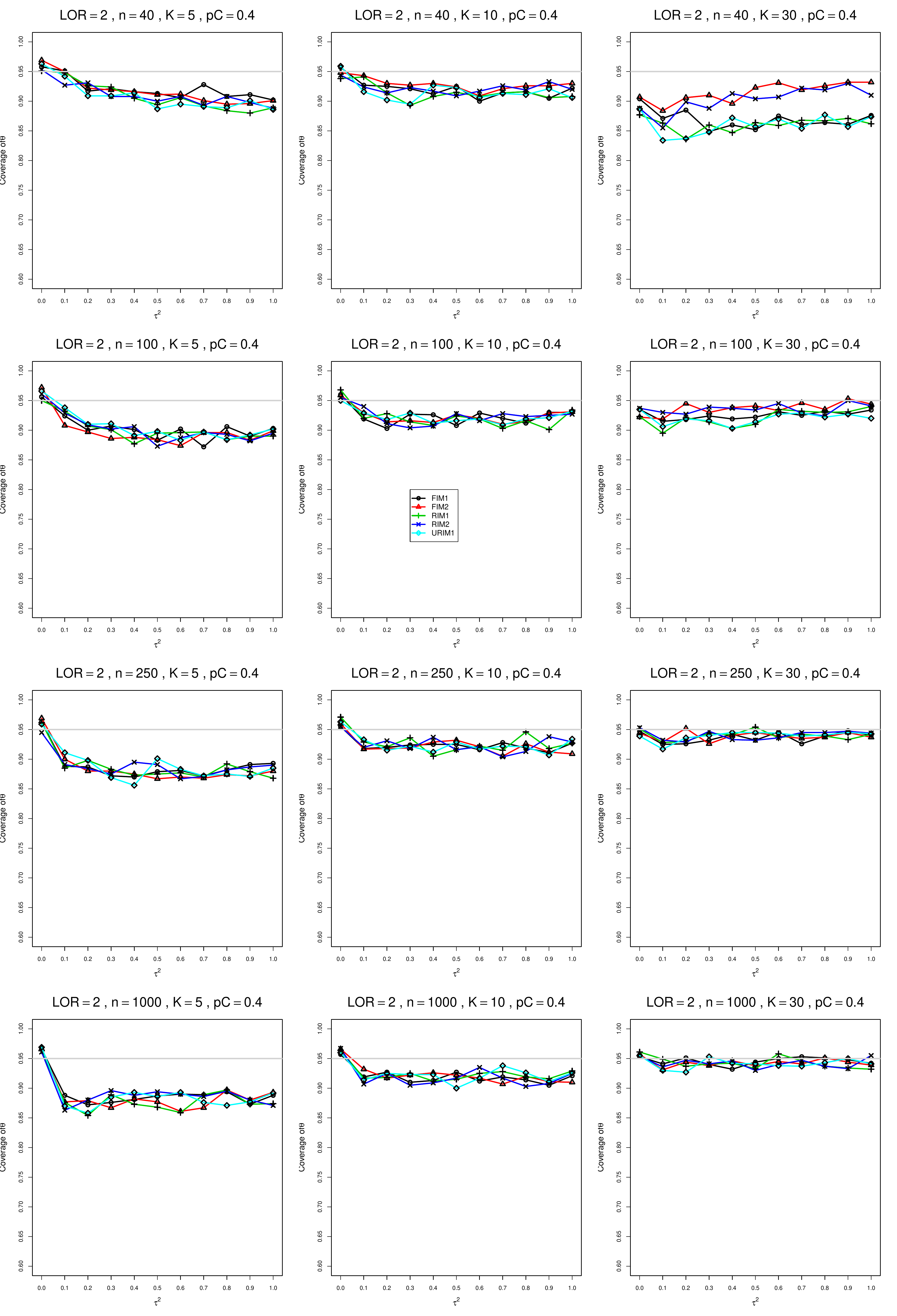}
	\caption{Coverage of the Kulinskaya-Dollinger confidence interval for $\theta=2$, $p_{C}=0.4$, $\sigma^2=0.4$, constant sample sizes $n=40,\;100,\;250,\;1000$.
The data-generation mechanisms are FIM1 ($\circ$), FIM2 ($\triangle$), RIM1 (+), RIM2 ($\times$), and URIM1 ($\diamond$).
		\label{PlotCovThetamu2andpC04LOR_KDsigma04}}
\end{figure}

\clearpage
\subsection*{A3.5 Coverage of $\hat{\theta}_{FIM2}$}
\renewcommand{\thefigure}{A3.5.\arabic{figure}}
\setcounter{figure}{0}

\begin{figure}[t]
	\centering
	\includegraphics[scale=0.33]{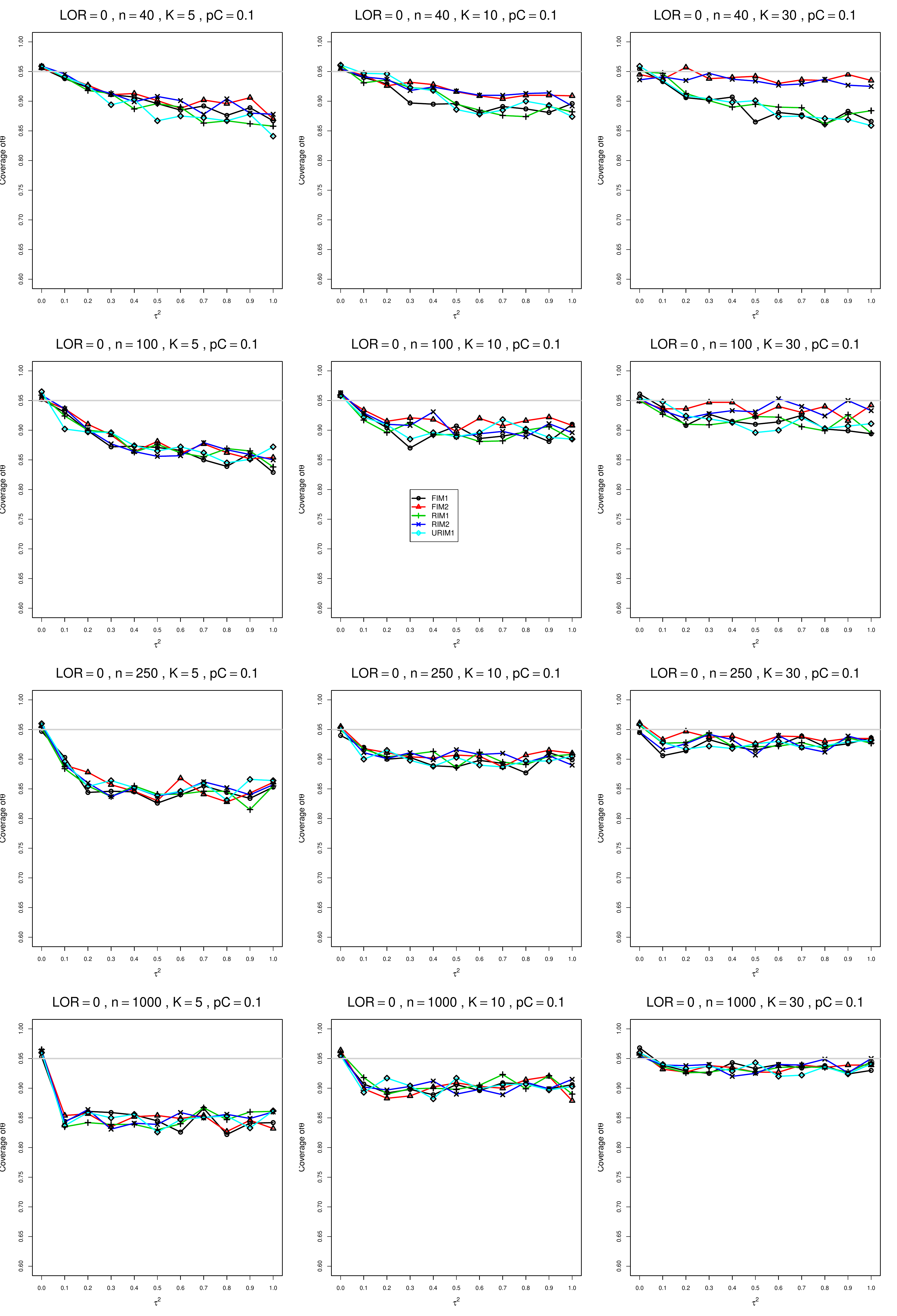}
	\caption{Coverage of the Fixed-intercept with $c=1/2$ confidence interval for $\theta=0$, $p_{C}=0.1$, $\sigma^2=0.1$, constant sample sizes $n=40,\;100,\;250,\;1000$.
The data-generation mechanisms are FIM1 ($\circ$), FIM2 ($\triangle$), RIM1 (+), RIM2 ($\times$), and URIM1 ($\diamond$).
		\label{PlotCovThetamu0andpC01LOR_UMFSsigma01}}
\end{figure}
\begin{figure}[t]
	\centering
	\includegraphics[scale=0.33]{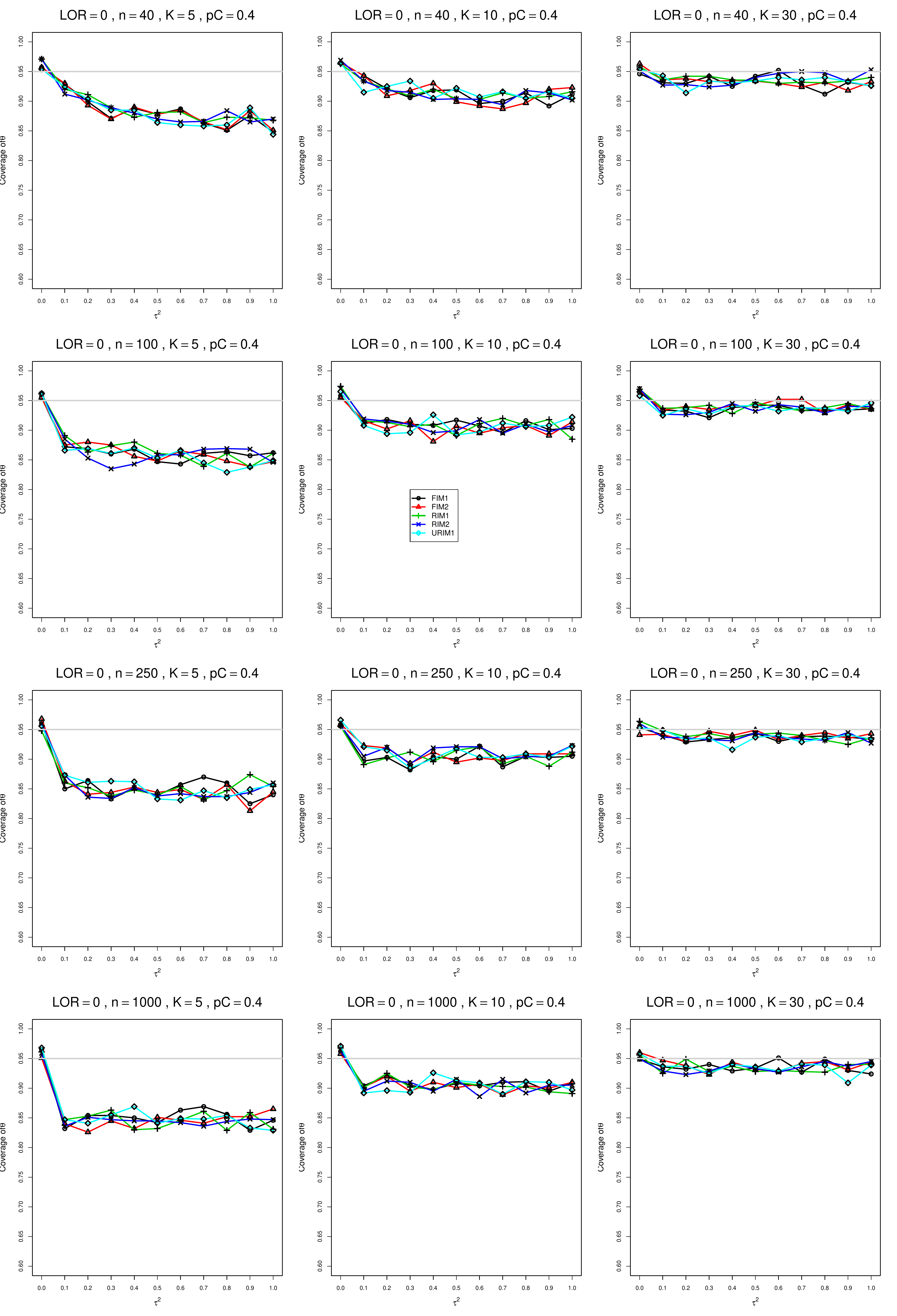}
	\caption{Coverage of the Fixed-intercept with $c=1/2$ confidence intervalfor $\theta=0$, $p_{C}=0.4$, $\sigma^2=0.1$, constant sample sizes $n=40,\;100,\;250,\;1000$.
The data-generation mechanisms are FIM1 ($\circ$), FIM2 ($\triangle$), RIM1 (+), RIM2 ($\times$), and URIM1 ($\diamond$).
		\label{PlotCovThetamu0andpC04LOR_UMFSsigma01}}
\end{figure}
\begin{figure}[t]
	\centering
	\includegraphics[scale=0.33]{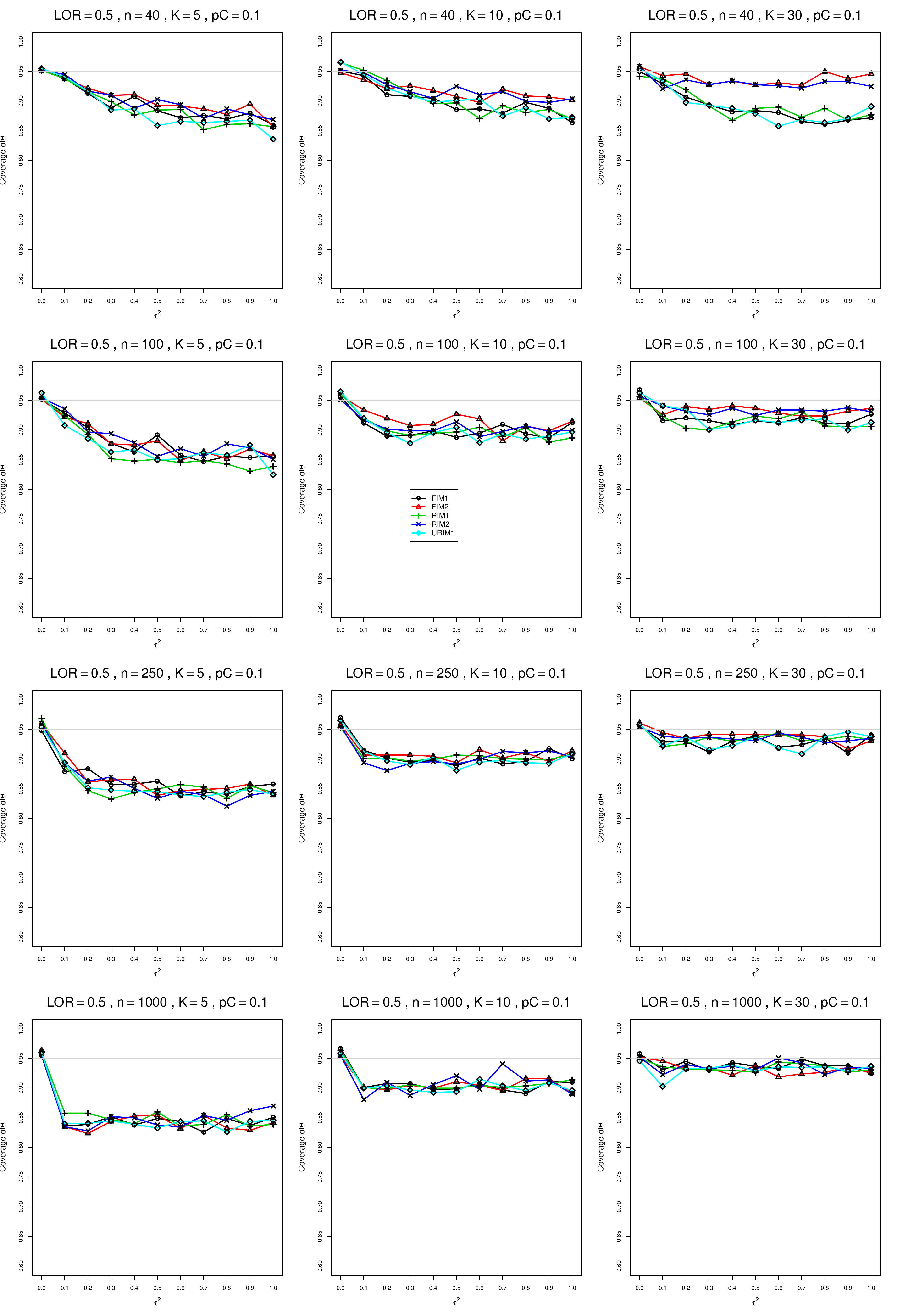}
	\caption{Coverage of the Fixed-intercept with $c=1/2$ confidence intervalfor $\theta=0.5$, $p_{C}=0.1$, $\sigma^2=0.1$, constant sample sizes $n=40,\;100,\;250,\;1000$.
The data-generation mechanisms are FIM1 ($\circ$), FIM2 ($\triangle$), RIM1 (+), RIM2 ($\times$), and URIM1 ($\diamond$).
		\label{PlotCovThetamu05andpC01LOR_UMFSsigma01}}
\end{figure}
\begin{figure}[t]
	\centering
	\includegraphics[scale=0.33]{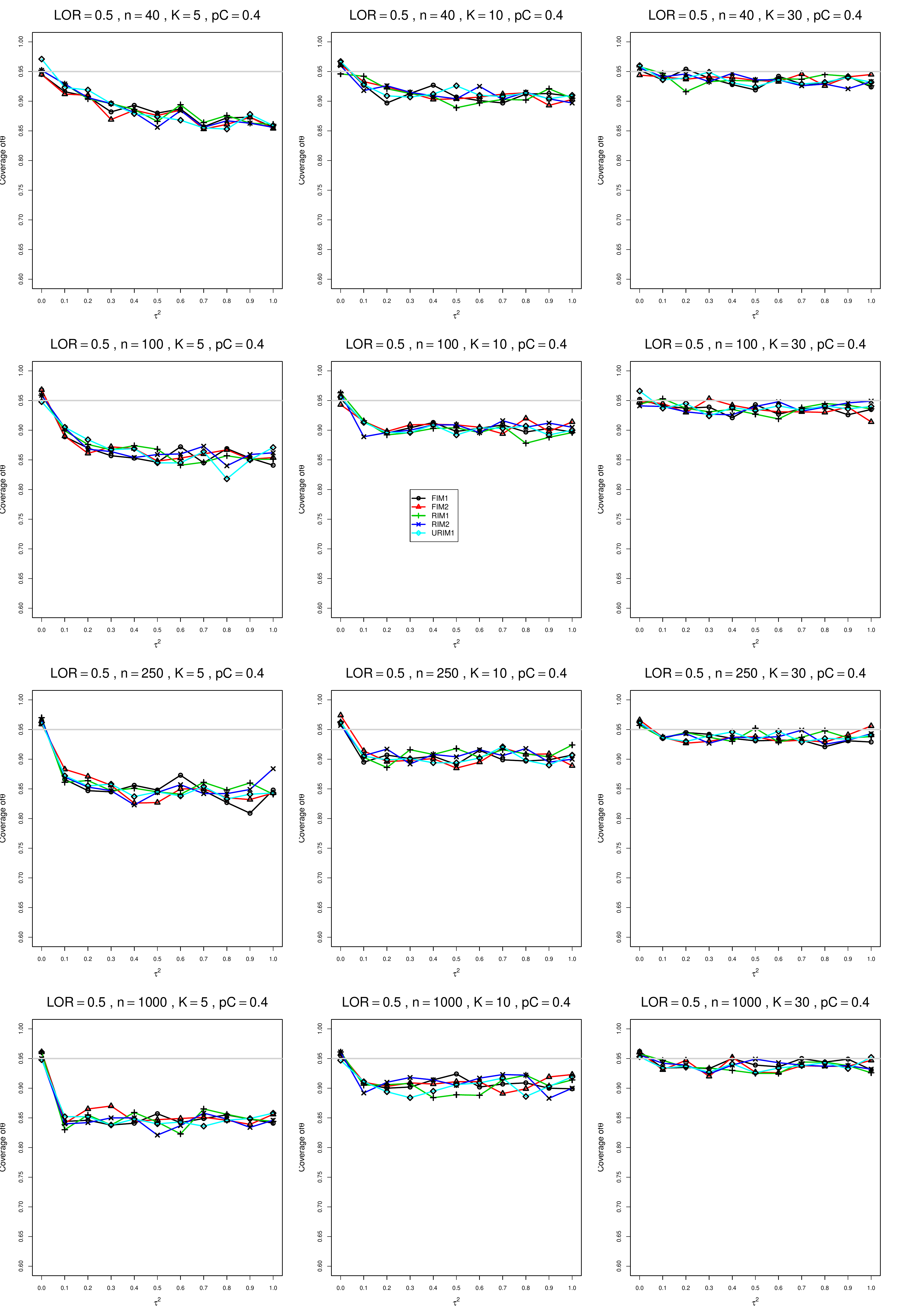}
	\caption{Coverage of the Fixed-intercept with $c=1/2$ confidence intervalfor $\theta=0.5$, $p_{C}=0.4$, $\sigma^2=0.1$, constant sample sizes $n=40,\;100,\;250,\;1000$.
The data-generation mechanisms are FIM1 ($\circ$), FIM2 ($\triangle$), RIM1 (+), RIM2 ($\times$), and URIM1 ($\diamond$).
		\label{PlotCovThetamu05andpC04LOR_UMFSsigma01}}
\end{figure}
\begin{figure}[t]
	\centering
	\includegraphics[scale=0.33]{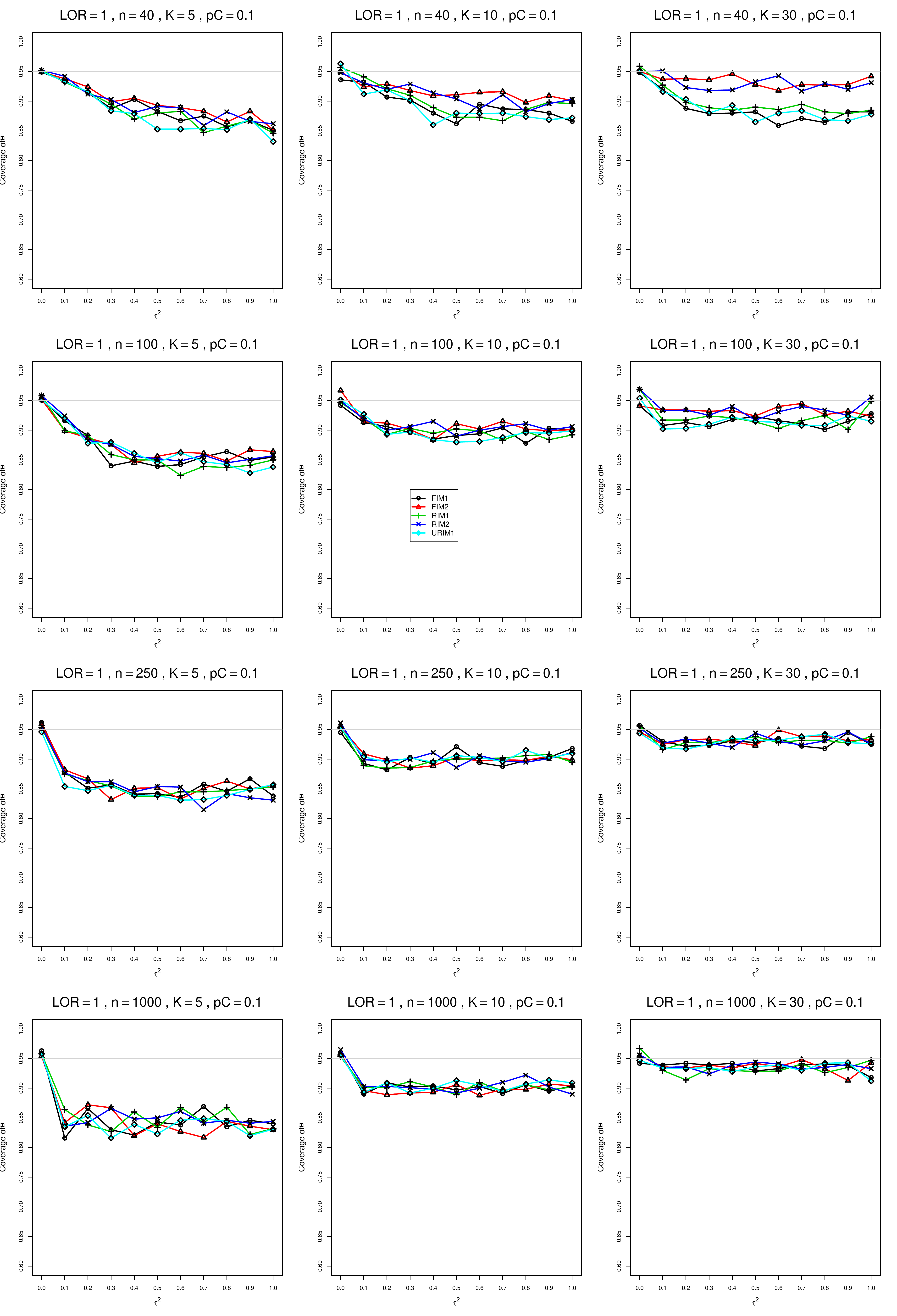}
	\caption{Coverage of the Fixed-intercept with $c=1/2$ confidence intervalfor $\theta=1$, $p_{C}=0.1$, $\sigma^2=0.1$, constant sample sizes $n=40,\;100,\;250,\;1000$.
The data-generation mechanisms are FIM1 ($\circ$), FIM2 ($\triangle$), RIM1 (+), RIM2 ($\times$), and URIM1 ($\diamond$).
		\label{PlotCovThetamu1andpC01LOR_UMFSsigma01}}
\end{figure}
\begin{figure}[t]
	\centering
	\includegraphics[scale=0.33]{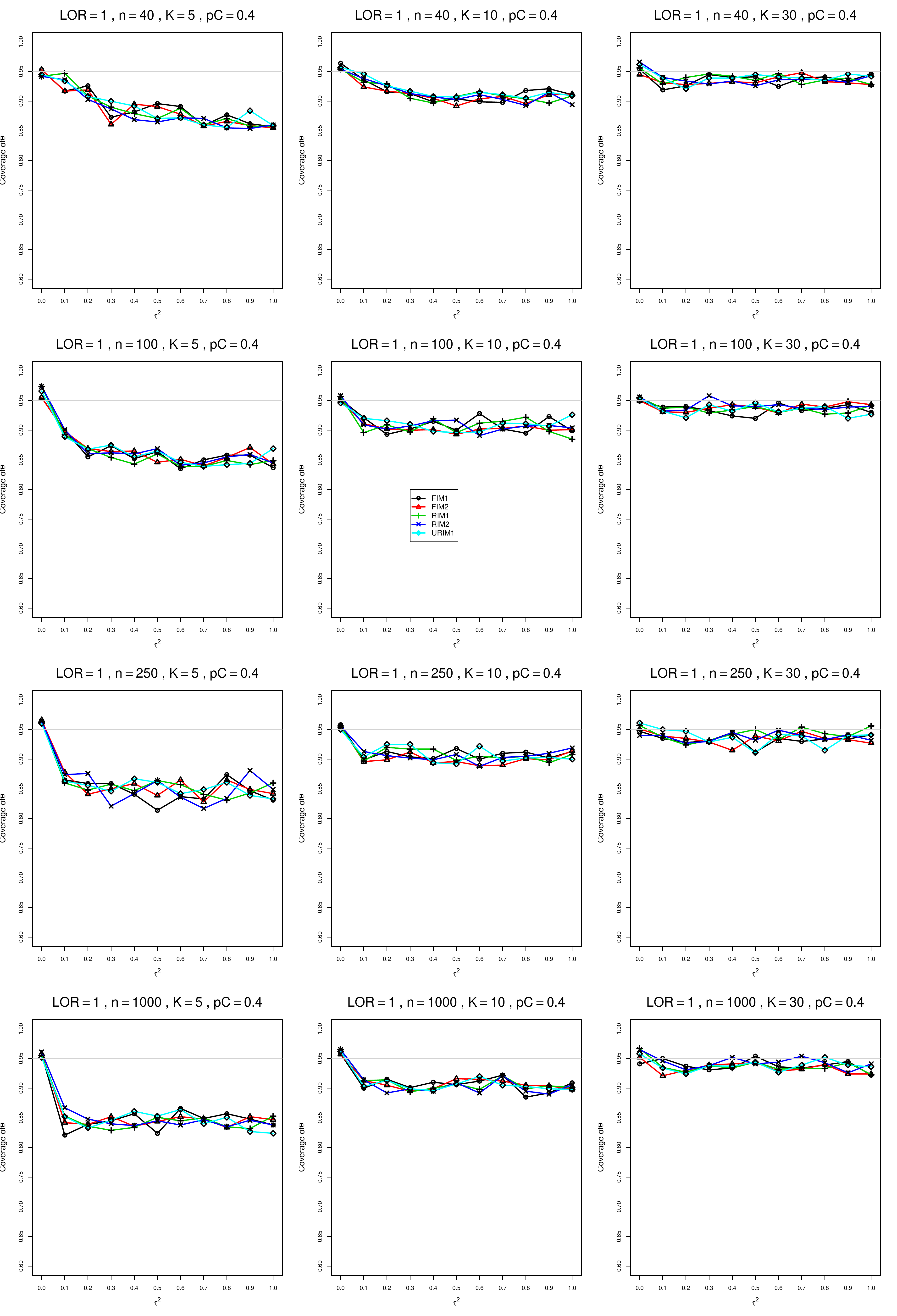}
	\caption{Coverage of the Fixed-intercept with $c=1/2$ confidence intervalfor $\theta=1$, $p_{C}=0.4$, $\sigma^2=0.1$, constant sample sizes $n=40,\;100,\;250,\;1000$.
The data-generation mechanisms are FIM1 ($\circ$), FIM2 ($\triangle$), RIM1 (+), RIM2 ($\times$), and URIM1 ($\diamond$).
		\label{PlotCovThetamu1andpC04LOR_UMFSsigma01}}
\end{figure}
\begin{figure}[t]
	\centering
	\includegraphics[scale=0.33]{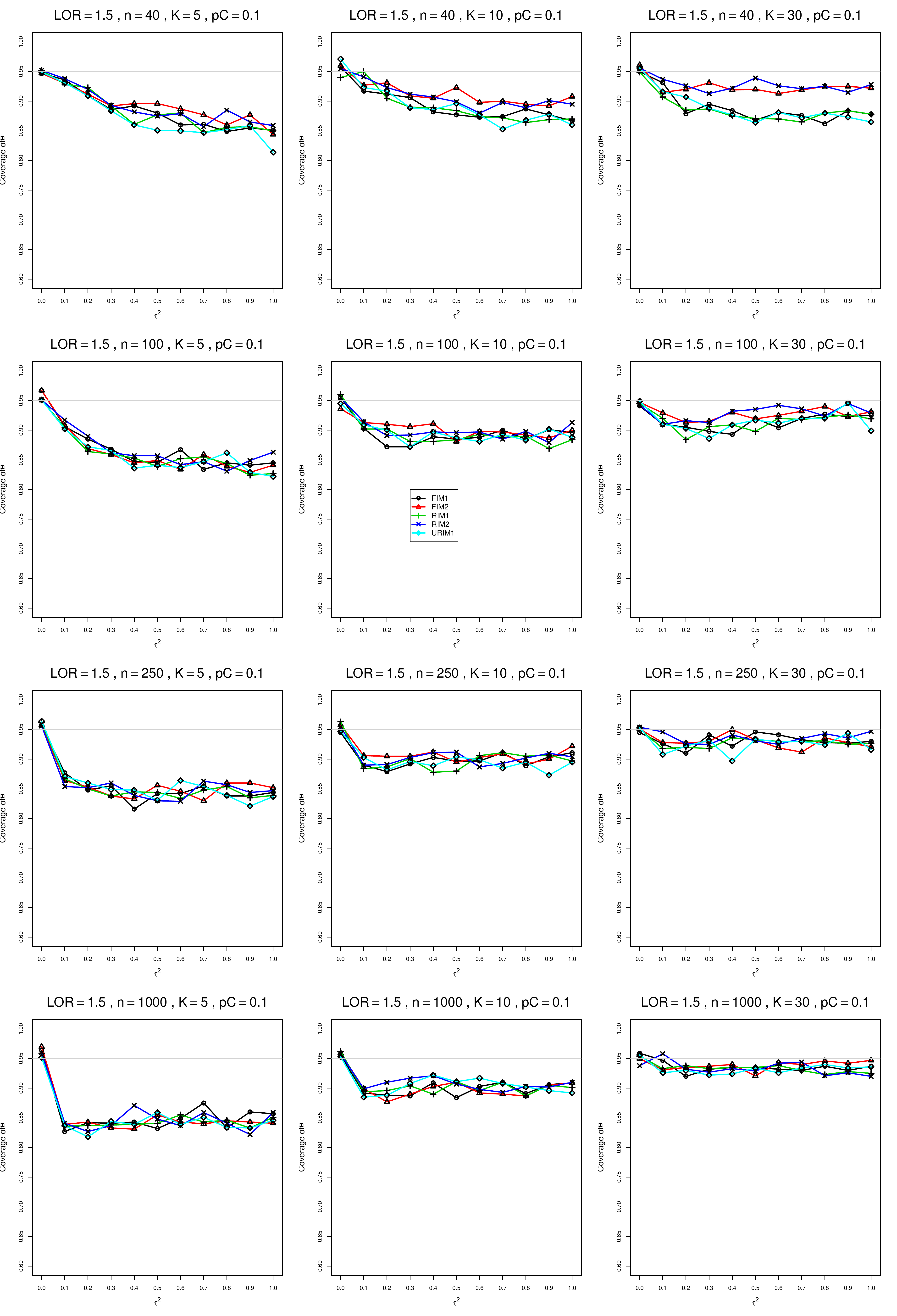}
	\caption{Coverage of the Fixed-intercept with $c=1/2$ confidence intervalfor $\theta=1.5$, $p_{C}=0.1$, $\sigma^2=0.1$, constant sample sizes $n=40,\;100,\;250,\;1000$.
The data-generation mechanisms are FIM1 ($\circ$), FIM2 ($\triangle$), RIM1 (+), RIM2 ($\times$), and URIM1 ($\diamond$).
		\label{PlotCovThetamu15andpC01LOR_UMFSsigma01}}
\end{figure}
\begin{figure}[t]
	\centering
	\includegraphics[scale=0.33]{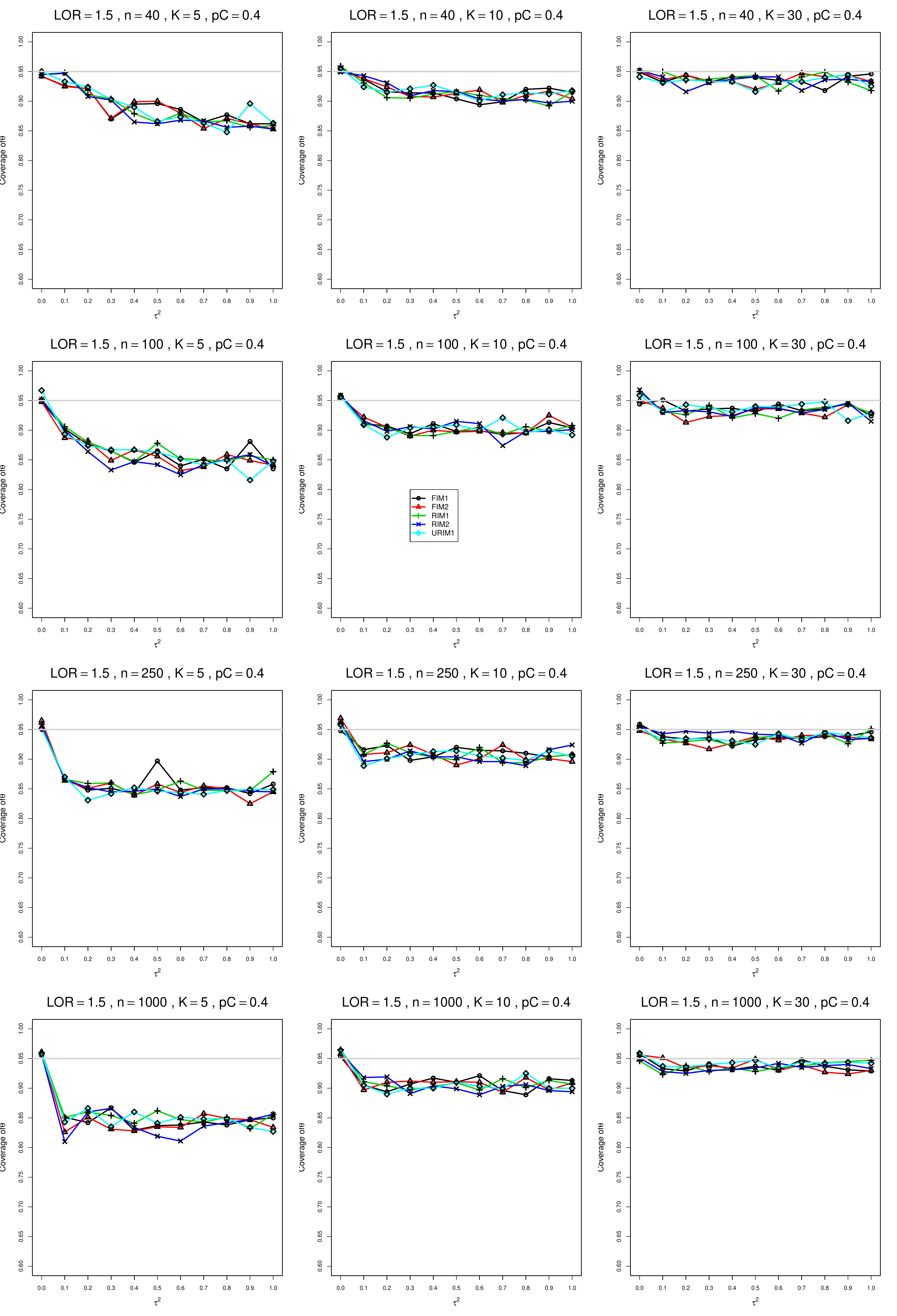}
	\caption{Coverage of the Fixed-intercept with $c=1/2$ confidence intervalfor $\theta=1.5$, $p_{C}=0.4$, $\sigma^2=0.1$, constant sample sizes $n=40,\;100,\;250,\;1000$.
The data-generation mechanisms are FIM1 ($\circ$), FIM2 ($\triangle$), RIM1 (+), RIM2 ($\times$), and URIM1 ($\diamond$).
		\label{PlotCovThetamu15andpC04LOR_UMFSsigma01}}
\end{figure}
\begin{figure}[t]
	\centering
	\includegraphics[scale=0.33]{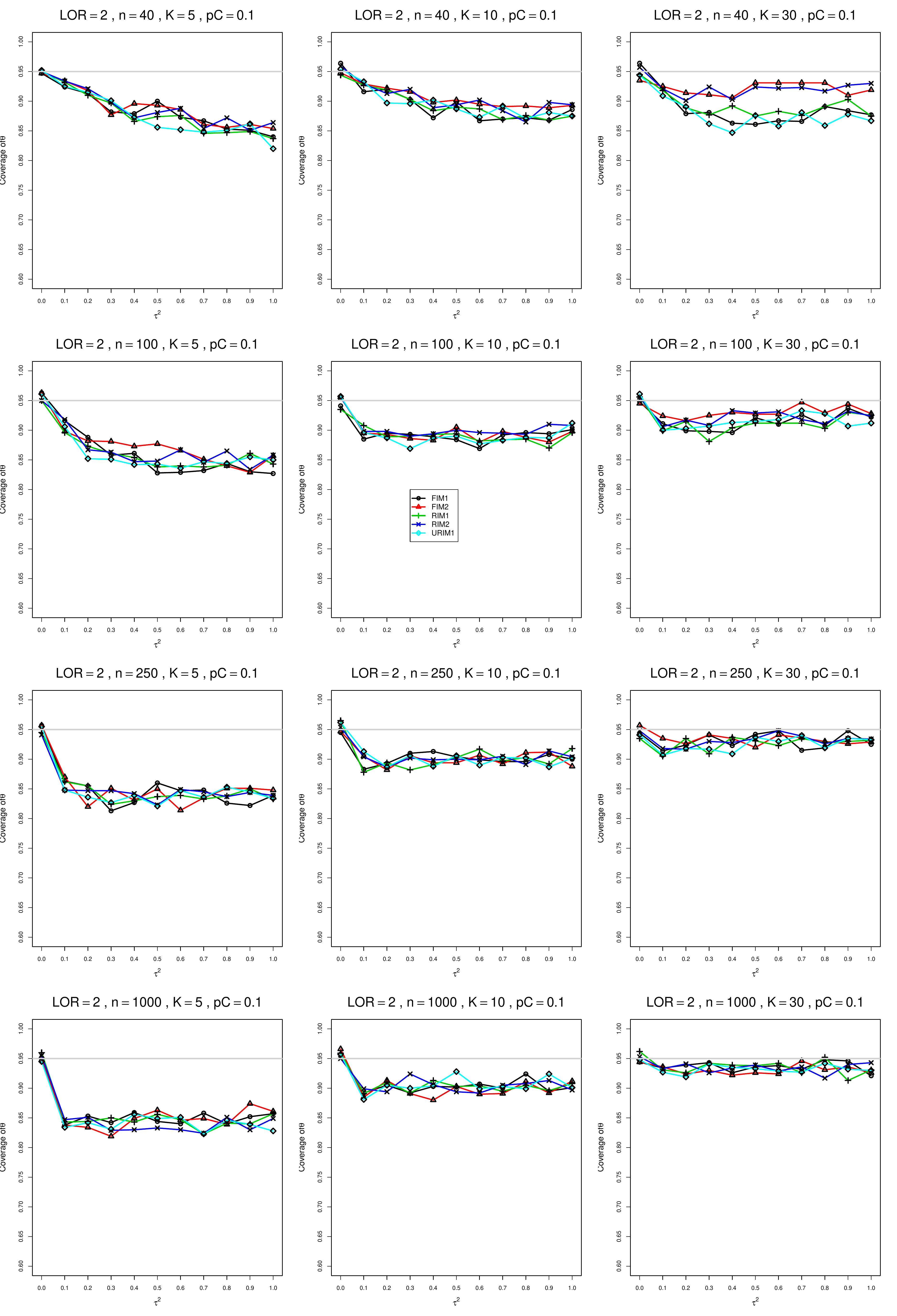}
	\caption{Coverage of the Fixed-intercept with $c=1/2$ confidence intervalfor $\theta=2$, $p_{C}=0.1$, $\sigma^2=0.1$, constant sample sizes $n=40,\;100,\;250,\;1000$.
The data-generation mechanisms are FIM1 ($\circ$), FIM2 ($\triangle$), RIM1 (+), RIM2 ($\times$), and URIM1 ($\diamond$).
		\label{PlotCovThetamu2andpC01LOR_UMFSsigma01}}
\end{figure}
\begin{figure}[t]
	\centering
	\includegraphics[scale=0.33]{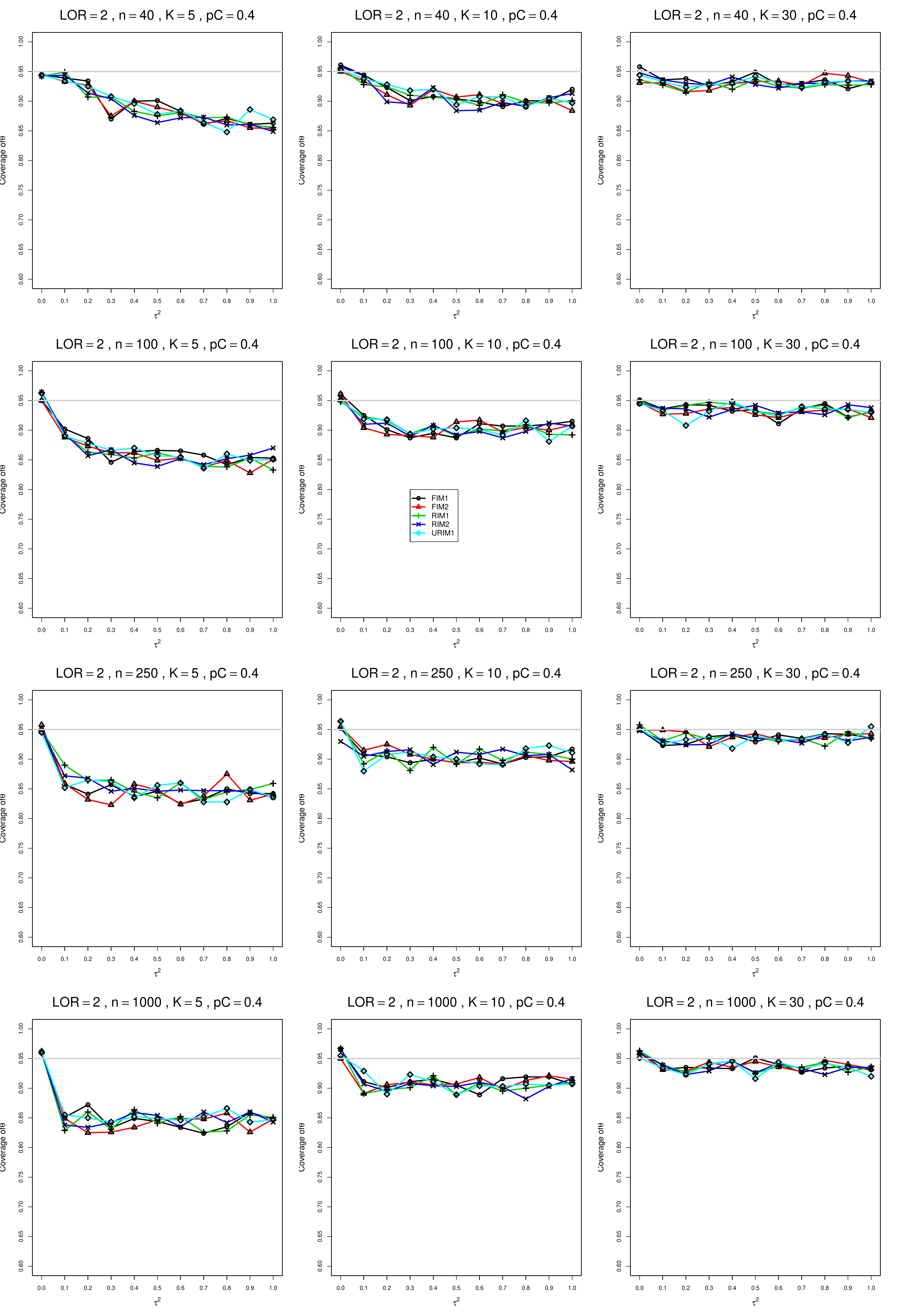}
	\caption{Coverage of the Fixed-intercept with $c=1/2$ confidence intervalfor $\theta=2$, $p_{C}=0.4$, $\sigma^2=0.1$, constant sample sizes $n=40,\;100,\;250,\;1000$.
The data-generation mechanisms are FIM1 ($\circ$), FIM2 ($\triangle$), RIM1 (+), RIM2 ($\times$), and URIM1 ($\diamond$).
		\label{PlotCovThetamu2andpC04LOR_UMFSsigma01}}
\end{figure}
\begin{figure}[t]
	\centering
	\includegraphics[scale=0.33]{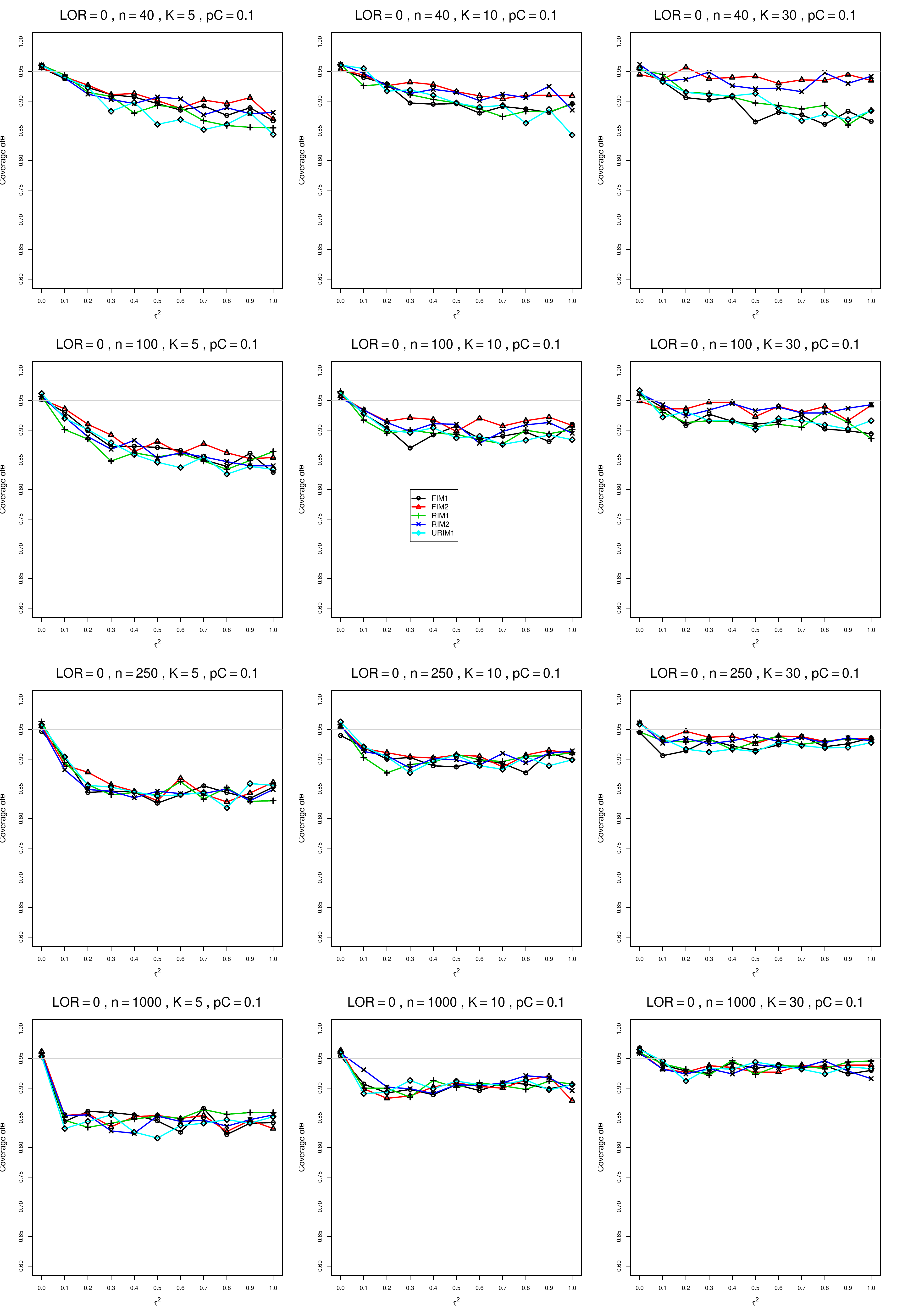}
	\caption{Coverage of the Fixed-intercept with $c=1/2$ confidence intervalfor $\theta=0$, $p_{C}=0.1$, $\sigma^2=0.4$, constant sample sizes $n=40,\;100,\;250,\;1000$.
The data-generation mechanisms are FIM1 ($\circ$), FIM2 ($\triangle$), RIM1 (+), RIM2 ($\times$), and URIM1 ($\diamond$).
		\label{PlotCovThetamu0andpC01LOR_UMFSsigma04}}
\end{figure}
\begin{figure}[t]
	\centering
	\includegraphics[scale=0.33]{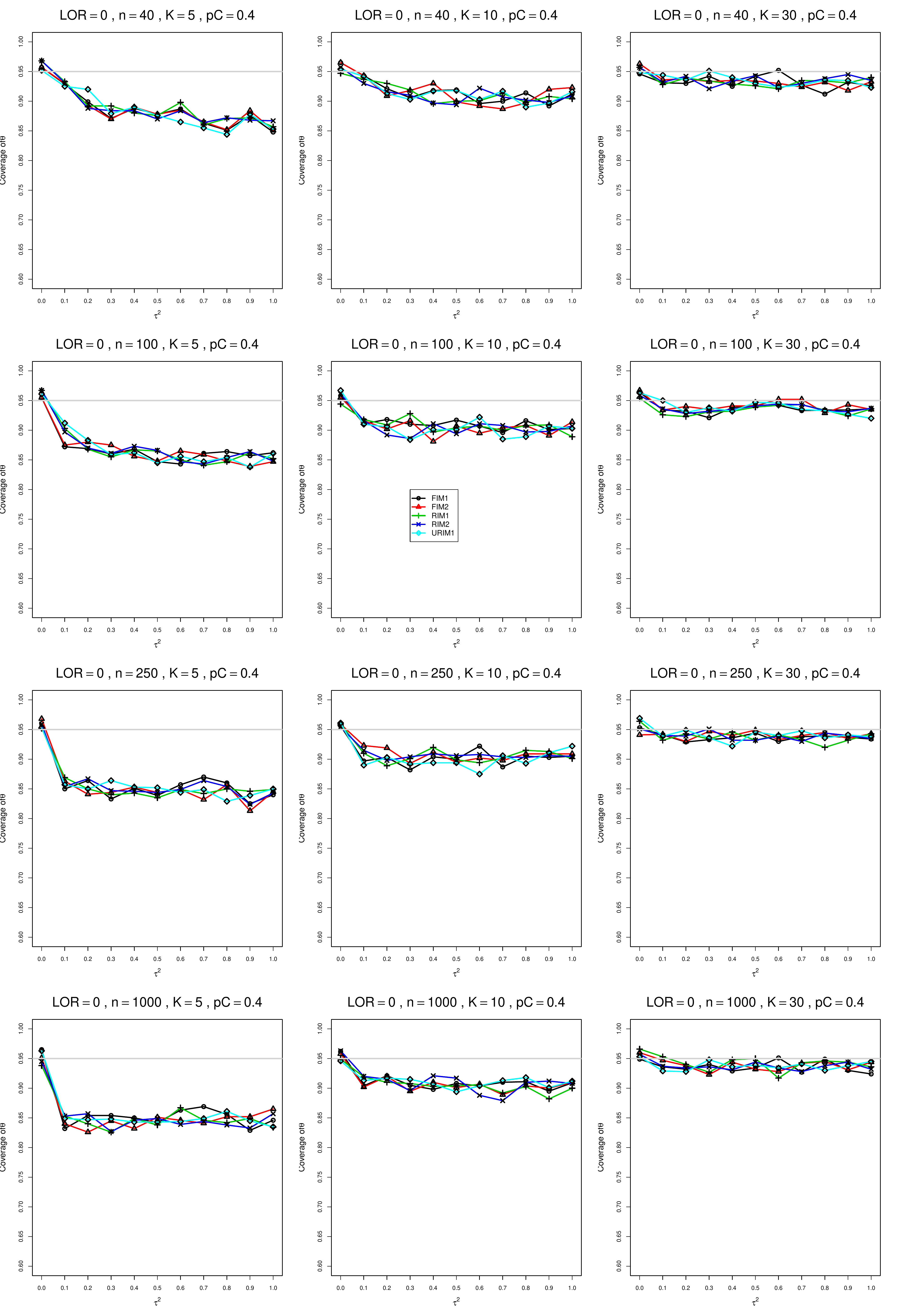}
	\caption{Coverage of the Fixed-intercept with $c=1/2$ confidence intervalfor $\theta=0$, $p_{C}=0.4$, $\sigma^2=0.4$, constant sample sizes $n=40,\;100,\;250,\;1000$.
The data-generation mechanisms are FIM1 ($\circ$), FIM2 ($\triangle$), RIM1 (+), RIM2 ($\times$), and URIM1 ($\diamond$).
		\label{PlotCovThetamu0andpC04LOR_UMFSsigma04}}
\end{figure}
\begin{figure}[t]
	\centering
	\includegraphics[scale=0.33]{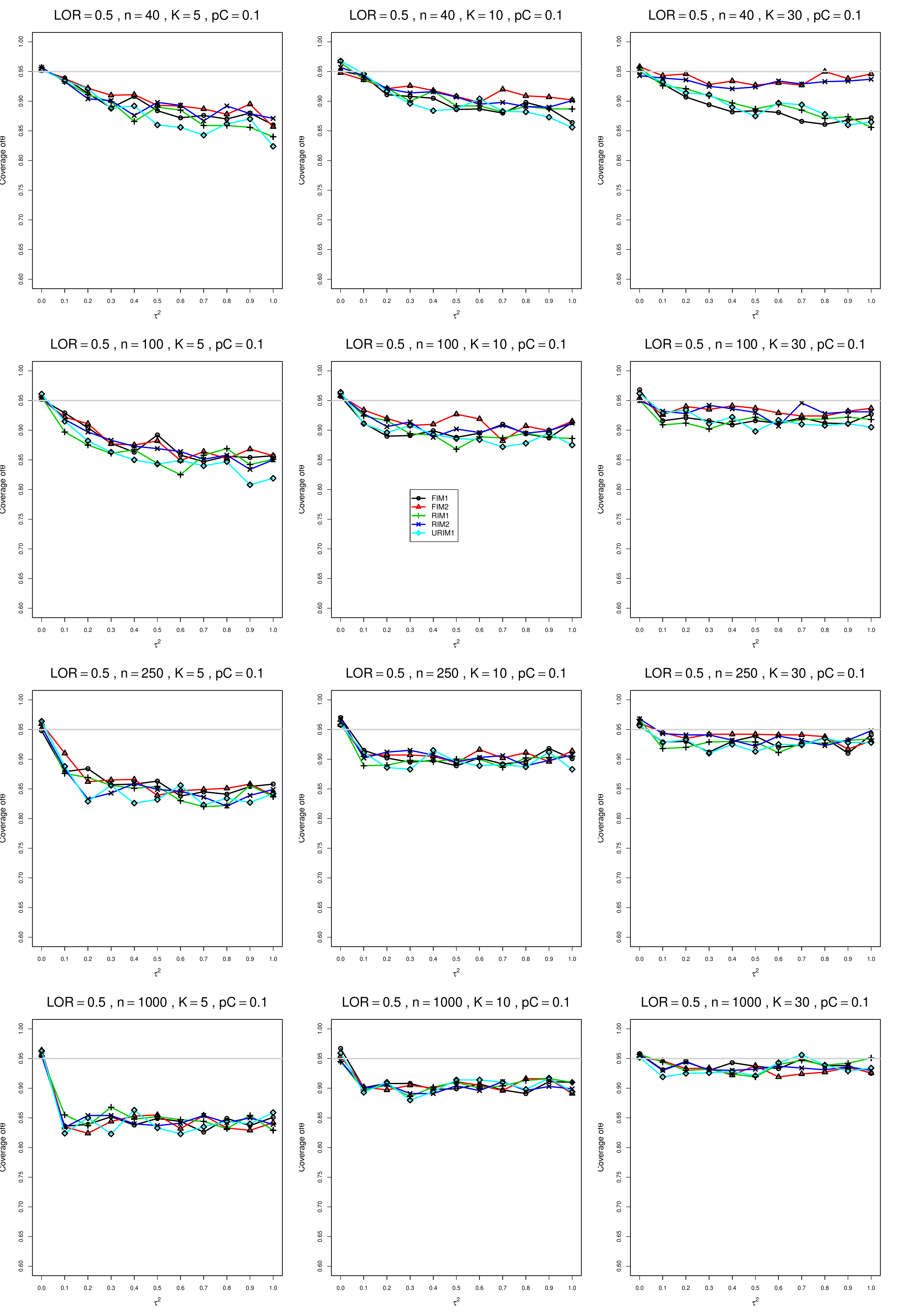}
	\caption{Coverage of the Fixed-intercept with $c=1/2$ confidence intervalfor $\theta=0.5$, $p_{C}=0.1$, $\sigma^2=0.4$, constant sample sizes $n=40,\;100,\;250,\;1000$.
The data-generation mechanisms are FIM1 ($\circ$), FIM2 ($\triangle$), RIM1 (+), RIM2 ($\times$), and URIM1 ($\diamond$).
		\label{PlotCovThetamu05andpC01LOR_UMFSsigma04}}
\end{figure}
\begin{figure}[t]
	\centering
	\includegraphics[scale=0.33]{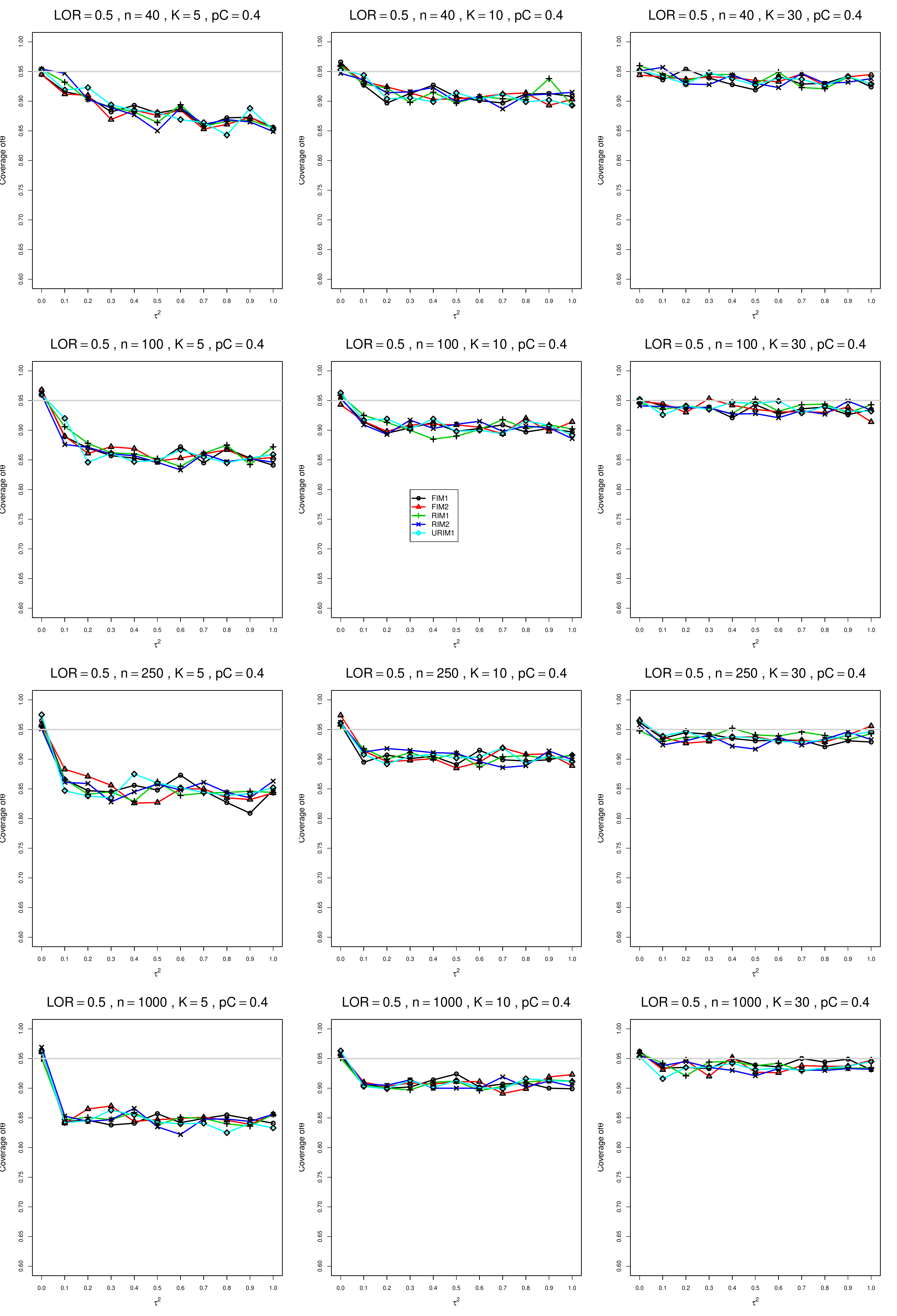}
	\caption{Coverage of the Fixed-intercept with $c=1/2$ confidence intervalfor $\theta=0.5$, $p_{C}=0.4$, $\sigma^2=0.4$, constant sample sizes $n=40,\;100,\;250,\;1000$.
The data-generation mechanisms are FIM1 ($\circ$), FIM2 ($\triangle$), RIM1 (+), RIM2 ($\times$), and URIM1 ($\diamond$).
		\label{PlotCovThetamu05andpC04LOR_UMFSsigma04}}
\end{figure}
\begin{figure}[t]
	\centering
	\includegraphics[scale=0.33]{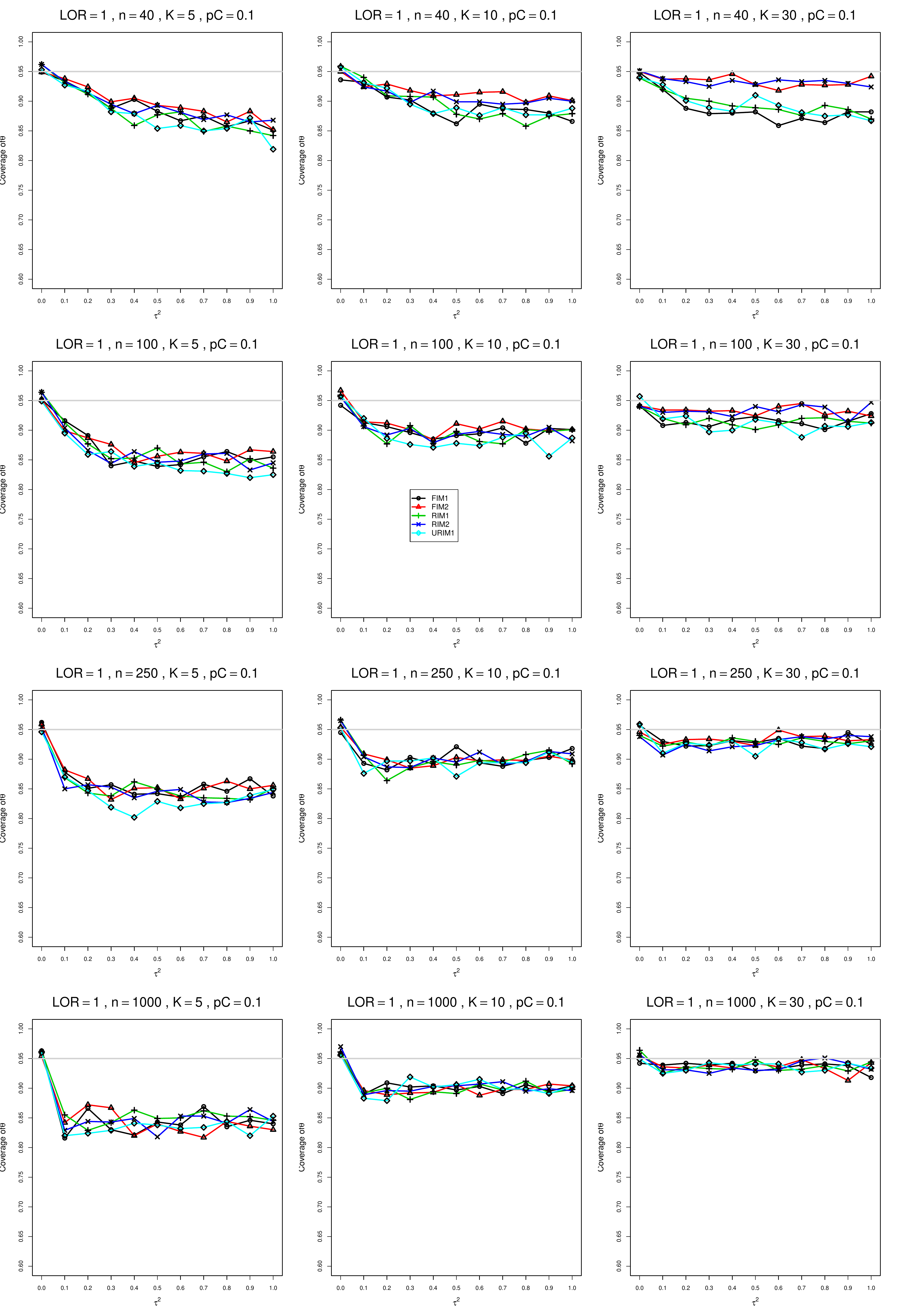}
	\caption{Coverage of the Fixed-intercept with $c=1/2$ confidence intervalfor $\theta=1$, $p_{C}=0.1$, $\sigma^2=0.4$, constant sample sizes $n=40,\;100,\;250,\;1000$.
The data-generation mechanisms are FIM1 ($\circ$), FIM2 ($\triangle$), RIM1 (+), RIM2 ($\times$), and URIM1 ($\diamond$).
		\label{PlotCovThetamu1andpC01LOR_UMFSsigma04}}
\end{figure}
\begin{figure}[t]
	\centering
	\includegraphics[scale=0.33]{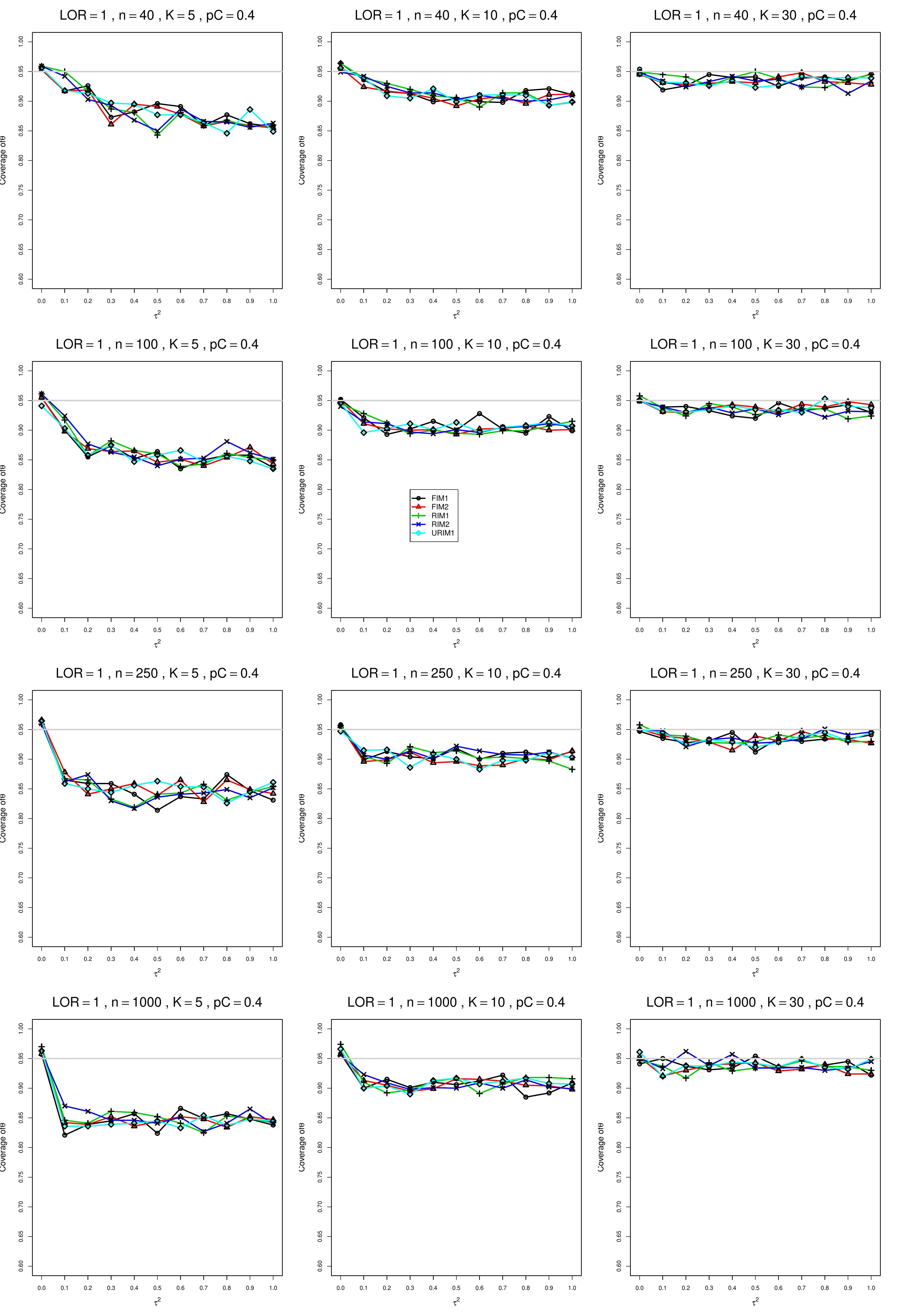}
	\caption{Coverage of the Fixed-intercept with $c=1/2$ confidence intervalfor $\theta=1$, $p_{C}=0.4$, $\sigma^2=0.4$, constant sample sizes $n=40,\;100,\;250,\;1000$.
The data-generation mechanisms are FIM1 ($\circ$), FIM2 ($\triangle$), RIM1 (+), RIM2 ($\times$), and URIM1 ($\diamond$).
		\label{PlotCovThetamu1andpC04LOR_UMFSsigma04}}
\end{figure}
\begin{figure}[t]
	\centering
	\includegraphics[scale=0.33]{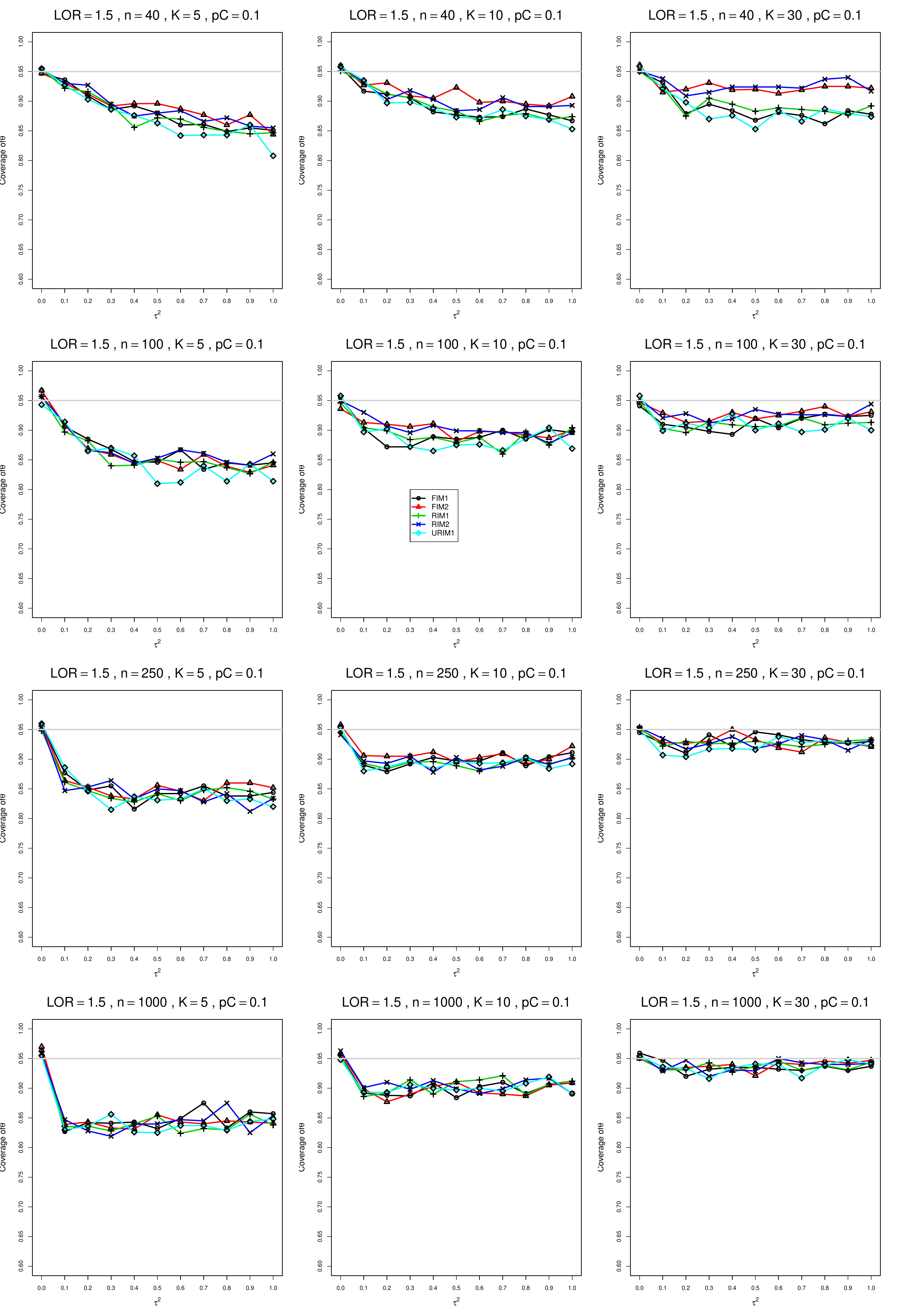}
	\caption{Coverage of the Fixed-intercept with $c=1/2$ confidence intervalfor $\theta=1.5$, $p_{C}=0.1$, $\sigma^2=0.4$, constant sample sizes $n=40,\;100,\;250,\;1000$.
The data-generation mechanisms are FIM1 ($\circ$), FIM2 ($\triangle$), RIM1 (+), RIM2 ($\times$), and URIM1 ($\diamond$).
		\label{PlotCovThetamu15andpC01LOR_UMFSsigma04}}
\end{figure}
\begin{figure}[t]
	\centering
	\includegraphics[scale=0.33]{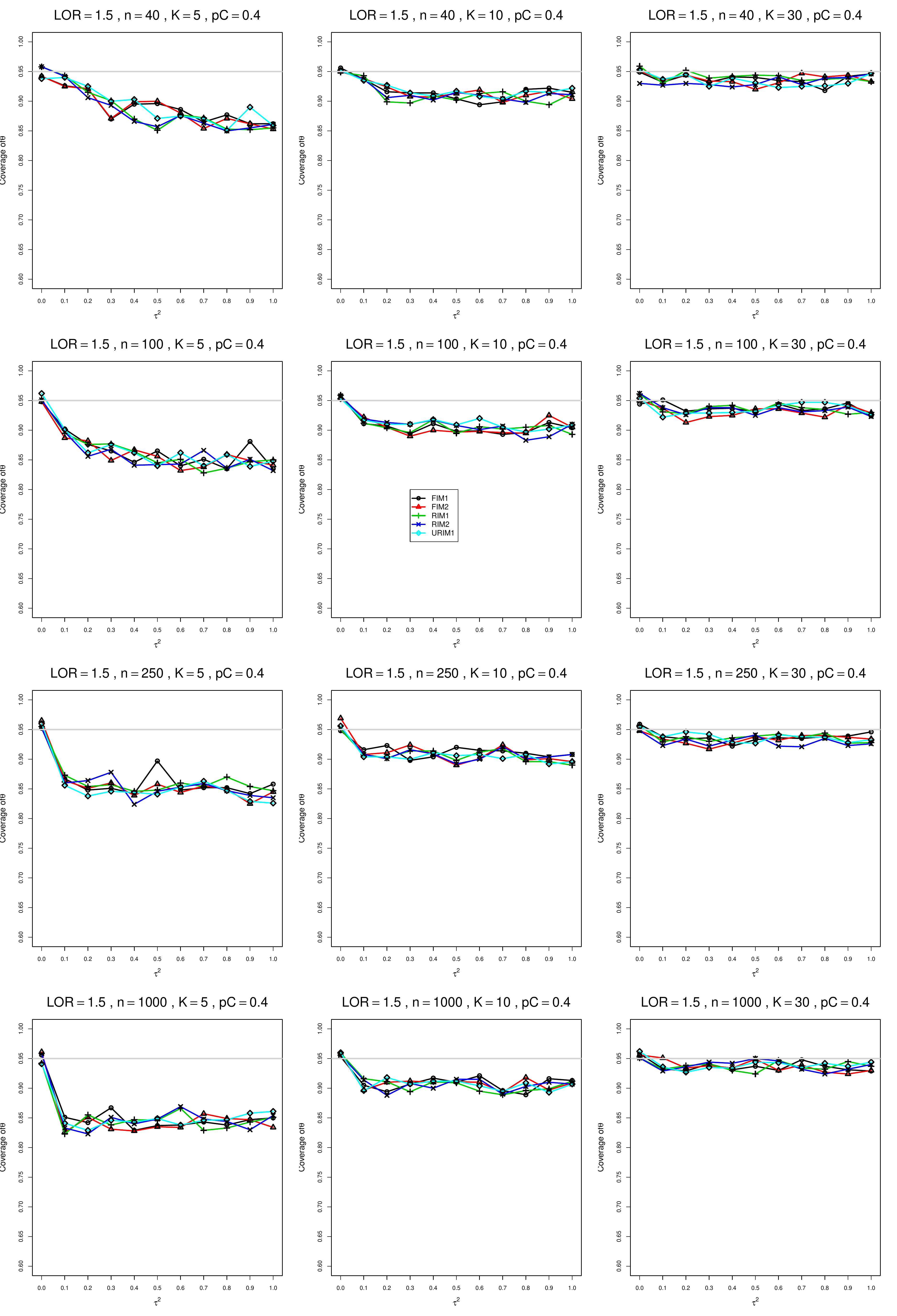}
	\caption{Coverage of the Fixed-intercept with $c=1/2$ confidence intervalfor $\theta=1.5$, $p_{C}=0.4$, $\sigma^2=0.4$, constant sample sizes $n=40,\;100,\;250,\;1000$.
The data-generation mechanisms are FIM1 ($\circ$), FIM2 ($\triangle$), RIM1 (+), RIM2 ($\times$), and URIM1 ($\diamond$).
		\label{PlotCovThetamu15andpC04LOR_UMFSsigma04}}
\end{figure}
\begin{figure}[t]
	\centering
	\includegraphics[scale=0.33]{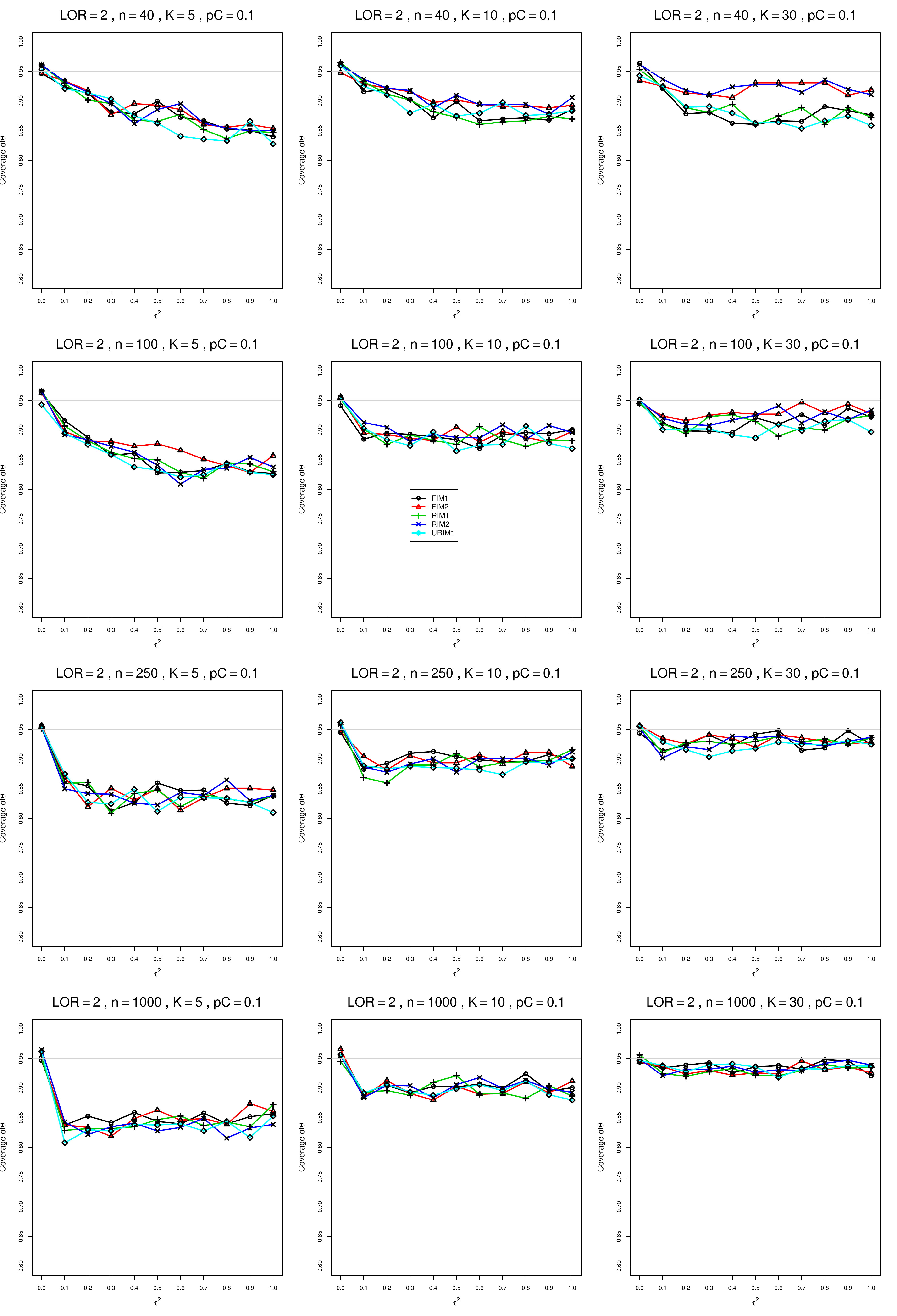}
	\caption{Coverage of the Fixed-intercept with $c=1/2$ confidence intervalfor $\theta=2$, $p_{C}=0.1$, $\sigma^2=0.4$, constant sample sizes $n=40,\;100,\;250,\;1000$.
The data-generation mechanisms are FIM1 ($\circ$), FIM2 ($\triangle$), RIM1 (+), RIM2 ($\times$), and URIM1 ($\diamond$).
		\label{PlotCovThetamu2andpC01LOR_UMFSsigma04}}
\end{figure}
\begin{figure}[t]
	\centering
	\includegraphics[scale=0.33]{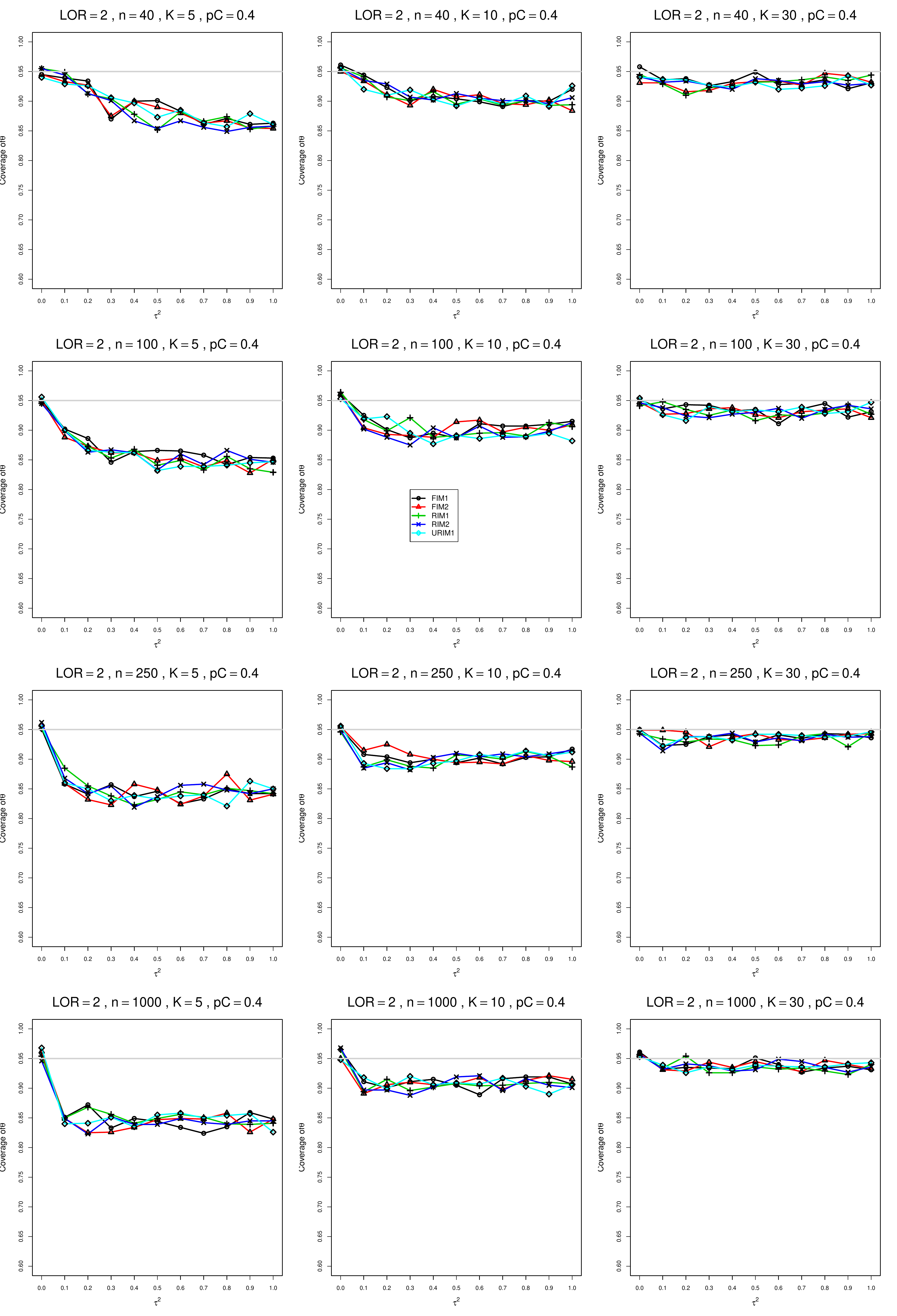}
	\caption{Coverage of the Fixed-intercept with $c=1/2$ confidence intervalfor $\theta=2$, $p_{C}=0.4$, $\sigma^2=0.4$, constant sample sizes $n=40,\;100,\;250,\;1000$.
The data-generation mechanisms are FIM1 ($\circ$), FIM2 ($\triangle$), RIM1 (+), RIM2 ($\times$), and URIM1 ($\diamond$).
		\label{PlotCovThetamu2andpC04LOR_UMFSsigma04}}
\end{figure}

\clearpage
\subsection*{A3.6 Coverage of $\hat{\theta}_{RIM2}$}
\renewcommand{\thefigure}{A3.6.\arabic{figure}}
\setcounter{figure}{0}

\begin{figure}[t]
	\centering
	\includegraphics[scale=0.33]{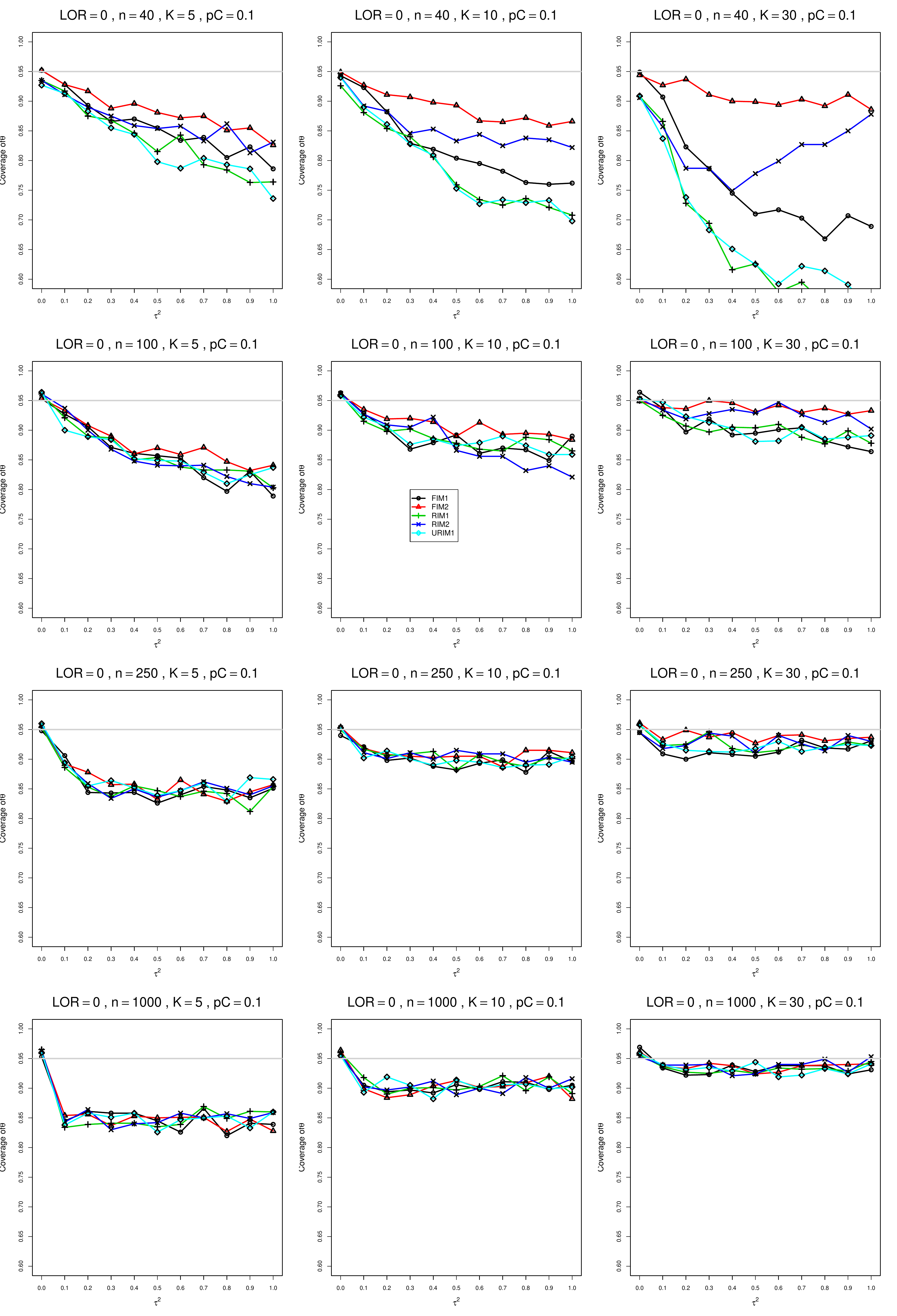}
	\caption{Coverage of the Random-intercept with $c=1/2$ confidence interval for $\theta=0$, $p_{C}=0.1$, $\sigma^2=0.1$, constant sample sizes $n=40,\;100,\;250,\;1000$.
The data-generation mechanisms are FIM1 ($\circ$), FIM2 ($\triangle$), RIM1 (+), RIM2 ($\times$), and URIM1 ($\diamond$).
		\label{PlotCovThetamu0andpC01LOR_UMRSsigma01}}
\end{figure}
\begin{figure}[t]
	\centering
	\includegraphics[scale=0.33]{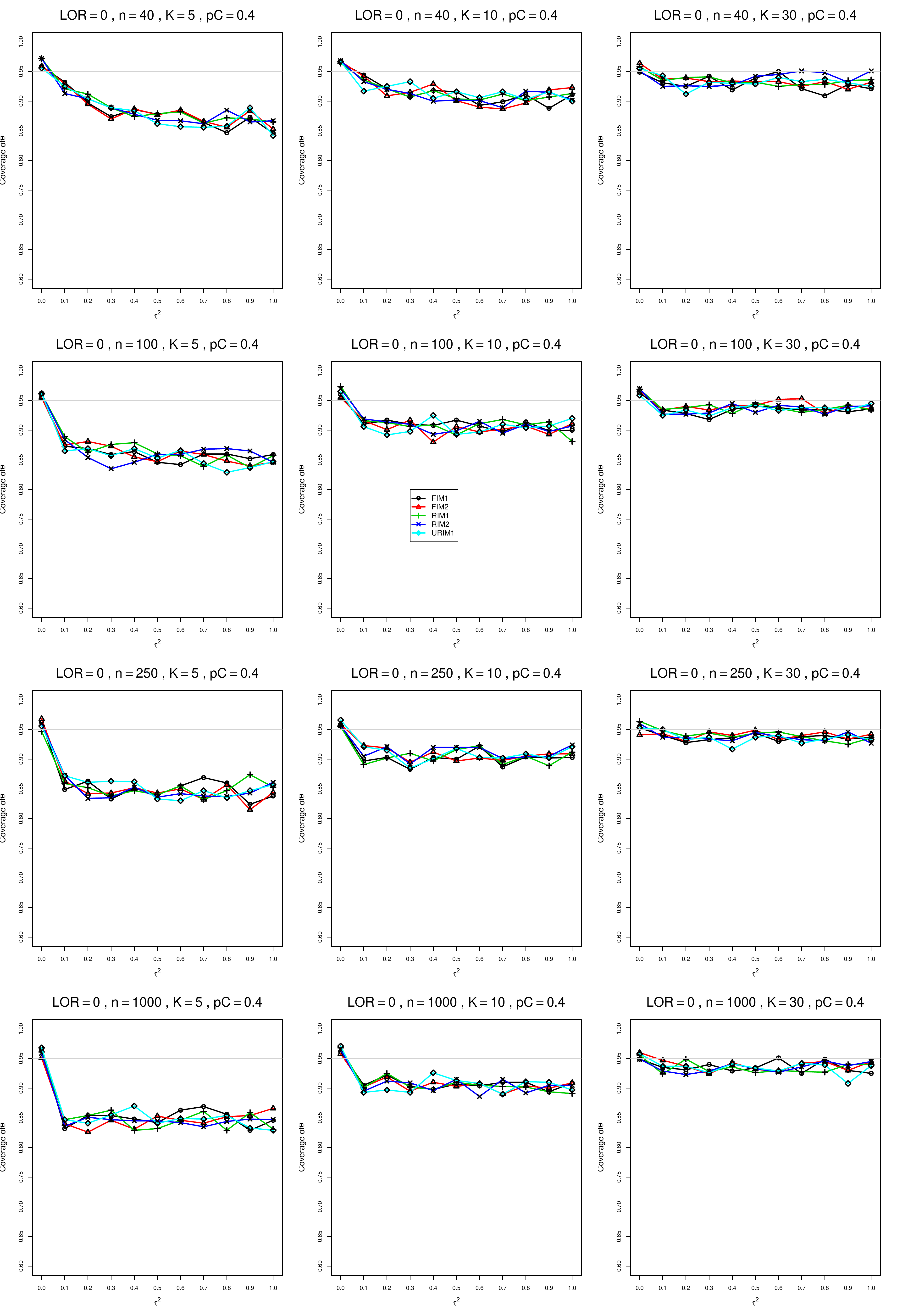}
	\caption{Coverage of the Random-intercept with $c=1/2$ confidence interval for $\theta=0$, $p_{C}=0.4$, $\sigma^2=0.1$, constant sample sizes $n=40,\;100,\;250,\;1000$.
The data-generation mechanisms are FIM1 ($\circ$), FIM2 ($\triangle$), RIM1 (+), RIM2 ($\times$), and URIM1 ($\diamond$).
		\label{PlotCovThetamu0andpC04LOR_UMRSsigma01}}
\end{figure}
\begin{figure}[t]
	\centering
	\includegraphics[scale=0.33]{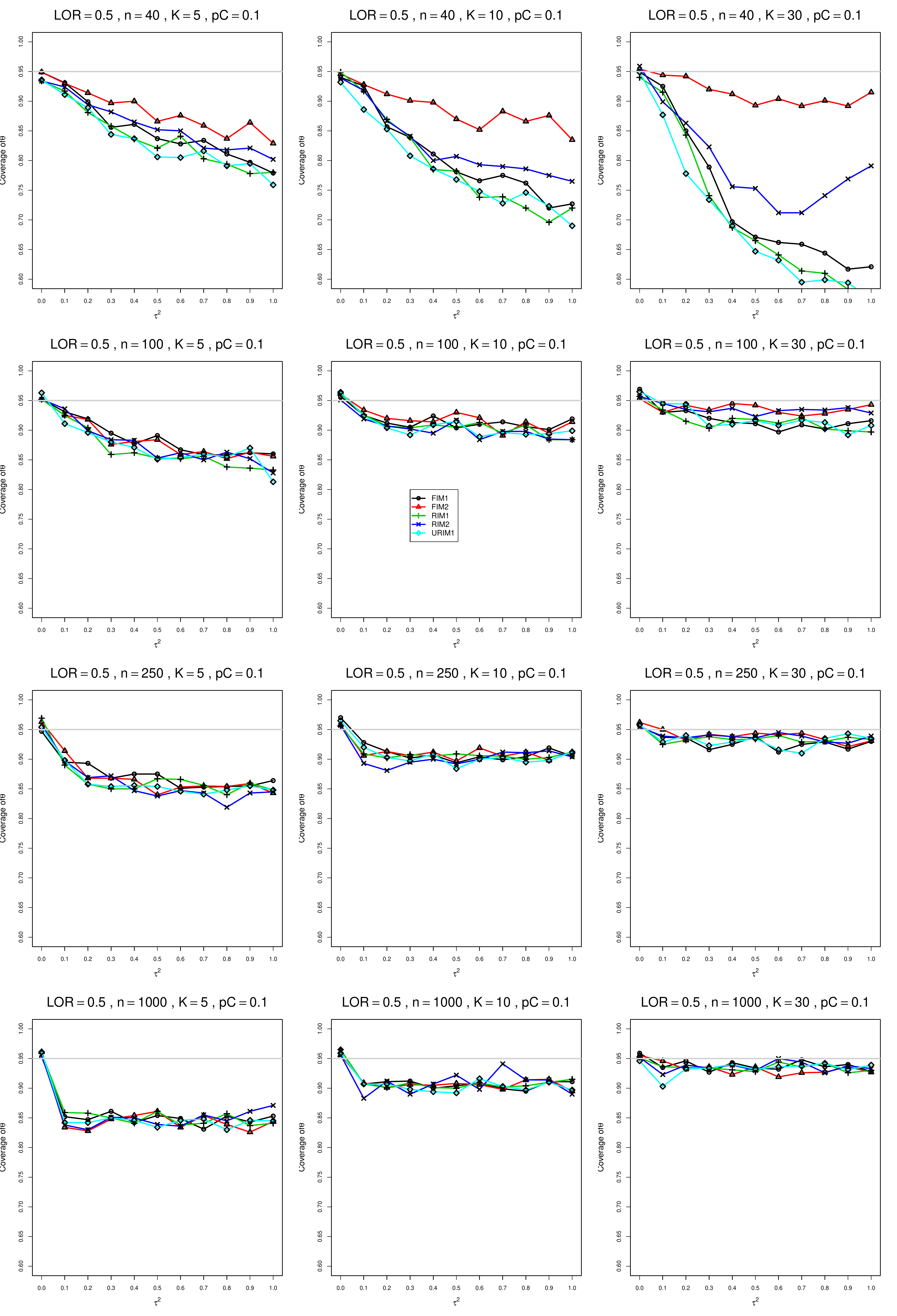}
	\caption{Coverage of the Random-intercept with $c=1/2$ confidence interval for $\theta=0.5$, $p_{C}=0.1$, $\sigma^2=0.1$, constant sample sizes $n=40,\;100,\;250,\;1000$.
The data-generation mechanisms are FIM1 ($\circ$), FIM2 ($\triangle$), RIM1 (+), RIM2 ($\times$), and URIM1 ($\diamond$).
		\label{PlotCovThetamu05andpC01LOR_UMRSsigma01}}
\end{figure}
\begin{figure}[t]
	\centering
	\includegraphics[scale=0.33]{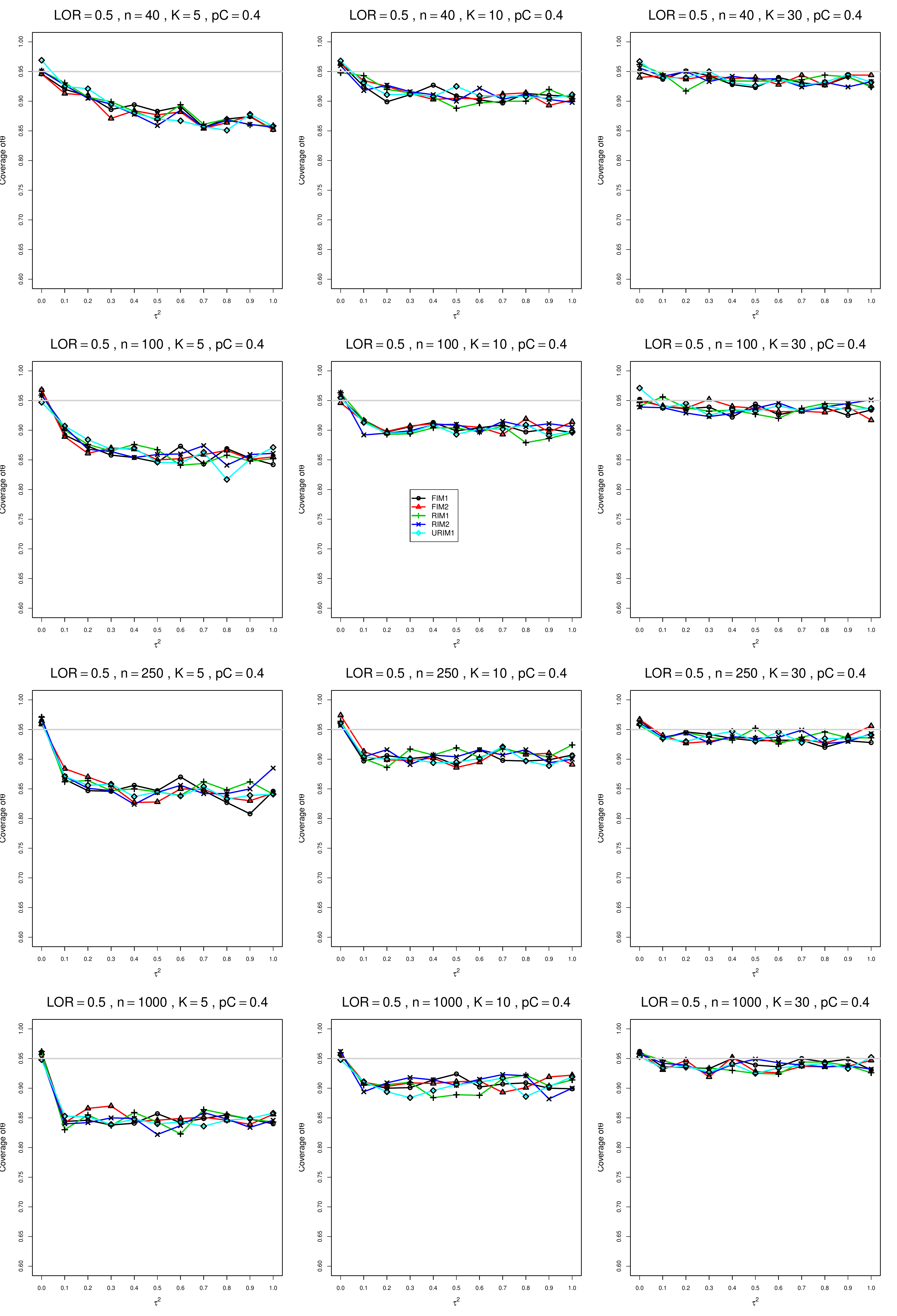}
	\caption{Coverage of the Random-intercept with $c=1/2$ confidence interval for $\theta=0.5$, $p_{C}=0.4$, $\sigma^2=0.1$, constant sample sizes $n=40,\;100,\;250,\;1000$.
The data-generation mechanisms are FIM1 ($\circ$), FIM2 ($\triangle$), RIM1 (+), RIM2 ($\times$), and URIM1 ($\diamond$).
		\label{PlotCovThetamu05andpC04LOR_UMRSsigma01}}
\end{figure}
\begin{figure}[t]
	\centering
	\includegraphics[scale=0.33]{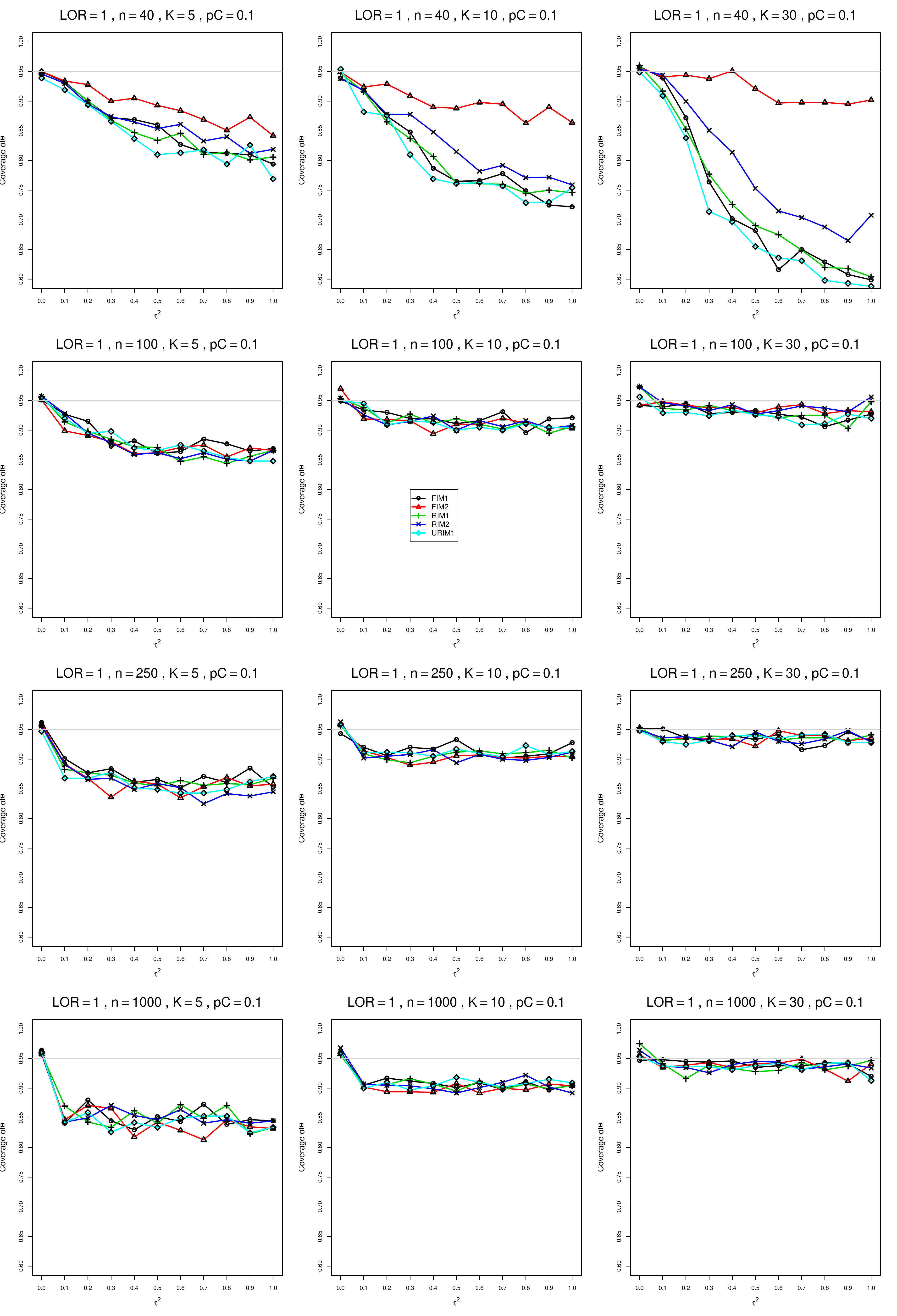}
	\caption{Coverage of the Random-intercept with $c=1/2$ confidence interval for $\theta=1$, $p_{C}=0.1$, $\sigma^2=0.1$, constant sample sizes $n=40,\;100,\;250,\;1000$.
The data-generation mechanisms are FIM1 ($\circ$), FIM2 ($\triangle$), RIM1 (+), RIM2 ($\times$), and URIM1 ($\diamond$).
		\label{PlotCovThetamu1andpC01LOR_UMRSsigma01}}
\end{figure}
\begin{figure}[t]
	\centering
	\includegraphics[scale=0.33]{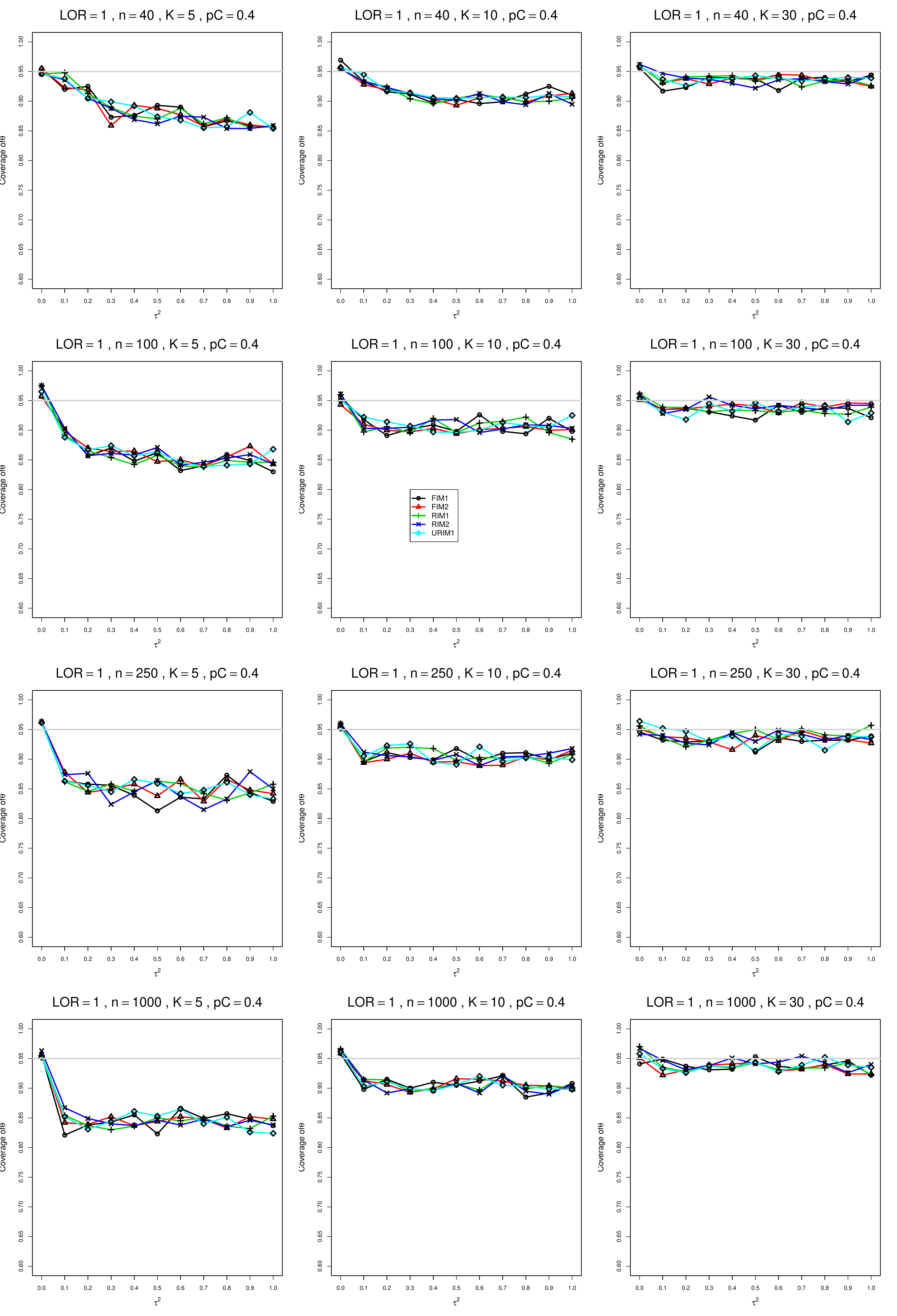}
	\caption{Coverage of the Random-intercept with $c=1/2$ confidence interval for $\theta=1$, $p_{C}=0.4$, $\sigma^2=0.1$, constant sample sizes $n=40,\;100,\;250,\;1000$.
The data-generation mechanisms are FIM1 ($\circ$), FIM2 ($\triangle$), RIM1 (+), RIM2 ($\times$), and URIM1 ($\diamond$).
		\label{PlotCovThetamu1andpC04LOR_UMRSsigma01}}
\end{figure}
\begin{figure}[t]
	\centering
	\includegraphics[scale=0.33]{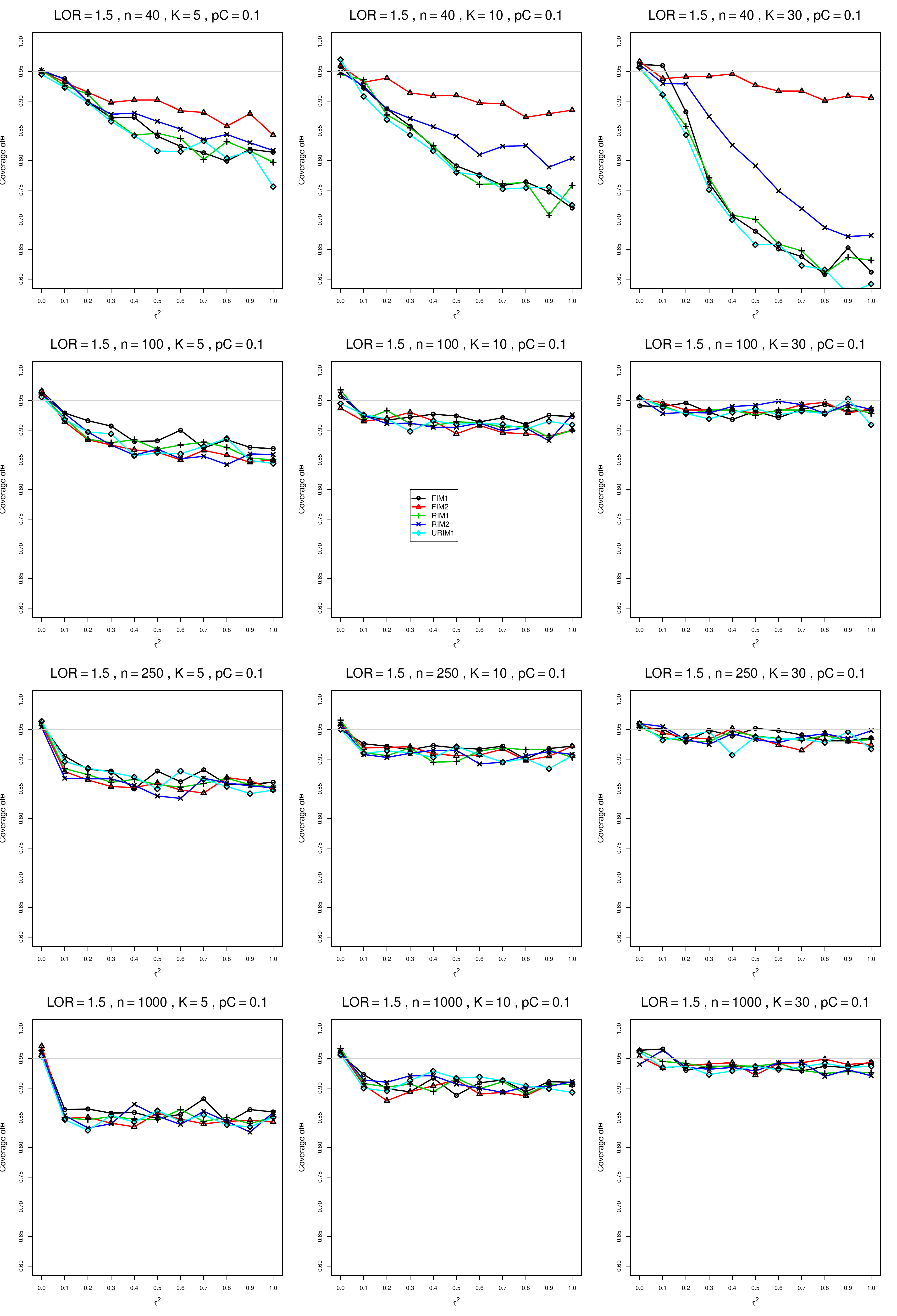}
	\caption{Coverage of the Random-intercept with $c=1/2$ confidence interval for $\theta=1.5$, $p_{C}=0.1$, $\sigma^2=0.1$, constant sample sizes $n=40,\;100,\;250,\;1000$.
The data-generation mechanisms are FIM1 ($\circ$), FIM2 ($\triangle$), RIM1 (+), RIM2 ($\times$), and URIM1 ($\diamond$).
		\label{PlotCovThetamu15andpC01LOR_UMRSsigma01}}
\end{figure}
\begin{figure}[t]
	\centering
	\includegraphics[scale=0.33]{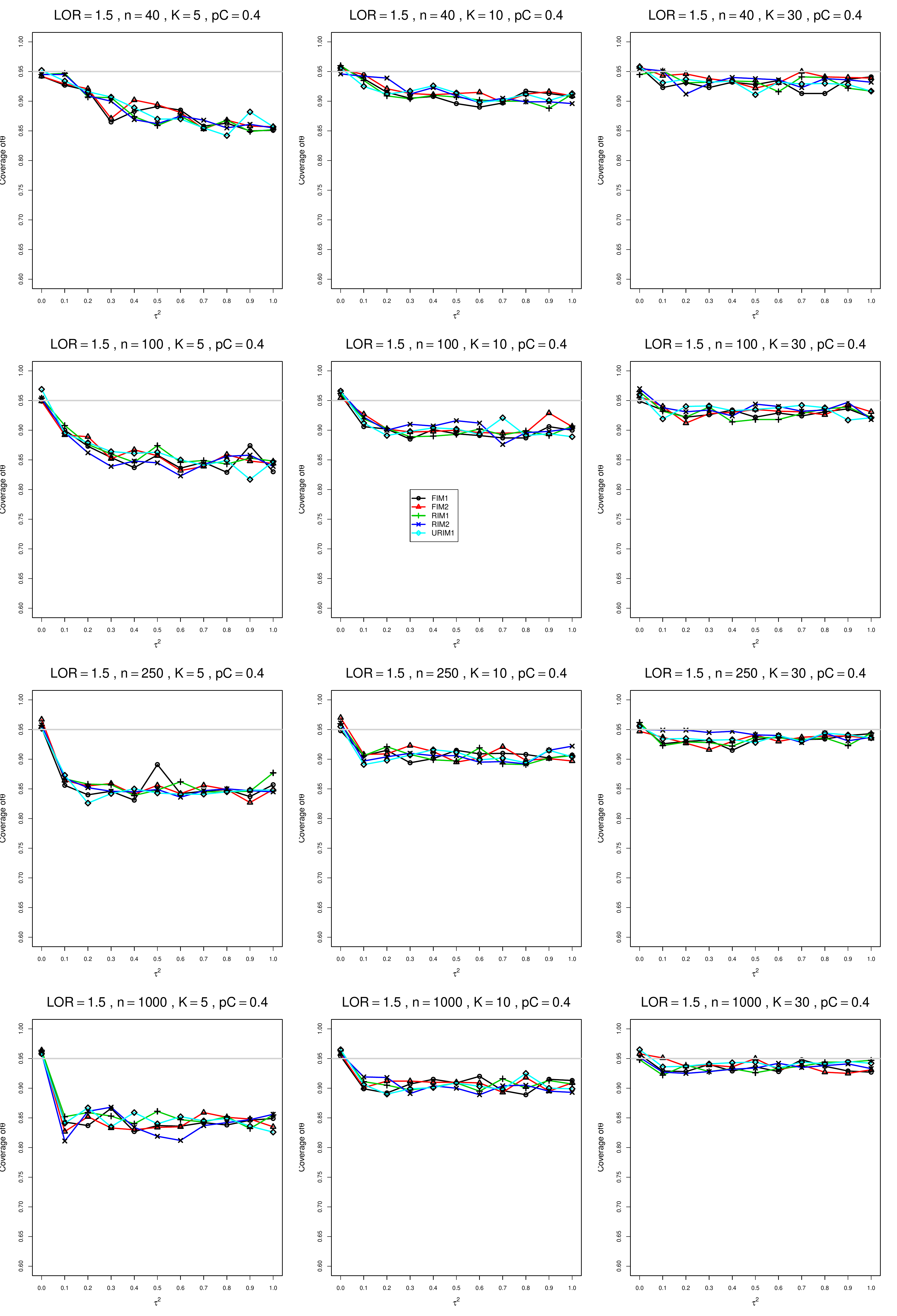}
	\caption{Coverage of the Random-intercept with $c=1/2$ confidence interval for $\theta=1.5$, $p_{C}=0.4$, $\sigma^2=0.1$, constant sample sizes $n=40,\;100,\;250,\;1000$.
The data-generation mechanisms are FIM1 ($\circ$), FIM2 ($\triangle$), RIM1 (+), RIM2 ($\times$), and URIM1 ($\diamond$).
		\label{PlotCovThetamu15andpC04LOR_UMRSsigma01}}
\end{figure}
\begin{figure}[t]
	\centering
	\includegraphics[scale=0.33]{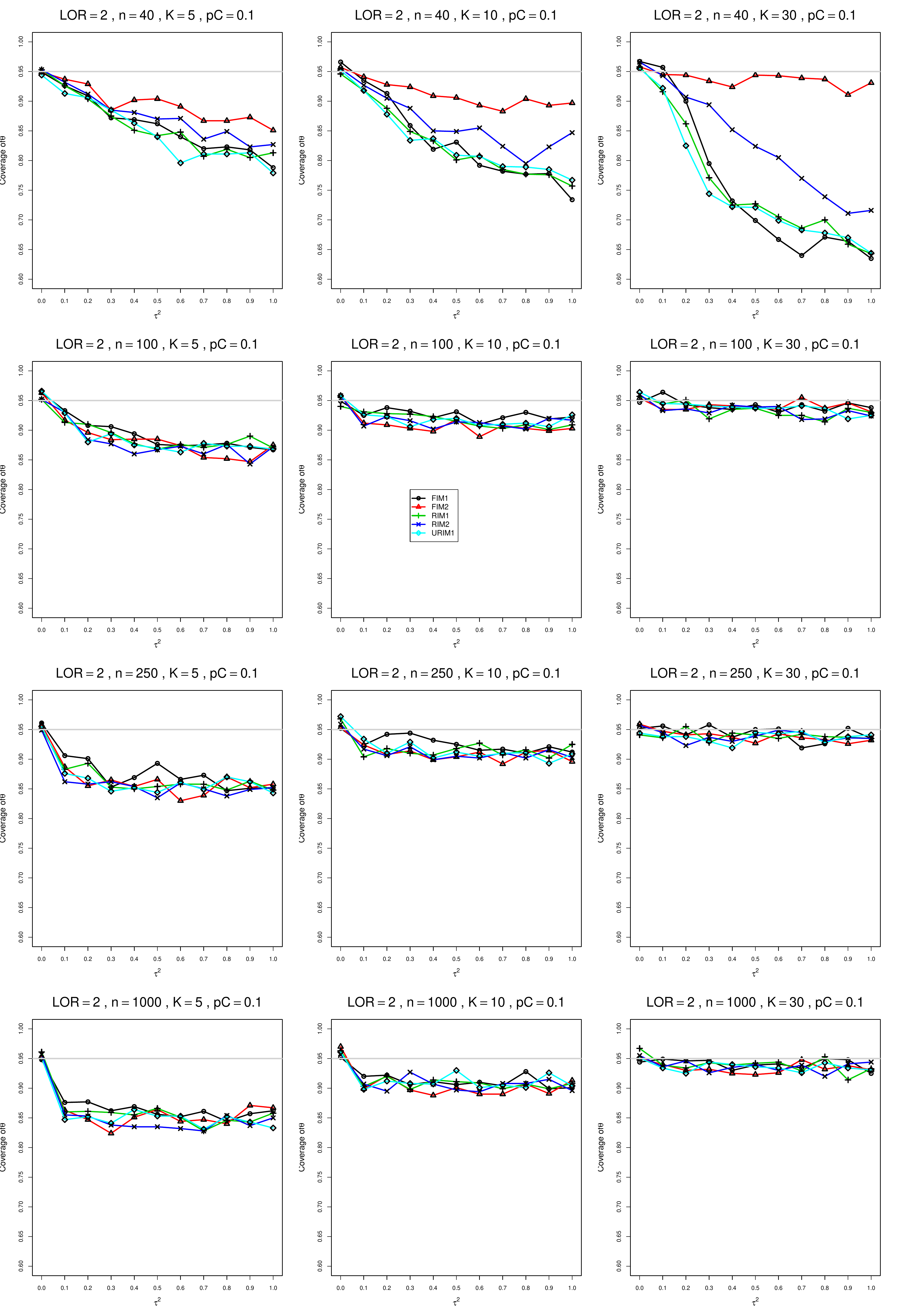}
	\caption{Coverage of the Random-intercept with $c=1/2$ confidence interval for $\theta=2$, $p_{C}=0.1$, $\sigma^2=0.1$, constant sample sizes $n=40,\;100,\;250,\;1000$.
The data-generation mechanisms are FIM1 ($\circ$), FIM2 ($\triangle$), RIM1 (+), RIM2 ($\times$), and URIM1 ($\diamond$).
		\label{PlotCovThetamu2andpC01LOR_UMRSsigma01}}
\end{figure}
\begin{figure}[t]
	\centering
	\includegraphics[scale=0.33]{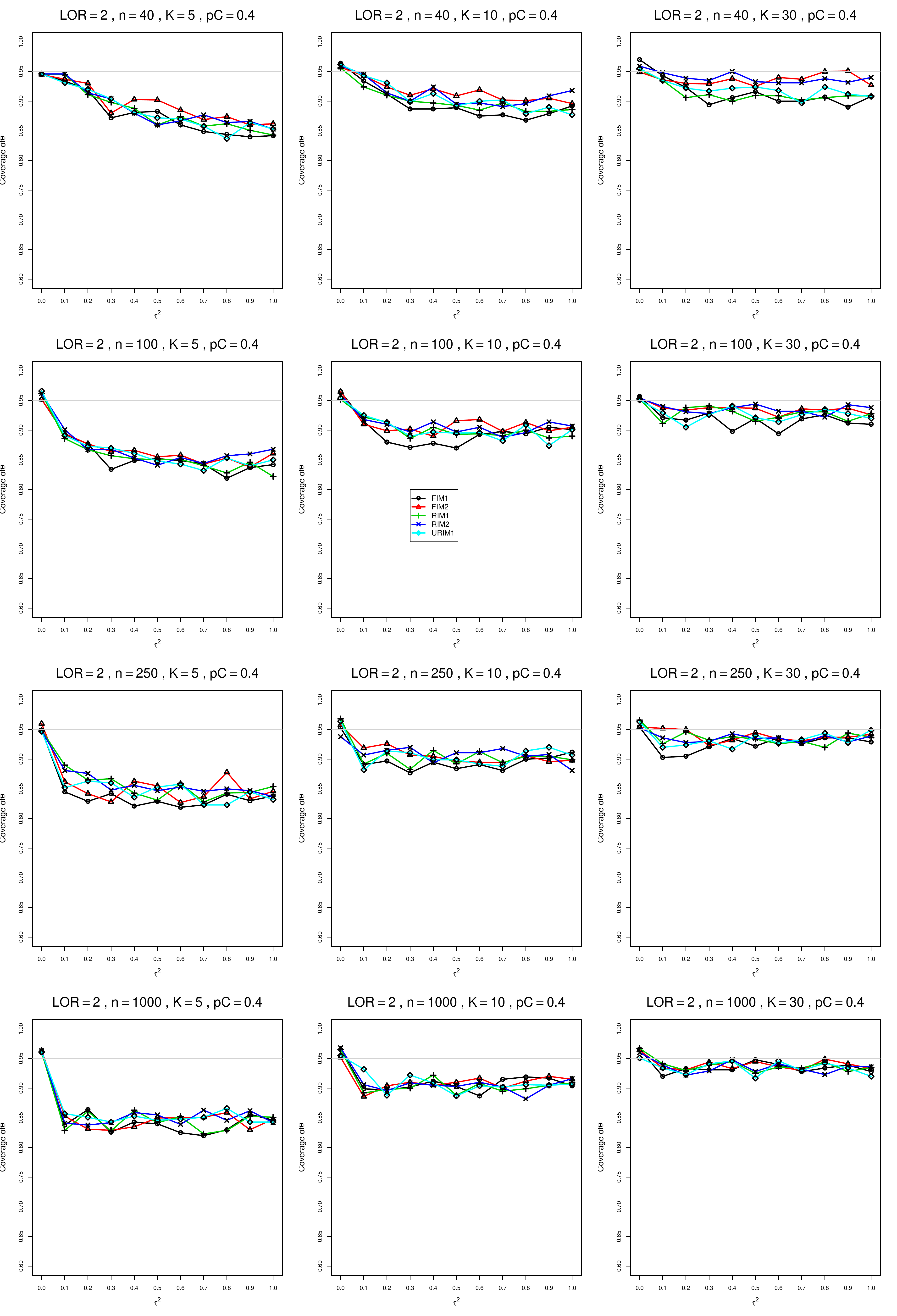}
	\caption{Coverage of the Random-intercept with $c=1/2$ confidence interval for $\theta=2$, $p_{C}=0.4$, $\sigma^2=0.1$, constant sample sizes $n=40,\;100,\;250,\;1000$.
The data-generation mechanisms are FIM1 ($\circ$), FIM2 ($\triangle$), RIM1 (+), RIM2 ($\times$), and URIM1 ($\diamond$).
		\label{PlotCovThetamu2andpC04LOR_UMRSsigma01}}
\end{figure}
\begin{figure}[t]
	\centering
	\includegraphics[scale=0.33]{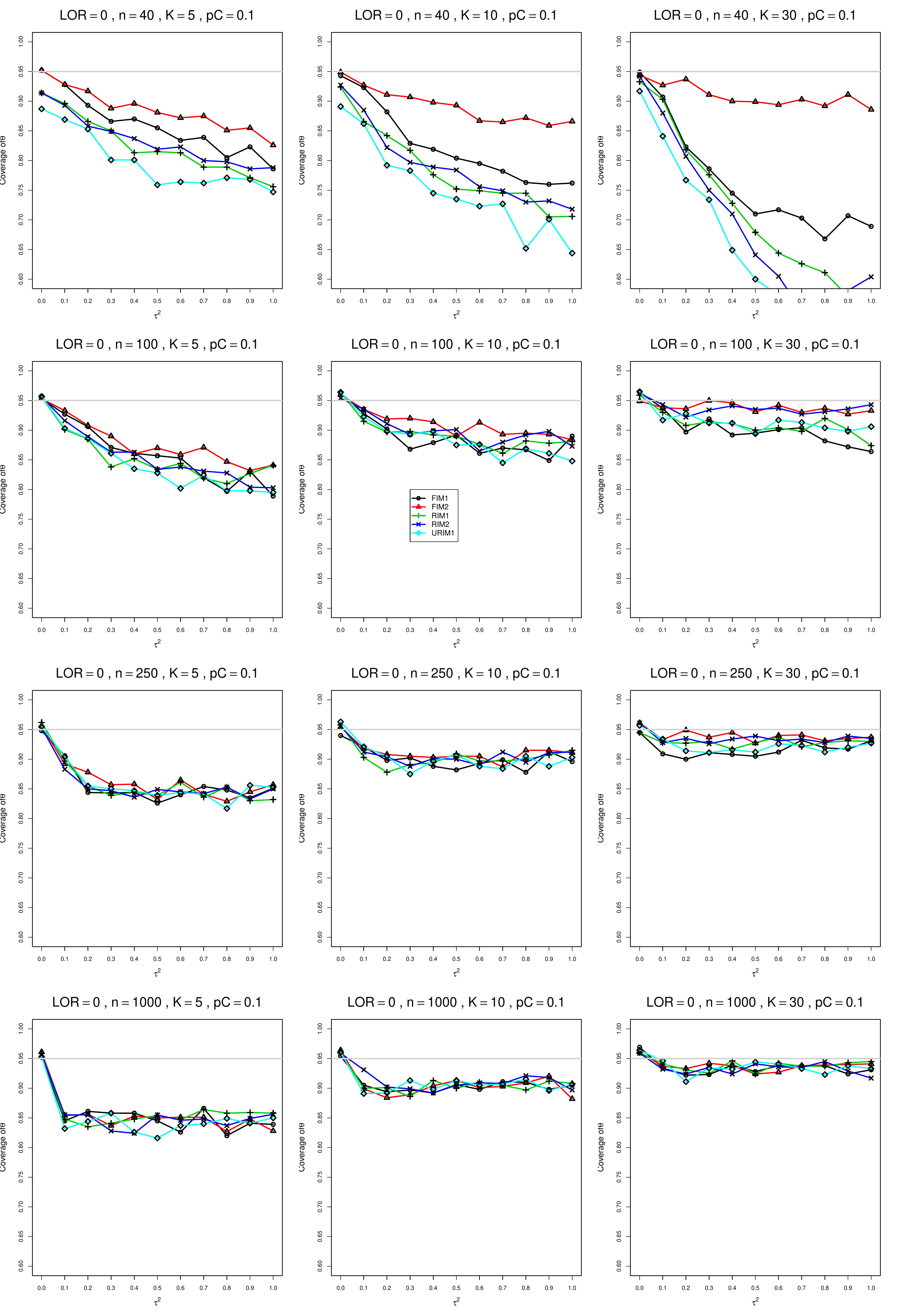}
	\caption{Coverage of the Random-intercept with $c=1/2$ confidence interval for $\theta=0$, $p_{C}=0.1$, $\sigma^2=0.4$, constant sample sizes $n=40,\;100,\;250,\;1000$.
The data-generation mechanisms are FIM1 ($\circ$), FIM2 ($\triangle$), RIM1 (+), RIM2 ($\times$), and URIM1 ($\diamond$).
		\label{PlotCovThetamu0andpC01LOR_UMRSsigma04}}
\end{figure}
\begin{figure}[t]
	\centering
	\includegraphics[scale=0.33]{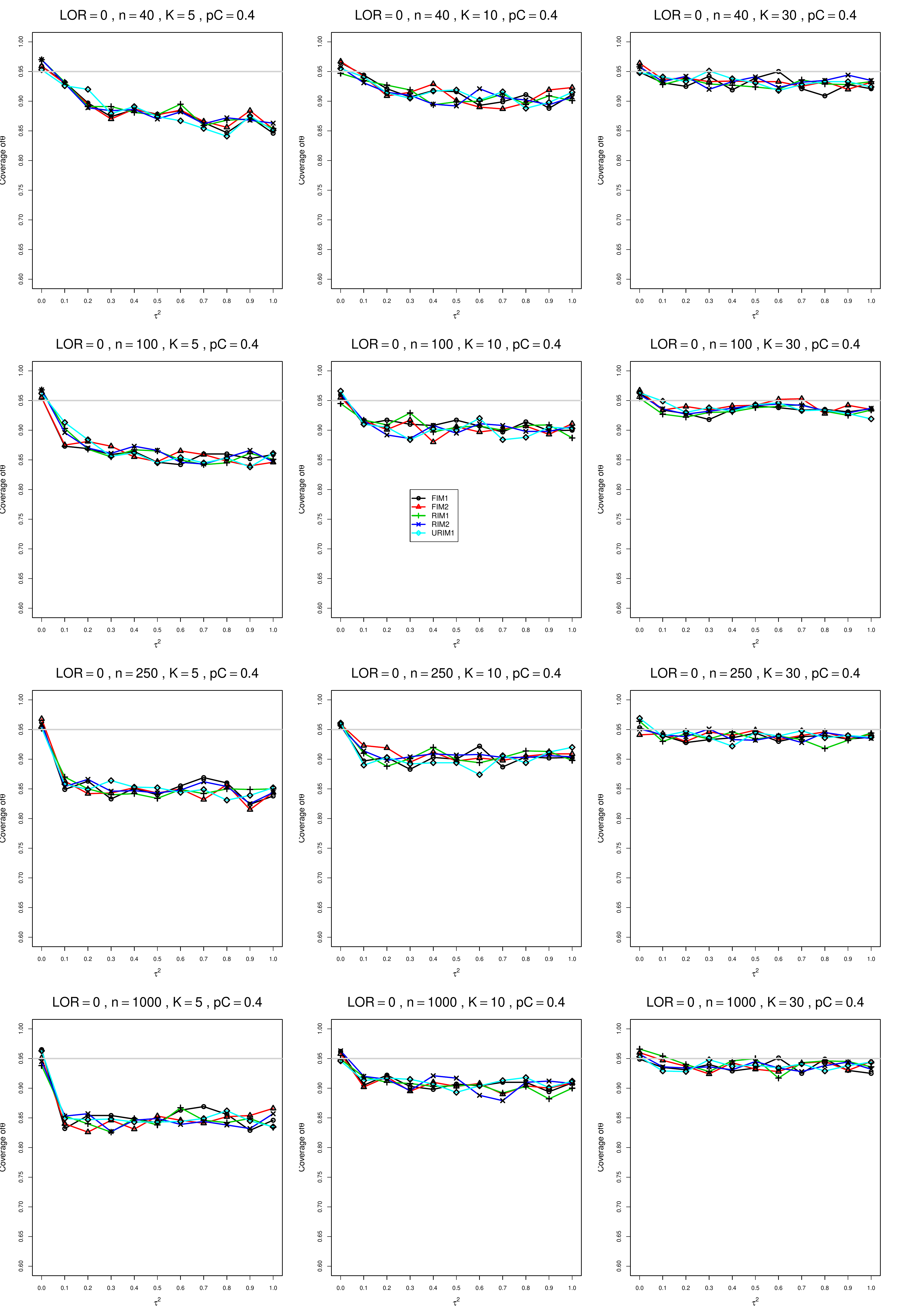}
	\caption{Coverage of the Random-intercept with $c=1/2$ confidence interval for $\theta=0$, $p_{C}=0.4$, $\sigma^2=0.4$, constant sample sizes $n=40,\;100,\;250,\;1000$.
The data-generation mechanisms are FIM1 ($\circ$), FIM2 ($\triangle$), RIM1 (+), RIM2 ($\times$), and URIM1 ($\diamond$).
		\label{PlotCovThetamu0andpC04LOR_UMRSsigma04}}
\end{figure}
\begin{figure}[t]
	\centering
	\includegraphics[scale=0.33]{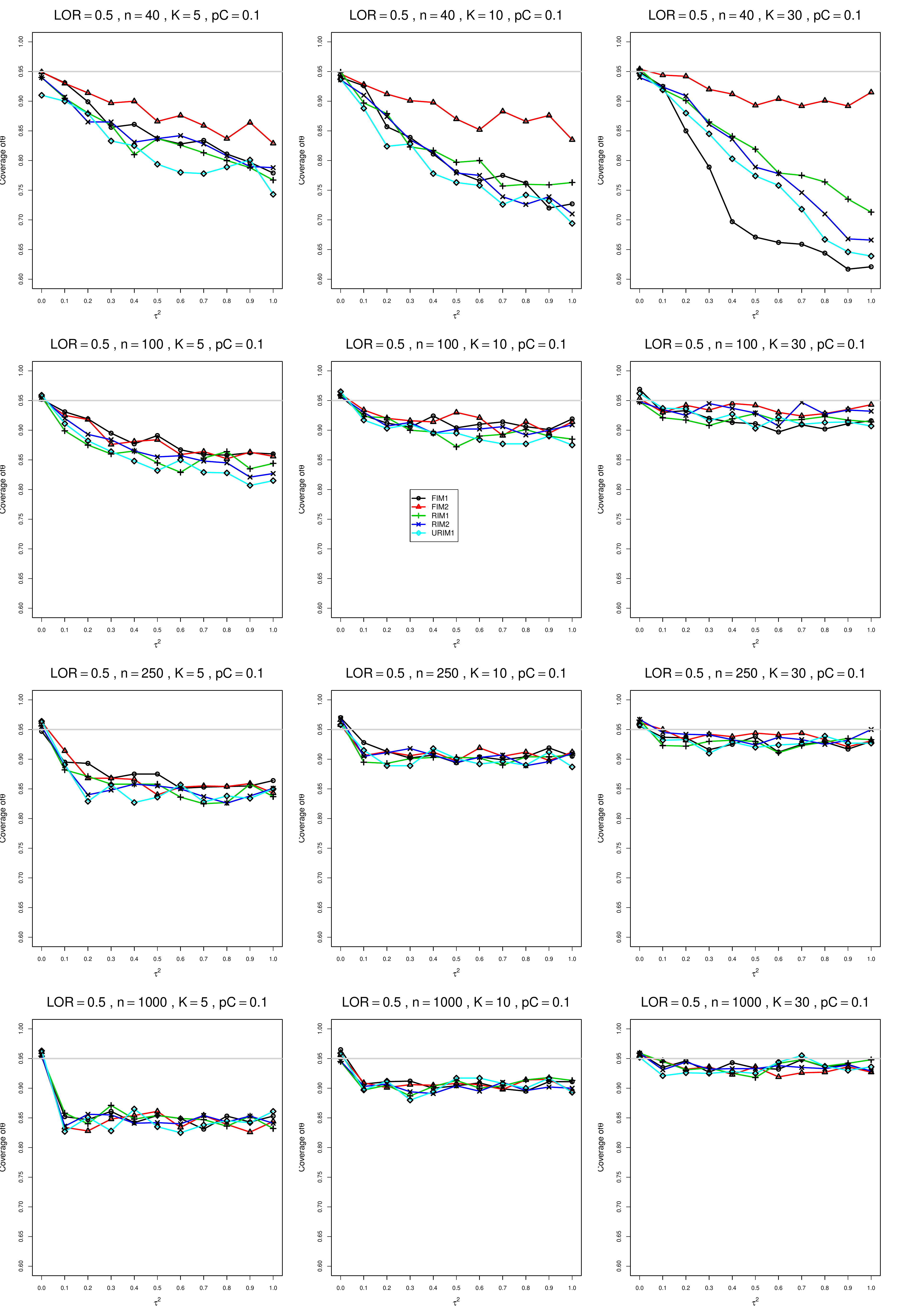}
	\caption{Coverage of the Random-intercept with $c=1/2$ confidence interval for $\theta=0.5$, $p_{C}=0.1$, $\sigma^2=0.4$, constant sample sizes $n=40,\;100,\;250,\;1000$.
The data-generation mechanisms are FIM1 ($\circ$), FIM2 ($\triangle$), RIM1 (+), RIM2 ($\times$), and URIM1 ($\diamond$).
		\label{PlotCovThetamu05andpC01LOR_UMRSsigma04}}
\end{figure}
\begin{figure}[t]
	\centering
	\includegraphics[scale=0.33]{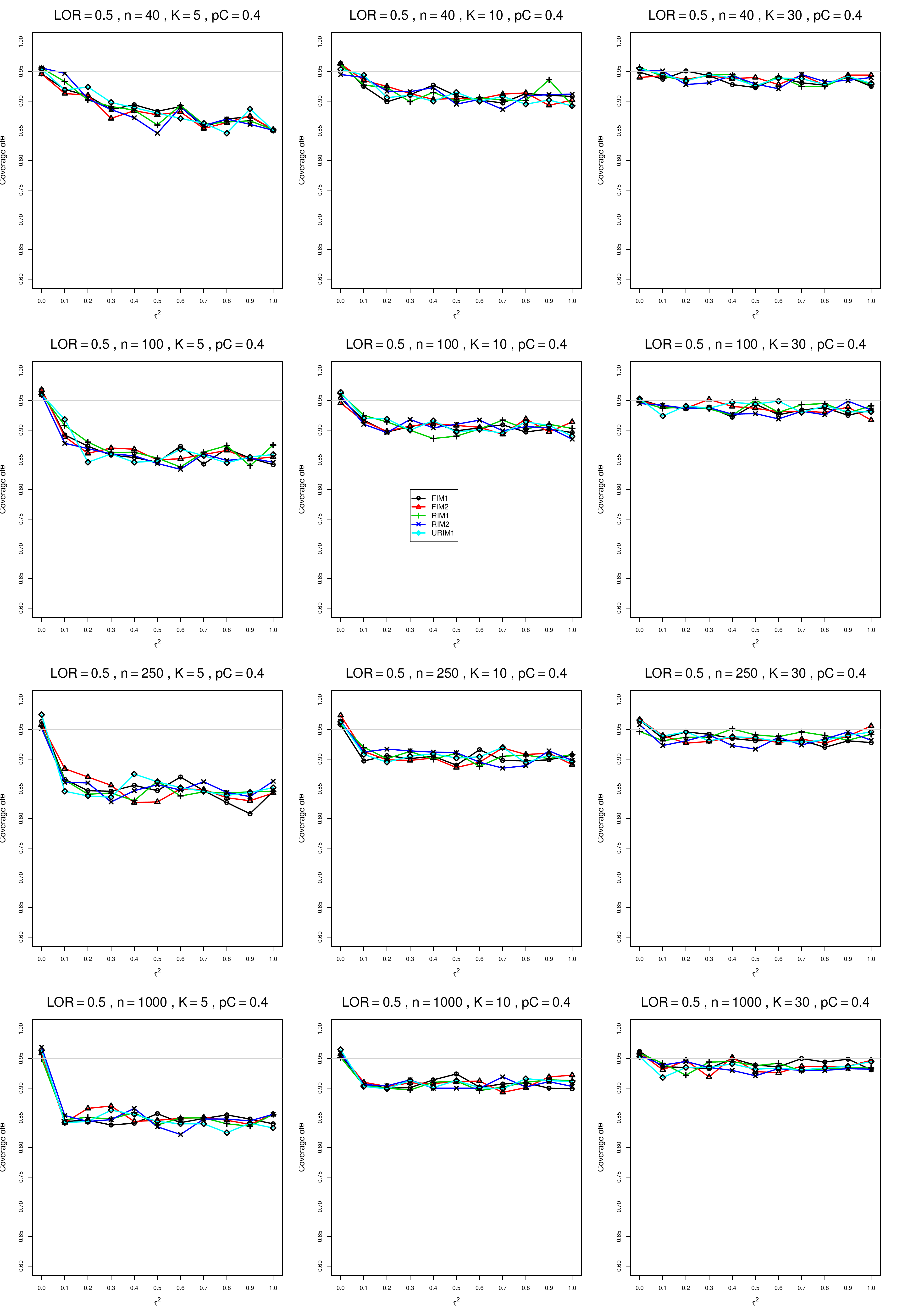}
	\caption{Coverage of the Random-intercept with $c=1/2$ confidence interval for $\theta=0.5$, $p_{C}=0.4$, $\sigma^2=0.4$, constant sample sizes $n=40,\;100,\;250,\;1000$.
The data-generation mechanisms are FIM1 ($\circ$), FIM2 ($\triangle$), RIM1 (+), RIM2 ($\times$), and URIM1 ($\diamond$).
		\label{PlotCovThetamu05andpC04LOR_UMRSsigma04}}
\end{figure}
\begin{figure}[t]
	\centering
	\includegraphics[scale=0.33]{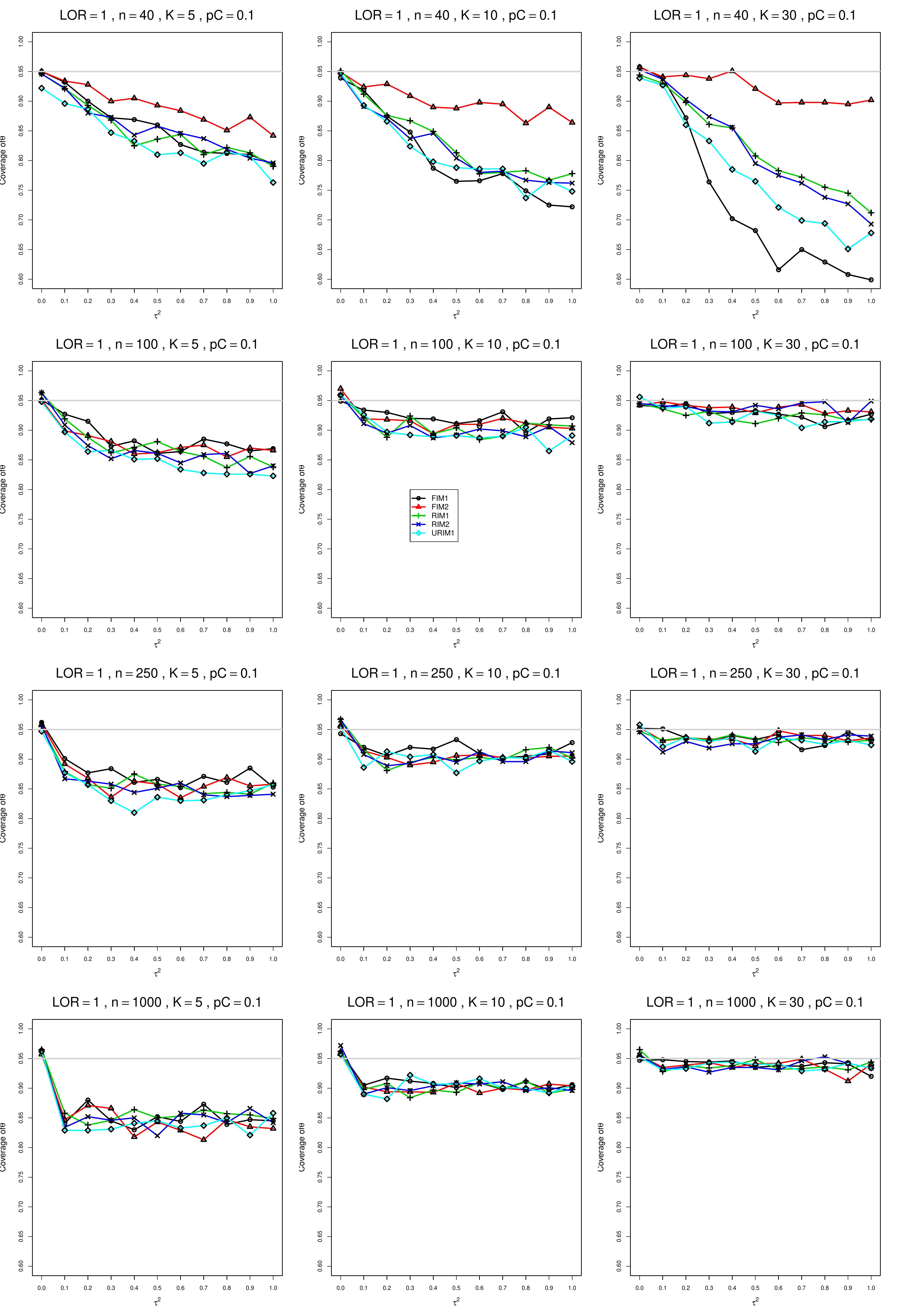}
	\caption{Coverage of the Random-intercept with $c=1/2$ confidence interval for $\theta=1$, $p_{C}=0.1$, $\sigma^2=0.4$, constant sample sizes $n=40,\;100,\;250,\;1000$.
The data-generation mechanisms are FIM1 ($\circ$), FIM2 ($\triangle$), RIM1 (+), RIM2 ($\times$), and URIM1 ($\diamond$).
		\label{PlotCovThetamu1andpC01LOR_UMRSsigma04}}
\end{figure}
\begin{figure}[t]
	\centering
	\includegraphics[scale=0.33]{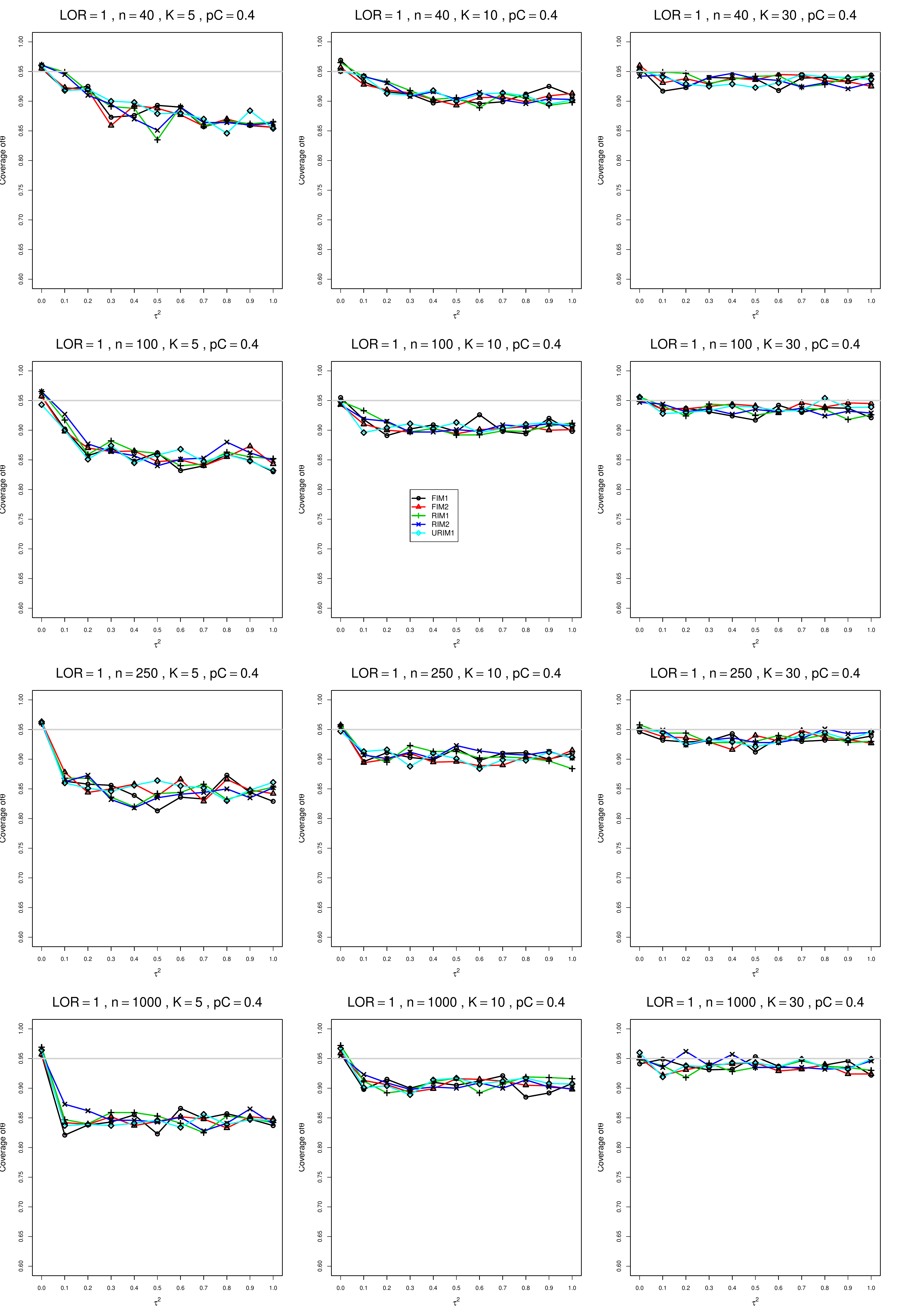}
	\caption{Coverage of the Random-intercept with $c=1/2$ confidence interval for $\theta=1$, $p_{C}=0.4$, $\sigma^2=0.4$, constant sample sizes $n=40,\;100,\;250,\;1000$.
The data-generation mechanisms are FIM1 ($\circ$), FIM2 ($\triangle$), RIM1 (+), RIM2 ($\times$), and URIM1 ($\diamond$).
		\label{PlotCovThetamu1andpC04LOR_UMRSsigma04}}
\end{figure}
\begin{figure}[t]
	\centering
	\includegraphics[scale=0.33]{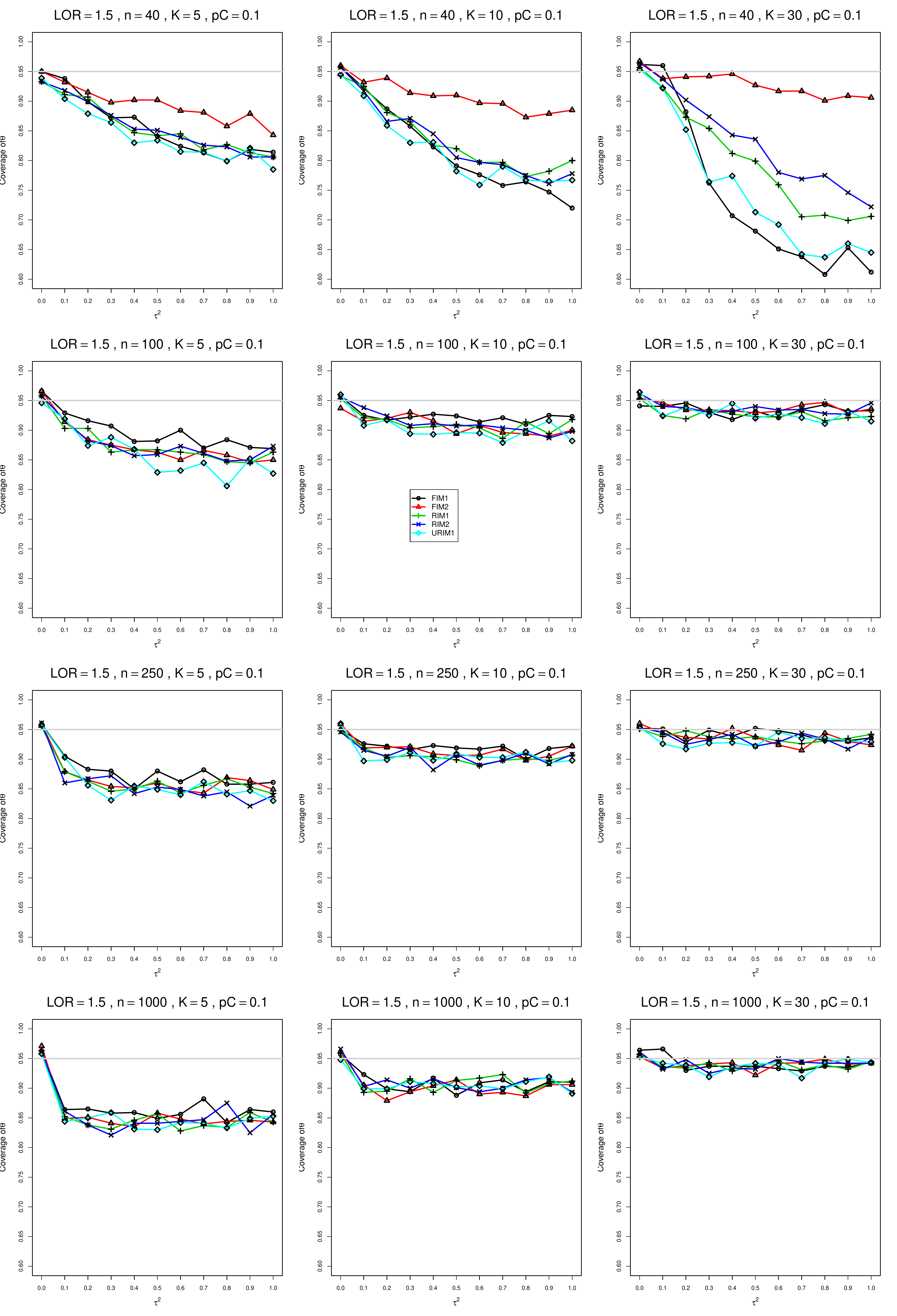}
	\caption{Coverage of the Random-intercept with $c=1/2$ confidence interval for $\theta=1.5$, $p_{C}=0.1$, $\sigma^2=0.4$, constant sample sizes $n=40,\;100,\;250,\;1000$.
The data-generation mechanisms are FIM1 ($\circ$), FIM2 ($\triangle$), RIM1 (+), RIM2 ($\times$), and URIM1 ($\diamond$).
		\label{PlotCovThetamu15andpC01LOR_UMRSsigma04}}
\end{figure}
\begin{figure}[t]
	\centering
	\includegraphics[scale=0.33]{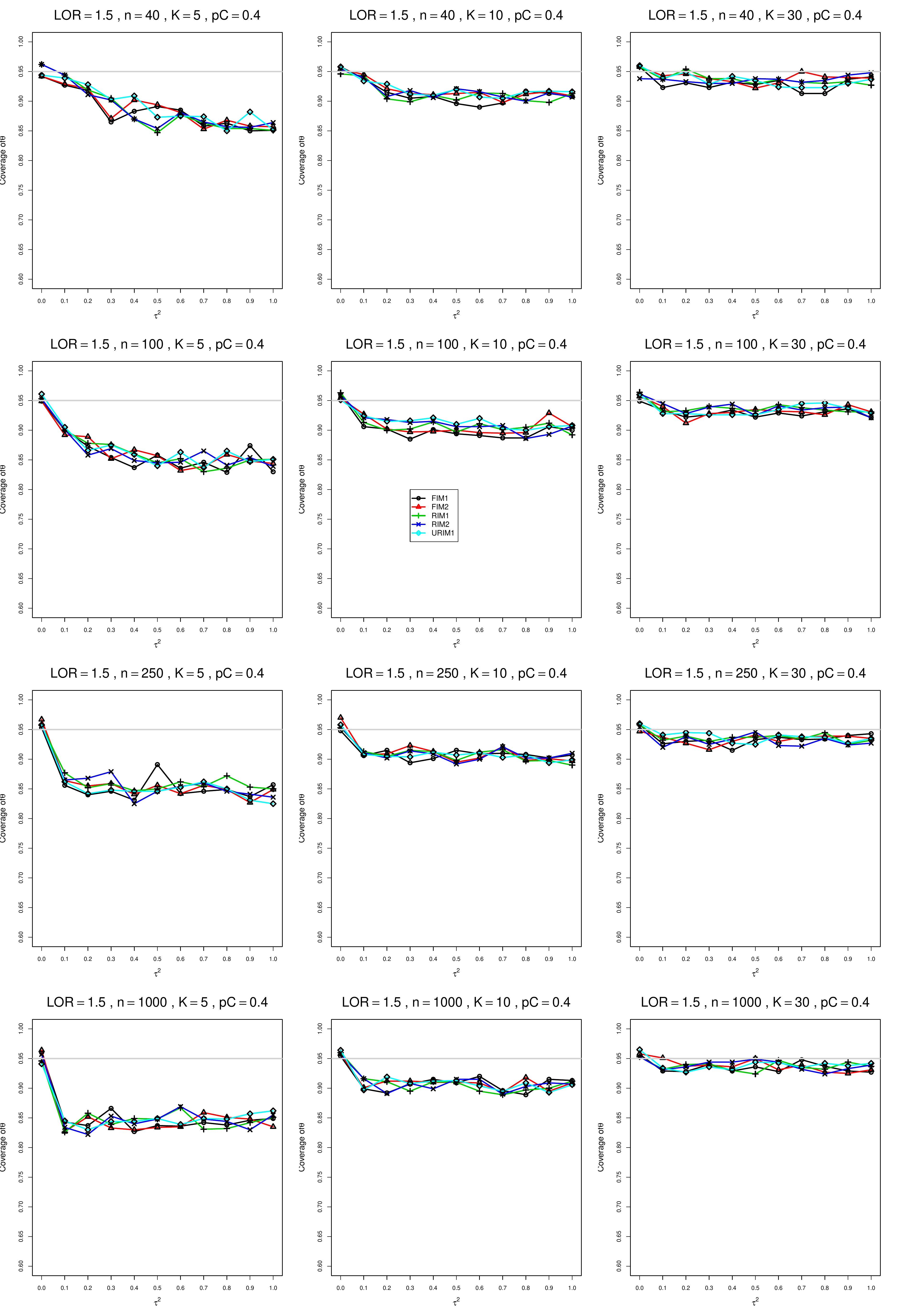}
	\caption{Coverage of the Random-intercept with $c=1/2$ confidence interval for $\theta=1.5$, $p_{C}=0.4$, $\sigma^2=0.4$, constant sample sizes $n=40,\;100,\;250,\;1000$.
The data-generation mechanisms are FIM1 ($\circ$), FIM2 ($\triangle$), RIM1 (+), RIM2 ($\times$), and URIM1 ($\diamond$).
		\label{PlotCovThetamu15andpC04LOR_UMRSsigma04}}
\end{figure}
\begin{figure}[t]
	\centering
	\includegraphics[scale=0.33]{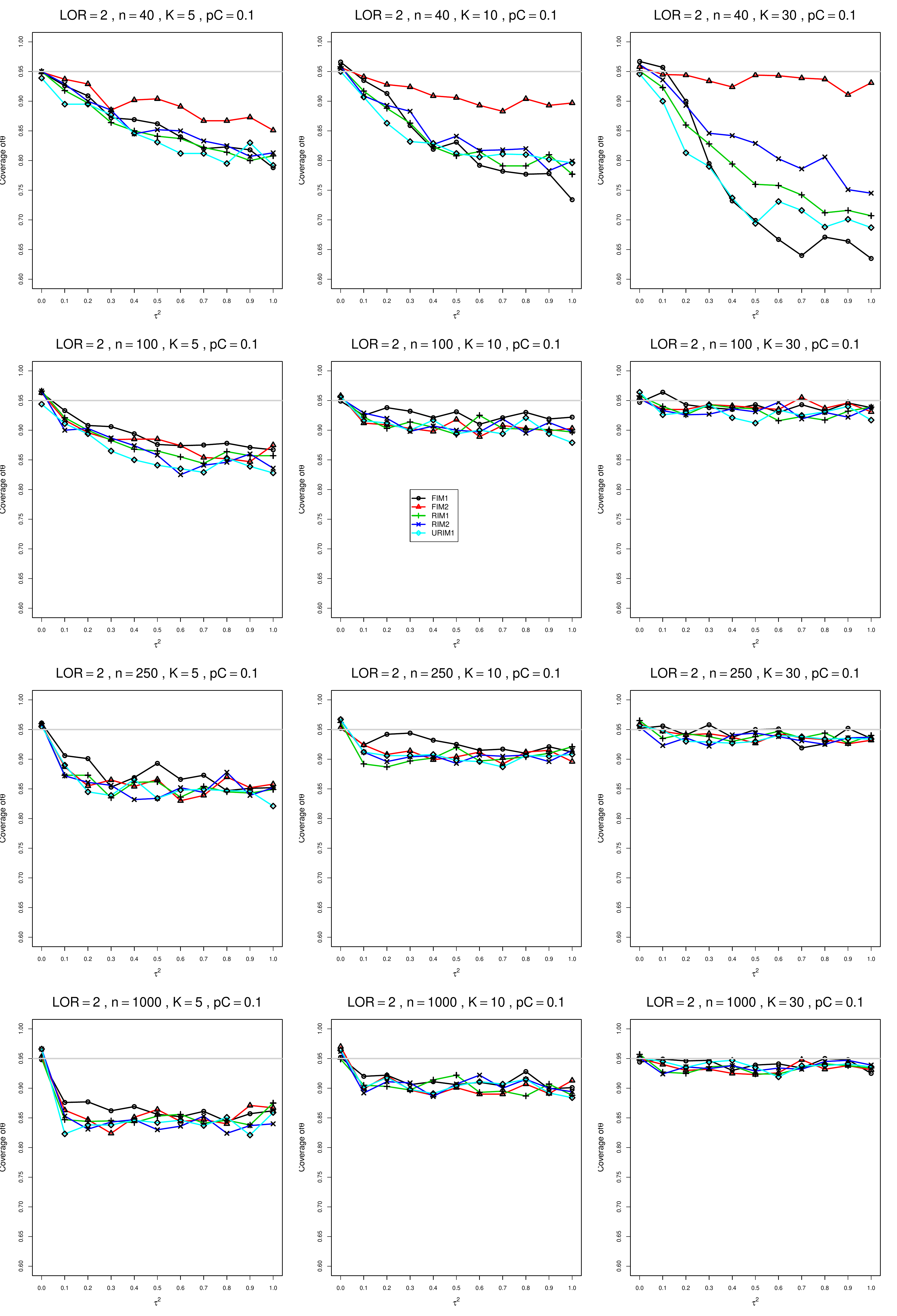}
	\caption{Coverage of the Random-intercept with $c=1/2$ confidence interval for $\theta=2$, $p_{C}=0.1$, $\sigma^2=0.4$, constant sample sizes $n=40,\;100,\;250,\;1000$.
The data-generation mechanisms are FIM1 ($\circ$), FIM2 ($\triangle$), RIM1 (+), RIM2 ($\times$), and URIM1 ($\diamond$).
		\label{PlotCovThetamu2andpC01LOR_UMRSsigma04}}
\end{figure}
\begin{figure}[t]
	\centering
	\includegraphics[scale=0.33]{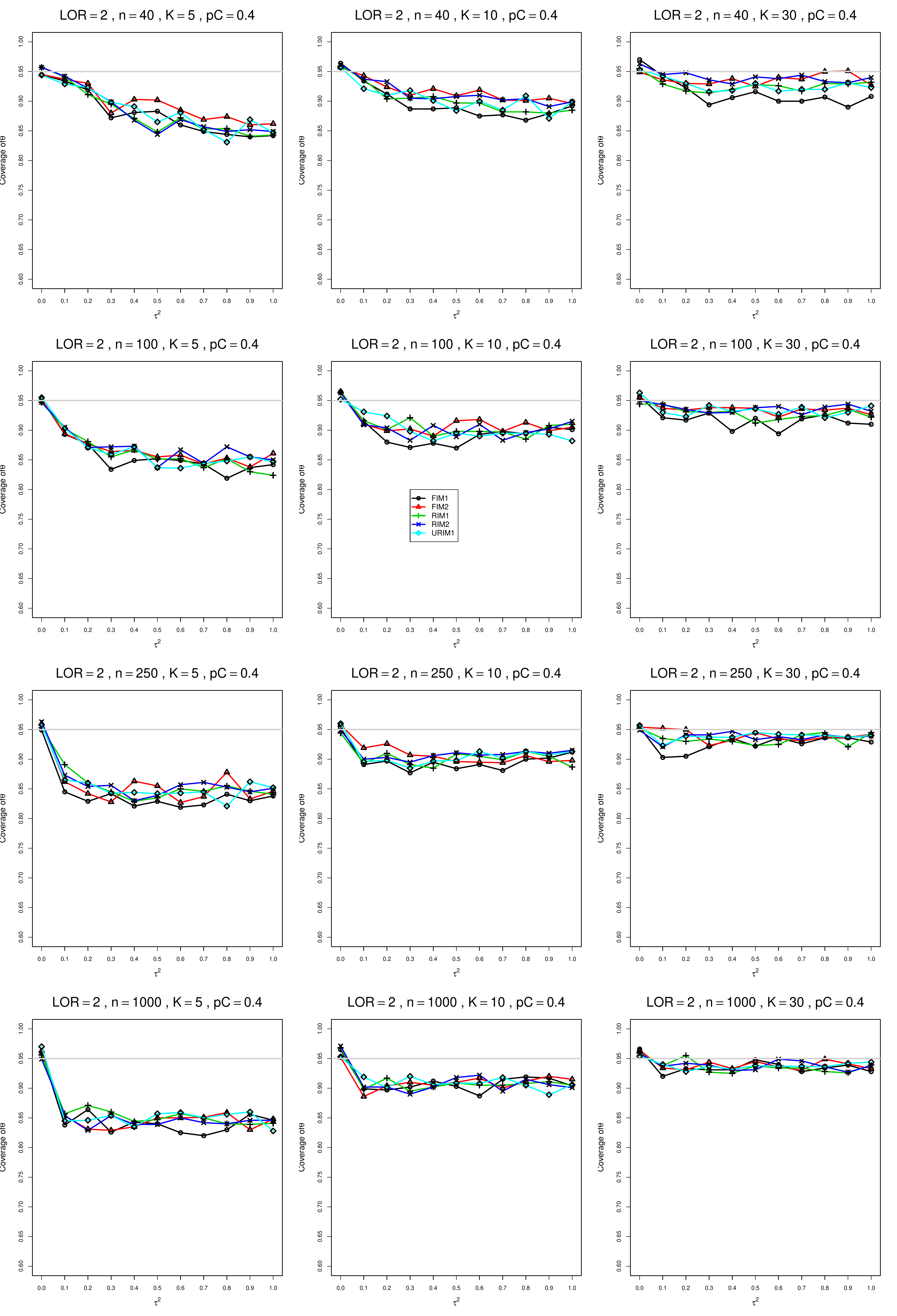}
	\caption{Coverage of the Random-intercept with $c=1/2$ confidence interval for $\theta=2$, $p_{C}=0.4$, $\sigma^2=0.4$, constant sample sizes $n=40,\;100,\;250,\;1000$.
The data-generation mechanisms are FIM1 ($\circ$), FIM2 ($\triangle$), RIM1 (+), RIM2 ($\times$), and URIM1 ($\diamond$).
		\label{PlotCovThetamu2andpC04LOR_UMRSsigma04}}
\end{figure}

\clearpage
\subsection*{A3.6 Coverage of $\hat{\theta}_{SSW}$ with HkSJ and $\hat{tau}_{KD}^2$}
\renewcommand{\thefigure}{A3.6.\arabic{figure}}
\setcounter{figure}{0}

\begin{figure}[t]
	\centering
	\includegraphics[scale=0.33]{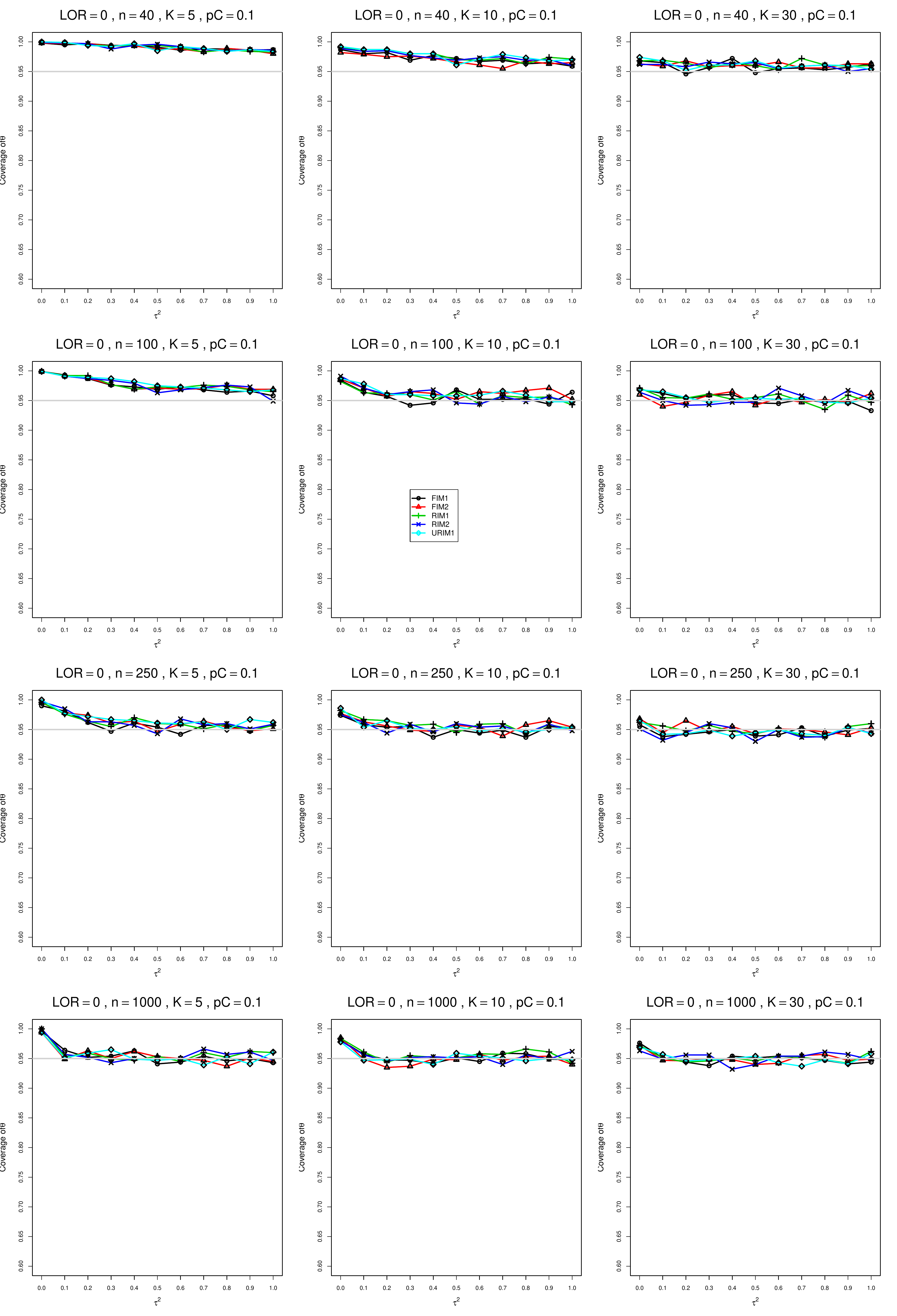}
	\caption{Coverage of the sample-size-weighted with Hartung-Knapp-Sidik-Jonkman (with KD) confidence interval for $\theta=0$, $p_{C}=0.1$, $\sigma^2=0.1$, constant sample sizes $n=40,\;100,\;250,\;1000$.
The data-generation mechanisms are FIM1 ($\circ$), FIM2 ($\triangle$), RIM1 (+), RIM2 ($\times$), and URIM1 ($\diamond$).
		\label{PlotCovThetamu0andpC01LOR_HkSJ_new_CMPsigma01}}
\end{figure}
\begin{figure}[t]
	\centering
	\includegraphics[scale=0.33]{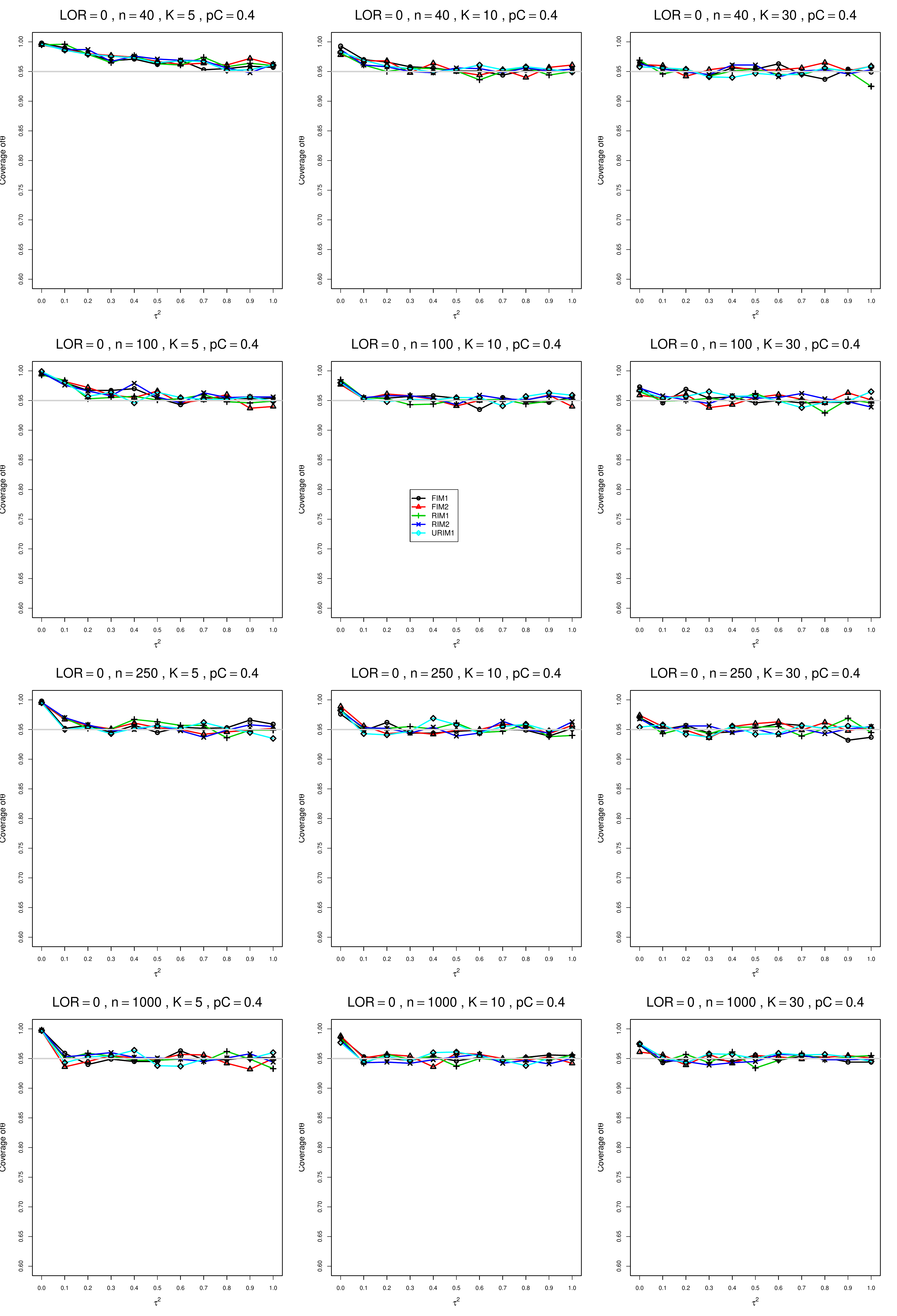}
	\caption{Coverage of the sample-size-weighted with Hartung-Knapp-Sidik-Jonkman (with KD) confidence interval for $\theta=0$, $p_{C}=0.4$, $\sigma^2=0.1$, constant sample sizes $n=40,\;100,\;250,\;1000$.
The data-generation mechanisms are FIM1 ($\circ$), FIM2 ($\triangle$), RIM1 (+), RIM2 ($\times$), and URIM1 ($\diamond$).
		\label{PlotCovThetamu0andpC04LOR_HkSJ_new_CMPsigma01}}
\end{figure}
\begin{figure}[t]
	\centering
	\includegraphics[scale=0.33]{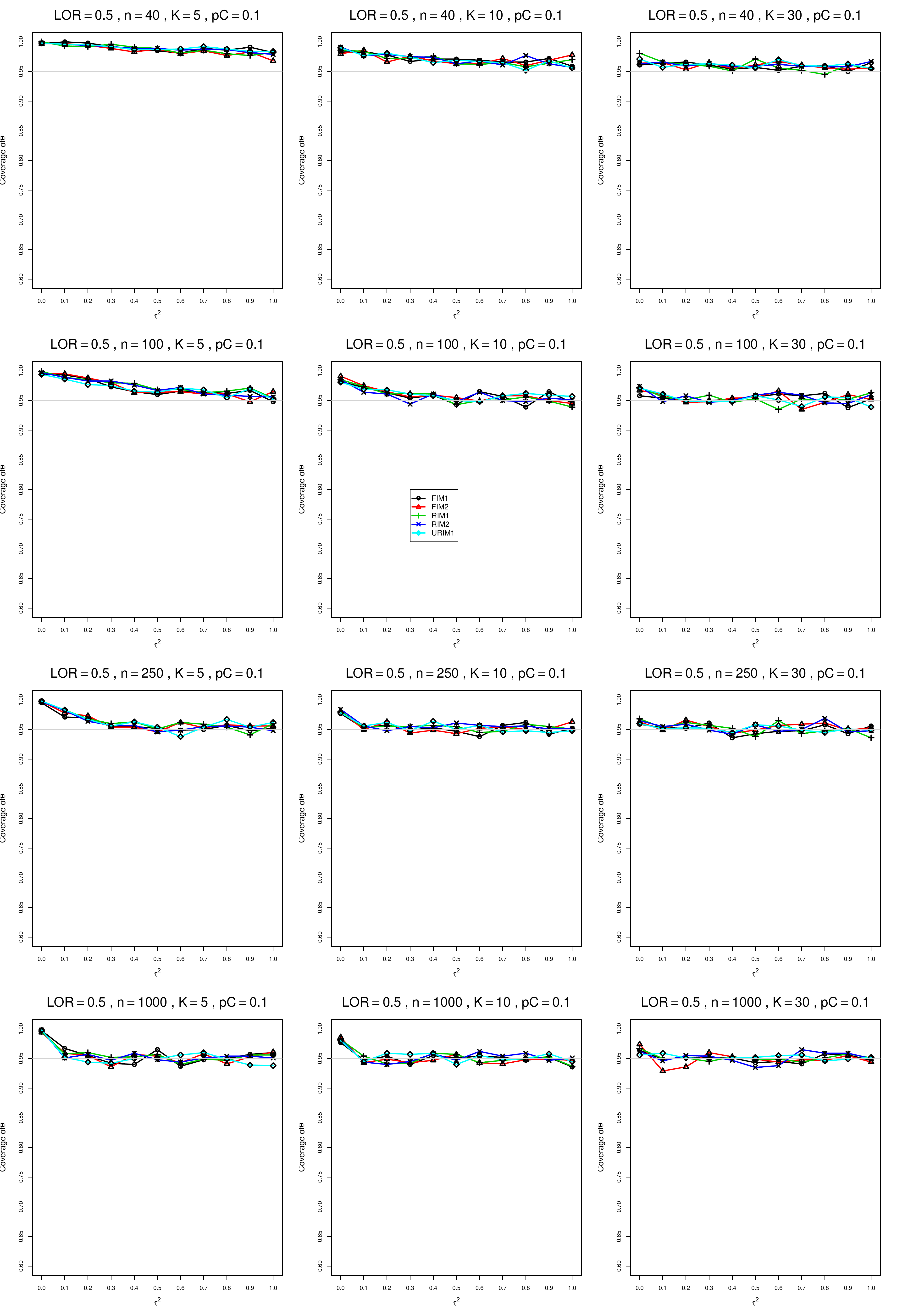}
	\caption{Coverage of the sample-size-weighted with Hartung-Knapp-Sidik-Jonkman (with KD) confidence interval for $\theta=0.5$, $p_{C}=0.1$, $\sigma^2=0.1$, constant sample sizes $n=40,\;100,\;250,\;1000$.
The data-generation mechanisms are FIM1 ($\circ$), FIM2 ($\triangle$), RIM1 (+), RIM2 ($\times$), and URIM1 ($\diamond$).
		\label{PlotCovThetamu05andpC01LOR_HkSJ_new_CMPsigma01}}
\end{figure}
\begin{figure}[t]
	\centering
	\includegraphics[scale=0.33]{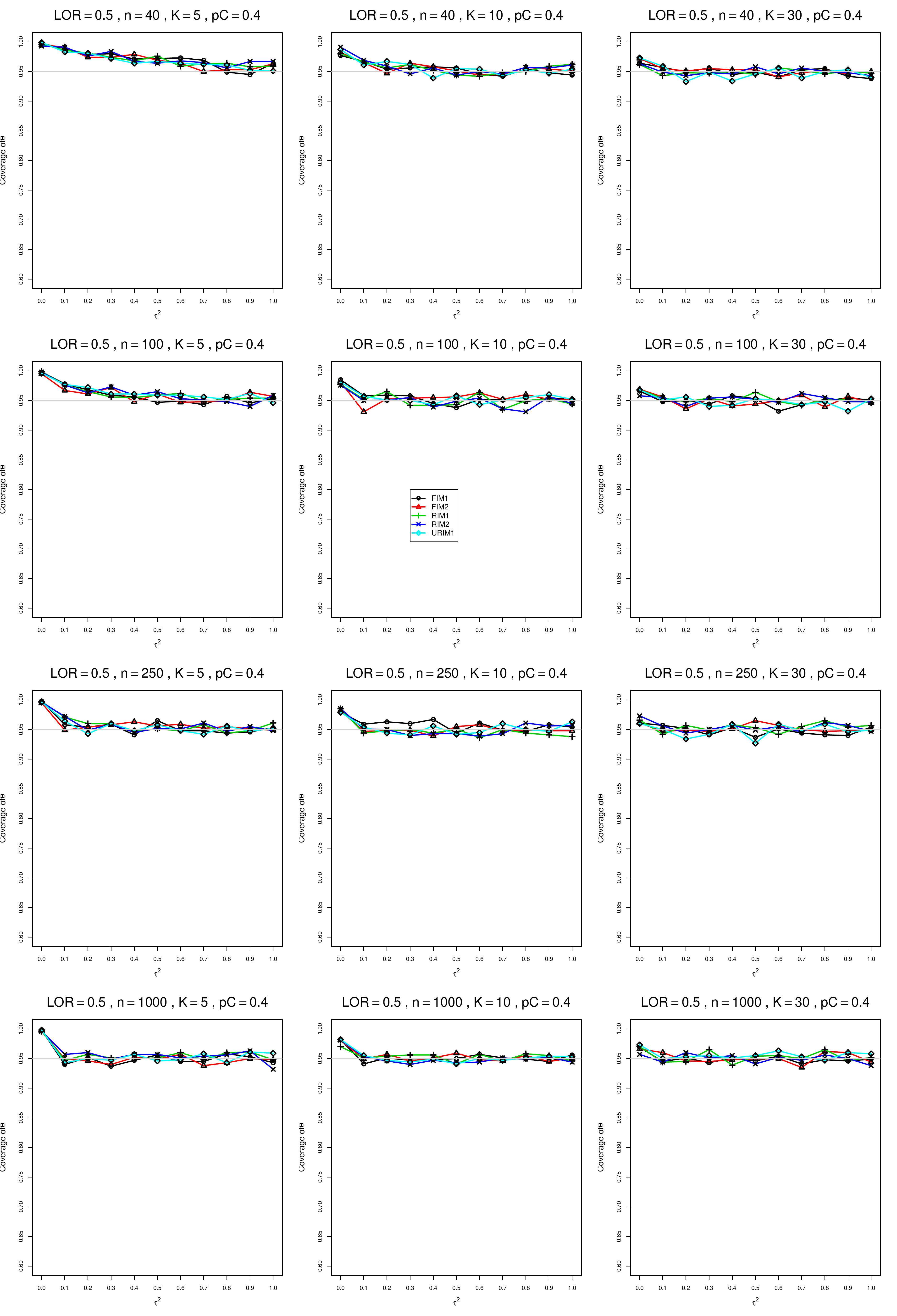}
	\caption{Coverage of the sample-size-weighted with Hartung-Knapp-Sidik-Jonkman (with KD) confidence interval for $\theta=0.5$, $p_{C}=0.4$, $\sigma^2=0.1$, constant sample sizes $n=40,\;100,\;250,\;1000$.
The data-generation mechanisms are FIM1 ($\circ$), FIM2 ($\triangle$), RIM1 (+), RIM2 ($\times$), and URIM1 ($\diamond$).
		\label{PlotCovThetamu05andpC04LOR_HkSJ_new_CMPsigma01}}
\end{figure}
\begin{figure}[t]
	\centering
	\includegraphics[scale=0.33]{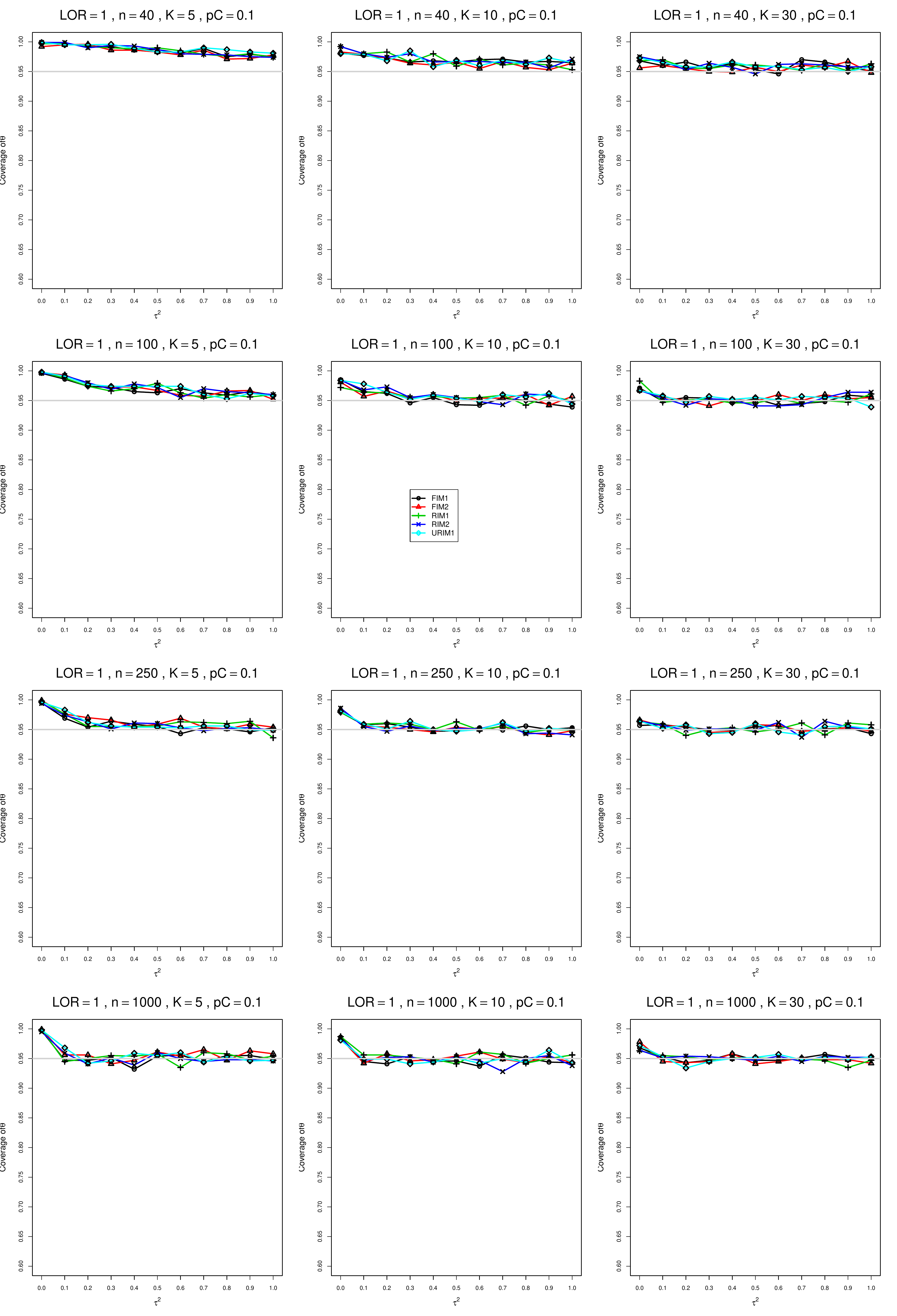}
	\caption{Coverage of the sample-size-weighted with Hartung-Knapp-Sidik-Jonkman (with KD) confidence interval for $\theta=1$, $p_{C}=0.1$, $\sigma^2=0.1$, constant sample sizes $n=40,\;100,\;250,\;1000$.
The data-generation mechanisms are FIM1 ($\circ$), FIM2 ($\triangle$), RIM1 (+), RIM2 ($\times$), and URIM1 ($\diamond$).
		\label{PlotCovThetamu1andpC01LOR_HkSJ_new_CMPsigma01}}
\end{figure}
\begin{figure}[t]
	\centering
	\includegraphics[scale=0.33]{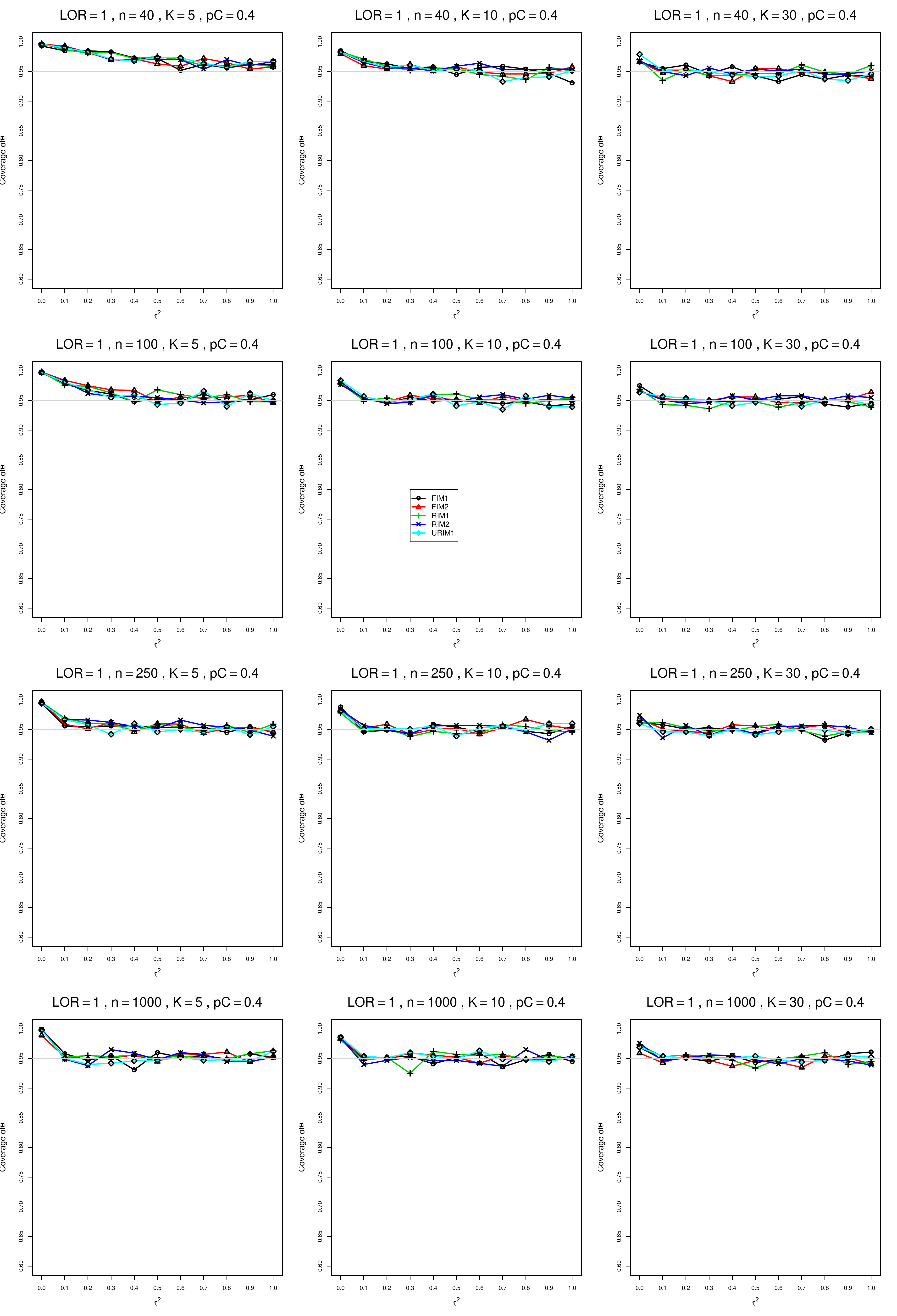}
	\caption{Coverage of the sample-size-weighted with Hartung-Knapp-Sidik-Jonkman (with KD) confidence interval for $\theta=1$, $p_{C}=0.4$, $\sigma^2=0.1$, constant sample sizes $n=40,\;100,\;250,\;1000$.
The data-generation mechanisms are FIM1 ($\circ$), FIM2 ($\triangle$), RIM1 (+), RIM2 ($\times$), and URIM1 ($\diamond$).
		\label{PlotCovThetamu1andpC04LOR_HkSJ_new_CMPsigma01}}
\end{figure}
\begin{figure}[t]
	\centering
	\includegraphics[scale=0.33]{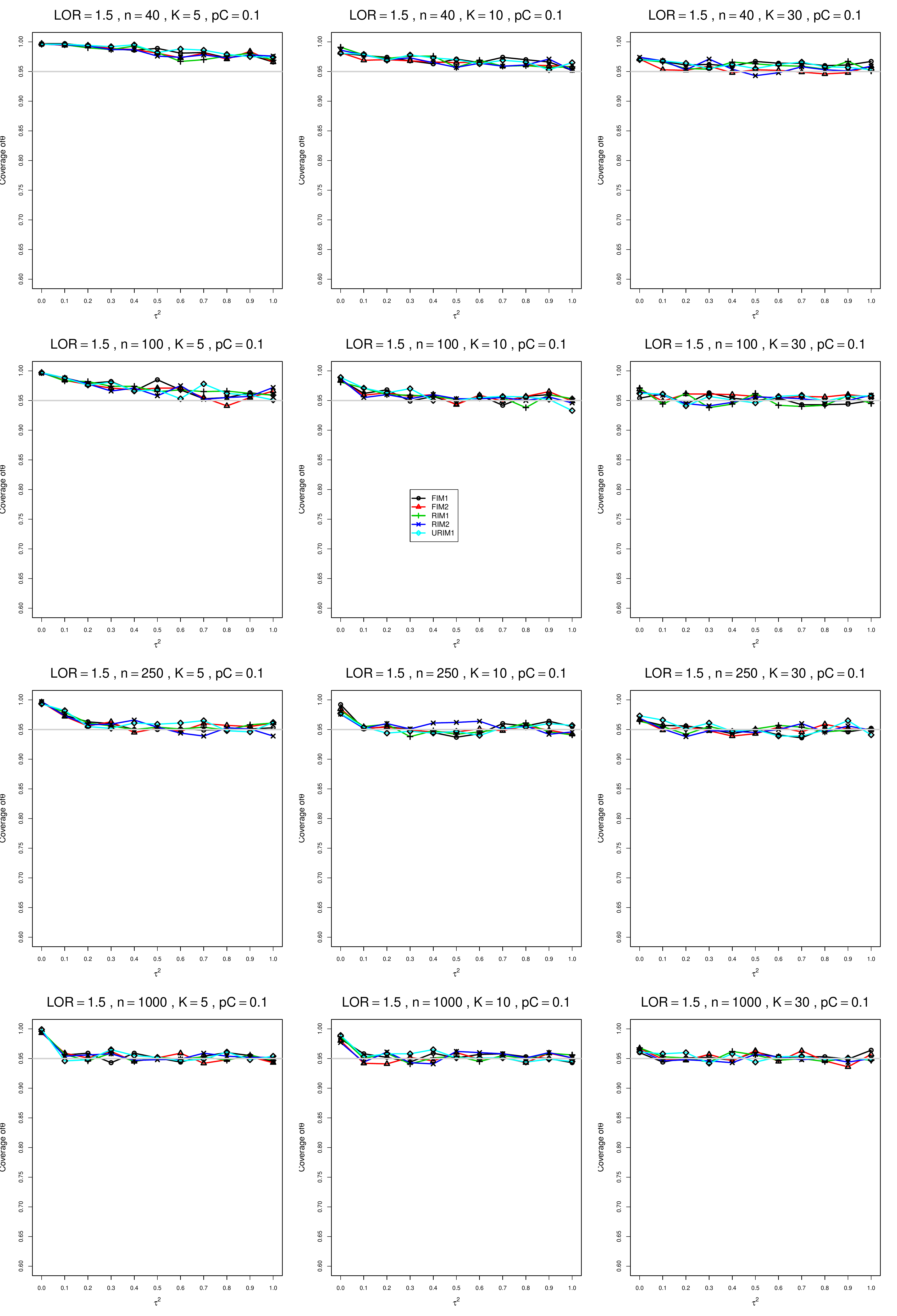}
	\caption{Coverage of the sample-size-weighted with Hartung-Knapp-Sidik-Jonkman (with KD) confidence interval for $\theta=1.5$, $p_{C}=0.1$, $\sigma^2=0.1$, constant sample sizes $n=40,\;100,\;250,\;1000$.
The data-generation mechanisms are FIM1 ($\circ$), FIM2 ($\triangle$), RIM1 (+), RIM2 ($\times$), and URIM1 ($\diamond$).
		\label{PlotCovThetamu15andpC01LOR_HkSJ_new_CMPsigma01}}
\end{figure}
\begin{figure}[t]
	\centering
	\includegraphics[scale=0.33]{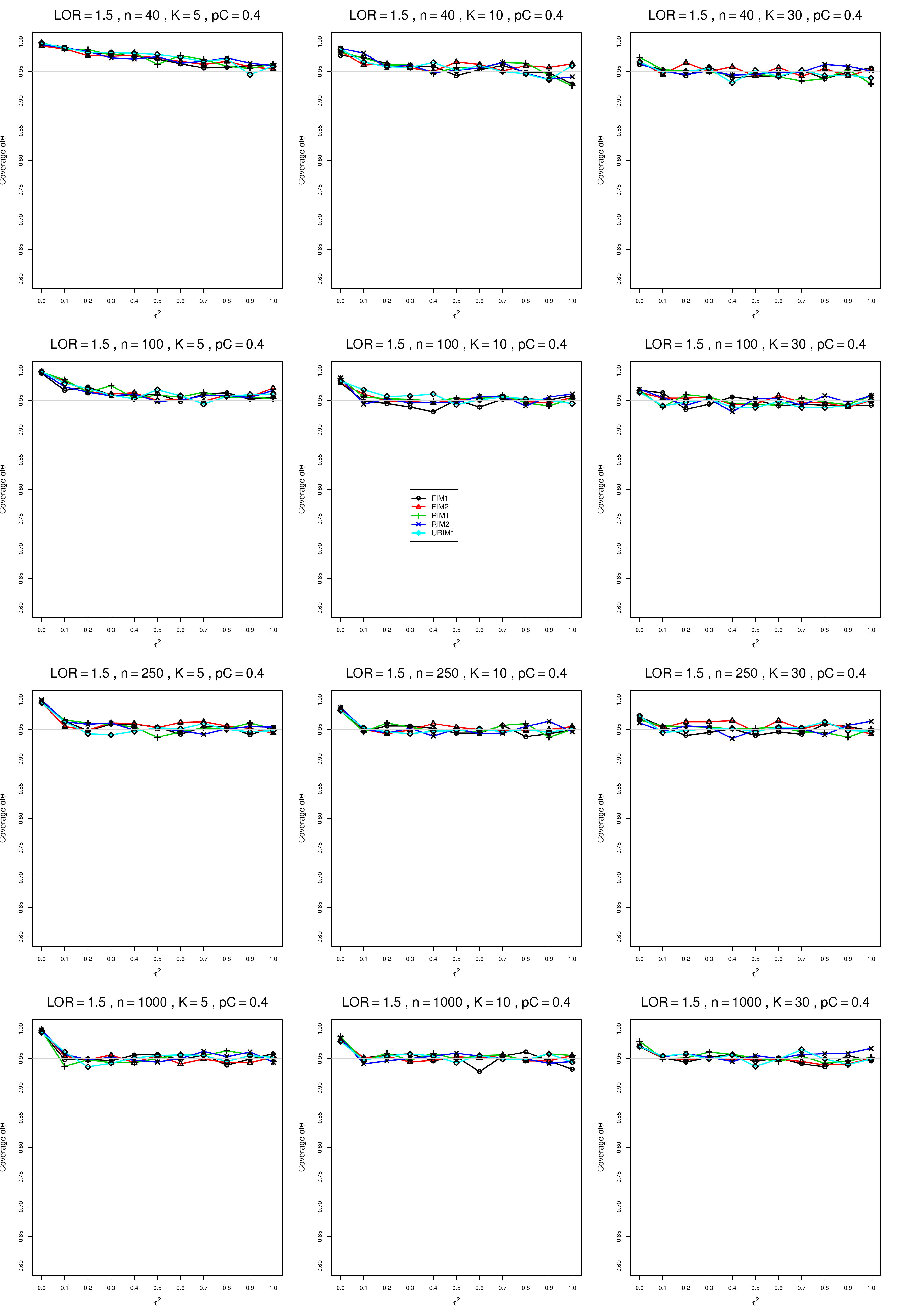}
	\caption{Coverage of the sample-size-weighted with Hartung-Knapp-Sidik-Jonkman (with KD) confidence interval for $\theta=1.5$, $p_{C}=0.4$, $\sigma^2=0.1$, constant sample sizes $n=40,\;100,\;250,\;1000$.
The data-generation mechanisms are FIM1 ($\circ$), FIM2 ($\triangle$), RIM1 (+), RIM2 ($\times$), and URIM1 ($\diamond$).
		\label{PlotCovThetamu15andpC04LOR_HkSJ_new_CMPsigma01}}
\end{figure}
\begin{figure}[t]
	\centering
	\includegraphics[scale=0.33]{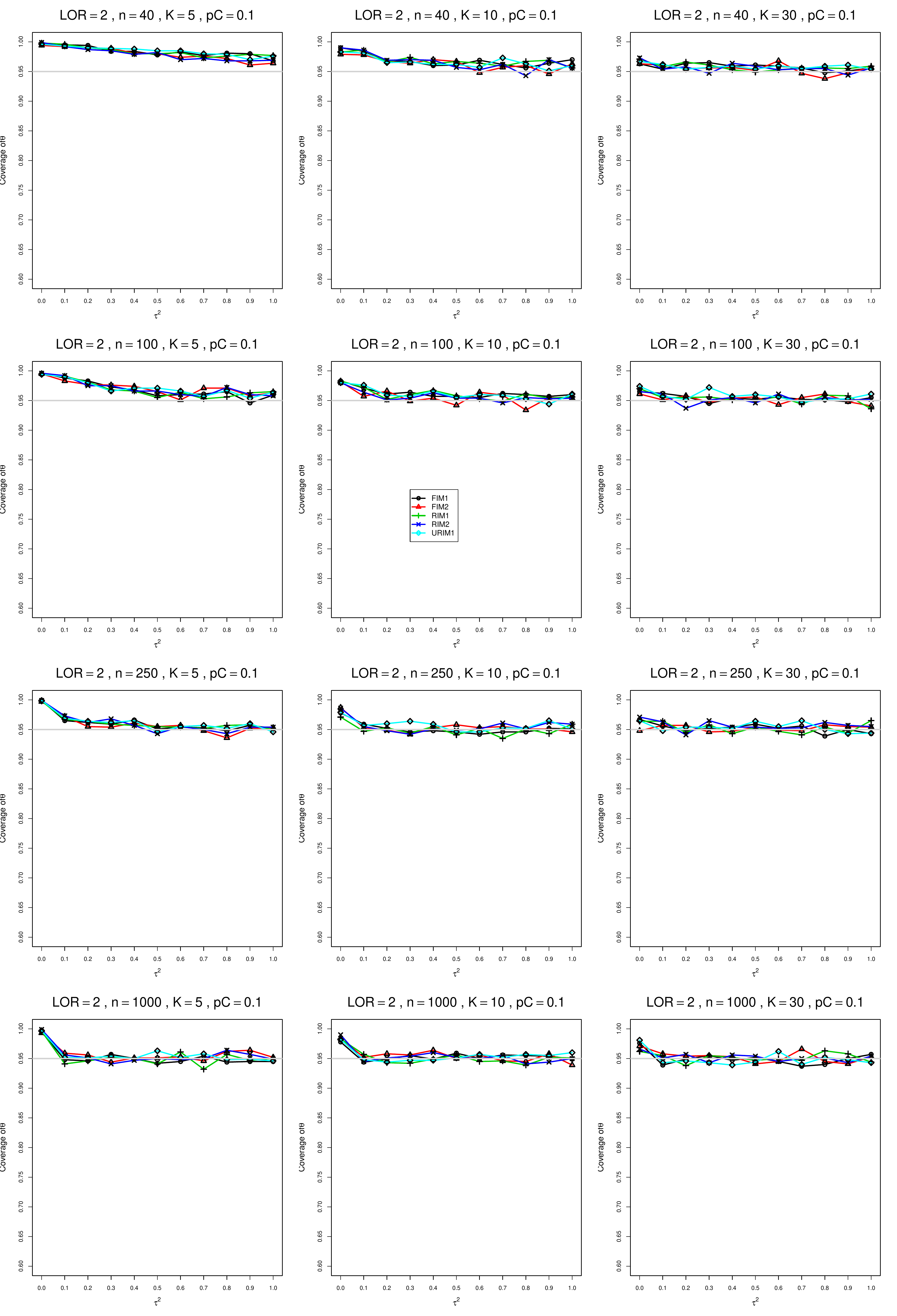}
	\caption{Coverage of the sample-size-weighted with Hartung-Knapp-Sidik-Jonkman (with KD) confidence interval for $\theta=2$, $p_{C}=0.1$, $\sigma^2=0.1$, constant sample sizes $n=40,\;100,\;250,\;1000$.
The data-generation mechanisms are FIM1 ($\circ$), FIM2 ($\triangle$), RIM1 (+), RIM2 ($\times$), and URIM1 ($\diamond$).
		\label{PlotCovThetamu2andpC01LOR_HkSJ_new_CMPsigma01}}
\end{figure}
\begin{figure}[t]
	\centering
	\includegraphics[scale=0.33]{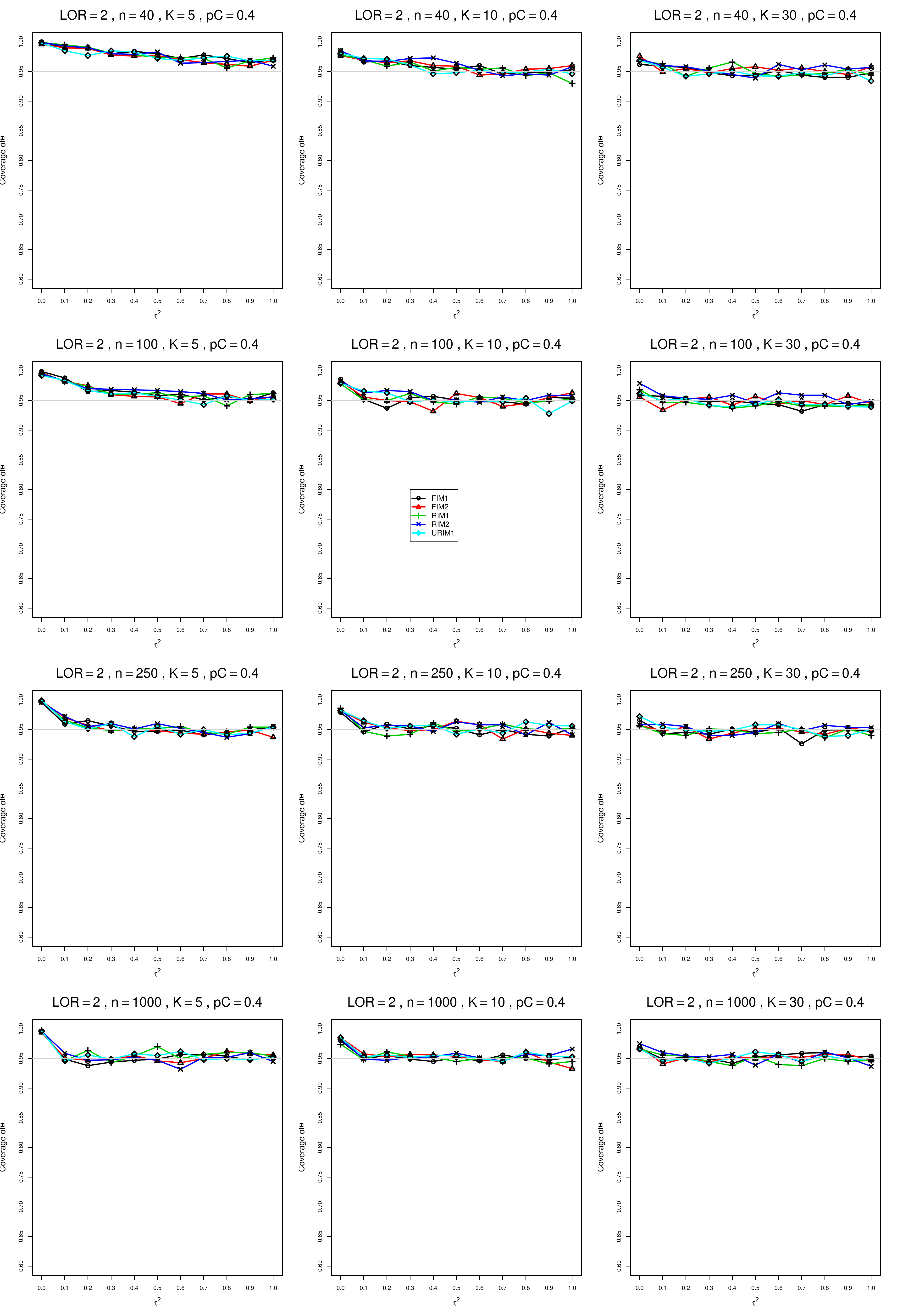}
	\caption{Coverage of the sample-size-weighted with Hartung-Knapp-Sidik-Jonkman (with KD) confidence interval for $\theta=2$, $p_{C}=0.4$, $\sigma^2=0.1$, constant sample sizes $n=40,\;100,\;250,\;1000$.
The data-generation mechanisms are FIM1 ($\circ$), FIM2 ($\triangle$), RIM1 (+), RIM2 ($\times$), and URIM1 ($\diamond$).
		\label{PlotCovThetamu2andpC04LOR_HkSJ_new_CMPsigma01}}
\end{figure}
\begin{figure}[t]
	\centering
	\includegraphics[scale=0.33]{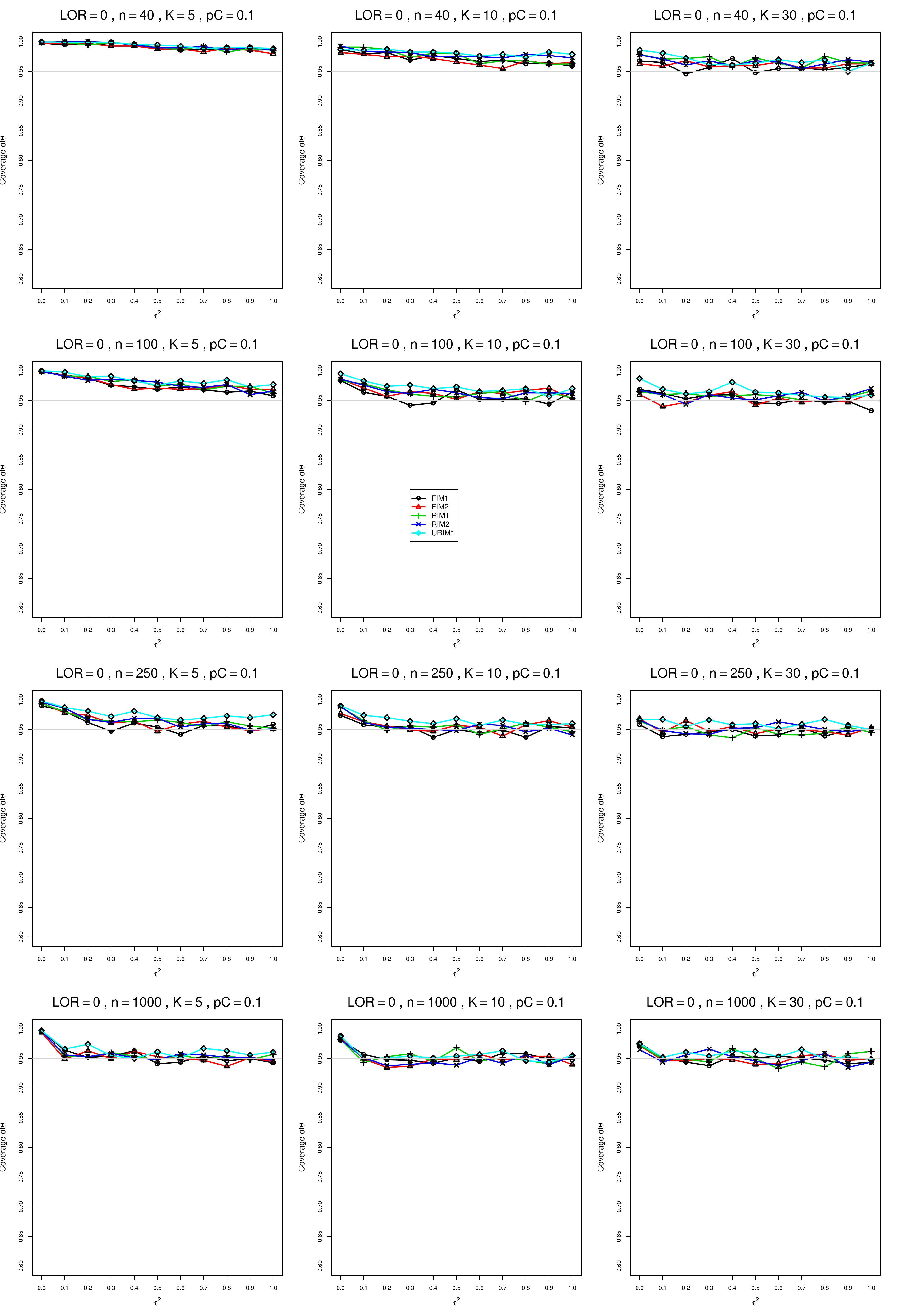}
	\caption{Coverage of the sample-size-weighted with Hartung-Knapp-Sidik-Jonkman (with KD) confidence interval for $\theta=0$, $p_{C}=0.1$, $\sigma^2=0.4$, constant sample sizes $n=40,\;100,\;250,\;1000$.
The data-generation mechanisms are FIM1 ($\circ$), FIM2 ($\triangle$), RIM1 (+), RIM2 ($\times$), and URIM1 ($\diamond$).
		\label{PlotCovThetamu0andpC01LOR_HkSJ_new_CMPsigma04}}
\end{figure}
\begin{figure}[t]
	\centering
	\includegraphics[scale=0.33]{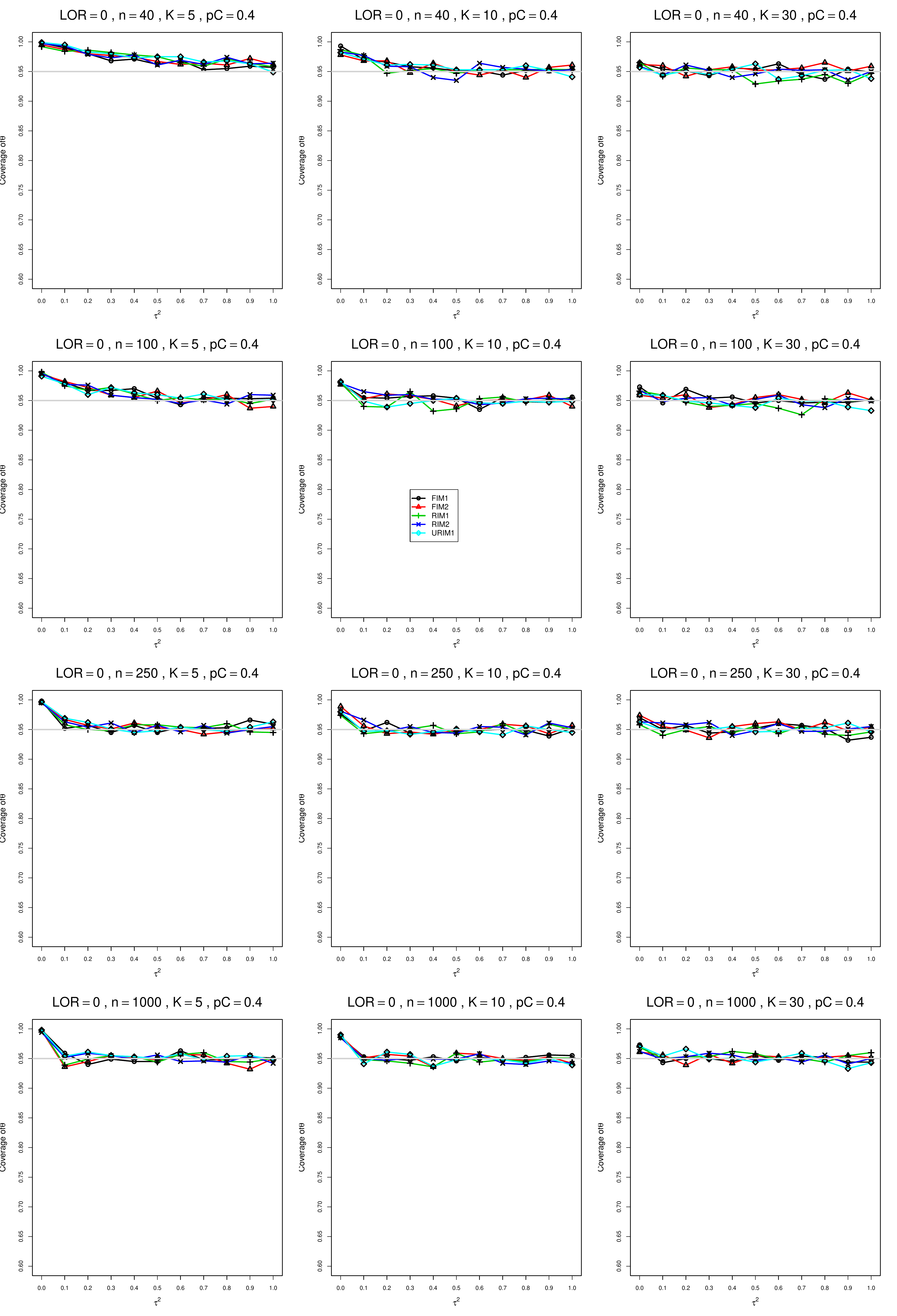}
	\caption{Coverage of the sample-size-weighted with Hartung-Knapp-Sidik-Jonkman (with KD) confidence interval for $\theta=0$, $p_{C}=0.4$, $\sigma^2=0.4$, constant sample sizes $n=40,\;100,\;250,\;1000$.
The data-generation mechanisms are FIM1 ($\circ$), FIM2 ($\triangle$), RIM1 (+), RIM2 ($\times$), and URIM1 ($\diamond$).
		\label{PlotCovThetamu0andpC04LOR_HkSJ_new_CMPsigma04}}
\end{figure}
\begin{figure}[t]
	\centering
	\includegraphics[scale=0.33]{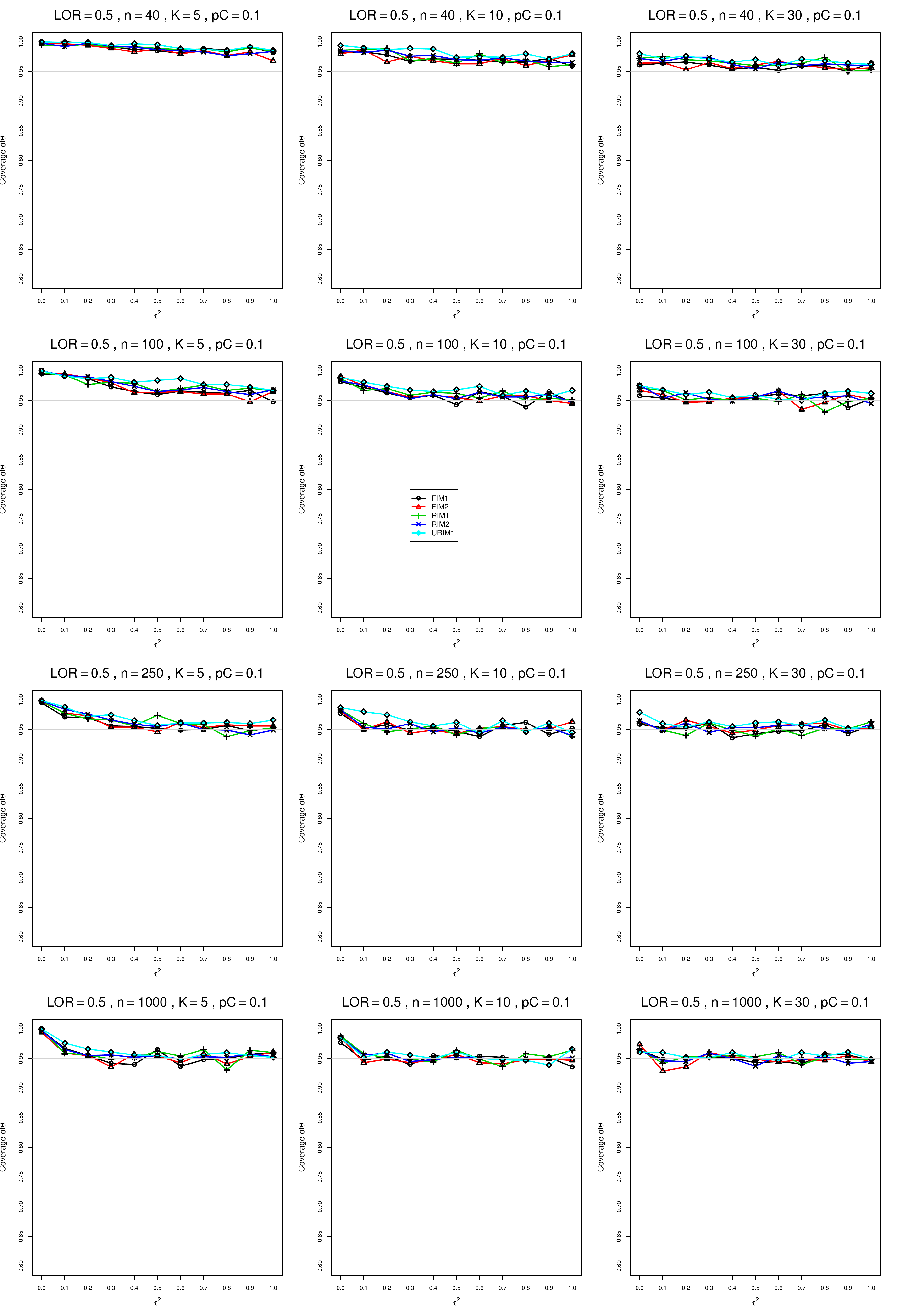}
	\caption{Coverage of the sample-size-weighted with Hartung-Knapp-Sidik-Jonkman (with KD) confidence interval for $\theta=0.5$, $p_{C}=0.1$, $\sigma^2=0.4$, constant sample sizes $n=40,\;100,\;250,\;1000$.
The data-generation mechanisms are FIM1 ($\circ$), FIM2 ($\triangle$), RIM1 (+), RIM2 ($\times$), and URIM1 ($\diamond$).
		\label{PlotCovThetamu05andpC01LOR_HkSJ_new_CMPsigma04}}
\end{figure}
\begin{figure}[t]
	\centering
	\includegraphics[scale=0.33]{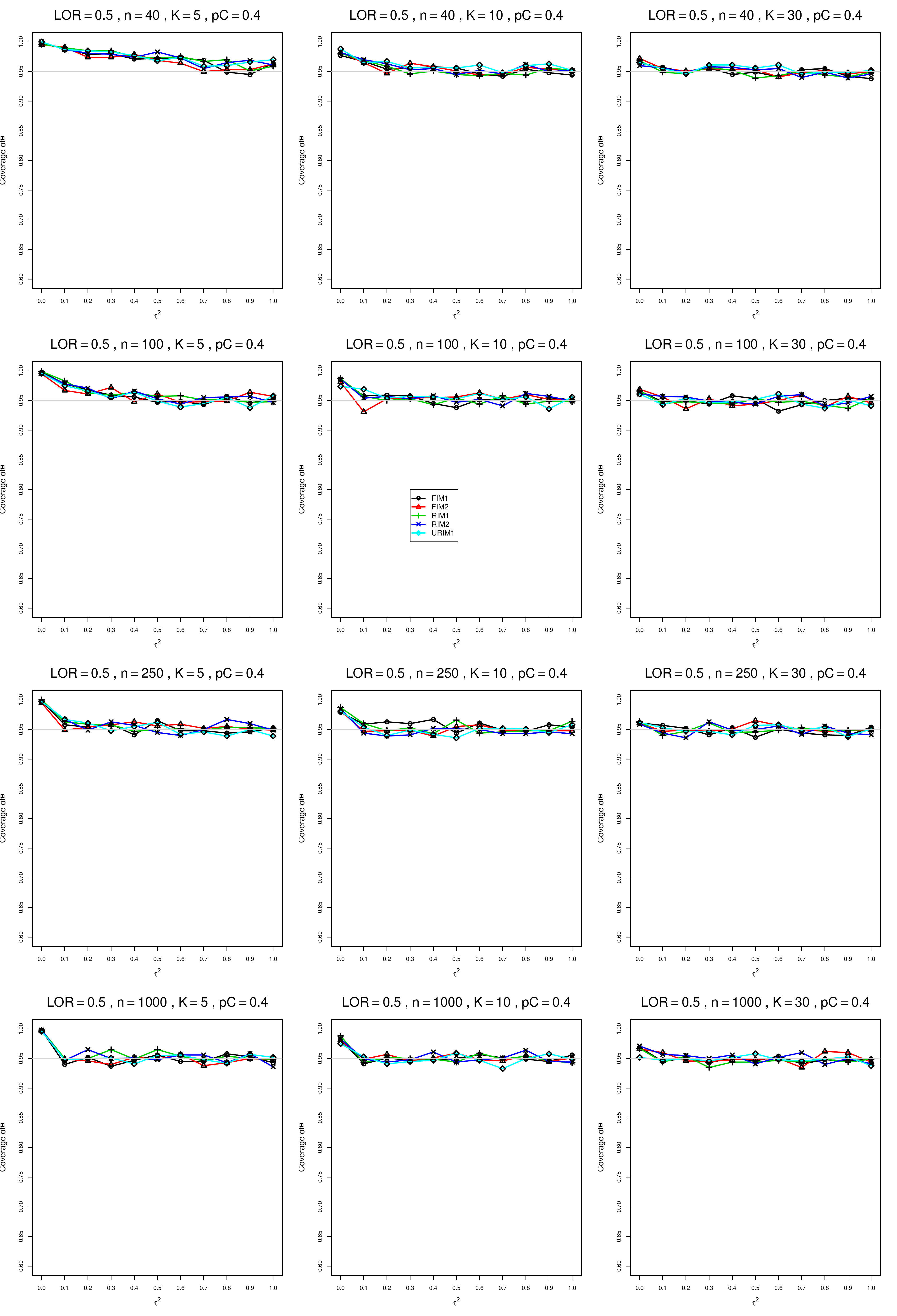}
	\caption{Coverage of the sample-size-weighted with Hartung-Knapp-Sidik-Jonkman (with KD) confidence interval for $\theta=0.5$, $p_{C}=0.4$, $\sigma^2=0.4$, constant sample sizes $n=40,\;100,\;250,\;1000$.
The data-generation mechanisms are FIM1 ($\circ$), FIM2 ($\triangle$), RIM1 (+), RIM2 ($\times$), and URIM1 ($\diamond$).
		\label{PlotCovThetamu05andpC04LOR_HkSJ_new_CMPsigma04}}
\end{figure}
\begin{figure}[t]
	\centering
	\includegraphics[scale=0.33]{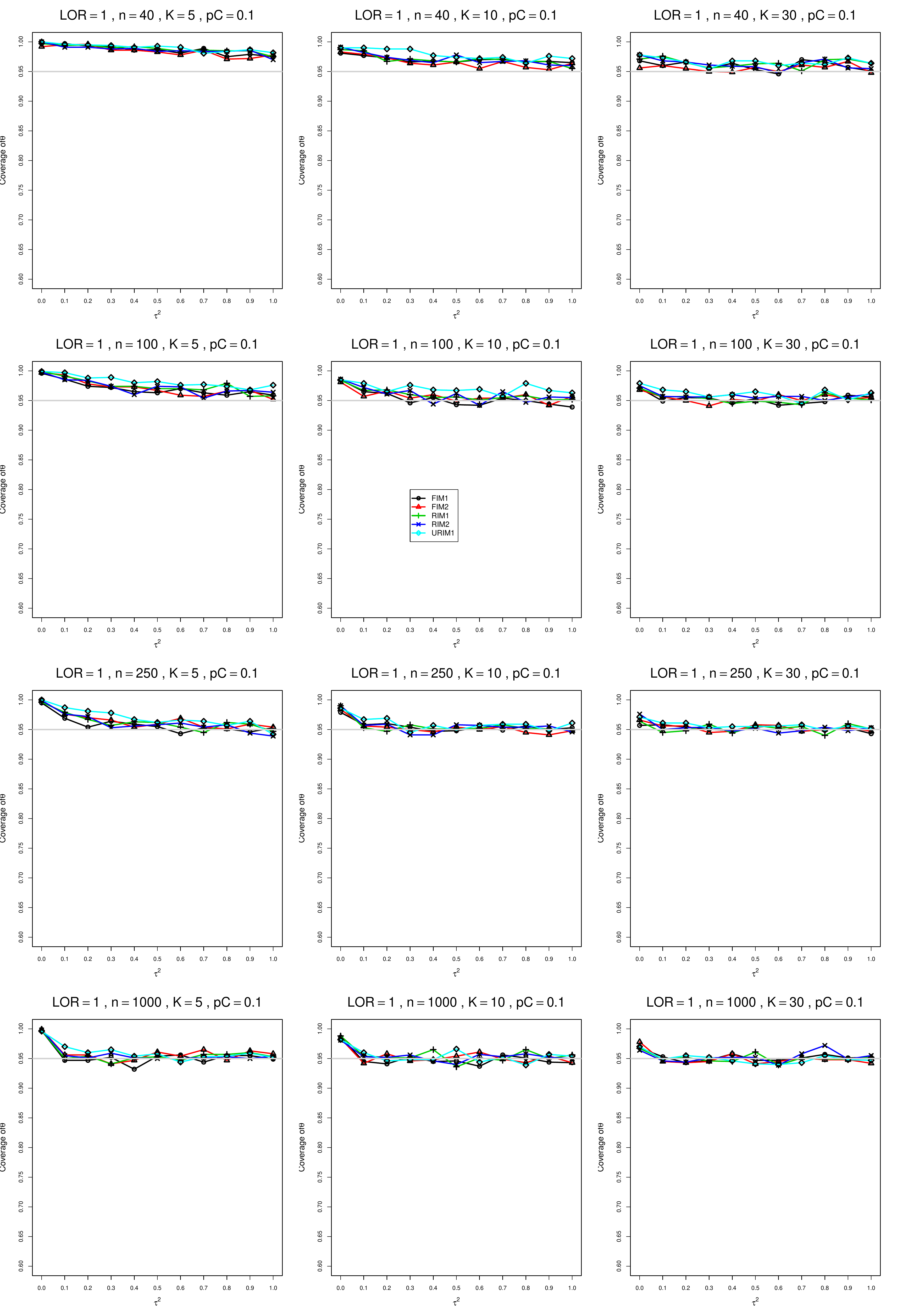}
	\caption{Coverage of the sample-size-weighted with Hartung-Knapp-Sidik-Jonkman (with KD) confidence interval for $\theta=1$, $p_{C}=0.1$, $\sigma^2=0.4$, constant sample sizes $n=40,\;100,\;250,\;1000$.
The data-generation mechanisms are FIM1 ($\circ$), FIM2 ($\triangle$), RIM1 (+), RIM2 ($\times$), and URIM1 ($\diamond$).
		\label{PlotCovThetamu1andpC01LOR_HkSJ_new_CMPsigma04}}
\end{figure}
\begin{figure}[t]
	\centering
	\includegraphics[scale=0.33]{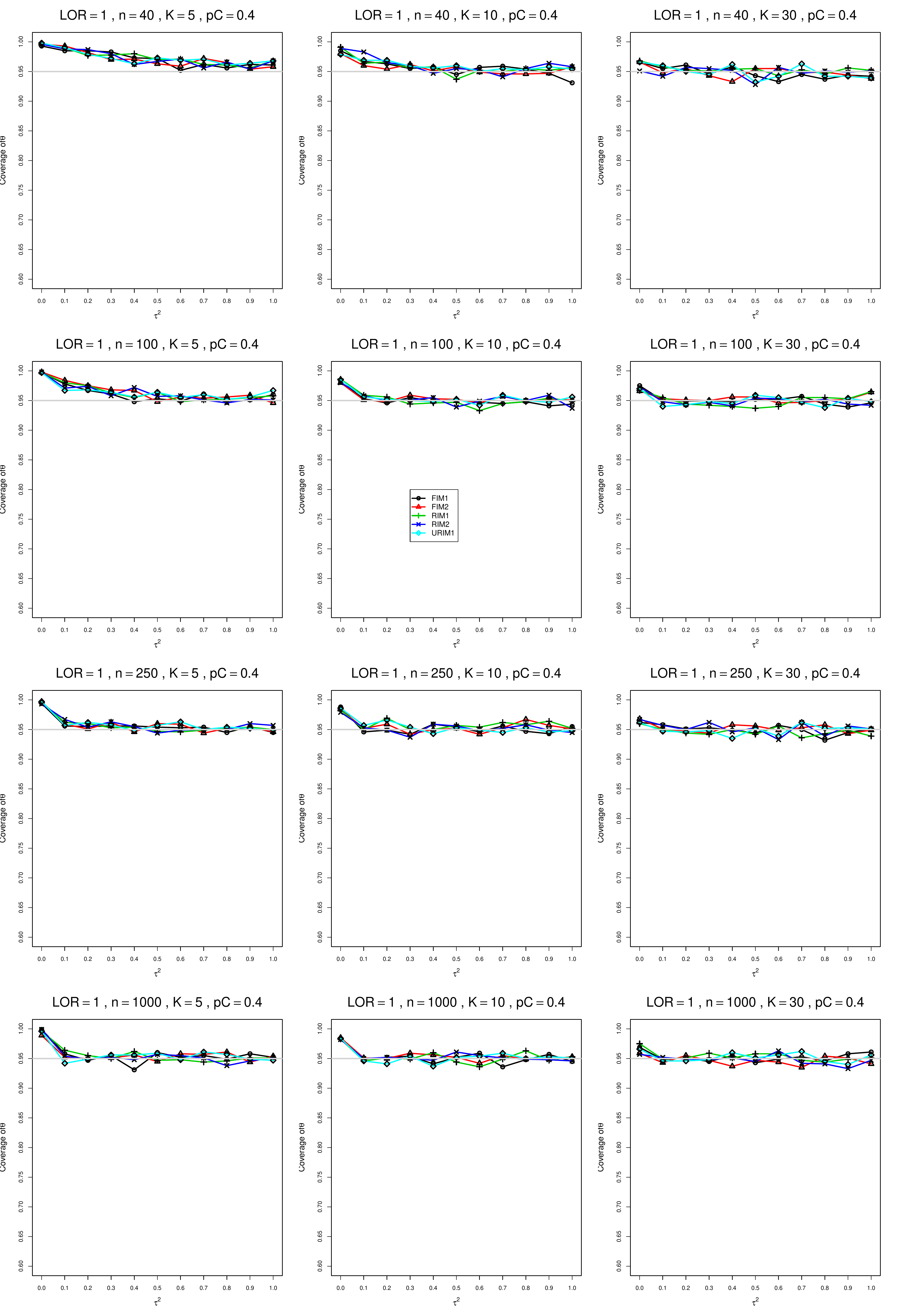}
	\caption{Coverage of the sample-size-weighted with Hartung-Knapp-Sidik-Jonkman (with KD) confidence interval for $\theta=1$, $p_{C}=0.4$, $\sigma^2=0.4$, constant sample sizes $n=40,\;100,\;250,\;1000$.
The data-generation mechanisms are FIM1 ($\circ$), FIM2 ($\triangle$), RIM1 (+), RIM2 ($\times$), and URIM1 ($\diamond$).
		\label{PlotCovThetamu1andpC04LOR_HkSJ_new_CMPsigma04}}
\end{figure}
\begin{figure}[t]
	\centering
	\includegraphics[scale=0.33]{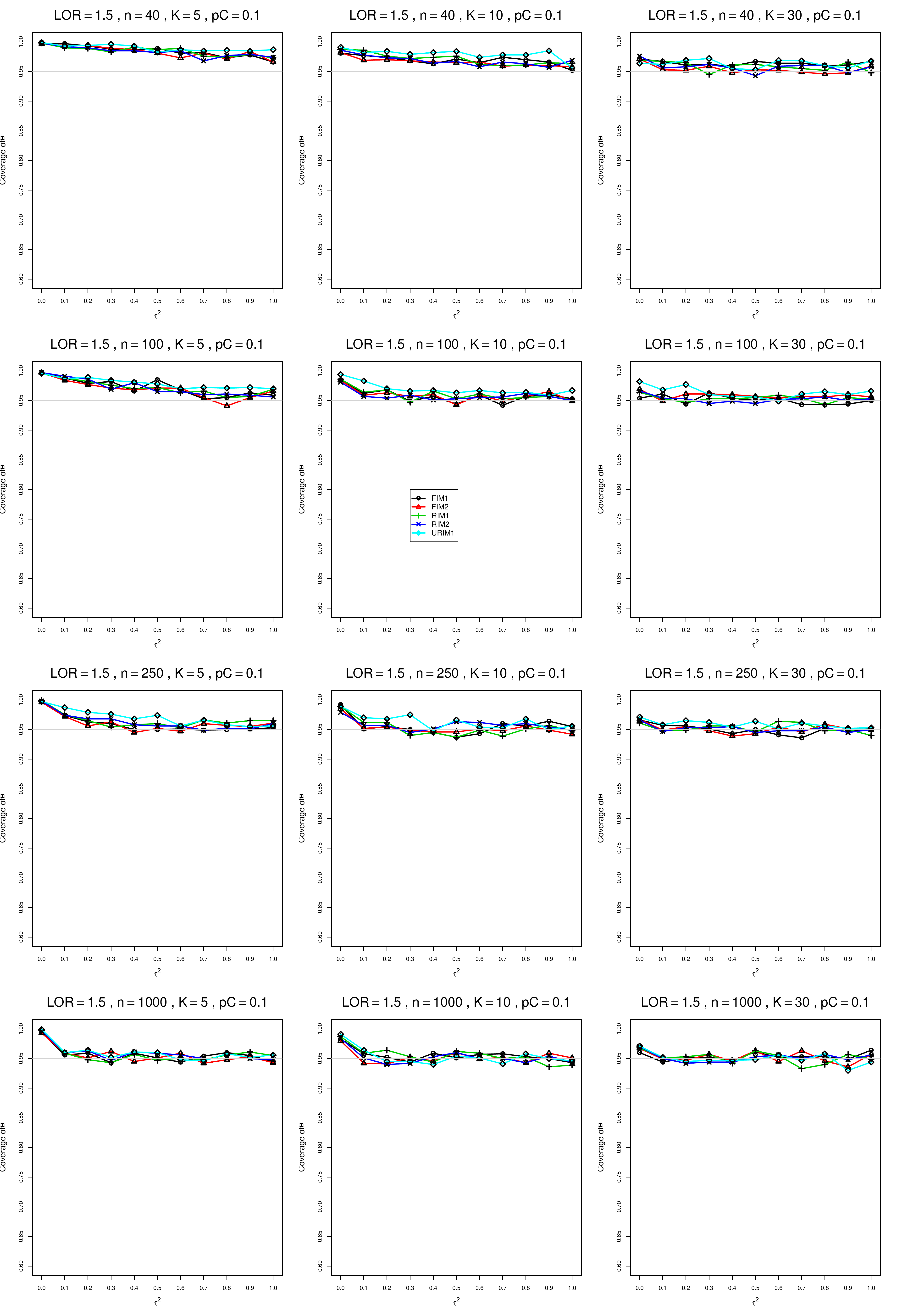}
	\caption{Coverage of the sample-size-weighted with Hartung-Knapp-Sidik-Jonkman (with KD) confidence interval for $\theta=1.5$, $p_{C}=0.1$, $\sigma^2=0.4$, constant sample sizes $n=40,\;100,\;250,\;1000$.
The data-generation mechanisms are FIM1 ($\circ$), FIM2 ($\triangle$), RIM1 (+), RIM2 ($\times$), and URIM1 ($\diamond$).
		\label{PlotCovThetamu15andpC01LOR_HkSJ_new_CMPsigma04}}
\end{figure}
\begin{figure}[t]
	\centering
	\includegraphics[scale=0.33]{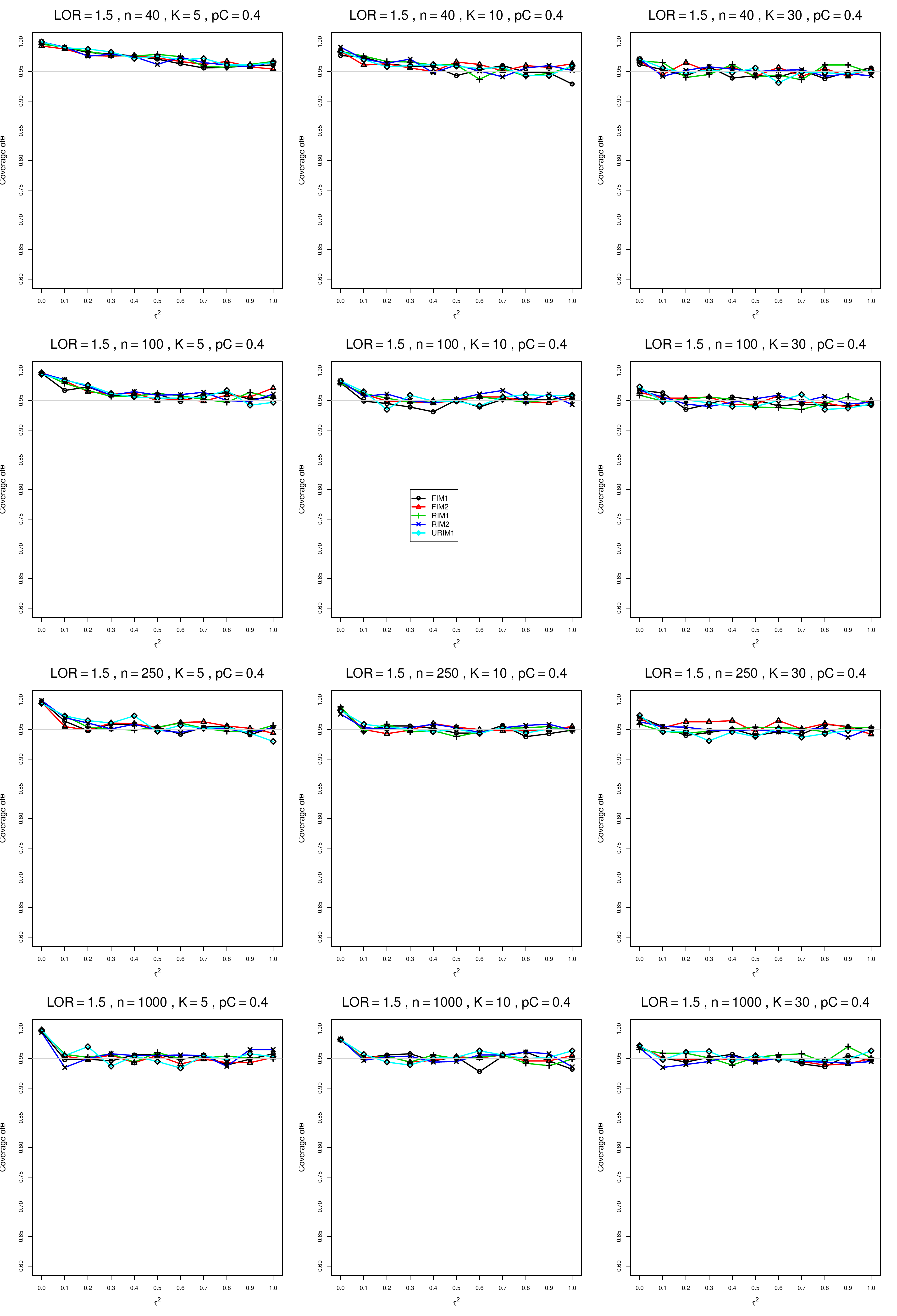}
	\caption{Coverage of the sample-size-weighted with Hartung-Knapp-Sidik-Jonkman (with KD) confidence interval for $\theta=1.5$, $p_{C}=0.4$, $\sigma^2=0.4$, constant sample sizes $n=40,\;100,\;250,\;1000$.
The data-generation mechanisms are FIM1 ($\circ$), FIM2 ($\triangle$), RIM1 (+), RIM2 ($\times$), and URIM1 ($\diamond$).
		\label{PlotCovThetamu15andpC04LOR_HkSJ_new_CMPsigma04}}
\end{figure}
\begin{figure}[t]
	\centering
	\includegraphics[scale=0.33]{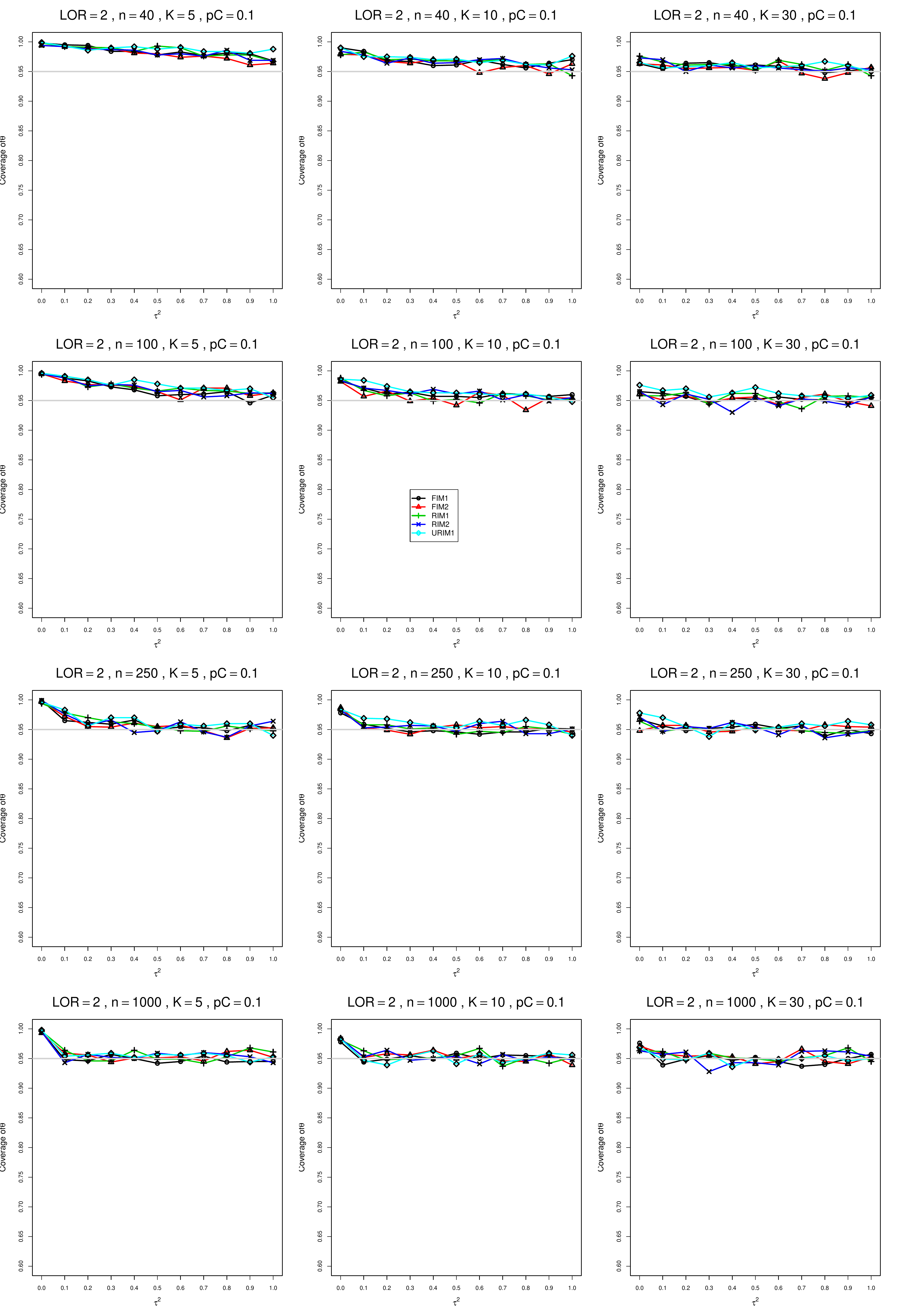}
	\caption{Coverage of the sample-size-weighted with Hartung-Knapp-Sidik-Jonkman (with KD) confidence interval for $\theta=2$, $p_{C}=0.1$, $\sigma^2=0.4$, constant sample sizes $n=40,\;100,\;250,\;1000$.
The data-generation mechanisms are FIM1 ($\circ$), FIM2 ($\triangle$), RIM1 (+), RIM2 ($\times$), and URIM1 ($\diamond$).
		\label{PlotCovThetamu2andpC01LOR_HkSJ_new_CMPsigma04}}
\end{figure}
\begin{figure}[t]
	\centering
	\includegraphics[scale=0.33]{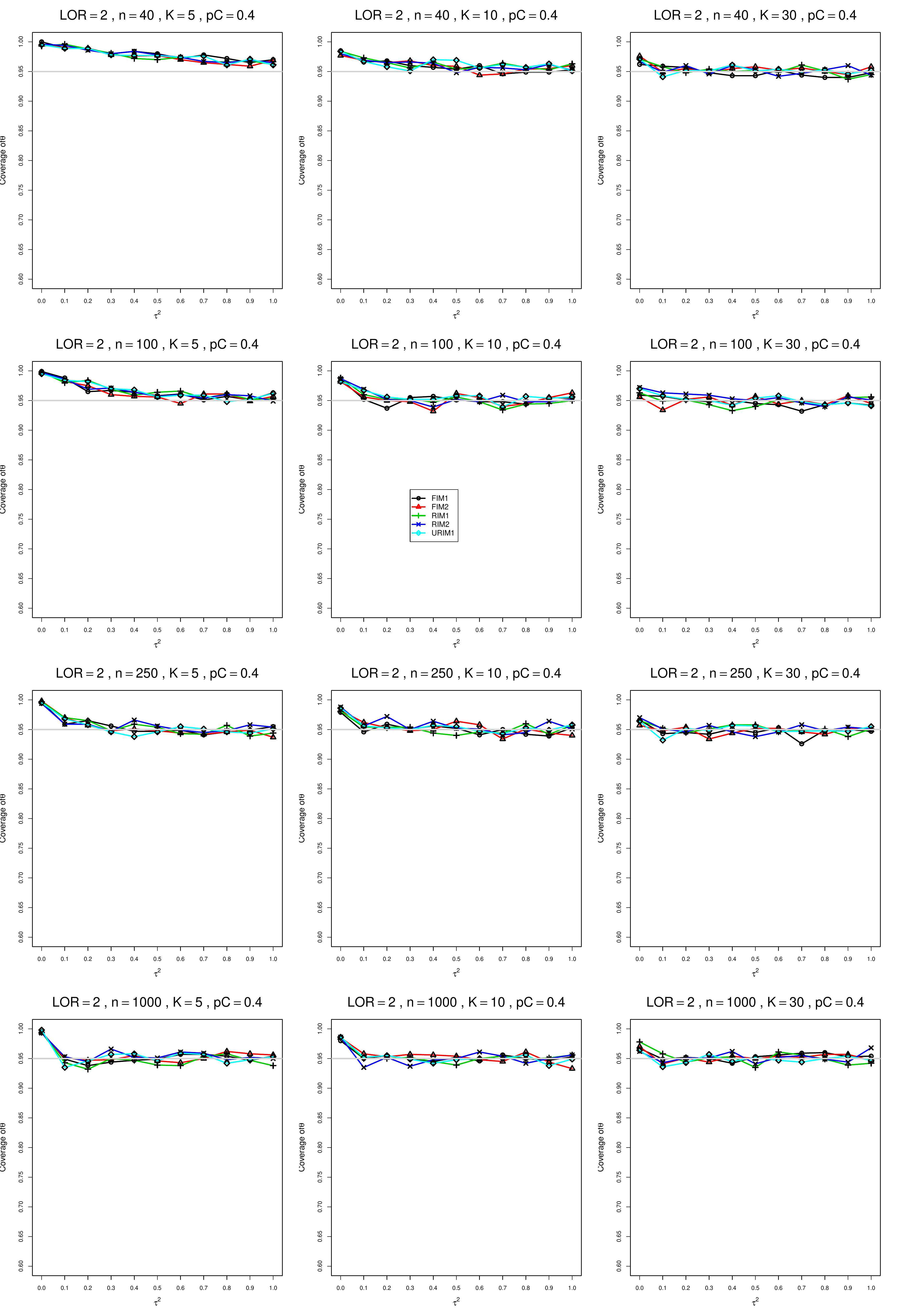}
	\caption{Coverage of the sample-size-weighted with Hartung-Knapp-Sidik-Jonkman (with KD) confidence interval for $\theta=2$, $p_{C}=0.4$, $\sigma^2=0.4$, constant sample sizes $n=40,\;100,\;250,\;1000$.
The data-generation mechanisms are FIM1 ($\circ$), FIM2 ($\triangle$), RIM1 (+), RIM2 ($\times$), and URIM1 ($\diamond$).
		\label{PlotCovThetamu2andpC04LOR_HkSJ_new_CMPsigma04}}
\end{figure}

\end{document}